\documentclass[a4paper,12pt]{article}
\pdfoutput=1
\usepackage{epsfig,amsmath,amsfonts,amsthm}
\usepackage[figuresright]{rotating}
\usepackage[left=28mm,right=20mm,top=30mm,bottom=20mm]{geometry}
\usepackage{subfig}
\usepackage{graphicx}
\usepackage{rotating}
\usepackage{booktabs}
\DeclareGraphicsExtensions{.pdf,.png,.jpg}
\usepackage{natbib}
\numberwithin{equation}{section}

\def\var{\hbox{Var}}

\def\max{\hbox{max}}

\usepackage[nokeyprefix]{refstyle}
\usepackage{varioref}
\usepackage{xr}
\usepackage{hyperref}
\begin{document}
\title{Simulation study of  estimating between-study variance and overall effect in meta-analysis of odds-ratios}

\author{Ilyas Bakbergenuly, David C. Hoaglin  and Elena Kulinskaya}
\date{\today}

\maketitle

\begin{center}
\textit{Abstract}
\end{center}
Random-effects meta-analysis requires an estimate of the between-study variance, $\tau^2$. We study methods of estimation of $\tau^2$  and its confidence interval in meta-analysis of odds ratio, and also the performance of related estimators of the overall effect.\\
We provide results of extensive simulations on  five point estimators of $\tau^2$ (the popular methods of DerSimonian-Laird, restricted maximum likelihood, and Mandel and Paule; the less-familiar method of Jackson; and the  new method (KD) based on the improved approximation to the distribution of the Q statistic by Kulinskaya and Dollinger (2015)); five interval estimators for $\tau^2$ (profile likelihood, Q-profile, Biggerstaff and Jackson, Jackson, and KD), six point estimators of the overall effect (the five inverse-variance estimators related to the point estimators of $\tau^2$ and an estimator  (SSW) whose weights use only study-level sample sizes), and eight interval estimators for the overall effect (five based on the point estimators for $\tau^2$; the Hartung-Knapp-Sidik-Jonkman (HKSJ) interval; a KD-based modification of HKSJ; and an interval based on the sample-size-weighted estimator).  Results of our simulations
show that none of the point estimators of $\tau^2$ can be recommended, however the new KD estimator provides a reliable coverage of $\tau^2$. Inverse-variance estimators of the overall effect are substantially biased. The SSW estimator of the overall effect and the related confidence interval provide the reliable point and interval estimation of log-odds-ratio.


\section{Introduction}
Meta-analysis is broadly used for combining estimates of a measure of effect from a set of studies in order to estimate an overall (pooled) effect. In studies with binary individual-level outcomes, the most common measure of treatment effect is the odds ratio. The standard method for combining study-level estimates uses a weighted average with inverse-variance weights. Our primary interest lies in meta-analysis of odds ratios via the random-effects model (REM), in which heterogeneity of the true study-level effects is usually  modelled through a study-level distribution with an unknown between-study variance $\tau^2$. Inverse-variance weights require an estimate of the between-study variance, which is also of interest  in assessing heterogeneity.

A number of methods provide estimates of between-study variance. \cite{veroniki2016methods} and \cite{Langan_2018_RSM_1316} provide comprehensive reviews. The most popular is the \cite{dersimonian1986meta} method. Recommended alternative point estimators include restricted maximum likelihood (REML), the method of \cite{mandel1970interlaboratory}, and the method of \cite{jackson2013confidence}. Interval estimators recommended by \cite{veroniki2016methods} include profile likelihood, the Q-profile interval (\cite{viechtbauer2007confidence}), and the generalized Q-profile intervals of  \cite{biggerstaff2008exact} and \cite{jackson2013confidence}. Quality of estimation varies with the effect measure; the simulation study {of estimating heterogeneity of odds-ratios} by \cite{Aert2019} found the last three methods lacking.


In meta-analyses that use inverse-variance weights, the actual measure of effect is the logarithm of the odds ratio (LOR), and the data are the logarithm of each study's sample odds ratio and the large-sample estimate of its variance.

Most moment-based methods of estimating heterogeneity use the moments of Cochran's $Q$ or its generalization (\cite{dersimonian2007random}). However, studies have shown (\cite{kulinskaya2015accurate, Aert2019}) that, for log-odds-ratio, these statistics do not follow the nominal chi-squared distribution or the mixture of chi-squared distributions derived by \cite{biggerstaff2008exact} and \cite{jackson2013confidence}. These departures result in biases and in undercoverage of the standard estimators of between-study variance. Also, in combination with inverse-variance weighting, they lead to biased point estimation of the overall effect and undercoverage of the associated confidence intervals (see \cite{Veroniki_2018_RSM_1319} for a review). Therefore, for estimating  between-study variance, we propose a method based on an improved approximation to the moments of Cochran's $Q$ statistic, suggested by \cite{kulinskaya2015accurate}. For the overall effect, we propose a weighted average in which the weights depend only on the effective sample sizes.

To compare our proposals with previous methods, we use simulation to study bias in five point estimators of the between-study variance, and coverage of five interval estimators of the between-study variance. We also study bias in six point estimators of the overall effect, and coverage of eight interval estimators of the overall effect.

\section{Estimation of study-level log-odds-ratio} \label{studyLOR}
Consider $K$ studies that used a particular individual-level binary outcome.
Each study $i$ reports a pair of independent binomial variables, $X_{i1}$ and $X_{i2}$, the numbers of events in $n_{i1}$ subjects in the Treatment arm ($j = 1$) and $n_{i2}$ subjects in the Control arm ($j = 2$) such that, for $i = 1, \ldots ,K$,
$$X_{i1}\sim {Binom}(n_{i1},p_{i1})\qquad \text{and}\qquad X_{i2}\sim {Binom}(n_{i2},p_{i2}).$$
The log-odds-ratio for Study $i$ is
\begin{equation}\label{eq:psi}
\theta_{i}=\log\left(\frac{p_{i1}(1-p_{i2})}{p_{i2}(1-p_{i1})}\right)\qquad\text{estimated by} \qquad
\hat\theta_{i}=\log\left(\frac{\hat p_{i1}(1-\hat p_{i2})}{\hat p_{i2}(1-\hat p_{i1})}\right).
\end{equation}
The large-sample variance of $\hat{\theta}_i$, derived by the delta method, is
\begin{equation}\label{eq:sigma}
{\sigma}_{i}^2=\var(\hat{\theta}_{i})=\frac{1}{n_{i1}{p}_{i1}(1-{p}_{i1})}+\frac{1}{n_{i2}{p}_{i2}(1-{p}_{i2})}.
\end{equation}
Estimation of $\theta$ and $\sigma^2_i$ requires estimates of the $p_{ij}$. The usual (and maximum-likelihood) estimate of $p_{ij}$ is $\hat{p}_{ij} = x_{ij} / n_{ij}$, but an adjustment is necessary when either of the observed counts is 0 or $n_{ij}$ (i.e., when the $2 \times 2$ table for Study $i$ contains a 0 cell). The standard approach adds $1/2$ to $x_{i1}$, $n_{i1} - x_{i1}$, $x_{i2}$, and $n_{i2} - x_{i2}$ when the $2 \times 2$ table contains exactly one 0 cell, and it omits Study $i$ when the $2 \times 2$ table contains two 0 cells. An alternative approach always adds $a$ $(>0)$ to all four cells of the $2 \times 2$ table for each of the $K$ studies; that is, it estimates $p_{ij}$ by $\hat{p}_{ij(a)}=(x_{ij}+a)/(n_{ij}+2a)$. The most common choice, $a = 1/2$, removes biases of order $n^{-1}$ in $\hat{\theta}_i$ and of order $n^{-2}$ in its estimated variance given by Equation~(\ref{eq:sigma}) (\cite{gart1985}).

\section{Standard random-effects model} \label{sec:StdREM}
The standard random-effects model assumes that each estimated study-level effect, $\hat{\theta}_i$, has an approximately normal distribution and that the true study-level effects, $\theta_{i}$, follow a normal distribution:
\begin{equation}\label{standardREM}
\hat{\theta}_{i} \sim N(\theta_{i}, \sigma_{i}^2) \quad \text{and} \quad \theta_{i} \sim N(\theta, \tau^2).
\end{equation}
Thus, the marginal distribution of $\hat{\theta}_{i}$ is $N(\theta, \sigma_{i}^2 + \tau^2)$. Although the $\sigma_{i}^2$ are generally unknown, they are routinely replaced by their estimates, $\hat{\sigma}_{i}^2$. A key step involves estimating the between-study variance, $\tau^{2}$; the standard random-effects model uses the DerSimonian-Laird estimate (\cite{dersimonian1986meta}). The estimate of the overall effect is then
\begin{equation}\label{thetahatRE}
\hat{\theta}_{RE} = {\sum \limits_{i=1}^{K} \hat{w}_{i} \hat{\theta}_{i}} / {\sum \limits_{i=1}^{K} \hat{w}_{i}},
\end{equation}
where $\hat{w}_{i} = \hat{w}_{i}(\hat{\tau}^2) = (\hat{\sigma}_{i}^2 + \hat{\tau}^2) ^ {-1}$ is the inverse-variance weight for Study $i$. If the $\sigma_i^2$ and $\tau^2$ were known, the variance of $\hat{\theta}_{RE}$ would be $[\sum w_{i}] ^ {-1}$ with $w_i = (\sigma_i^2 + \tau^2) ^ {-1}$. In practice, the variance of $\hat{\theta}_{RE}$ is traditionally estimated by $[\sum \hat{w}_{i}(\hat{\tau}^2)] ^ {-1}$, and a confidence interval for $\theta$ uses critical values from the normal distribution.

\section{Point and interval estimation of $\tau^2$ by Kulinskaya-Dollinger method (KD)}
The chi-squared approximation for $Q$ is inaccurate, and the actual distribution of $Q$ depends on the effect measure.   Under the null hypothesis of homogeneity of the log-odds-ratio, \cite{kulinskaya2015accurate}  derive corrected approximations for the mean and variance of $Q$ and match those corrected moments to obtain a gamma distribution that (as their simulations confirm) closely fits the null distribution of $Q$. 

We propose a new method of estimating $\tau^2$ based on this improved approximation.
Let $E_{KD}({Q})$ denote the corrected expected value of $Q$.  Then one obtains the KD estimate $\hat{\tau}_{KD}^2$ by iteratively solving
\begin{equation}
Q(\tau^2)=\sum\limits_{i=1}^{K}\frac{(\theta_{i}-\hat{\theta}_{RE})^{2}}{\hat{\sigma}_{i}^2+\tau^2}=E_{KD}({Q}).
\end{equation}

We also propose a new KD confidence interval for the between-study variance. This  interval for $\tau^2$ combines the Q-profile approach and the improved approximation by \cite{kulinskaya2015accurate}. This corrected Q-profile confidence interval can be estimated from the lower and upper quantiles of $F_Q$, the cumulative distribution function for the corrected distribution of $Q$:
\begin{equation}
Q(\tau_{L}^2)=F_{Q;0.975}\qquad Q(\tau_{U}^2)=F_{Q;0.025}
\end{equation}
The upper and lower confidence limits for $\tau^2$ can be calculated iteratively.

\section{Sample size weighted (SSW)  point and interval  estimation  of $\theta$}

In an attempt to avoid the bias in the inverse-variance-weighted estimators, we included a point estimator whose weights depend only on the studies' sample sizes (\cite{hedges1985statistical, hunter1990methods}). For this estimator (SSW),
$w_{i} = \tilde{n}_i = n_{iT}n_{iC}/(n_{iT} + n_{iC})$; $\tilde{n}_i$ is the effective sample size in Study $i$. These weights would coincide with the inverse-variance weights if all the probabilities across studies were equal.

The interval estimator corresponding to SSW (SSW KD) uses the SSW point estimator as its center, and its  half-width equals the estimated standard deviation of SSW under the random-effects model times the critical value from the $t$ distribution on $K - 1$ degrees of freedom.  The estimator of the variance of SSW is
\begin{equation}\label{eq:varianceOfSSW}
\widehat{\var}(\hat{\theta}_{\mathit{SSW}})= \frac{\sum \tilde{n}_i^2 (v_i^2 + \hat{\tau}^2)} {(\sum \tilde{n}_i)^2},
\end{equation}
in which $v_i^2$ comes from Equation (\ref{eq:sigma})  and $\hat{\tau}^2 = \hat{\tau}_{\mathit{KD}}^2$.

\section{Simulation study}
In a simulation study with log-odds-ratio as the effect measure, we varied six  parameters: the number of studies $K$, the total sample size of each study $n$, the proportion of observations in the control arm $q$, the overall true LOR $\theta$,  the between-study variance $\tau^2$, and the probability of an event in the control arm.

The number of studies $K = (5, 10, 30)$.

We included sample sizes that were equal for all $K$ studies and sample sizes that varied among studies. The total sample sizes were $n = (40, 100, 250, 1000)$ for equal sample sizes and $\bar{n} = (30, 60, 100, 160)$ for unequal sample sizes. In choosing sample sizes that varied among studies, we followed a suggestion of \cite{sanchez2000testing}, who selected study sizes having skewness $1.464$, which they considered typical in behavioral and health sciences. The average study sizes were $\bar{n} = (30, 60, 100, 160)$,  where $\bar{n}=30$ corresponds to $K=5$ studies of sizes $(12,16,18,20,84)$, $\bar{n}=60$ corresponds to studies of sizes $(24,32,36,40,168)$, $\bar{n}=100$ corresponds to $(64,72,76,80,208)$, and $\bar{n}=160$ corresponds to $(124,132,136,140,268)$. Table \ref{unequal sample sizes} summarizes these sample sizes. For $K = 10$ and $K = 30$, the same set of sample sizes was used twice and six times, respectively.

The values of $q$ were .5 and .75. The sample sizes of the treatment and control arms were $n_{iT}=\lceil{(1 - q_i)n_{i}}\rceil$ and $n_{iC}=n_{i}-n_{iT}$, $i=1,\ldots,K$.

The values of the overall true LOR $\theta$ were $0(0.5)2$.

The values of the between-study variance $\tau^2$ were $0(0.1)1$, corresponding to small to moderate heterogeneity, and $1((1)10$ for moderate to large heterogeneity.

The probability in the control arm, $p_{iC}$, was $0.1,\; 0.2,\; 0.4$.

Altogether, the simulations comprised 7,920 combinations of the six parameters. We generated 10,000 meta-analyses for each combination.

The true values of LOR ($\theta_{i}$) in the $K$ studies were generated from normal distributions with mean $\theta$ and variance $\tau^2$.

For a given probability  $p_{iC}$, the number of events in the control group $X_{iC}$ was generated from the Binomial $(n_{iC}, p_{iC})$ distribution. The number of events in the treatment group $X_{iT}$ was  generated  from the Binomial $(n_{iT}, p_{iT})$ distribution with
$p_{iT}=p_{iC}\exp(\theta_{i})/(1 - p_{iC} + p_{iC}\exp(\theta_{i}))$.

The estimate of effect size in Study $i$, $\hat\theta_i$ is calculated as in Equation~(\ref{eq:psi}),
and its sampling variance is estimated by substitution of $\hat p_{ij}$ in Equation~(\ref{eq:sigma}).
The methods differ, however, in the way they obtain $\hat{p}_{ij}$ from $x_{ij}$ and $n_{ij}$.
For all standard methods, we added $1/2$ to each cell of the $2 \times 2$ table only when the table had at least one cell equal to 0. This approach corresponds to the default values of the arguments \textbf{add}, \textbf{to} and  \textbf{drop00} of the \textit{escalc} procedure from \textit{metafor}, \cite{viechtbauer2015package}.  

For the KD methods, we corrected for bias by adding $a=1/2$ to each cell of all $K$ tables, and we dropped the double zero studies. We also tried always adding $1/2$ in standard methods, but the results were worse.

\begin{table}
	\centering
	\caption{Unequal sample sizes for simulations}
	\label{unequal sample sizes}
	\begin{tabular}{|l|l|l|l|l|l|}
		\hline
		$\bar{n}\setminus{K}$&$1$ & $2$ & $3$ &$4$  & $5$                       \\
		\hline
		30   & 12  & 16  & 18  & 20  & 84 \\
		60   & 24  & 32  & 36  & 40  & 168  \\
		100   & 64  & 72  & 76   & 80   & 208          \\
		160   & 124   & 132  & 136   & 140   &268         \\
		\hline
	\end{tabular}
\end{table}

\section {Methods of estimation of $\tau^2$ and $\theta$ used  in simulations}
\subsection*{Point estimators of $\tau^2$}
\begin{itemize}
\item DL - DerSimonian and Laird method by \cite{dersimonian1986meta}
\item J - method by \cite{jackson2013confidence}
\item KD - new improved moment method based on \cite{kulinskaya2015accurate}
\item MP - Mandel-Paule method \cite{mandel1970interlaboratory}
\item REML - Restricted Maximum Likelihood method
\end{itemize}
\subsection*{Interval estimators of $\tau^2$}
\begin{itemize}
\item BJ - method by \cite{biggerstaff2008exact}
\item J - method by \cite{jackson2013confidence}
\item KD - new improved Q-profile method  based on \cite{kulinskaya2015accurate}
\item PL - profile likelihood confidence interval based on $\tau_{REML}^2$
\item QP - Q-profile confidence interval method \cite{viechtbauer2007confidence}
\end{itemize}

\subsection*{Point estimators of $\theta$ }
Inverse variance methods with $\tau^2$ estimated by:
\begin{itemize}
\item DL - DerSimonian and Laird method by \cite{dersimonian1986meta}
\item J -  method by \cite{jackson2013confidence}\item REML-Restricted Maximum Likelihood Method
\item KD - improved moment method based on  \cite{kulinskaya2015accurate}
\item MP - Mandel Paule method \cite{mandel1970interlaboratory}
\item REML - Restricted Maximum Likelihood method
\end{itemize}
and
\begin{itemize}
\item SSW - fixed weights estimator of $\theta$
\end{itemize}

\subsection*{Interval estimators of $\theta$}
Standard inverse-variance methods  using normal quantiles, with  $\tau^2$ estimated by:
\begin{itemize}
\item DL - DerSimonian and Laird method by \cite{dersimonian1986meta}
\item J -  method by \cite{jackson2013confidence}\item REML-Restricted Maximum Likelihood Method
\item KD - improved moment method based on  \cite{kulinskaya2015accurate}
\item MP - Mandel Paule method \cite{mandel1970interlaboratory}
\item REML - Restricted Maximum Likelihood method
\end{itemize}
Inverse-variance methods with modified variance of $\theta$ and t-quantiles as in  \cite{hartung2001refined} and \cite{sidik2002simple}
\begin{itemize}
\item HKSJ (DL) -   $\tau^2$ estimated by DL
\item HKSJ KD -   $\tau^2$ estimated by KD
\end{itemize}
and
\begin{itemize}
\item SSW KD - fixed weights estimator of $\theta$ with the variance given by (\ref{eq:varianceOfSSW}) and t-quantiles
\end{itemize}

\subsection*{Studies with  zero events in one or both arms}
\begin{itemize}
\item J - adds continuity correction $1/2$ to each cell in case of zeros only
\item DL - adds continuity correction $1/2$  to each cell in case of zeros only
\item REML - adds continuity correction $1/2$  to each cell in case of zeros only
\item MP - adds continuity correction $1/2$  to each cell in case of zeros only
\item KD - always adds continuity correction $1/2$  to each cell; excludes double zeros

\end{itemize}

\subsection{Results of simulation studies }

Our full simulation results, comprising $300$ figures, each presenting $12$ combinations of $K$ by $n$ values,  are provided in Appendices A and B.  The short  summary is given below.

\subsubsection*{Bias in estimation of $\tau^2$ (Web Appendix A1)}

None of the point estimators of $\tau^2$ has bias consistently close enough to 0 to be recommended, but among the  existing estimators, MP and KD provide better choices for small and large $K$, respectively.

\subsubsection*{Coverage in estimation of $\tau^2$ (Web Appendix A2)}

Coverage of $\tau^2$ is generally good for  $K=5$, but is considerably worse for larger number of studies, especially so for large values of $\theta$. All methods are somewhat conservative at $\tau^2=0$. Overall, KD performs the best.
The large number of studies $K$ presents the greatest challenge for the standard methods. PL is the most affected, with considerable undercoverage up to $n=100$ for medium to large values of $\tau^2$. The other methods also have low coverage for small $n$, but they improve faster with increasing  $n$. KD provides reliable coverage except for  small sample sizes combined with $p_C=0.4$ and  $\theta\geq 1.5$, where its undecoverage worsens  with  increasing $\tau^2$, though it is still considerably  better than all the competitors.

\subsubsection*{Bias in estimation of $\theta$ (Web Appendix B1)}
In the results for the bias of the point estimators of $\theta$, a common pattern is that the bias is roughly linearly related to $\tau^2$ with a positive slope. 

As expected, in the vast majority of situations, SSW avoids most, if not all, of the bias in the IV-weighted estimators. The bias of the inverse variance estimators affects their efficiency, so that SSW is sometimes more efficient (it terms of its mean squared error) than its competitors.

\subsubsection*{Coverage in estimation of $\theta$ (Web Appendix B2)}
Because of the undercoverage of the standard CIs based on the IV-weighted estimators, we do not recommend them. HKSJ and HKSJ KD often have coverage close to 95\%, but they sometimes have serious undercoverage. All problems are typically exacerbated for the unbalanced sample sizes.  The newly proposed SSW KD interval often has coverage somewhat greater than 95\%, but its coverage is at least 93\% (except for a few cases involving $K = 30$ and unequal sample sizes with $\bar{n} = 30$).

\section{Summary}
Our extensive simulations demonstrate that the existing methods of meta-analysis of odds ratio often present a  biased view  both on  the heterogeneity and the overall effect. In brief:\\
small sample sizes  are rather problematic, and meta-analyses that involve numerous small studies are especially challenging. Because the study-level effects and their variances are related,  estimates of the overall effects are biased, and the coverage of confidence intervals is too low, especially for small sample sizes and  large number of studies.

The between-study variance, $\tau^2$, is typically estimated by generic methods which assume normality of the estimated effects $\hat\theta_i$. It is usually overestimated near zero,  but the standard methods are  negatively biased for larger values of $\tau^2$.
 Our findings agree with those by \cite{Aert2019} that the standard interval estimation of $\tau^2$ is often too liberal. The behavior of the profile likelihood method is especially erratic.  

Therefore we   proposed and studied by simulation the new moment method of estimation of $\tau^2$ based on the improved approximation to distribution of Cochran's $Q$ for odds ratios by \cite{kulinskaya2015accurate}.  The KD method provides  reliable interval estimation of $\tau^2$  across all values of $\tau^2$, $n$, and $K$. The point estimation of $\tau^2$  is more challenging, and even though KD provides better point estimation   for $K=30$, it is positively biased for small values of $K$.

Arguably, the main purpose of a meta-analysis is to provide point and interval estimates of an overall effect.
Our results show that the inverse-variance-weighted estimates of the overall effect are biased.
These biases (and even their sign), depend on the $\tau^2$ and  the true value of $\theta$, worsen for the unbalanced studies, and may be considerable even for reasonably large sample sizes such as $n=250$. The coverage of the overall effect follows the same patterns because the centering of confidence intervals is biased. Additionally, traditional intervals using normal quantiles are too narrow, and the use of t-quantiles as in HKSJ methods, brings noticeable though not sufficient improvement.

A pragmatic approach to unbiased estimation of $\theta$ uses weights that do not involve estimated variances of study-level estimates, for example, weights proportional to the study sizes $n_i$. \cite{hedges1985statistical}, \cite{hunter1990methods} and \cite{Shuster-2010}, among others, have proposed such weights.
We propose  to use weights proportional  to an effective sample size,  $\tilde{n}_i=n_{iT}n_{iC}/n_i$; these are the optimal inverse-variance weights for LOR  when all the probabilities are equal. 

A reasonable estimator of $\tau^2$, such as MP or KD   can be used as $\hat{\tau}^2$. Further, confidence intervals for $\theta$ centered at $\hat{\theta}_{\mathit{SSW}}$ with $\hat{\tau}_{\mathit{KD}}^2$ in Equation~(\ref{eq:varianceOfSSW}) can be used. In our simulations, this is by far the best interval estimator of $\theta$, providing near nominal coverage under all studied conditions.

\section*{Funding}
The work by E. Kulinskaya was supported by the Economic and Social Research Council [grant number ES/L011859/1].
\section*{Appendices description}
\begin{itemize}
	\item Appendix A: Plots for bias and coverage of $\tau^2$.
	\item Appendix B: Plots for bias, mean squared error, and coverage of estimators of the log-odds-ratio 
\end{itemize}

\bibliographystyle{plainnat}
\bibliography{LOR_biblio_13Feb19}
\clearpage

\setcounter{section}{0}
\renewcommand{\thefigure}{A1.1.\arabic{figure}}
\renewcommand{\thesection}{A.\arabic{section}}
\setcounter{figure}{0}

\text{\LARGE{\bf{Appendices}}}
\section{Plots for bias of between-study variance.}
Subsections A1.1, A1.2 and A1.3 correspond to $p_{C}=0.1,\; 0.2,\; 0.4$ respectively. 
For a given $p_{C}$ value, each figure corresponds to a value of $\theta (= 0, 0.5, 1, 1.5, 2)$, a value of $q (= 0.5, 0.75)$, a value of $\tau^2 = 0.0(0.1)1.0$, and a set of values of $n$ (= 40, 100, 250, 1000) or $\bar{n}$ (= 30, 60, 100, 160).\\
Each figure contains a panel (with $\tau^2$ on the horizontal axis) for each combination of n (or $\bar{n}$) and $K (=5, 10, 30)$.\\
The point estimators of $\tau^2$ are
\begin{itemize}
	\item DL (DerSimonian-Laird)
	\item REML (Restricted Maximum Likelihood )
	\item MP (Mandel-Paule)
	\item KD (new improved moment estimator based on Kulinskaya and Dollinger (2015)) 
	\item J (Jackson)
\end{itemize}

\clearpage
\subsection*{A1.1 Probability in the control arm $p_{C}=0.1$}
\begin{figure}[t]
	\centering
	\includegraphics[scale=0.33]{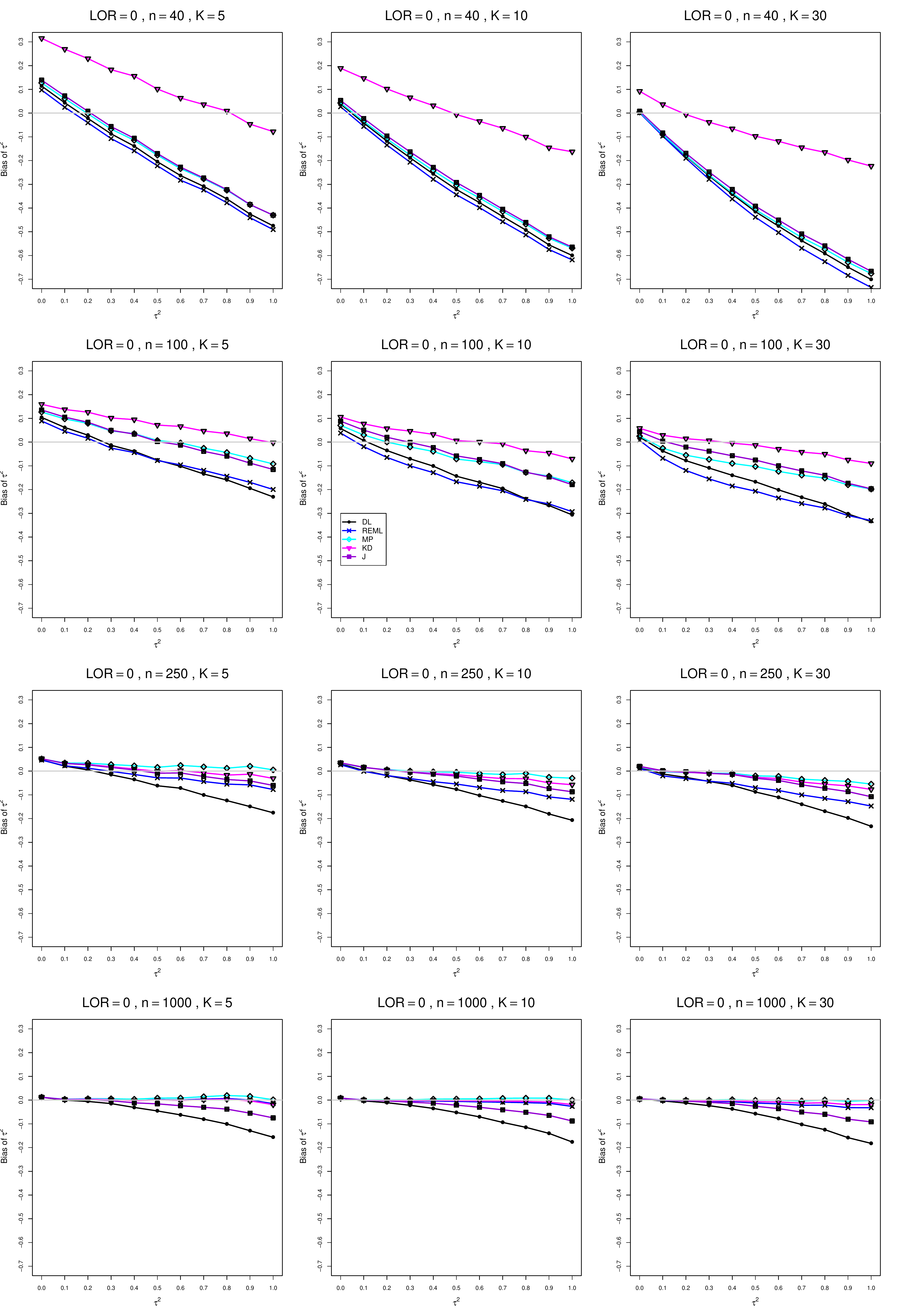}
	\caption{Bias of the estimation of  between-studies variance $\tau^2$ for $\theta=0$, $p_{iC}=0.1$, $q=0.5$, equal sample sizes $n=40,\;100,\;250,\;1000$. 
		\label{BiasTauLOR0q05piC01}}
\end{figure}

\begin{figure}[t]
	\centering
	\includegraphics[scale=0.33]{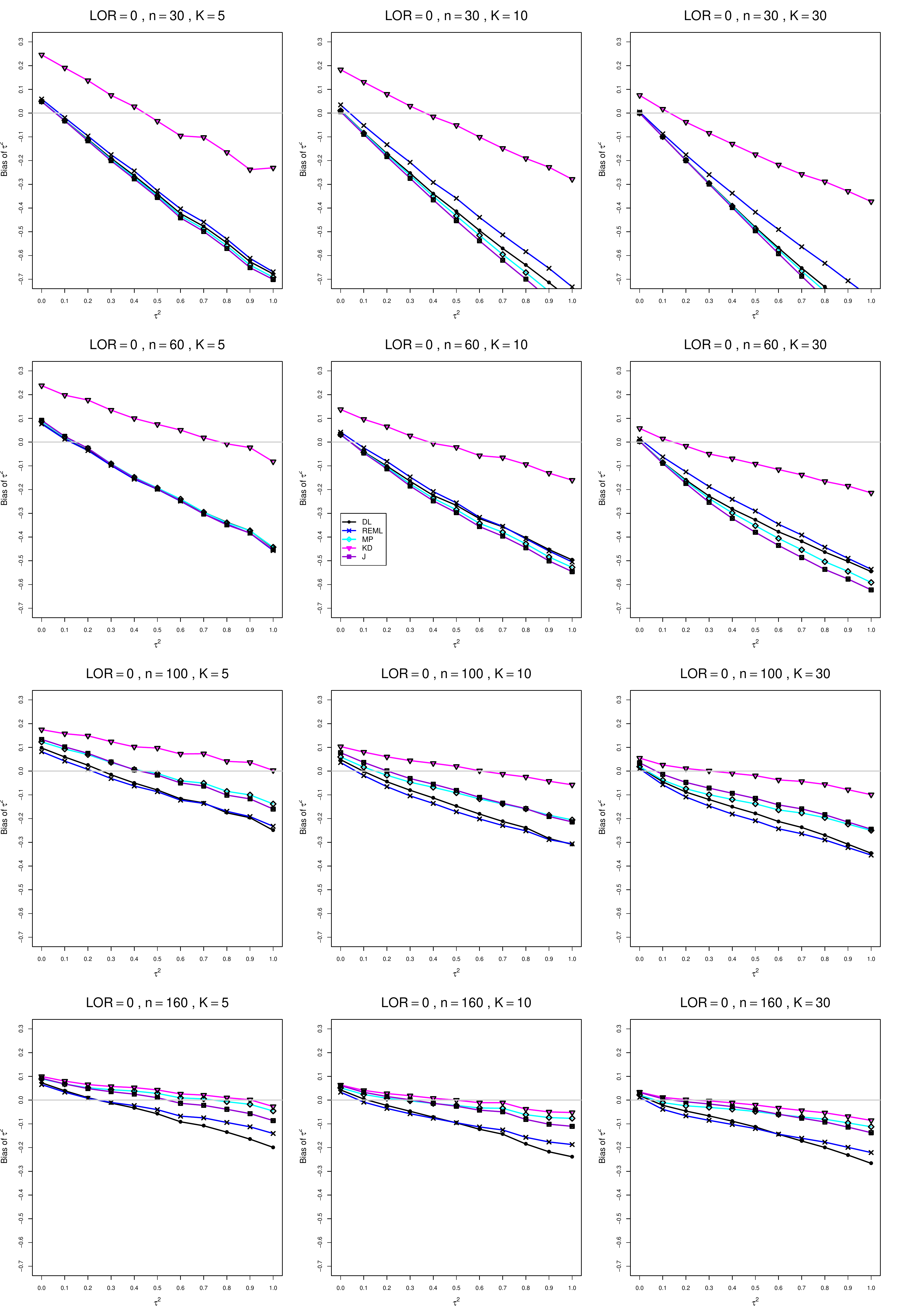}
	\caption{Bias of the estimation of  between-studies variance $\tau^2$ for $\theta=0$, $p_{iC}=0.1$, $q=0.5$, 
		unequal sample sizes $n=30,\; 60,\;100,\;160$.
		\label{BiasTauLOR0q05piC01_unequal_sample_sizes}}
\end{figure}

\begin{figure}[t]
	\centering
	\includegraphics[scale=0.33]{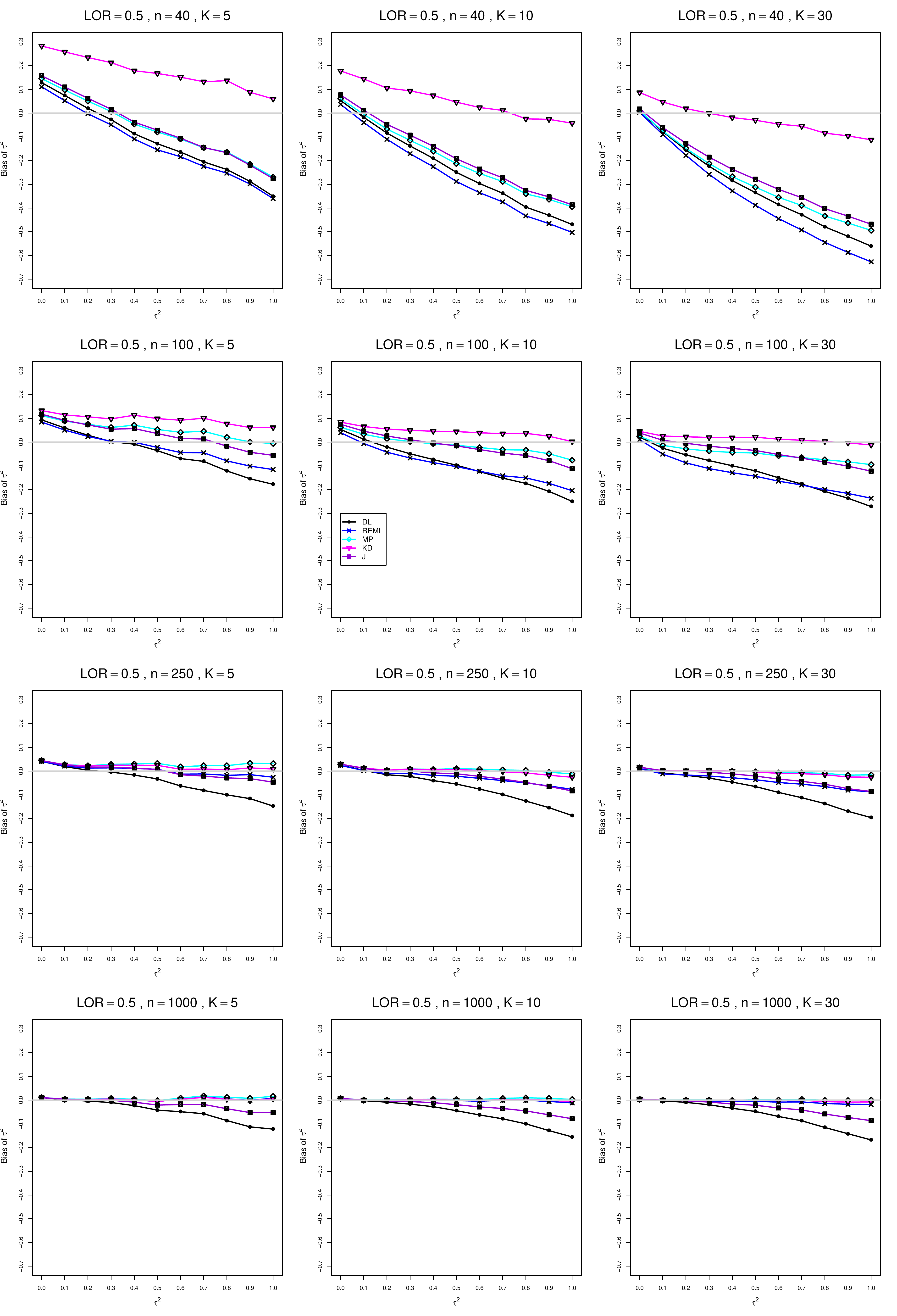}
	\caption{Bias of the estimation of  between-studies variance $\tau^2$ for $\theta=0.5$, $p_{iC}=0.1$, $q=0.5$, equal sample sizes $n=40,\;100,\;250,\;1000$.  
		\label{BiasTauLOR05q05piC01}}
\end{figure}

\begin{figure}[t]
	\centering
	\includegraphics[scale=0.33]{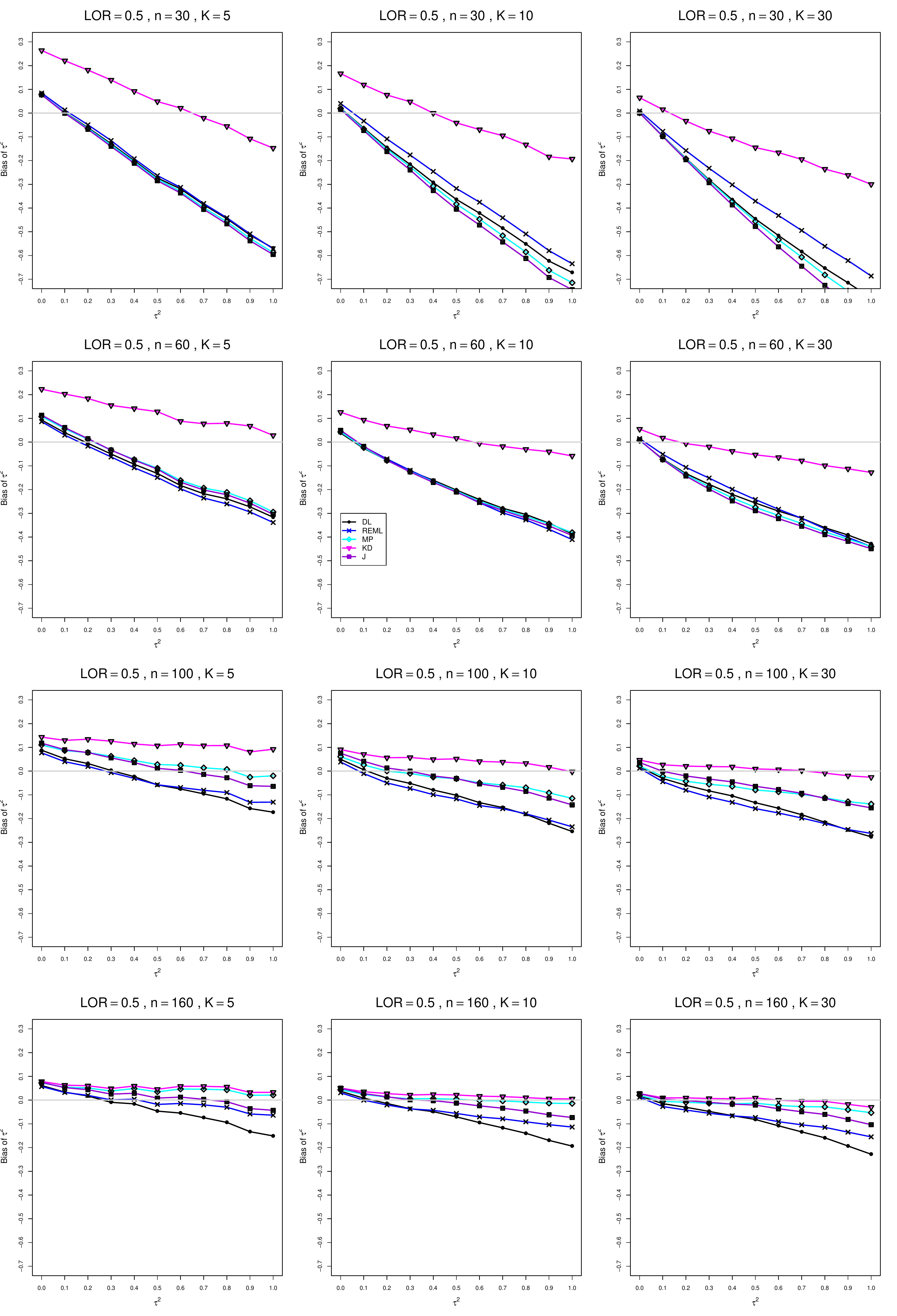}
	\caption{Bias of the estimation of  between-studies variance $\tau^2$ for $\theta=0.5$, $p_{iC}=0.1$, $q=0.5$, 
		unequal sample sizes $n=30,\; 60,\;100,\;160$. 
		\label{BiasTauLOR05q05piC01_unequal_sample_sizes}}
\end{figure}

\begin{figure}[t]
	\centering
	\includegraphics[scale=0.33]{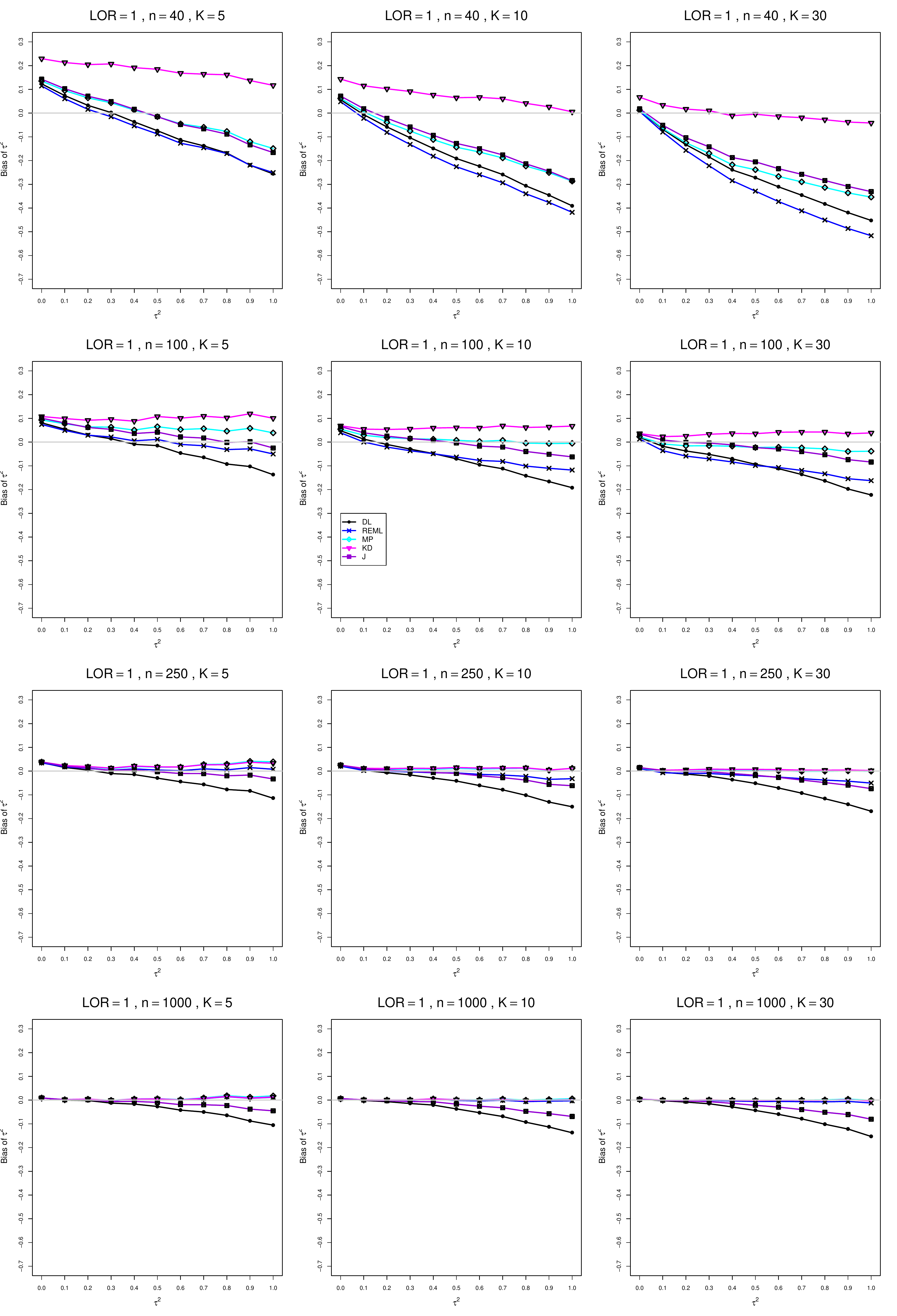}
	\caption{Bias of the estimation of  between-studies variance $\tau^2$ for $\theta=1$, $p_{iC}=0.1$, $q=0.5$, equal sample sizes $n=40,\;100,\;250,\;1000$.
		\label{BiasTauLOR1q05piC01}}
\end{figure}

\begin{figure}[t]
	\centering
	\includegraphics[scale=0.33]{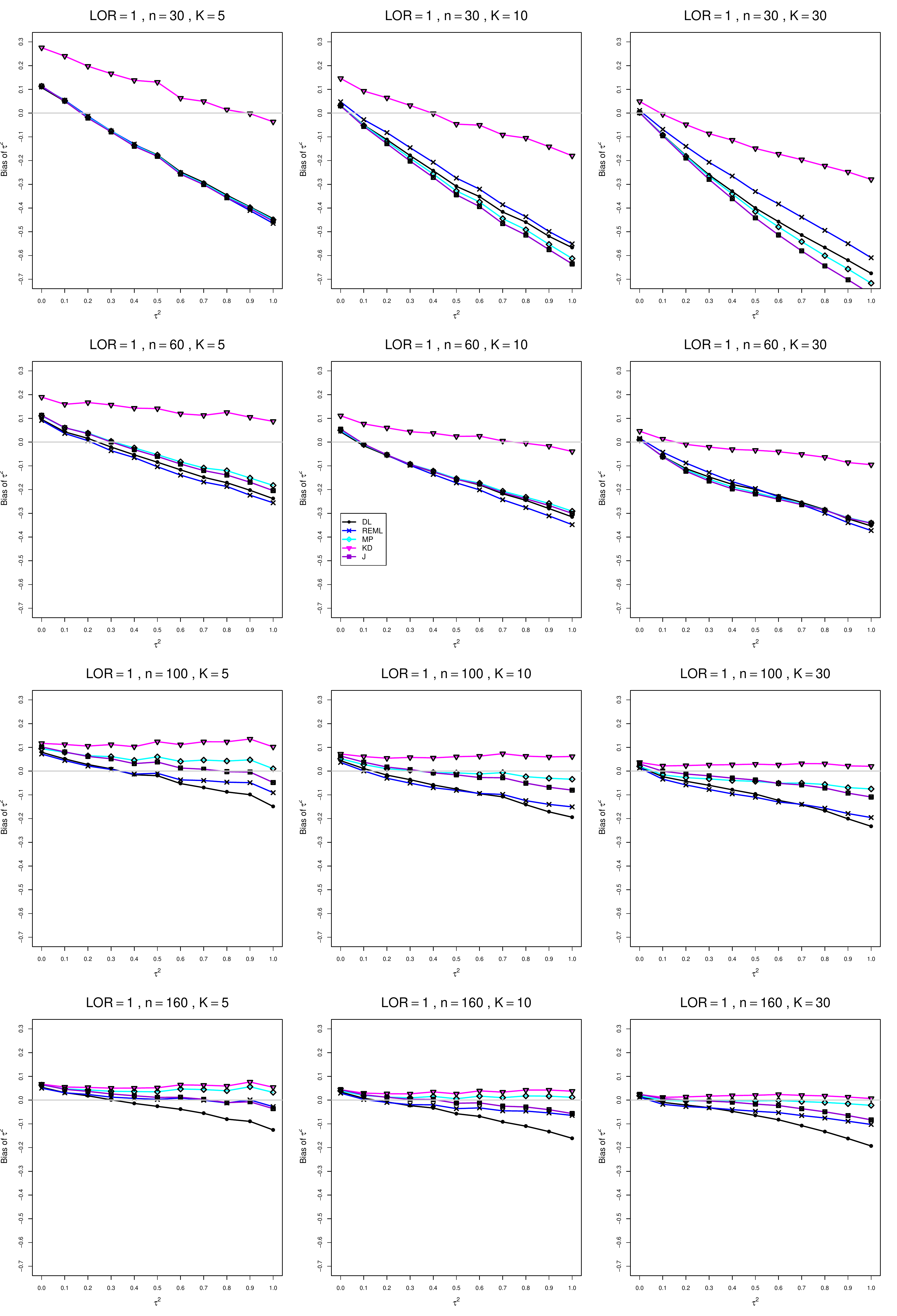}
	\caption{Bias of the estimation of  between-studies variance $\tau^2$ for $\theta=1$, $p_{iC}=0.1$, $q=0.5$, 
		unequal sample sizes $n=30,\; 60,\;100,\;160$.
		\label{BiasTauLOR1q05piC01_unequal_sample_sizes}}
\end{figure}

\begin{figure}[t]
	\centering
	\includegraphics[scale=0.33]{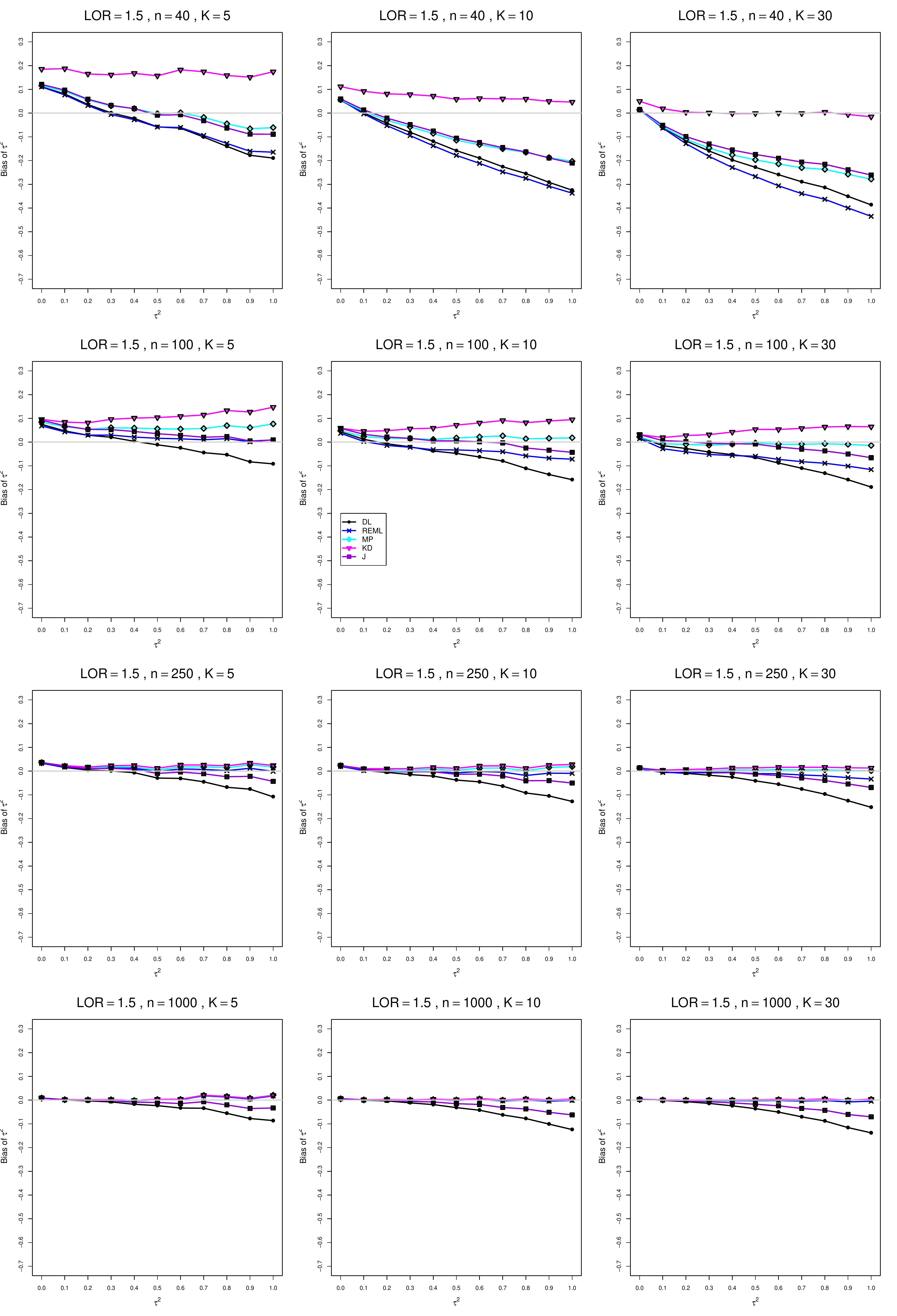}
	\caption{Bias of the estimation of  between-studies variance $\tau^2$ for $\theta=1.5$, $p_{iC}=0.1$, $q=0.5$, equal sample sizes $n=40,\;100,\;250,\;1000$.  
		\label{BiasTauLOR15q05piC01}}
\end{figure}

\begin{figure}[t]
	\centering
	\includegraphics[scale=0.33]{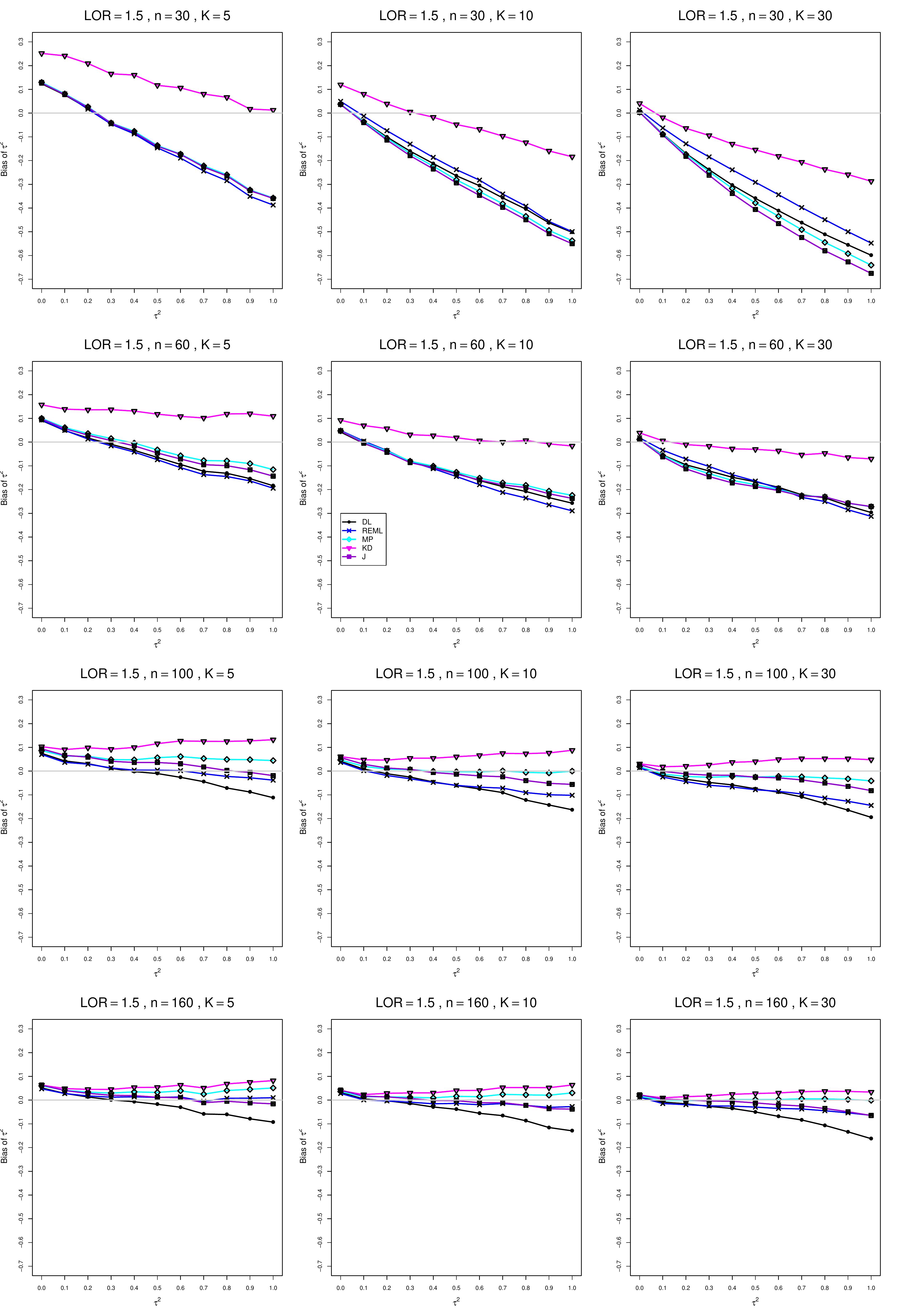}
	\caption{Bias of the estimation of  between-studies variance $\tau^2$ for $\theta=1.5$, $p_{iC}=0.1$, $q=0.5$, 
		unequal sample sizes $n=30,\; 60,\;100,\;160$.
		\label{BiasTauLOR15q05piC01_unequal_sample_sizes}}
\end{figure}

\begin{figure}[t]
	\centering
	\includegraphics[scale=0.33]{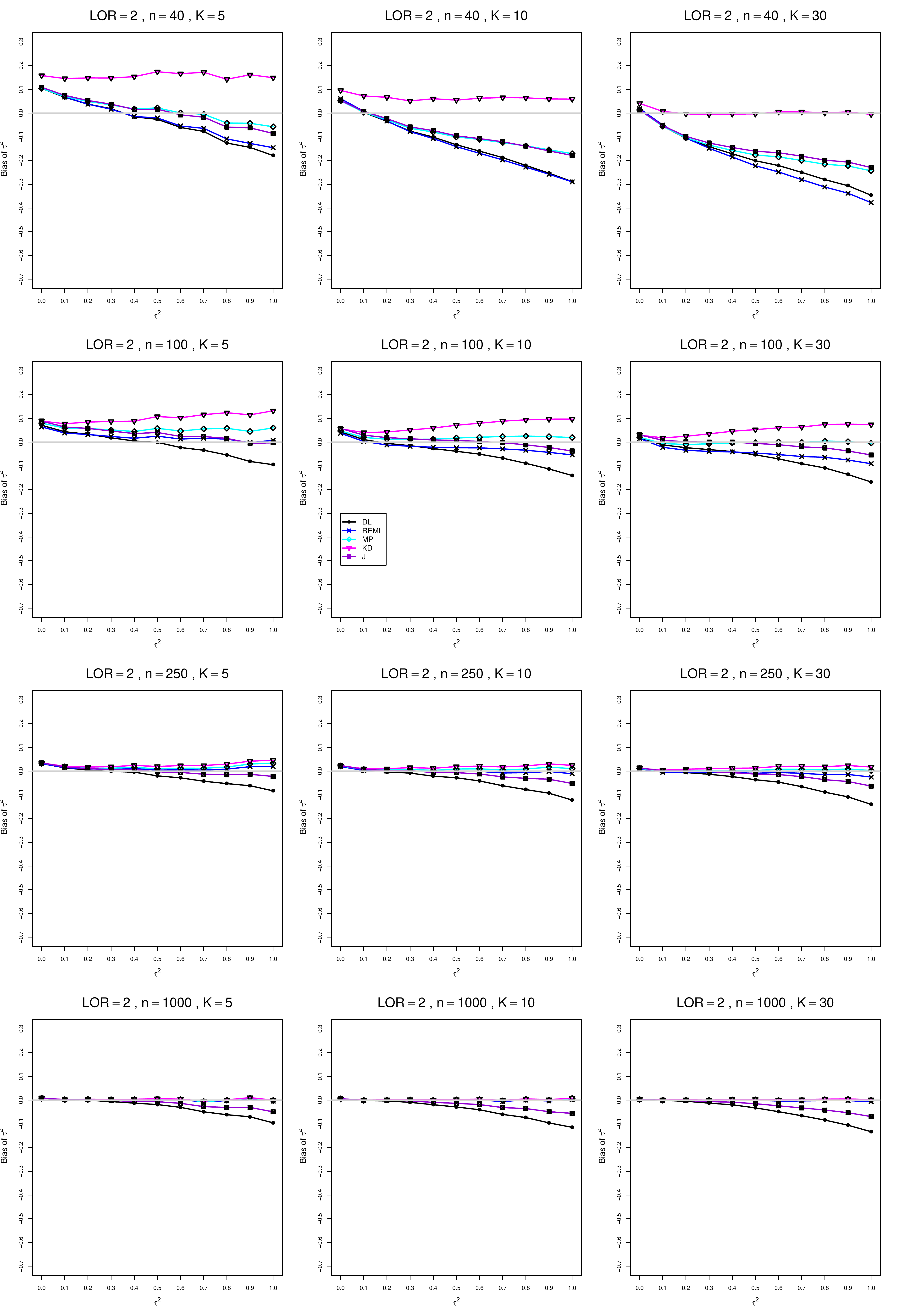}
	\caption{Bias of the estimation of  between-studies variance $\tau^2$ for $\theta=2$, $p_{iC}=0.1$, $q=0.5$, equal sample sizes $n=40,\;100,\;250,\;1000$.
		\label{BiasTauLOR2q05piC01}}
\end{figure}

\begin{figure}[t]
	\centering
	\includegraphics[scale=0.33]{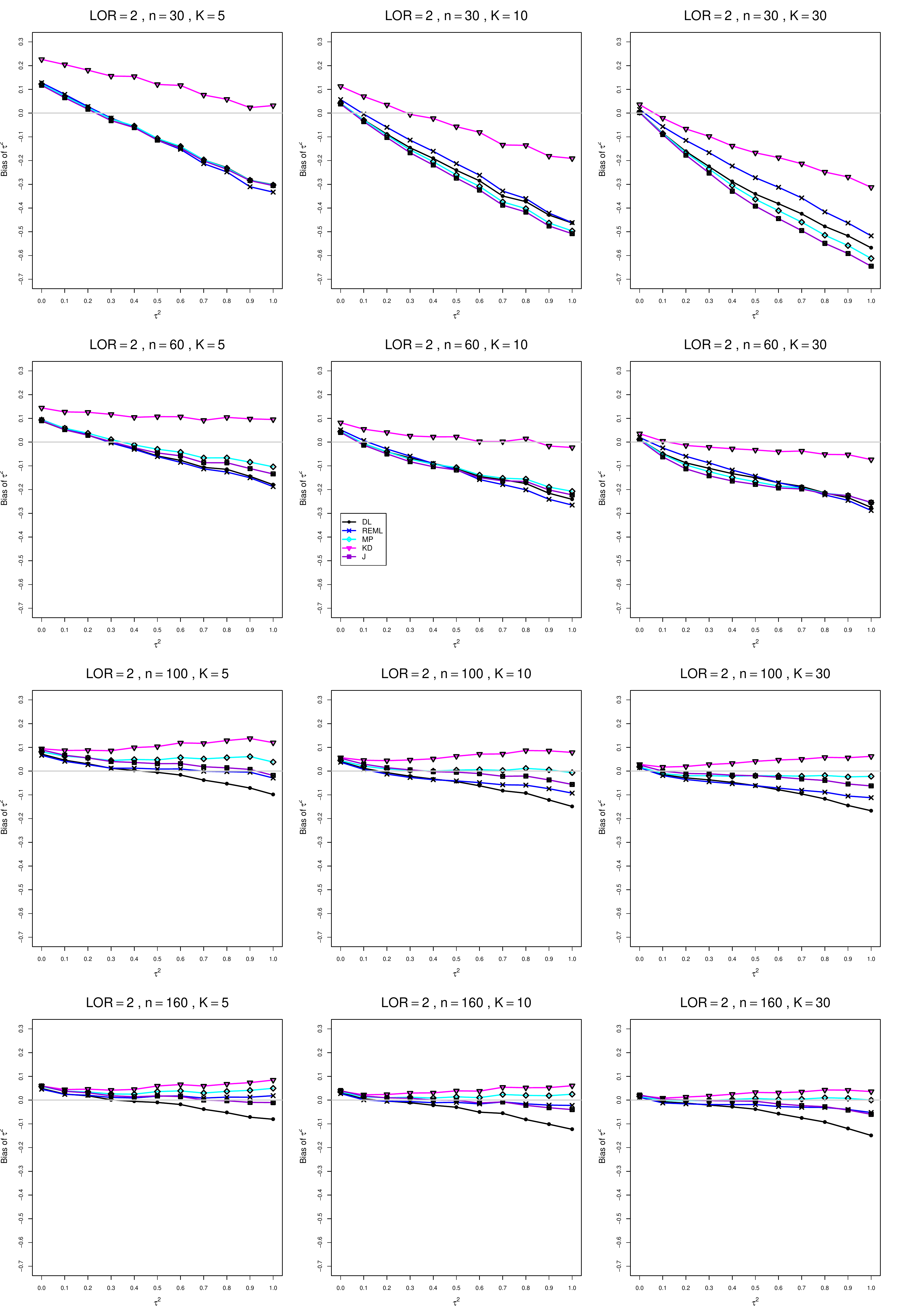}
	\caption{Bias of the estimation of  between-studies variance $\tau^2$ for $\theta=2$, $p_{iC}=0.1$, $q=0.5$, 
		unequal sample sizes $n=30,\; 60,\;100,\;160$. 
		\label{BiasTauLOR2q05piC01_unequal_sample_sizes}}
\end{figure}


\begin{figure}[t]
	\centering
	\includegraphics[scale=0.33]{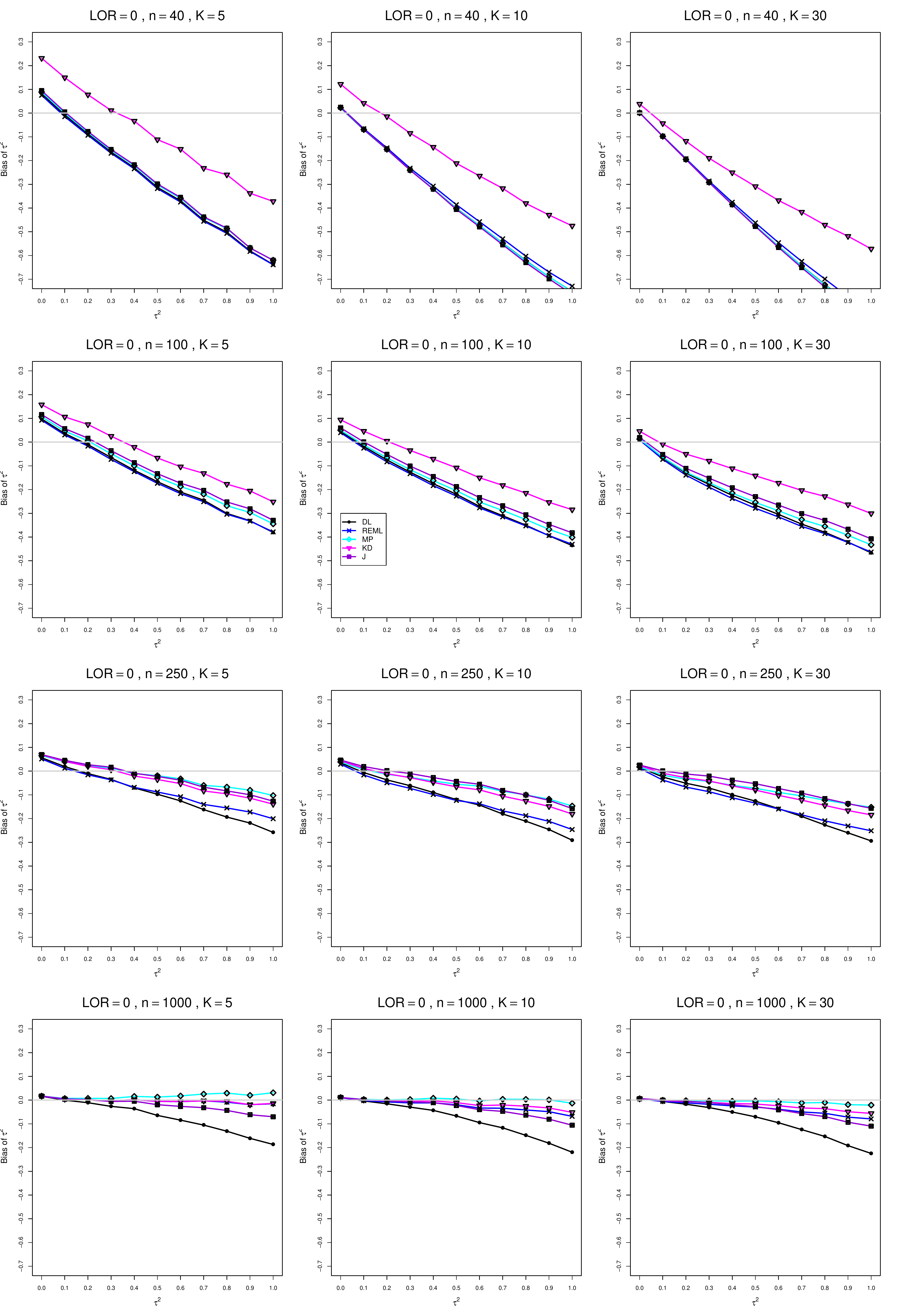}
	\caption{Bias of the estimation of  between-studies variance $\tau^2$ for $\theta=0$, $p_{iC}=0.1$, $q=0.75$, equal sample sizes $n=40,\;100,\;250,\;1000$.  
		\label{BiasTauLOR0q075piC01}}
\end{figure}

\begin{figure}[t]
	\centering
	\includegraphics[scale=0.33]{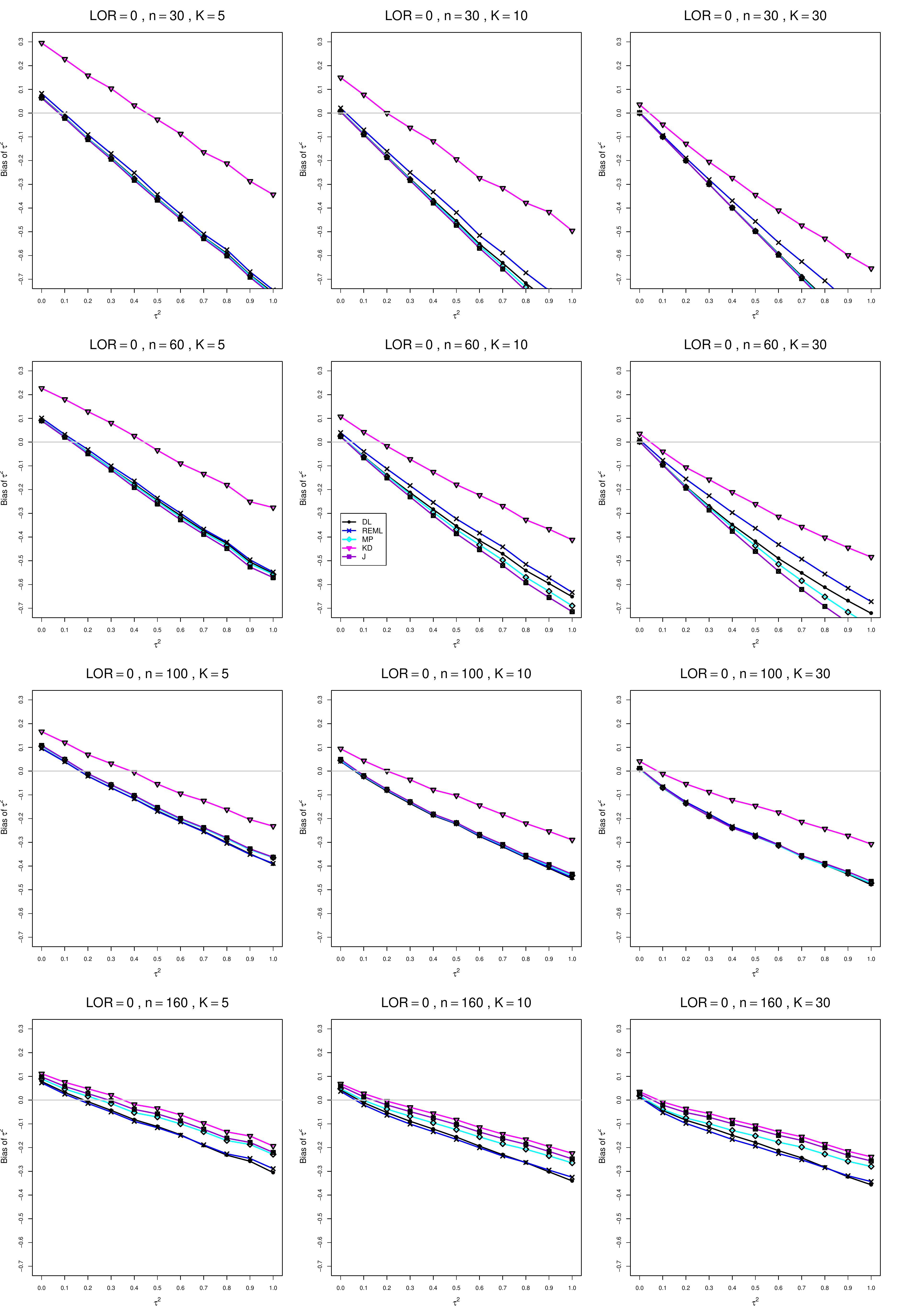}
	\caption{Bias of the estimation of  between-studies variance $\tau^2$ for $\theta=0$, $p_{iC}=0.1$, $q=0.75$, 
		unequal sample sizes $n=30,\; 60,\;100,\;160$.
		\label{BiasTauLOR0q075piC01_unequal_sample_sizes}}
\end{figure}

\begin{figure}[t]
	\centering
	\includegraphics[scale=0.33]{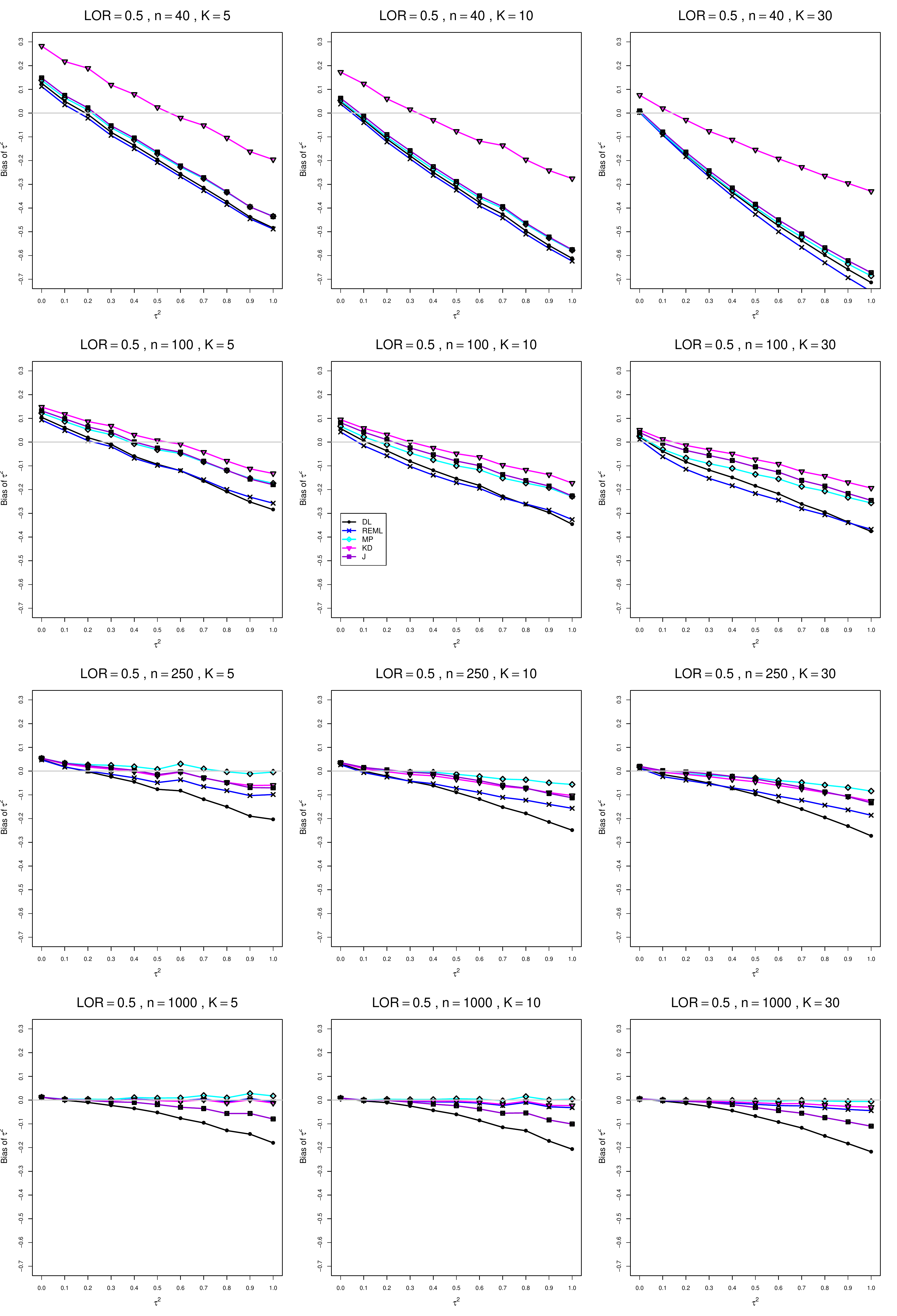}
	\caption{Bias of the estimation of  between-studies variance $\tau^2$ for $\theta=0.5$, $p_{iC}=0.1$, $q=0.75$, equal sample sizes $n=40,\;100,\;250,\;1000$.  
		\label{BiasTauLOR05q075piC01}}
\end{figure}

\begin{figure}[t]
	\centering
	\includegraphics[scale=0.33]{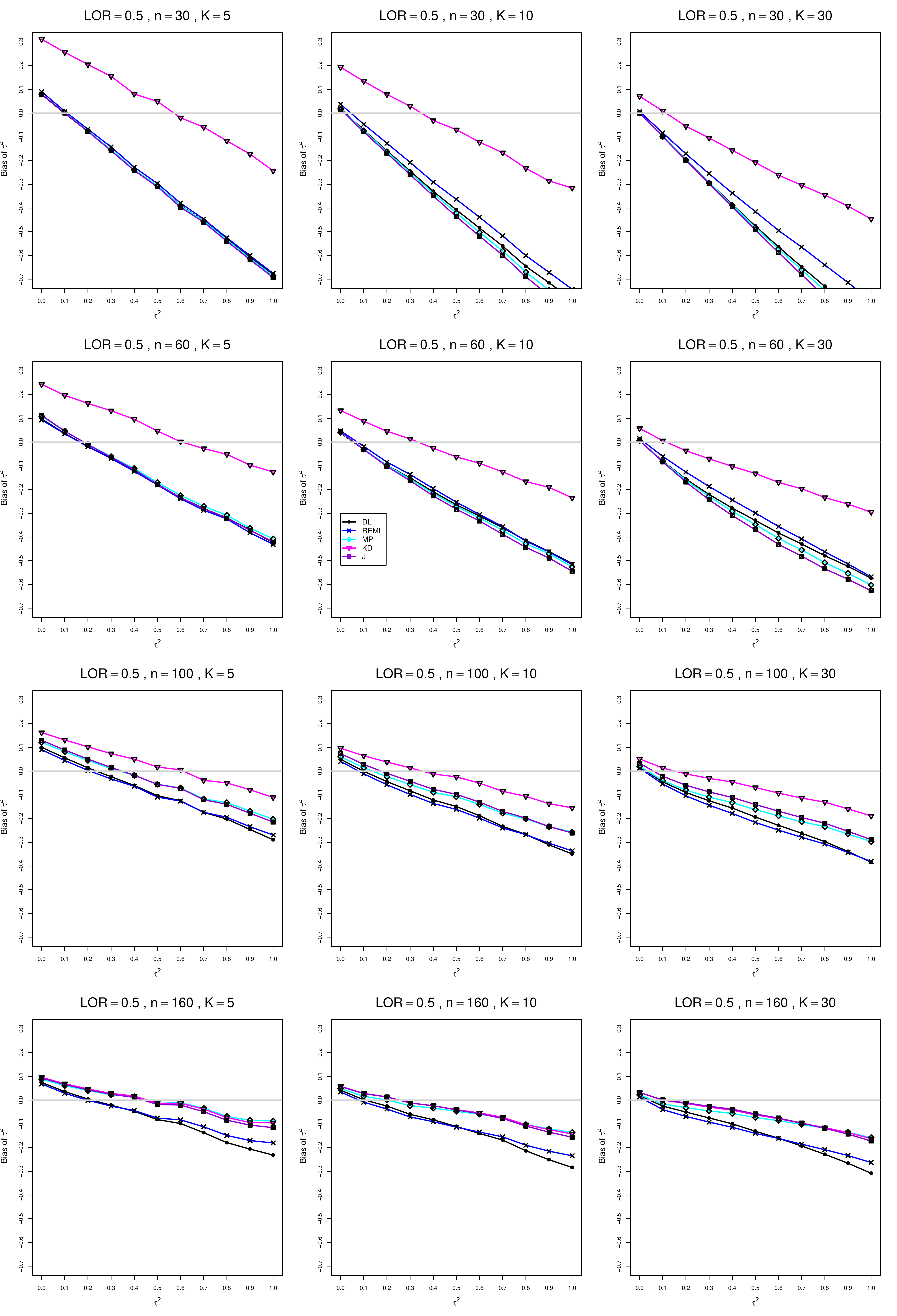}
	\caption{Bias of the estimation of  between-studies variance $\tau^2$ for $\theta=0.5$, $p_{iC}=0.1$, $q=0.75$, 
		unequal sample sizes $n=30,\; 60,\;100,\;160$.
		\label{BiasTauLOR05q075piC01_unequal_sample_sizes}}
\end{figure}

\begin{figure}[t]
	\centering
	\includegraphics[scale=0.33]{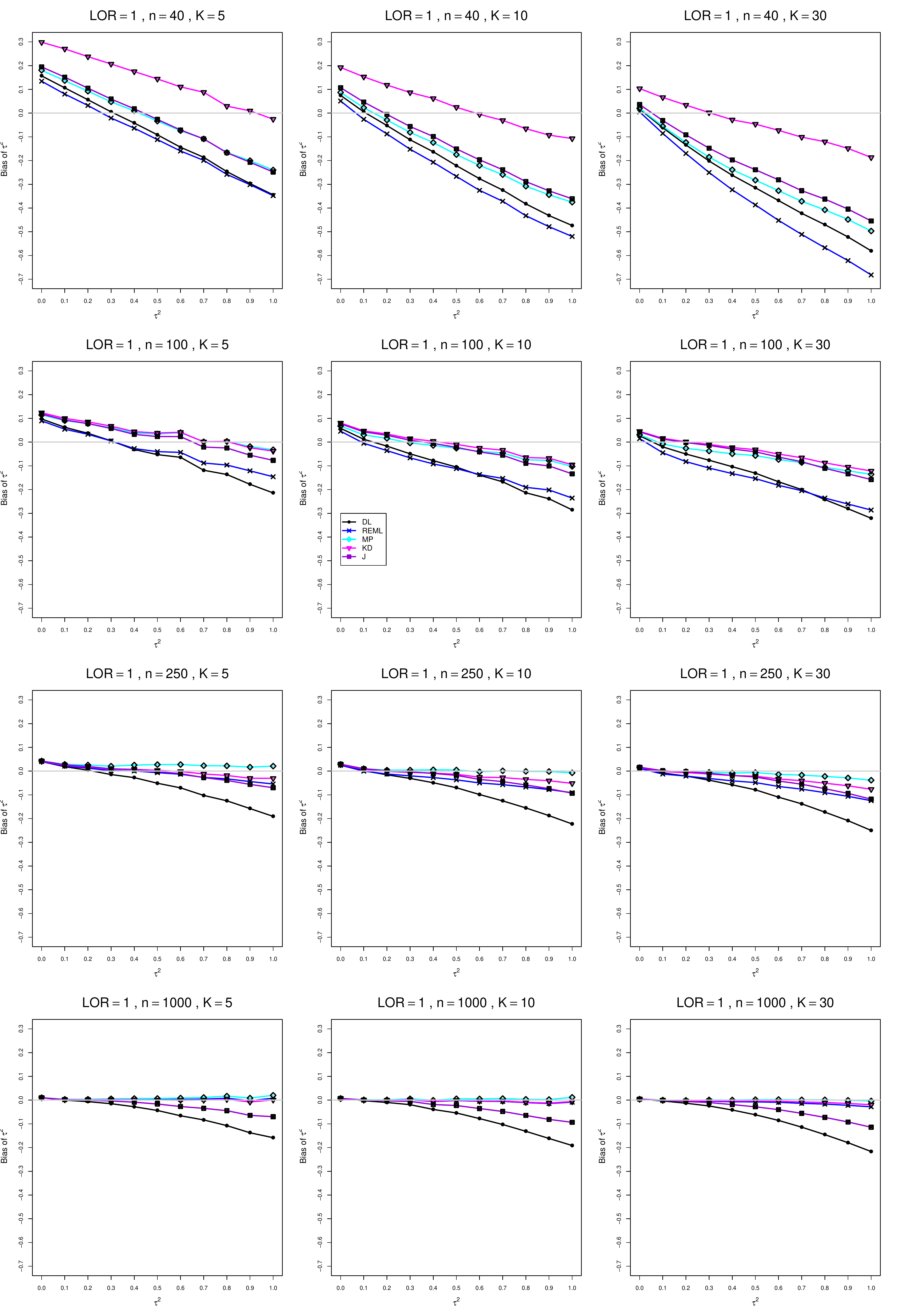}
	\caption{Bias of the estimation of  between-studies variance $\tau^2$ for $\theta=1$, $p_{iC}=0.1$, $q=0.75$, equal sample sizes $n=40,\;100,\;250,\;1000$. 
		\label{BiasTauLOR1q075piC01}}
\end{figure}

\begin{figure}[t]
	\centering
	\includegraphics[scale=0.33]{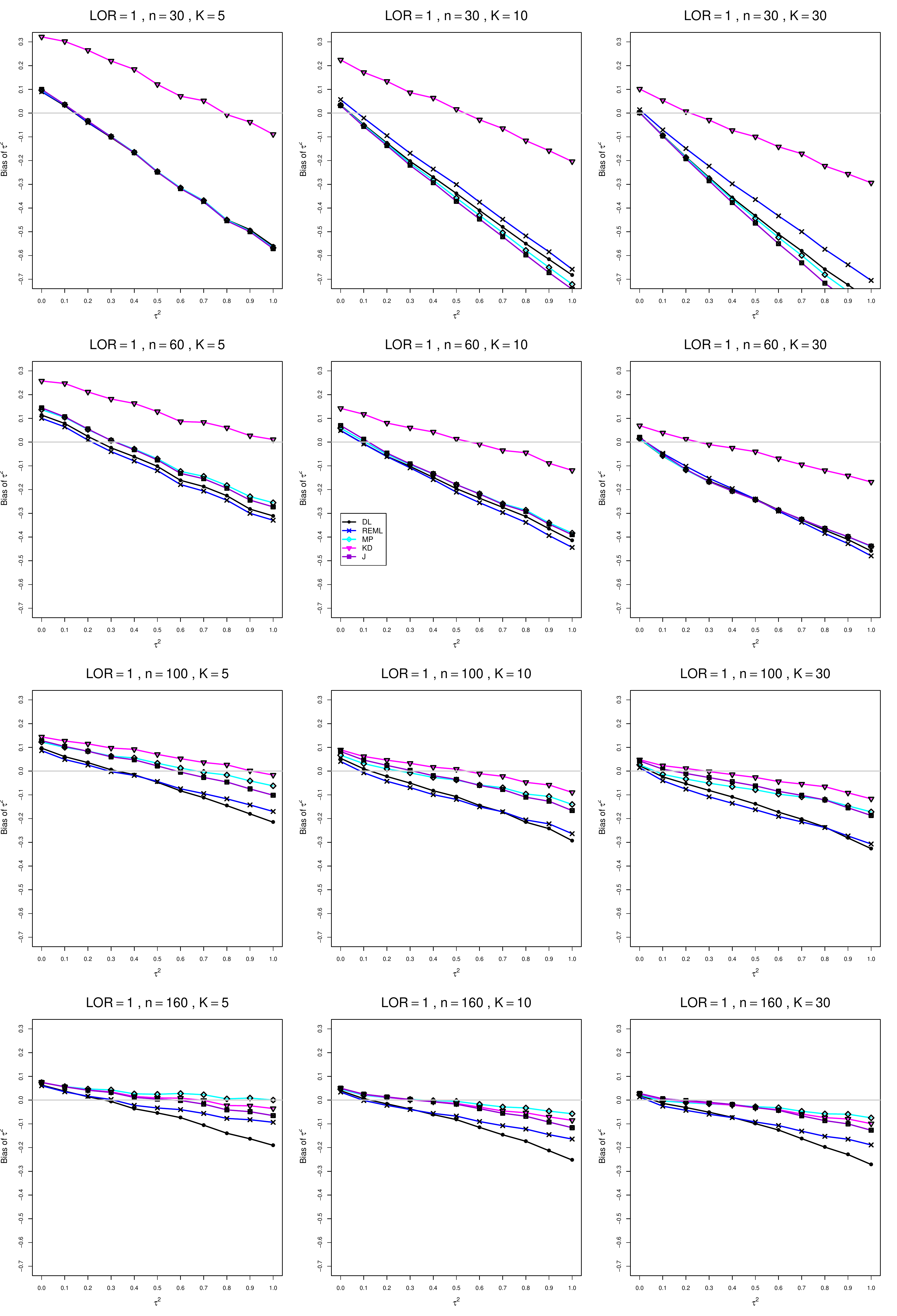}
	\caption{Bias of the estimation of  between-studies variance $\tau^2$ for $\theta=1$, $p_{iC}=0.1$, $q=0.75$, 
		unequal sample sizes $n=30,\; 60,\;100,\;160$.
		\label{BiasTauLOR1q075piC01_unequal_sample_sizes}}
\end{figure}

\begin{figure}[t]
	\centering
	\includegraphics[scale=0.33]{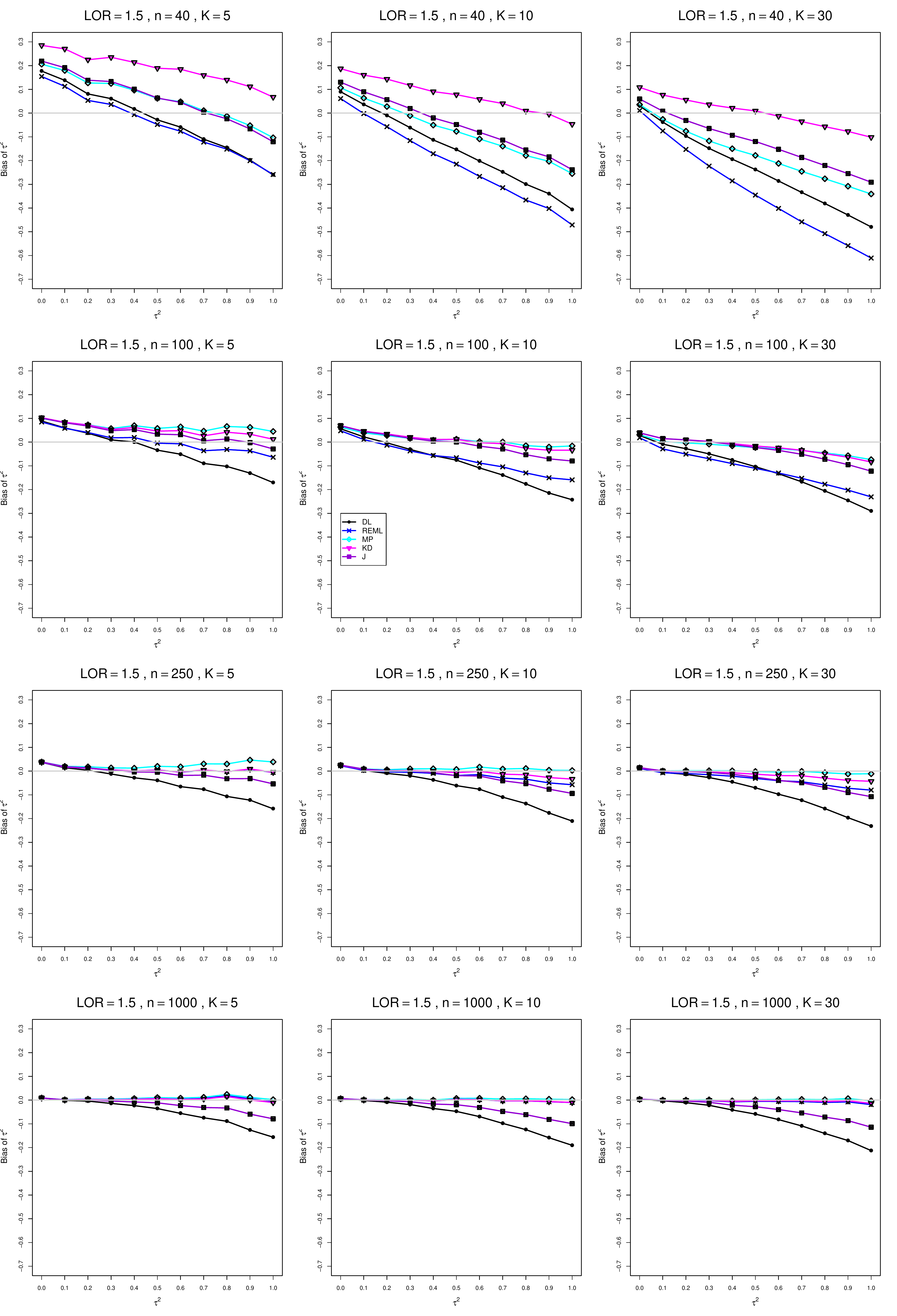}
	\caption{Bias of the estimation of  between-studies variance $\tau^2$ for $\theta=1.5$, $p_{iC}=0.1$, $q=0.75$, equal sample sizes $n=40,\;100,\;250,\;1000$.  
		\label{BiasTauLOR15q075piC01}}
\end{figure}

\begin{figure}[t]
	\centering
	\includegraphics[scale=0.33]{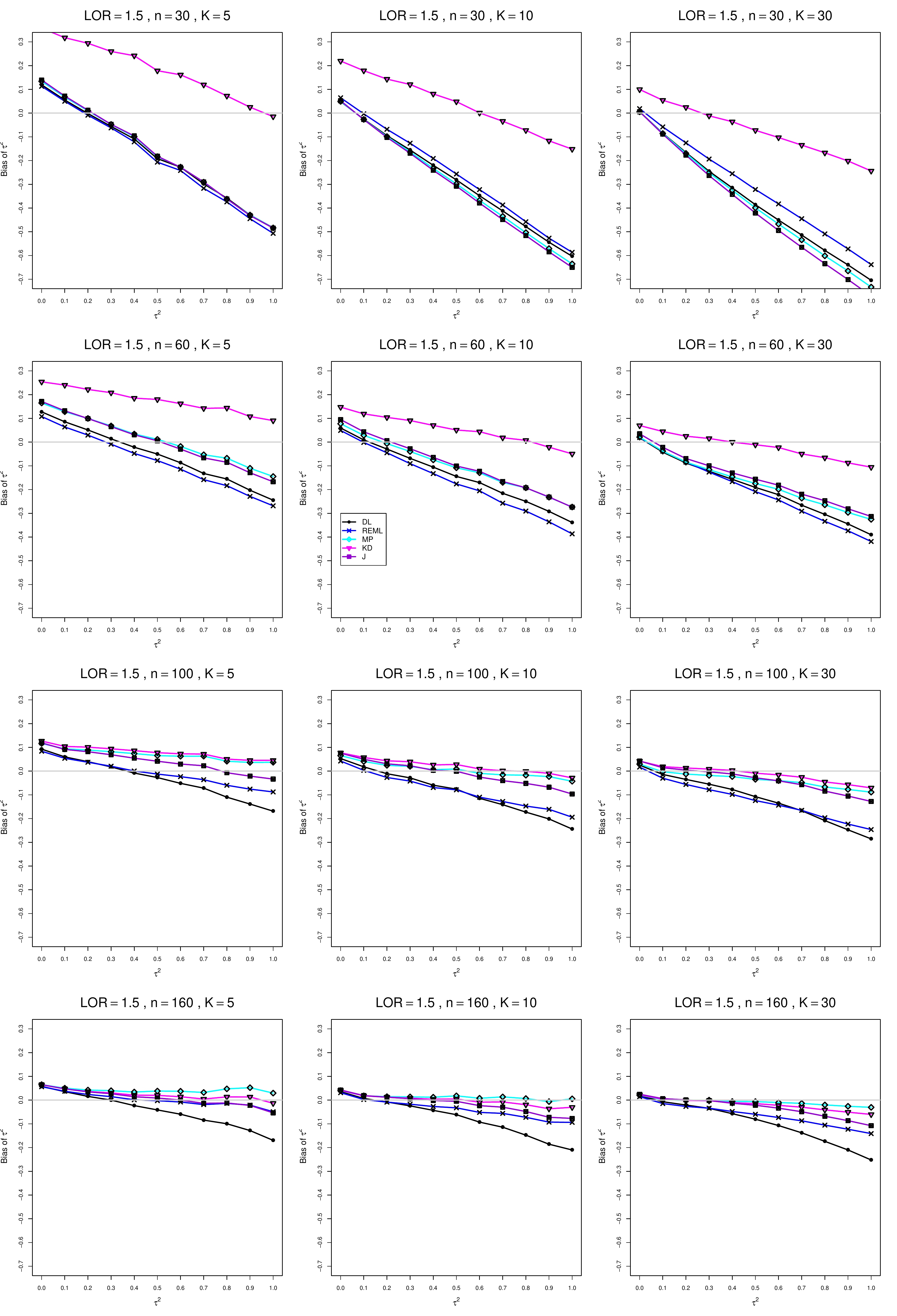}
	\caption{Bias of the estimation of  between-studies variance $\tau^2$ for $\theta=1.5$, $p_{iC}=0.1$, $q=0.75$, 
		unequal sample sizes $n=30,\; 60,\;100,\;160$. 
		\label{BiasTauLOR15q075piC01_unequal_sample_sizes}}
\end{figure}

\clearpage

\begin{figure}[t]
	\centering
	\includegraphics[scale=0.33]{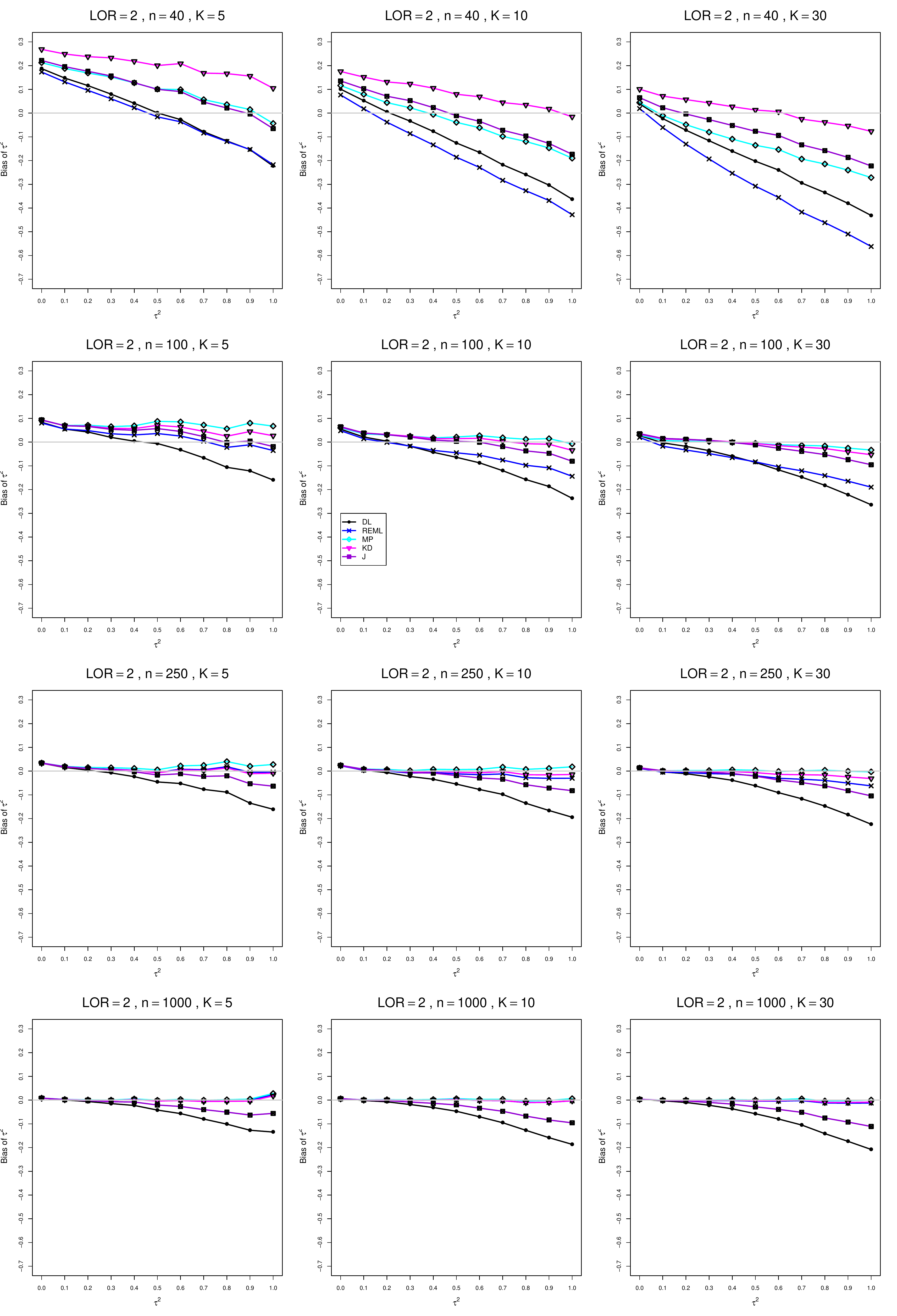}
	\caption{Bias of the estimation of  between-studies variance $\tau^2$ for $\theta=2$, $p_{iC}=0.1$, $q=0.75$, equal sample sizes $n=40,\;100,\;250,\;1000$.  
		\label{BiasTauLOR2q075piC01}}
\end{figure}

\begin{figure}[t]
	\centering
	\includegraphics[scale=0.33]{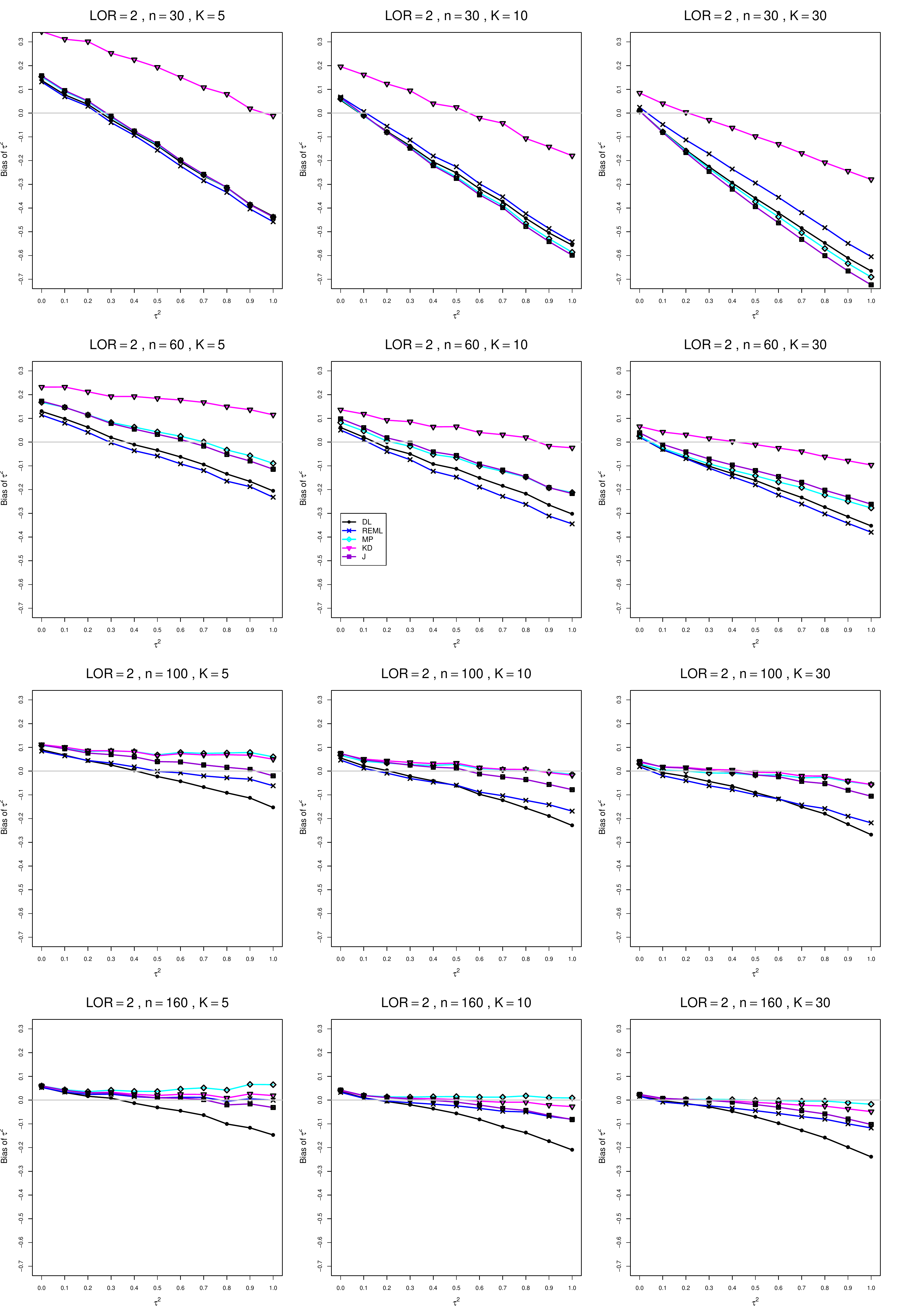}
	\caption{Bias of the estimation of  between-studies variance $\tau^2$ for $\theta=2$, $p_{iC}=0.1$, $q=0.75$, 
		unequal sample sizes $n=30,\; 60,\;100,\;160$.
		\label{BiasTauLOR2q075piC01_unequal_sample_sizes}}
\end{figure}

\clearpage
\renewcommand{\thefigure}{A1.2.\arabic{figure}}
\setcounter{figure}{0}
\subsection*{A1.2 Probability in the control arm $p_{C}=0.2$}
\begin{figure}[t]
	\centering
	\includegraphics[scale=0.33]{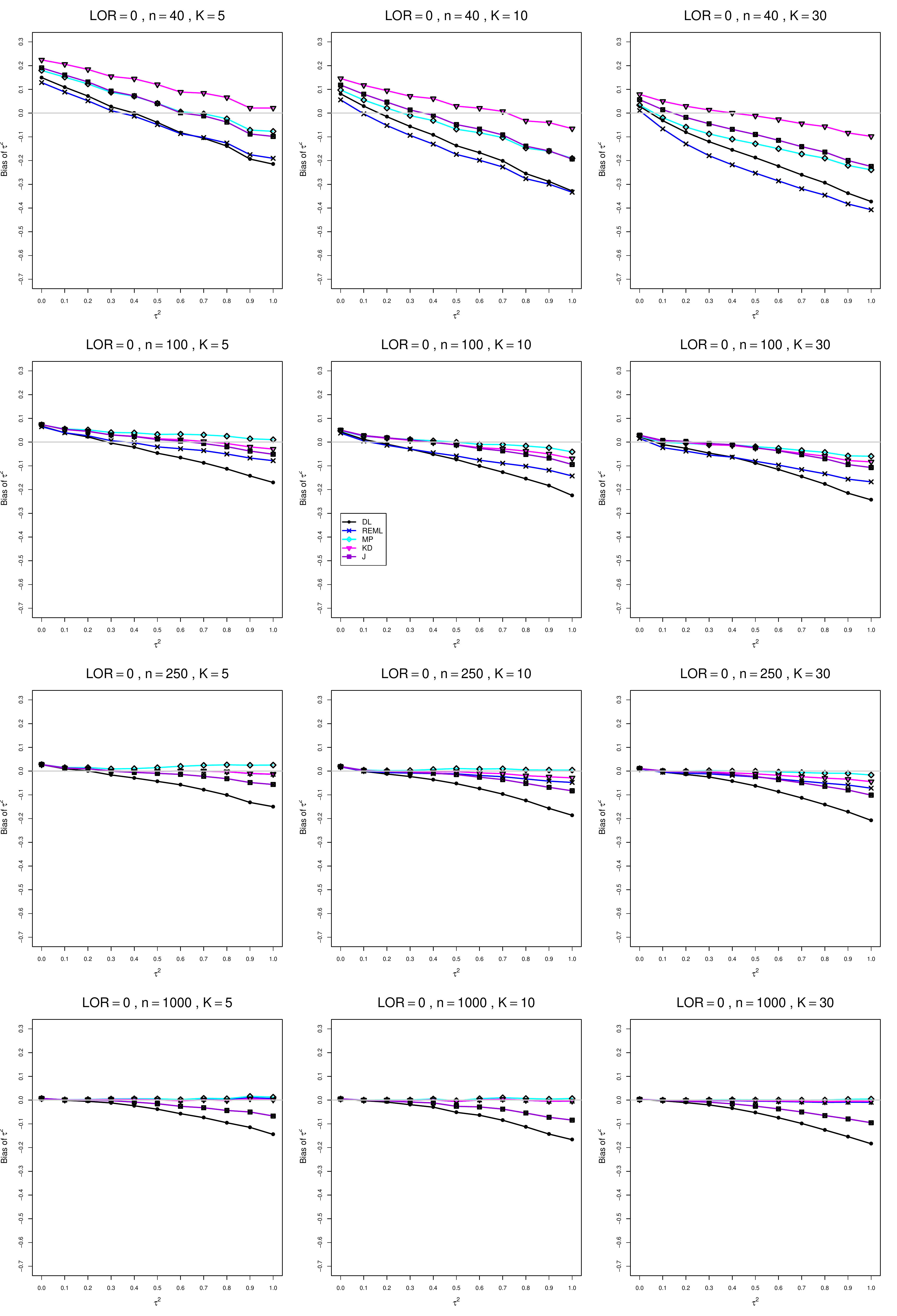}
	\caption{Bias of the estimation of  between-studies variance $\tau^2$ for $\theta=0$, $p_{iC}=0.2$, $q=0.5$, equal sample sizes $n=40,\;100,\;250,\;1000$. 
		\label{BiasTauLOR0q05piC02}}
\end{figure}

\begin{figure}[t]
	\centering
	\includegraphics[scale=0.33]{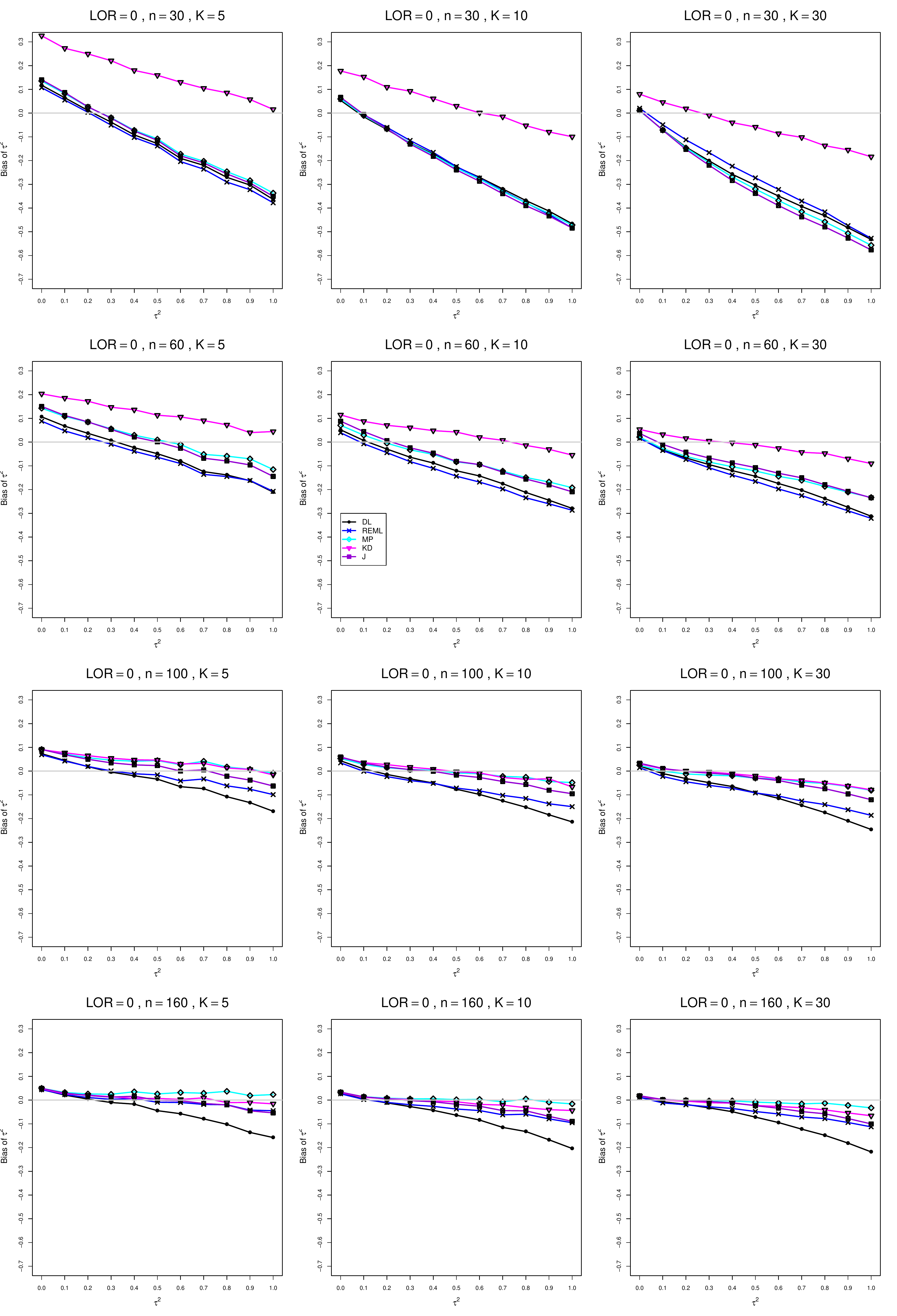}
	\caption{Bias of the estimation of  between-studies variance $\tau^2$ for $\theta=0$, $p_{iC}=0.2$, $q=0.5$,
		unequal sample sizes $n=30,\; 60,\;100,\;160$. 
		\label{BiasTauLOR0q05piC02_unequal_sample_sizes}}
\end{figure}

\begin{figure}[t]
	\centering
	\includegraphics[scale=0.33]{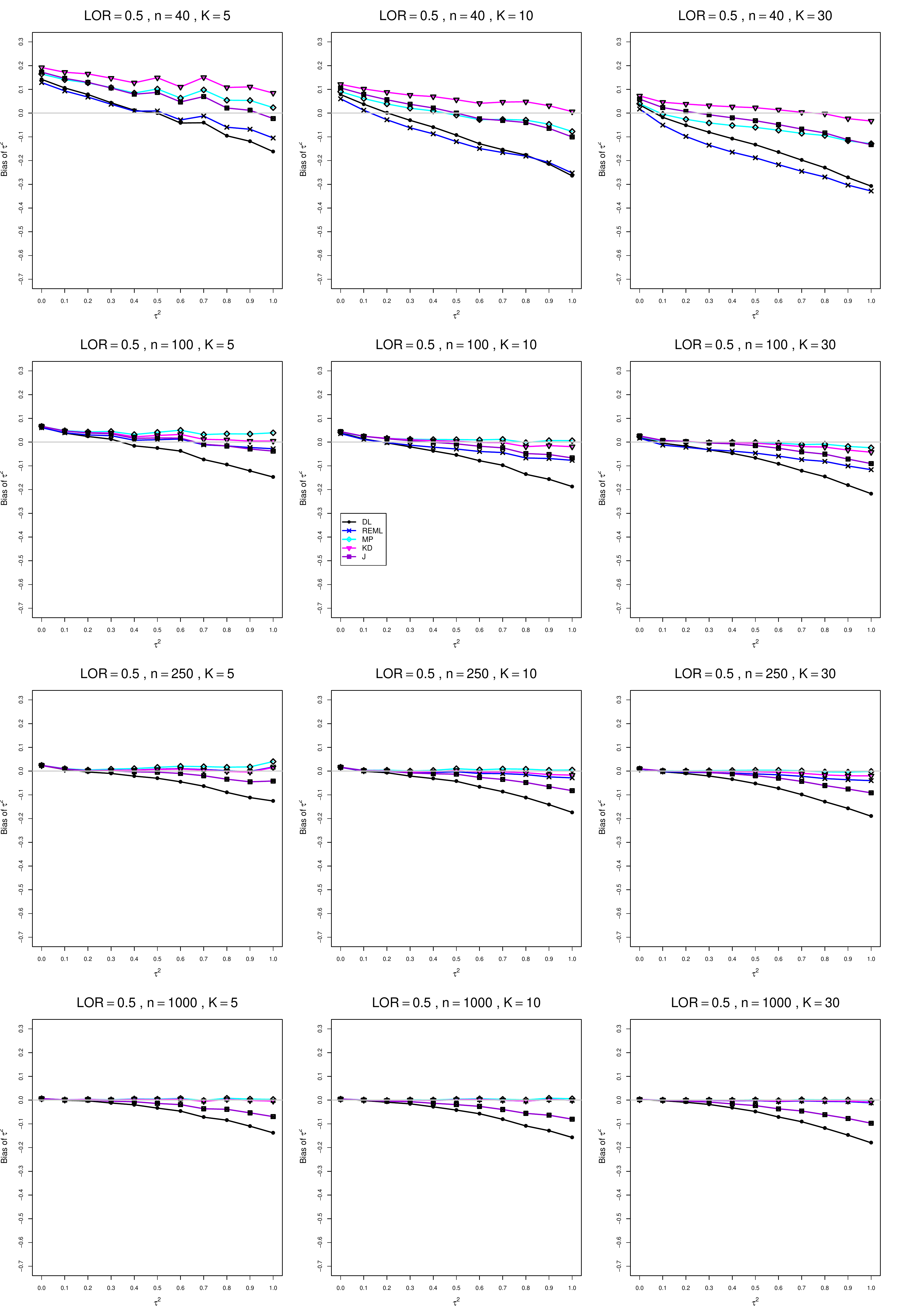}
	\caption{Bias of the estimation of  between-studies variance $\tau^2$ for $\theta=0.5$, $p_{iC}=0.2$, $q=0.5$, equal sample sizes $n=40,\;100,\;250,\;1000$.  
		\label{BiasTauLOR05q05piC02}}
\end{figure}

\begin{figure}[t]
	\centering
	\includegraphics[scale=0.33]{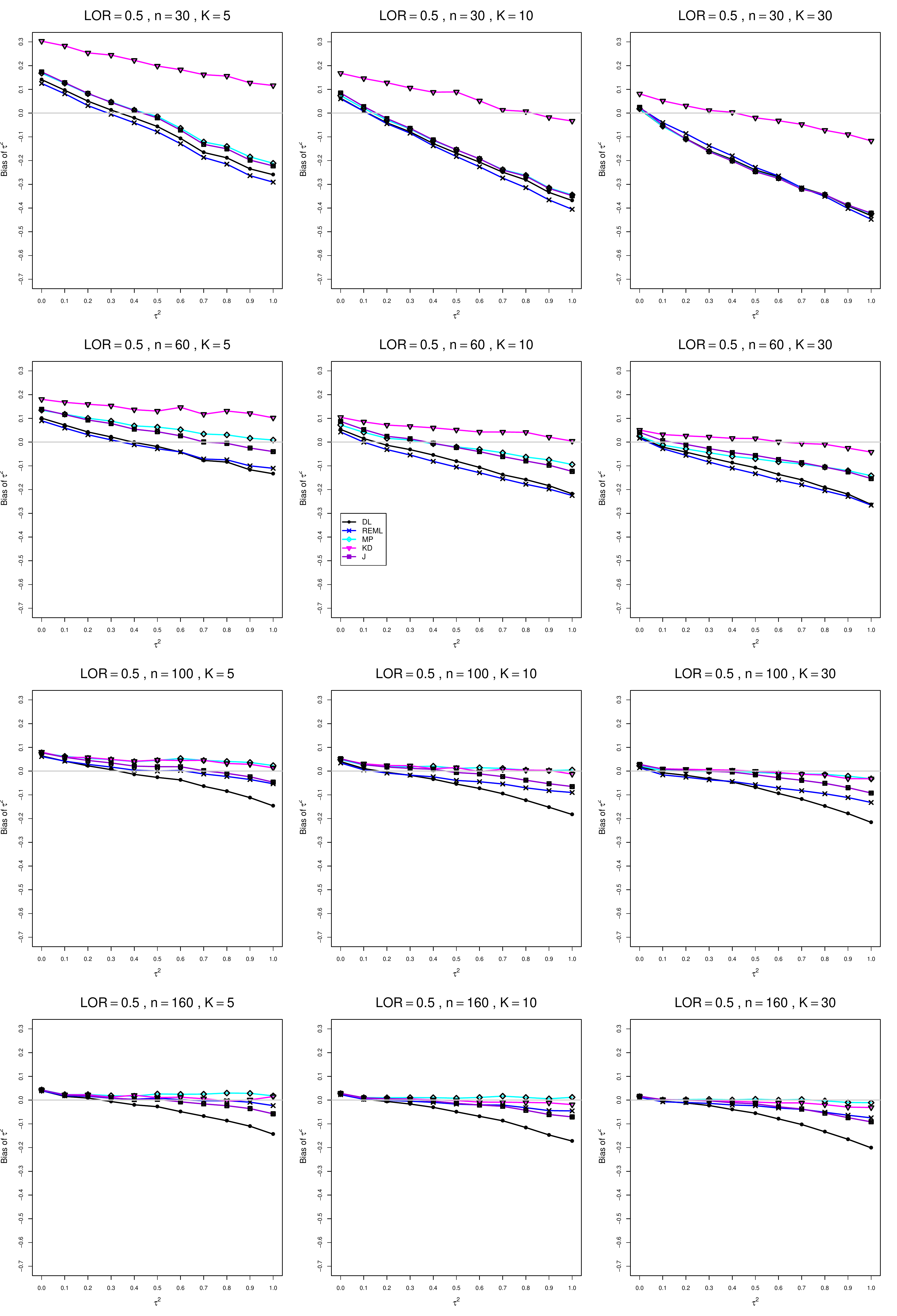}
	\caption{Bias of the estimation of  between-studies variance $\tau^2$ for $\theta=0.5$, $p_{iC}=0.2$, $q=0.5$,
		unequal sample sizes $n=30,\; 60,\;100,\;160$. 
		\label{BiasTauLOR05q05piC02_unequal_sample_sizes}}
\end{figure}

\begin{figure}[t]
	\centering
	\includegraphics[scale=0.33]{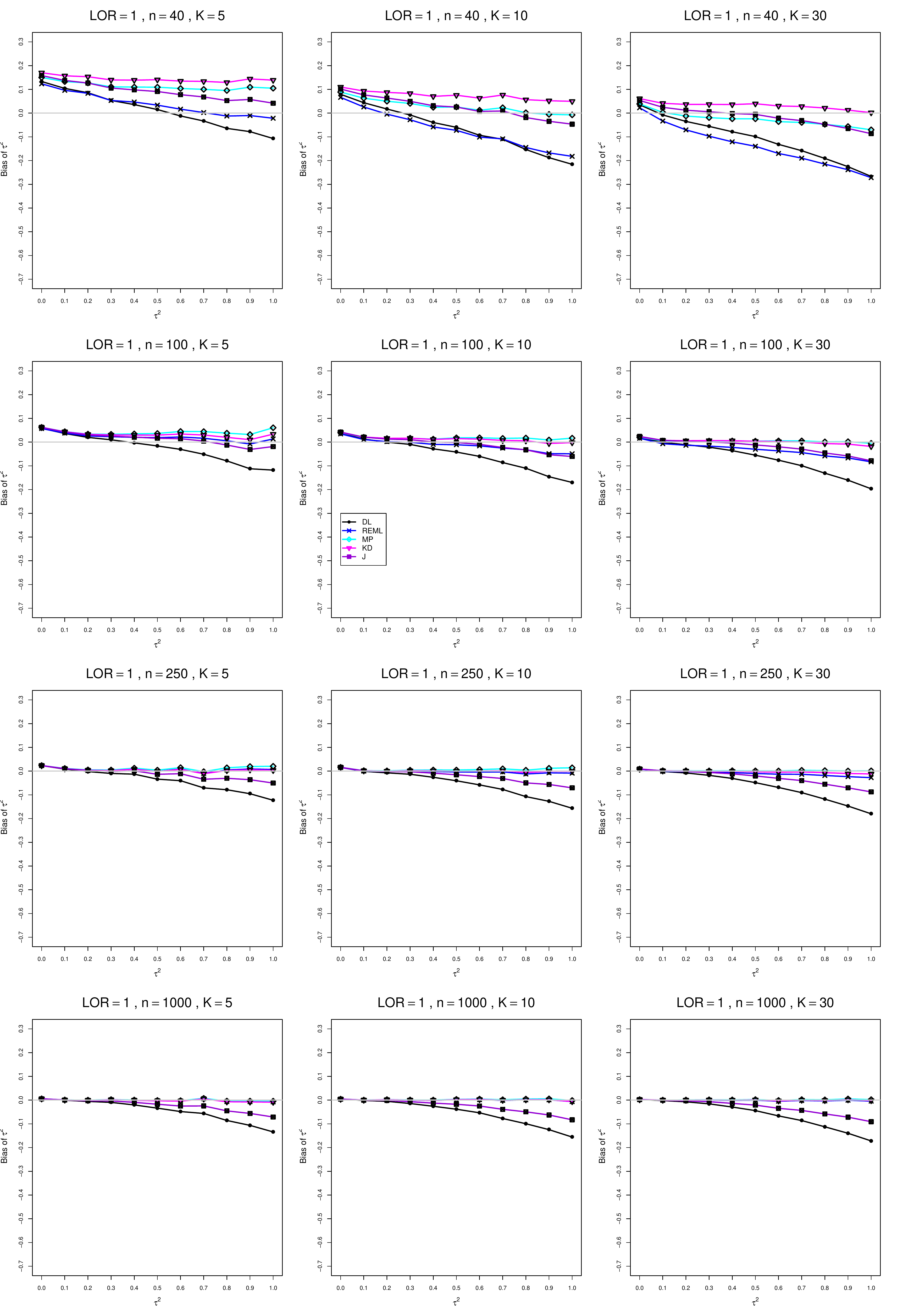}
	\caption{Bias of the estimation of  between-studies variance $\tau^2$ for $\theta=1$, $p_{iC}=0.2$, $q=0.5$, equal sample sizes $n=40,\;100,\;250,\;1000$. 
		\label{BiasTauLOR1q05piC02}}
\end{figure}

\begin{figure}[t]
	\centering
	\includegraphics[scale=0.33]{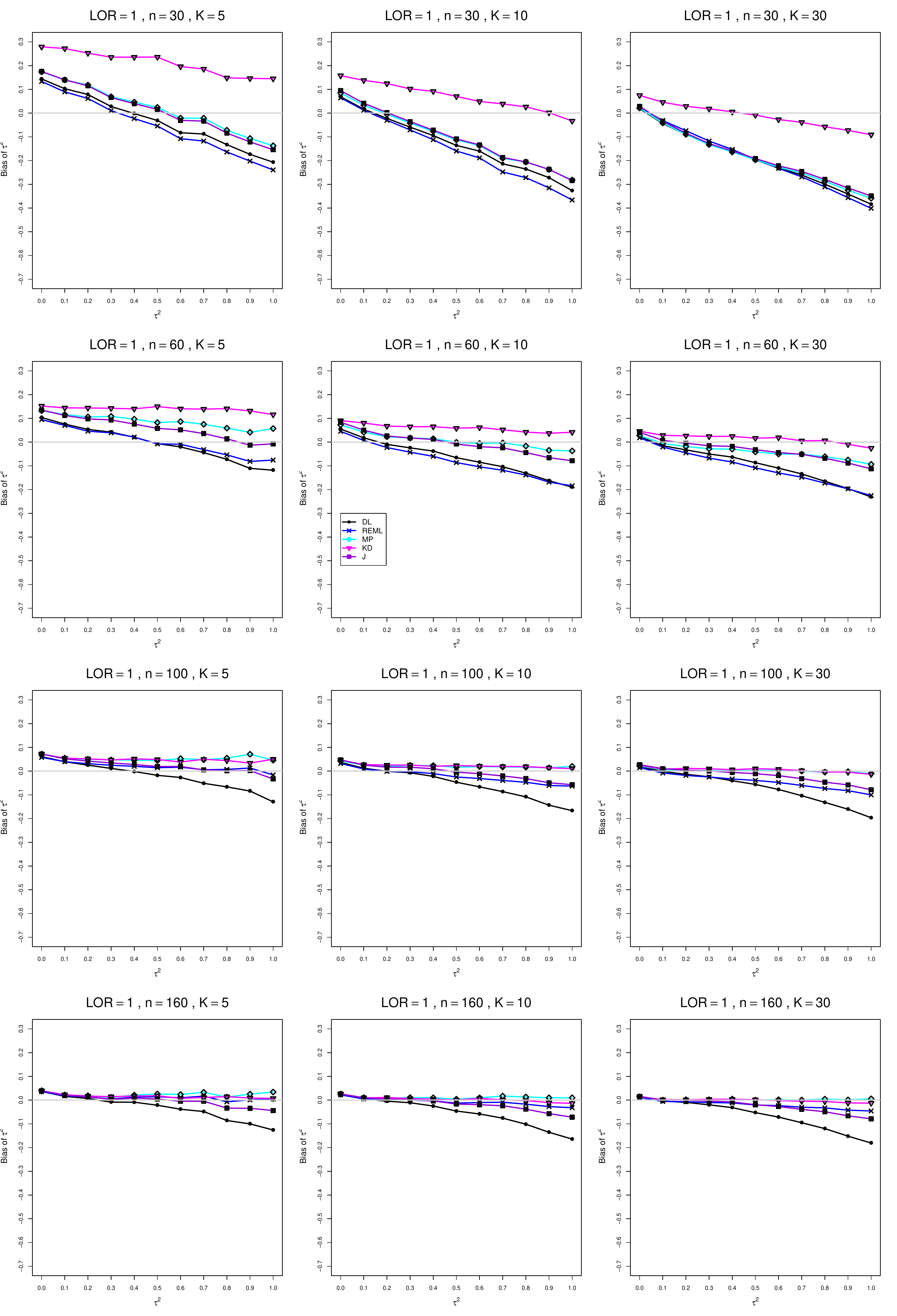}
	\caption{Bias of the estimation of  between-studies variance $\tau^2$ for $\theta=1$, $p_{iC}=0.2$, $q=0.5$, equal sample sizes $n=60,\;100,\;160$. 
		\label{BiasTauLOR1q05piC02_unequal_sample_sizes}}
\end{figure}

\begin{figure}[t]
	\centering
	\includegraphics[scale=0.33]{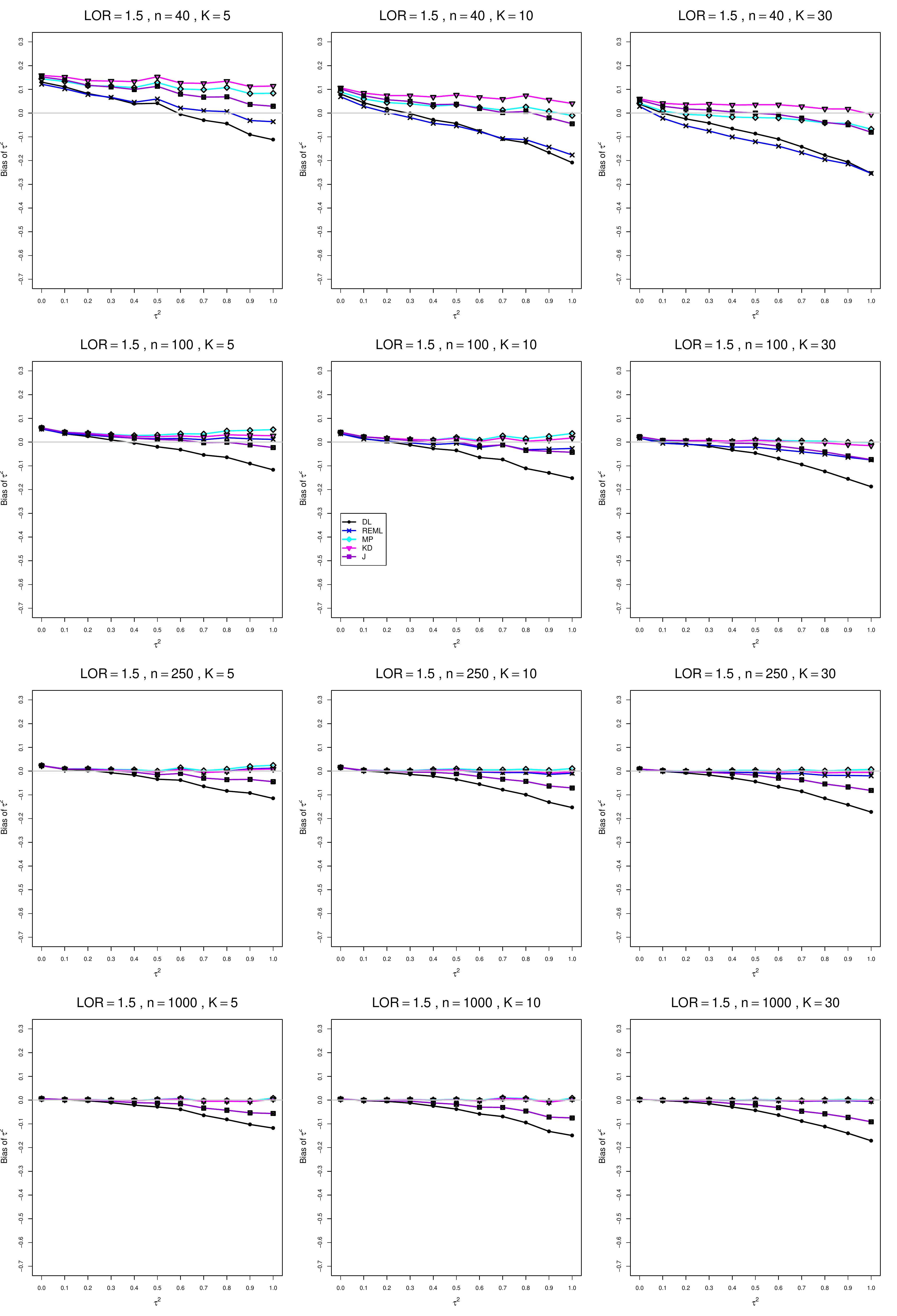}
	\caption{Bias of the estimation of  between-studies variance $\tau^2$ for $\theta=1.5$, $p_{iC}=0.2$, $q=0.5$, equal sample sizes $n=40,\;100,\;250,\;1000$. 
		\label{BiasTauLOR15q05piC02}}
\end{figure}
\begin{figure}[t]
	\centering
	\includegraphics[scale=0.33]{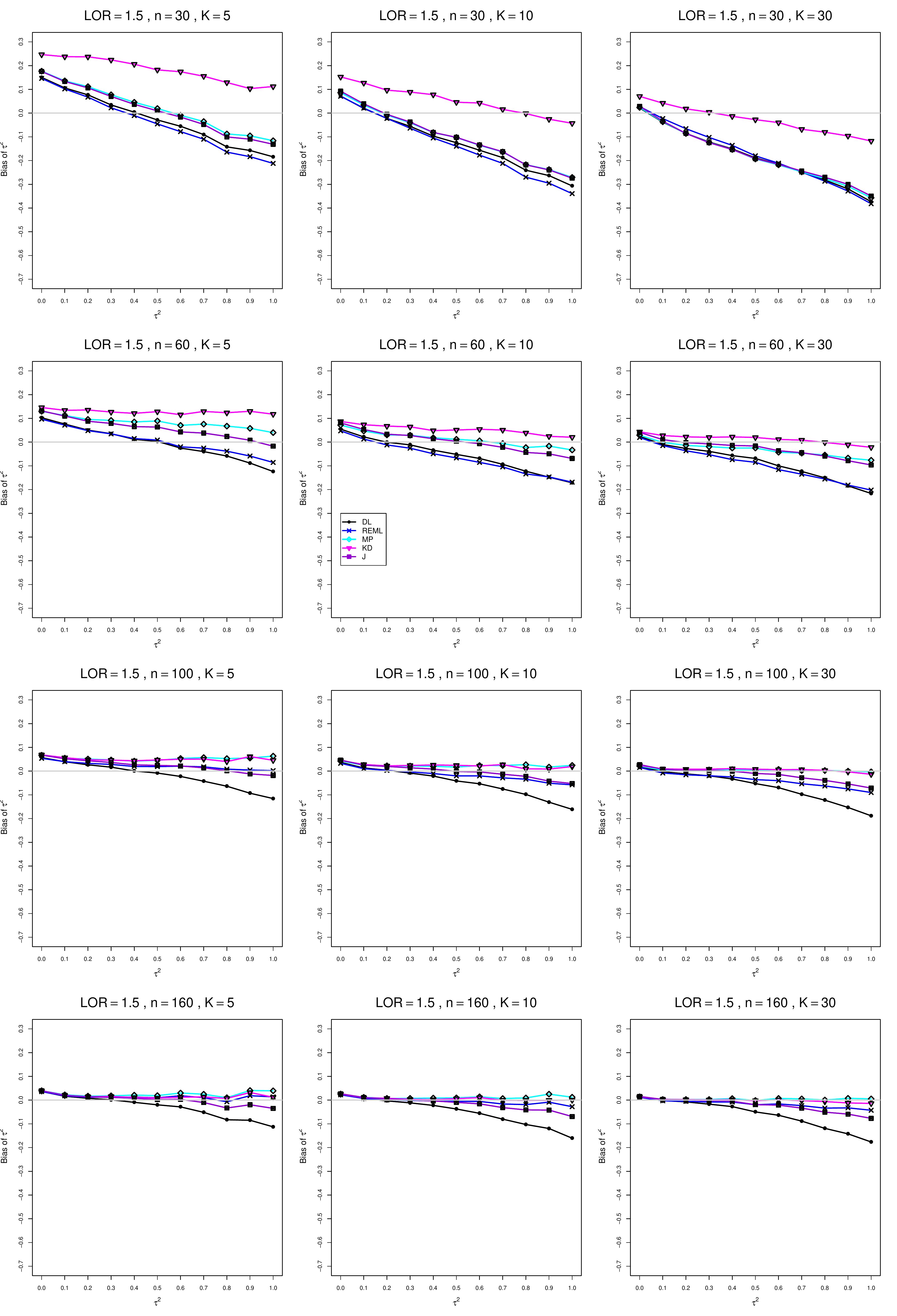}
	\caption{Bias of the estimation of  between-studies variance $\tau^2$ for $\theta=1.5$, $p_{iC}=0.2$, $q=0.5$, 
		unequal sample sizes $n=30,\; 60,\;100,\;160$. 
		\label{BiasTauLOR15q05piC02_unequal_sample_sizes}}
\end{figure}

\begin{figure}[t]
	\centering
	\includegraphics[scale=0.33]{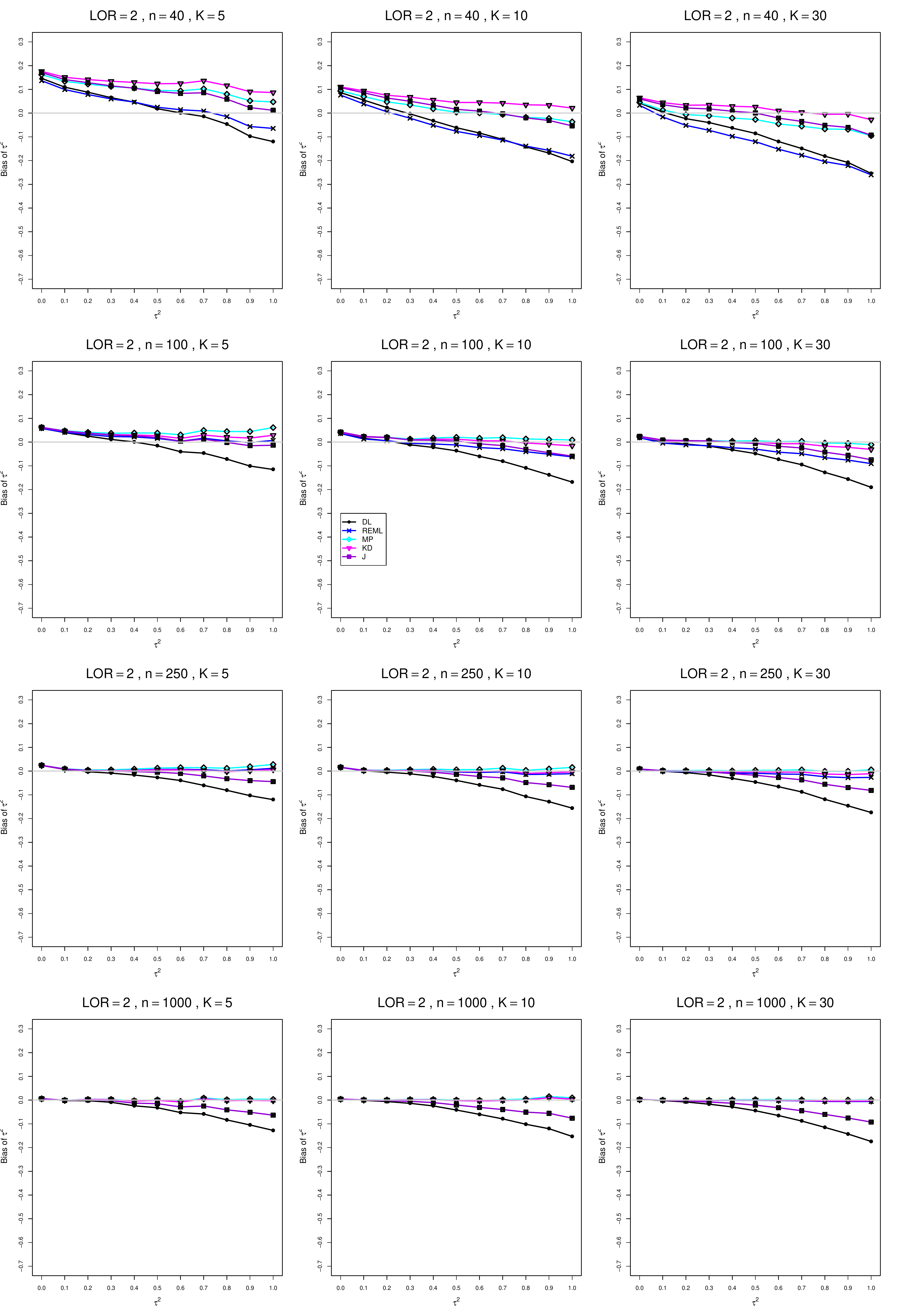}
	\caption{Bias of the estimation of  between-studies variance $\tau^2$ for $\theta=2$, $p_{iC}=0.2$, $q=0.5$, equal sample sizes $n=40,\;100,\;250,\;1000$. 
		\label{BiasTauLOR2q05piC02}}
\end{figure}
\begin{figure}[t]
	\centering
	\includegraphics[scale=0.33]{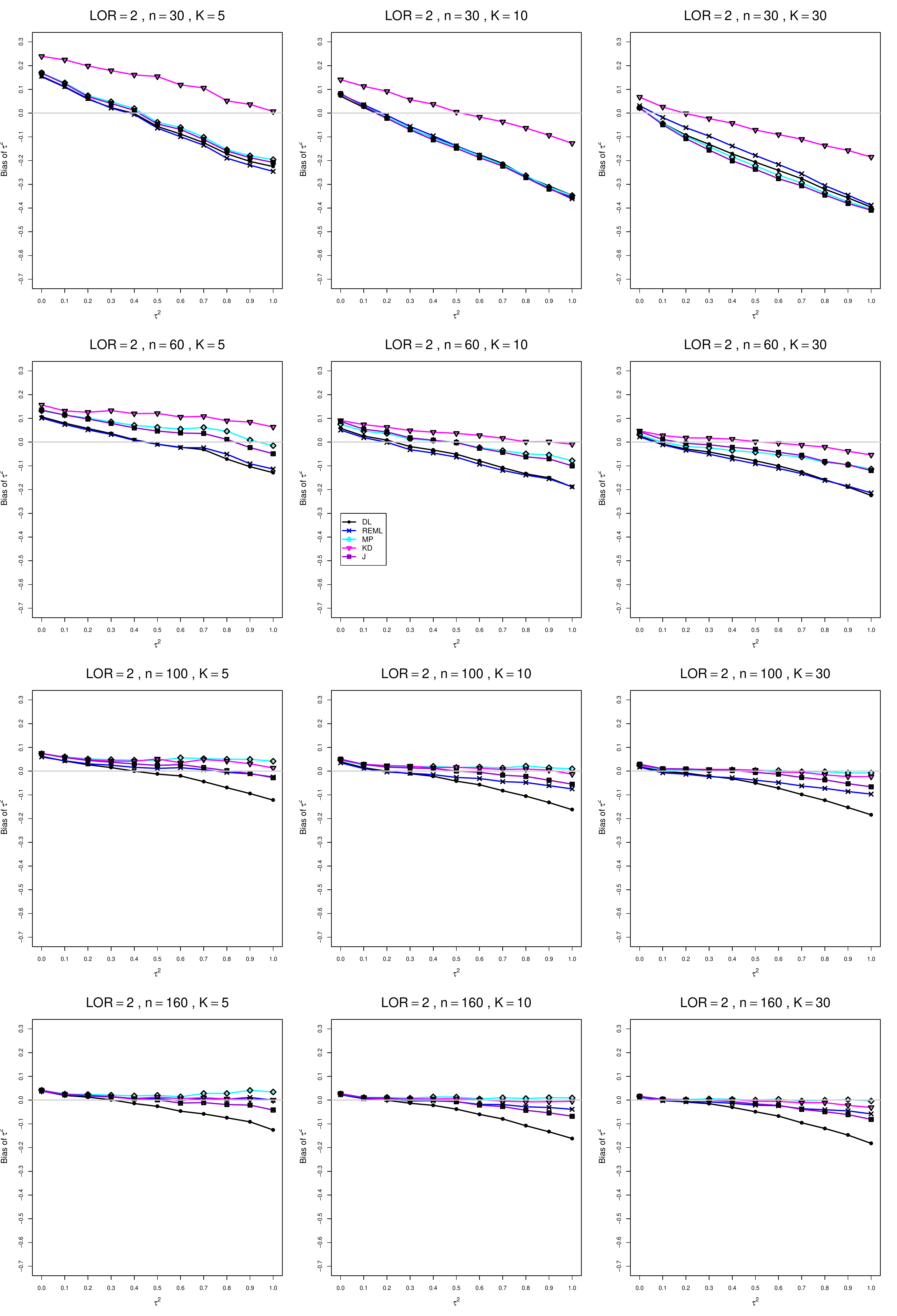}
	\caption{Bias of the estimation of  between-studies variance $\tau^2$ for $\theta=2$, $p_{iC}=0.2$, $q=0.5$, 
		unequal sample sizes $n=30,\; 60,\;100,\;160$. 
		\label{BiasTauLOR2q05piC02_unequal_sample_sizes}}
\end{figure}


\begin{figure}[t]
	\centering
	\includegraphics[scale=0.33]{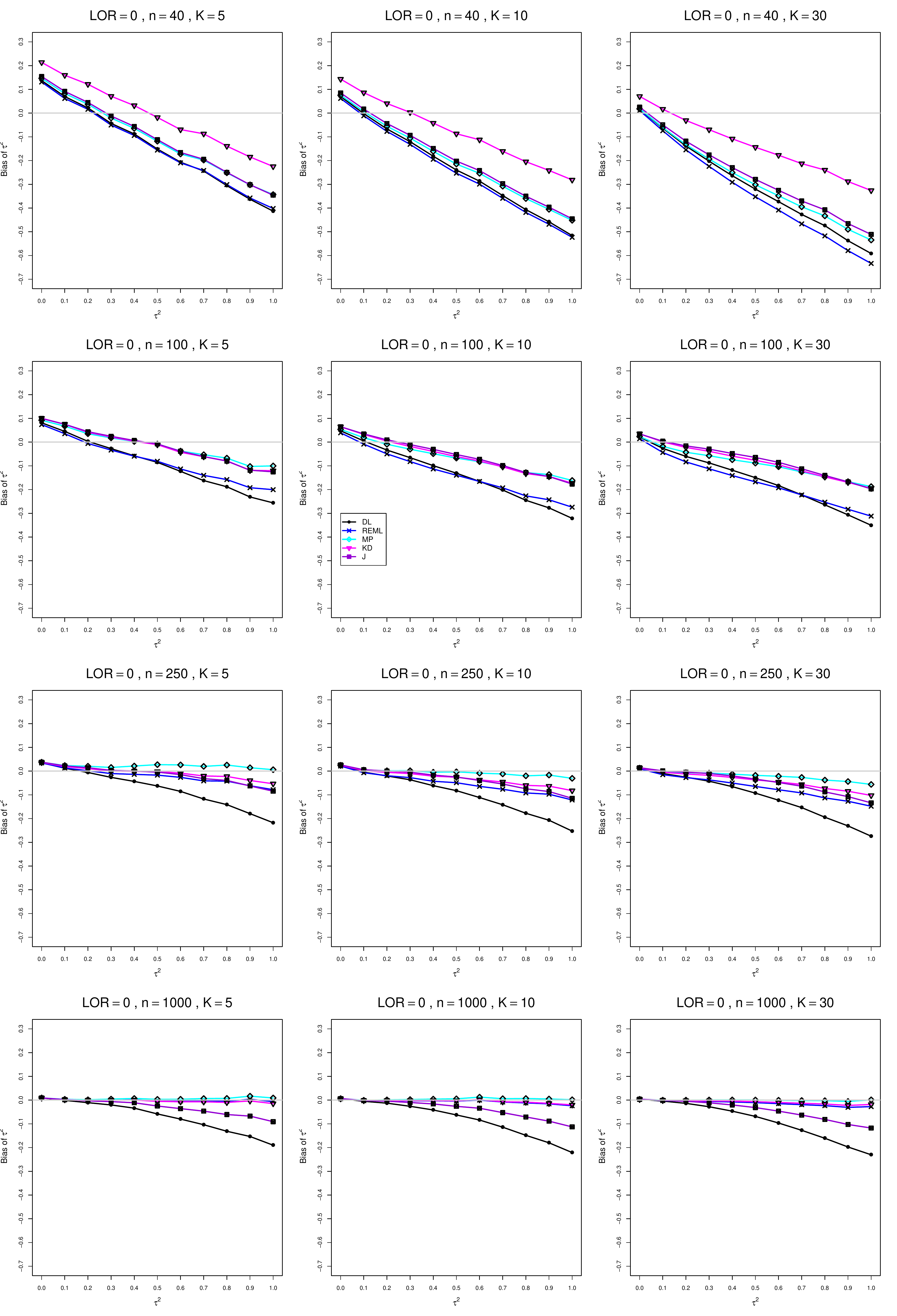}
	\caption{Bias of the estimation of  between-studies variance $\tau^2$ for $\theta=0$, $p_{iC}=0.2$, $q=0.75$, equal sample sizes $n=40,\;100,\;250,\;1000$. 
		\label{BiasTauLOR0q075piC02}}
\end{figure}

\begin{figure}[t]
	\centering
	\includegraphics[scale=0.33]{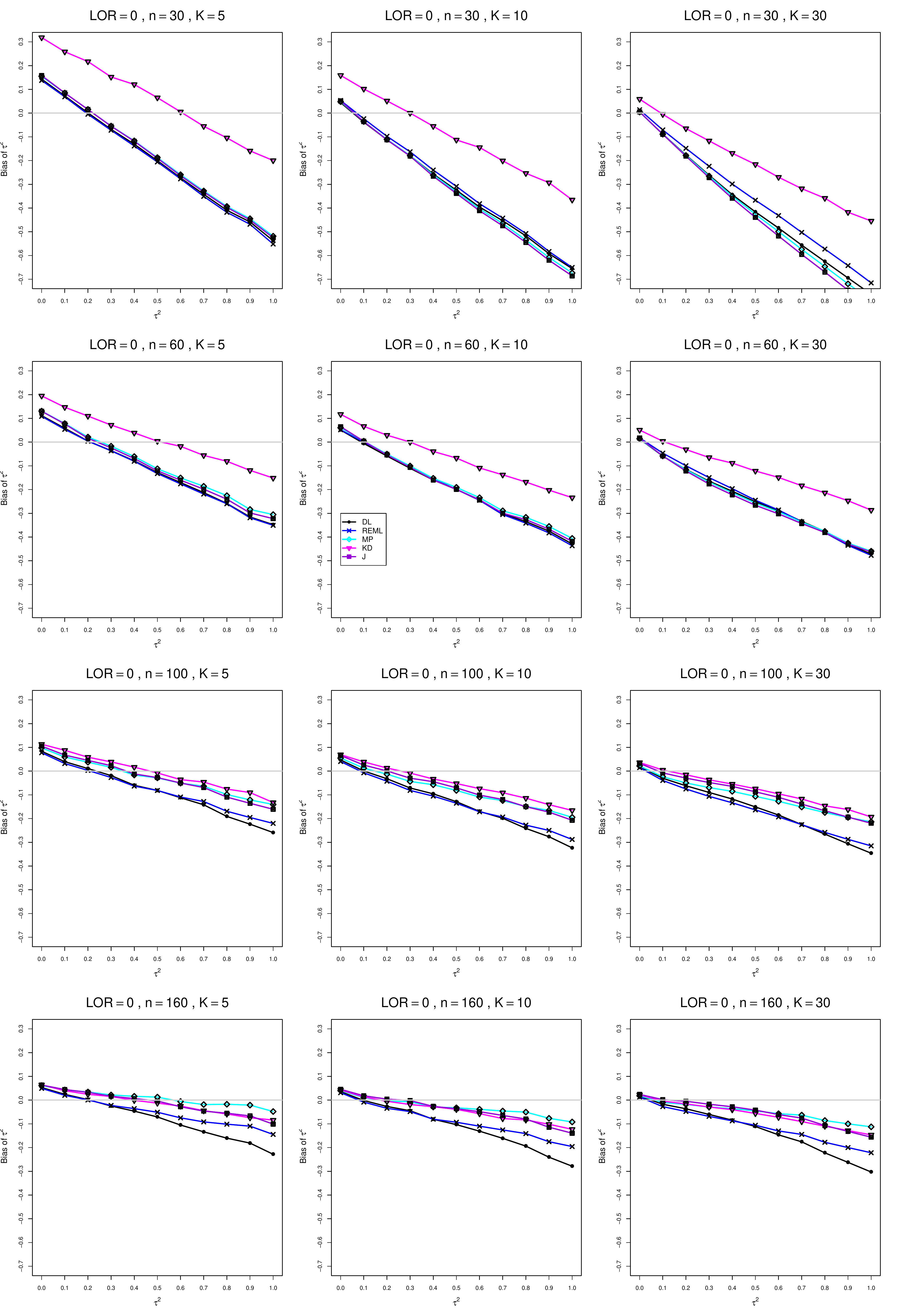}
	\caption{Bias of the estimation of  between-studies variance $\tau^2$ for $\theta=0$, $p_{iC}=0.2$, $q=0.75$, 
		unequal sample sizes $n=30,\; 60,\;100,\;160$. 
		\label{BiasTauLOR0q075piC02_unequal_sample_sizes}}
\end{figure}

\begin{figure}[t]
	\centering
	\includegraphics[scale=0.33]{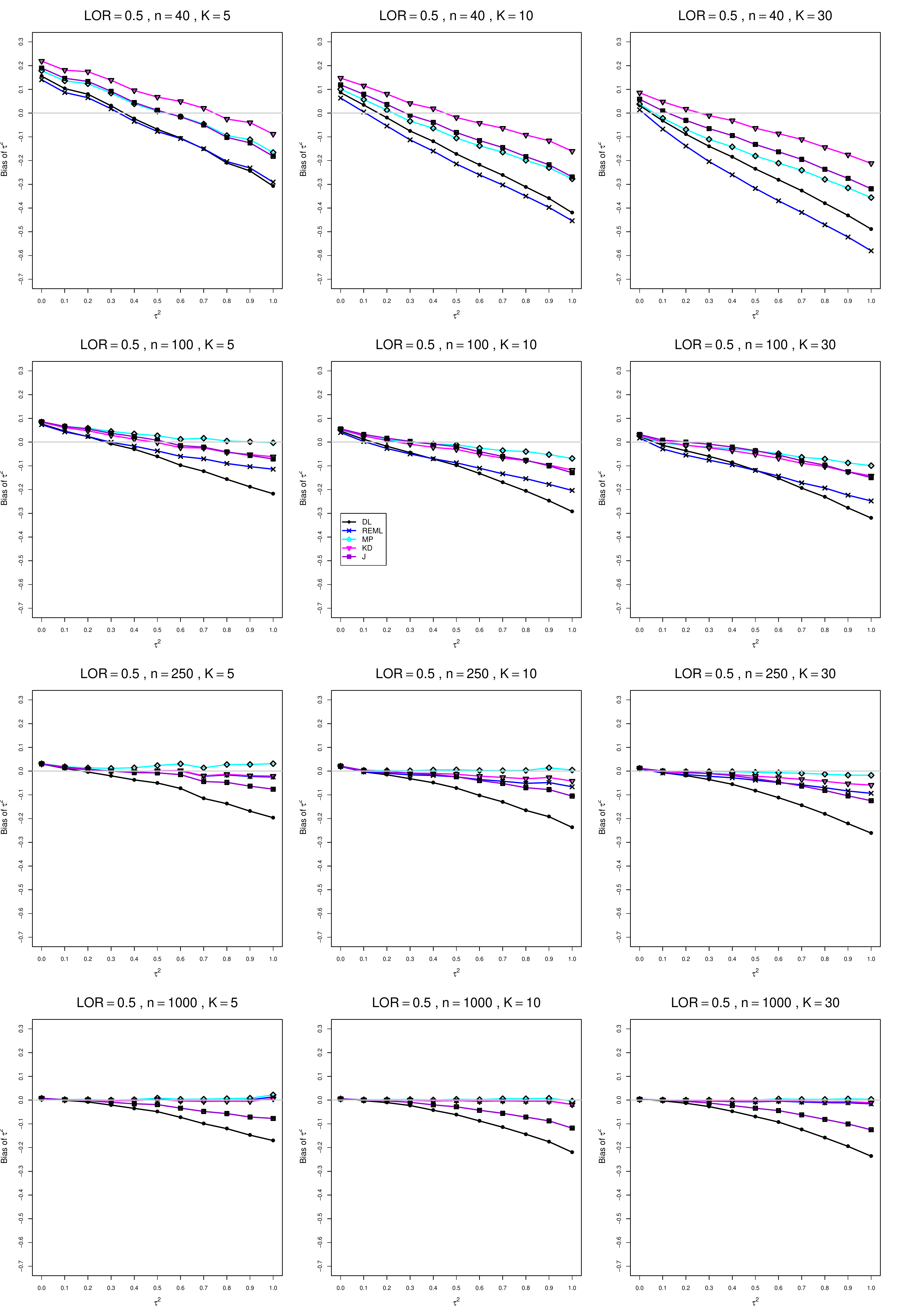}
	\caption{Bias of the estimation of  between-studies variance $\tau^2$ for $\theta=0.5$, $p_{iC}=0.2$, $q=0.75$, equal sample sizes $n=40,\;100,\;250,\;1000$. 
		\label{BiasTauLOR05q075piC02}}
\end{figure}

\begin{figure}[t]
	\centering
	\includegraphics[scale=0.33]{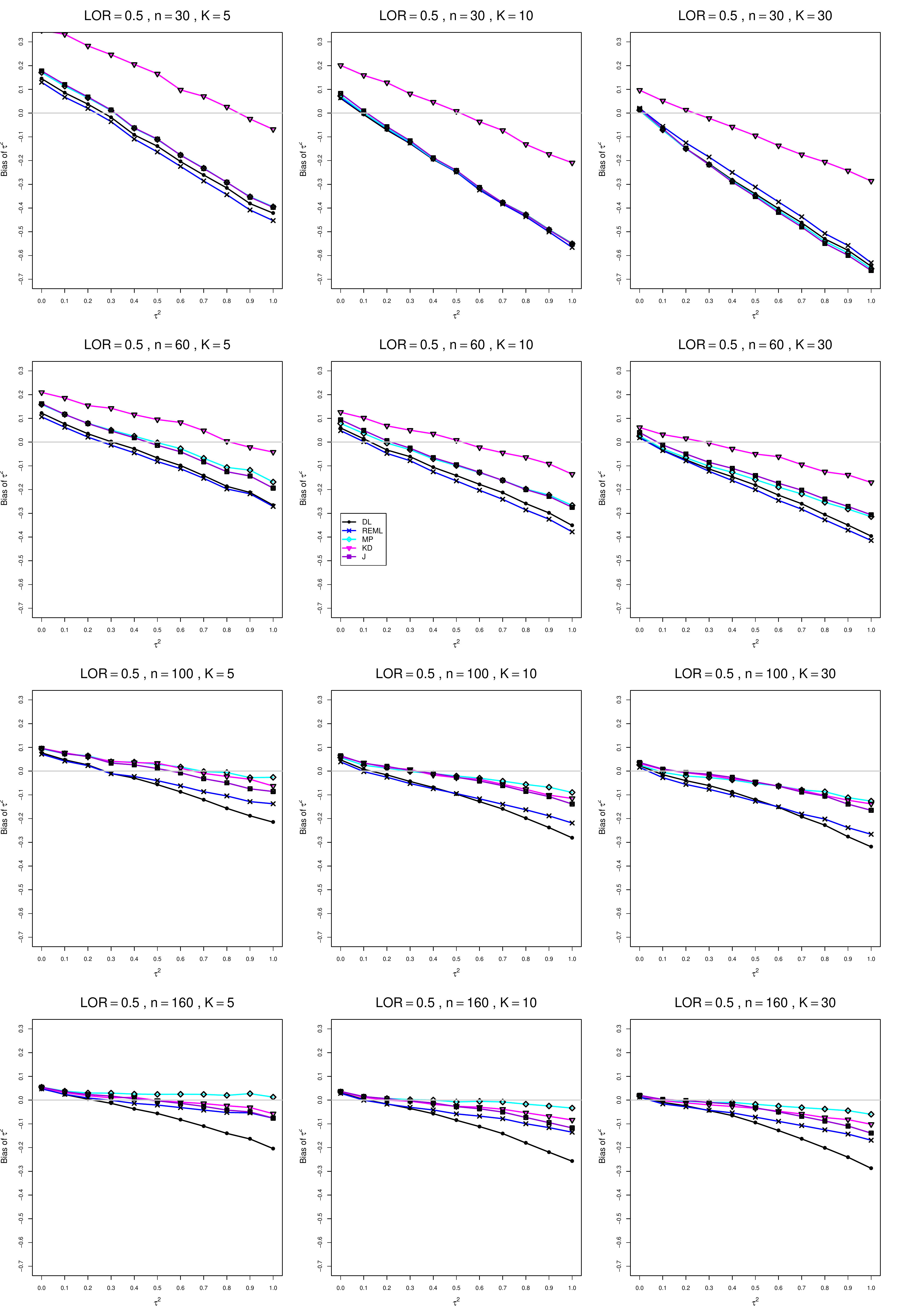}
	\caption{Bias of the estimation of  between-studies variance $\tau^2$ for $\theta=0.5$, $p_{iC}=0.2$, $q=0.75$,
		unequal sample sizes $n=30,\; 60,\;100,\;160$. 
		\label{BiasTauLOR05q075piC02_unequal_sample_sizes}}
\end{figure}
\clearpage

\begin{figure}[t]
	\centering
	\includegraphics[scale=0.33]{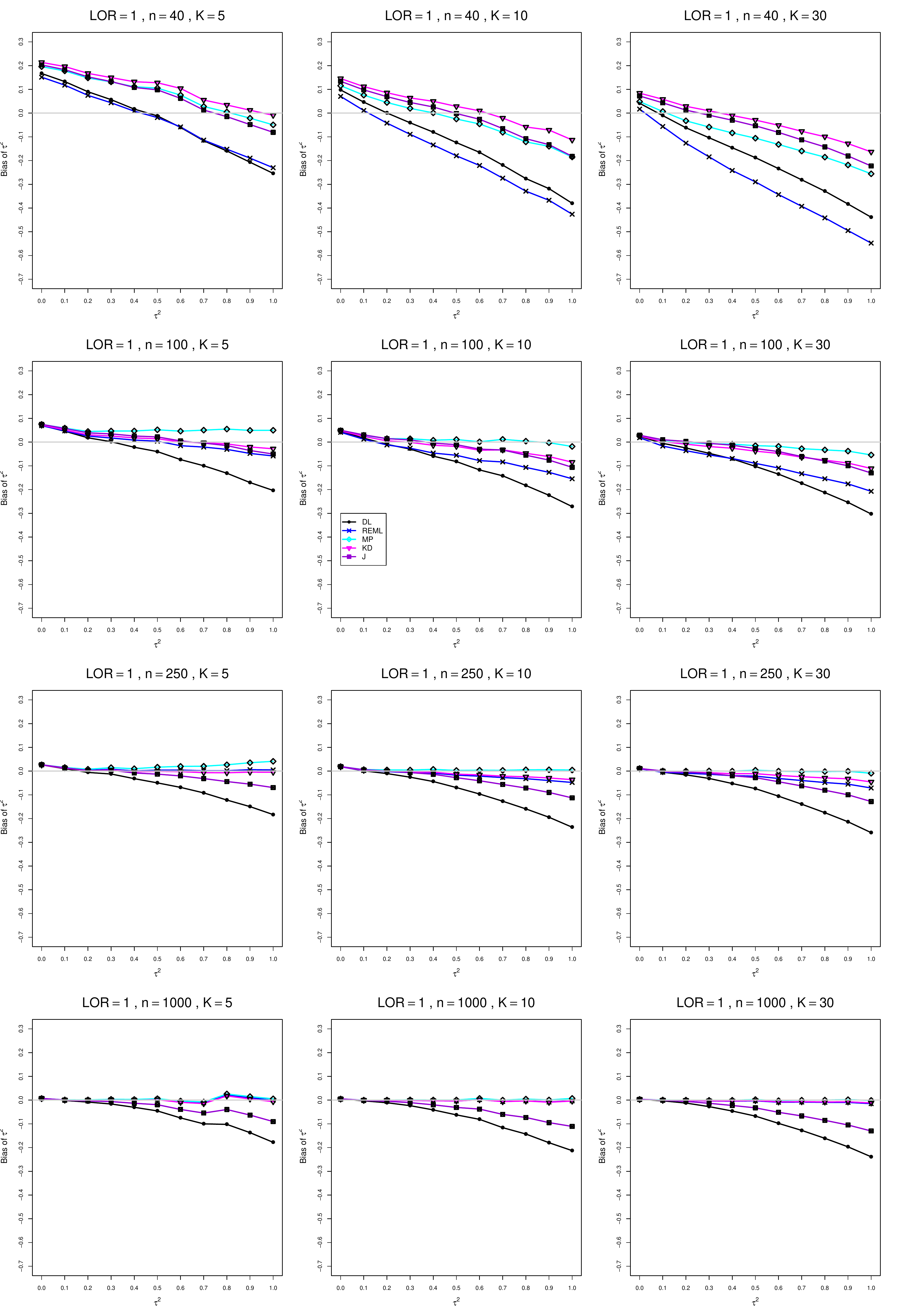}
	\caption{Bias of the estimation of  between-studies variance $\tau^2$ for $\theta=1$, $p_{iC}=0.2$, $q=0.75$, equal sample sizes $n=40,\;100,\;250,\;1000$. 
		\label{BiasTauLOR1q075piC02}}
\end{figure}

\begin{figure}[t]
	\centering
	\includegraphics[scale=0.33]{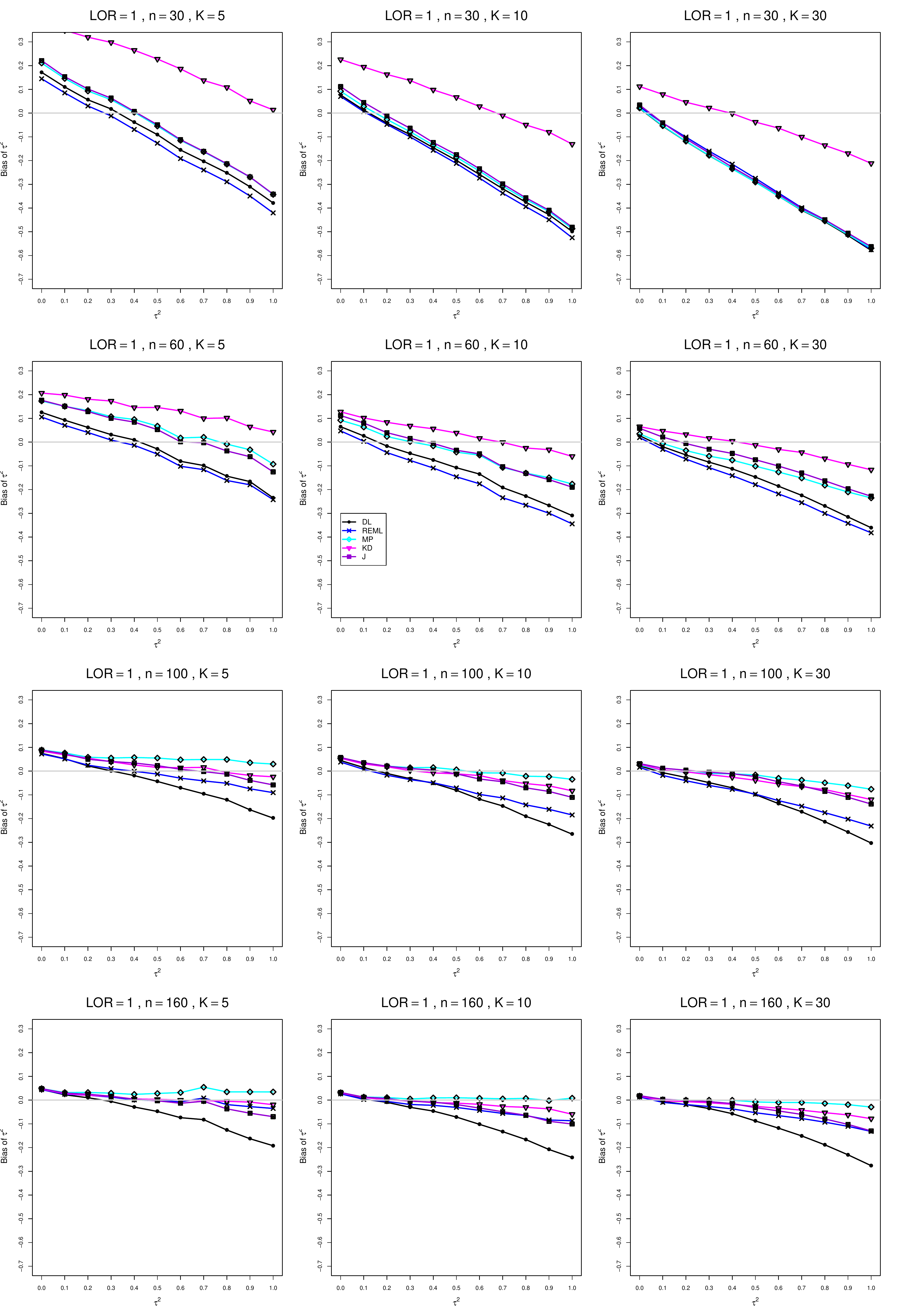}
	\caption{Bias of the estimation of  between-studies variance $\tau^2$ for $\theta=1$, $p_{iC}=0.2$, $q=0.75$, 
		unequal sample sizes $n=30,\; 60,\;100,\;160$. 
		\label{BiasTauLOR1q075piC02_unequal_sample_sizes}}
\end{figure}

\begin{figure}[t]
	\centering
	\includegraphics[scale=0.33]{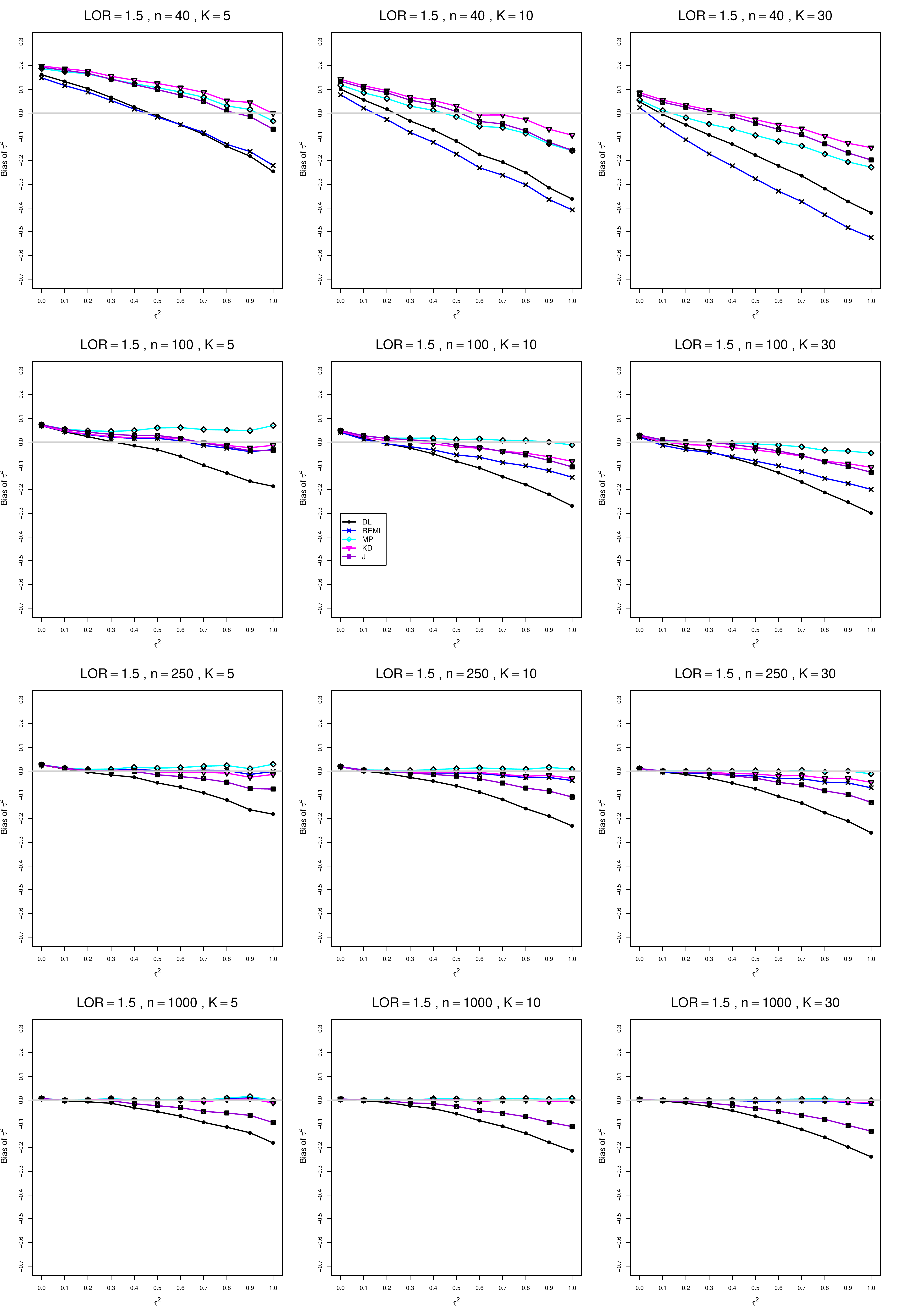}
	\caption{Bias of the estimation of  between-studies variance $\tau^2$ for $\theta=1.5$, $p_{iC}=0.2$, $q=0.75$, equal sample sizes $n=40,\;100,\;250,\;1000$. 
		\label{BiasTauLOR15q075piC02}}
\end{figure}

\begin{figure}[t]
	\centering
	\includegraphics[scale=0.33]{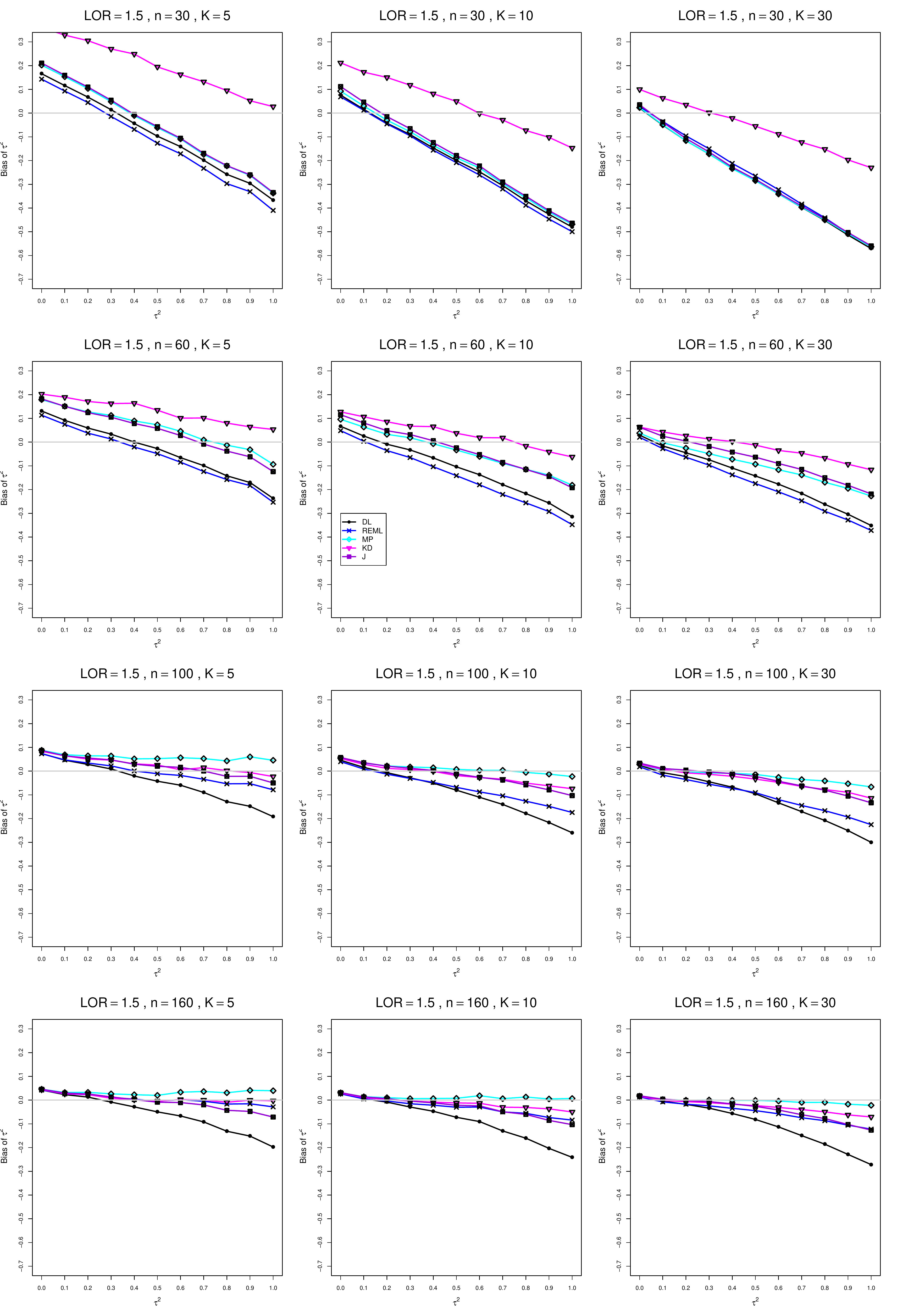}
	\caption{Bias of the estimation of  between-studies variance $\tau^2$ for $\theta=1.5$, $p_{iC}=0.2$, $q=0.75$, 
		unequal sample sizes $n=30,\; 60,\;100,\;160$. 
		\label{BiasTauLOR15q075piC02_unequal_sample_sizes}}
\end{figure}

\begin{figure}[t]
	\centering
	\includegraphics[scale=0.33]{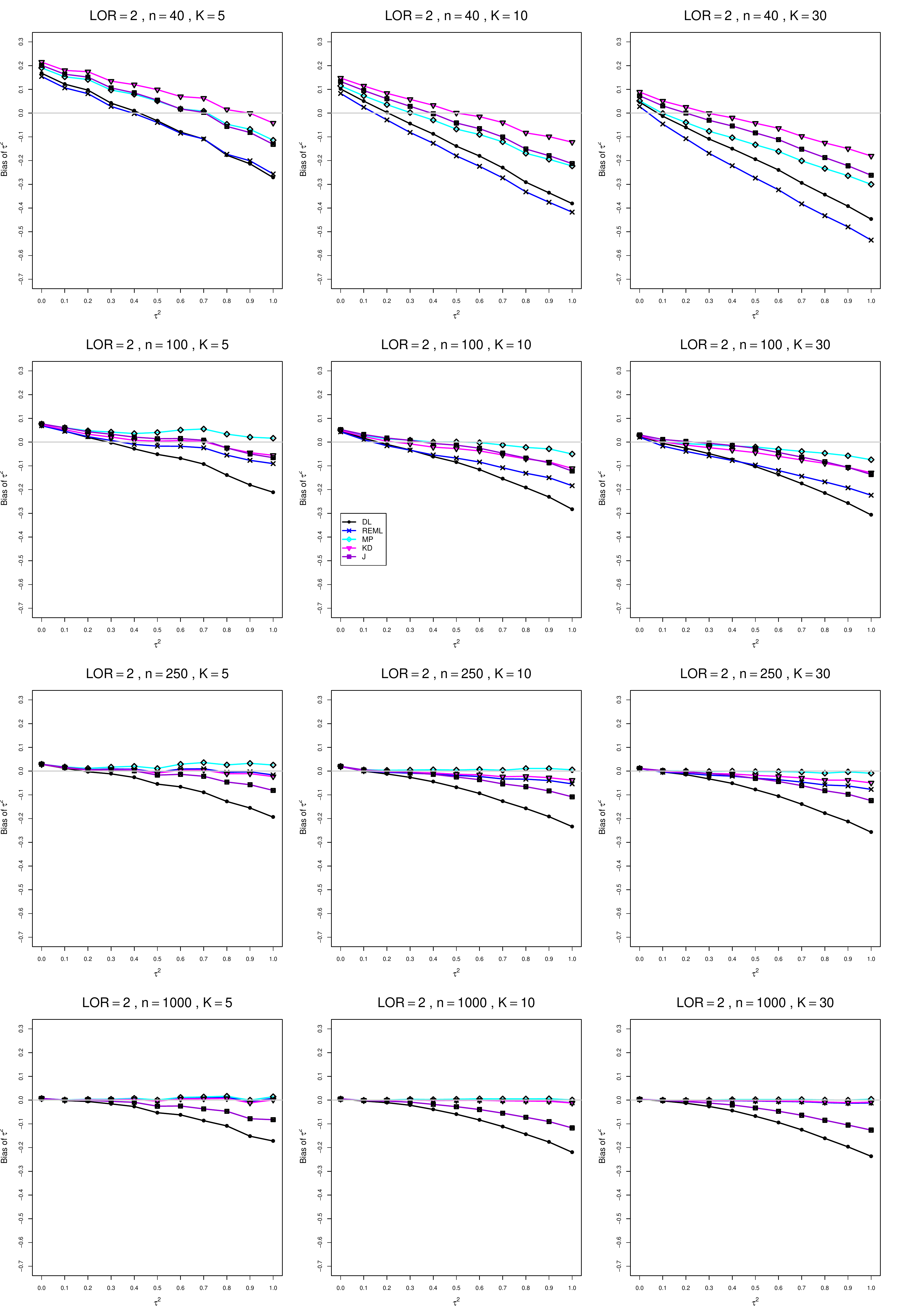}
	\caption{Bias of the estimation of  between-studies variance $\tau^2$ for $\theta=2$, $p_{iC}=0.2$, $q=0.75$, equal sample sizes $n=40,\;100,\;250,\;1000$. 
		\label{BiasTauLOR2q075piC02}}
\end{figure}

\begin{figure}[t]
	\centering
	\includegraphics[scale=0.33]{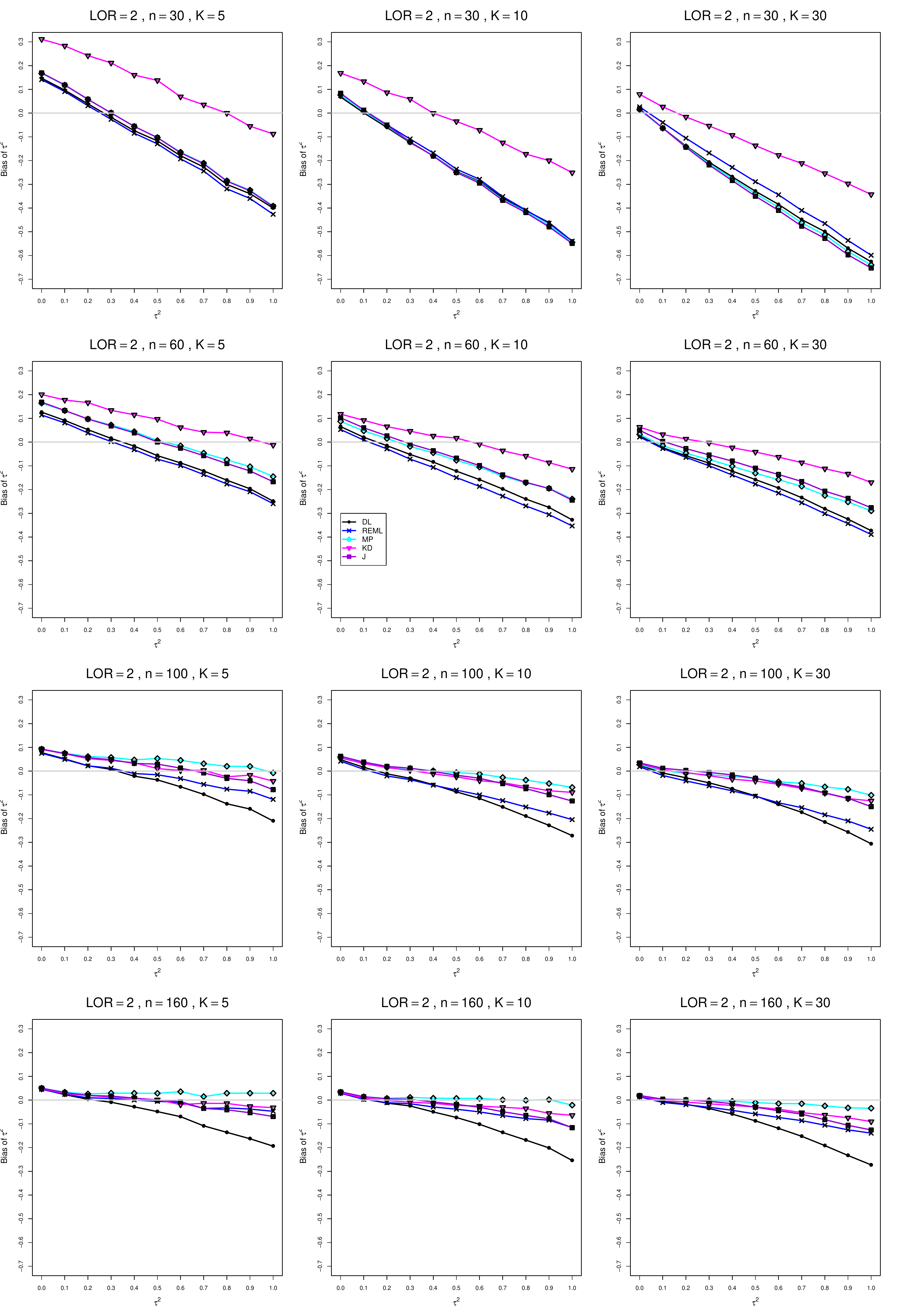}
	\caption{Bias of the estimation of  between-studies variance $\tau^2$ for $\theta=2$, $p_{iC}=0.2$, $q=0.75$, 
		unequal sample sizes $n=30,\; 60,\;100,\;160$. 
		\label{BiasTauLOR2q075piC02_unequal_sample_sizes}}
\end{figure}
\clearpage
\renewcommand{\thefigure}{A1.3.\arabic{figure}}
\setcounter{figure}{0}
\subsection*{A1.3 Probability in the control arm $p_{C}=0.4$}
\begin{figure}[t]
	\centering
	\includegraphics[scale=0.33]{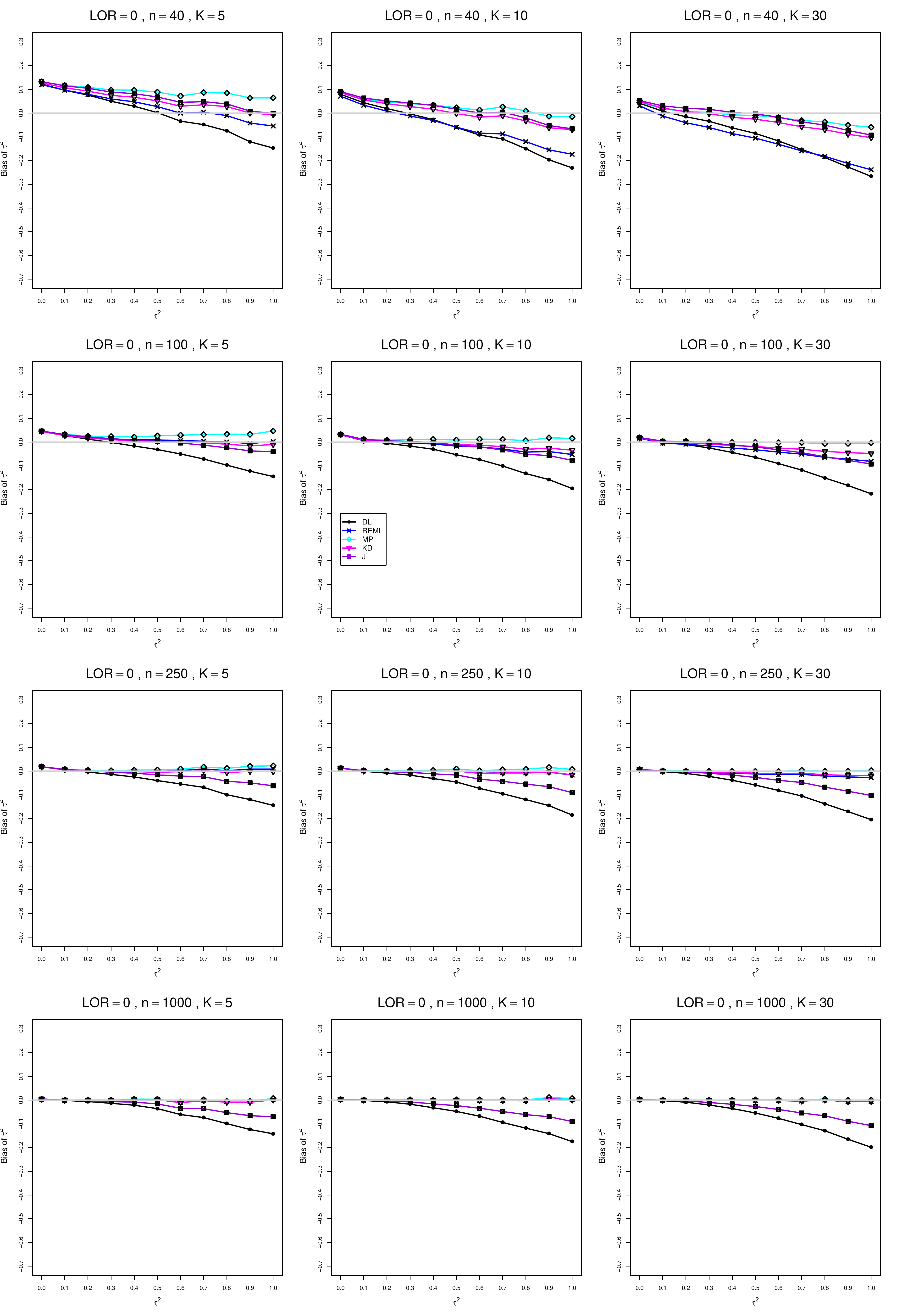}
	\caption{Bias of the estimation of  between-studies variance $\tau^2$ for $\theta=0$, $p_{iC}=0.4$, $q=0.5$, equal sample sizes $n=40,\;100,\;250,\;1000$. 
		\label{BiasTauLOR0q05piC04}}
\end{figure}

\begin{figure}[t]
	\centering
	\includegraphics[scale=0.33]{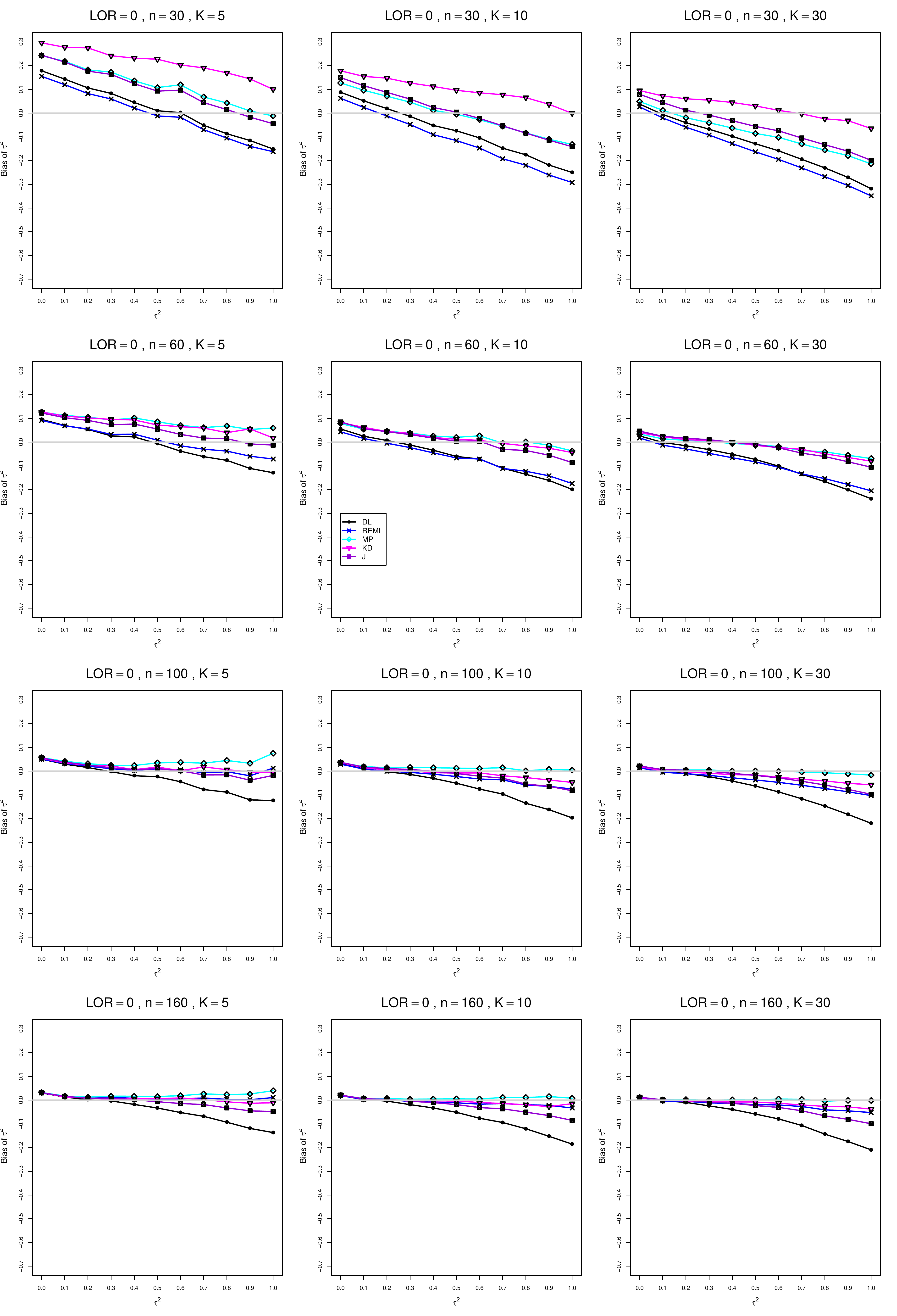}
	\caption{Bias of the estimation of  between-studies variance $\tau^2$ for $\theta=0$, $p_{iC}=0.4$, $q=0.5$, 
		unequal sample sizes $n=30,\; 60,\;100,\;160$. 
		\label{BiasTauLOR0q05piC04_unequal_sample_sizes}}
\end{figure}

\begin{figure}[t]
	\centering
	\includegraphics[scale=0.33]{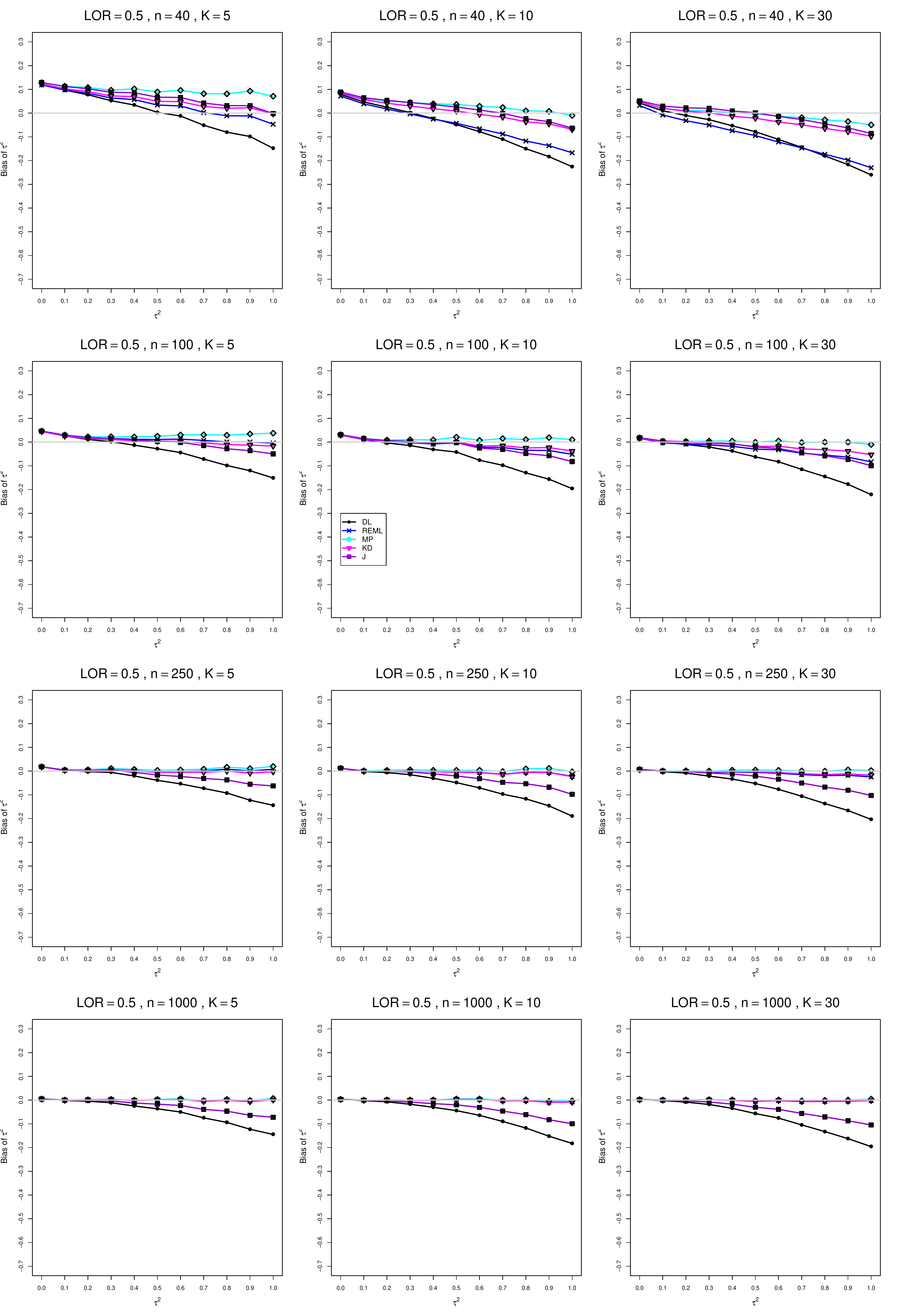}
	\caption{Bias of the estimation of  between-studies variance $\tau^2$ for $\theta=0.5$, $p_{iC}=0.4$, $q=0.5$, equal sample sizes $n=40,\;100,\;250,\;1000$. 
		\label{BiasTauLOR05q05piC04}}
\end{figure}

\begin{figure}[t]
	\centering
	\includegraphics[scale=0.33]{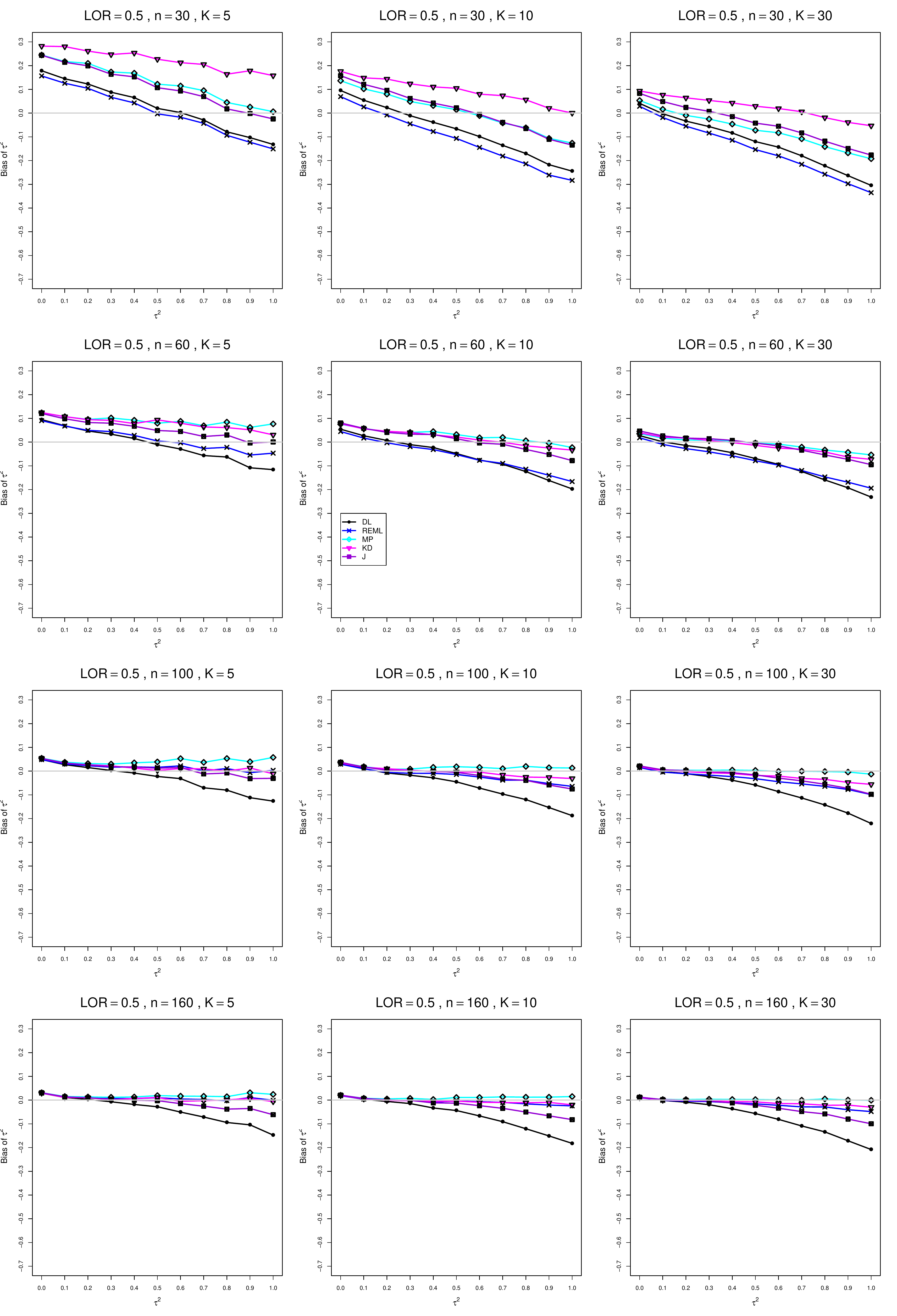}
	\caption{Bias of the estimation of  between-studies variance $\tau^2$ for $\theta=0.5$, $p_{iC}=0.4$, $q=0.5$, 
		unequal sample sizes $n=30,\; 60,\;100,\;160$. 
		\label{BiasTauLOR05q05piC04_unequal_sample_sizes}}
\end{figure}

\begin{figure}[t]
	\centering
	\includegraphics[scale=0.33]{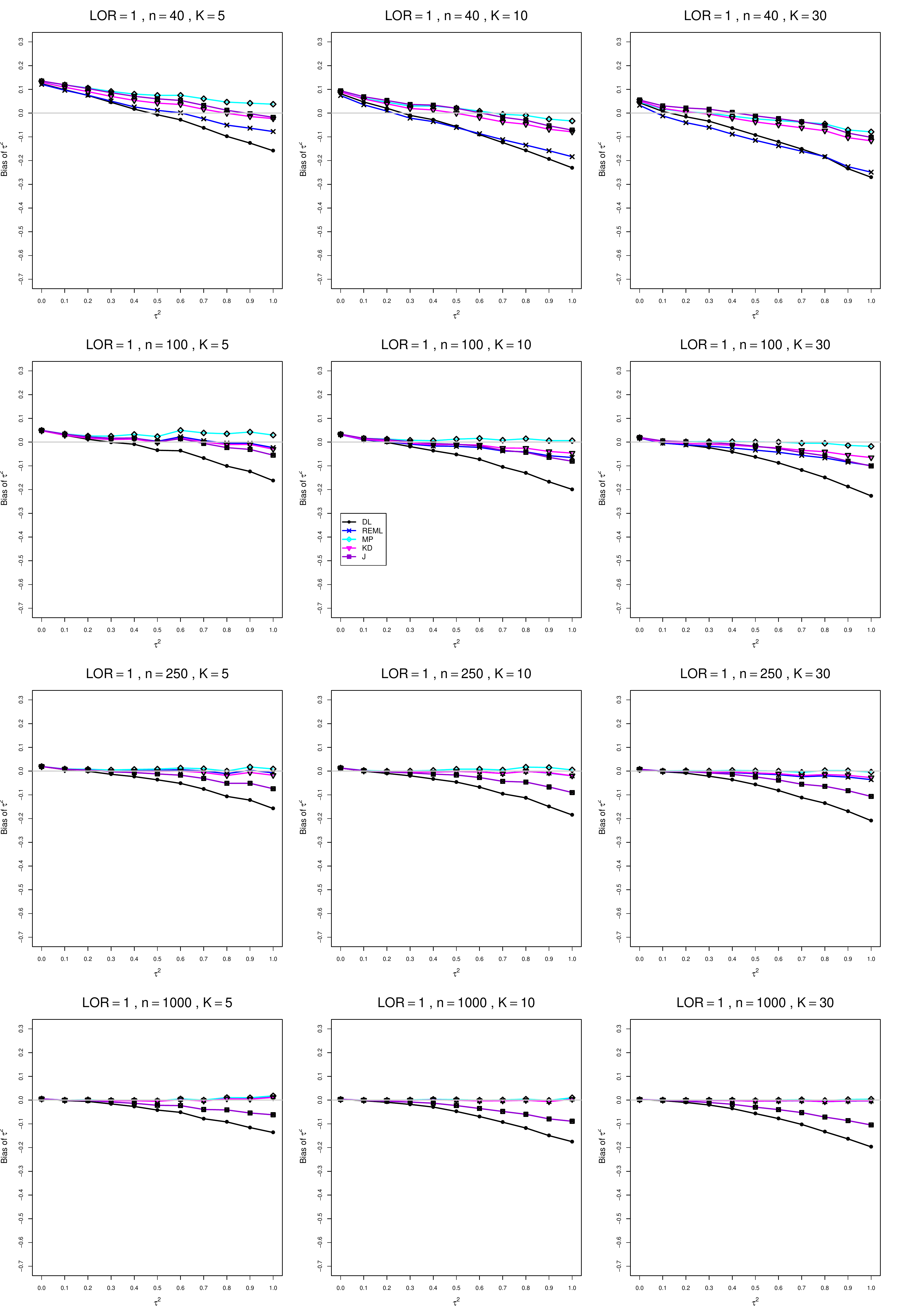}
	\caption{Bias of the estimation of  between-studies variance $\tau^2$ for $\theta=1$, $p_{iC}=0.4$, $q=0.5$, equal sample sizes $n=40,\;100,\;250,\;1000$. 
		\label{BiasTauLOR1q05piC04}}
\end{figure}
\begin{figure}[t]
	\centering
	\includegraphics[scale=0.33]{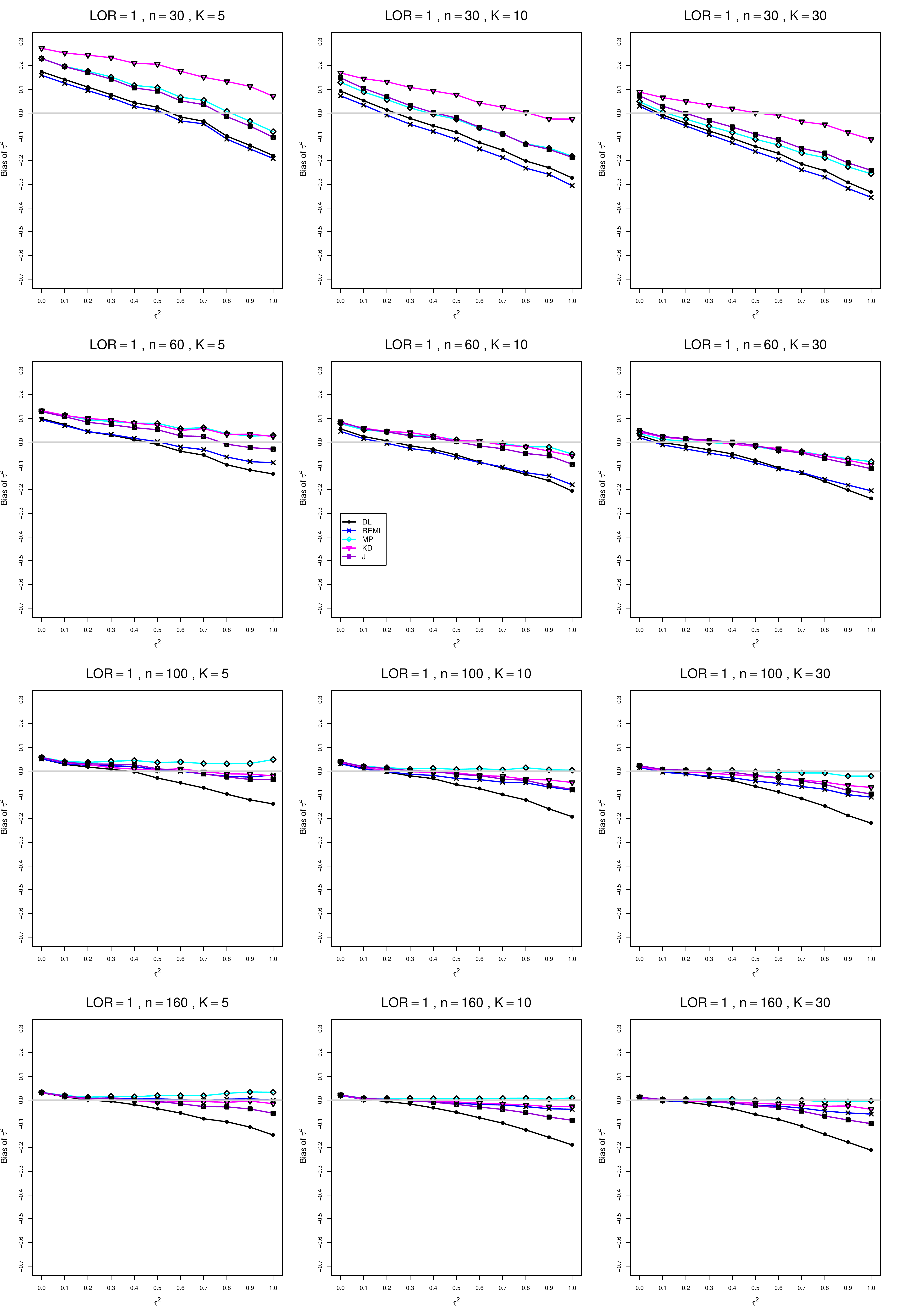}
	\caption{Bias of the estimation of  between-studies variance $\tau^2$ for $\theta=1$, $p_{iC}=0.4$, $q=0.5$, 
		unequal sample sizes $n=30,\; 60,\;100,\;160$. 
		\label{BiasTauLOR1q05piC04_unequal_sample_sizes}}
\end{figure}

\begin{figure}[t]
	\centering
	\includegraphics[scale=0.33]{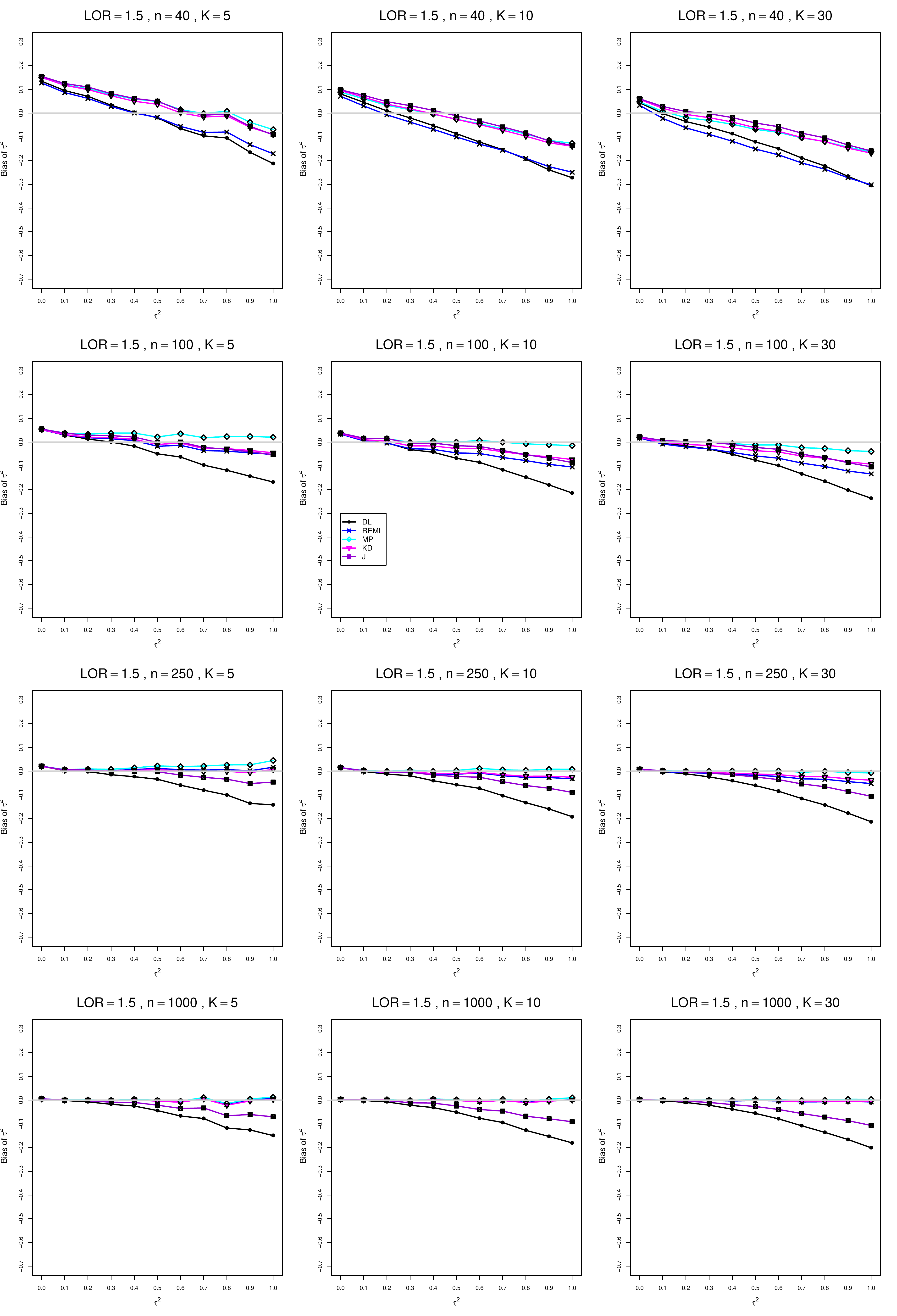}
	\caption{Bias of the estimation of  between-studies variance $\tau^2$ for $\theta=1.5$, $p_{iC}=0.4$, $q=0.5$, equal sample sizes $n=40,\;100,\;250,\;1000$. 
		\label{BiasTauLOR15q05piC04}}
\end{figure}
\begin{figure}[t]
	\centering
	\includegraphics[scale=0.33]{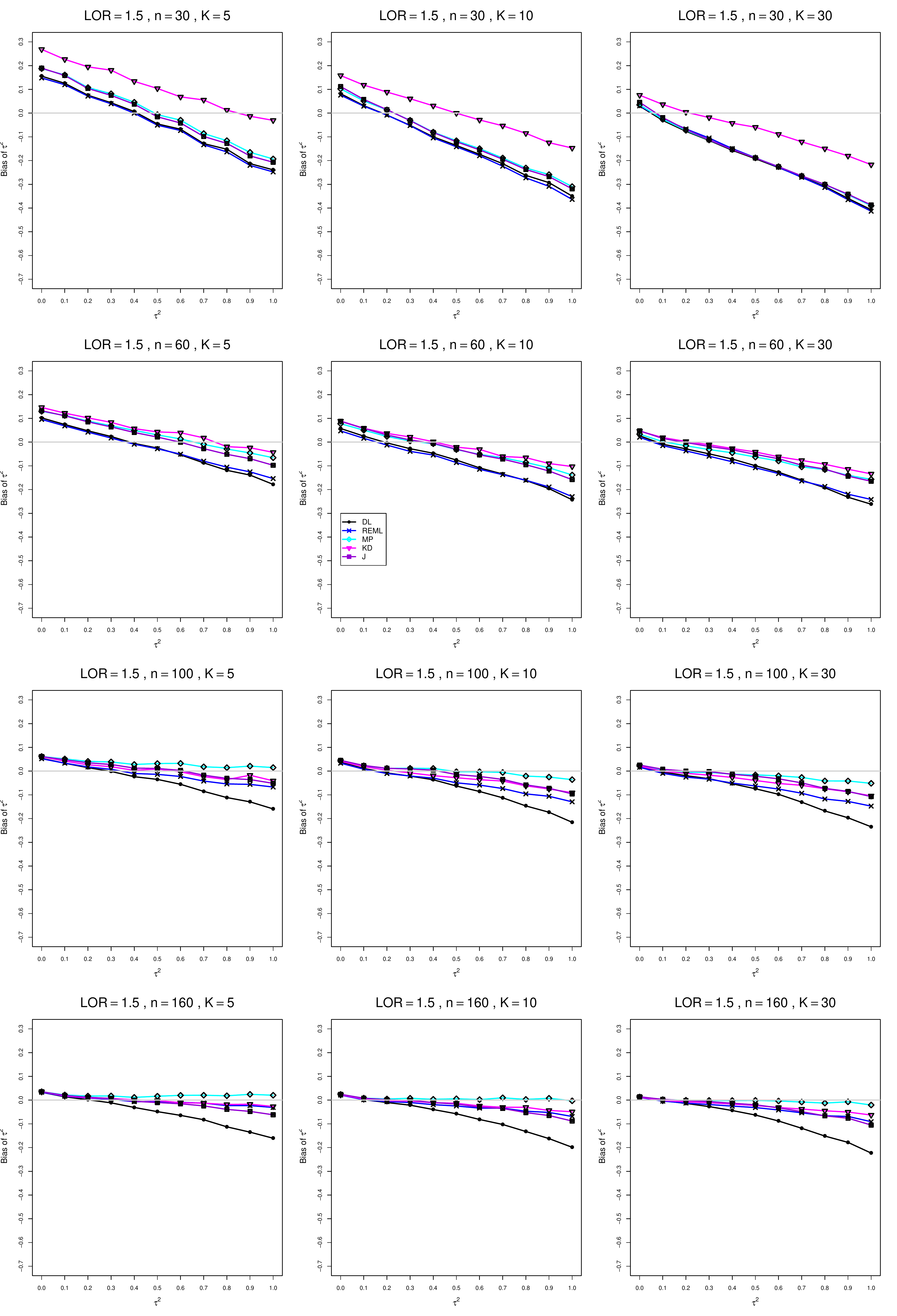}
	\caption{Bias of the estimation of  between-studies variance $\tau^2$ for $\theta=1.5$, $p_{iC}=0.4$, $q=0.5$, 
		unequal sample sizes $n=30,\; 60,\;100,\;160$. 
		\label{BiasTauLOR15q05piC04_unequal_sample_sizes}}
\end{figure}

\begin{figure}[t]
	\centering
	\includegraphics[scale=0.33]{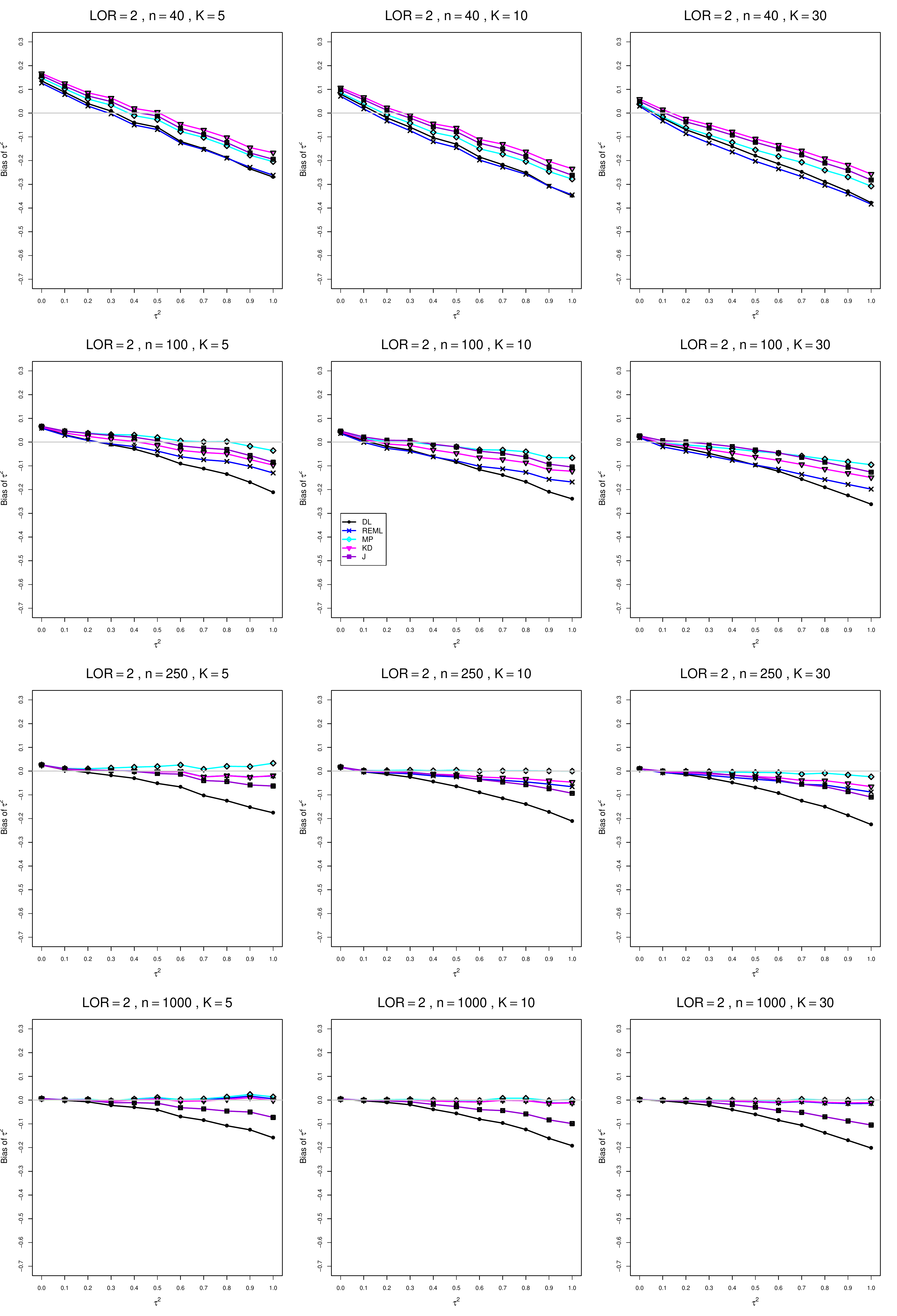}
	\caption{Bias of the estimation of  between-studies variance $\tau^2$ for $\theta=2$, $p_{iC}=0.4$, $q=0.5$, equal sample sizes $n=40,\;100,\;250,\;1000$. 
		\label{BiasTauLOR2q05piC04}}
\end{figure}
\begin{figure}[t]
	\centering
	\includegraphics[scale=0.33]{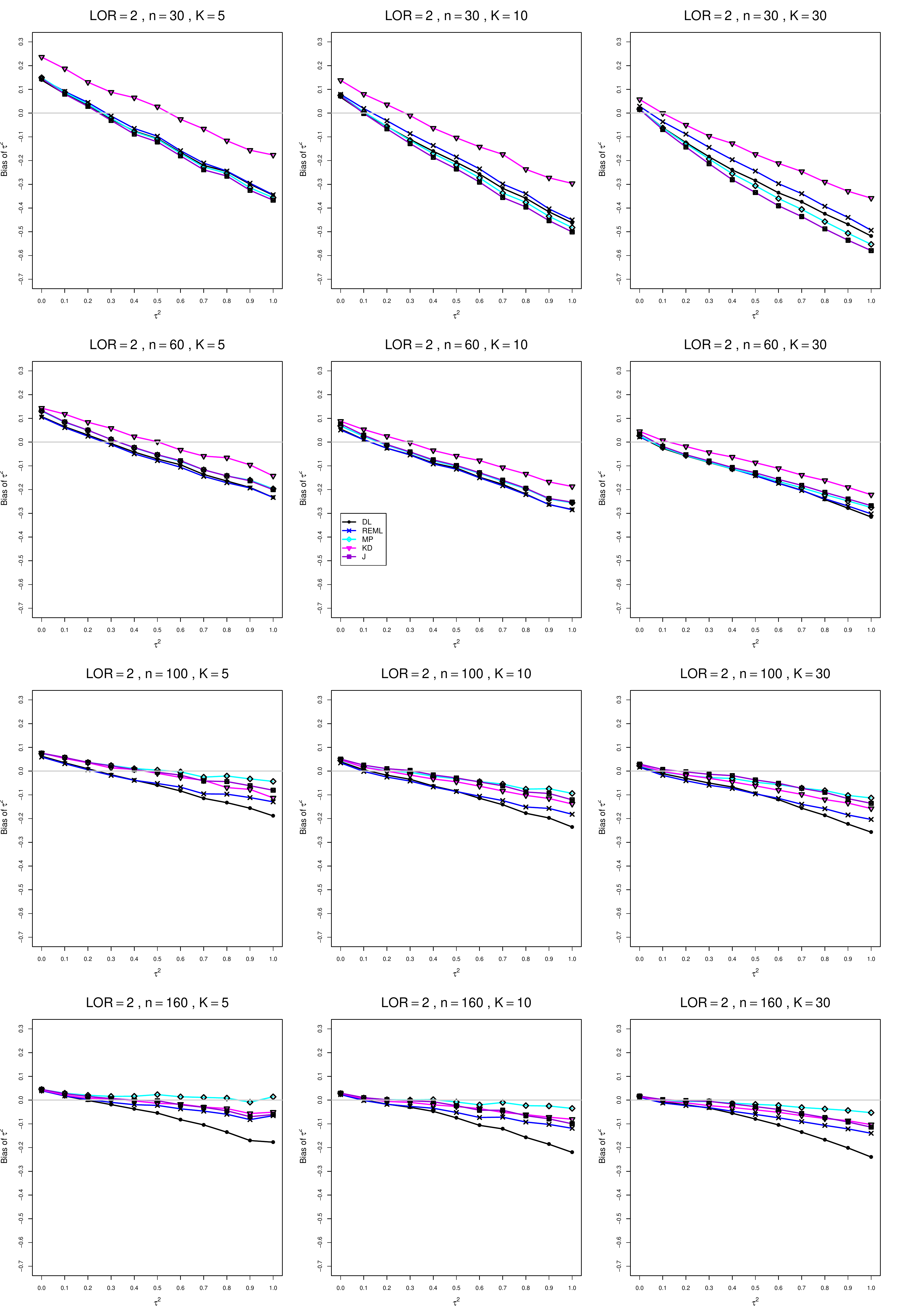}
	\caption{Bias of the estimation of  between-studies variance $\tau^2$ for $\theta=2$, $p_{iC}=0.4$, $q=0.5$, 
		unequal sample sizes $n=30,\; 60,\;100,\;160$. 
		\label{BiasTauLOR2q05piC04_unequal_sample_sizes}}
\end{figure}


\begin{figure}[t]
	\centering
	\includegraphics[scale=0.33]{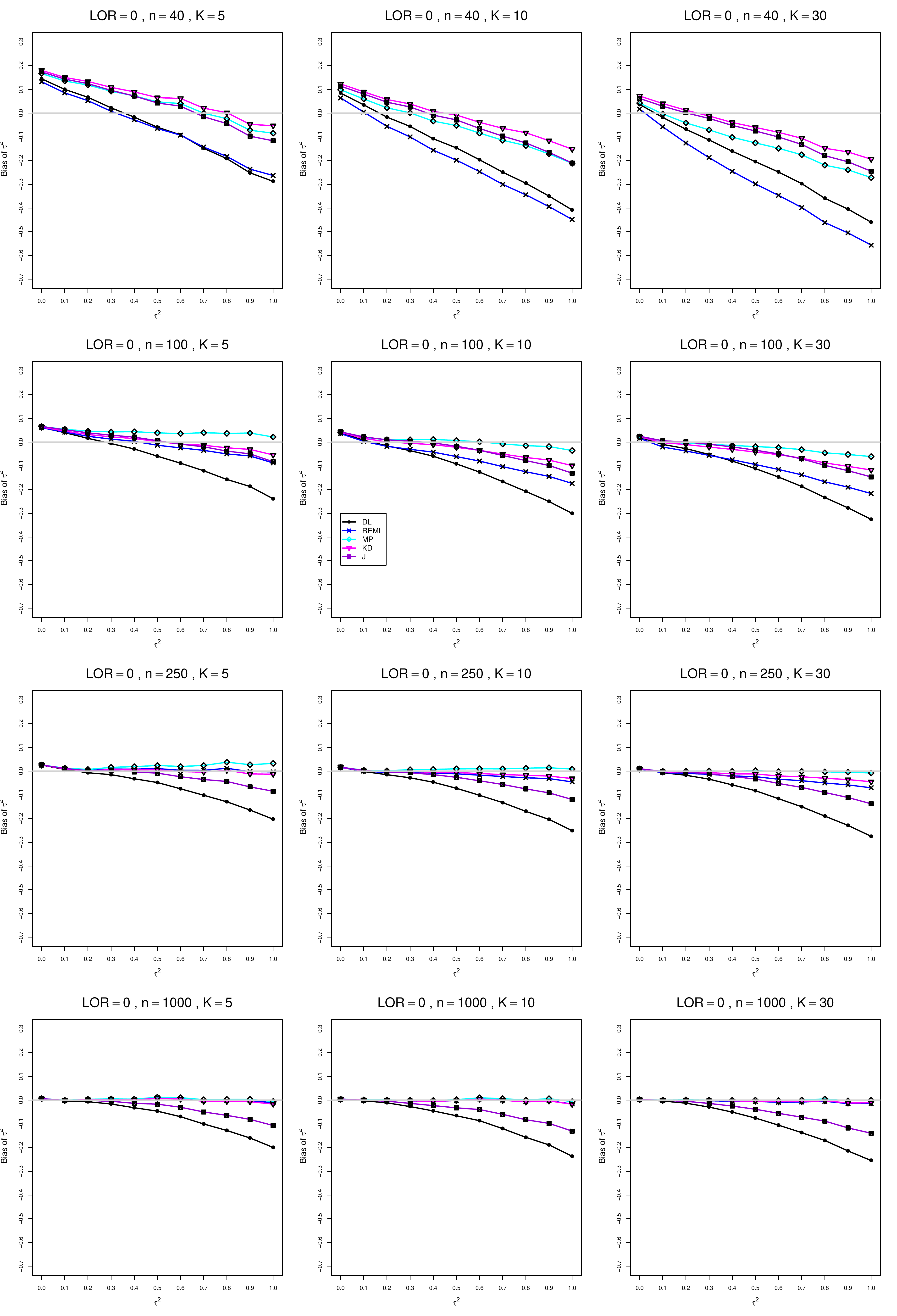}
	\caption{Bias of the estimation of  between-studies variance $\tau^2$ for $\theta=0$, $p_{iC}=0.4$, $q=0.75$, equal sample sizes $n=40,\;100,\;250,\;1000$. 
		\label{BiasTauLOR0q075piC04}}
\end{figure}
\begin{figure}[t]
	\centering
	\includegraphics[scale=0.33]{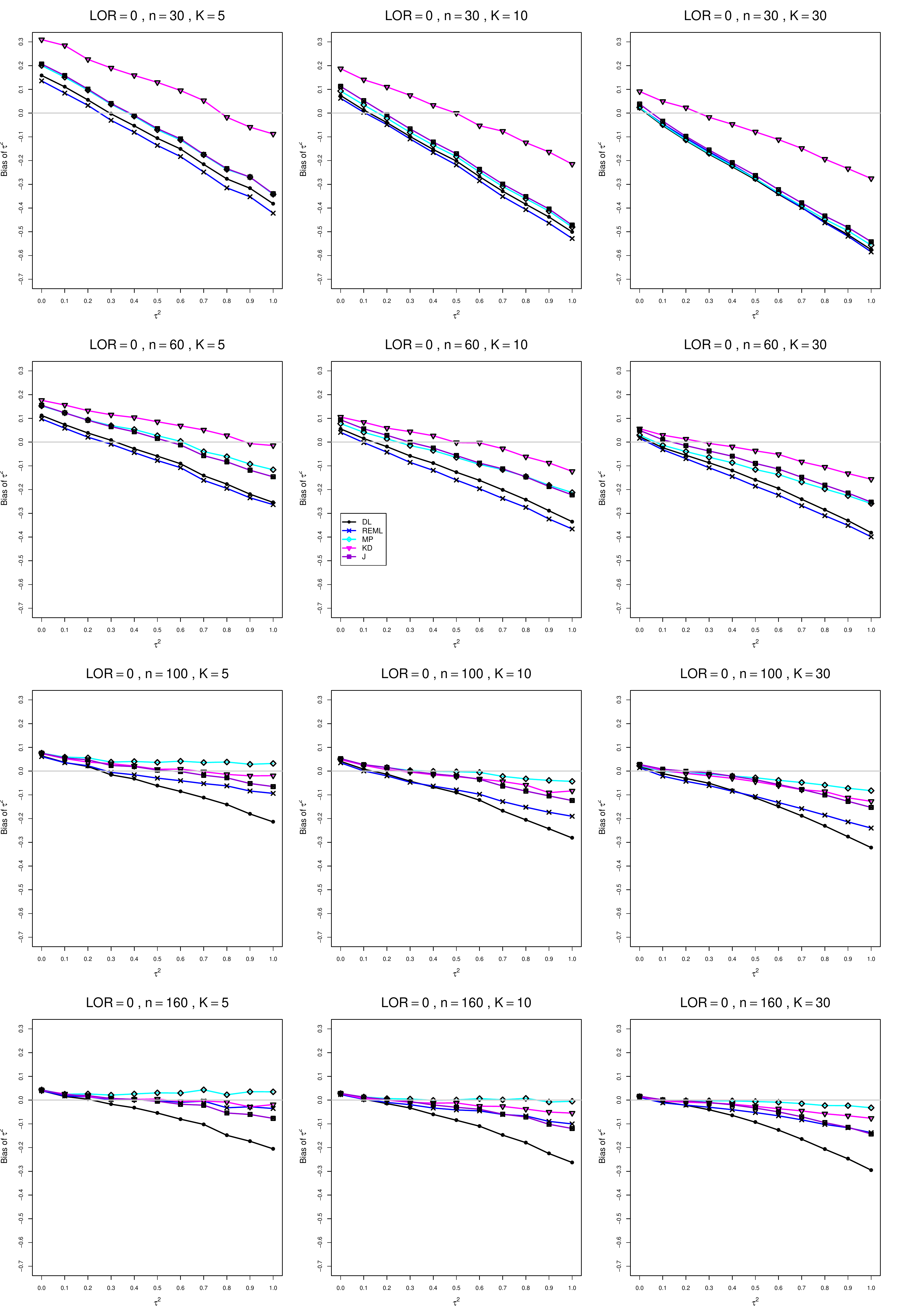}
	\caption{Bias of the estimation of  between-studies variance $\tau^2$ for $\theta=0$, $p_{iC}=0.4$, $q=0.75$, 
		unequal sample sizes $n=30,\; 60,\;100,\;160$. 
		\label{BiasTauLOR0q075piC04_unequal_sample_sizes}}
\end{figure}

\begin{figure}[t]
	\centering
	\includegraphics[scale=0.33]{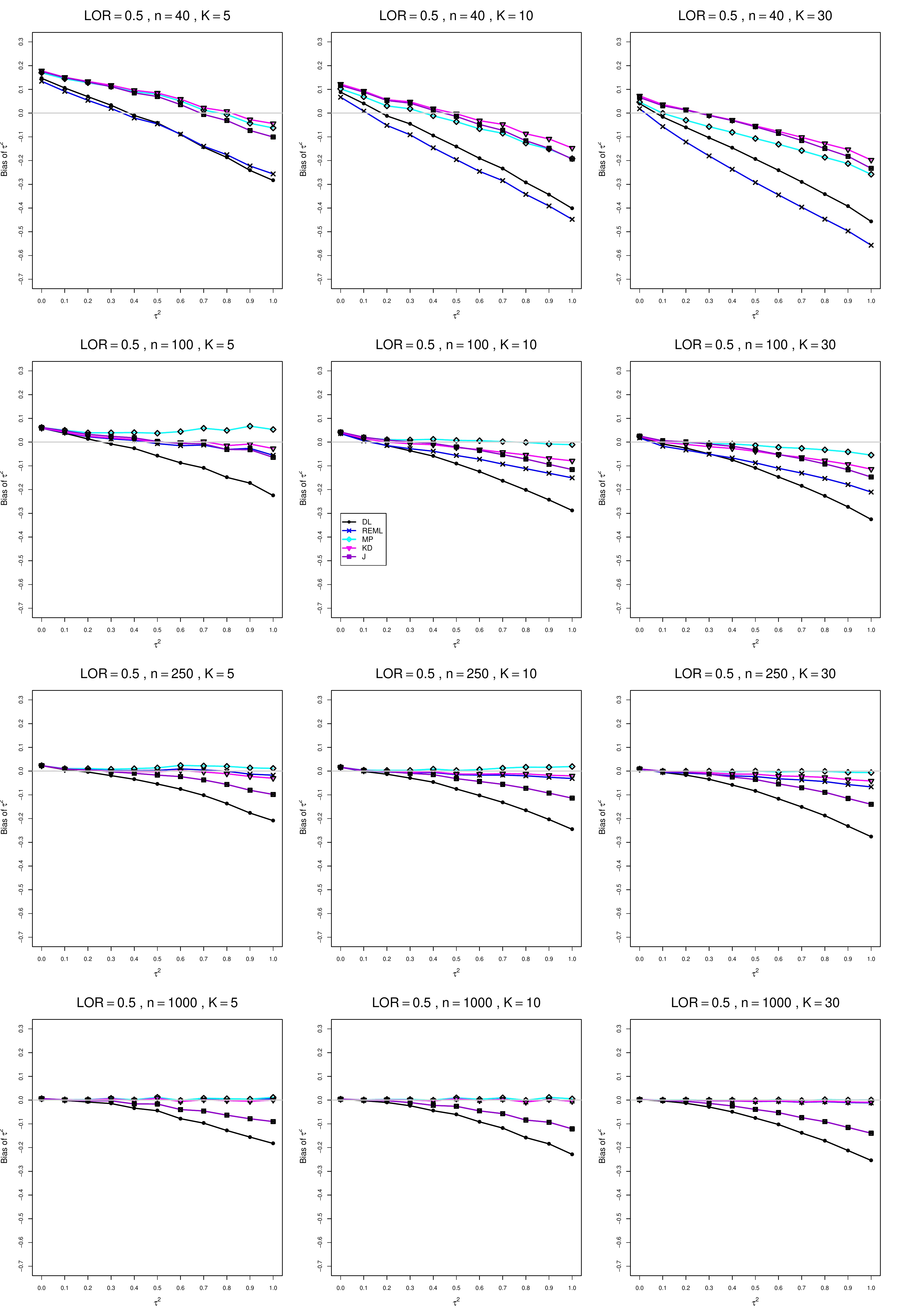}
	\caption{Bias of the estimation of  between-studies variance $\tau^2$ for $\theta=0.5$, $p_{iC}=0.4$, $q=0.75$, $n=40,\;100,\;250,\;1000$. 
		\label{BiasTauLOR05q075piC04}}
\end{figure}
\begin{figure}[t]
	\centering
	\includegraphics[scale=0.33]{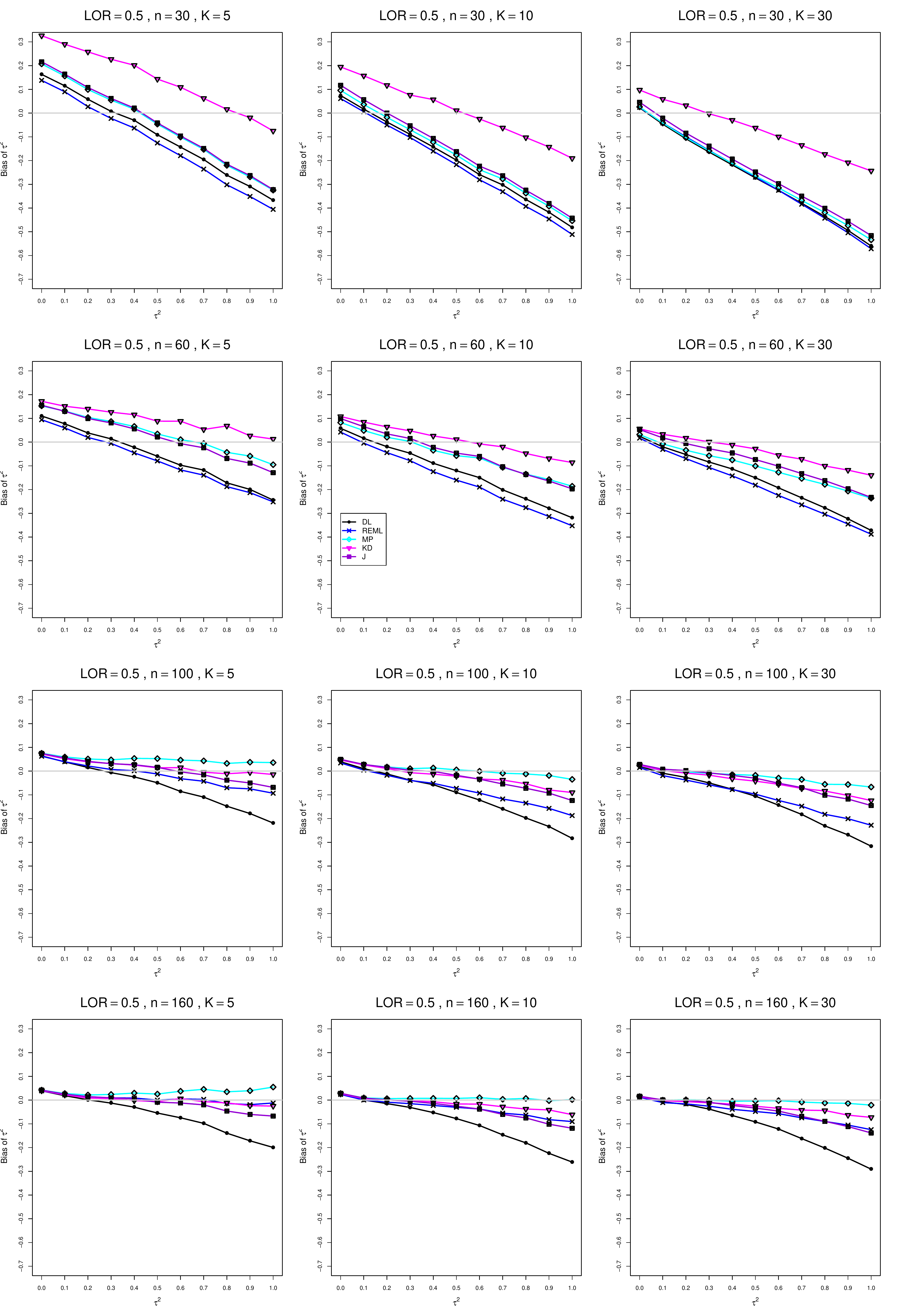}
	\caption{Bias of the estimation of  between-studies variance $\tau^2$ for $\theta=0.5$, $p_{iC}=0.4$, $q=0.75$, 
		unequal sample sizes $n=30,\; 60,\;100,\;160$. 
		\label{BiasTauLOR05q075piC04_unequal_sample_sizes}}
\end{figure}
\clearpage

\begin{figure}[t]
	\centering
	\includegraphics[scale=0.33]{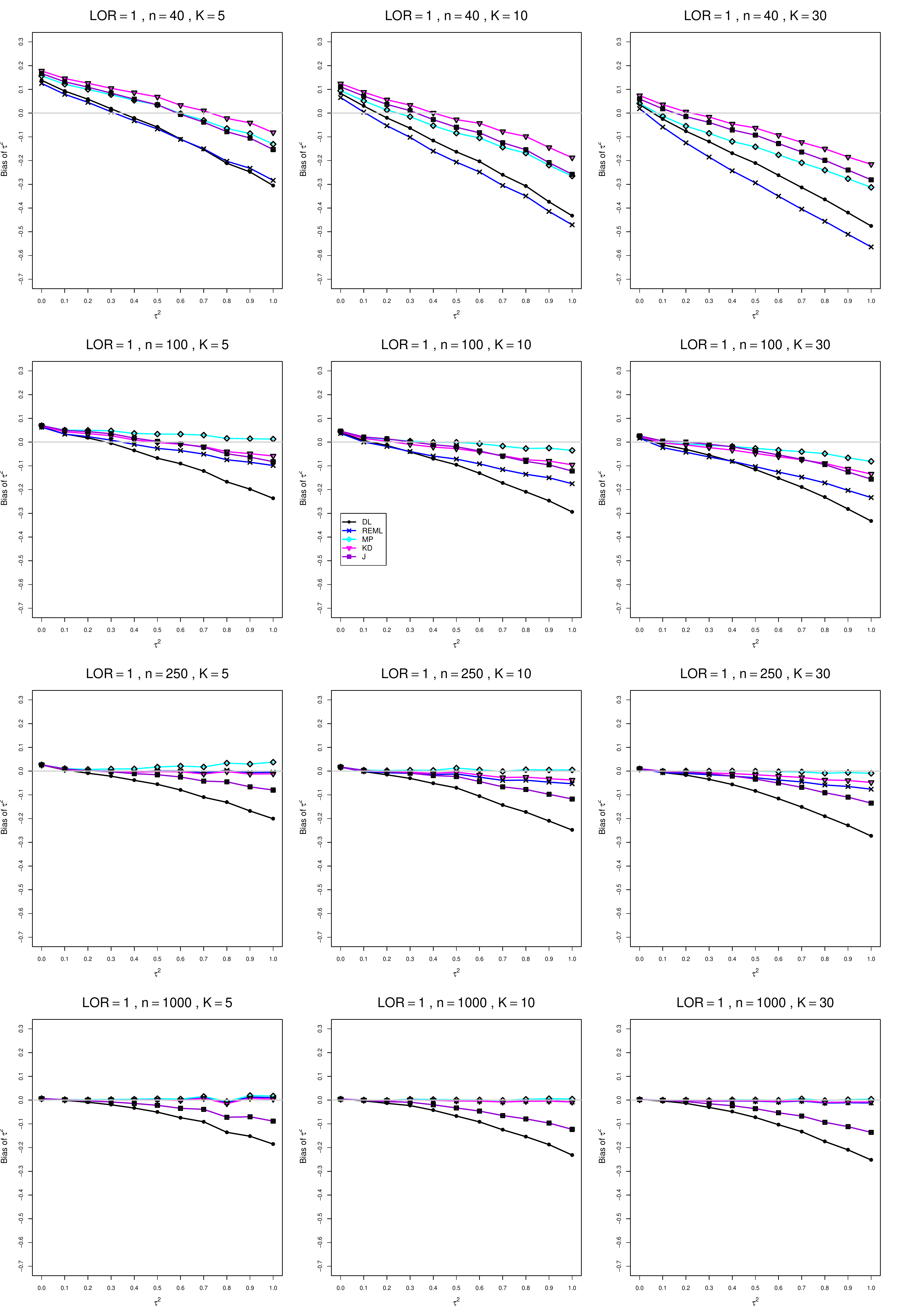}
	\caption{Bias of the estimation of  between-studies variance $\tau^2$ for $\theta=1$, $p_{iC}=0.4$, $q=0.75$, equal sample sizes $n=40,\;100,\;250,\;1000$. 
		\label{BiasTauLOR1q075piC04}}
\end{figure}

\begin{figure}[t]
	\centering
	\includegraphics[scale=0.33]{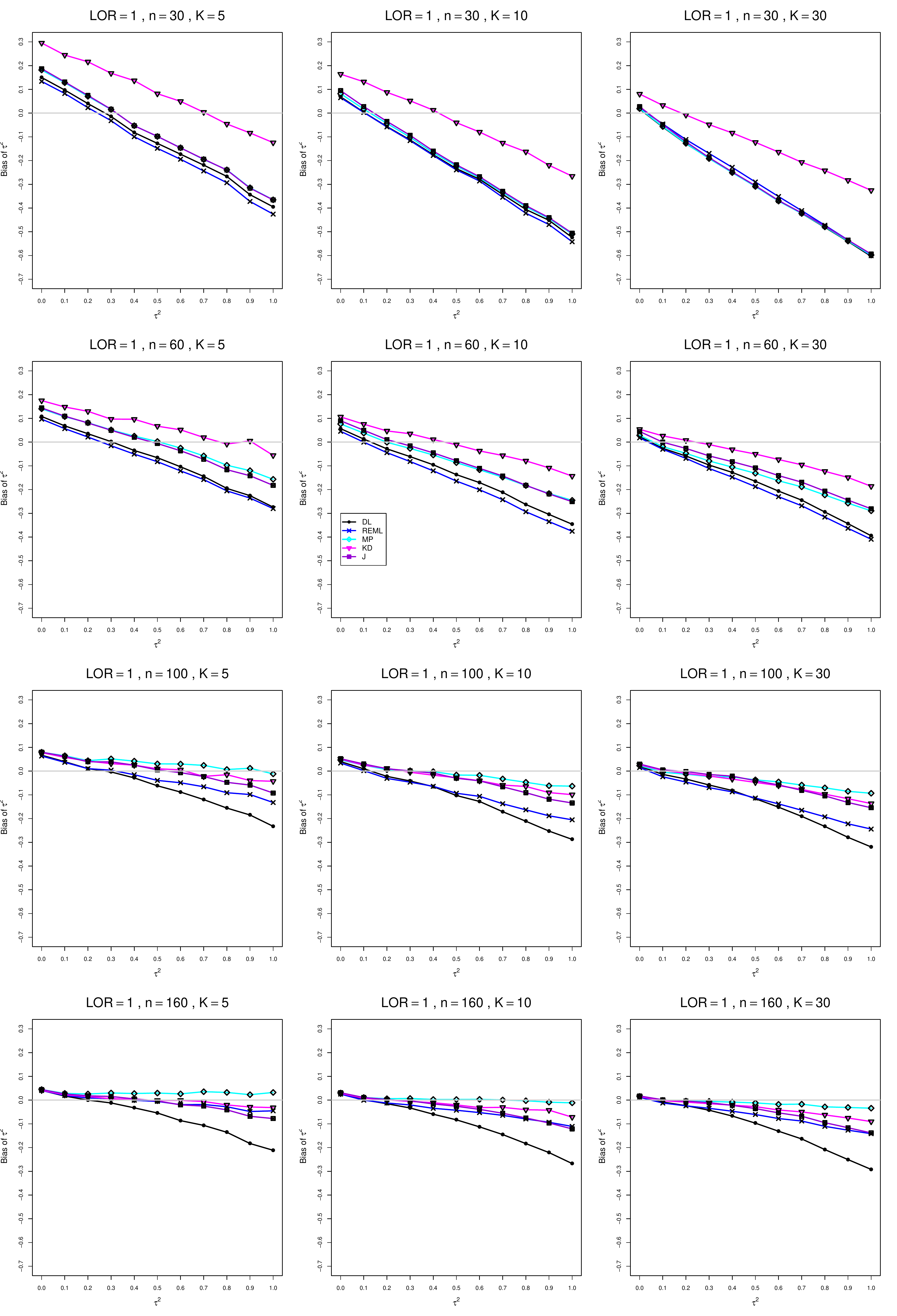}
	\caption{Bias of the estimation of  between-studies variance $\tau^2$ for $\theta=1$, $p_{iC}=0.4$, $q=0.75$, 
		unequal sample sizes $n=30,\; 60,\;100,\;160$. 
		\label{BiasTauLOR1q075piC04_unequal_sample_sizes}}
\end{figure}

\begin{figure}[t]
	\centering
	\includegraphics[scale=0.33]{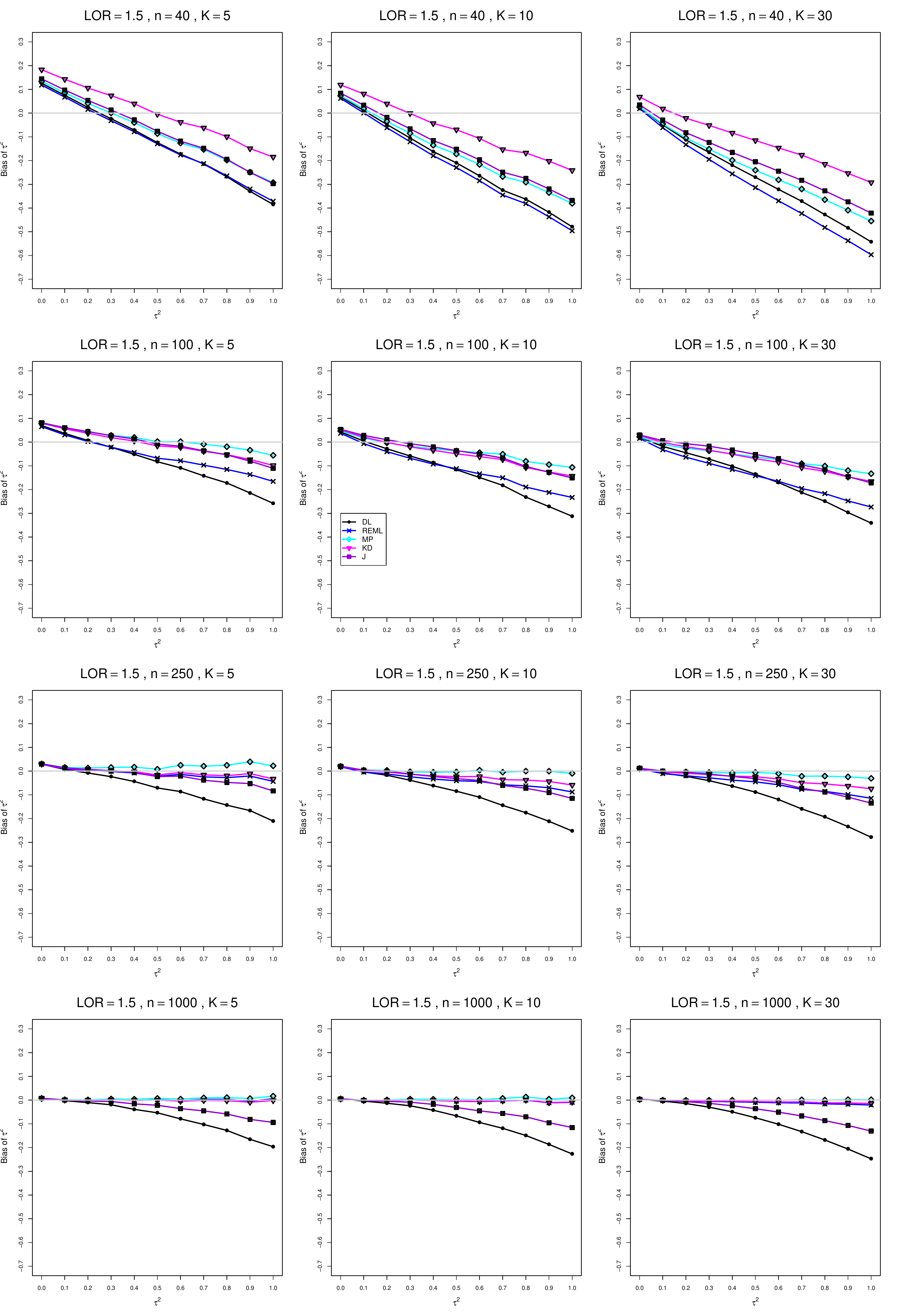}
	\caption{Bias of the estimation of  between-studies variance $\tau^2$ for $\theta=1.5$, $p_{iC}=0.4$, $q=0.75$, equal sample sizes $n=40,\;100,\;250,\;1000$. 
		\label{BiasTauLOR15q075piC04}}
\end{figure}

\begin{figure}[t]
	\centering
	\includegraphics[scale=0.33]{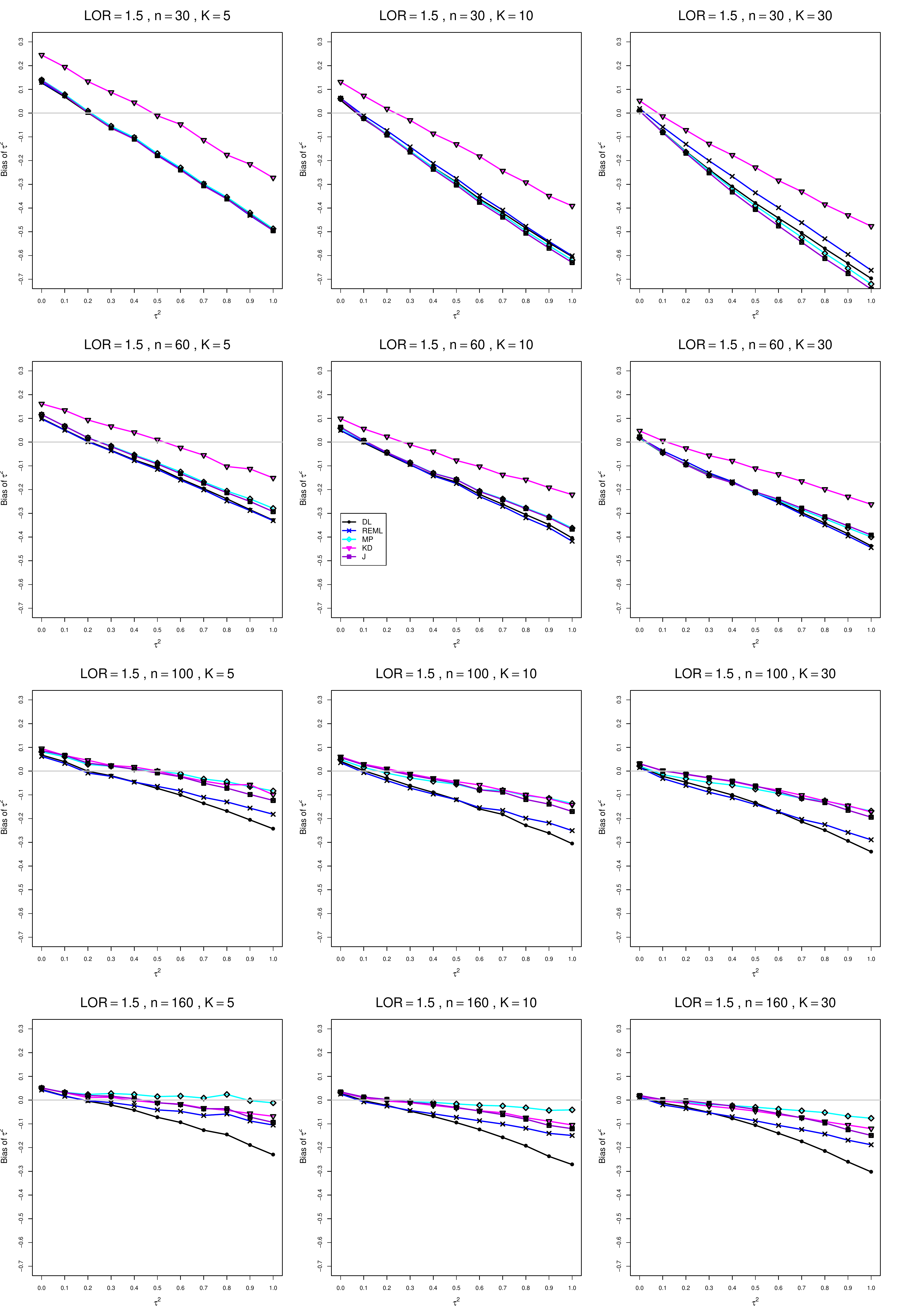}
	\caption{Bias of the estimation of  between-studies variance $\tau^2$ for $\theta=1.5$, $p_{iC}=0.4$, $q=0.75$, 
		unequal sample sizes $n=30,\; 60,\;100,\;160$. 
		\label{BiasTauLOR15q075piC04_unequal_sample_sizes}}
\end{figure}

\begin{figure}[t]
	\centering
	\includegraphics[scale=0.33]{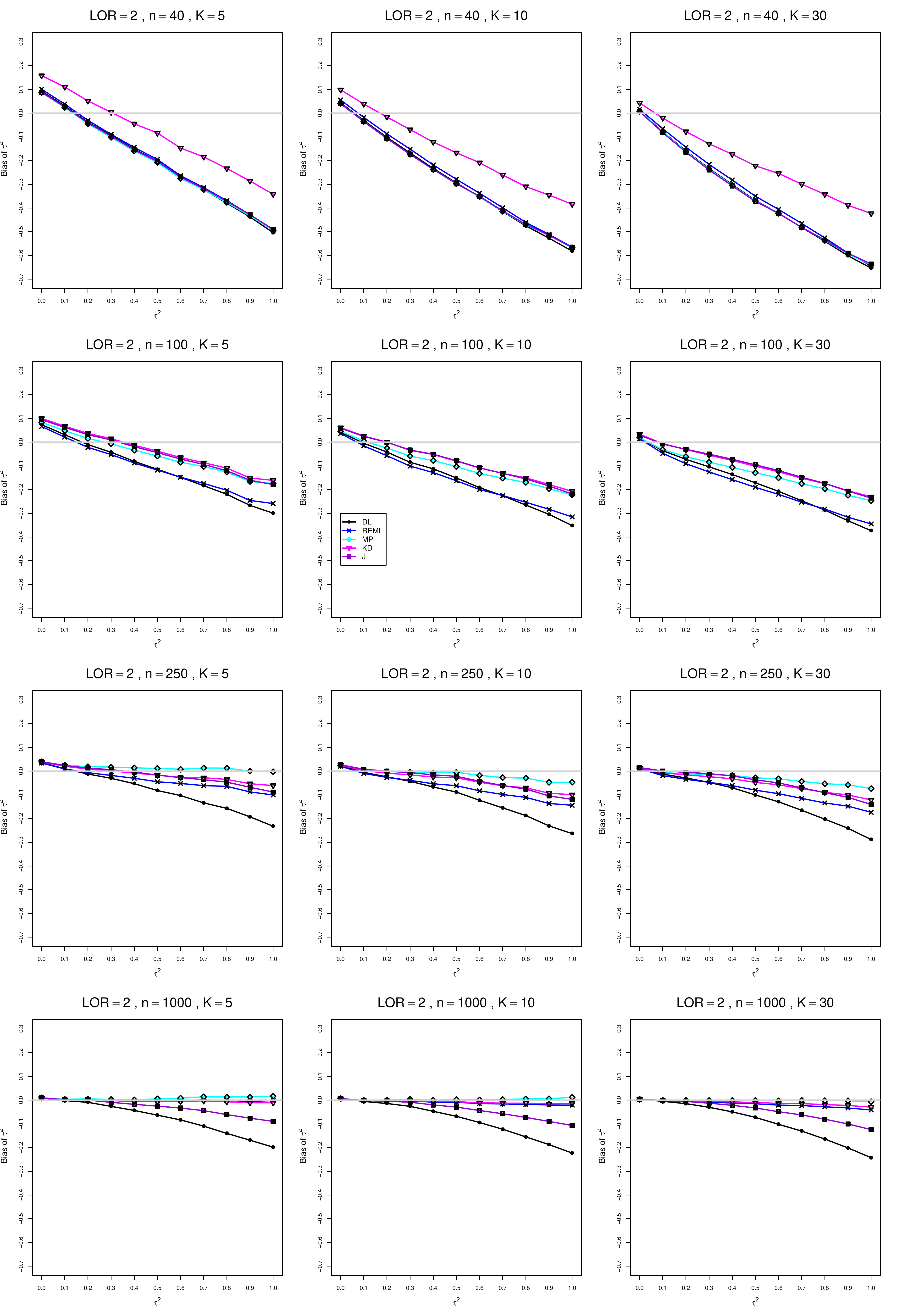}
	\caption{Bias of the estimation of  between-studies variance $\tau^2$ for $\theta=2$, $p_{iC}=0.4$, $q=0.75$, equal sample sizes $n=40,\;100,\;250,\;1000$. 
		\label{BiasTauLOR2q075piC04}}
\end{figure}

\begin{figure}[t]
	\centering
	\includegraphics[scale=0.33]{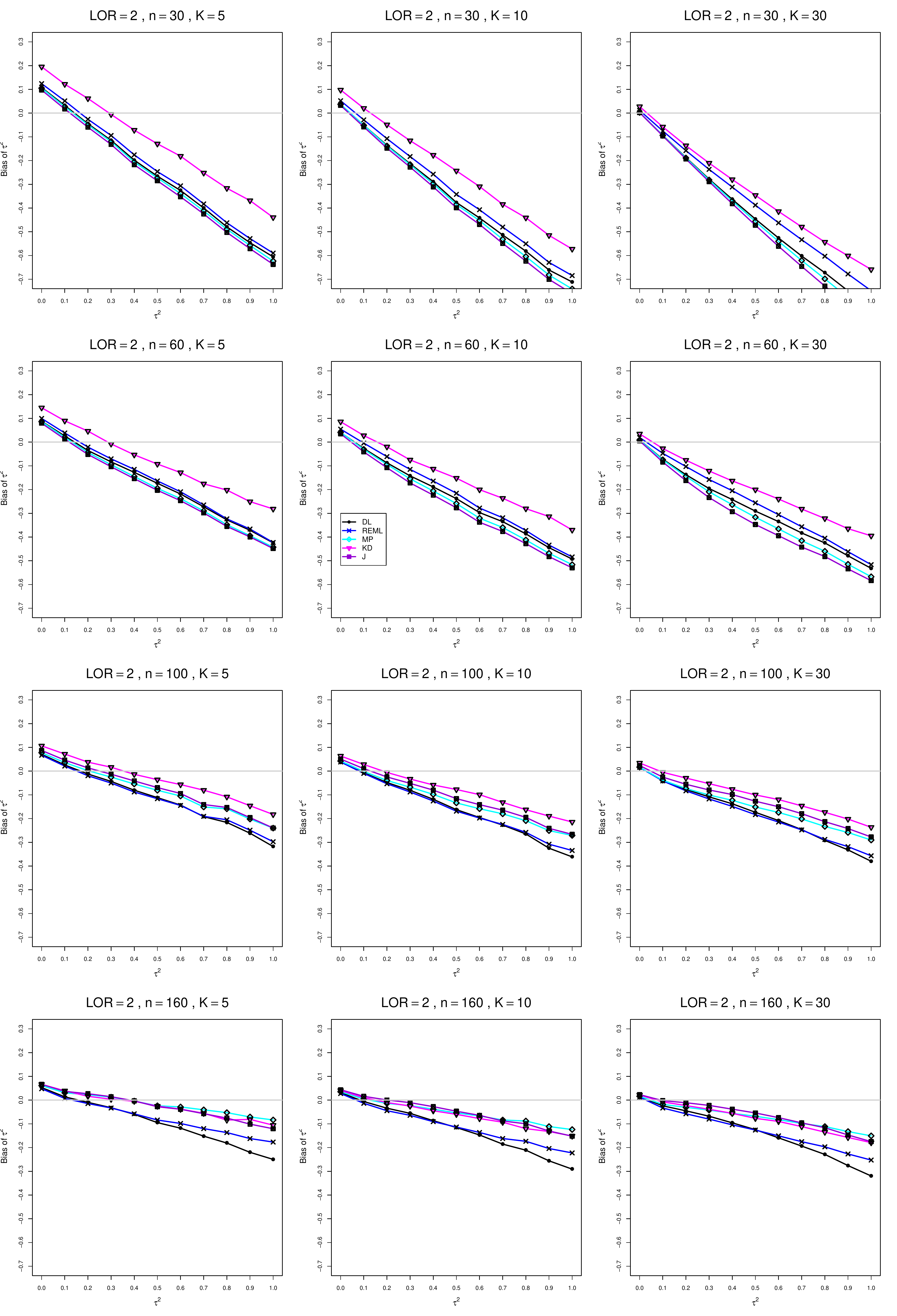}
	\caption{Bias of the estimation of  between-studies variance $\tau^2$ for $\theta=2$, $p_{iC}=0.4$, $q=0.75$,
		unequal sample sizes $n=30,\; 60,\;100,\;160$. 
		\label{BiasTauLOR2q075piC04_unequal_sample_sizes}}
\end{figure}

\clearpage
\renewcommand{\thefigure}{A2.1.\arabic{figure}}
\setcounter{figure}{0}
\setcounter{section}{1}
\section{Coverage of between-study variance.}
Subsections A2.1, A2.2 and A2.3 correspond to $p_{C}=0.1,\; 0.2,\; 0.4$ respectively. 
For a given $p_{C}$ value, each figure corresponds to a value of $\theta (= 0, 0.5, 1, 1.5, 2)$, a value of $q (= 0.5, 0.75)$, a value of $\tau^2 = 0.0(0.1)1.0$, and a set of values of $n$ (= 40, 100, 250, 1000) or $\bar{n} (= 30, 60, 100, 160)$.\\
Each figure contains a panel (with $\tau^2$ on the horizontal axis) for each combination of n (or $\bar{n}$) and $K (=5, 10, 30)$.\\
The interval estimators of $\tau^2$ are
\begin{itemize}
	\item QP (Q-profile confidence interval)
	\item BJ (Biggerstaff and Jackson interval )
	\item PL (Profile likelihood interval)
	\item KD (Improved Q-profile confidence interval based on Kulinskaya and Dollinger (2015)) 
	\item J (Jacksons interval)
\end{itemize}
\clearpage

\renewcommand{\thefigure}{A2.1.\arabic{figure}}
\setcounter{figure}{0}
\subsection*{A2.1 Probability in the control arm $p_{C}=0.1$}
\clearpage
\begin{figure}[t]
	\centering
	\includegraphics[scale=0.33]{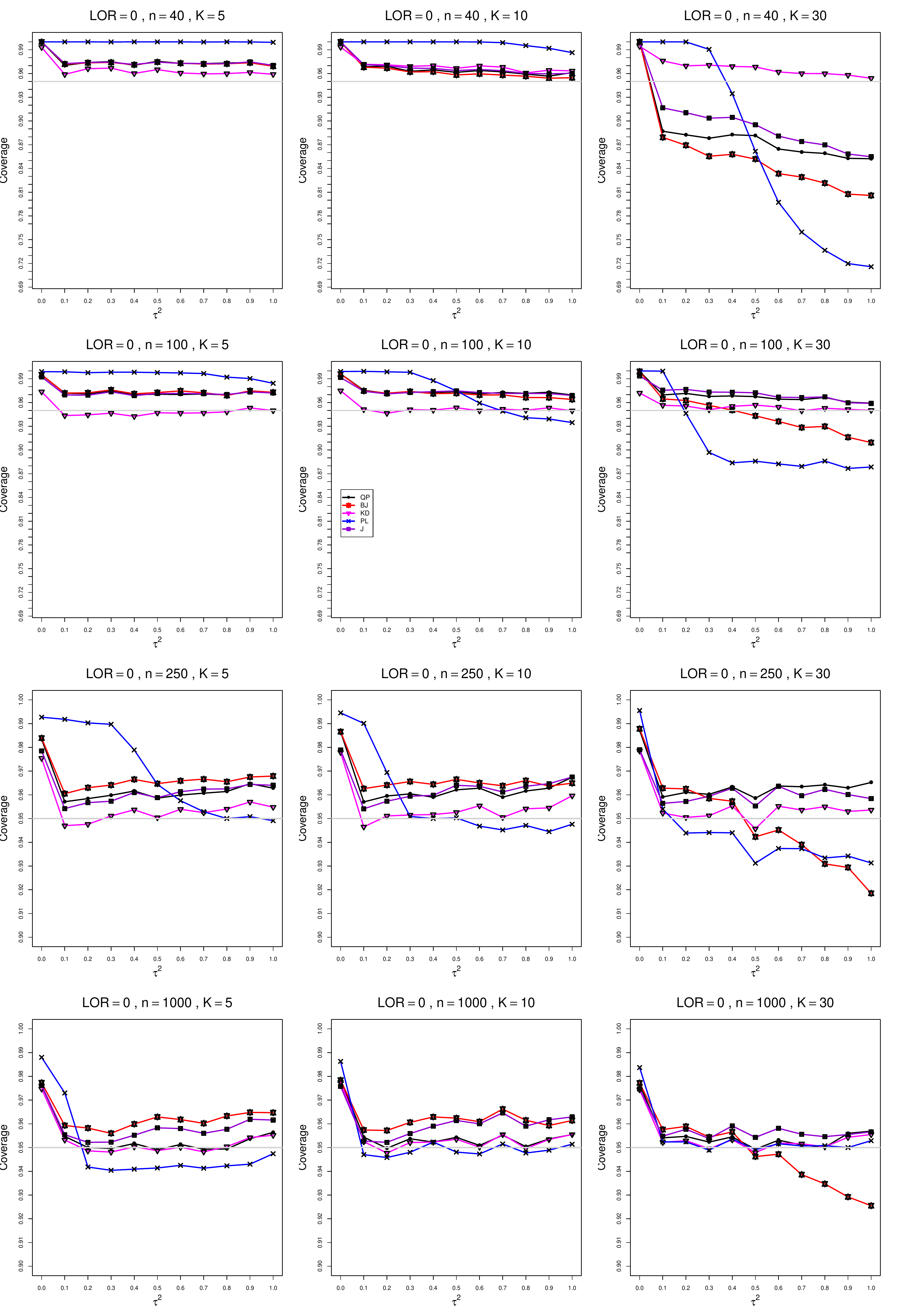}
	\caption{Coverage of  between-studies variance $\tau^2$ for $\theta=0$, $p_{iC}=0.1$, $q=0.5$, equal sample sizes $n=40,\;100,\;250,\;1000$. 
		\label{CovTauLOR0q05piC01}}
\end{figure}

\begin{figure}[t]
	\centering
	\includegraphics[scale=0.33]{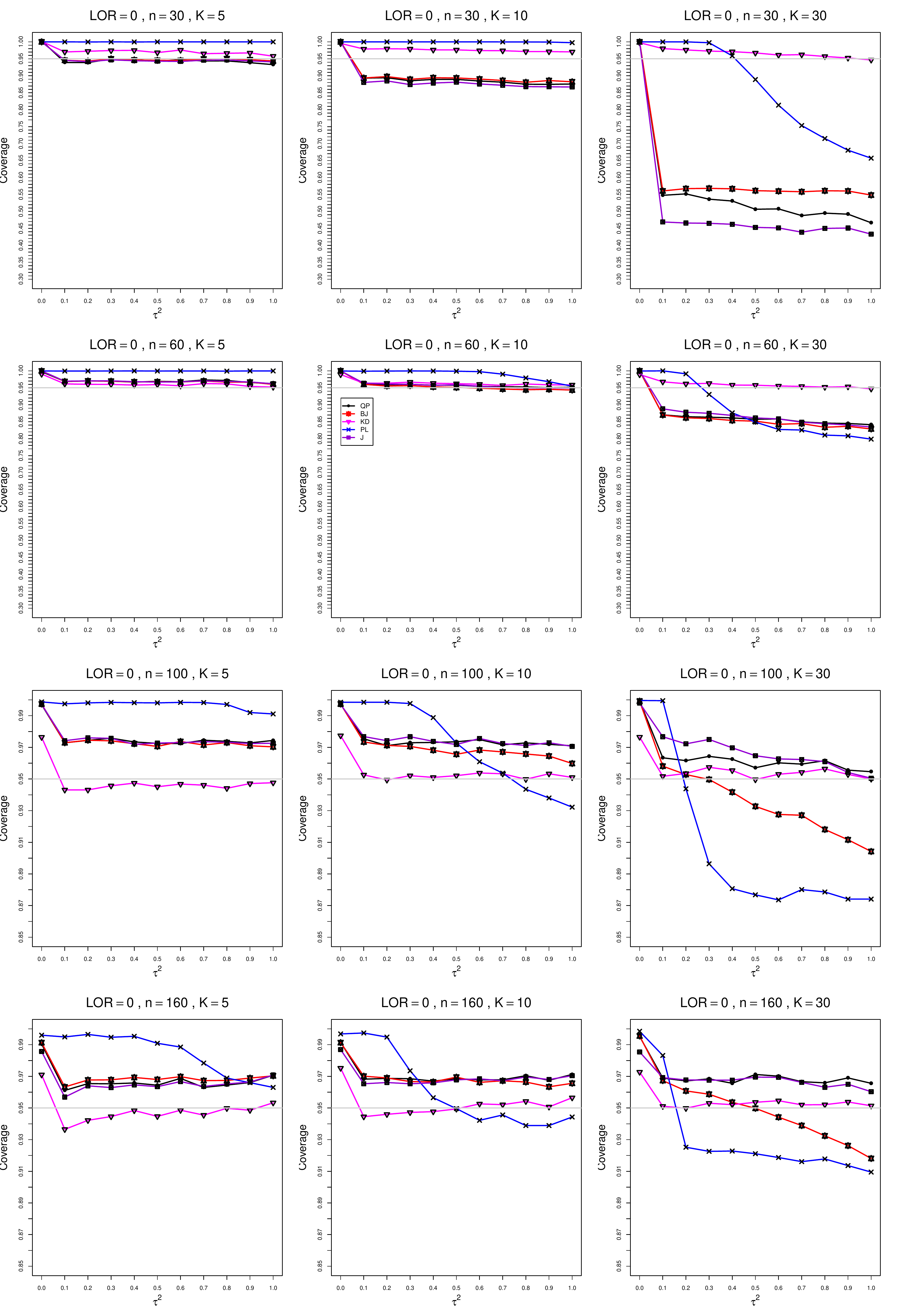}
	\caption{Coverage of  between-studies variance $\tau^2$ for $\theta=0$, $p_{iC}=0.1$, $q=0.5$, 
		unequal sample sizes $n=30,\; 60,\;100,\;160$. 
		\label{CovTauLOR0q05piC01_unequal_sample_sizes}}
\end{figure}

\begin{figure}[t]
	\centering
	\includegraphics[scale=0.33]{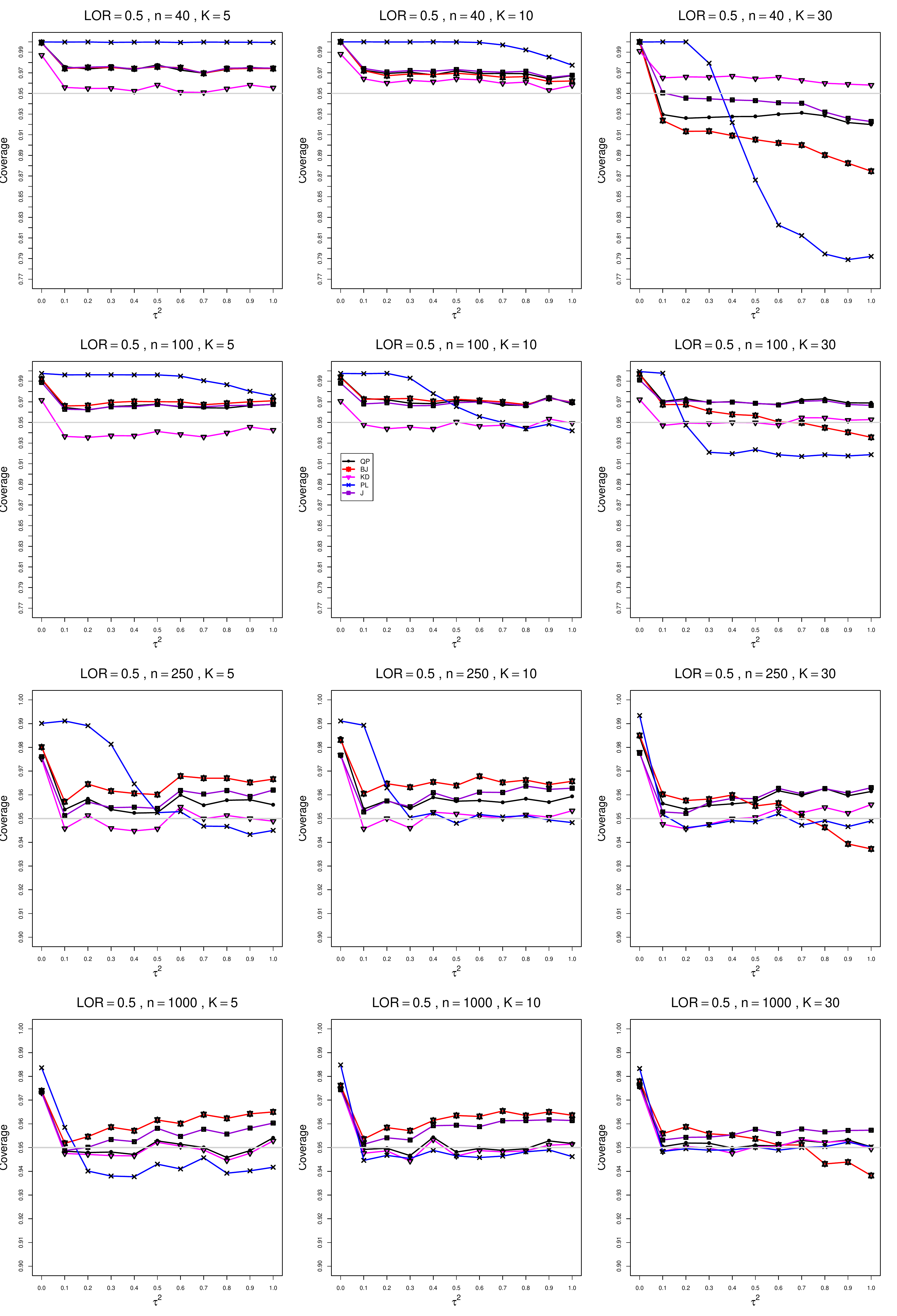}
	\caption{Coverage of  between-studies variance $\tau^2$ for $\theta=0.5$, $p_{iC}=0.1$, $q=0.5$, equal sample sizes $n=40,\;100,\;250,\;1000$. 
		\label{CovTauLOR05q05piC01}}
\end{figure}

\begin{figure}[t]
	\centering
	\includegraphics[scale=0.33]{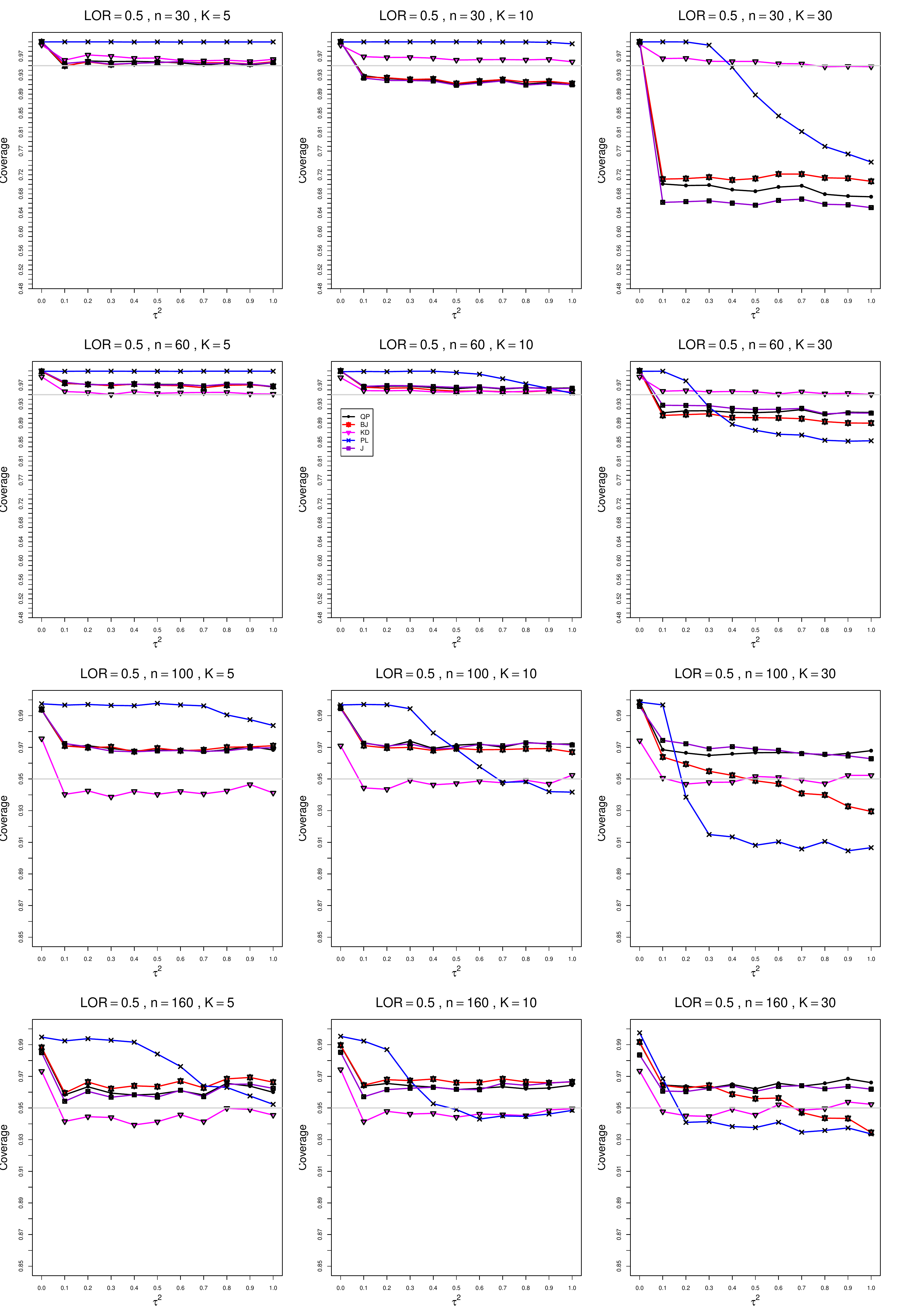}
	\caption{Coverage of  between-studies variance $\tau^2$ for $\theta=0.5$, $p_{iC}=0.1$, $q=0.5$,
		unequal sample sizes $n=30,\; 60,\;100,\;160$. 
		\label{CovTauLOR05q05piC01_unequal_sample_sizes}}
\end{figure}

\begin{figure}[t]
	\centering
	\includegraphics[scale=0.33]{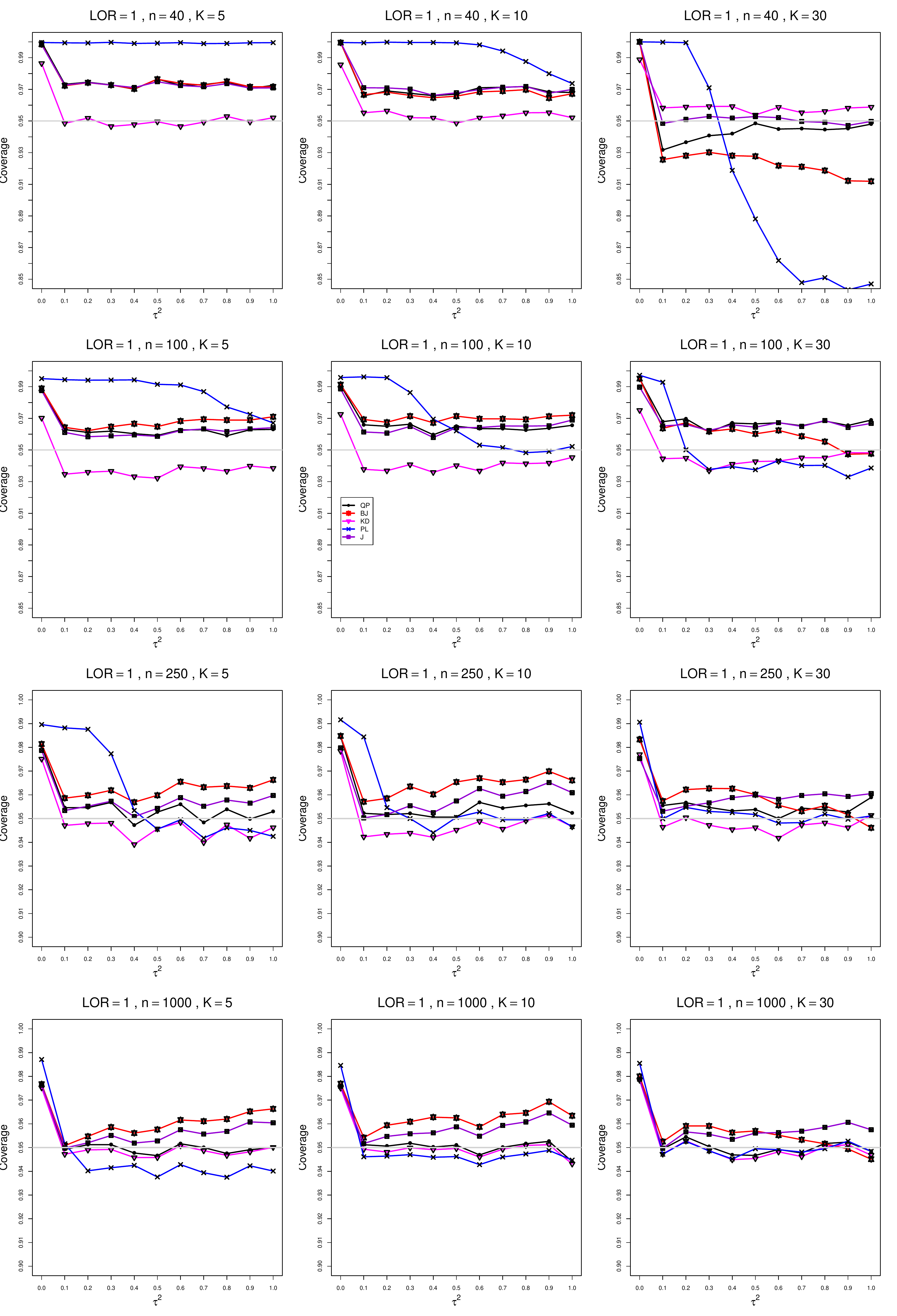}
	\caption{Coverage of  between-studies variance $\tau^2$ for $\theta=1$, $p_{iC}=0.1$, $q=0.5$, equal sample sizes $n=40,\;100,\;250,\;1000$. 
		\label{CovTauLOR1q05piC01}}
\end{figure}

\begin{figure}[t]
	\centering
	\includegraphics[scale=0.33]{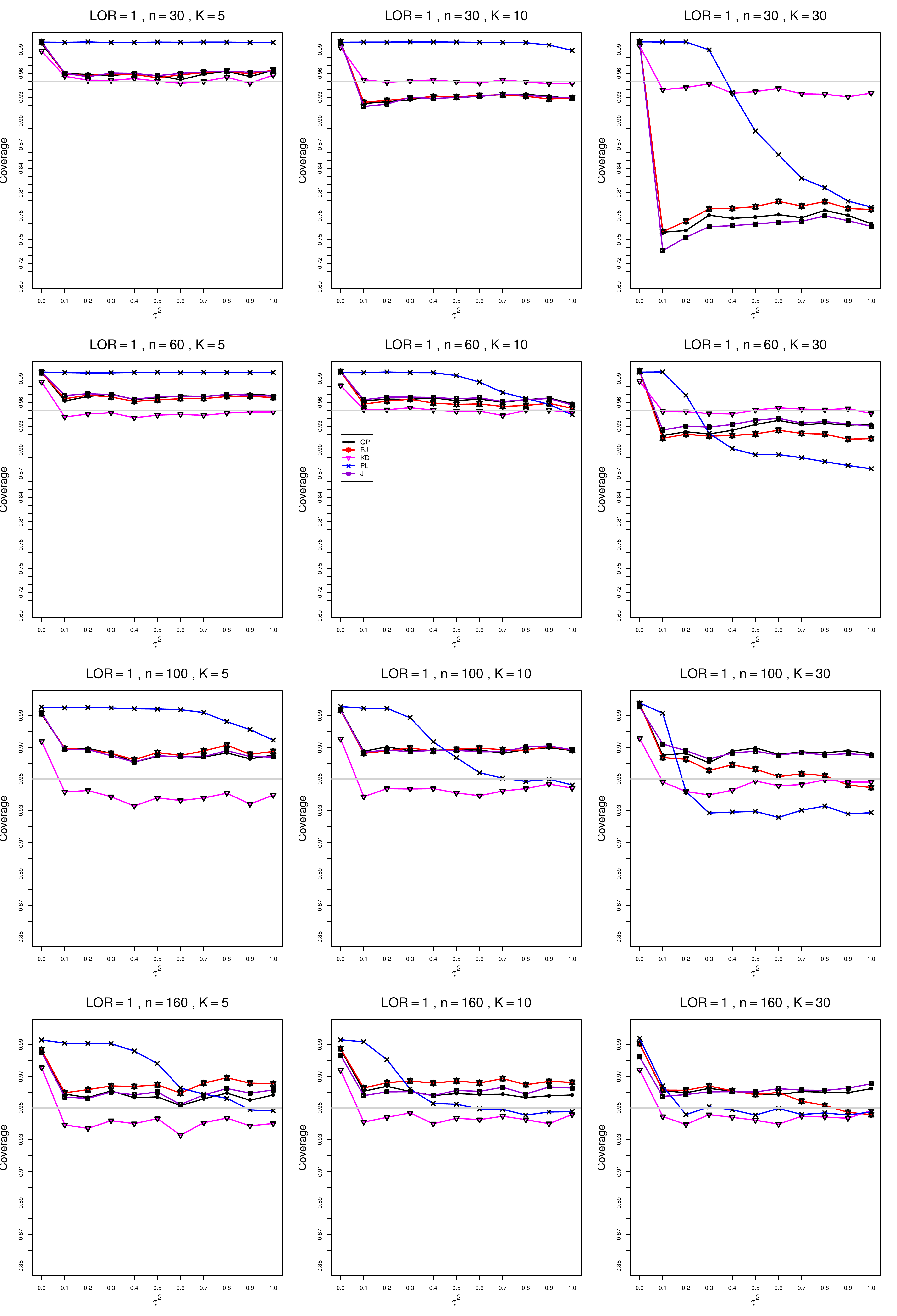}
	\caption{Coverage of  between-studies variance $\tau^2$ for $\theta=1$, $p_{iC}=0.1$, $q=0.5$,
		unequal sample sizes $n=30,\; 60,\;100,\;160$. 
		\label{CovTauLOR1q05piC01_unequal_sample_sizes}}
\end{figure}

\begin{figure}[t]
	\centering
	\includegraphics[scale=0.33]{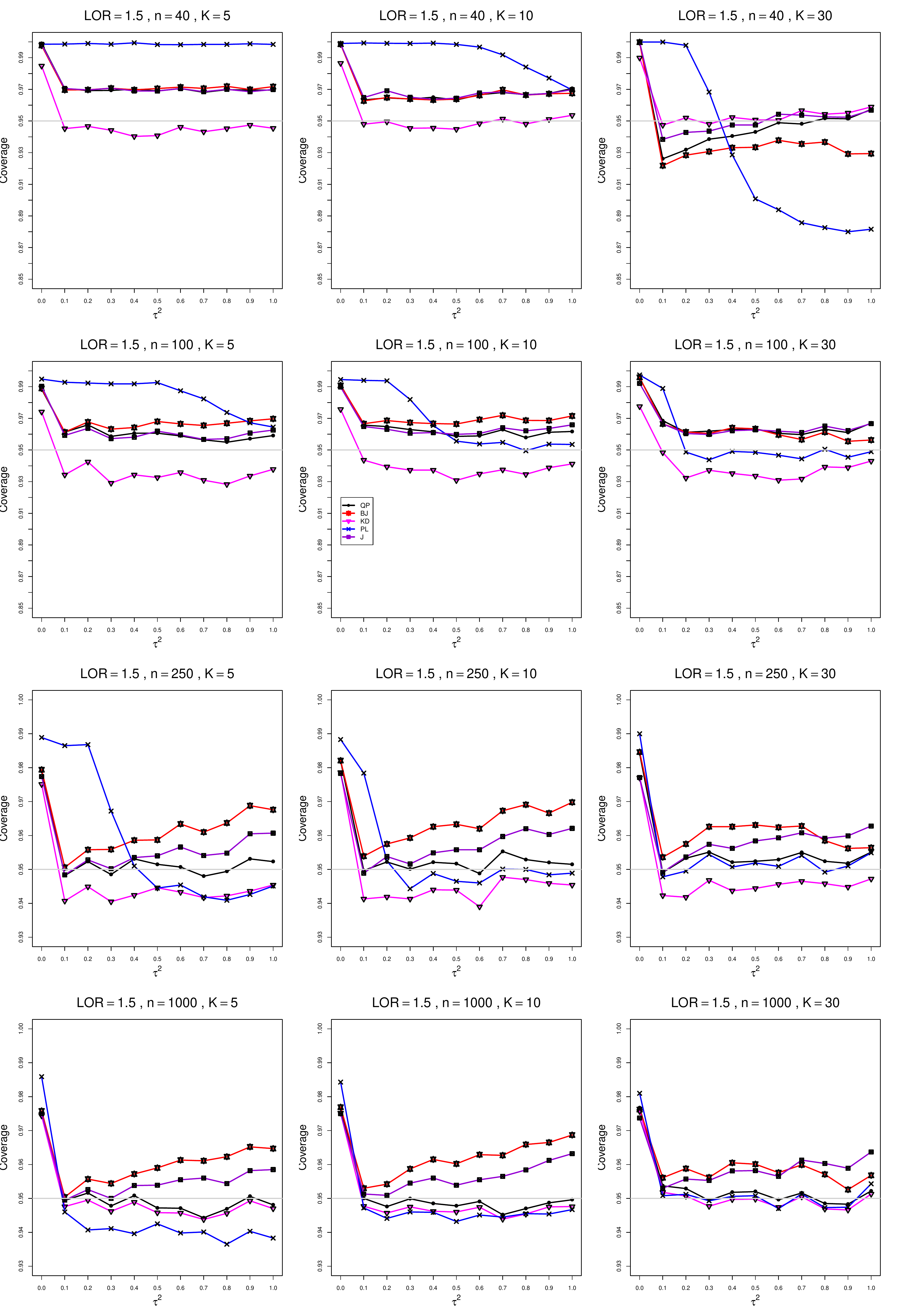}
	\caption{Coverage of  between-studies variance $\tau^2$ for $\theta=1.5$, $p_{iC}=0.1$, $q=0.5$, equal sample sizes $n=40,\;100,\;250,\;1000$. 
		\label{CovTauLOR15q05piC01}}
\end{figure}

\begin{figure}[t]
	\centering
	\includegraphics[scale=0.33]{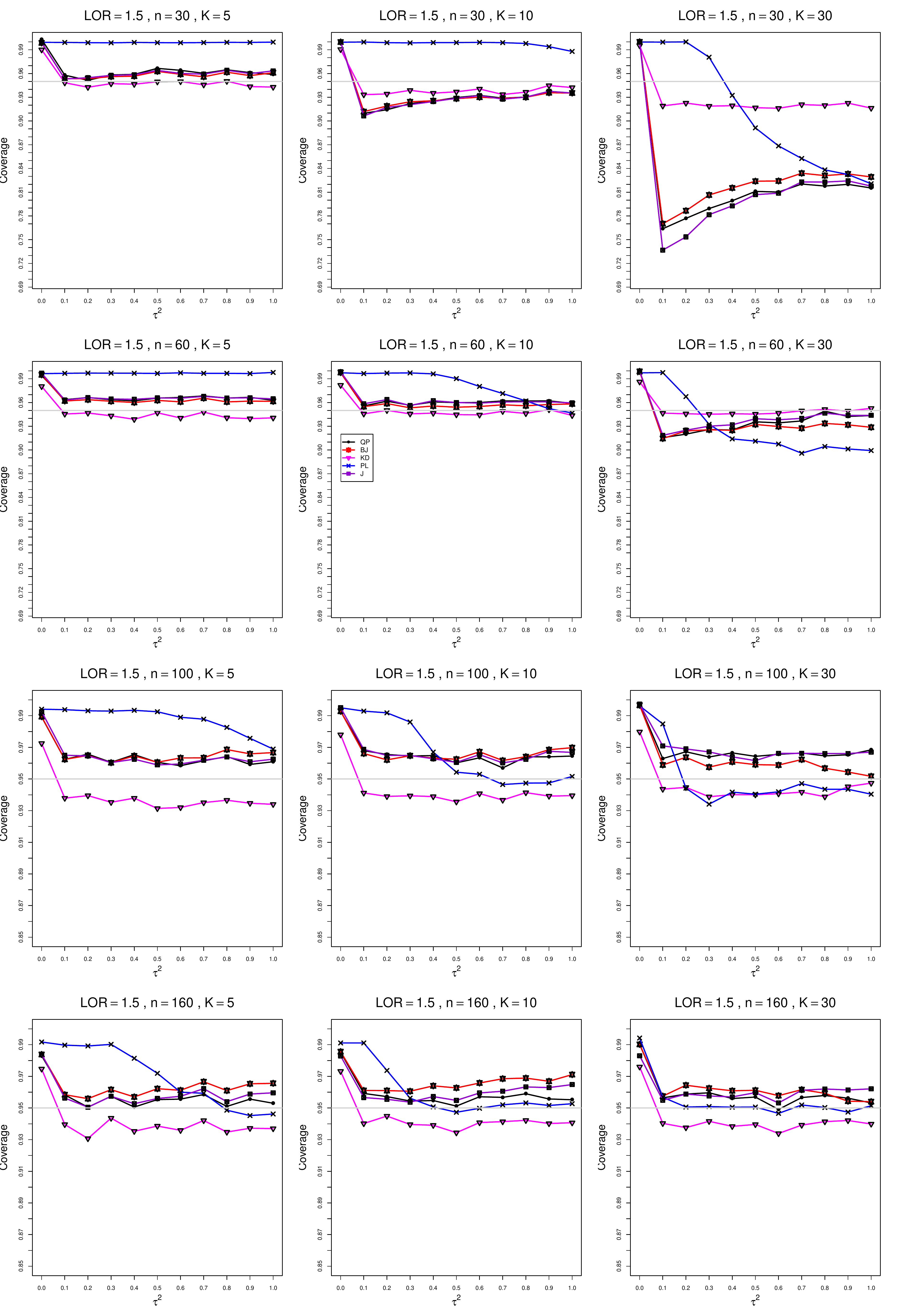}
	\caption{Coverage of  between-studies variance $\tau^2$ for $\theta=1.5$, $p_{iC}=0.1$, $q=0.5$,
		unequal sample sizes $n=30,\; 60,\;100,\;160$. 
		\label{CovTauLOR15q05piC01_unequal_sample_sizes}}
\end{figure}

\begin{figure}[t]
	\centering
	\includegraphics[scale=0.33]{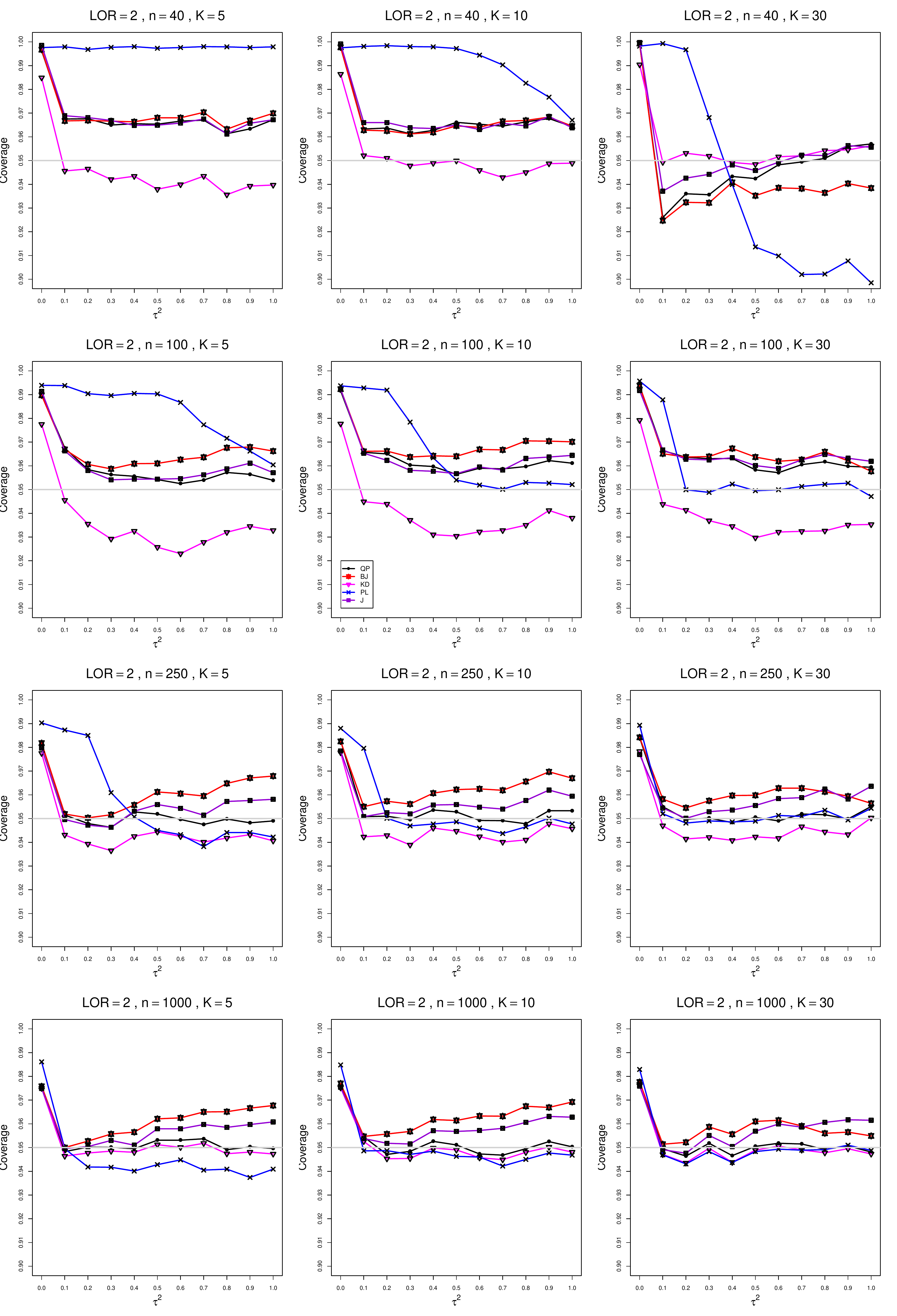}
	\caption{Coverage of  between-studies variance $\tau^2$ for $\theta=2$, $p_{iC}=0.1$, $q=0.5$, equal sample sizes $n=40,\;100,\;250,\;1000$. 
		\label{CovTauLOR2q05piC01}}
\end{figure}

\begin{figure}[t]
	\centering
	\includegraphics[scale=0.33]{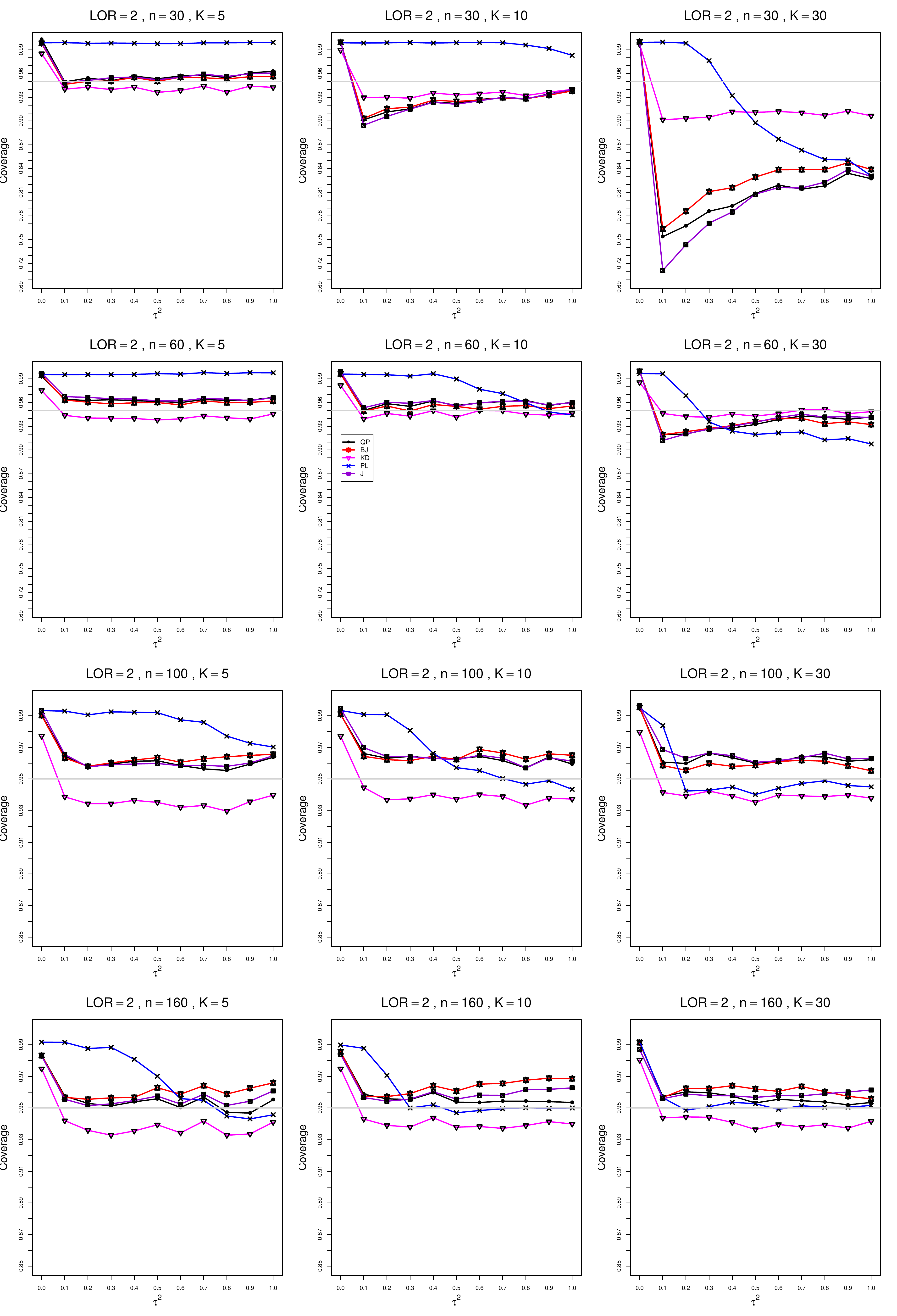}
	\caption{Coverage of  between-studies variance $\tau^2$ for $\theta=2$, $p_{iC}=0.1$, $q=0.5$,
		unequal sample sizes $n=30,\; 60,\;100,\;160$. 
		\label{CovTauLOR2q05piC01_unequal_sample_sizes}}
\end{figure}


\begin{figure}[t]
	\centering
	\includegraphics[scale=0.33]{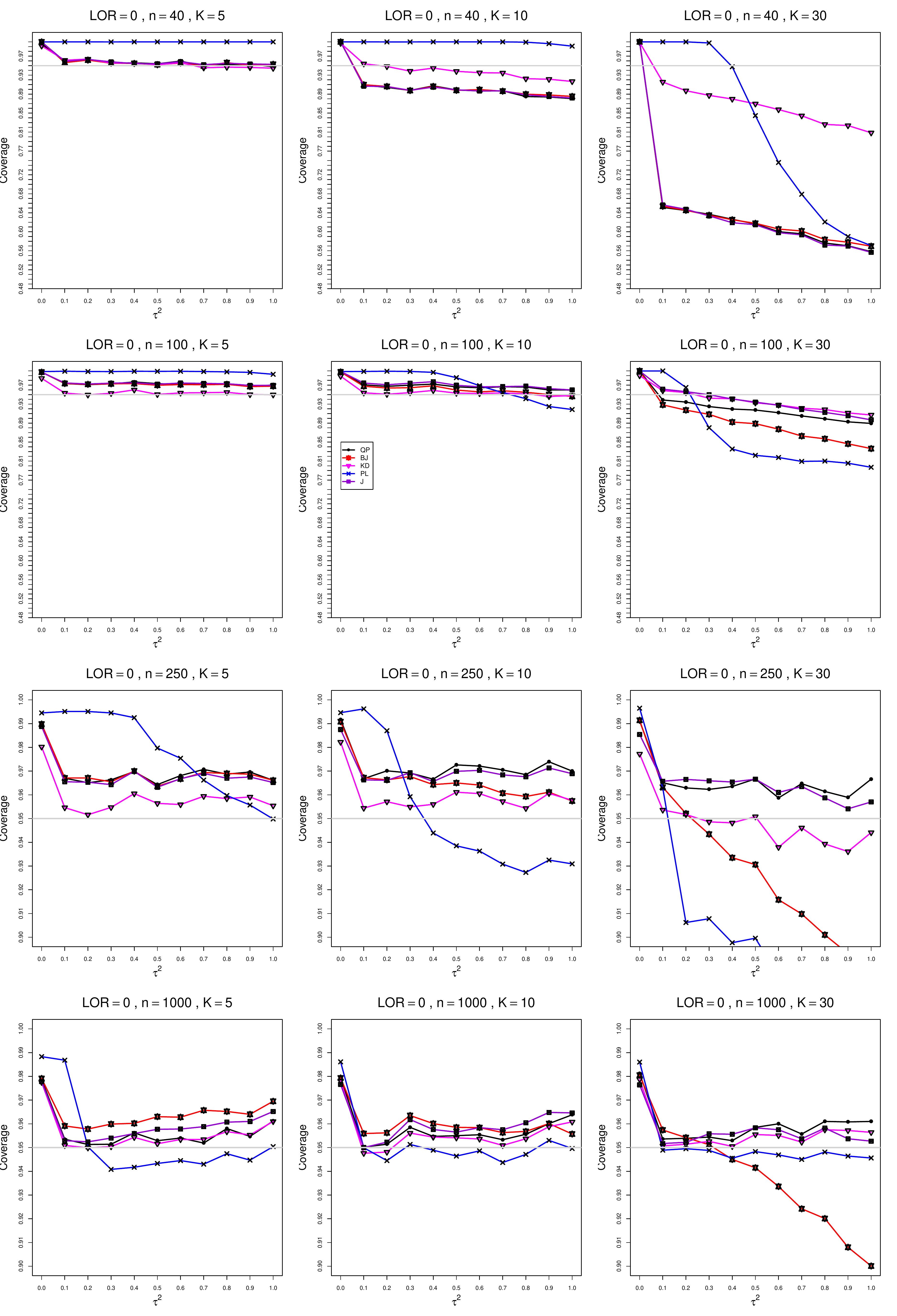}
	\caption{Coverage of  between-studies variance $\tau^2$ for $\theta=0$, $p_{iC}=0.1$, $q=0.75$, equal sample sizes $n=40,\;100,\;250,\;1000$. 
		\label{CovTauLOR0q075piC01}}
\end{figure}

\begin{figure}[t]
	\centering
	\includegraphics[scale=0.33]{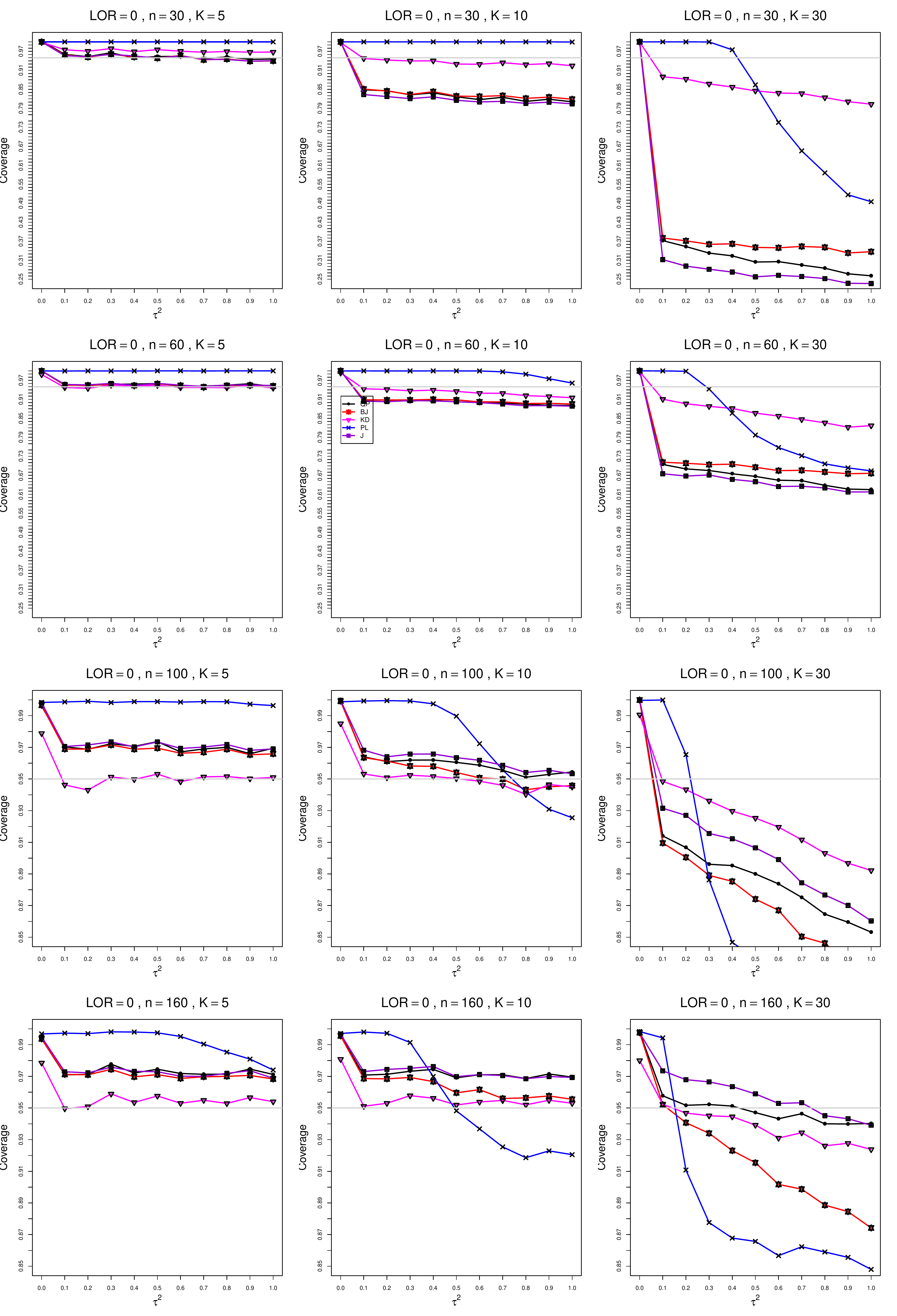}
	\caption{Coverage of  between-studies variance $\tau^2$ for $\theta=0$, $p_{iC}=0.1$, $q=0.75$, 
		unequal sample sizes $n=30,\; 60,\;100,\;160$. 
		\label{CovTauLOR0q075piC01_unequal_sample_sizes}}
\end{figure}

\begin{figure}[t]
	\centering
	\includegraphics[scale=0.33]{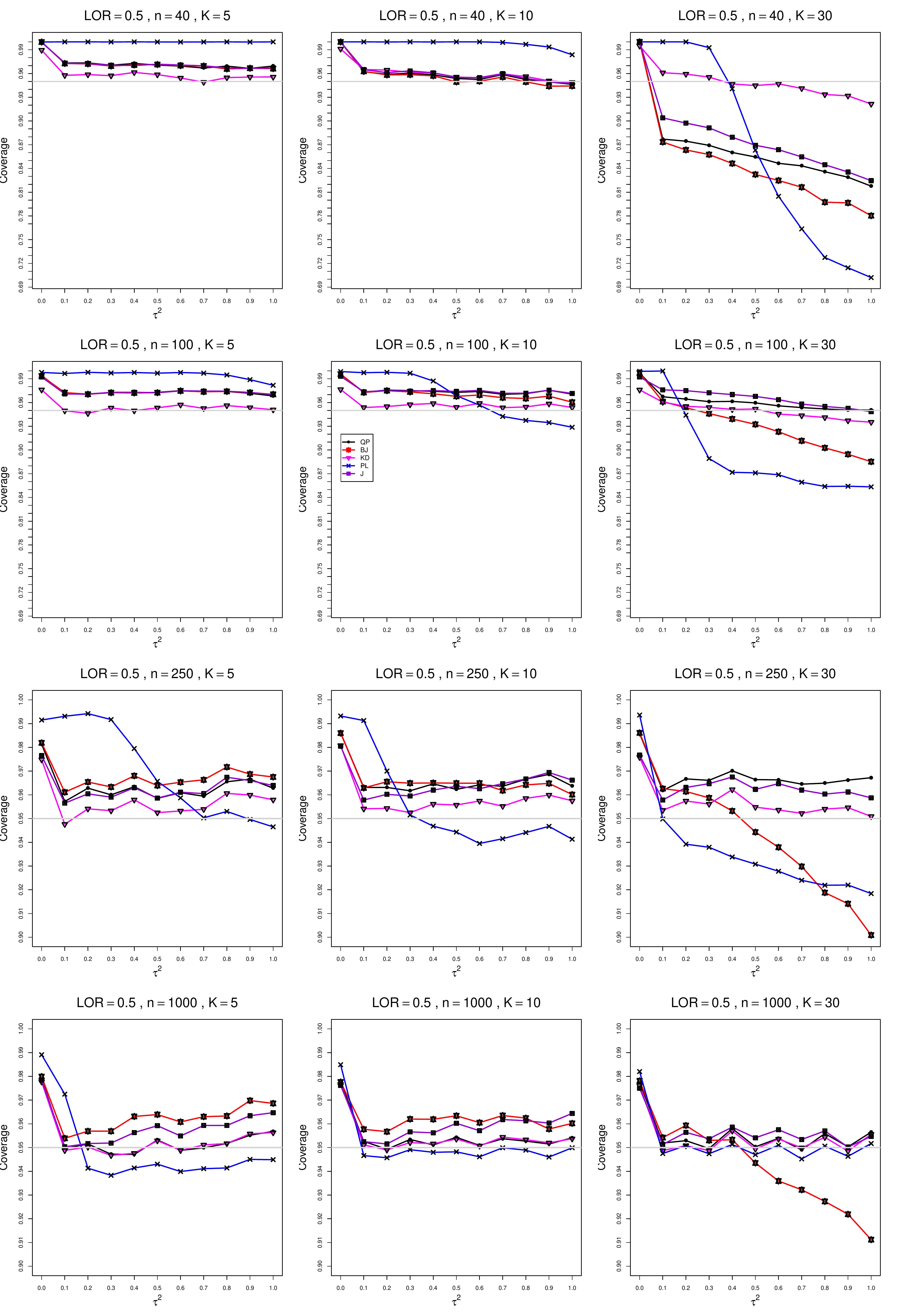}
	\caption{Coverage of  between-studies variance $\tau^2$ for $\theta=0.5$, $p_{iC}=0.1$, $q=0.75$, equal sample sizes $n=40,\;100,\;250,\;1000$. 
		\label{CovTauLOR05q075piC01}}
\end{figure}
\begin{figure}[t]
	\centering
	\includegraphics[scale=0.33]{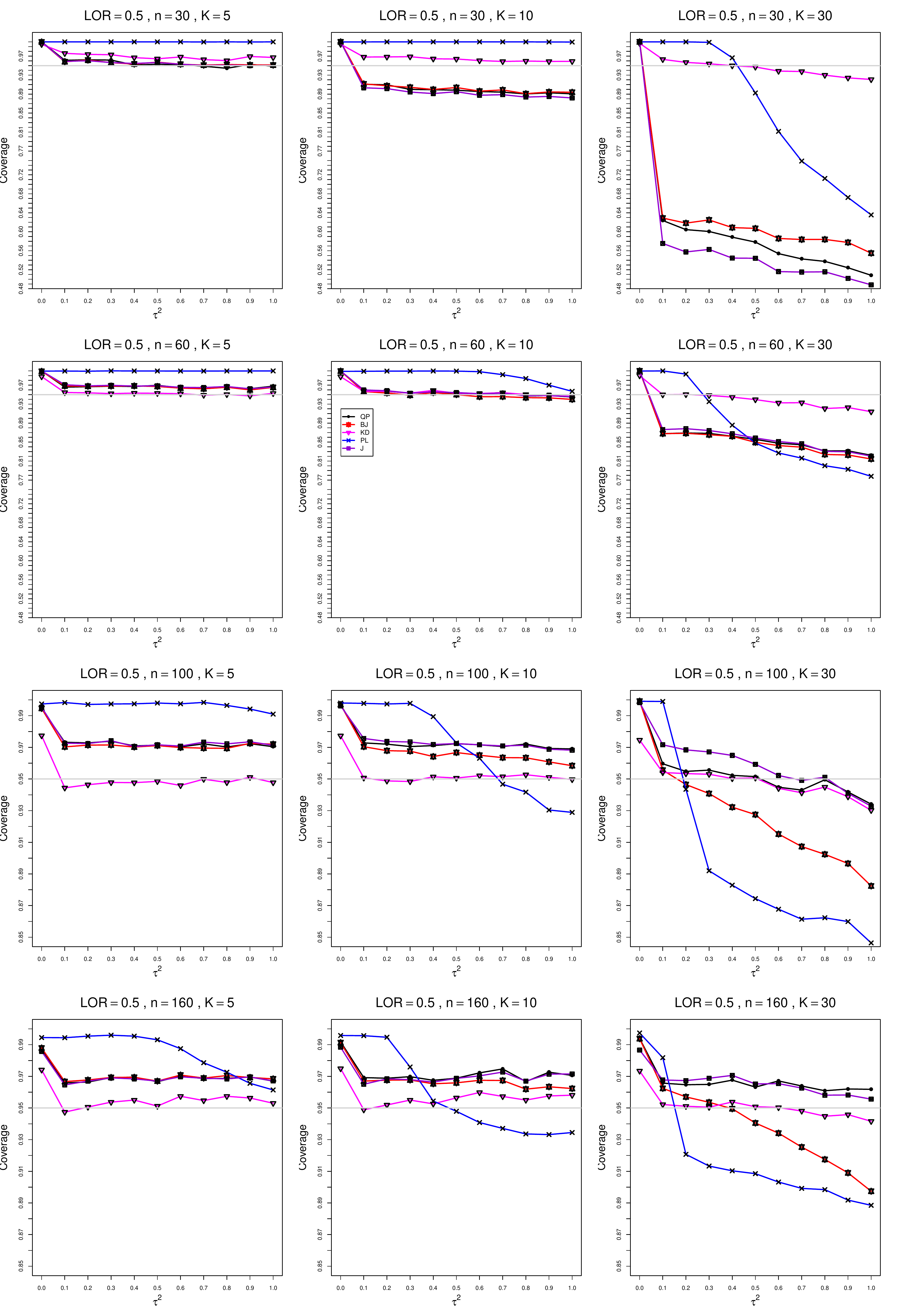}
	\caption{Coverage of  between-studies variance $\tau^2$ for $\theta=0.5$, $p_{iC}=0.1$, $q=0.75$,
		unequal sample sizes $n=30,\; 60,\;100,\;160$. 
		\label{CovTauLOR05q075piC01_unequal_sample_sizes}}
\end{figure}

\begin{figure}[t]
	\centering
	\includegraphics[scale=0.33]{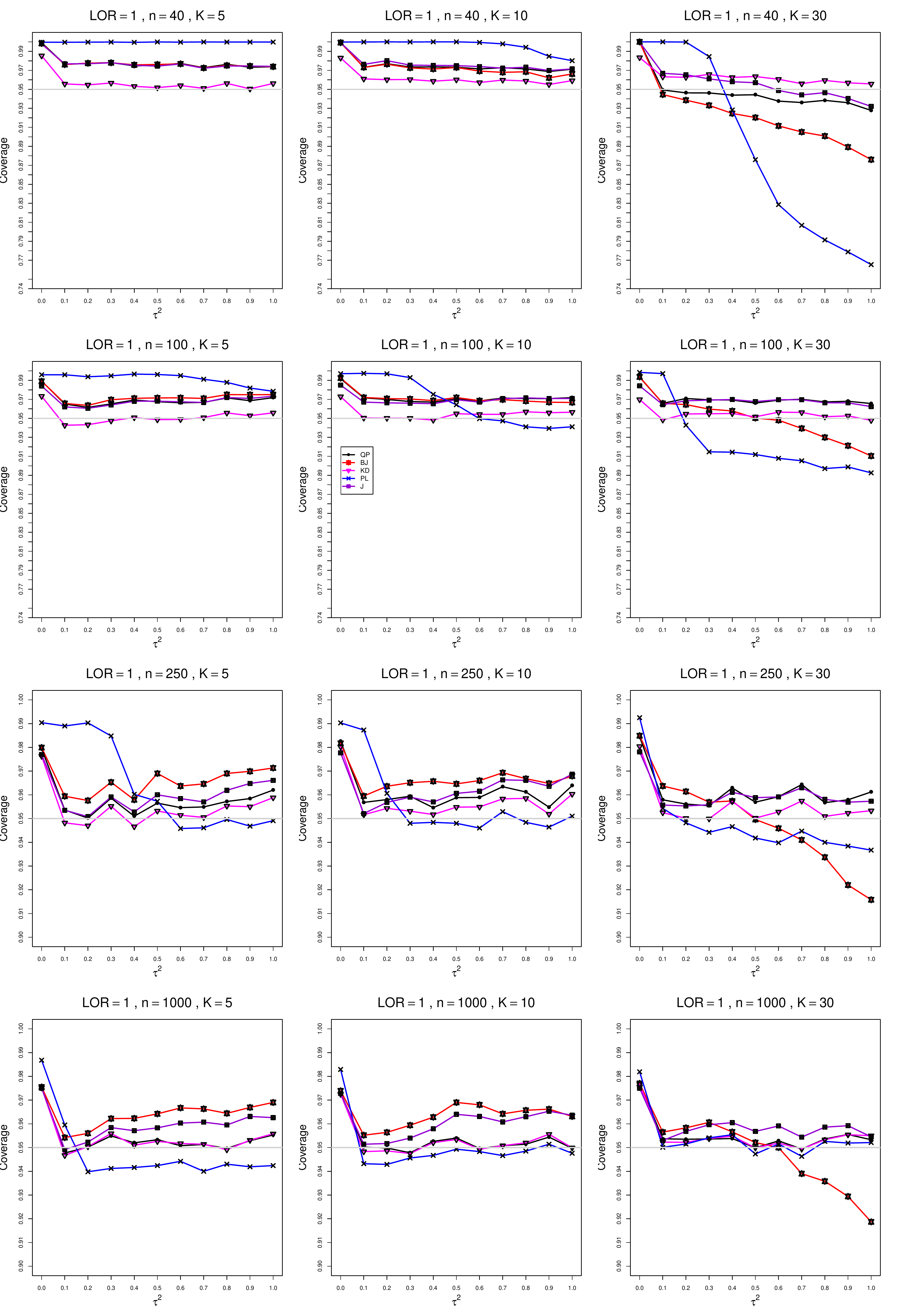}
	\caption{Coverage of  between-studies variance $\tau^2$ for $\theta=1$, $p_{iC}=0.1$, $q=0.75$, equal sample sizes $n=40,\;100,\;250,\;1000$. 
		\label{CovTauLOR1q075piC01}}
\end{figure}

\begin{figure}[t]
	\centering
	\includegraphics[scale=0.33]{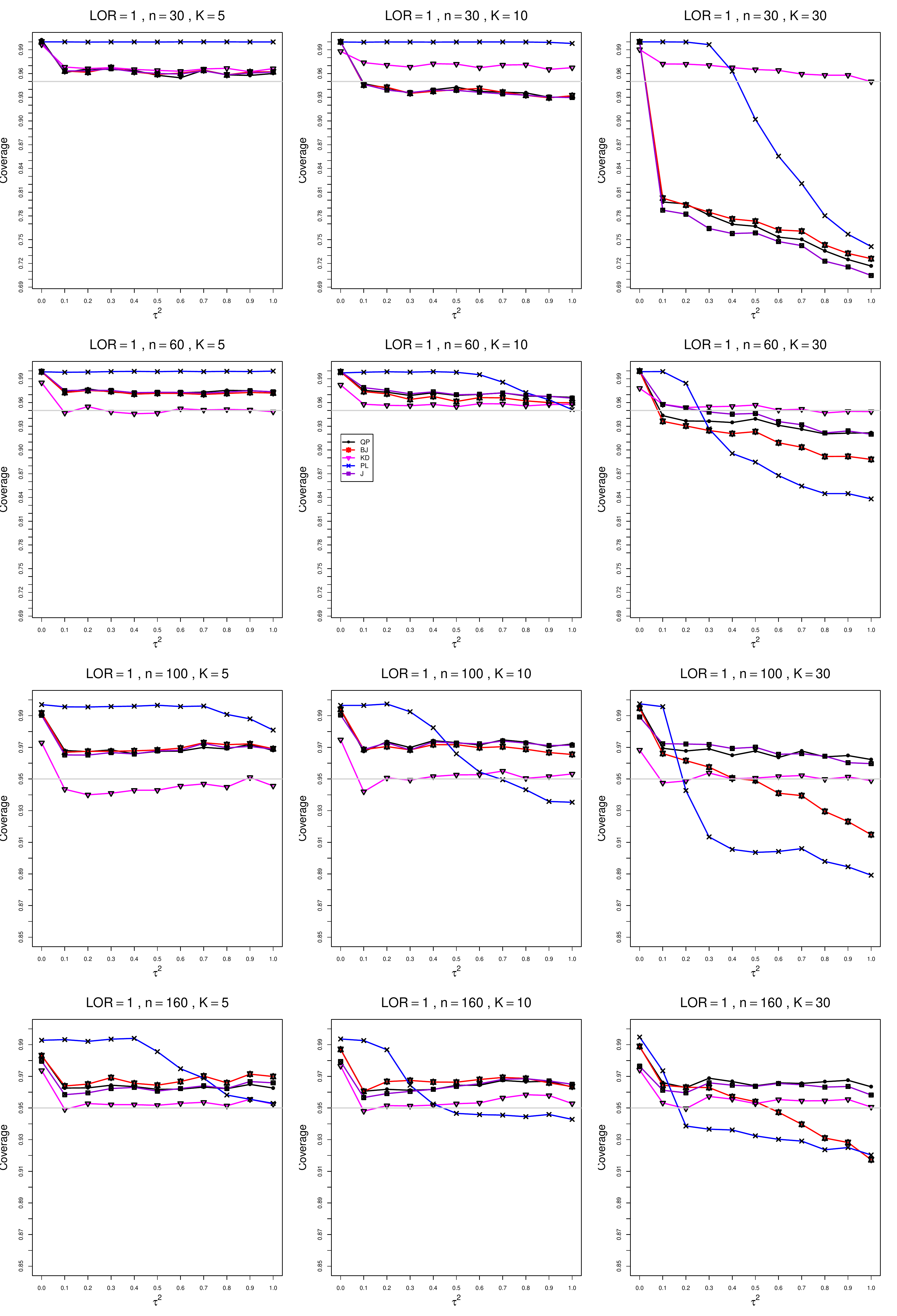}
	\caption{Coverage of  between-studies variance $\tau^2$ for $\theta=1$, $p_{iC}=0.1$, $q=0.75$,
		unequal sample sizes $n=30,\; 60,\;100,\;160$. 
		\label{CovTauLOR1q075piC01_unequal_sample_sizes}}
\end{figure}

\begin{figure}[t]
	\centering
	\includegraphics[scale=0.33]{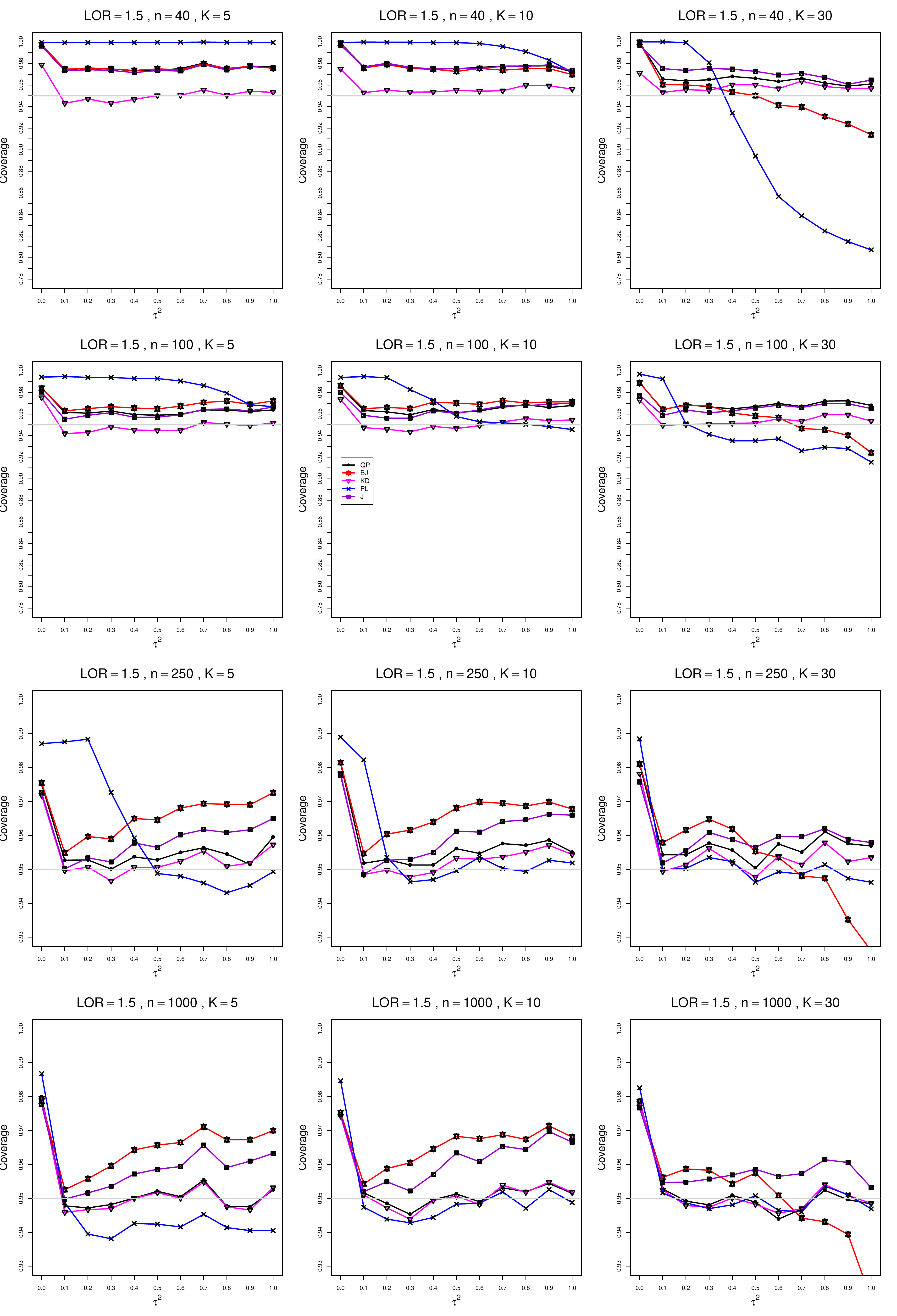}
	\caption{Coverage of  between-studies variance $\tau^2$ for $\theta=1.5$, $p_{iC}=0.1$, $q=0.75$, equal sample sizes $n=40,\;100,\;250,\;1000$. 
		\label{CovTauLOR15q075piC01}}
\end{figure}

\begin{figure}[t]
	\centering
	\includegraphics[scale=0.33]{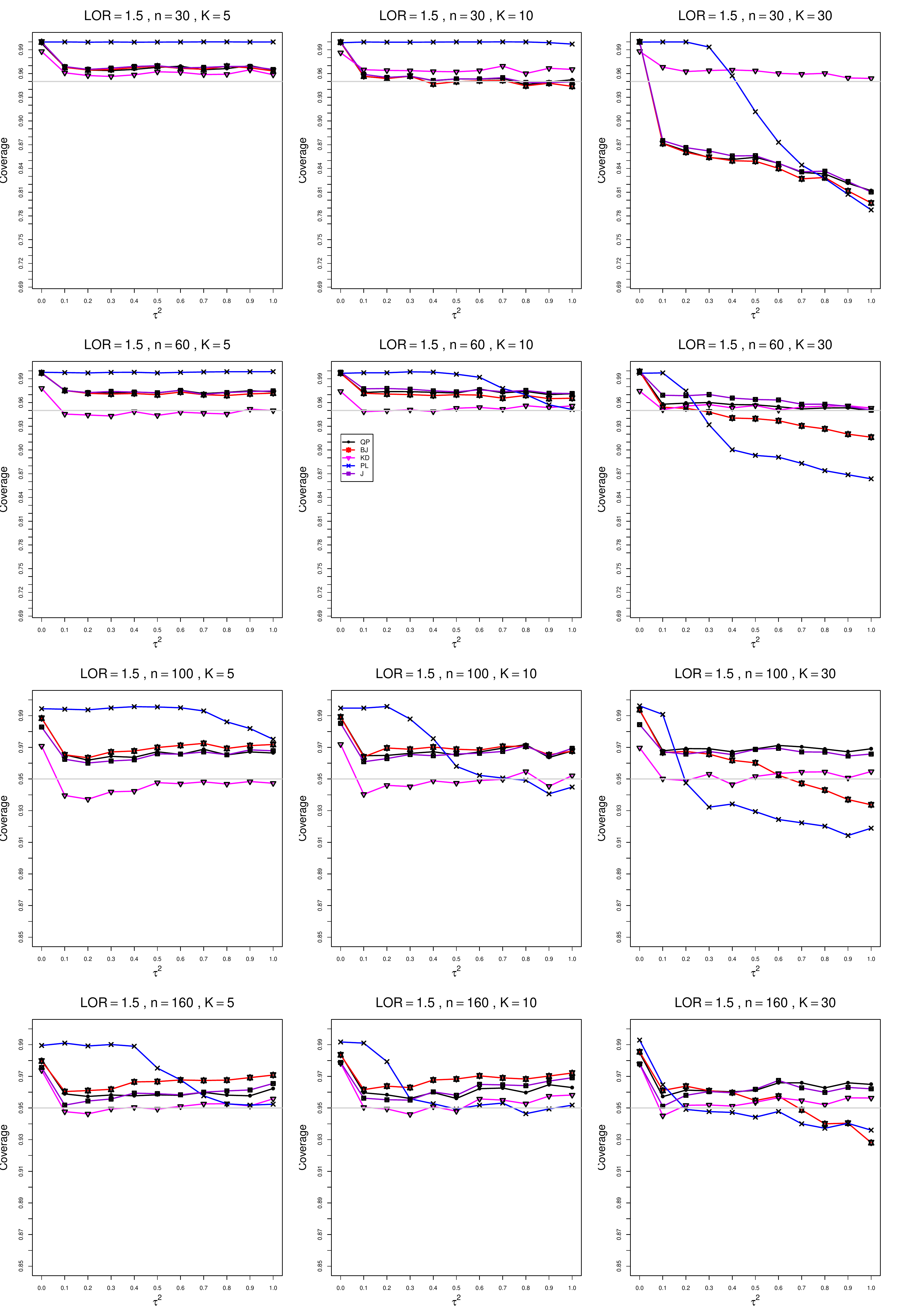}
	\caption{Coverage of  between-studies variance $\tau^2$ for $\theta=1.5$, $p_{iC}=0.1$, $q=0.75$,
		unequal sample sizes $n=30,\; 60,\;100,\;160$. 
		\label{CovTauLOR15q075piC01_unequal_sample_sizes}}
\end{figure}

\clearpage
\begin{figure}[t]
	\centering
	\includegraphics[scale=0.33]{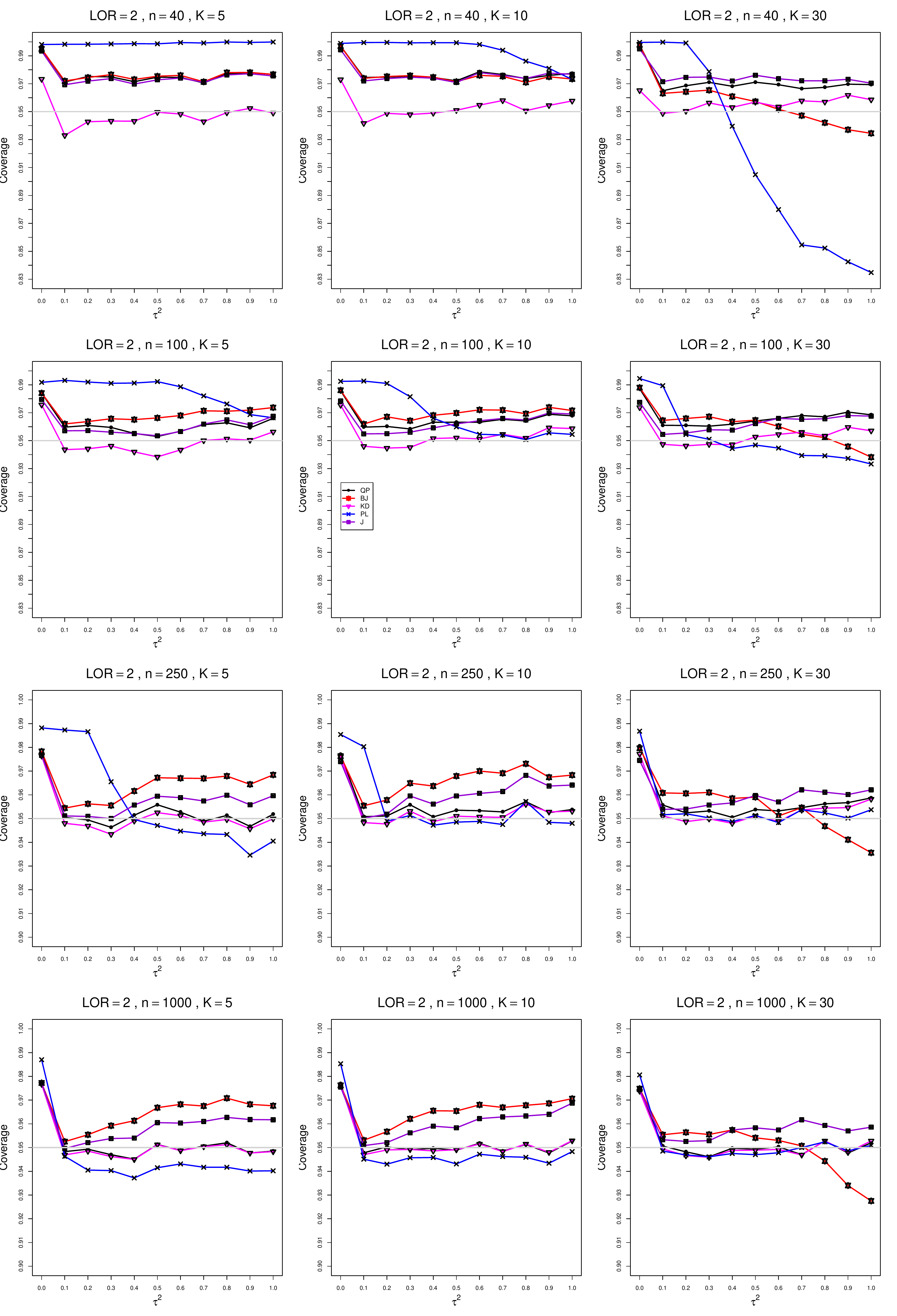}
	\caption{Coverage of  between-studies variance $\tau^2$ for $\theta=2$, $p_{iC}=0.1$, $q=0.75$, equal sample sizes $n=40,\;100,\;250,\;1000$. 
		\label{CovTauLOR2q075piC01}}
\end{figure}

\begin{figure}[t]
	\centering
	\includegraphics[scale=0.33]{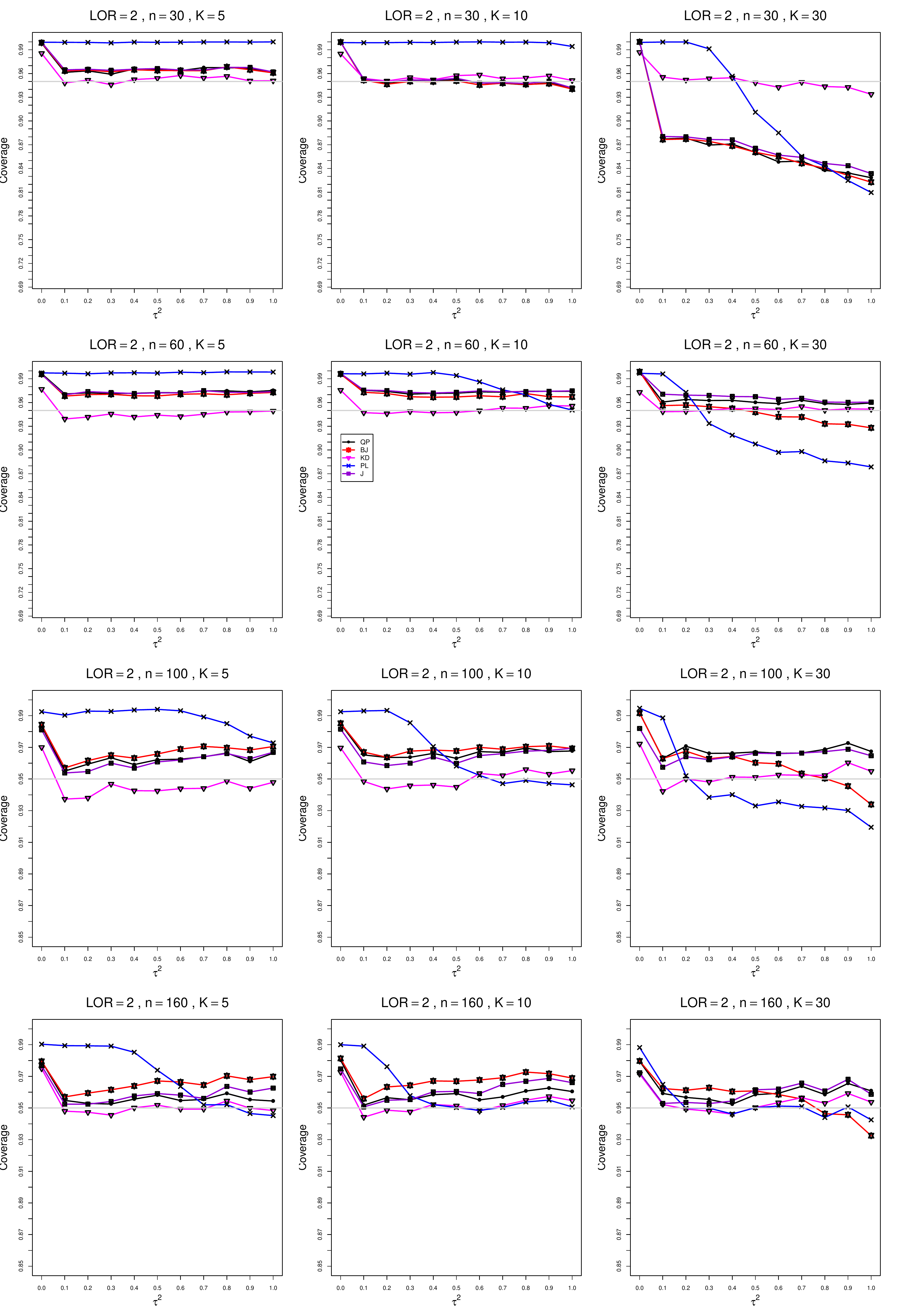}
	\caption{Coverage of  between-studies variance $\tau^2$ for $\theta=2$, $p_{iC}=0.1$, $q=0.75$, 
		unequal sample sizes $n=30,\; 60,\;100,\;160$. 
		\label{CovTauLOR2q075piC01_unequal_sample_sizes}}
\end{figure}
\clearpage
\renewcommand{\thefigure}{A2.2.\arabic{figure}}
\setcounter{figure}{0}
\subsection*{A2.2 Probability in the control arm $p_{C}=0.2$}
\begin{figure}[t]
	\centering
	\includegraphics[scale=0.33]{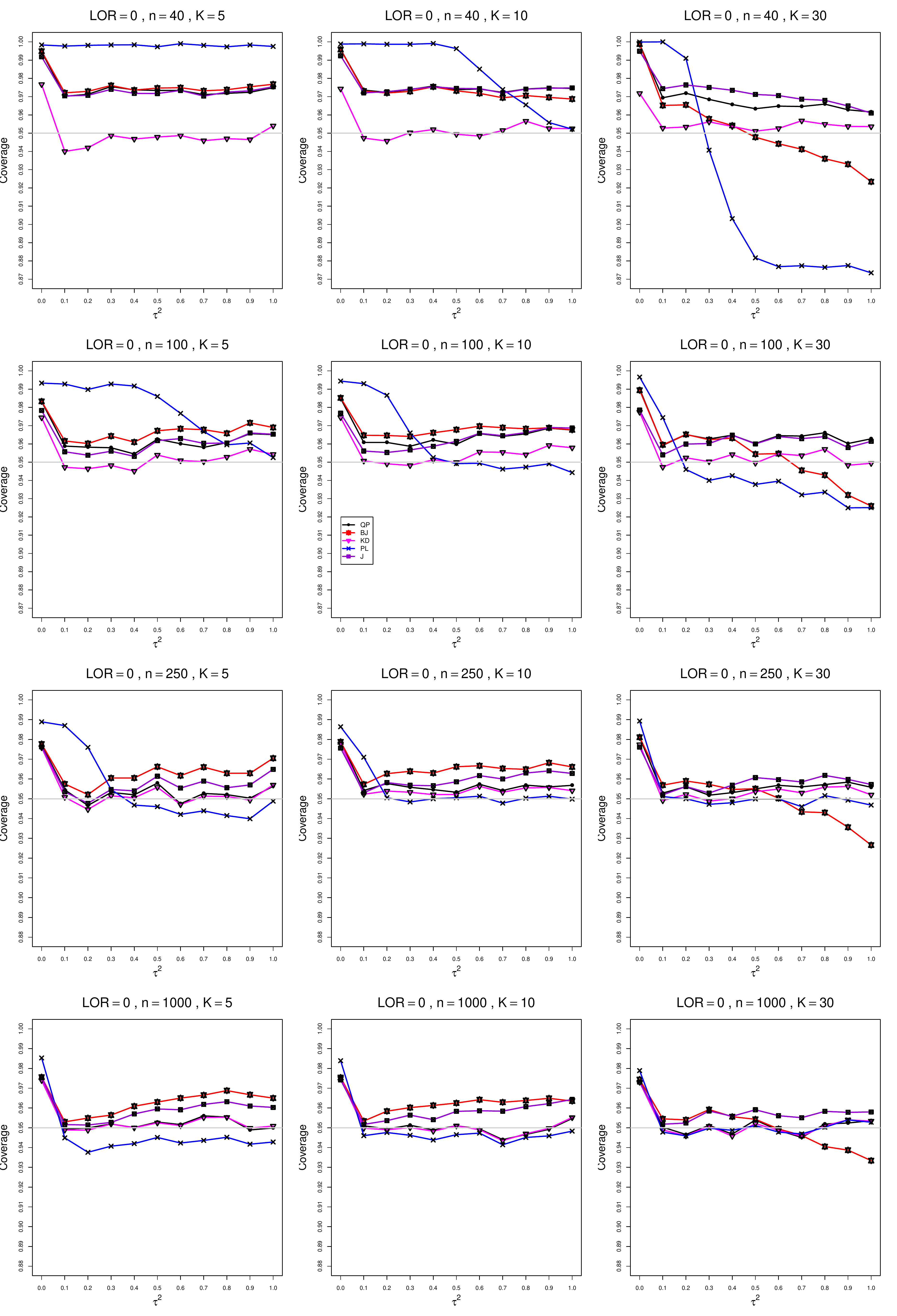}
	\caption{Coverage of  between-studies variance $\tau^2$ for $\theta=0$, $p_{iC}=0.2$, $q=0.5$, equal sample sizes $n=40,\;100,\;250,\;1000$. 
		\label{CovTauLOR0q05piC02}}
\end{figure}

\begin{figure}[t]
	\centering
	\includegraphics[scale=0.33]{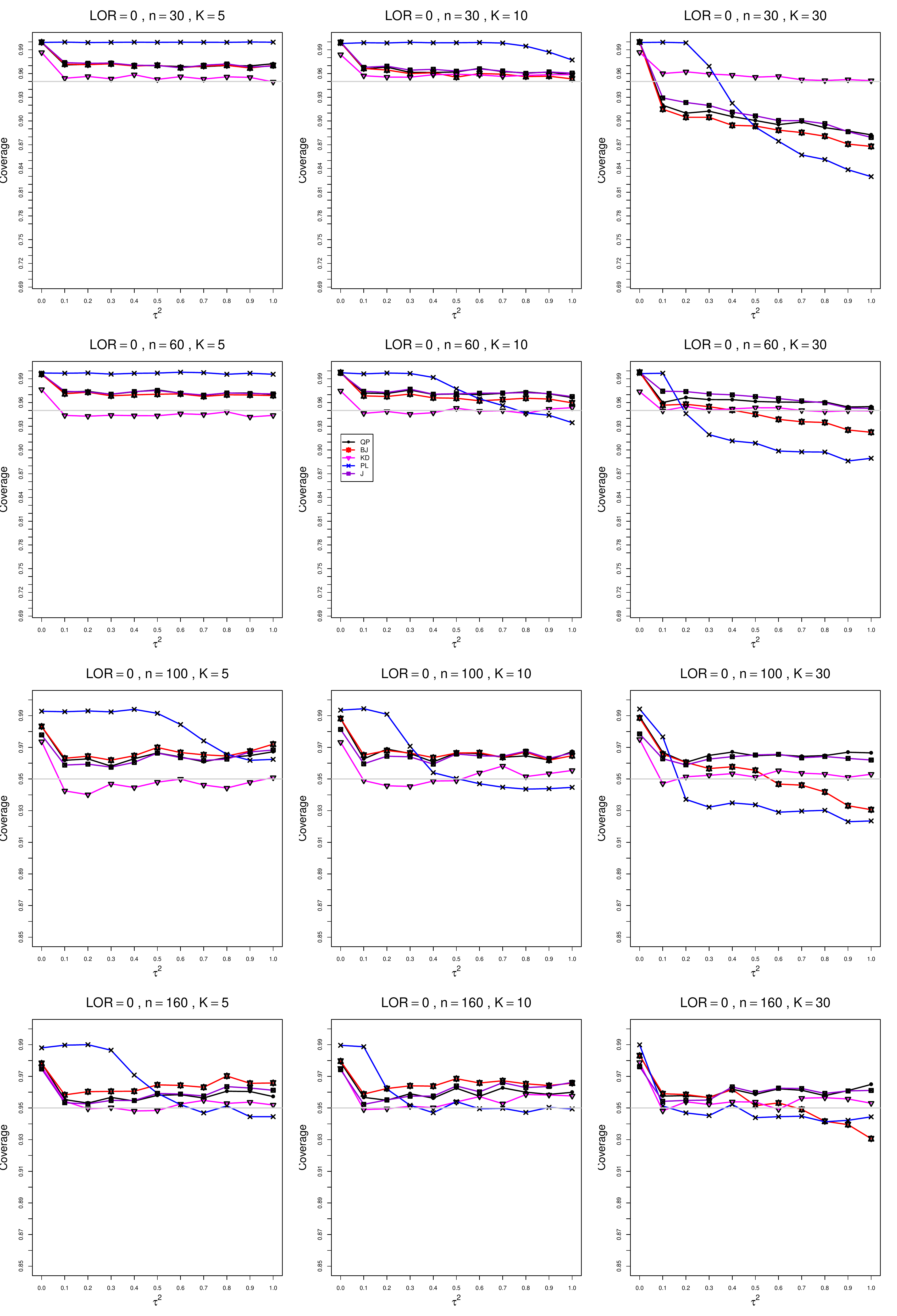}
	\caption{Coverage of  between-studies variance $\tau^2$ for $\theta=0$, $p_{iC}=0.2$, $q=0.5$,
		unequal sample sizes $n=30,\; 60,\;100,\;160$. 
		\label{CovTauLOR0q05piC02_unequal_sample_sizes}}
\end{figure}

\begin{figure}[t]
	\centering
	\includegraphics[scale=0.33]{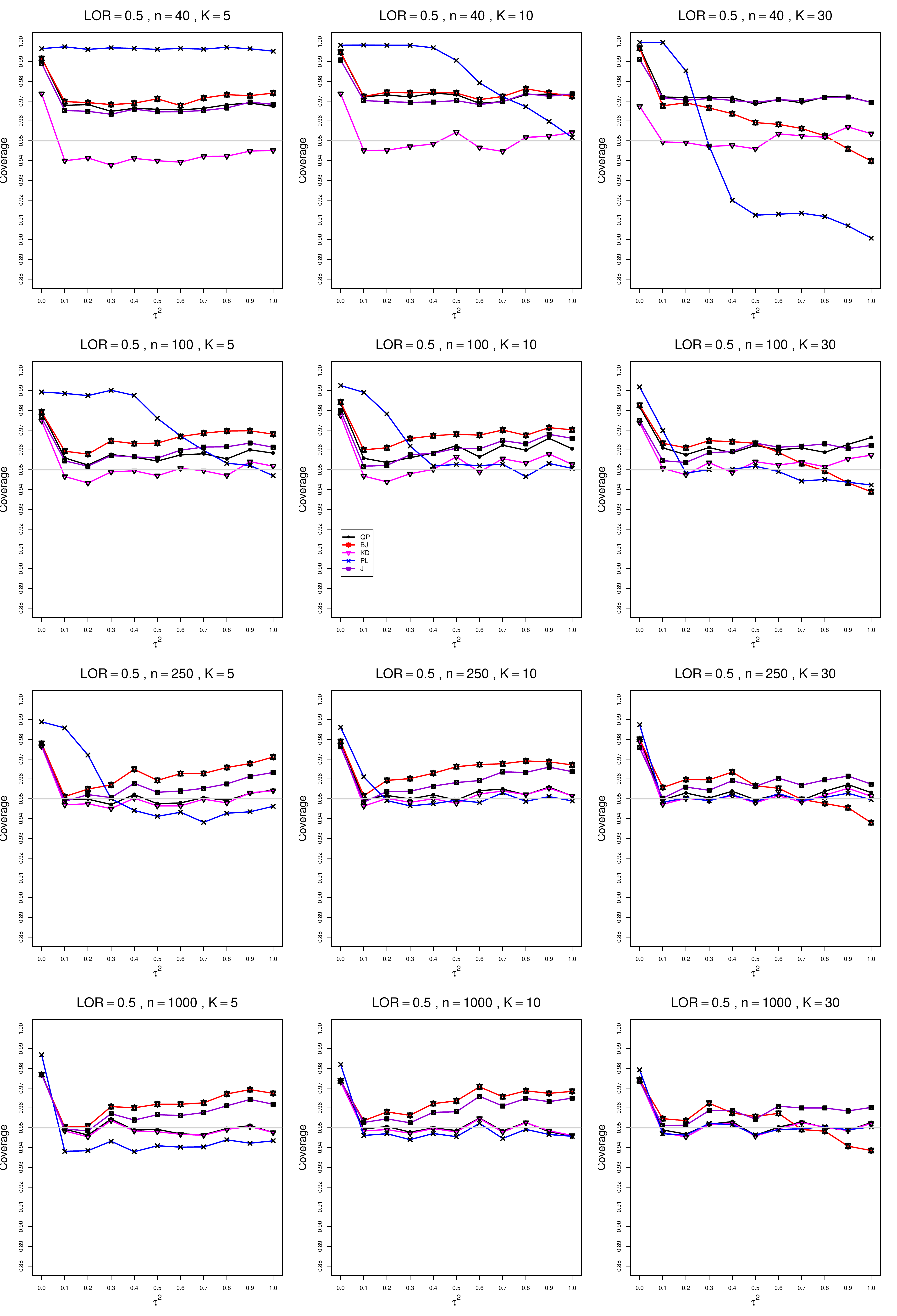}
	\caption{Coverage of  between-studies variance $\tau^2$ for $\theta=0.5$, $p_{iC}=0.2$, $q=0.5$, equal sample sizes $n=40,\;100,\;250,\;1000$. 
		\label{CovTauLOR05q05piC02}}
\end{figure}

\begin{figure}[t]
	\centering
	\includegraphics[scale=0.33]{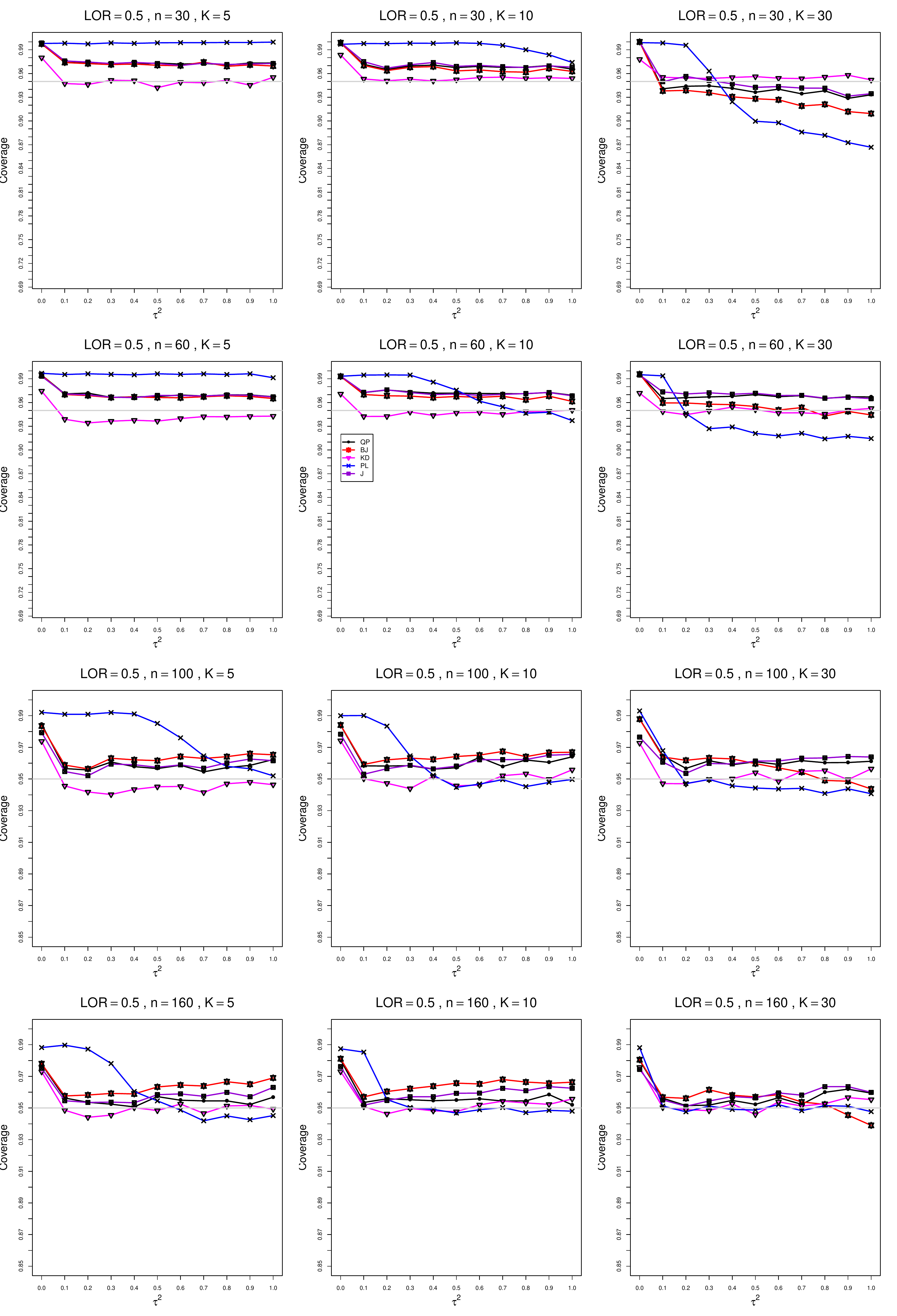}
	\caption{Coverage of  between-studies variance $\tau^2$ for $\theta=0.5$, $p_{iC}=0.2$, $q=0.5$,
		unequal sample sizes $n=30,\; 60,\;100,\;160$. 
		\label{CovTauLOR05q05piC02_unequal_sample_sizes}}
\end{figure}

\begin{figure}[t]
	\centering
	\includegraphics[scale=0.33]{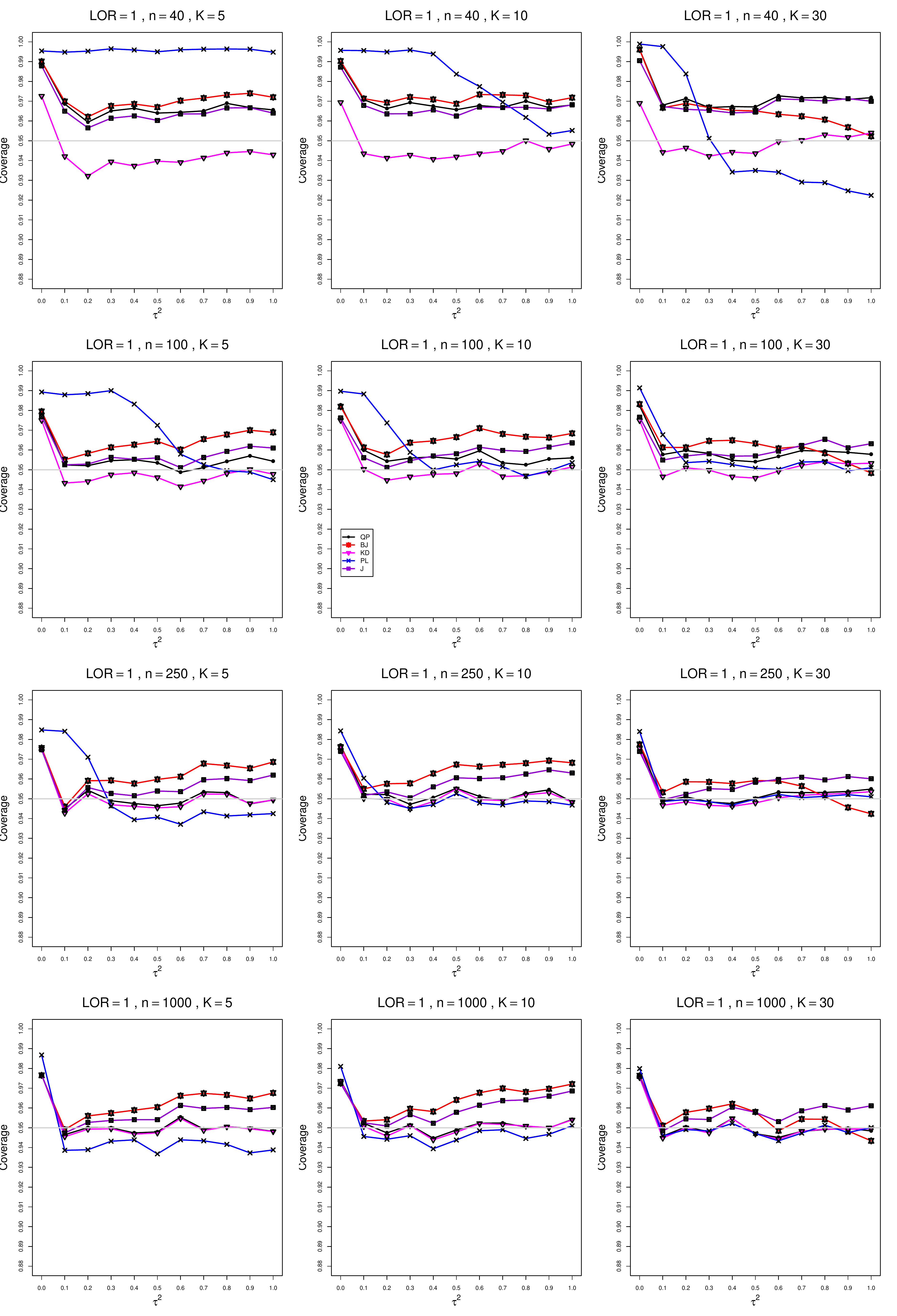}
	\caption{Coverage of  between-studies variance $\tau^2$ for $\theta=1$, $p_{iC}=0.2$, $q=0.5$, equal sample sizes $n=40,\;100,\;250,\;1000$. 
		\label{CovTauLOR1q05piC02}}
\end{figure}

\begin{figure}[t]
	\centering
	\includegraphics[scale=0.33]{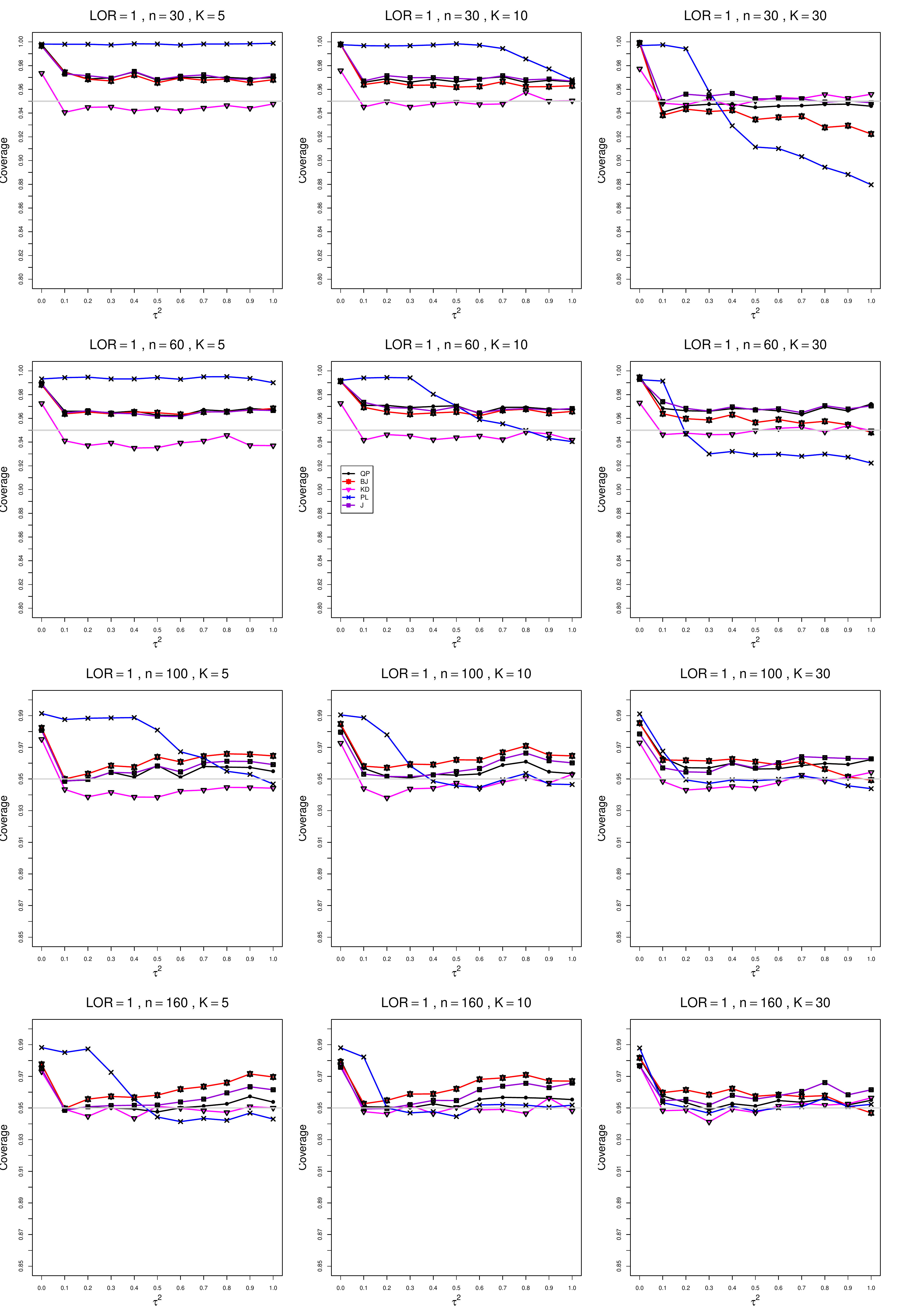}
	\caption{Coverage of  between-studies variance $\tau^2$ for $\theta=1$, $p_{iC}=0.2$, $q=0.5$,
		unequal sample sizes $n=30,\; 60,\;100,\;160$. 
		\label{CovTauLOR1q05piC02_unequal_sample_sizes}}
\end{figure}

\begin{figure}[t]
	\centering
	\includegraphics[scale=0.33]{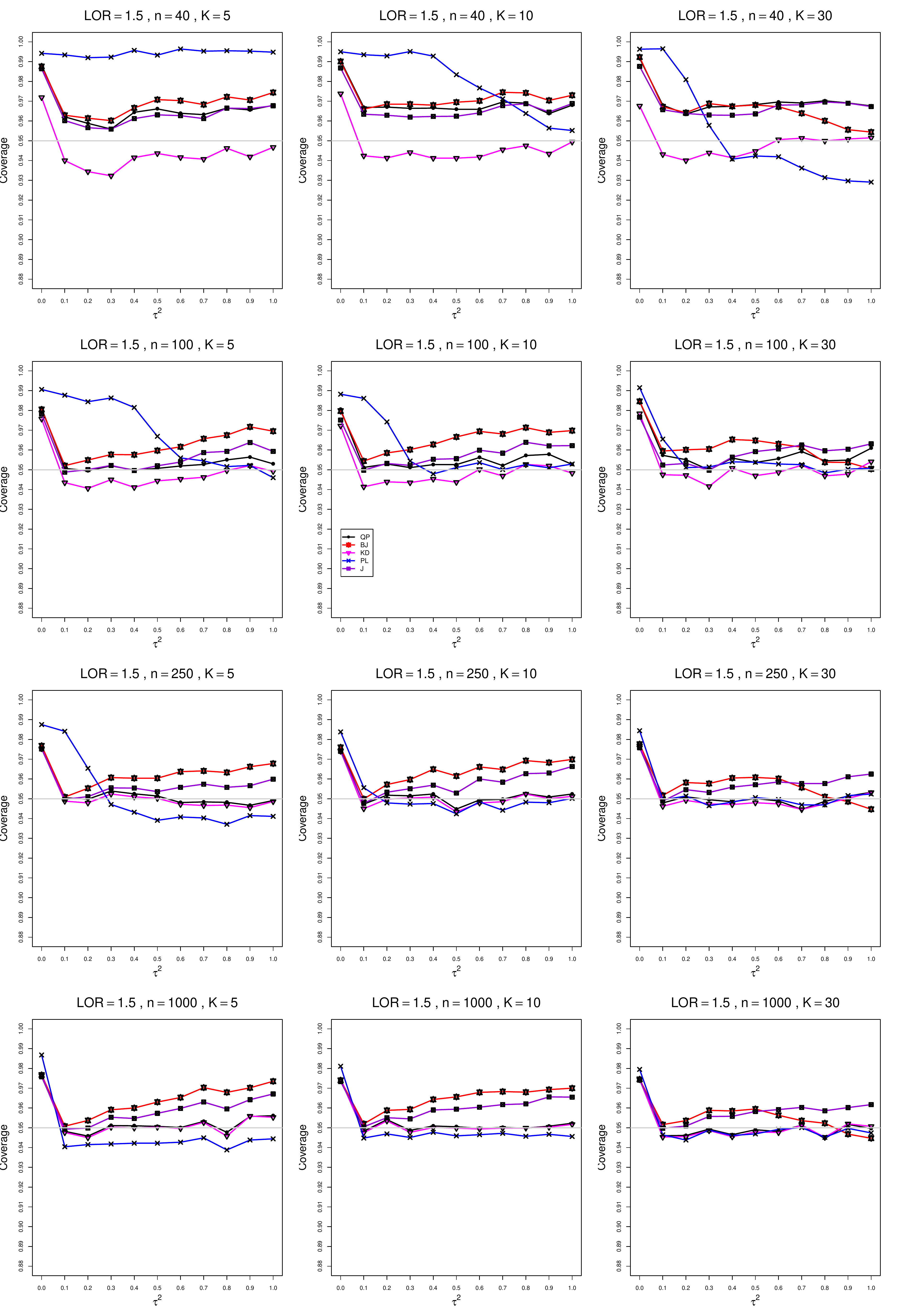}
	\caption{Coverage of  between-studies variance $\tau^2$ for $\theta=1.5$, $p_{iC}=0.2$, $q=0.5$, equal sample sizes $n=40,\;100,\;250,\;1000$. 
		\label{CovTauLOR15q05piC02}}
\end{figure}

\begin{figure}[t]
	\centering
	\includegraphics[scale=0.33]{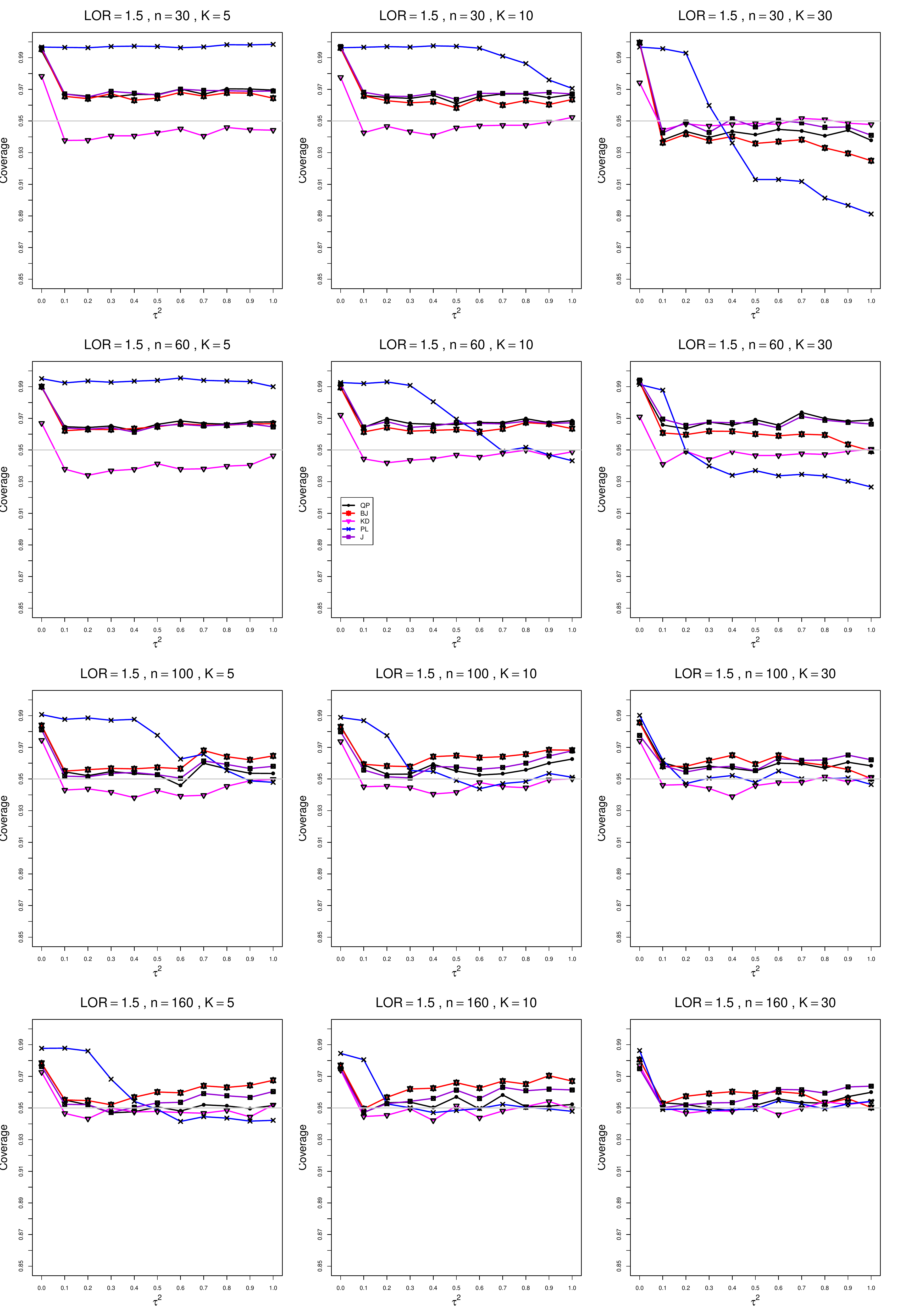}
	\caption{Coverage of  between-studies variance $\tau^2$ for $\theta=1.5$, $p_{iC}=0.2$, $q=0.5$,
		unequal sample sizes $n=30,\; 60,\;100,\;160$. 
		\label{CovTauLOR15q05piC02_unequal_sample_sizes}}
\end{figure}

\begin{figure}[t]
	\centering
	\includegraphics[scale=0.33]{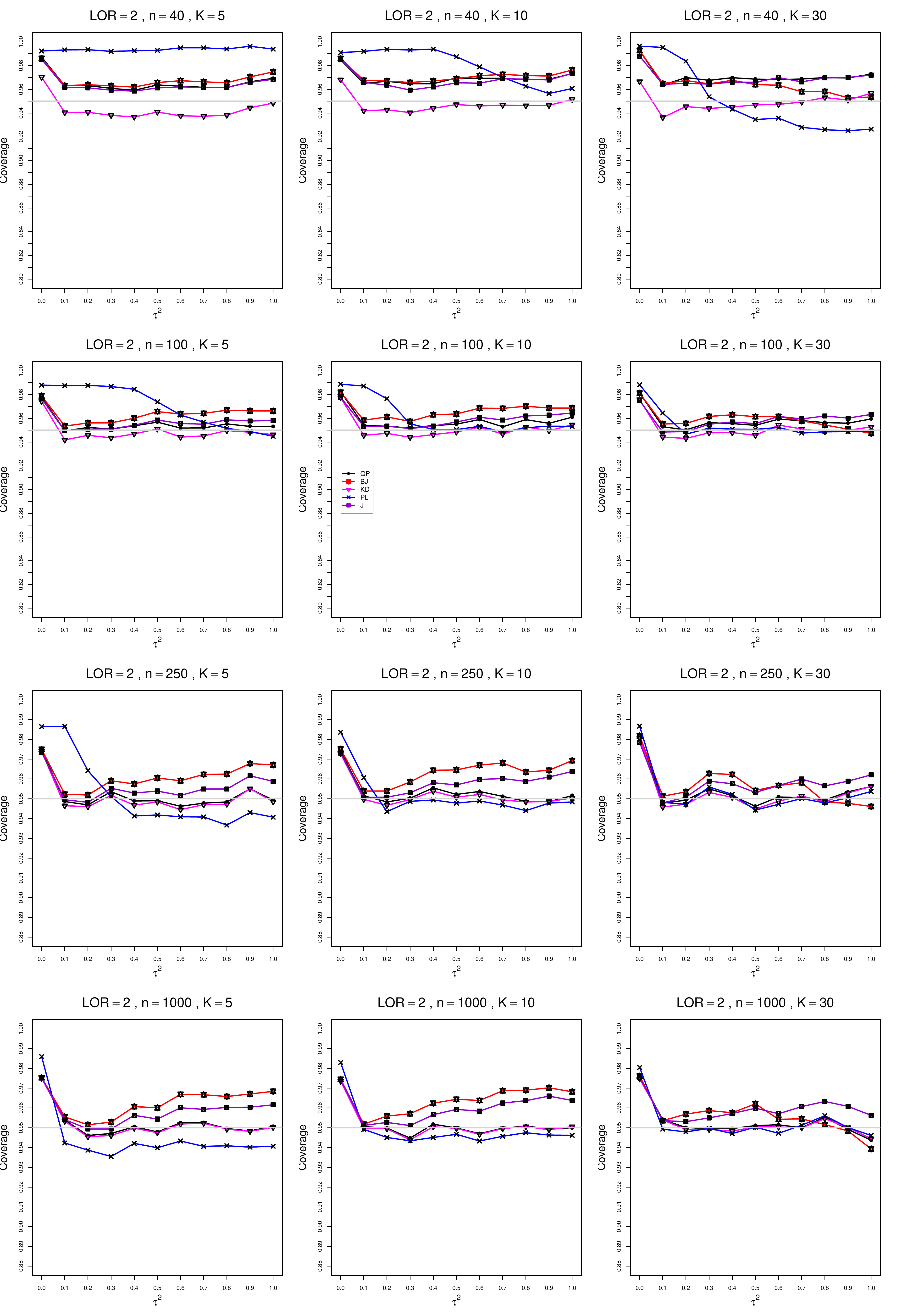}
	\caption{Coverage of  between-studies variance $\tau^2$ for $\theta=2$, $p_{iC}=0.2$, $q=0.5$, equal sample sizes $n=40,\;100,\;250,\;1000$. 
		\label{CovTauLOR2q05piC02}}
\end{figure}

\begin{figure}[t]
	\centering
	\includegraphics[scale=0.33]{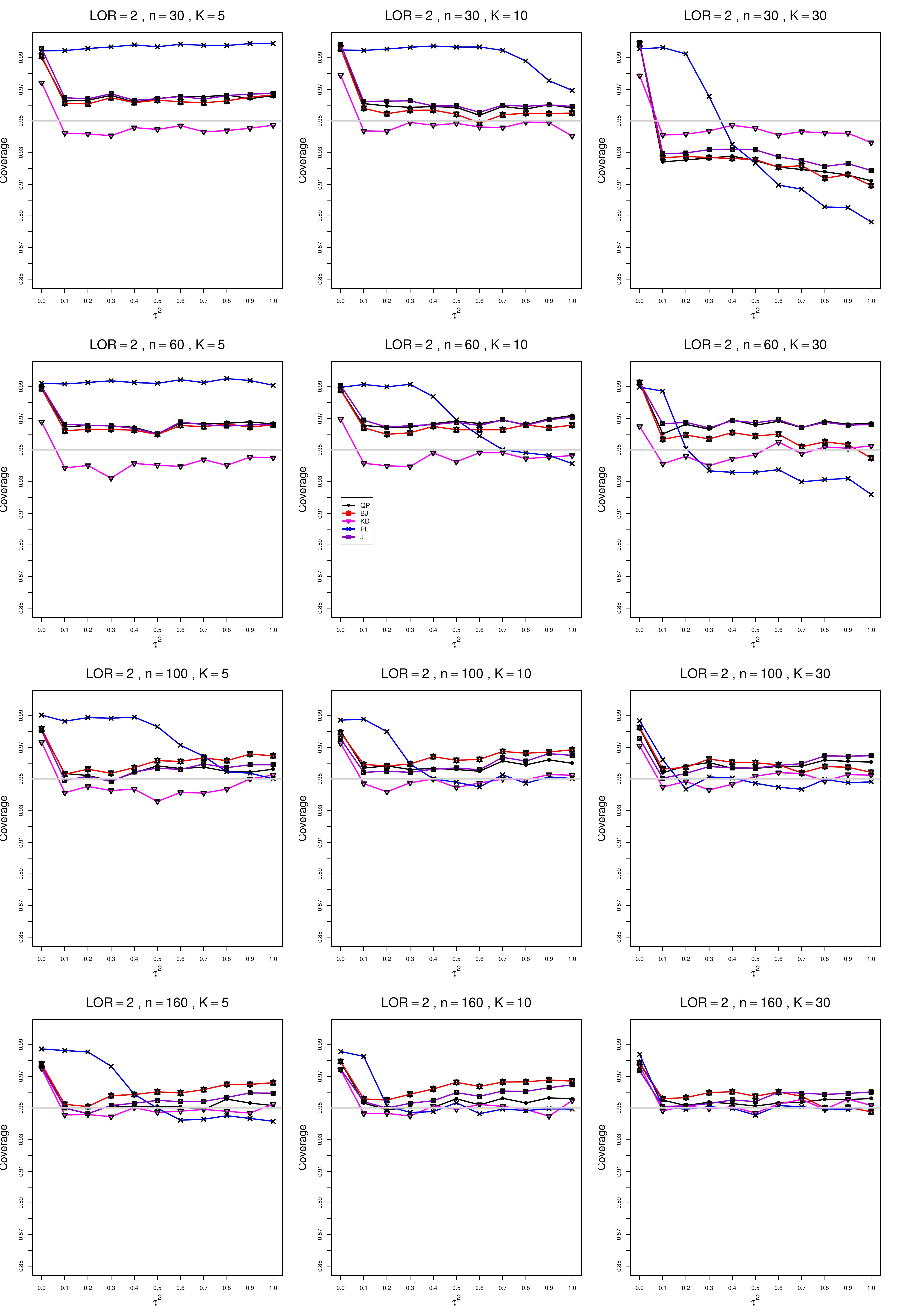}
	\caption{Coverage of  between-studies variance $\tau^2$ for $\theta=2$, $p_{iC}=0.2$, $q=0.5$,
		unequal sample sizes $n=30,\; 60,\;100,\;160$. 
		\label{CovTauLOR2q05piC02_unequal_sample_sizes}}
\end{figure}


\begin{figure}[t]
	\centering
	\includegraphics[scale=0.33]{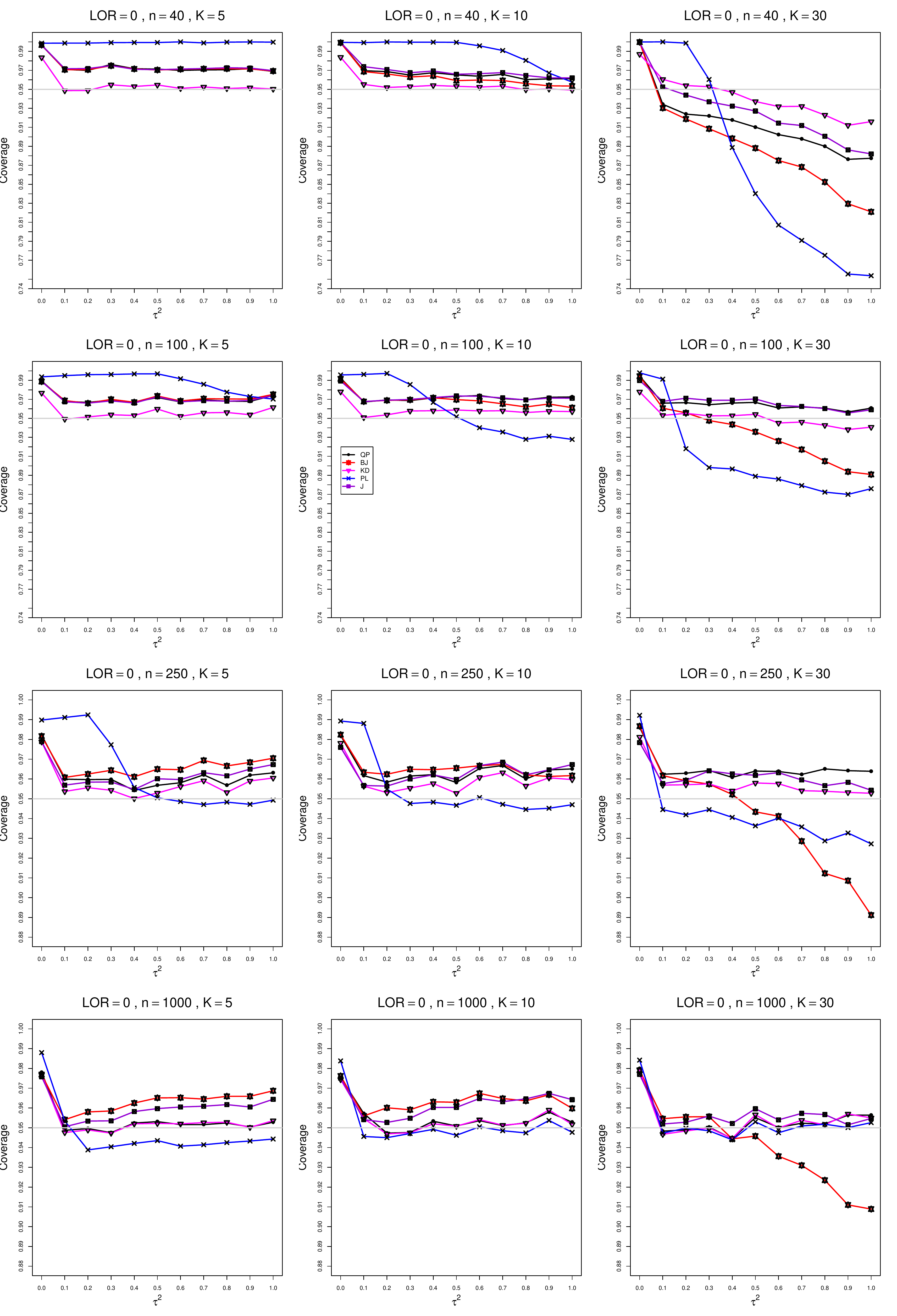}
	\caption{Coverage of  between-studies variance $\tau^2$ for $\theta=0$, $p_{iC}=0.2$, $q=0.75$, equal sample sizes $n=40,\;100,\;250,\;1000$. 
		\label{CovTauLOR0q075piC02}}
\end{figure}

\begin{figure}[t]
	\centering
	\includegraphics[scale=0.33]{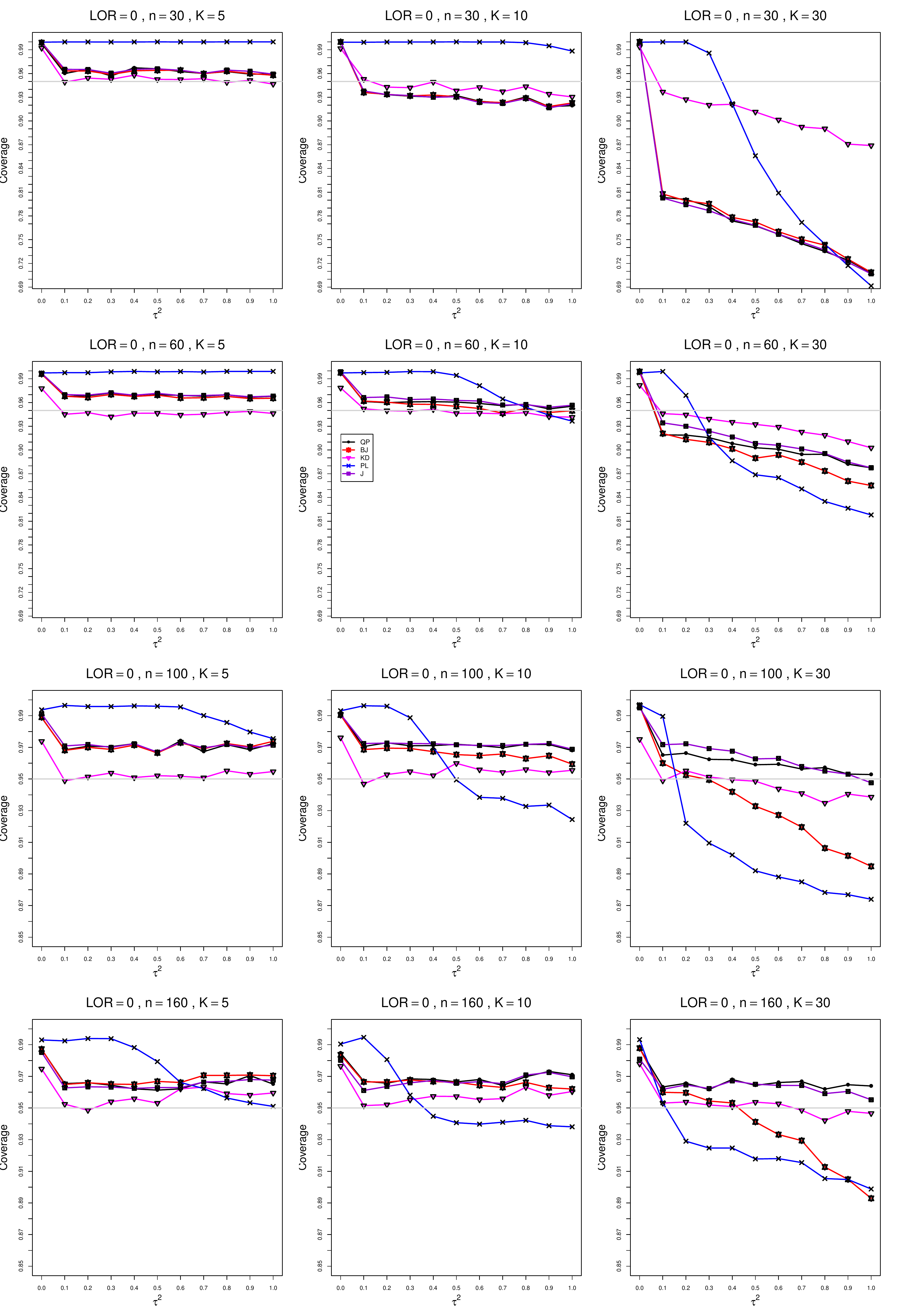}
	\caption{Coverage of  between-studies variance $\tau^2$ for $\theta=0$, $p_{iC}=0.2$, $q=0.75$,
		unequal sample sizes $n=30,\; 60,\;100,\;160$. 
		\label{CovTauLOR0q075piC02_unequal_sample_sizes}}
\end{figure}

\begin{figure}[t]
	\centering
	\includegraphics[scale=0.33]{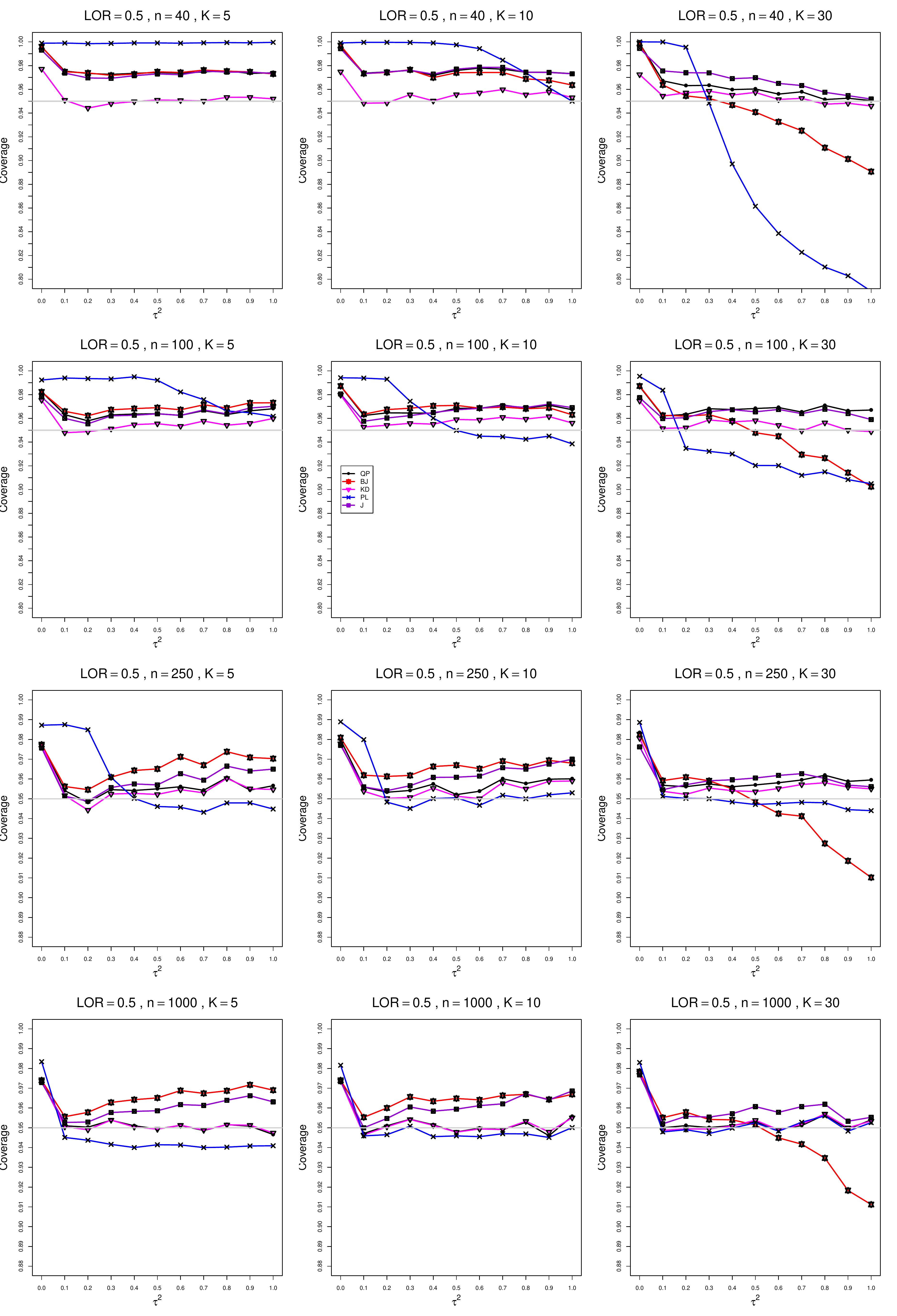}
	\caption{Coverage of  between-studies variance $\tau^2$ for $\theta=0.5$, $p_{iC}=0.2$, $q=0.75$, equal sample sizes $n=40,\;100,\;250,\;1000$. 
		\label{CovTauLOR05q075piC02}}
\end{figure}

\begin{figure}[t]
	\centering
	\includegraphics[scale=0.33]{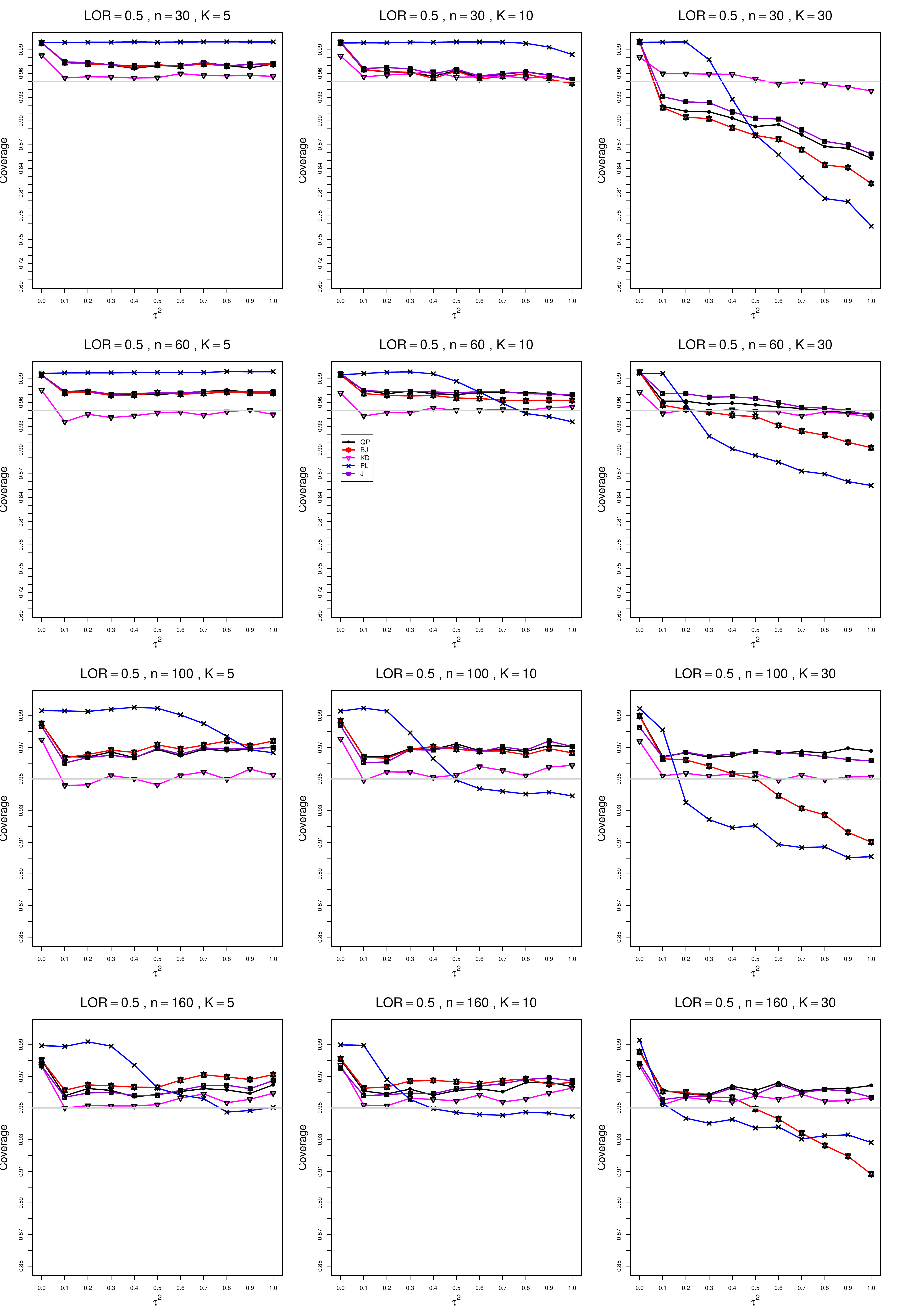}
	\caption{Coverage of  between-studies variance $\tau^2$ for $\theta=0.5$, $p_{iC}=0.2$, $q=0.75$,
		unequal sample sizes $n=30,\; 60,\;100,\;160$. 
		\label{CovTauLOR05q075piC02_unequal_sample_sizes}}
\end{figure}
\clearpage

\begin{figure}[t]
	\centering
	\includegraphics[scale=0.33]{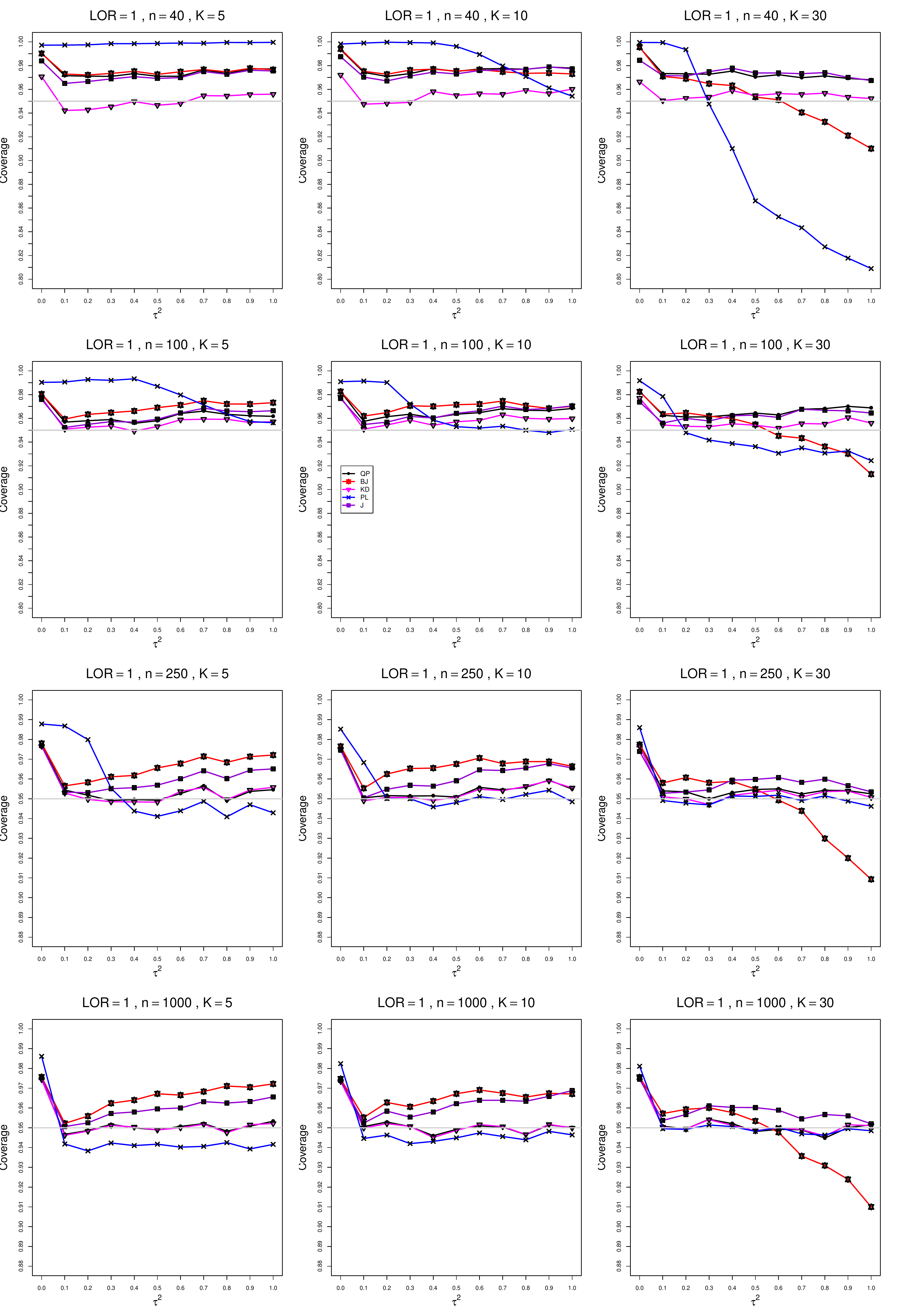}
	\caption{Coverage of  between-studies variance $\tau^2$ for $\theta=1$, $p_{iC}=0.2$, $q=0.75$, equal sample sizes $n=40,\;100,\;250,\;1000$. 
		\label{CovTauLOR1q075piC02}}
\end{figure}

\begin{figure}[t]
	\centering
	\includegraphics[scale=0.33]{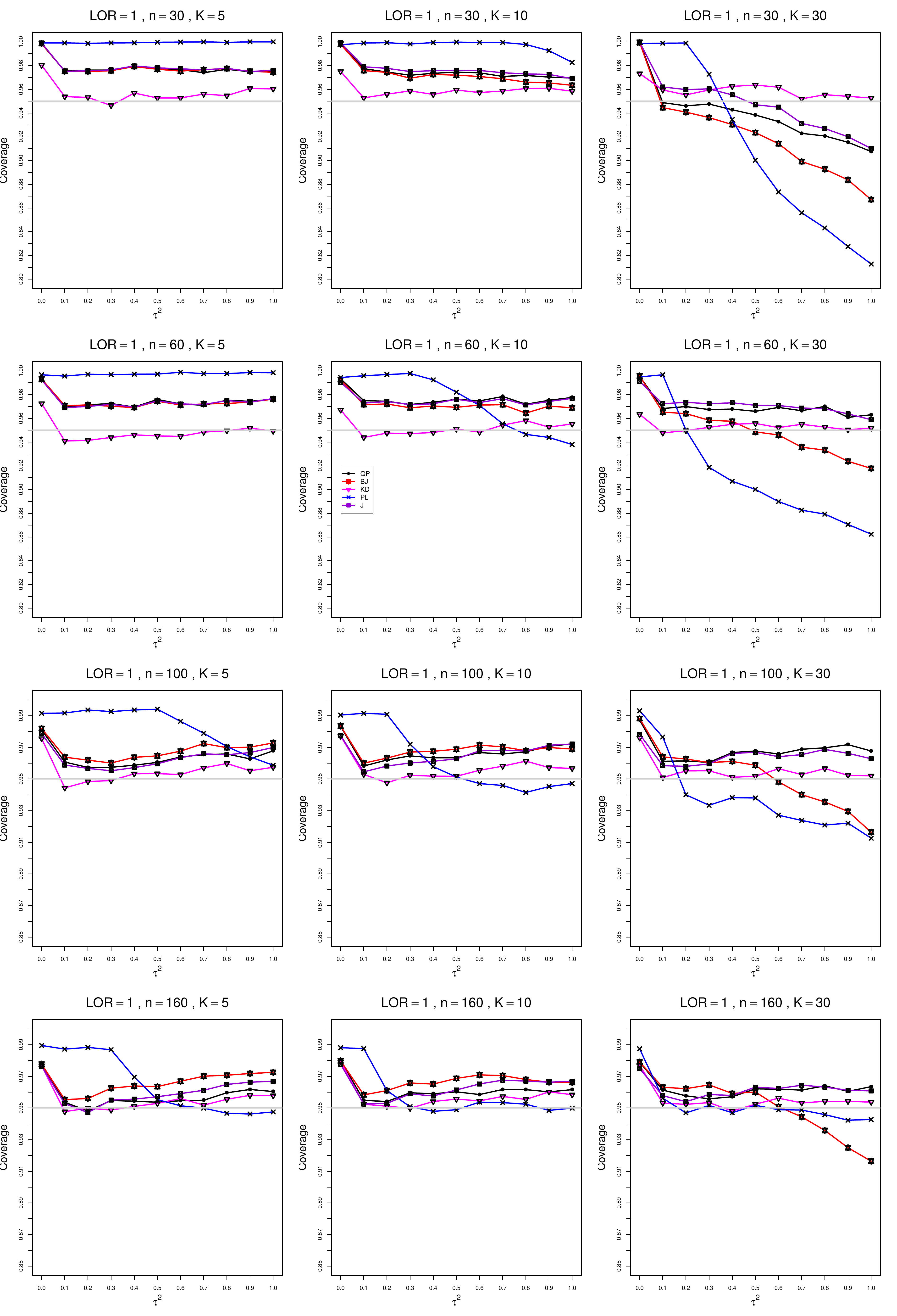}
	\caption{Coverage of  between-studies variance $\tau^2$ for $\theta=1$, $p_{iC}=0.2$, $q=0.75$,
		unequal sample sizes $n=30,\; 60,\;100,\;160$. 
		\label{CovTauLOR1q075piC02_unequal_sample_sizes}}
\end{figure}

\begin{figure}[t]
	\centering
	\includegraphics[scale=0.33]{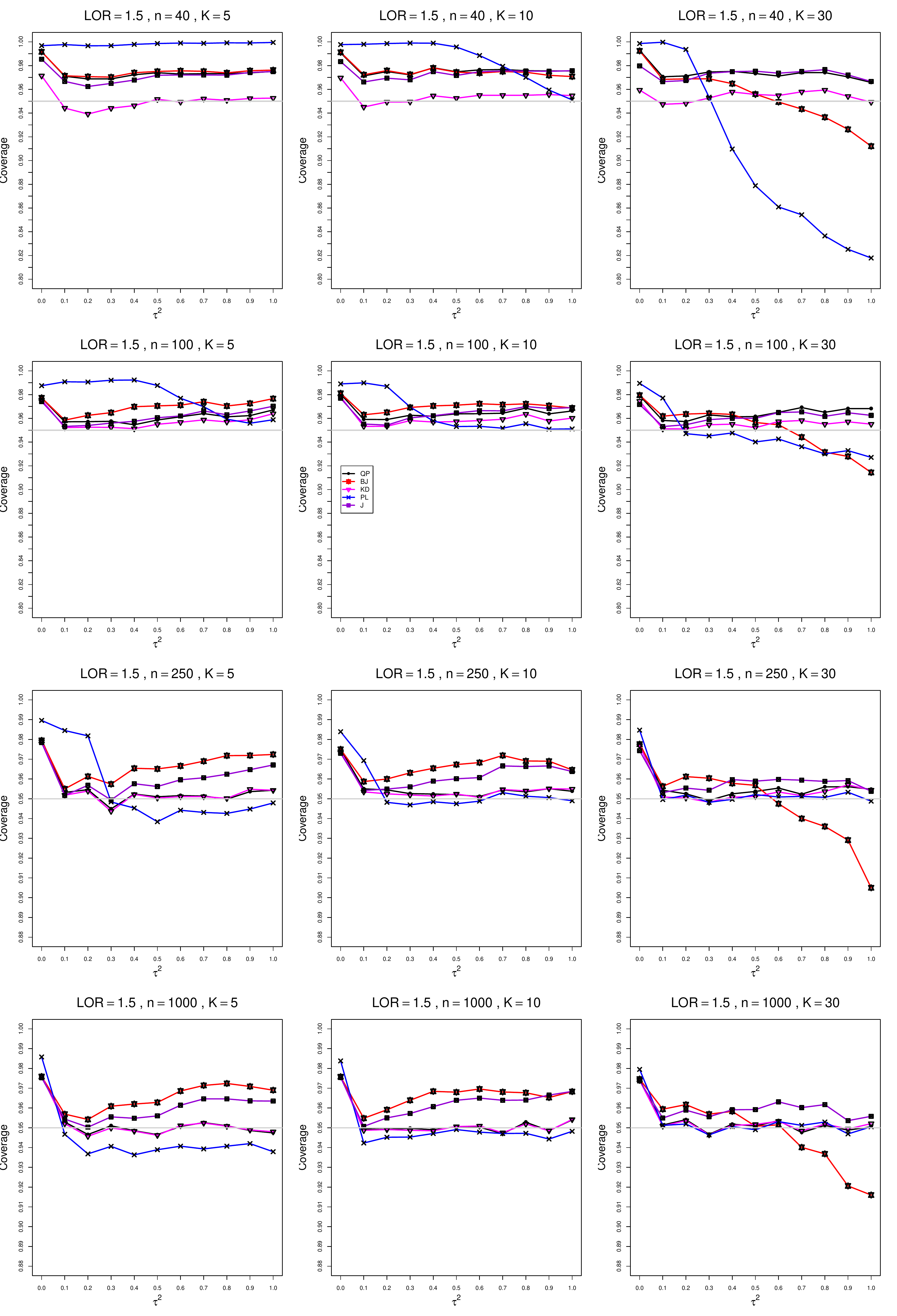}
	\caption{Coverage of  between-studies variance $\tau^2$ for $\theta=1.5$, $p_{iC}=0.2$, $q=0.75$, equal sample sizes $n=40,\;100,\;250,\;1000$. 
		\label{CovTauLOR15q075piC02}}
\end{figure}

\begin{figure}[t]
	\centering
	\includegraphics[scale=0.33]{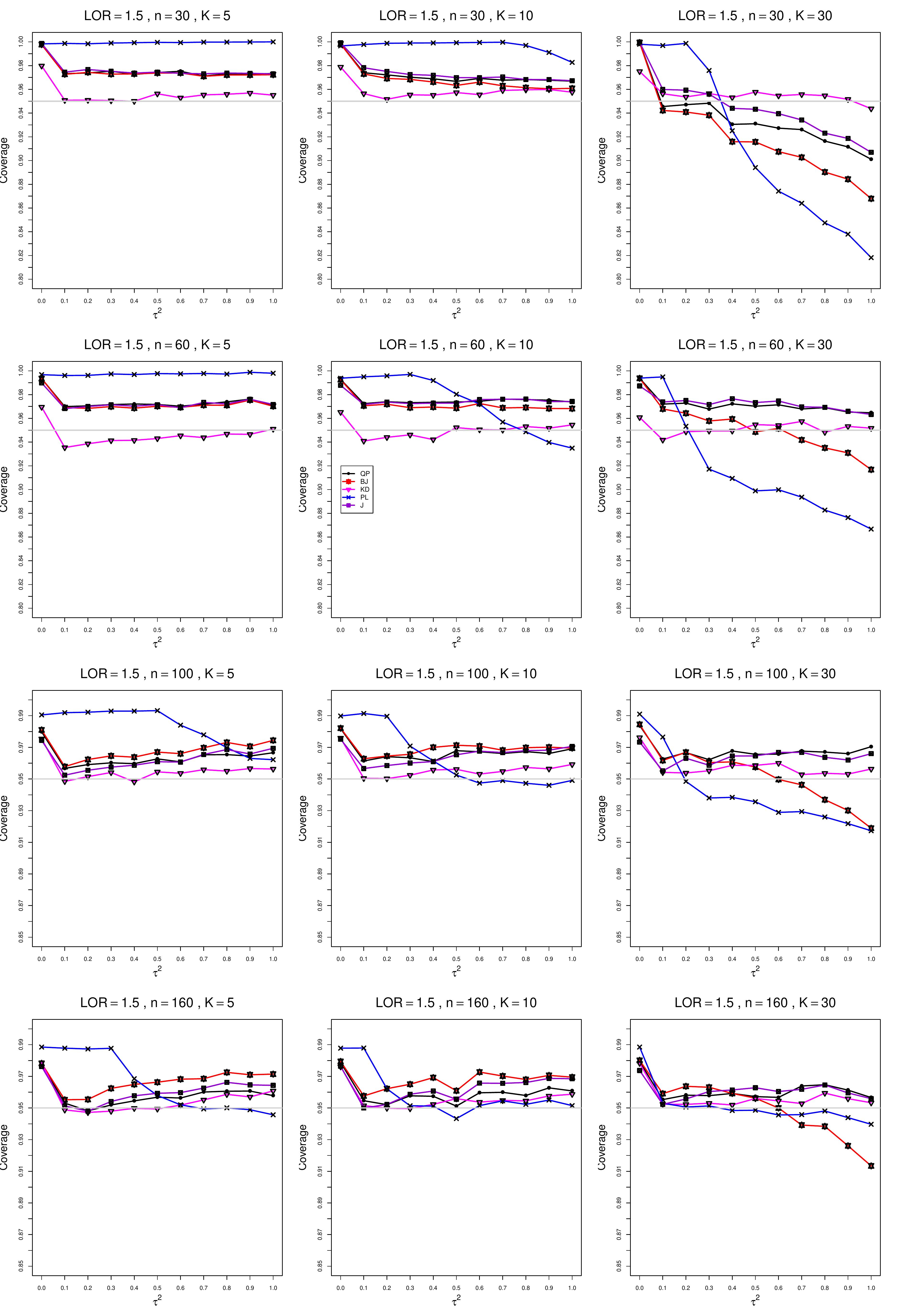}
	\caption{Coverage of  between-studies variance $\tau^2$ for $\theta=1.5$, $p_{iC}=0.2$, $q=0.75$,
		unequal sample sizes $n=30,\; 60,\;100,\;160$. 
		\label{CovTauLOR15q075piC02_unequal_sample_sizes}}
\end{figure}

\begin{figure}[t]
	\centering
	\includegraphics[scale=0.33]{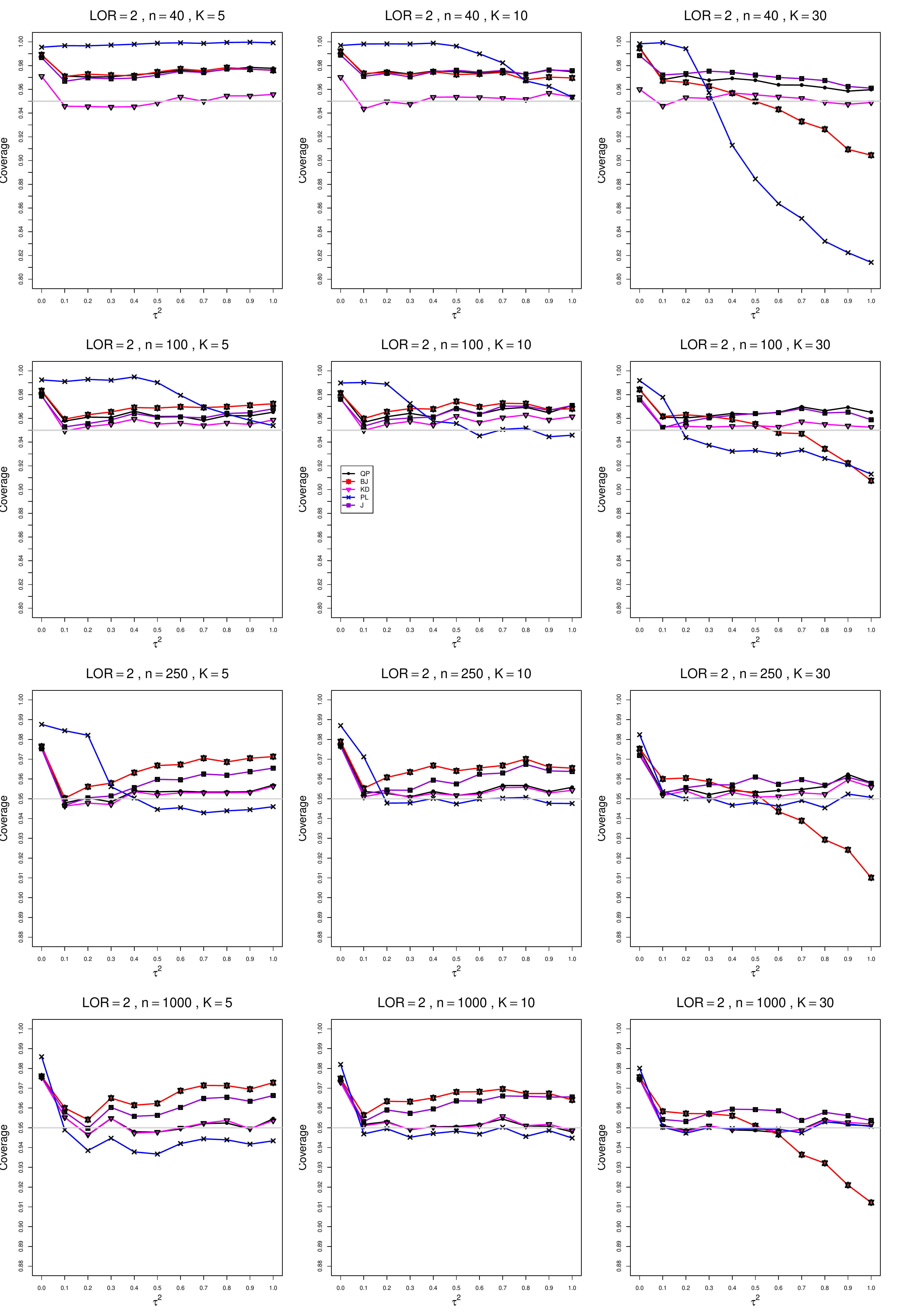}
	\caption{Coverage of  between-studies variance $\tau^2$ for $\theta=2$, $p_{iC}=0.2$, $q=0.75$, equal sample sizes  $n=40,\;100,\;250,\;1000$. 
		\label{CovTauLOR2q075piC02}}
\end{figure}

\begin{figure}[t]
	\centering
	\includegraphics[scale=0.33]{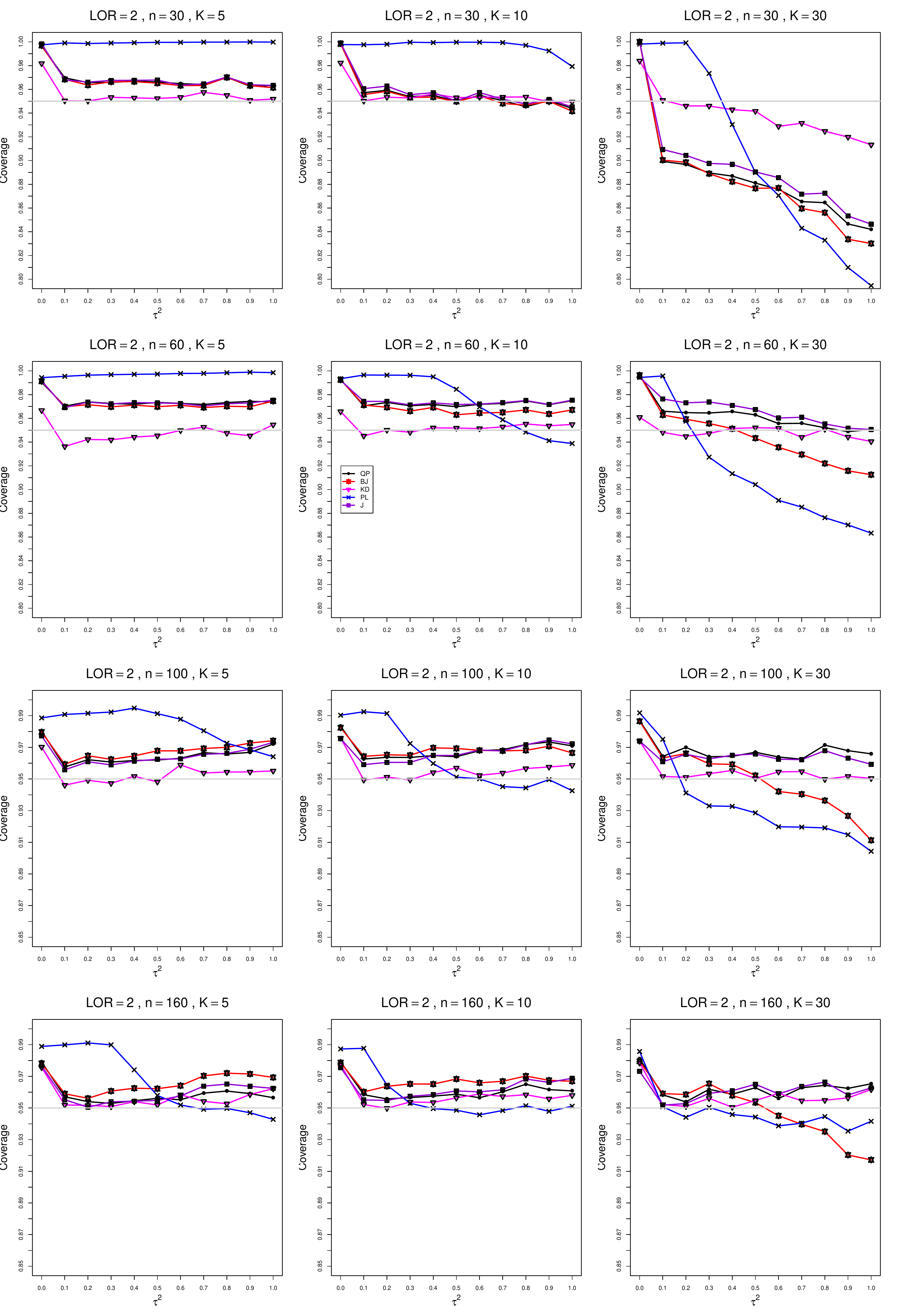}
	\caption{Coverage of  between-studies variance $\tau^2$ for $\theta=2$, $p_{iC}=0.2$, $q=0.75$,
		unequal sample sizes $n=30,\; 60,\;100,\;160$. 
		\label{CovTauLOR2q075piC02_unequal_sample_sizes}}
\end{figure}
\clearpage
\renewcommand{\thefigure}{A2.3.\arabic{figure}}
\setcounter{figure}{0}
\subsection*{A2.3 Probability in the control arm $p_{C}=0.4$}
\begin{figure}[t]
	\centering
	\includegraphics[scale=0.33]{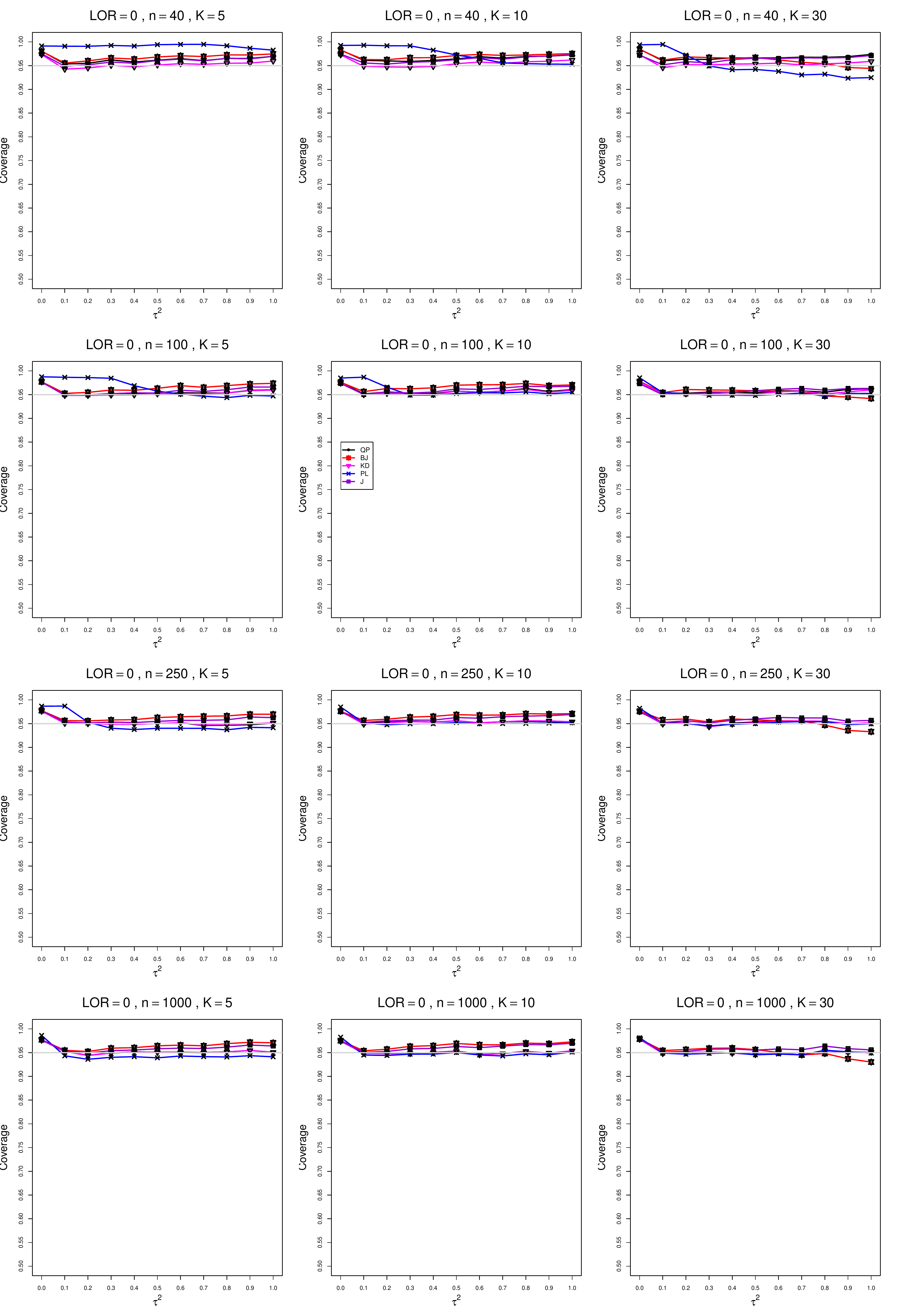}
	\caption{Coverage of  between-studies variance $\tau^2$ for $\theta=0$, $p_{iC}=0.4$, $q=0.5$, equal sample sizes $n=40,\;100,\;250,\;1000$. 
		\label{CovTauLOR0q05piC04}}
\end{figure}

\begin{figure}[t]
	\centering
	\includegraphics[scale=0.33]{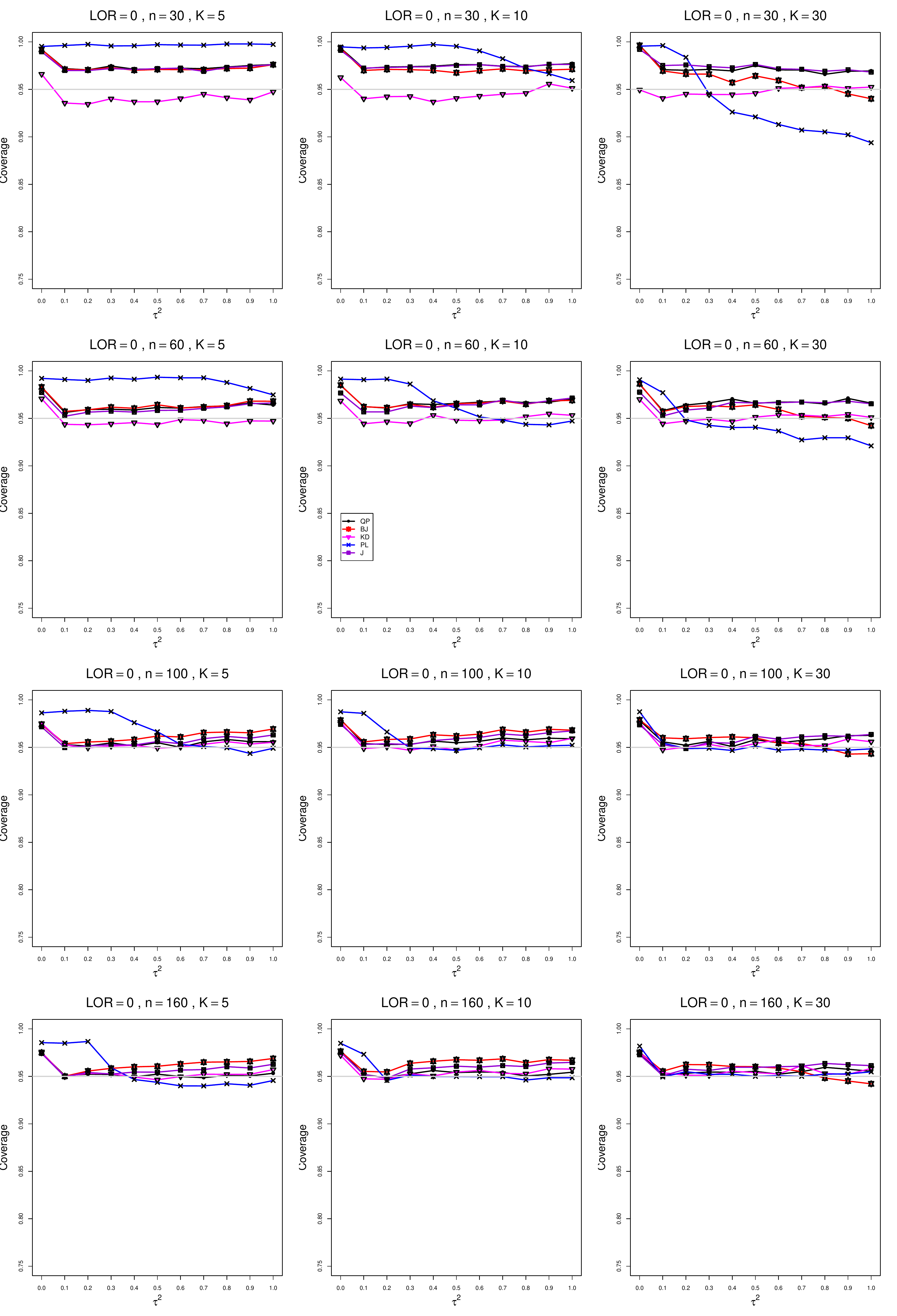}
	\caption{Coverage of  between-studies variance $\tau^2$ for $\theta=0$, $p_{iC}=0.4$, $q=0.5$,
		unequal sample sizes $n=30,\; 60,\;100,\;160$. 
		\label{CovTauLOR0q05piC04_unequal_sample_sizes}}
\end{figure}

\begin{figure}[t]
	\centering
	\includegraphics[scale=0.33]{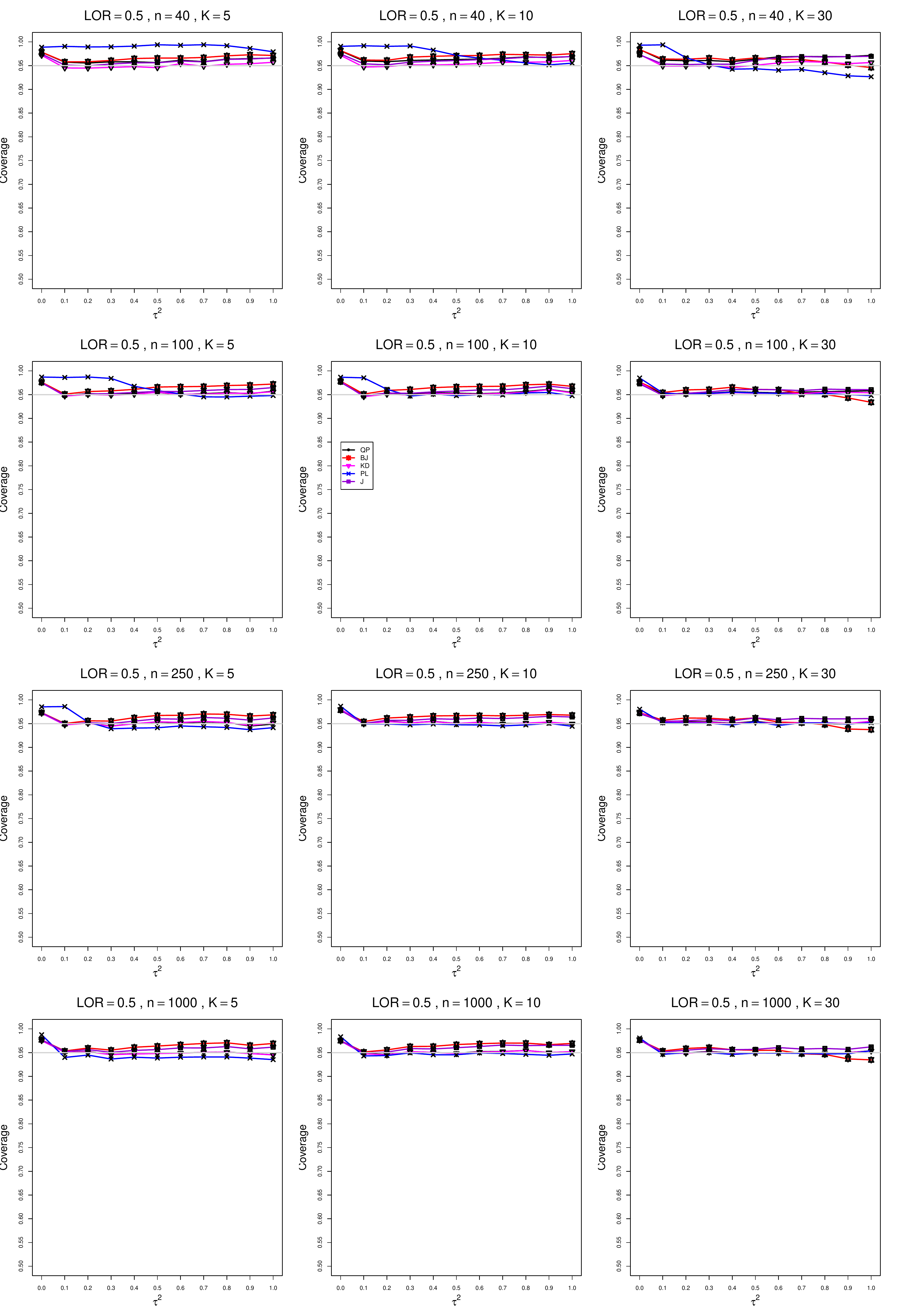}
	\caption{Coverage of  between-studies variance $\tau^2$ for $\theta=0.5$, $p_{iC}=0.4$, $q=0.5$, equal sample sizes  $n=40,\;100,\;250,\;1000$. 
		\label{CovTauLOR05q05piC04}}
\end{figure}

\begin{figure}[t]
	\centering
	\includegraphics[scale=0.33]{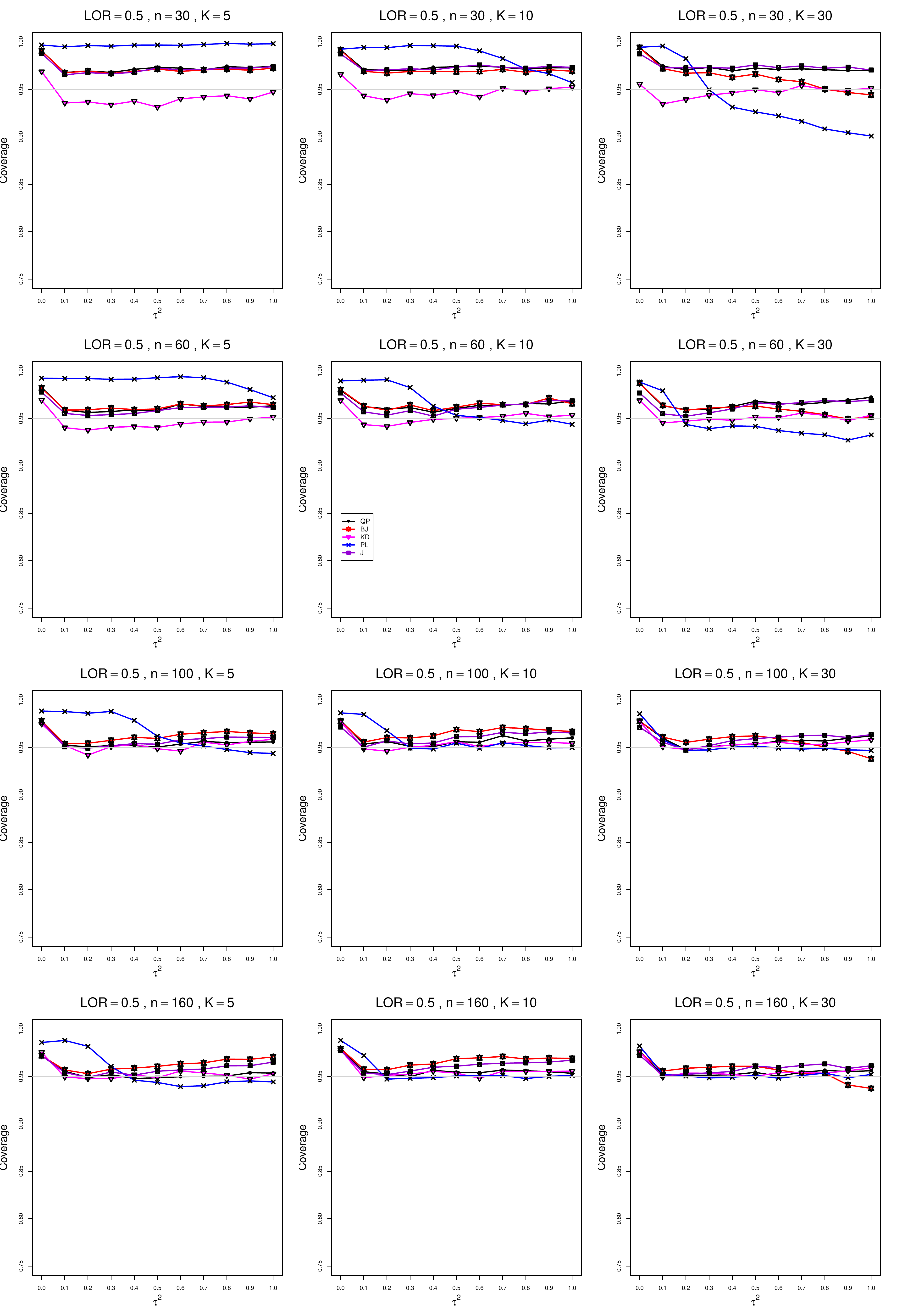}
	\caption{Coverage of  between-studies variance $\tau^2$ for $\theta=0.5$, $p_{iC}=0.4$, $q=0.5$,
		unequal sample sizes $n=30,\; 60,\;100,\;160$. 
		\label{CovTauLOR05q05piC04_unequal_sample_sizes}}
\end{figure}

\begin{figure}[t]
	\centering
	\includegraphics[scale=0.33]{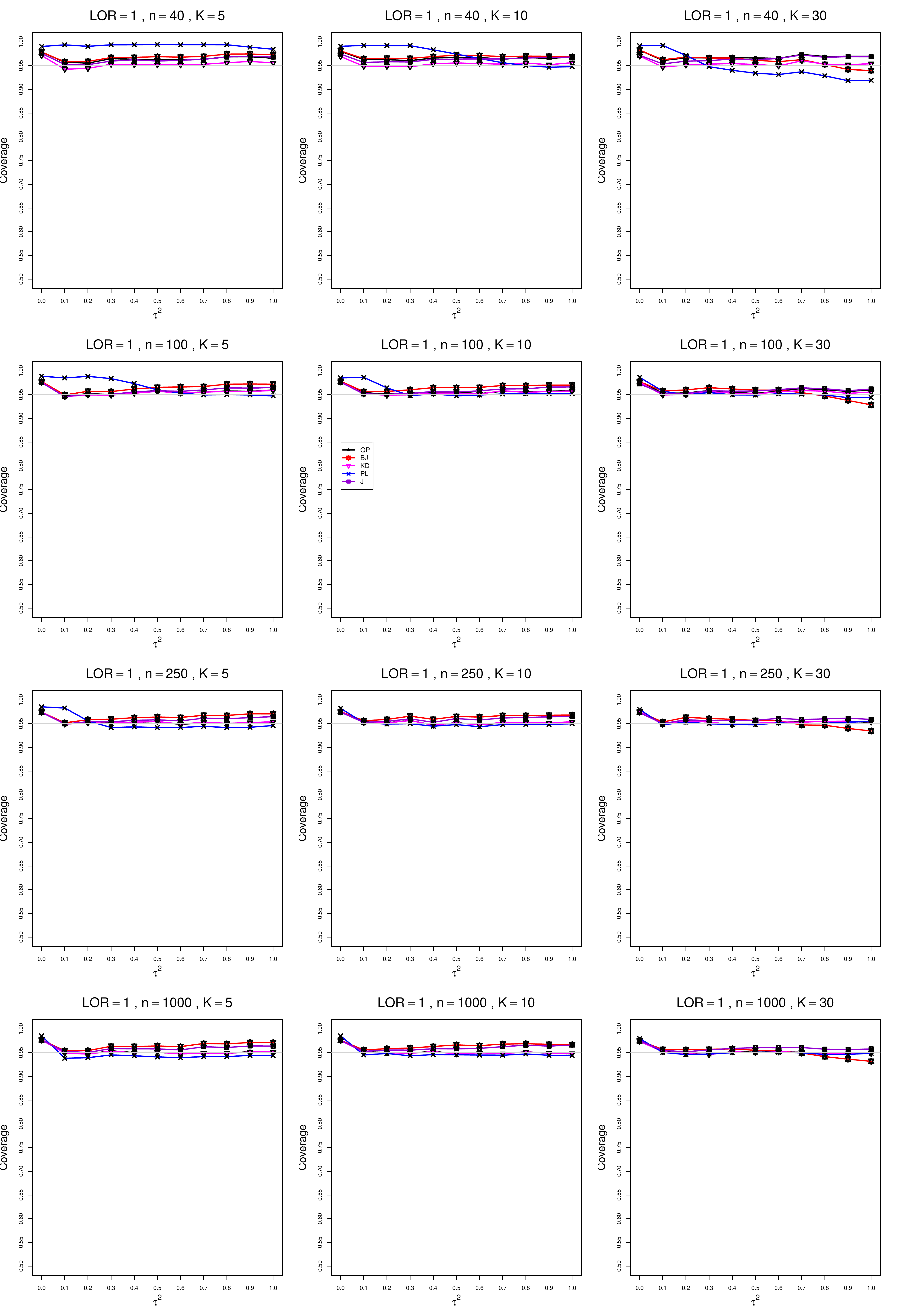}
	\caption{Coverage of  between-studies variance $\tau^2$ for $\theta=1$, $p_{iC}=0.4$, $q=0.5$, equal sample sizes $n=40,\;100,\;250,\;1000$. 
		\label{CovTauLOR1q05piC04}}
\end{figure}

\begin{figure}[t]
	\centering
	\includegraphics[scale=0.33]{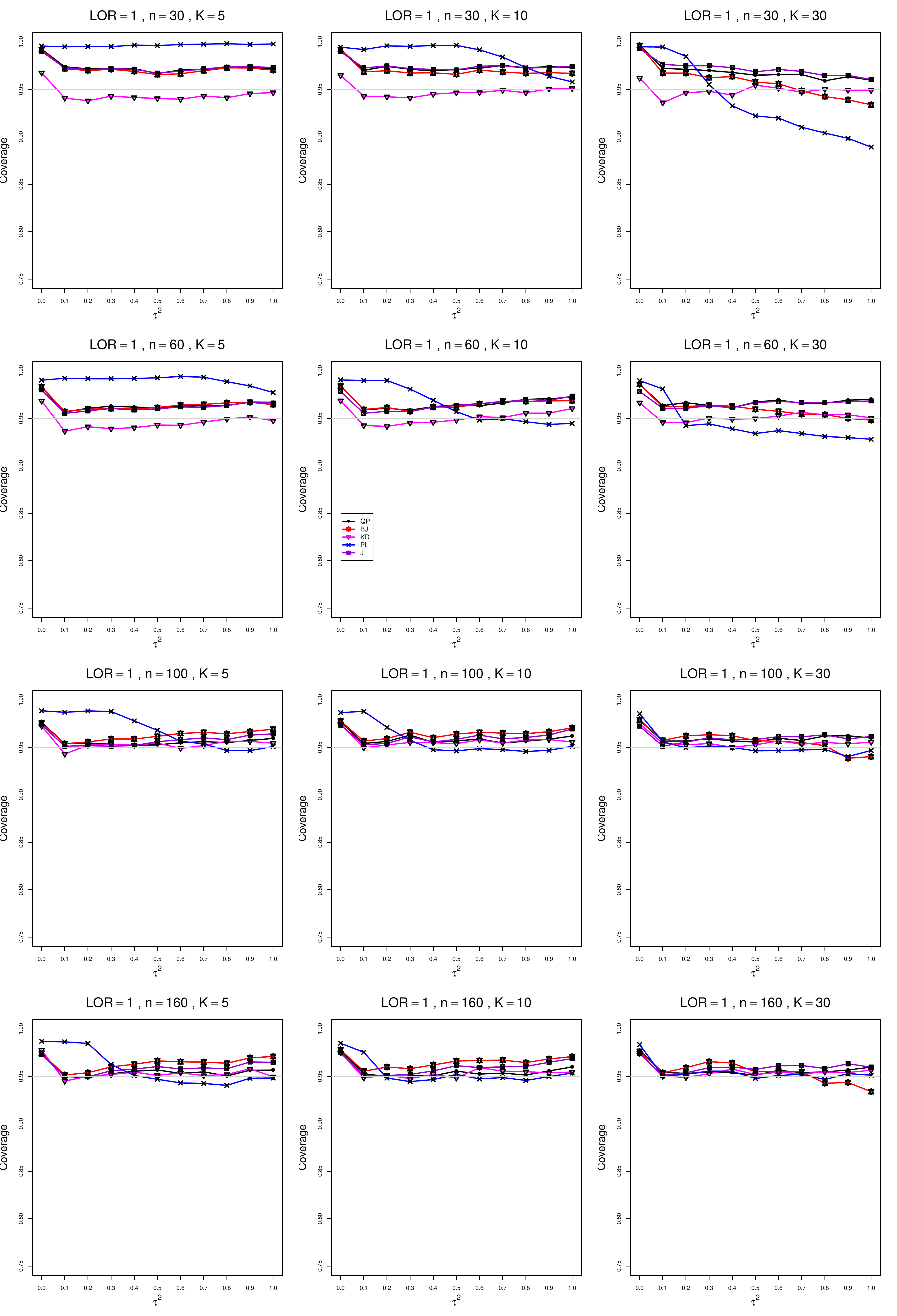}
	\caption{Coverage of  between-studies variance $\tau^2$ for $\theta=1$, $p_{iC}=0.4$, $q=0.5$,
		unequal sample sizes $n=30,\; 60,\;100,\;160$. 
		\label{CovTauLOR1q05piC04_unequal_sample_sizes}}
\end{figure}

\begin{figure}[t]
	\centering
	\includegraphics[scale=0.33]{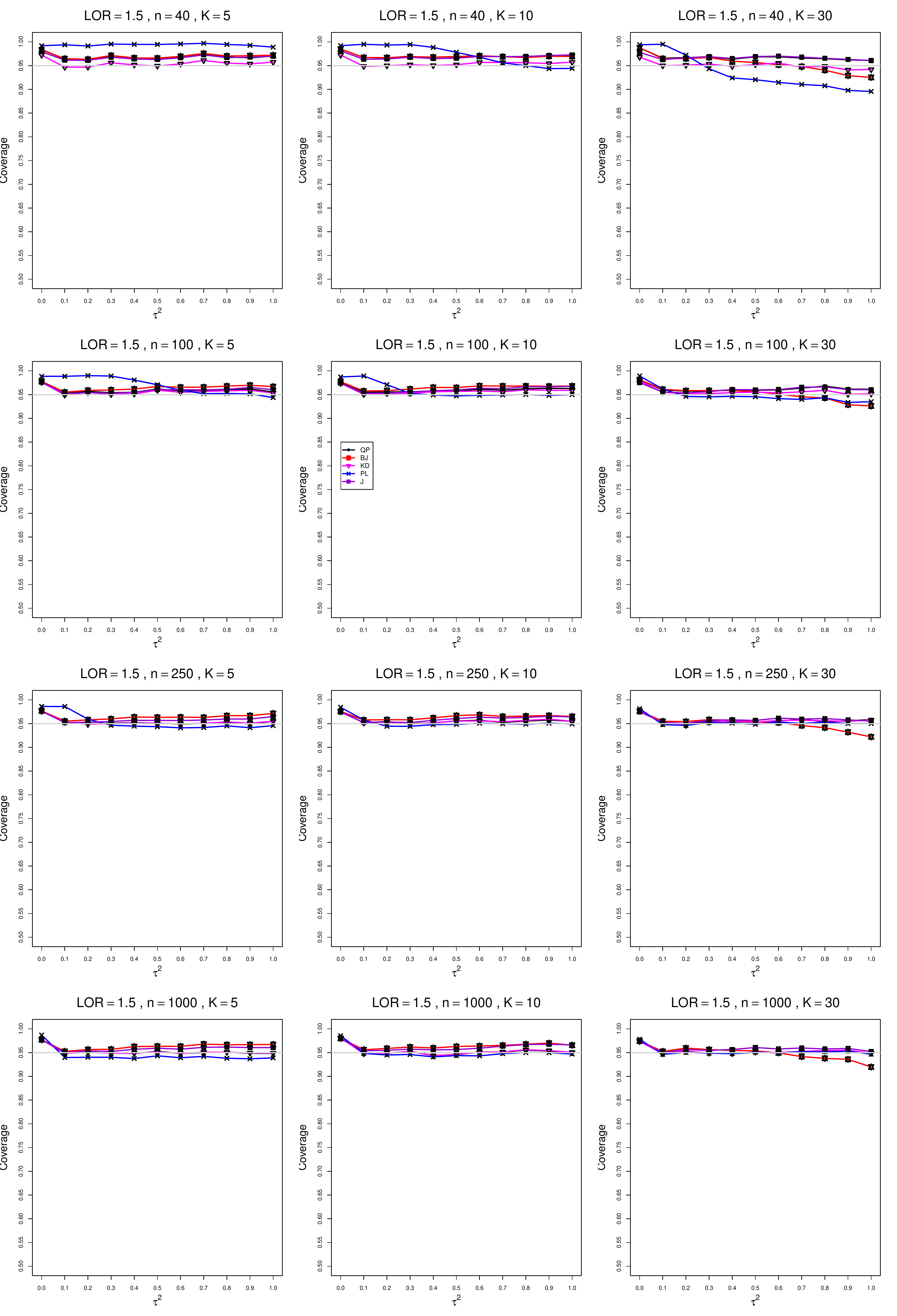}
	\caption{Coverage of  between-studies variance $\tau^2$ for $\theta=1.5$, $p_{iC}=0.4$, $q=0.5$, equal sample sizes $n=40,\;100,\;250,\;1000$. 
		\label{CovTauLOR15q05piC04}}
\end{figure}

\begin{figure}[t]
	\centering
	\includegraphics[scale=0.33]{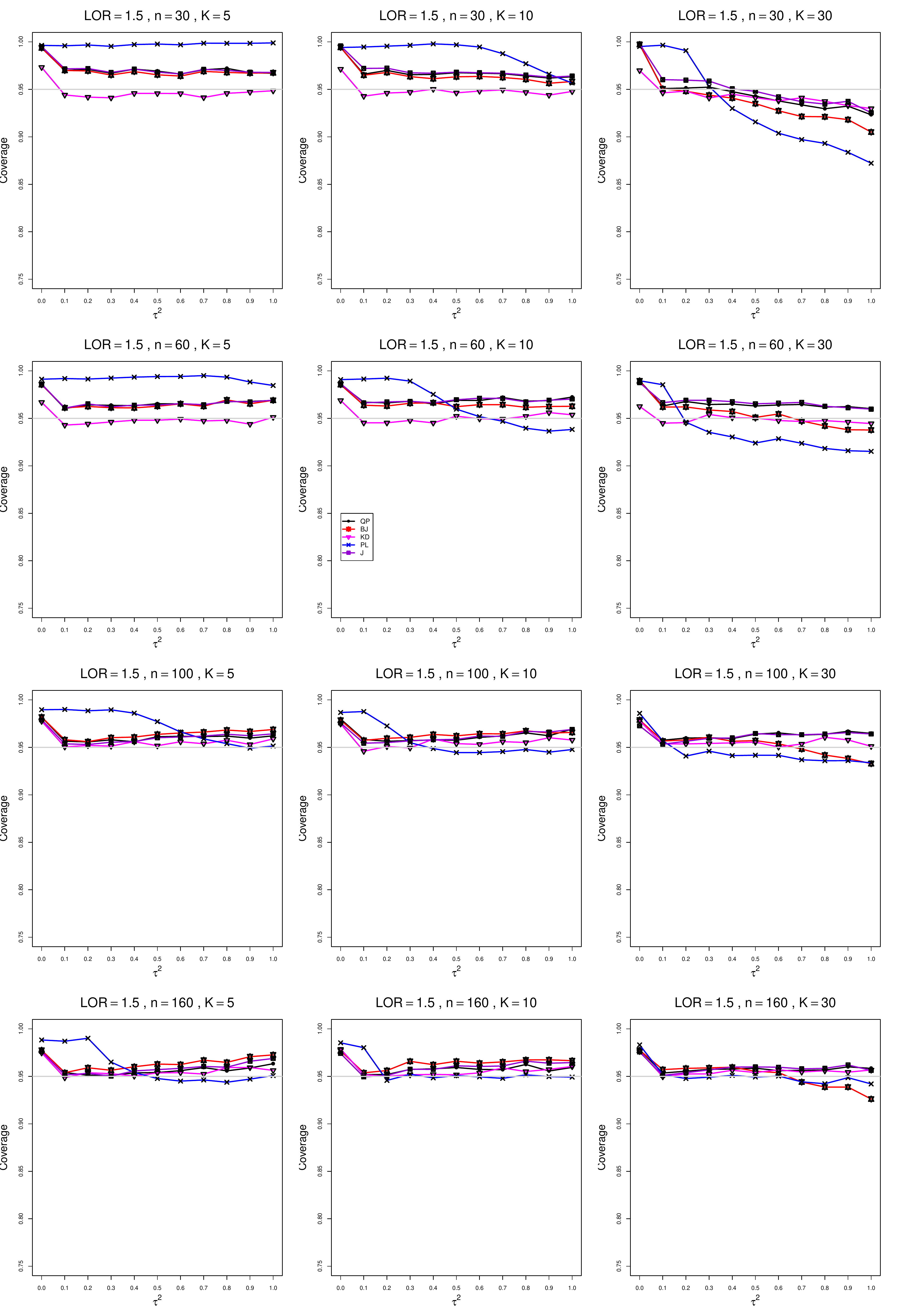}
	\caption{Coverage of  between-studies variance $\tau^2$ for $\theta=1.5$, $p_{iC}=0.4$, $q=0.5$,
		unequal sample sizes $n=30,\; 60,\;100,\;160$. 
		\label{CovTauLOR15q05piC04_unequal_sample_sizes}}
\end{figure}

\begin{figure}[t]
	\centering
	\includegraphics[scale=0.33]{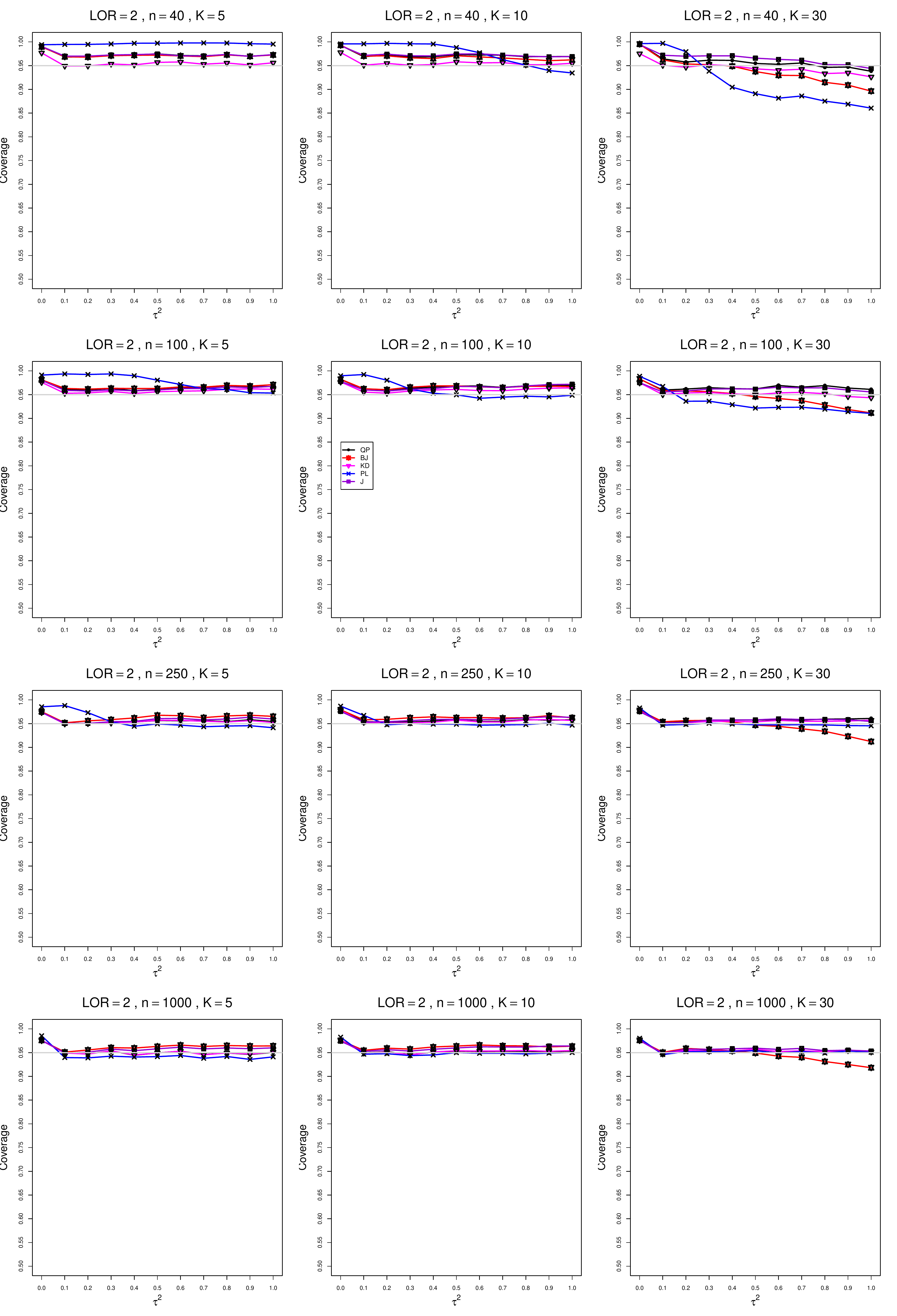}
	\caption{Coverage of  between-studies variance $\tau^2$ for $\theta=2$, $p_{iC}=0.4$, $q=0.5$, equal sample sizes $n=40,\;100,\;250,\;1000$. 
		\label{CovTauLOR2q05piC04}}
\end{figure}

\begin{figure}[t]
	\centering
	\includegraphics[scale=0.33]{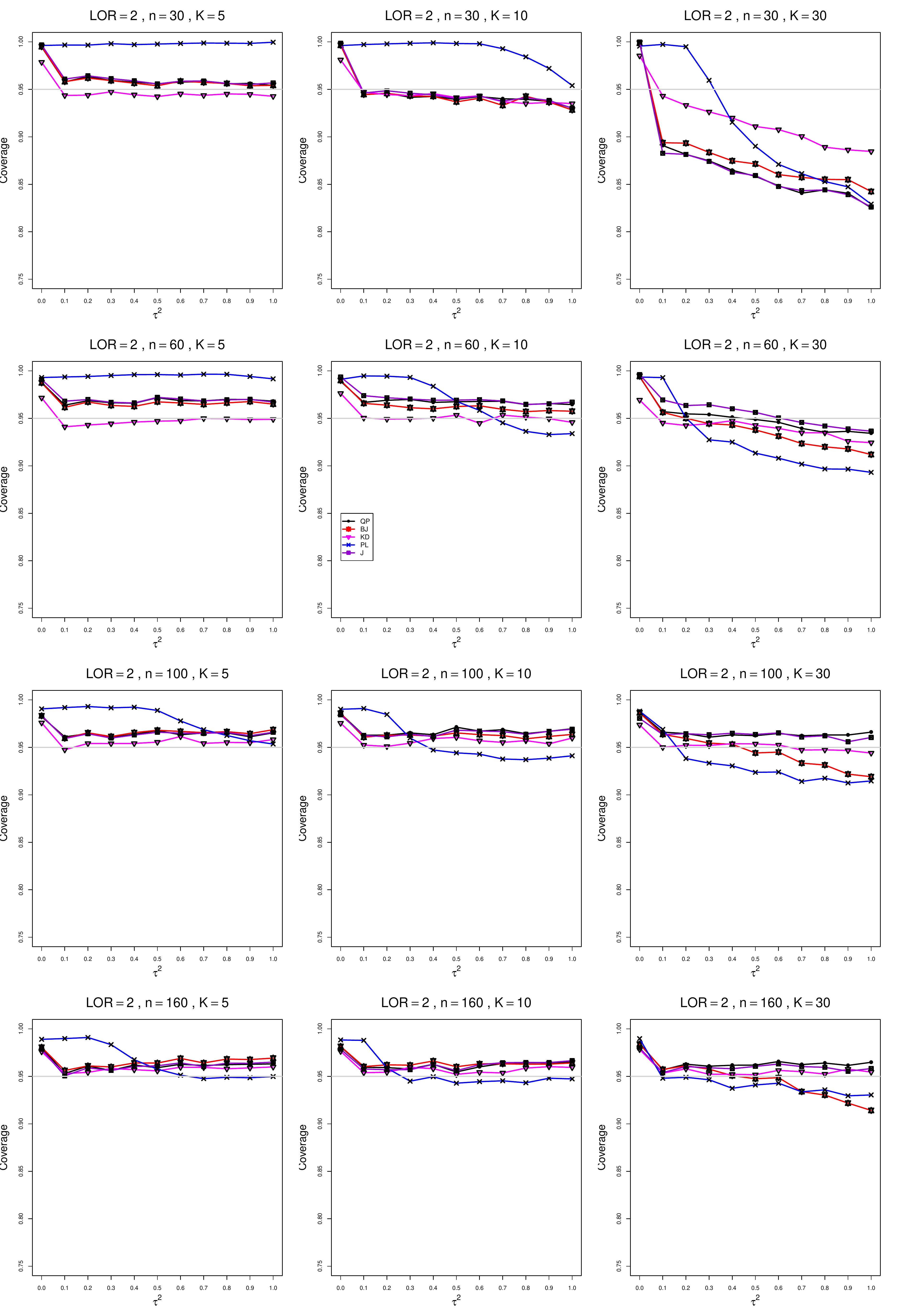}
	\caption{Coverage of  between-studies variance $\tau^2$ for $\theta=2$, $p_{iC}=0.4$, $q=0.5$,
		unequal sample sizes $n=30,\; 60,\;100,\;160$. 
		\label{CovTauLOR2q05piC04_unequal_sample_sizes}}
\end{figure}


\begin{figure}[t]
	\centering
	\includegraphics[scale=0.33]{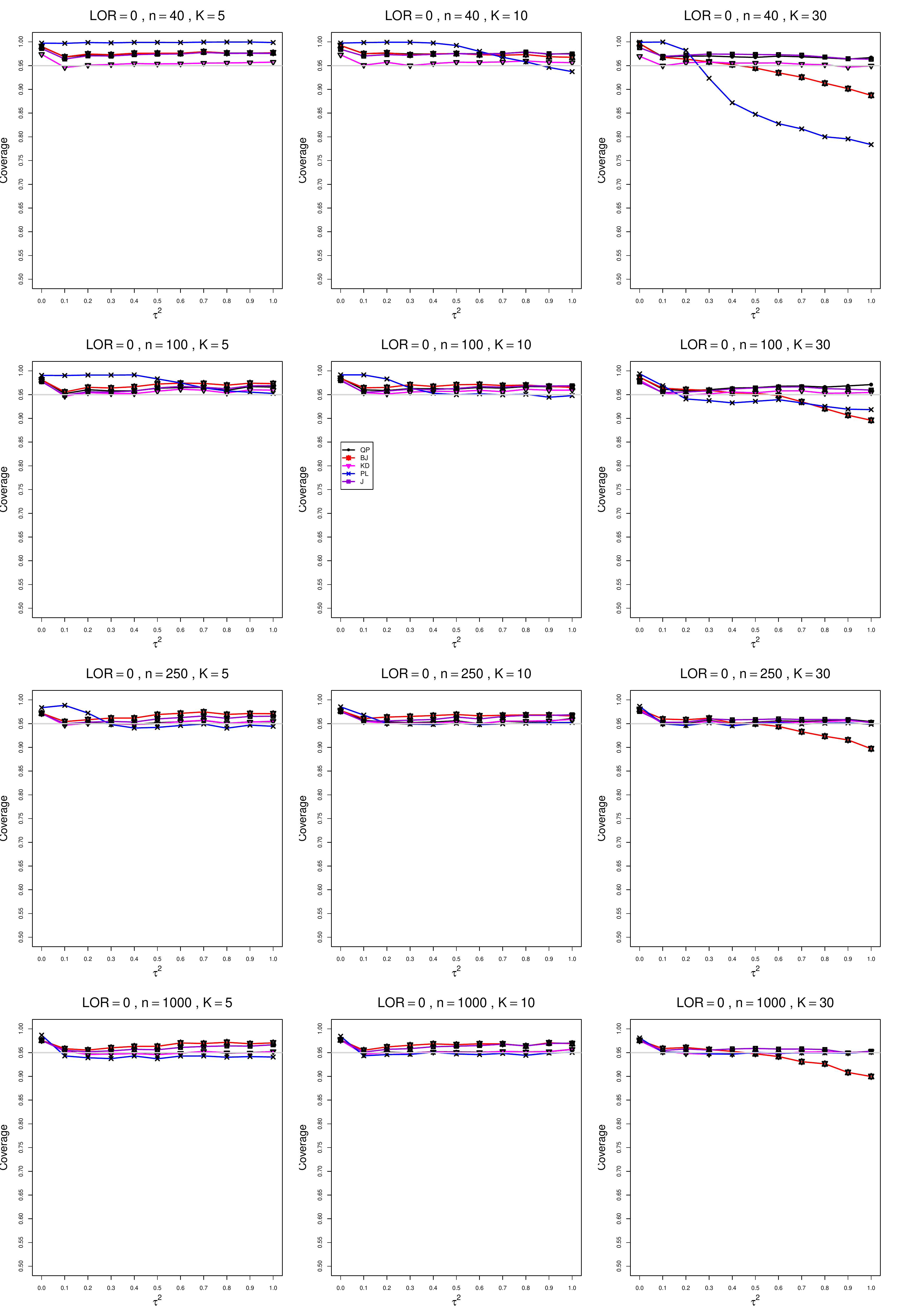}
	\caption{Coverage of  between-studies variance $\tau^2$ for $\theta=0$, $p_{iC}=0.4$, $q=0.75$, equal sample sizes $n=40,\;100,\;250,\;1000$. 
		\label{CovTauLOR0q075piC04}}
\end{figure}

\begin{figure}[t]
	\centering
	\includegraphics[scale=0.33]{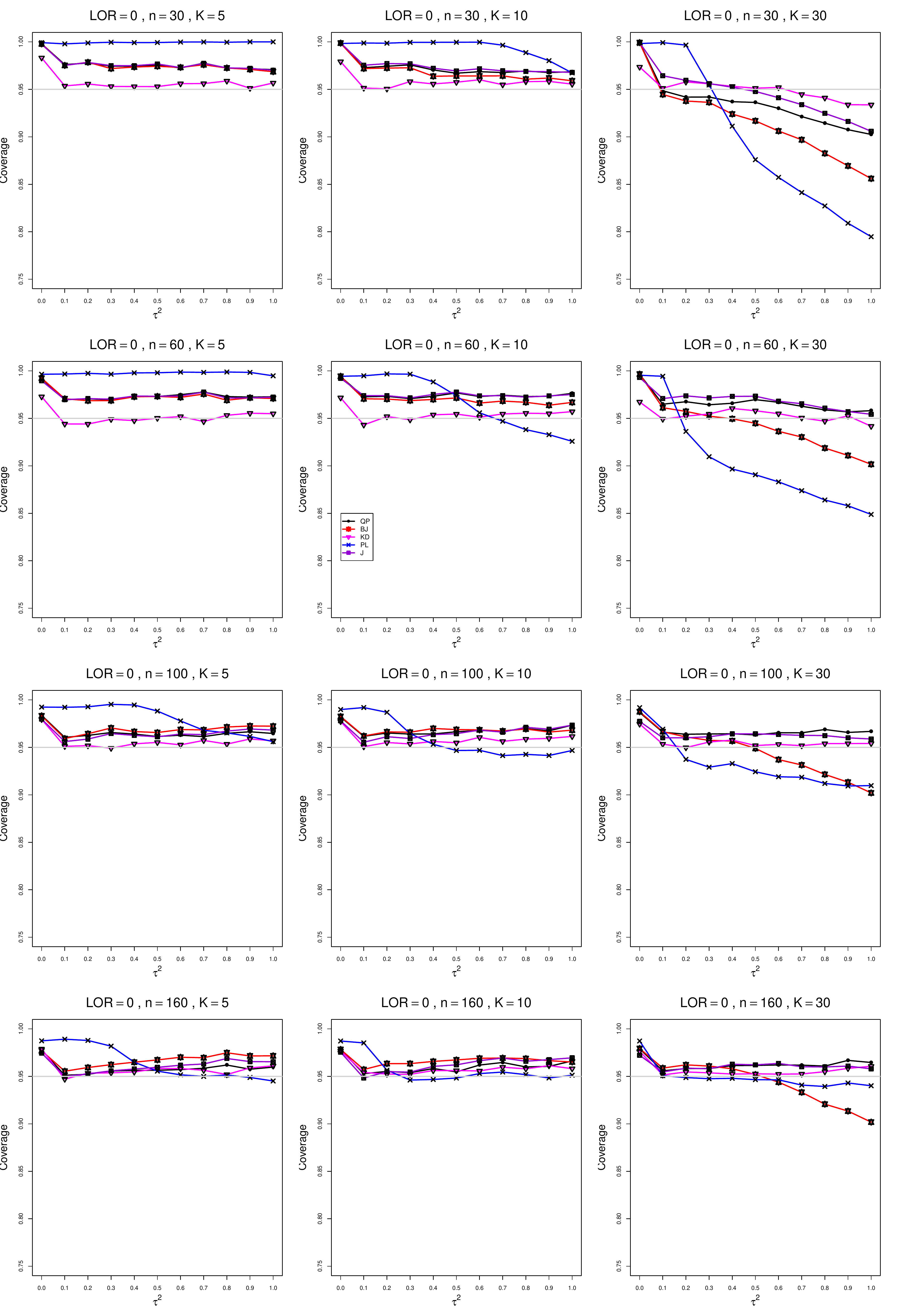}
	\caption{Coverage of  between-studies variance $\tau^2$ for $\theta=0$, $p_{iC}=0.4$, $q=0.75$,
		unequal sample sizes $n=30,\; 60,\;100,\;160$. 
		\label{CovTauLOR0q075piC04_unequal_sample_sizes}}
\end{figure}

\begin{figure}[t]
	\centering
	\includegraphics[scale=0.33]{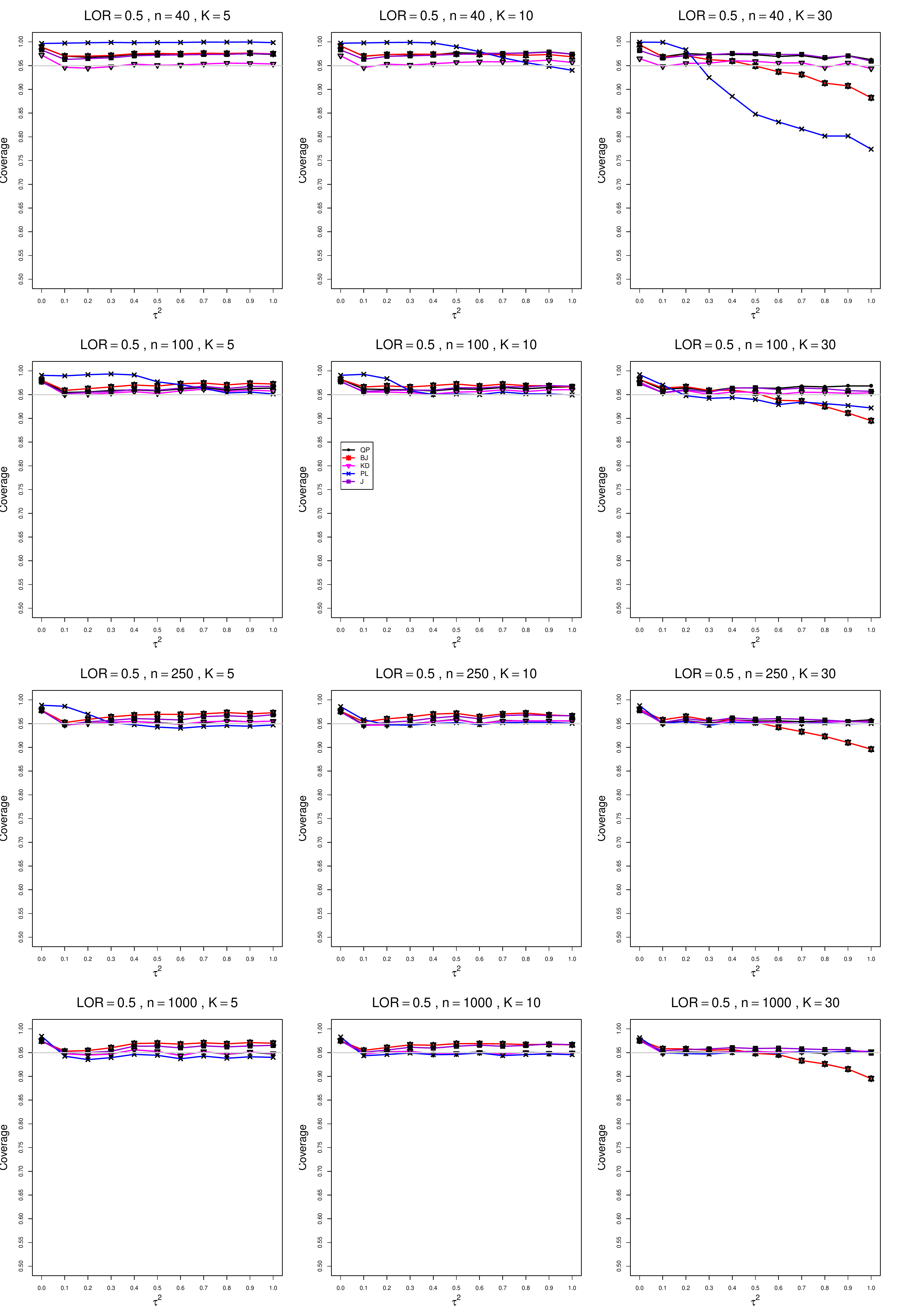}
	\caption{Coverage of  between-studies variance $\tau^2$ for $\theta=0.5$, $p_{iC}=0.4$, $q=0.75$, equal sample sizes $n=40,\;100,\;250,\;1000$. 
		\label{CovTauLOR05q075piC04}}
\end{figure}

\begin{figure}[t]
	\centering
	\includegraphics[scale=0.33]{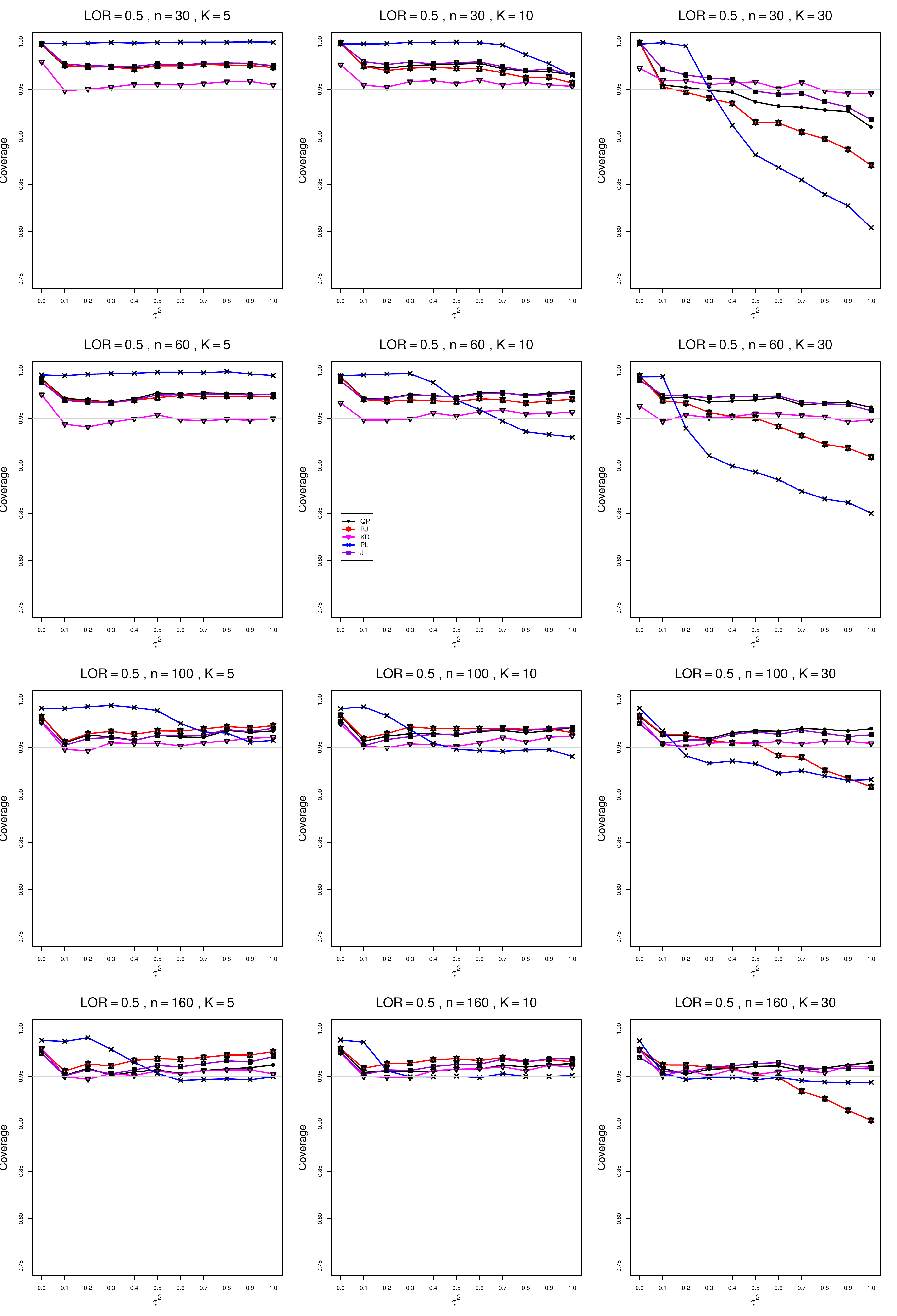}
	\caption{Coverage of  between-studies variance $\tau^2$ for $\theta=0.5$, $p_{iC}=0.4$, $q=0.75$,
		unequal sample sizes $n=30,\; 60,\;100,\;160$. 
		\label{CovTauLOR05q075piC04_unequal_sample_sizes}}
\end{figure}
\clearpage

\begin{figure}[t]
	\centering
	\includegraphics[scale=0.33]{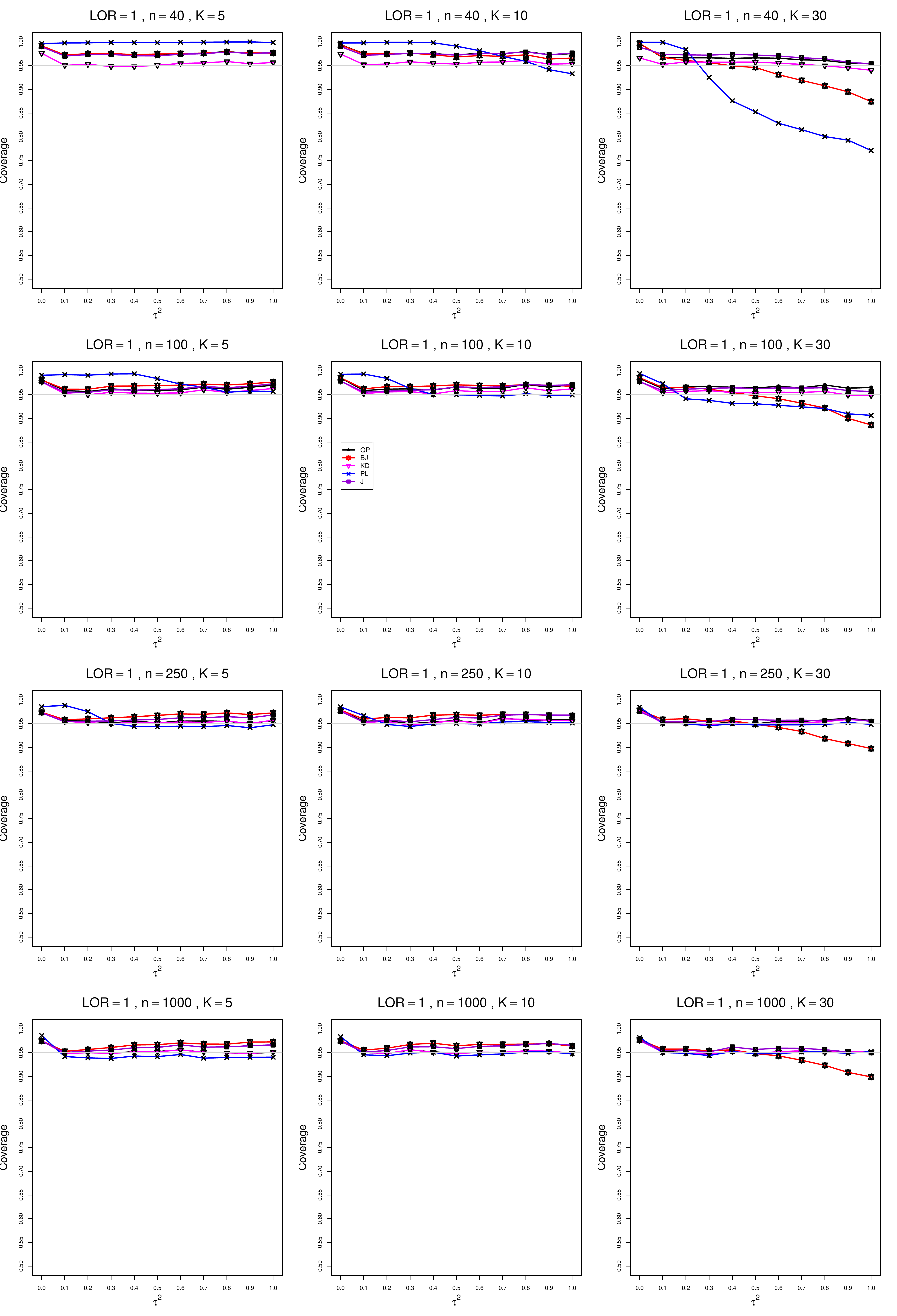}
	\caption{Coverage of  between-studies variance $\tau^2$ for $\theta=1$, $p_{iC}=0.4$, $q=0.75$, equal sample sizes $n=40,\;100,\;250,\;1000$. 
		\label{CovTauLOR1q075piC04}}
\end{figure}

\begin{figure}[t]
	\centering
	\includegraphics[scale=0.33]{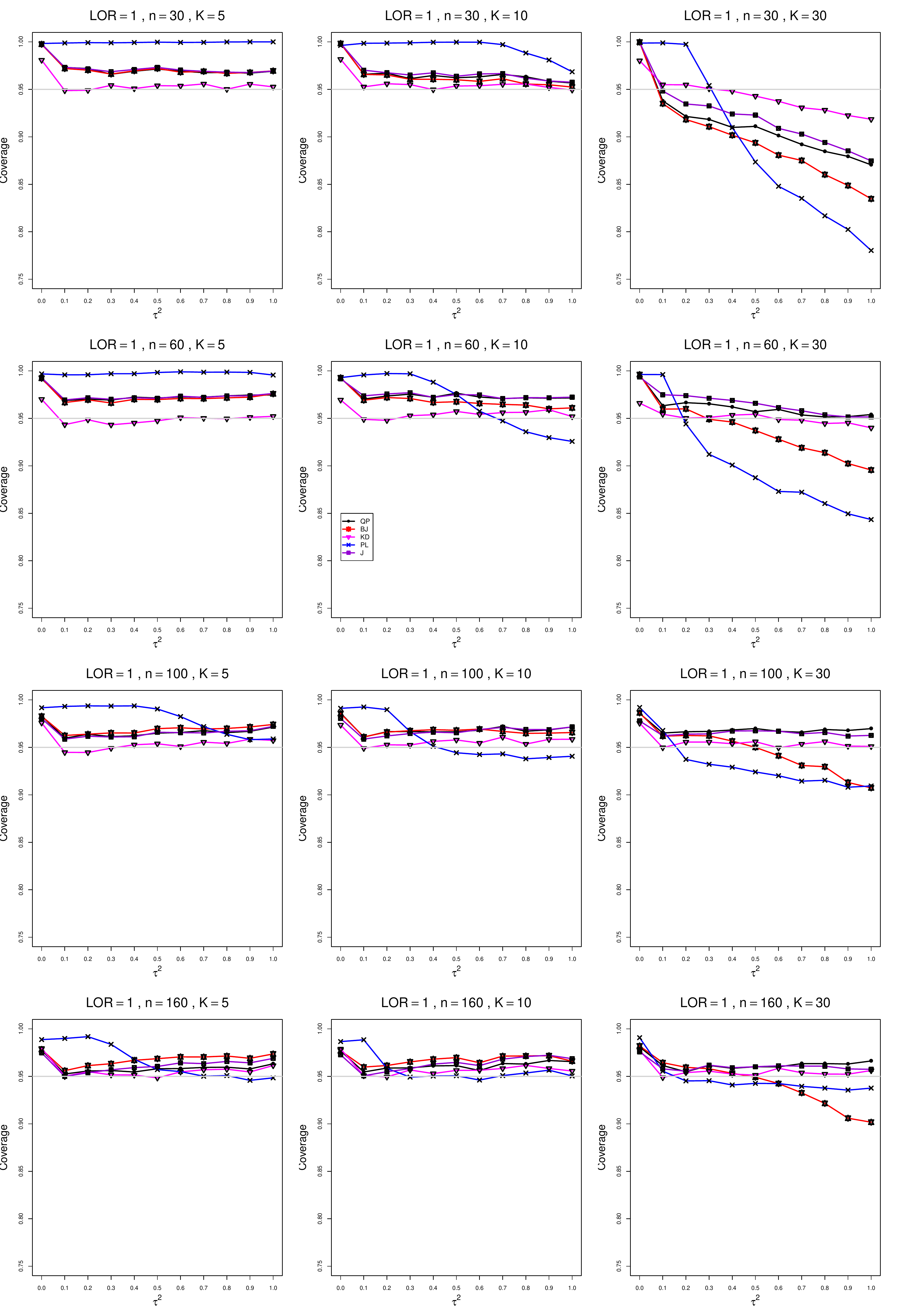}
	\caption{Coverage of  between-studies variance $\tau^2$ for $\theta=1$, $p_{iC}=0.4$, $q=0.75$,
		unequal sample sizes $n=30,\; 60,\;100,\;160$. 
		\label{CovTauLOR1q075piC04_unequal_sample_sizes}}
\end{figure}

\begin{figure}[t]
	\centering
	\includegraphics[scale=0.33]{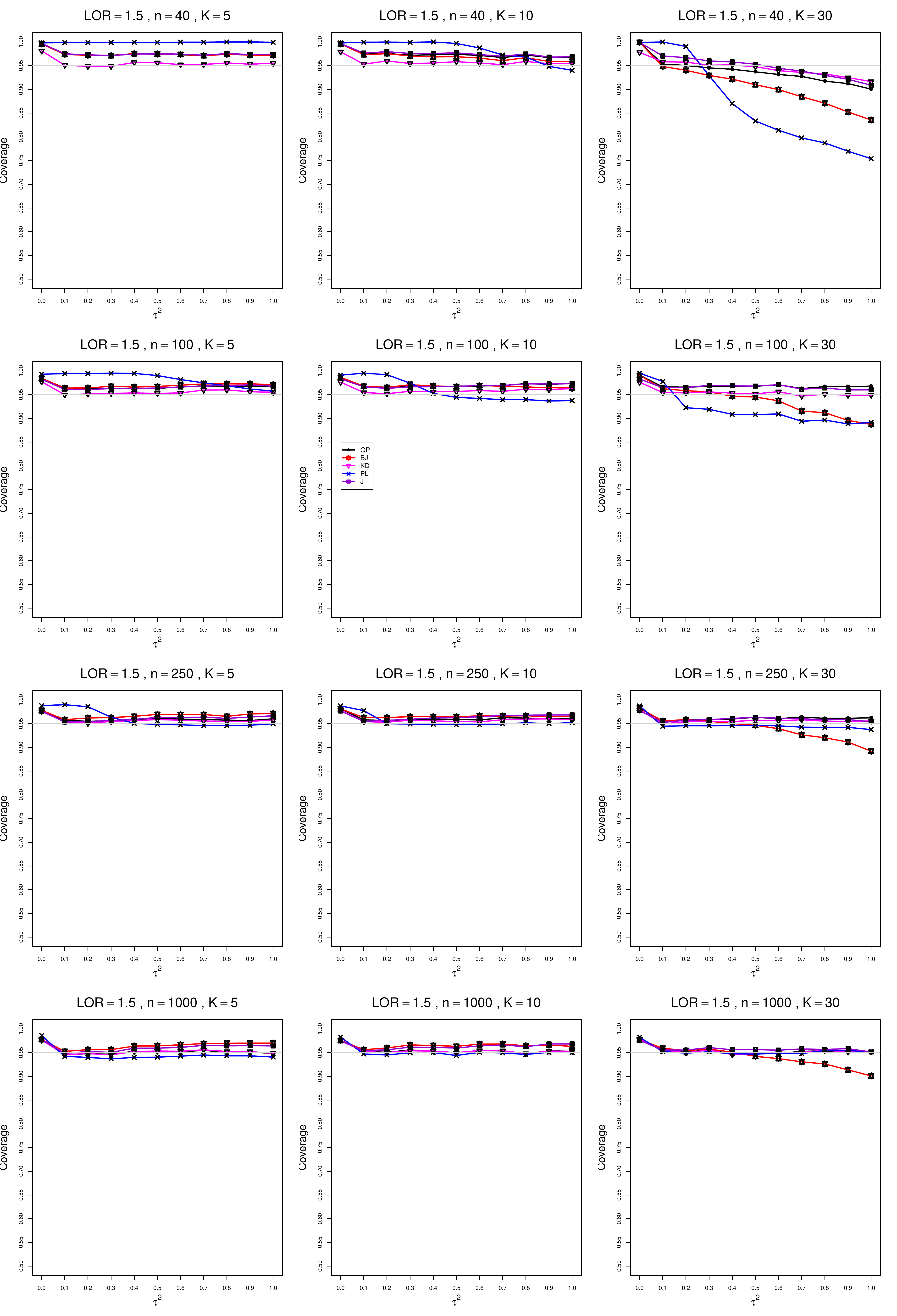}
	\caption{Coverage of  between-studies variance $\tau^2$ for $\theta=1.5$, $p_{iC}=0.4$, $q=0.75$, equal sample sizes $n=40,\;100,\;250,\;1000$. 
		\label{CovTauLOR15q075piC04}}
\end{figure}

\begin{figure}[t]
	\centering
	\includegraphics[scale=0.33]{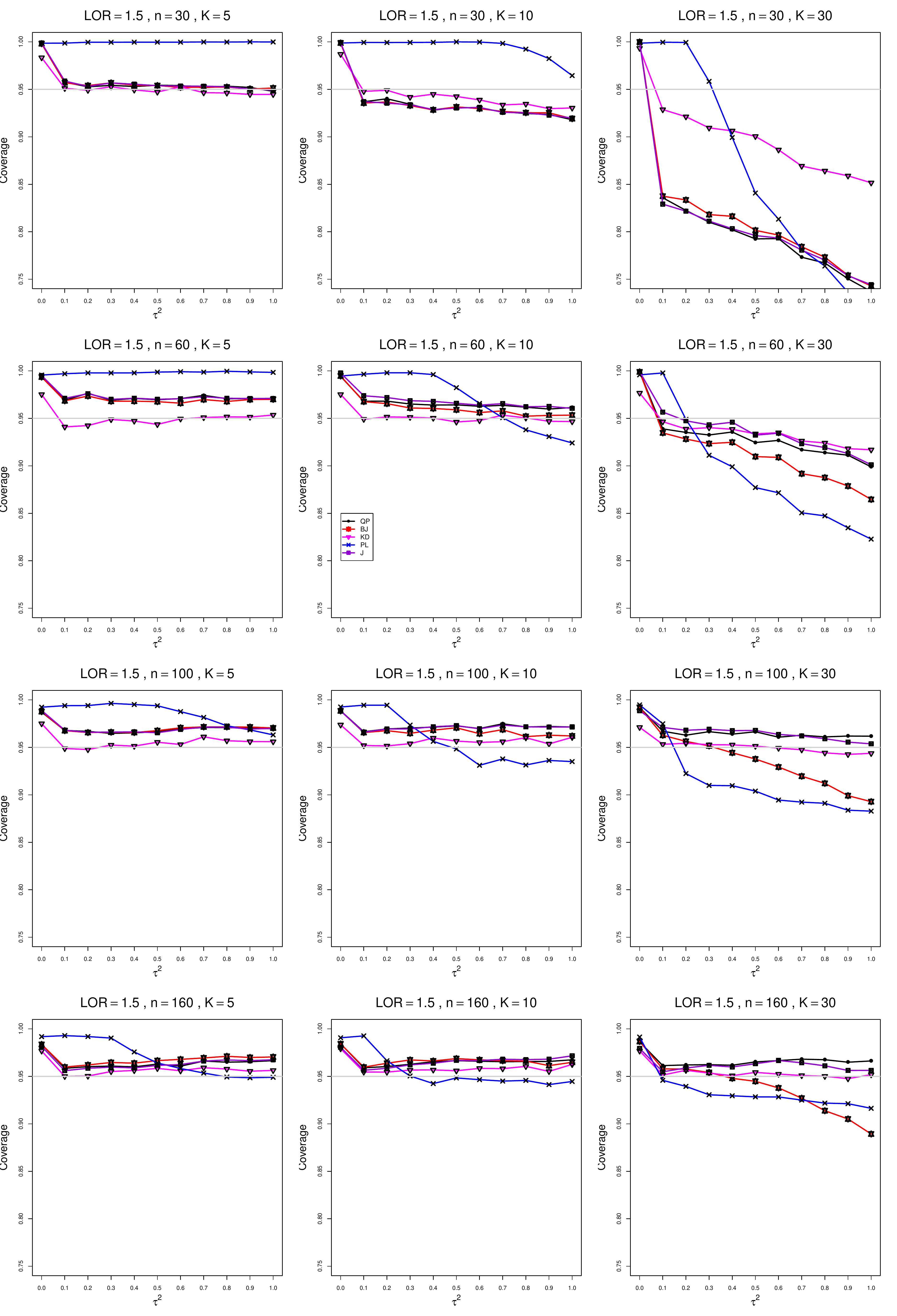}
	\caption{Coverage of  between-studies variance $\tau^2$ for $\theta=1.5$, $p_{iC}=0.4$, $q=0.75$,
		unequal sample sizes $n=30,\; 60,\;100,\;160$. 
		\label{CovTauLOR15q075piC04_unequal_sample_sizes}}
\end{figure}

\begin{figure}[t]
	\centering
	\includegraphics[scale=0.33]{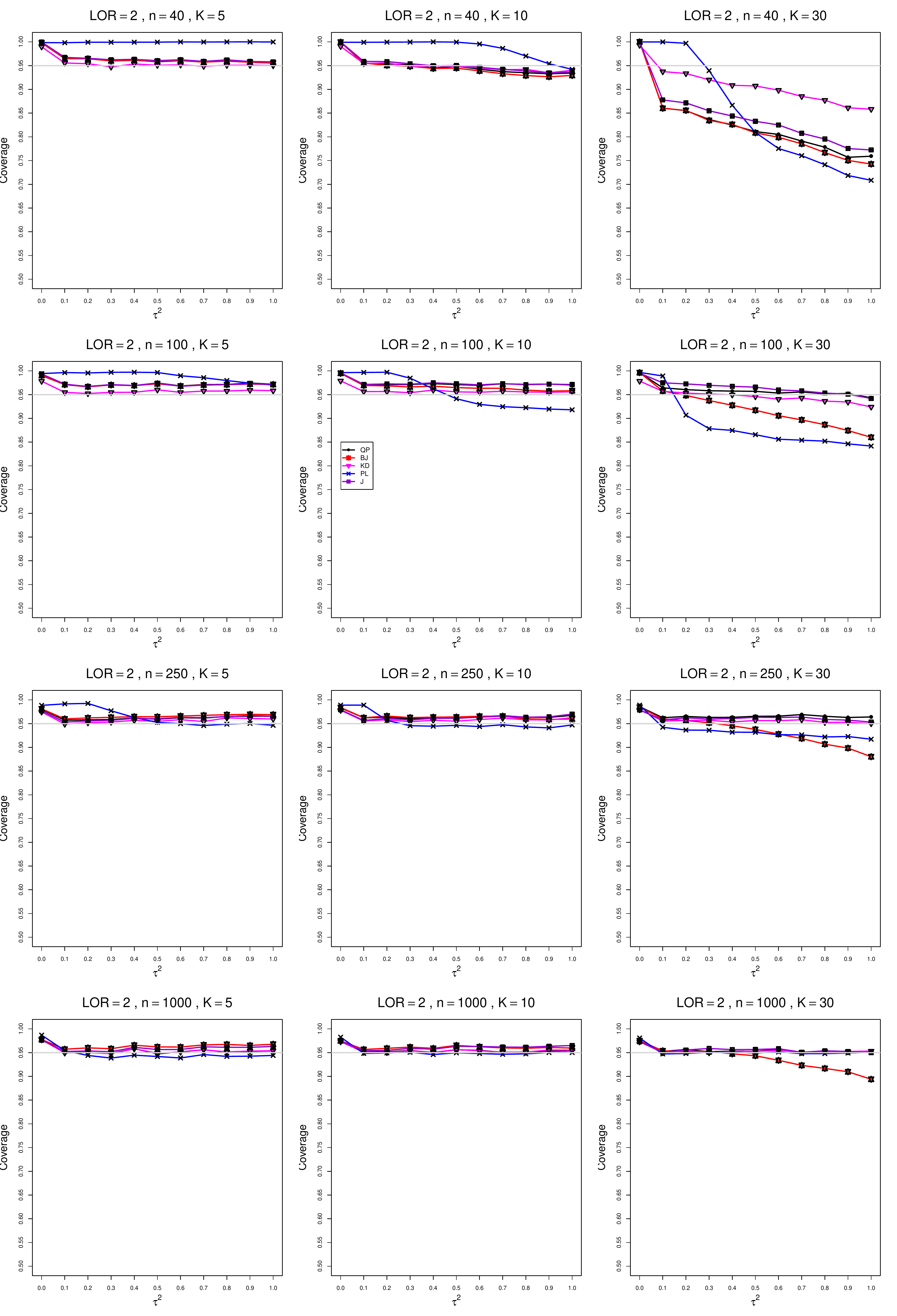}
	\caption{Coverage of  between-studies variance $\tau^2$ for $\theta=2$, $p_{iC}=0.4$, $q=0.75$, equal sample sizes $n=40,\;100,\;250,\;1000$. 
		\label{CovTauLOR2q075piC04}}
\end{figure}

\begin{figure}[t]
	\centering
	\includegraphics[scale=0.33]{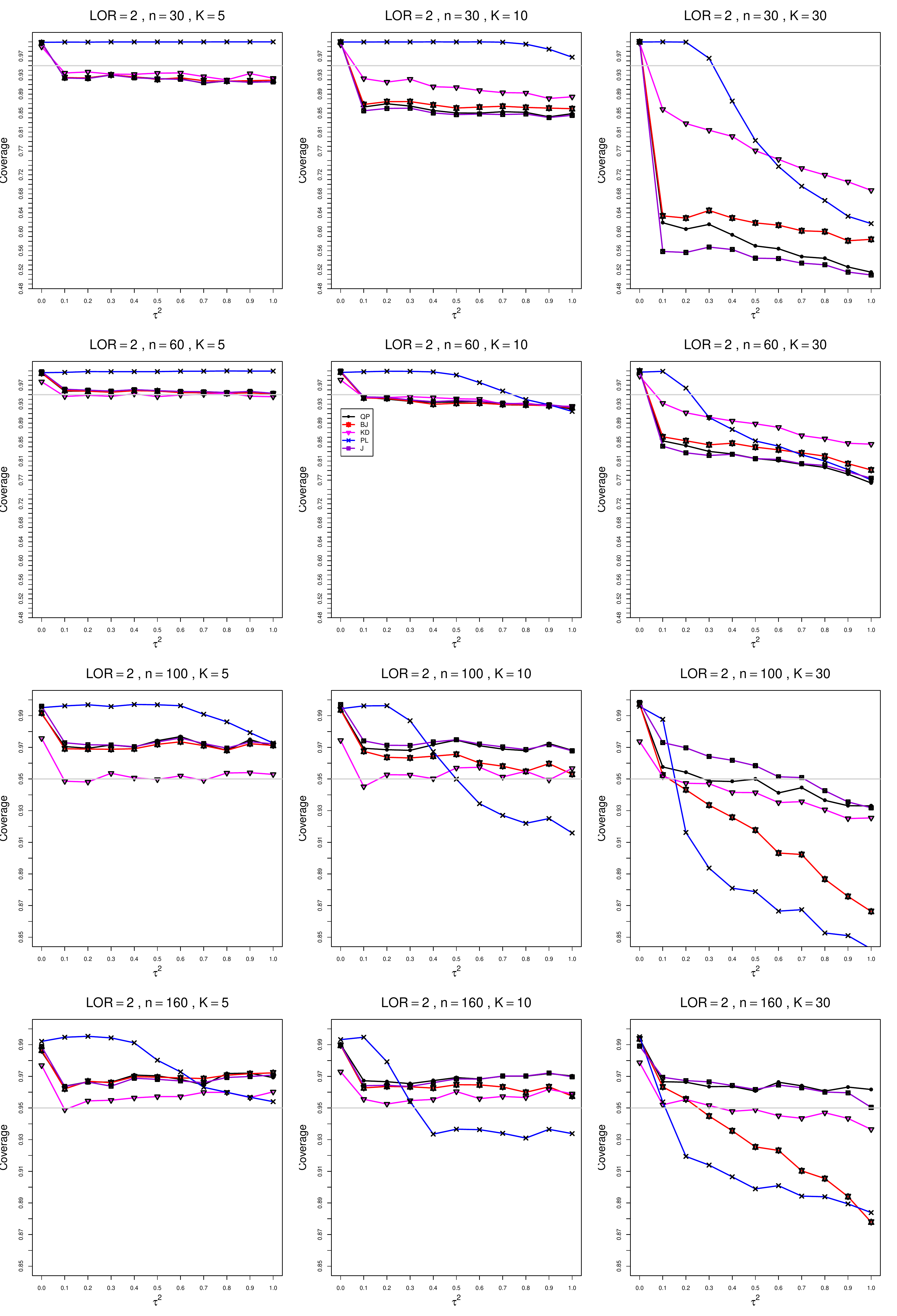}
	\caption{Coverage of  between-studies variance $\tau^2$ for $\theta=2$, $p_{iC}=0.4$, $q=0.75$,
		unequal sample sizes $n=30,\; 60,\;100,\;160$. 
		\label{CovTauLOR2q075piC04_unequal_sample_sizes}}
\end{figure}

\clearpage
\renewcommand{\thefigure}{B1.1.\arabic{figure}}
\renewcommand{\thesection}{B\arabic{section}}
\setcounter{figure}{0}
\setcounter{section}{0}

\section{Bias and mean squared error of point estimators of log-odds-ratio.}

Subsections B1.1, B1.2 and B1.3 correspond to $p_{C}=0.1,\; 0.2,\; 0.4$ respectively. 
For a given $p_{C}$ value, each figure corresponds to a value of $\theta (= 0, 0.5, 1, 1.5, 2)$, a value of $q (= 0.5, 0.75)$, a value of $\tau^2 = 0.0(0.1)1.0$, and a set of values of $n$ (= 40, 100, 250, 1000) or $\bar{n}$ (= 30, 60, 100, 160).\\
Figures for mean squared error (expressed as the ratio of the MSE of SSW to the MSEs of the inverse-variance-weighted estimators that use the MP or KD estimator of $\tau^2$) use the above values of $\theta$ and q but only n = 40, 100, 250, 1000.\\
Each figure contains a panel (with $\tau^2$ on the horizontal axis) for each combination of n (or $\bar{n}$) and $K (=5, 10, 30)$.\\
The point estimators of $\theta$ are
\begin{itemize}
	\item DL (DerSimonian-Laird)
	\item REML (restricted maximum likelihood)
	\item MP (Mandel-Paule)
	\item KD (Improved moment estimator based on Kulinskaya and Dollinger (2015)) 
	\item J (Jackson)
	\item SSW (sample-size weighted)
\end{itemize}
\clearpage
\subsection*{B1.1 Probability in the control arm $p_{C}=0.1$}
\clearpage
\begin{figure}[t]
	\centering
	\includegraphics[scale=0.33]{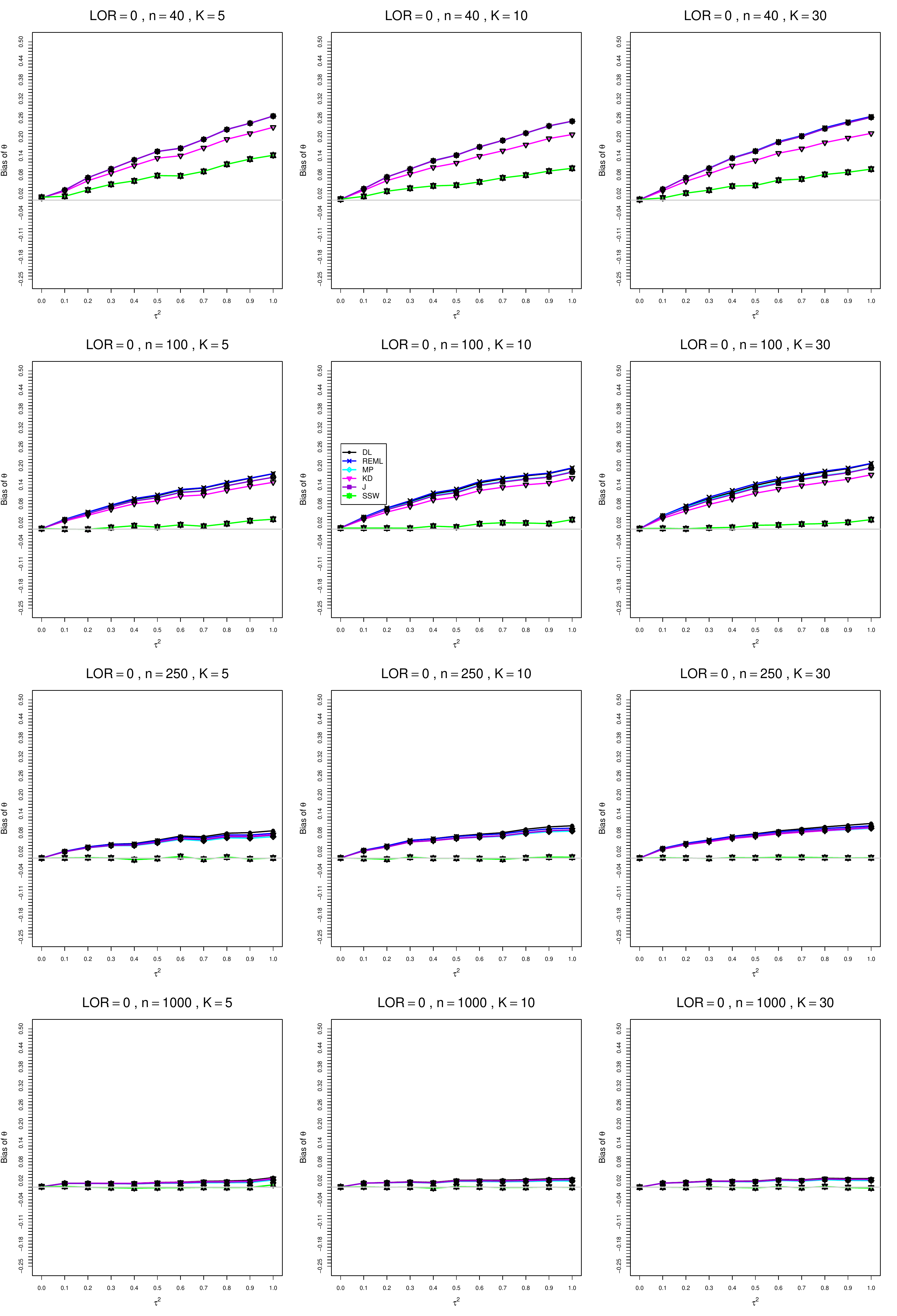}
	\caption{Bias of the estimation of  overall effect measure $\theta$ for $\theta=0$, $p_{iC}=0.1$, $q=0.5$,  equal sample sizes $n=40,\;100,\;250,\;1000$. 
		\label{BiasThetaLOR0q05piC01}}
\end{figure}

\begin{figure}[t]
	\centering
	\includegraphics[scale=0.33]{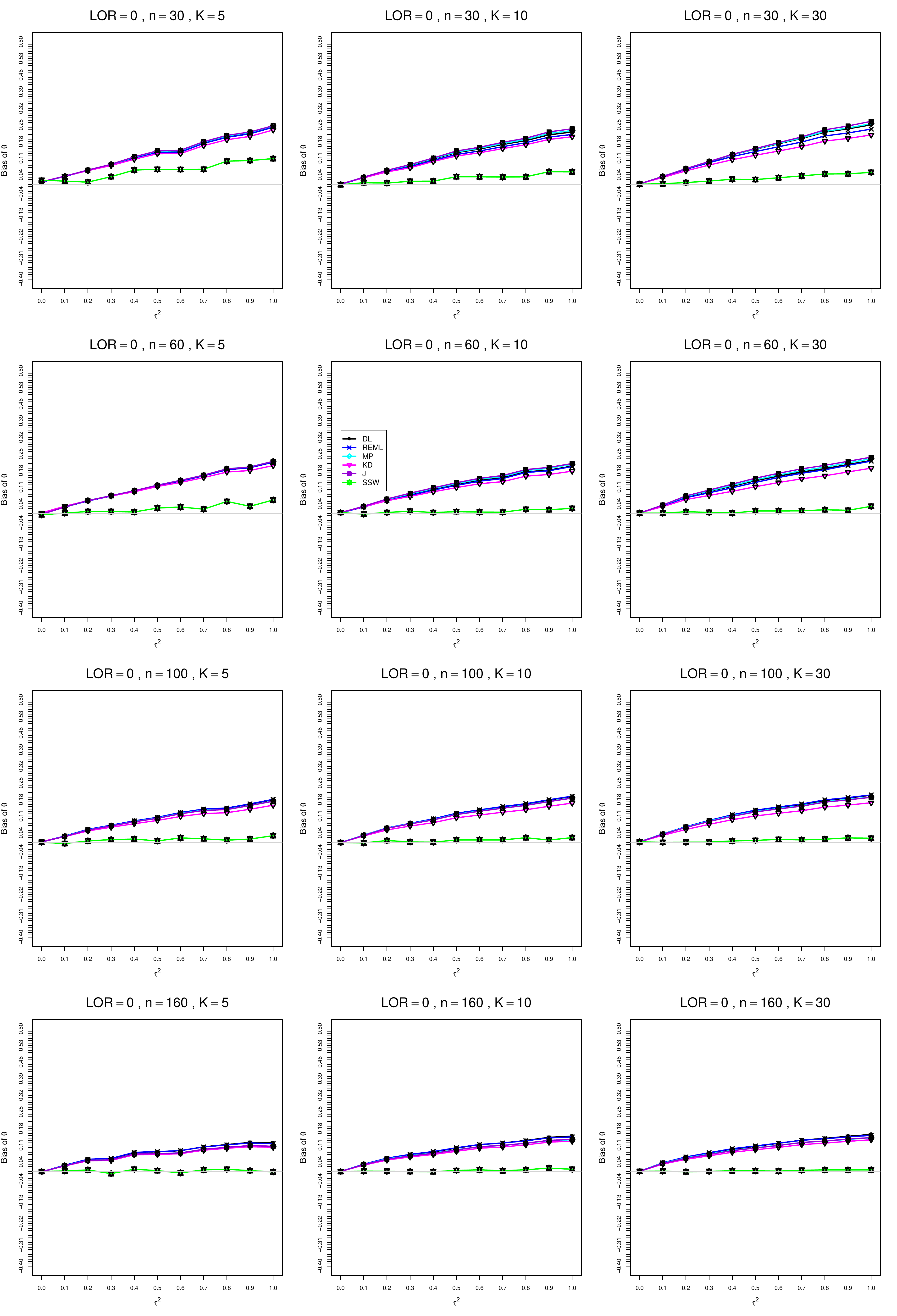}
	\caption{Bias of the estimation of  overall effect measure $\theta$ for $\theta=0$, $p_{iC}=0.1$, $q=0.5$, 
		unequal sample sizes $n=30,\; 60,\;100,\;160$. 
		\label{BiasThetaLOR0q05piC01_unequal_sample_sizes}}
\end{figure}

\begin{figure}[t]\centering
	\includegraphics[scale=0.35]{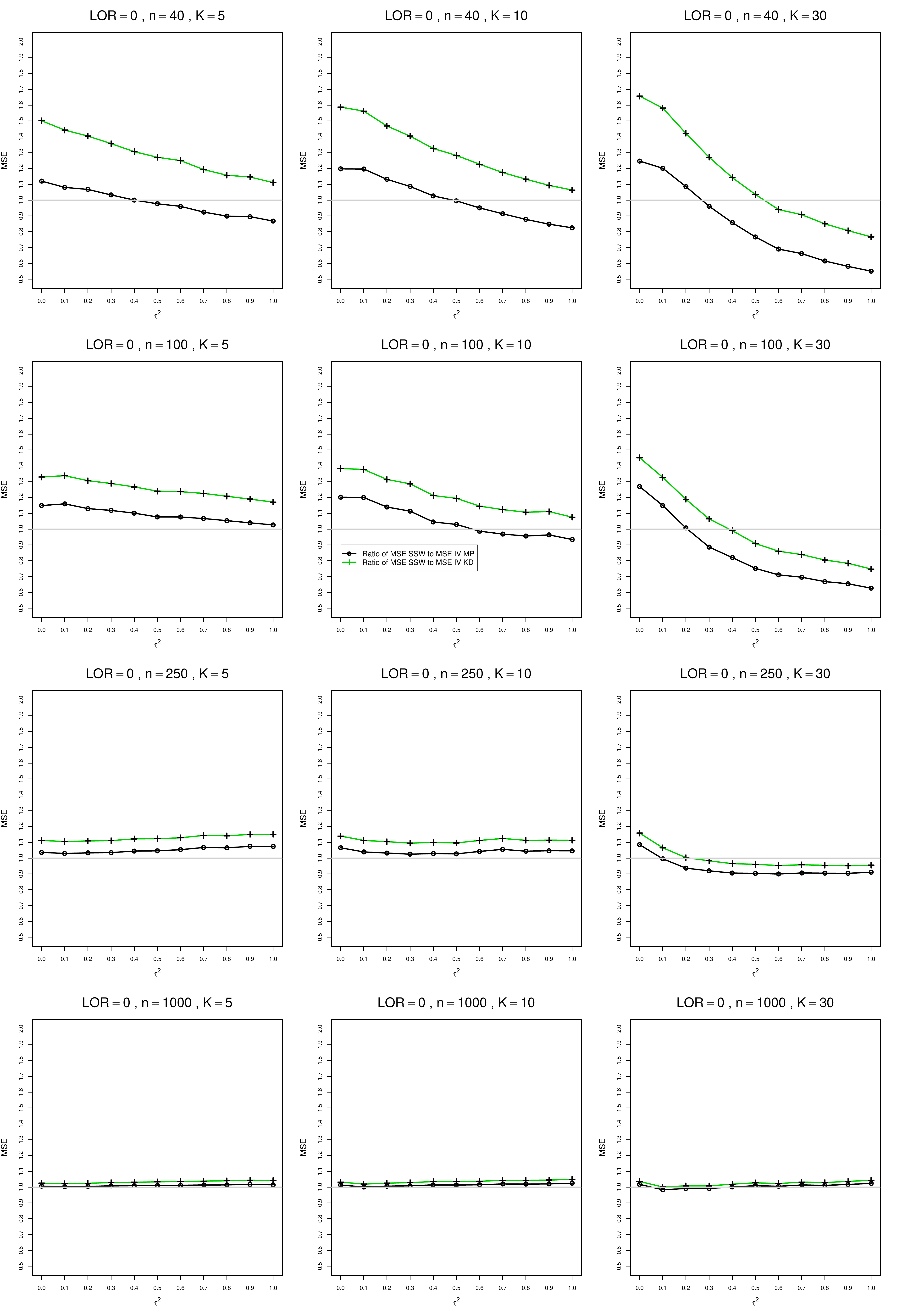}
	\caption{Ratio of mean squared errors of the fixed-weights to mean squared errors of inverse-variance estimator for $\theta=0$,$p_{iC}=0.1$, $q=0.5$, equal sample sizes $n=40,\;100,\;250,\;1000$. 
		\label{RatioOfMSEwithLOR0q05piC01fromMPandCMP}}
\end{figure}

\begin{figure}[t]\centering
	\includegraphics[scale=0.35]{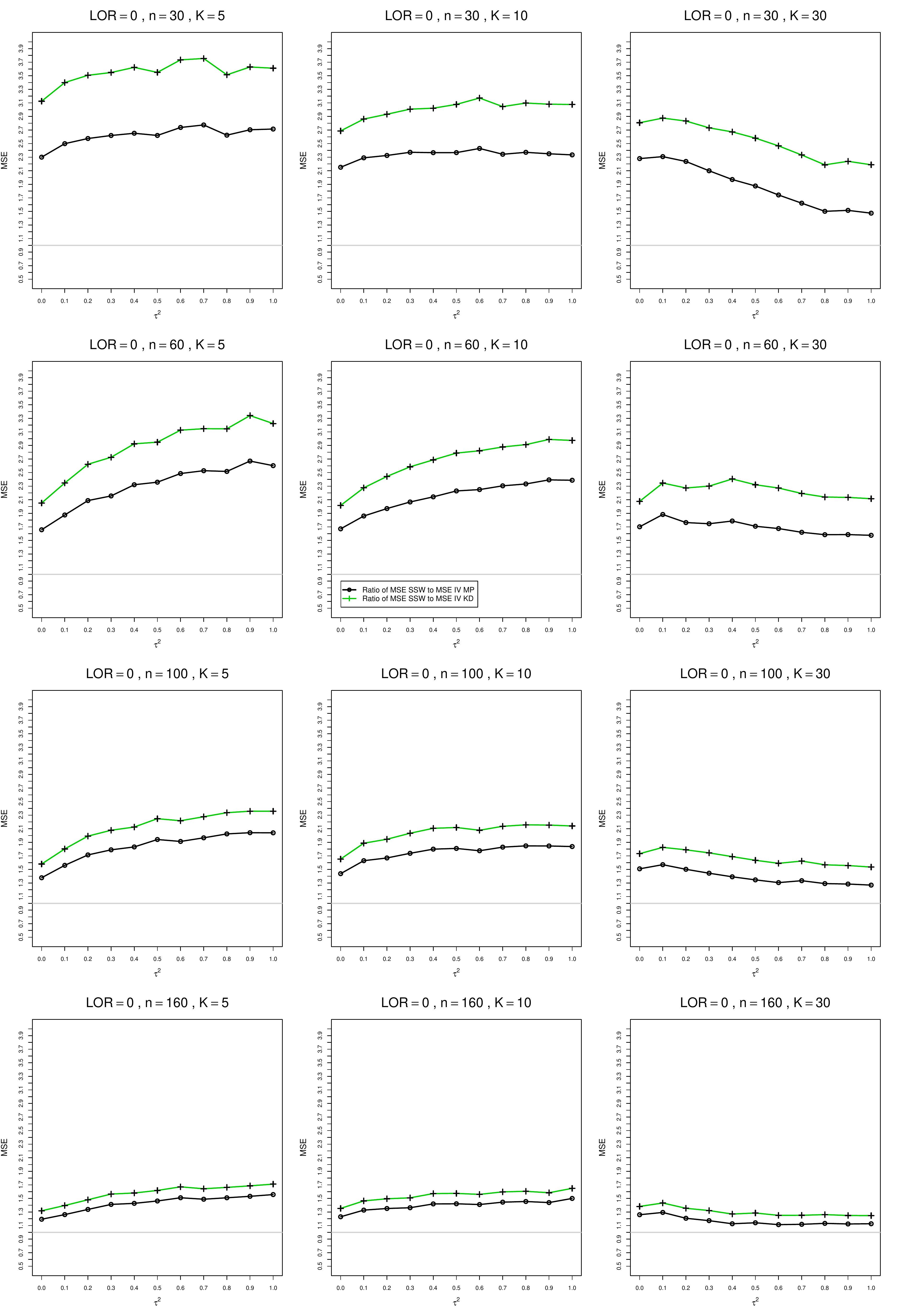}
	\caption{Ratio of mean squared errors of the fixed-weights to mean squared errors of inverse-variance estimator for $\theta=0$,$p_{iC}=0.1$, $q=0.5$, unequal sample sizes $n=30,\;60,\;100,\;160$. 
		\label{RatioOfMSEwithLOR0q05piC01fromMPandCMP_unequal_sample_sizes}}
\end{figure}


\begin{figure}[t]
	\centering
	\includegraphics[scale=0.33]{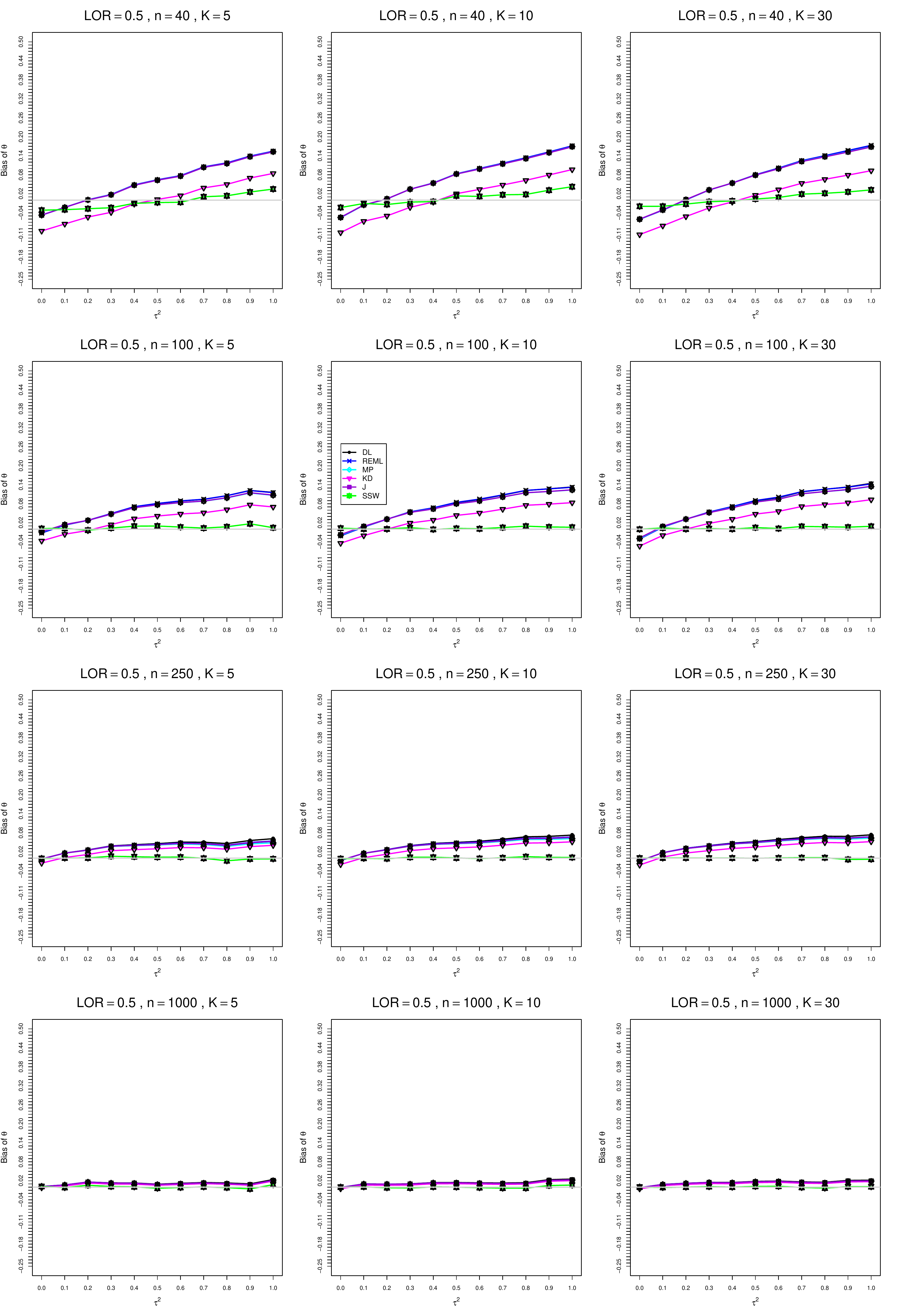}
	\caption{Bias of the estimation of  overall effect measure $\theta$ for $\theta=0.5$, $p_{iC}=0.1$, $q=0.5$, equal sample sizes $n=40,\;100,\;250,\;1000$. 
		\label{BiasThetaLOR05q05piC01}}
\end{figure}

\begin{figure}[t]
	\centering
	\includegraphics[scale=0.33]{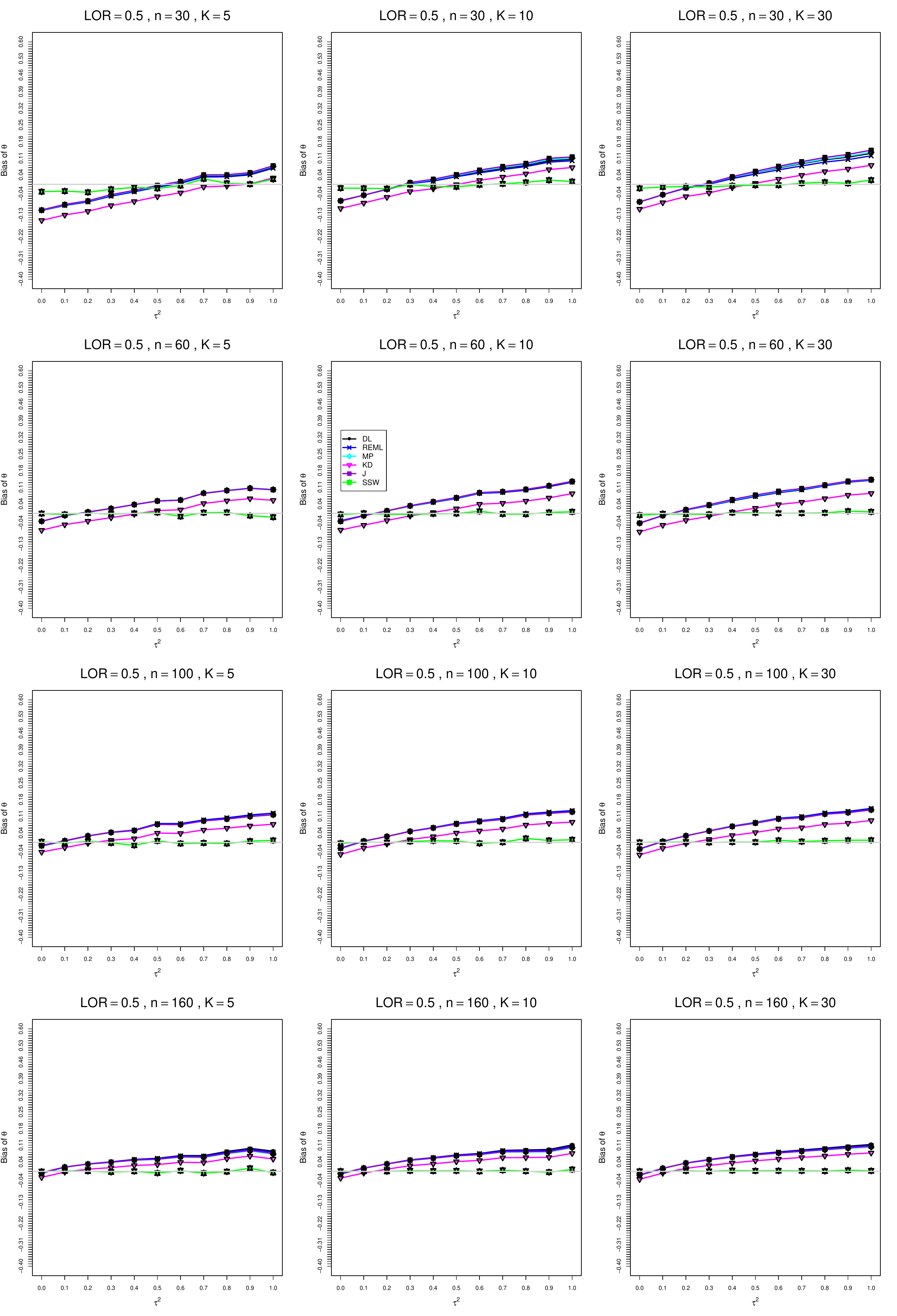}
	\caption{Bias of the estimation of  overall effect measure $\theta$ for $\theta=0.5$, $p_{iC}=0.1$, $q=0.5$, 
		unequal sample sizes $n=30,\; 60,\;100,\;160$. 
		\label{BiasThetaLOR05q05piC01_unequal_sample_sizes}}
\end{figure}

\begin{figure}[t]\centering
	\includegraphics[scale=0.35]{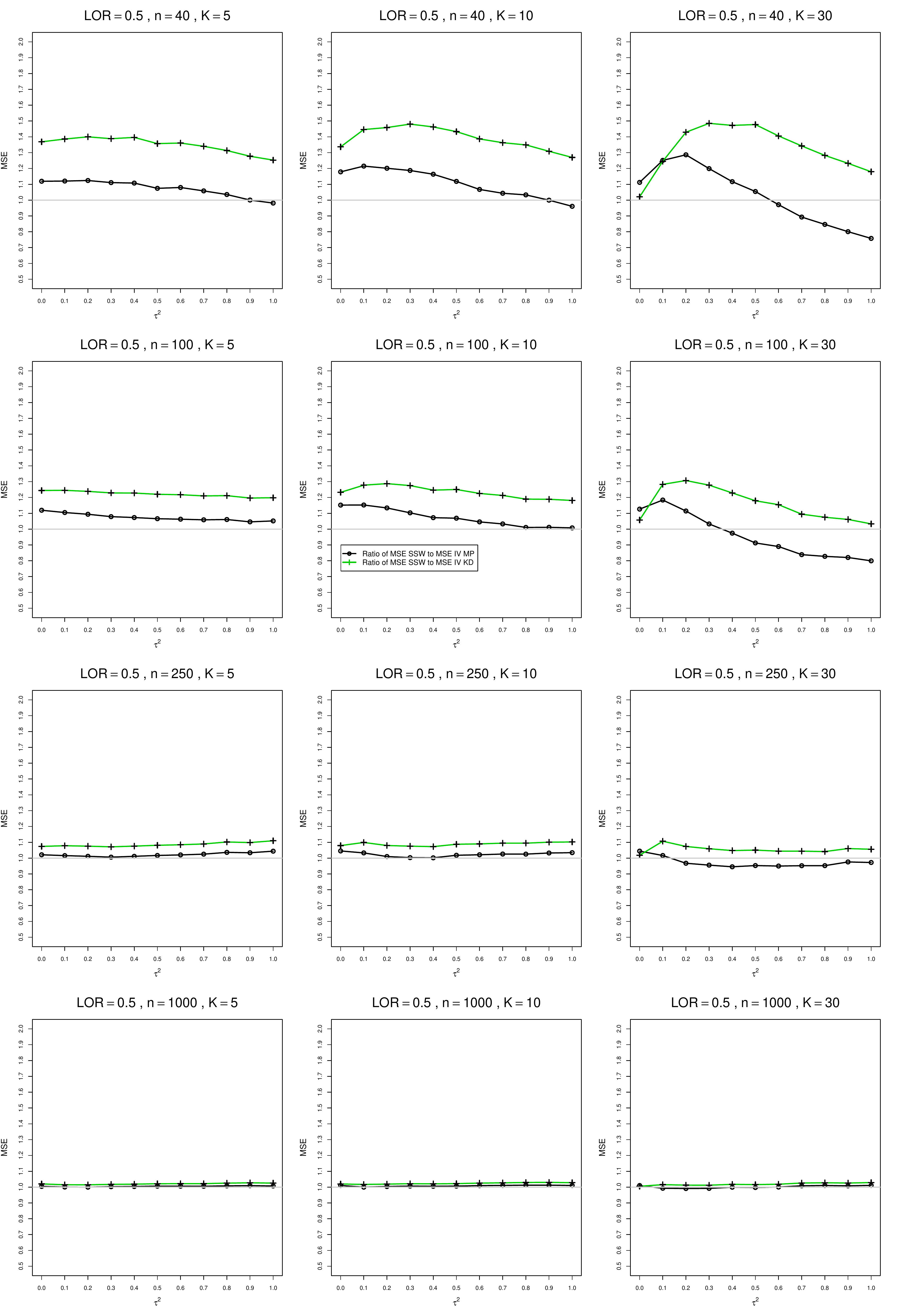}
	\caption{Ratio of mean squared errors of the fixed-weights to mean squared errors of inverse-variance estimator for $\theta=0.5$,$p_{iC}=0.1$, $q=0.5$, equal sample sizes $n=40,\;100,\;250,\;1000$. 
		\label{RatioOfMSEwithLOR05q05piC01fromMPandCMP}}
\end{figure}

\begin{figure}[t]\centering
	\includegraphics[scale=0.35]{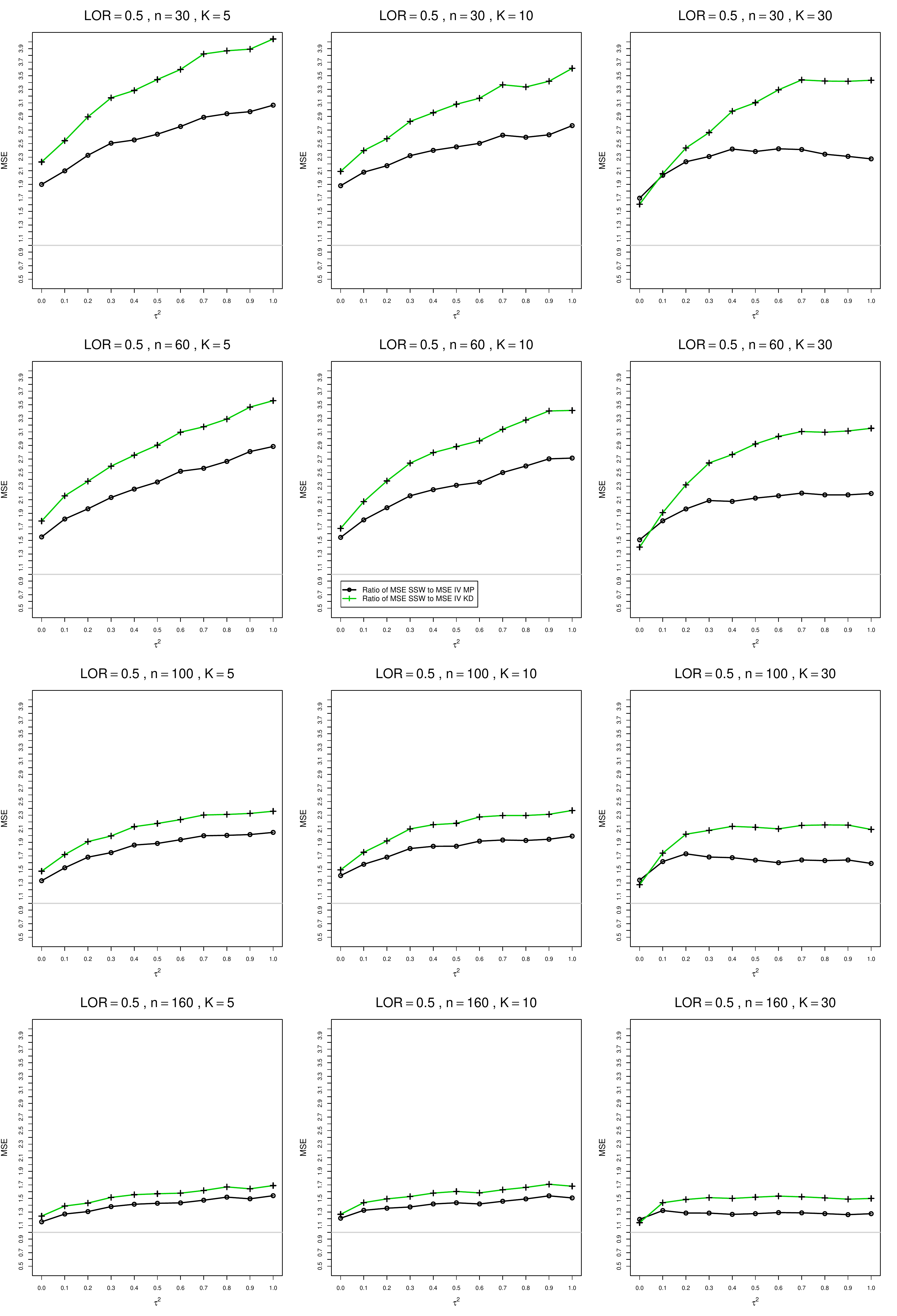}
	\caption{Ratio of mean squared errors of the fixed-weights to mean squared errors of inverse-variance estimator for $\theta=0.5$,$p_{iC}=0.1$, $q=0.5$, unequal sample sizes $n=30,\;60,\;100,\;160$. 
		\label{RatioOfMSEwithLOR05q05piC01fromMPandCMP_unequal_sample_sizes}}
\end{figure}


\begin{figure}[t]
	\centering
	\includegraphics[scale=0.33]{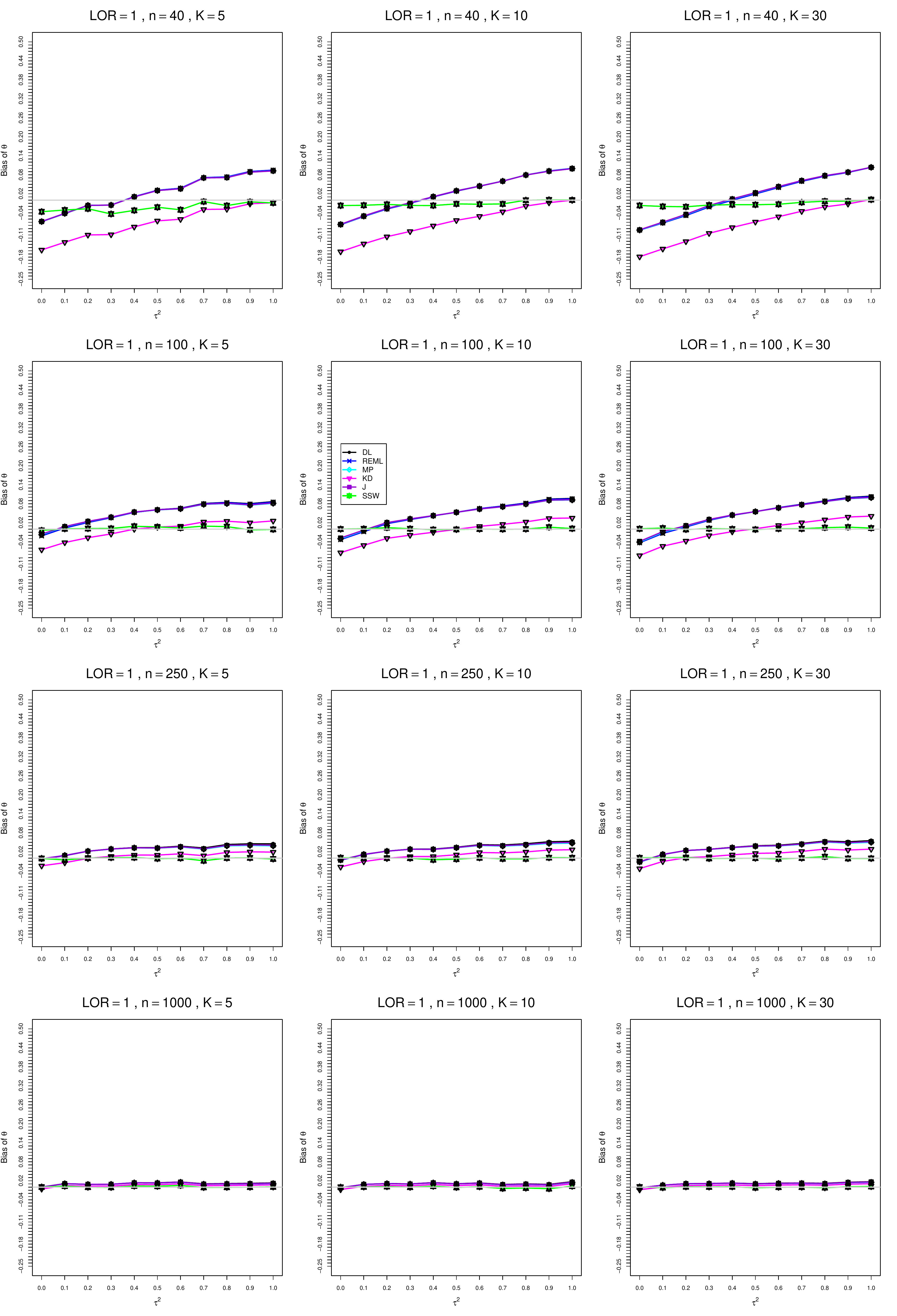}
	\caption{Bias of the estimation of  overall effect measure $\theta$ for $\theta=1$, $p_{iC}=0.1$, $q=0.5$, equal sample sizes $n=40,\;100,\;250,\;1000$. 
		\label{BiasThetaLOR1q05piC01}}
\end{figure}

\begin{figure}[t]
	\centering
	\includegraphics[scale=0.33]{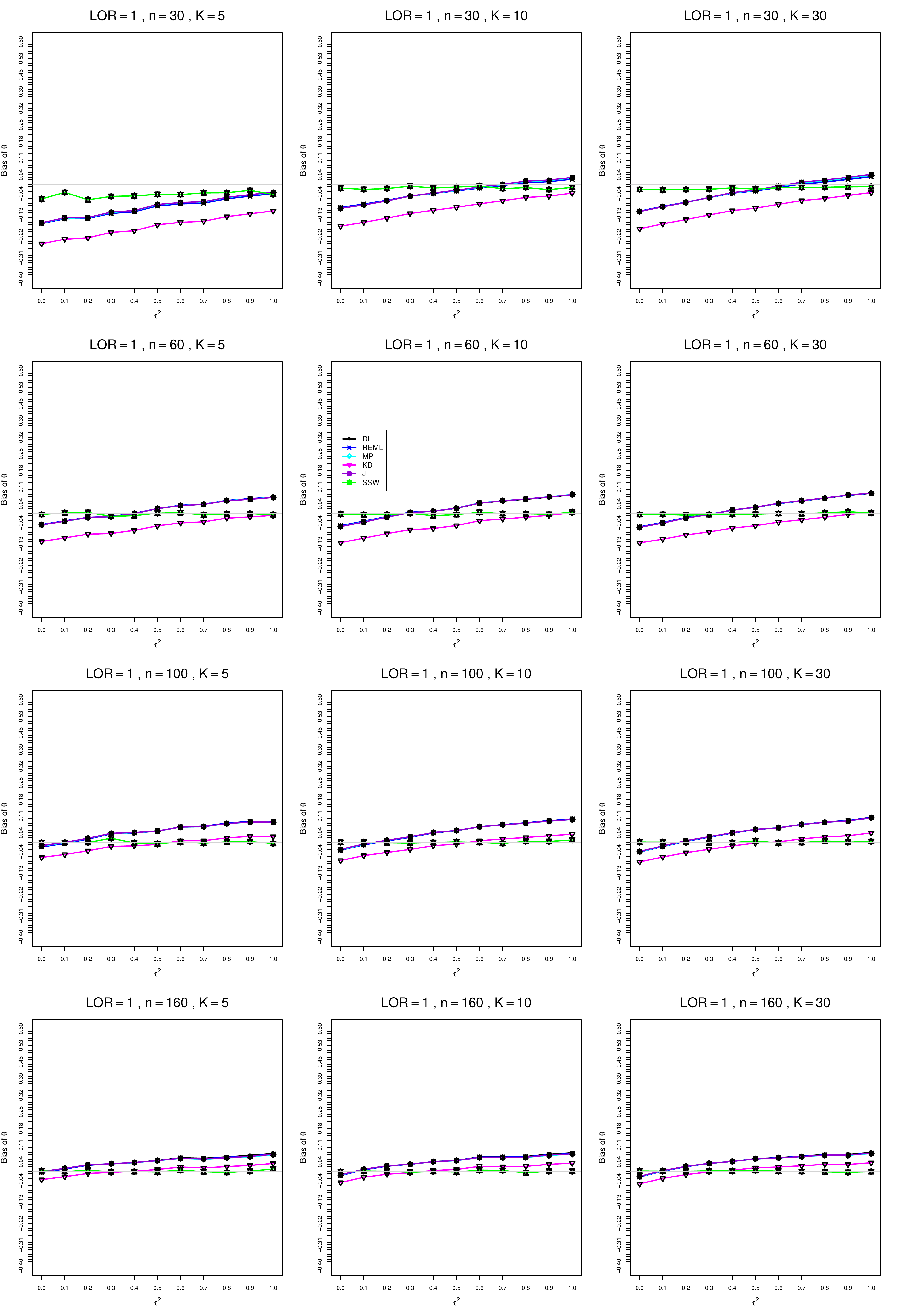}
	\caption{Bias of the estimation of  overall effect measure $\theta$ for LOR=1, $p_{iC}=0.1$, $q=0.5$, 
		unequal sample sizes $n=30,\; 60,\;100,\;160$. 
		\label{BiasThetaLOR1q05piC01_unequal_sample_sizes}}
\end{figure}

\begin{figure}[t]\centering
	\includegraphics[scale=0.35]{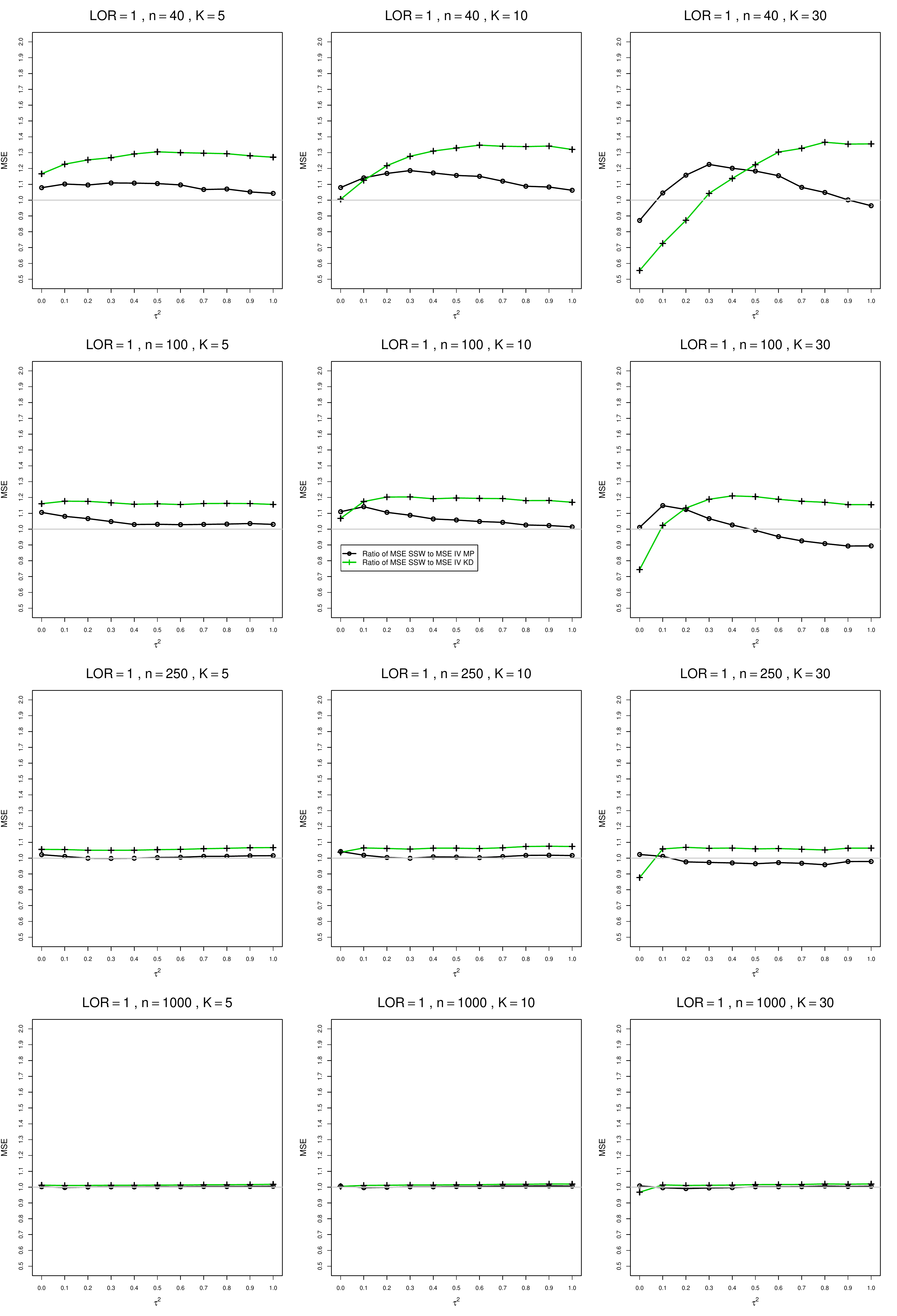}
	\caption{Ratio of mean squared errors of the fixed-weights to mean squared errors of inverse-variance estimator for $\theta=1$,$p_{iC}=0.1$, $q=0.5$, equal sample sizes $n=40,\;100,\;250,\;1000$. 
		\label{RatioOfMSEwithLOR1q05piC01fromMPandCMP}}
\end{figure}

\begin{figure}[t]\centering
	\includegraphics[scale=0.35]{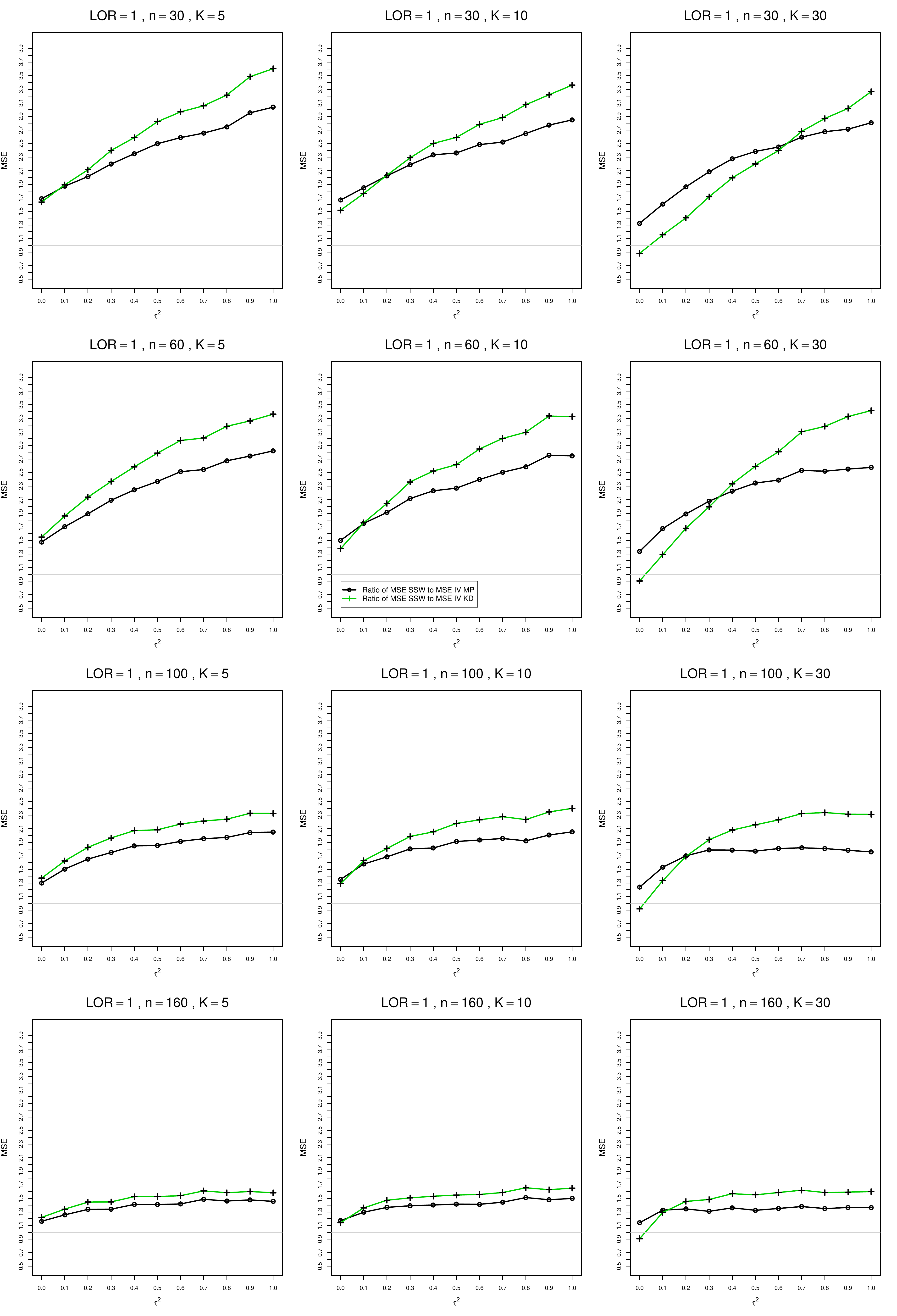}
	\caption{Ratio of mean squared errors of the fixed-weights to mean squared errors of inverse-variance estimator for $\theta=1$,$p_{iC}=0.1$, $q=0.5$, unequal sample sizes $n=30,\;60,\;100,\;160$. 
		\label{RatioOfMSEwithLOR1q05piC01fromMPandCMP_unequal_sample_sizes}}
\end{figure}

\begin{figure}[t]
	\centering
	\includegraphics[scale=0.33]{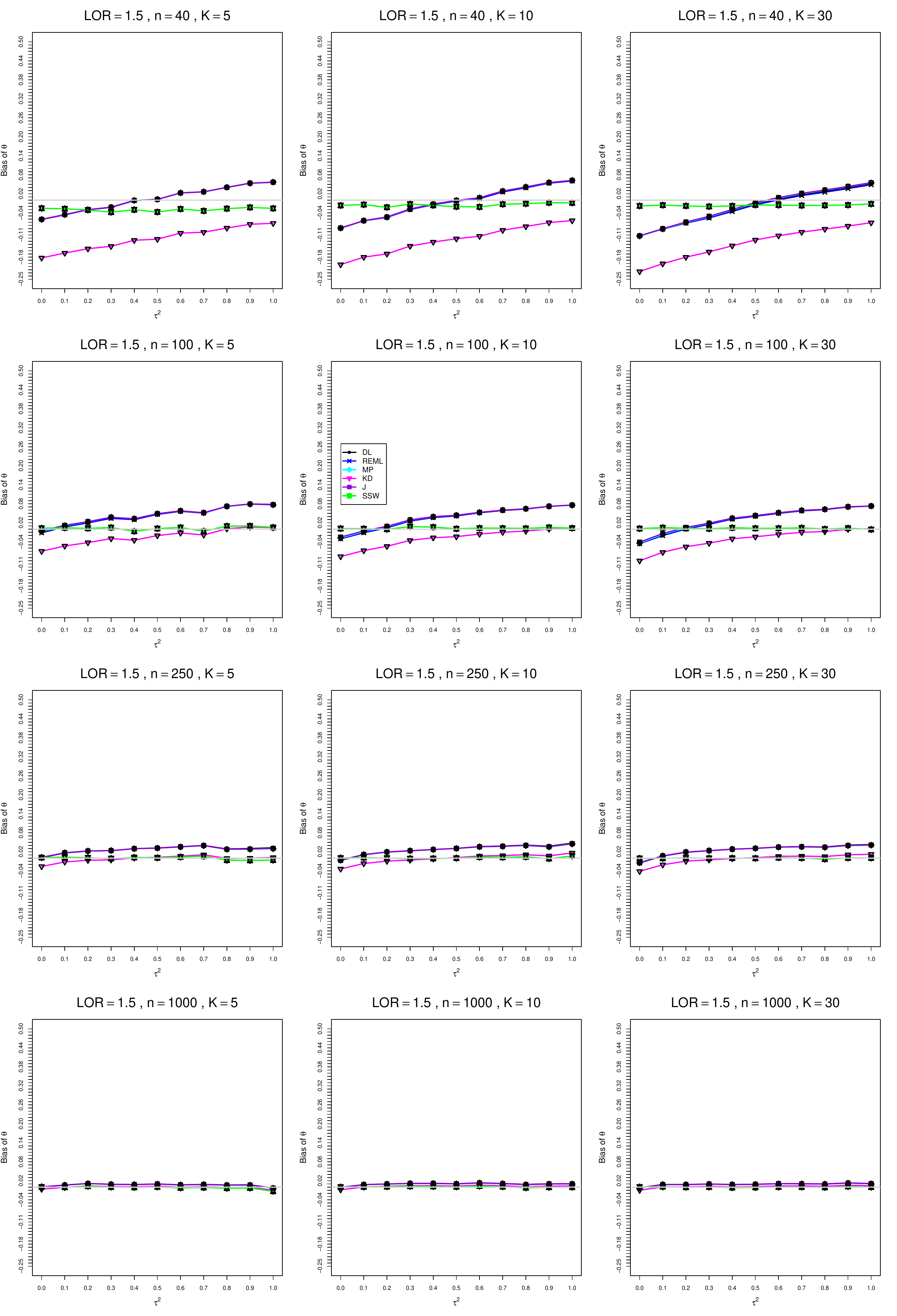}
	\caption{Bias of the estimation of  overall effect measure $\theta$ for $\theta=1.5$, $p_{iC}=0.1$, $q=0.5$, equal sample sizes $n=40,\;100,\;250,\;1000$. 
		\label{BiasThetaLOR15q05piC01}}
\end{figure}

\begin{figure}[t]
	\centering
	\includegraphics[scale=0.33]{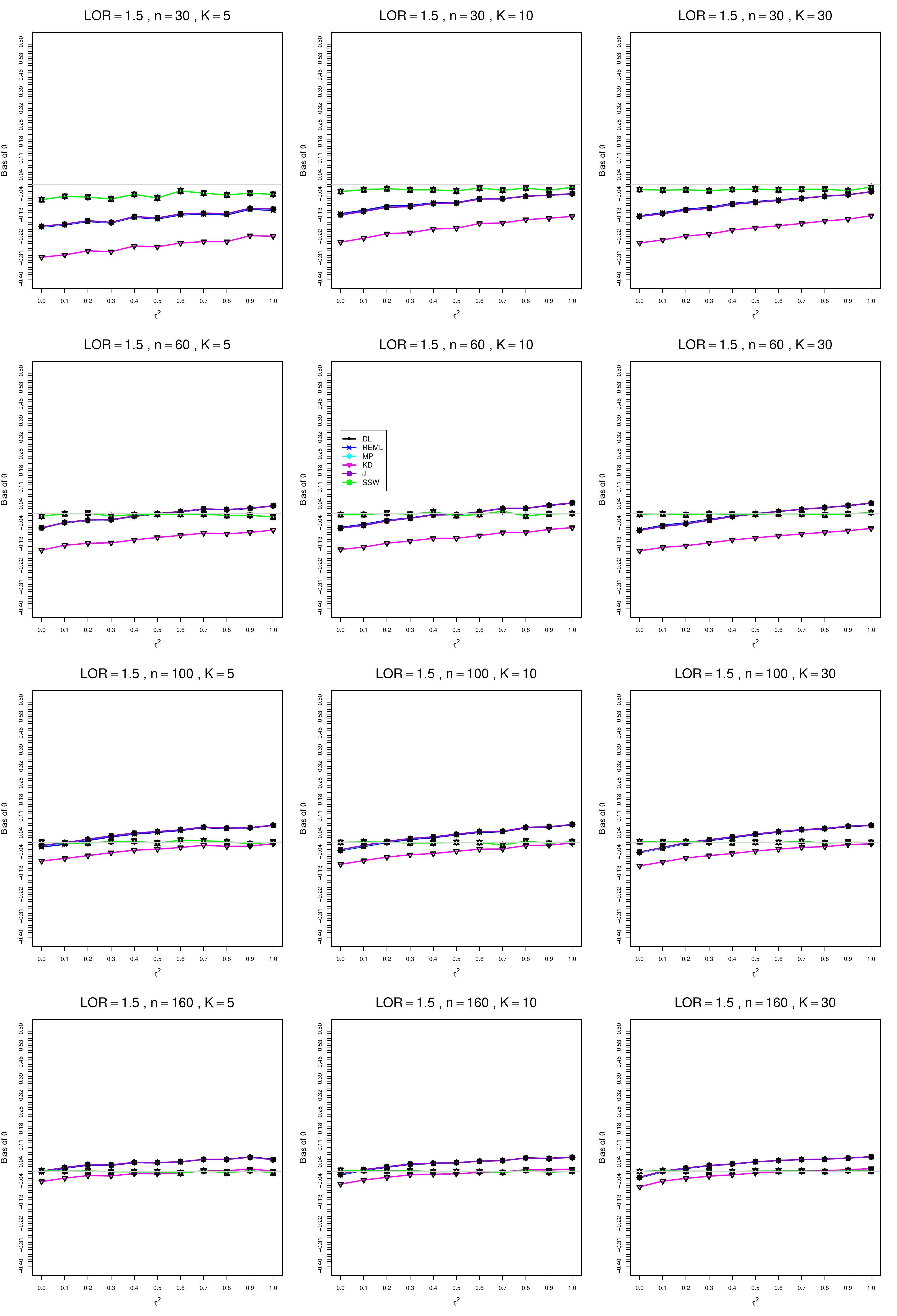}
	\caption{Bias of the estimation of  overall effect measure $\theta$ for $\theta=1.5$, $p_{iC}=0.1$, $q=0.5$, 
		unequal sample sizes $n=30,\; 60,\;100,\;160$. 
		\label{BiasThetaLOR15q05piC01_unequal_sample_sizes}}
\end{figure}

\begin{figure}[t]\centering
	\includegraphics[scale=0.35]{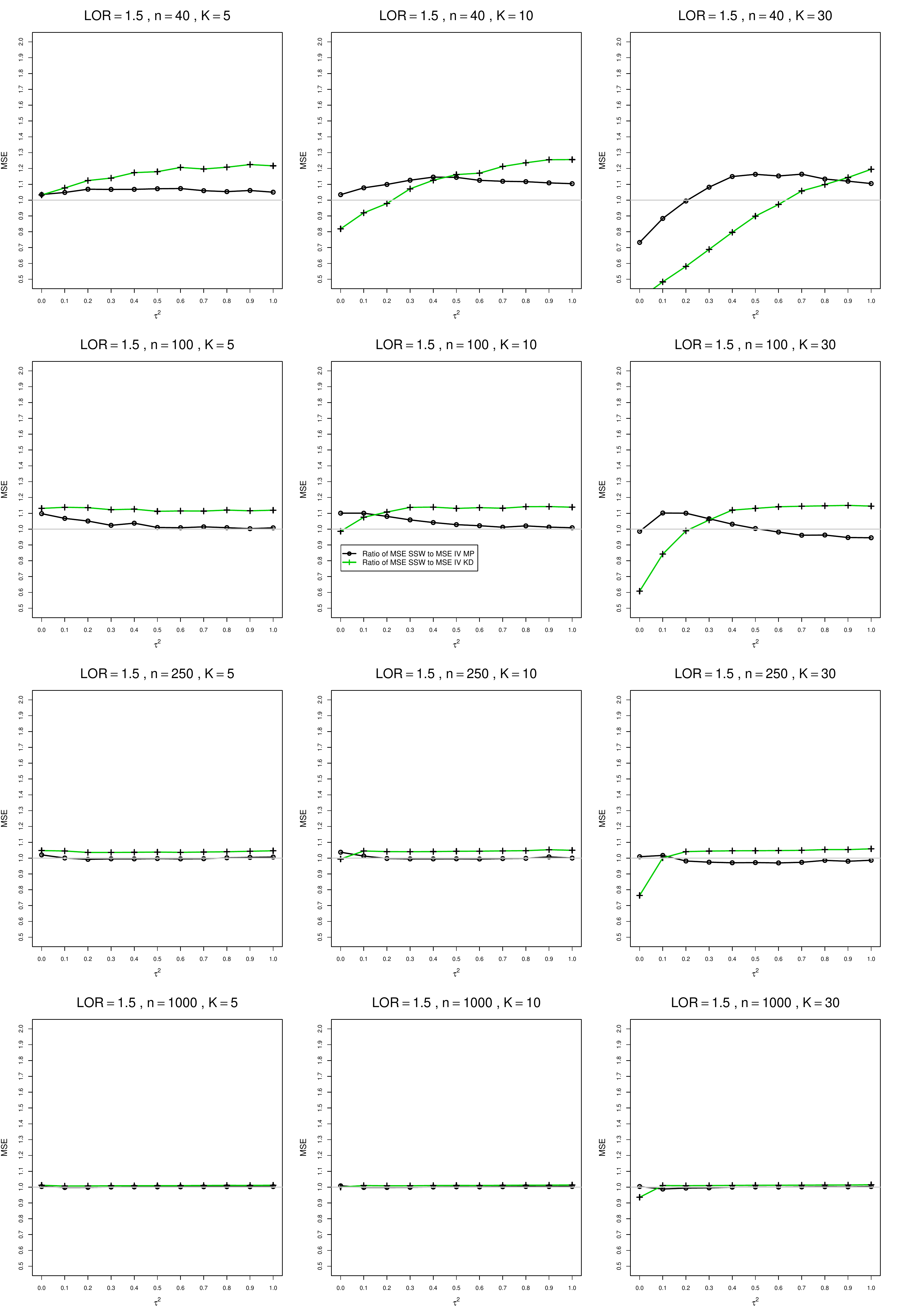}
	\caption{Ratio of mean squared errors of the fixed-weights to mean squared errors of inverse-variance estimator for $\theta=1.5$,$p_{iC}=0.1$, $q=0.5$, equal sample sizes $n=40,\;100,\;250,\;1000$. 
		\label{RatioOfMSEwithLOR15q05piC01fromMPandCMP}}
\end{figure}

\begin{figure}[t]\centering
	\includegraphics[scale=0.35]{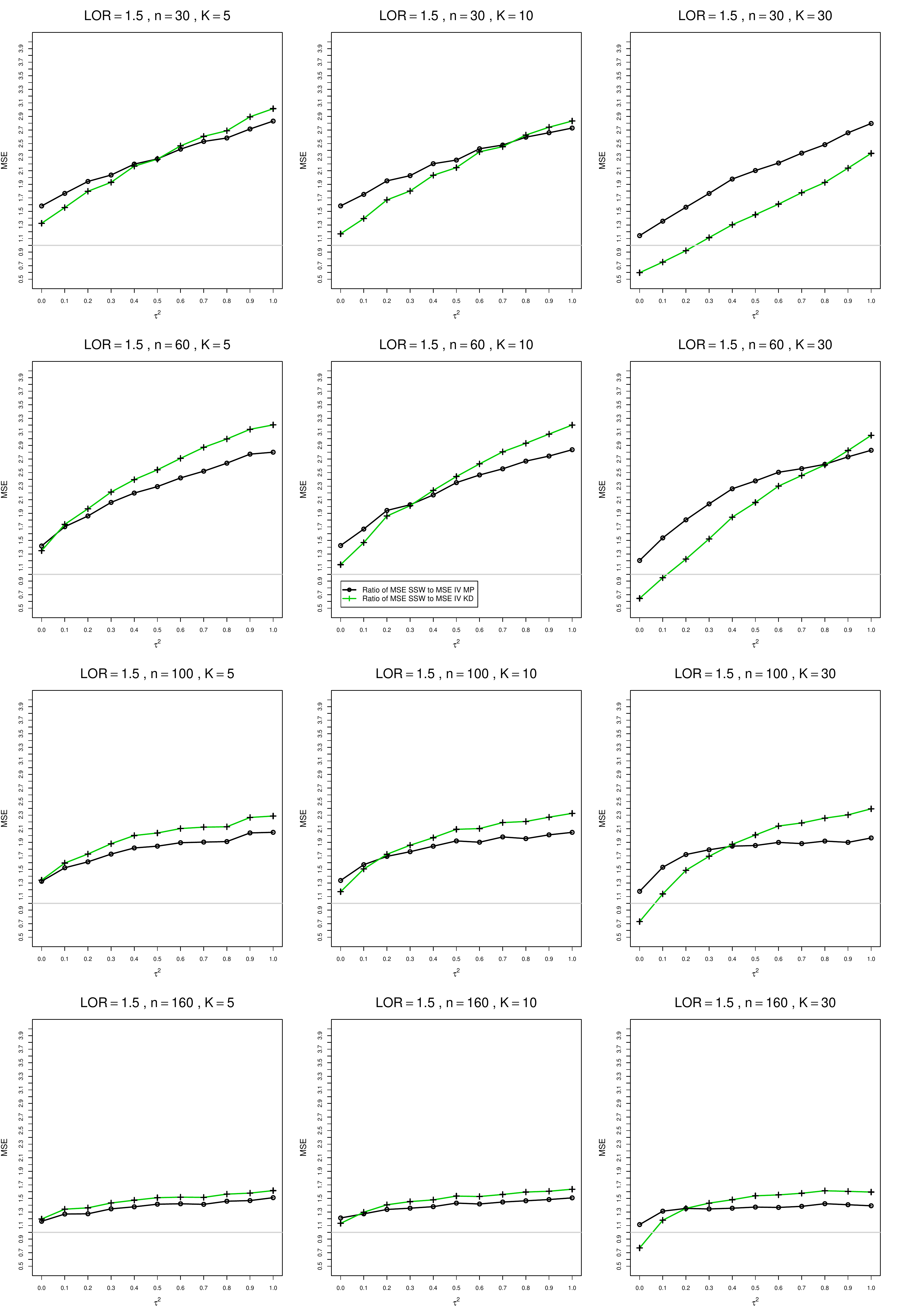}
	\caption{Ratio of mean squared errors of the fixed-weights to mean squared errors of inverse-variance estimator for $\theta=1.5$,$p_{iC}=0.1$, $q=0.5$, unequal sample sizes $n=30,\;60,\;100,\;160$. 
		\label{RatioOfMSEwithLOR15q05piC01fromMPandCMP_unequal_sample_sizes}}
\end{figure}

\begin{figure}[t]
	\centering
	\includegraphics[scale=0.33]{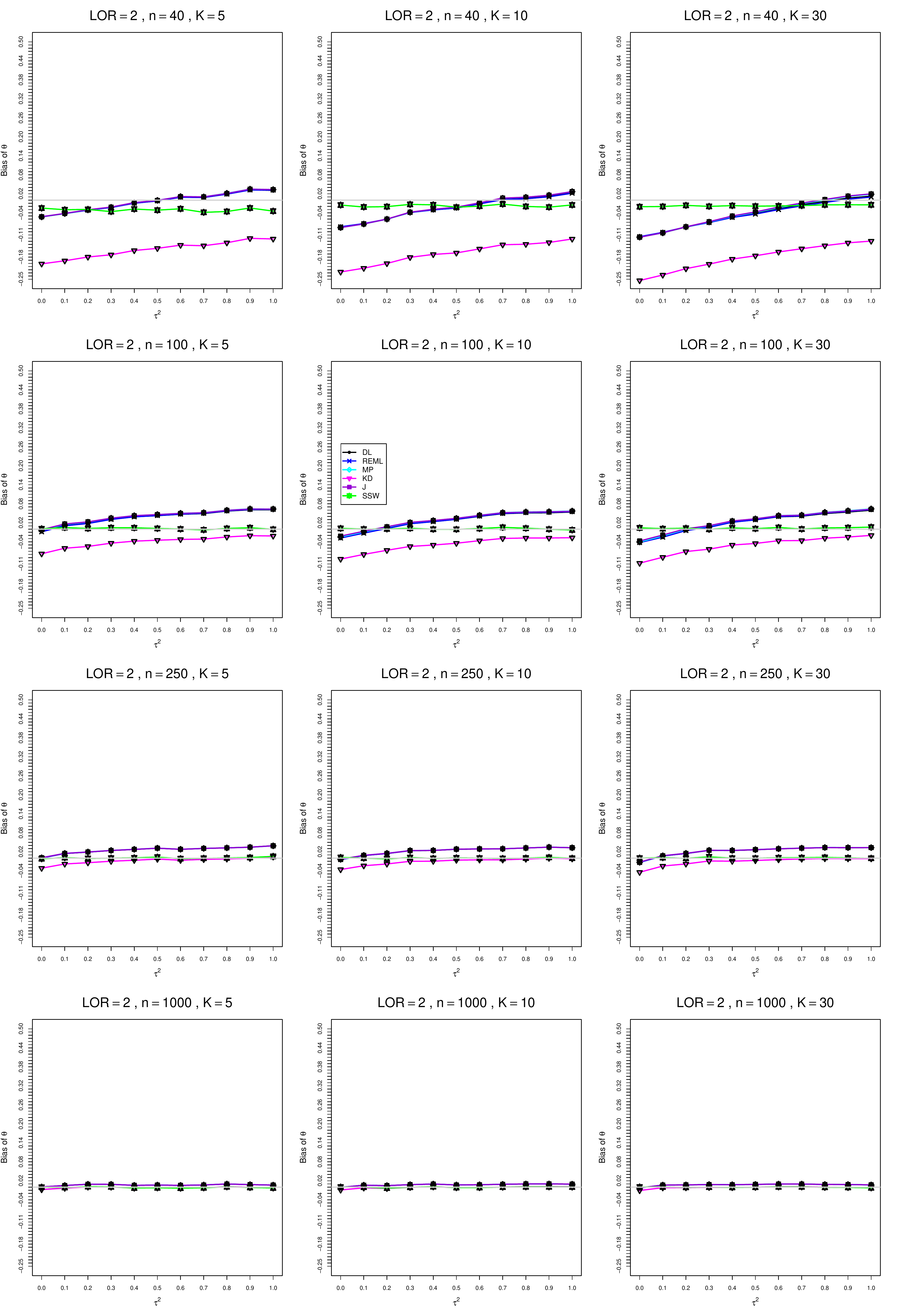}
	\caption{Bias of the estimation of  overall effect measure $\theta$ for $\theta=2$, $p_{iC}=0.1$, $q=0.5$, equal sample sizes $n=40,\;100,\;250,\;1000$. 
		\label{BiasThetaLOR2q05piC01}}
\end{figure}

\begin{figure}[t]
	\centering
	\includegraphics[scale=0.33]{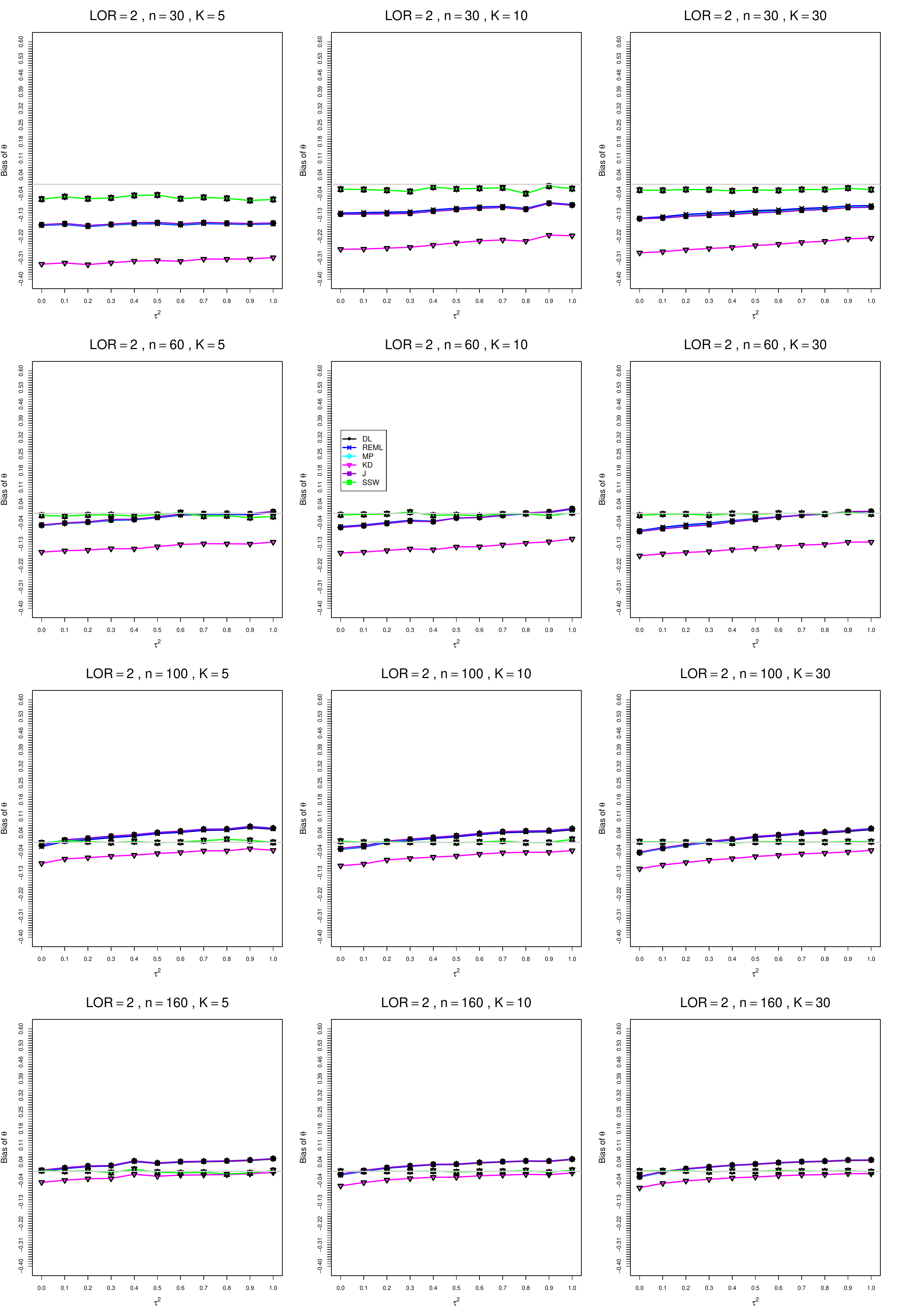}
	\caption{Bias of the estimation of  overall effect measure $\theta$ for $\theta=2$, $p_{iC}=0.1$, $q=0.5$, 
		unequal sample sizes $n=30,\; 60,\;100,\;160$. 
		\label{BiasThetaLOR2q05piC01_unequal_sample_sizes}}
\end{figure}

\begin{figure}[t]\centering
	\includegraphics[scale=0.35]{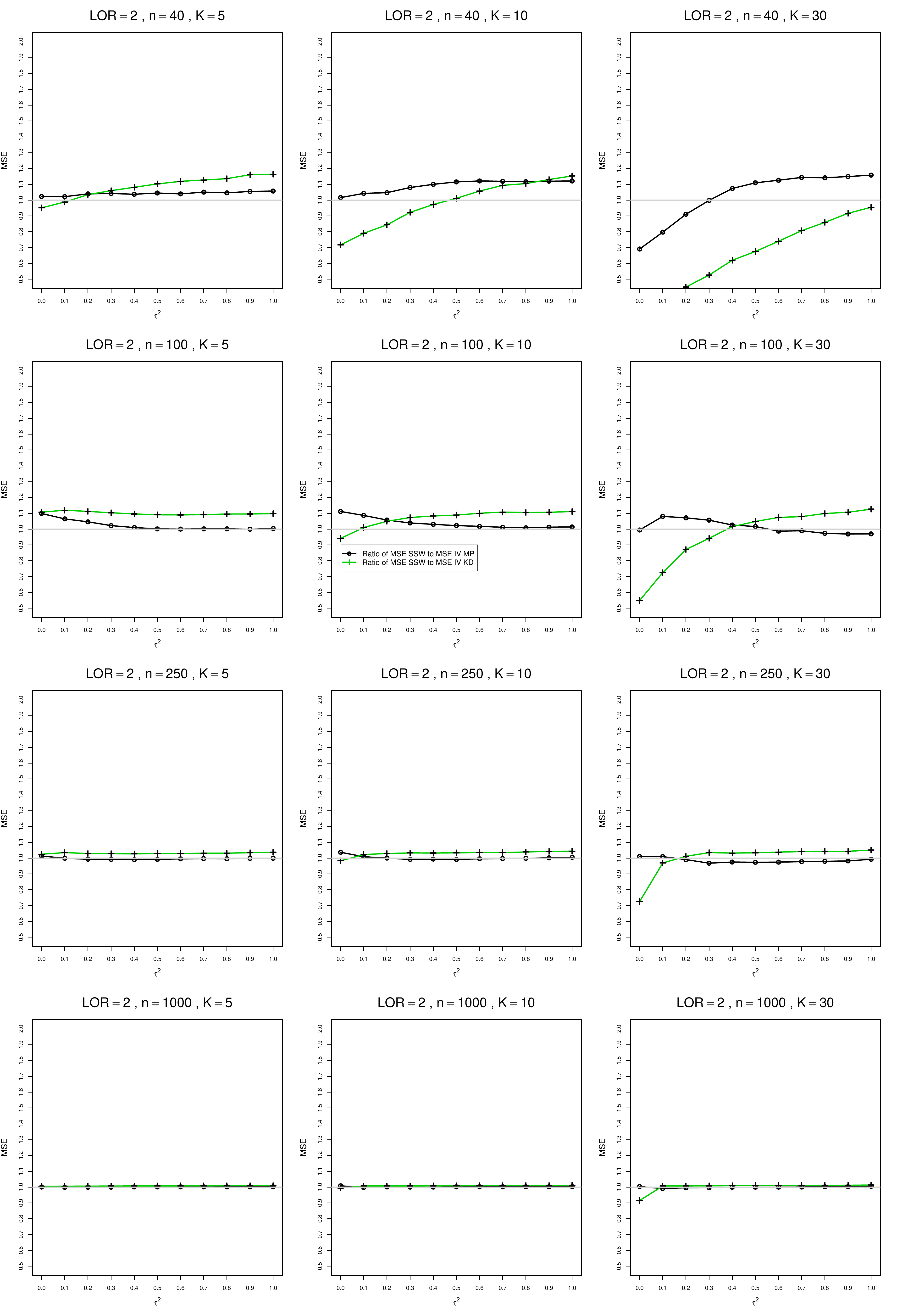}
	\caption{Ratio of mean squared errors of the fixed-weights to mean squared errors of inverse-variance estimator for $\theta=2$,$p_{iC}=0.1$, $q=0.5$, equal sample sizes $n=40,\;100,\;250,\;1000$. 
		\label{RatioOfMSEwithLOR2q05piC01fromMPandCMP}}
\end{figure}

\begin{figure}[t]\centering
	\includegraphics[scale=0.35]{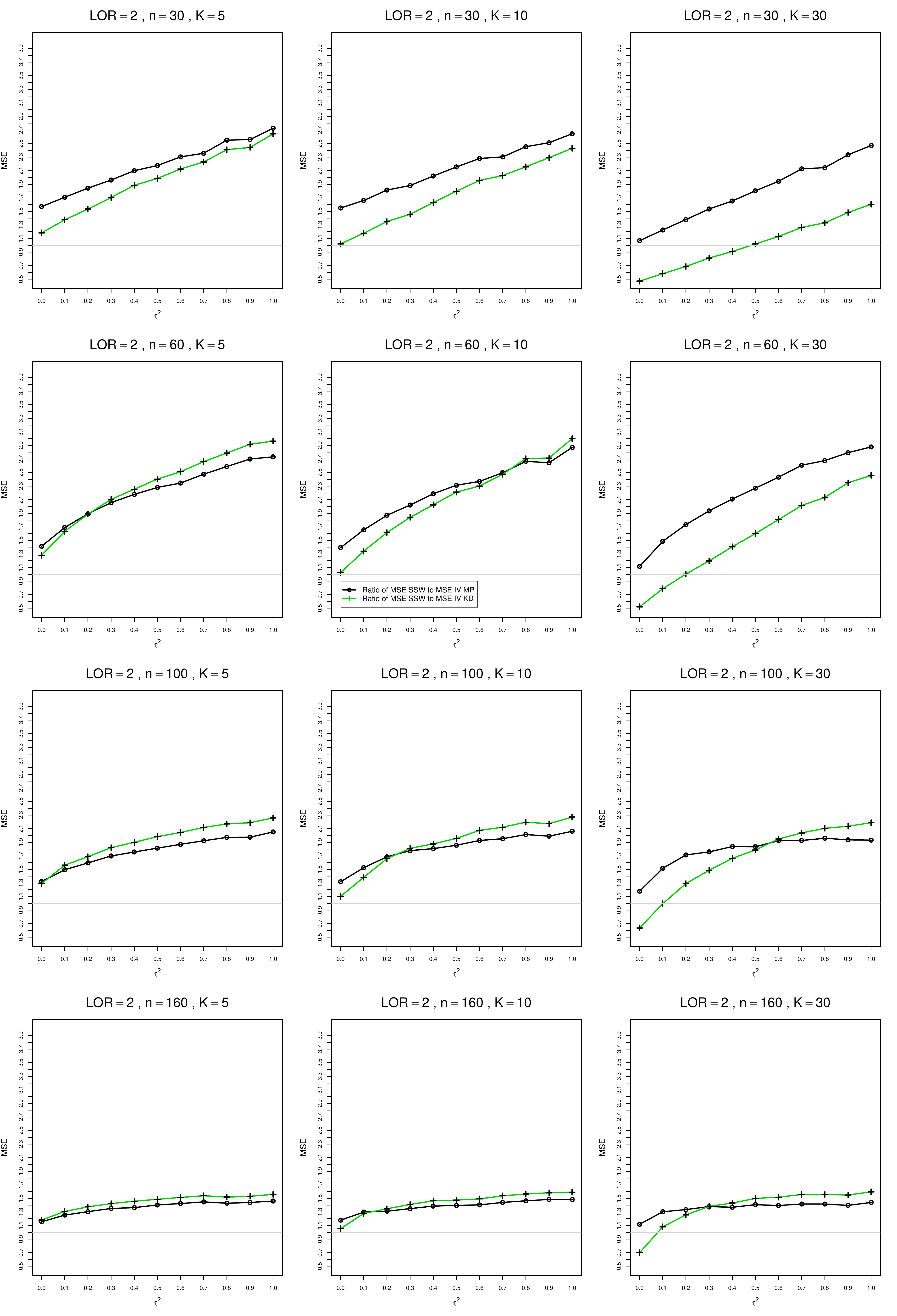}
	\caption{Ratio of mean squared errors of the fixed-weights to mean squared errors of inverse-variance estimator for $\theta=2$,$p_{iC}=0.1$, $q=0.5$, unequal sample sizes $n=30,\;60,\;100,\;160$. 
		\label{RatioOfMSEwithLOR2q05piC01fromMPandCMP_unequal_sample_sizes}}
\end{figure}


\begin{figure}[t]
	\centering
	\includegraphics[scale=0.33]{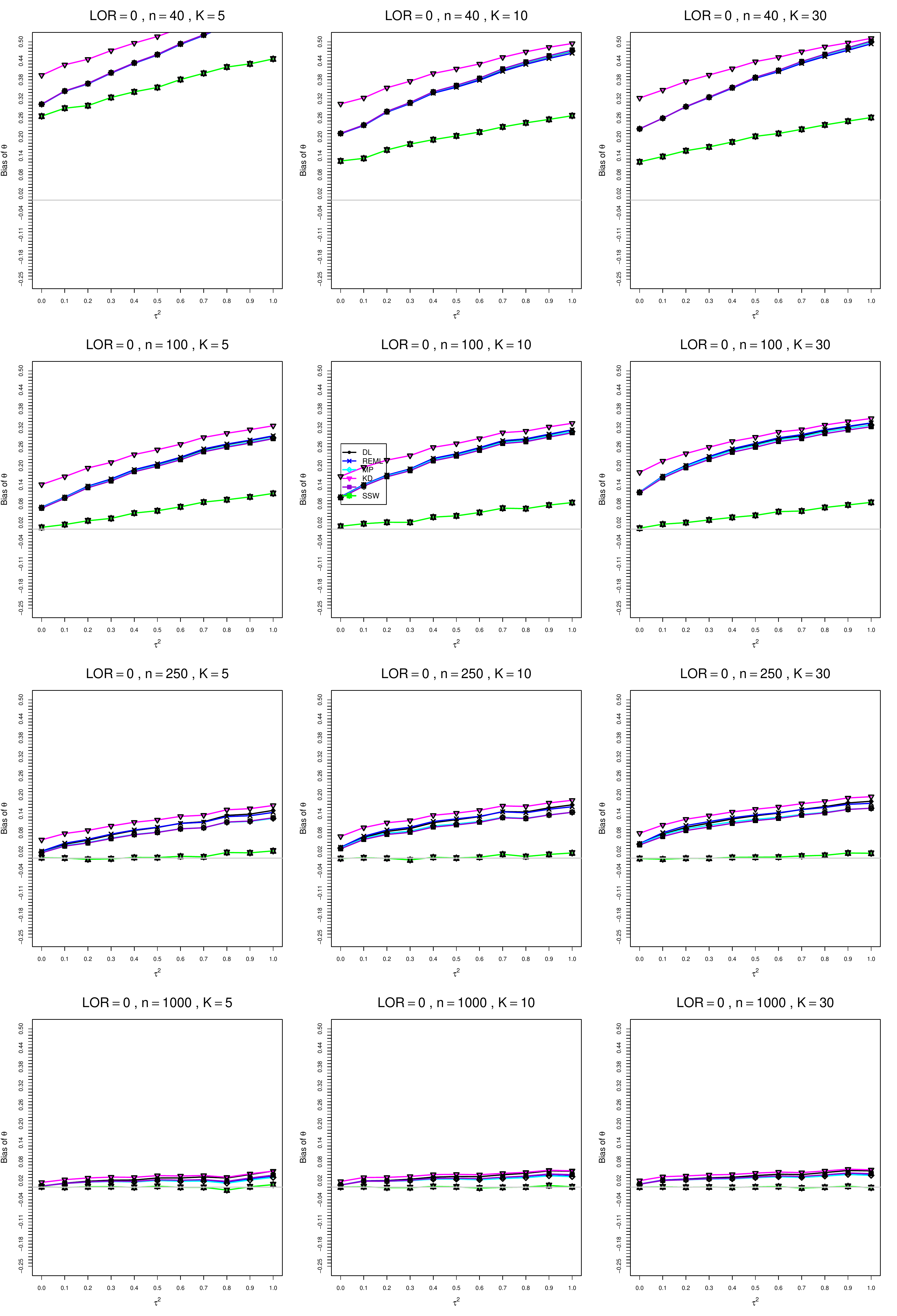}
	\caption{Bias of the estimation of  overall effect measure $\theta$ for $\theta=0$, $p_{iC}=0.1$, $q=0.75$, equal sample sizes $n=40,\;100,\;250,\;1000$. 
		\label{BiasThetaLOR0q075piC01}}
\end{figure}

\begin{figure}[t]
	\centering
	\includegraphics[scale=0.33]{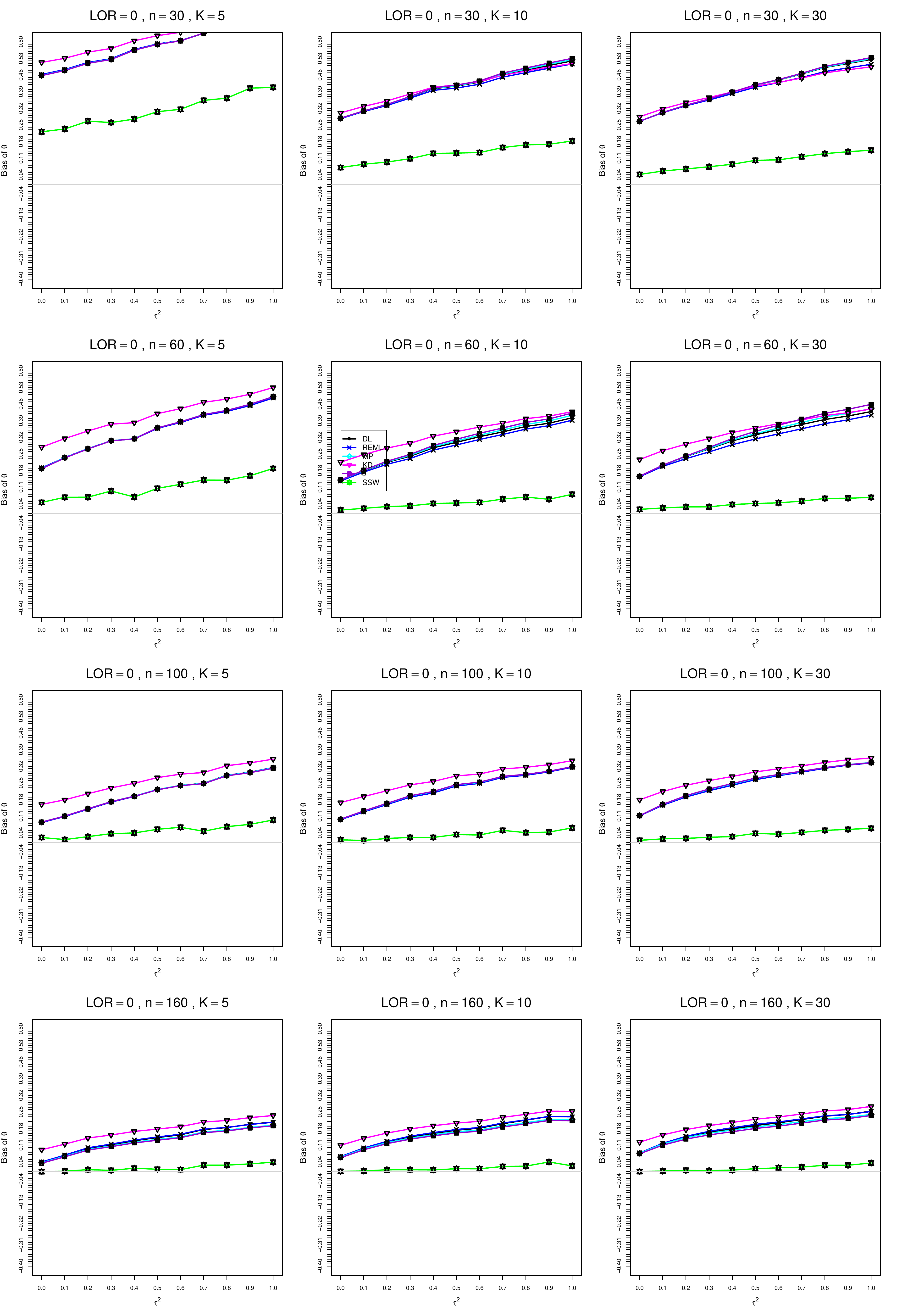}
	\caption{Bias of the estimation of  overall effect measure $\theta$ for $\theta=0$, $p_{iC}=0.1$, $q=0.75$, 
		unequal sample sizes $n=30,\; 60,\;100,\;160$. 
		\label{BiasThetaLOR0q075piC01_unequal_sample_sizes}}
\end{figure}

\begin{figure}[t]\centering
	\includegraphics[scale=0.35]{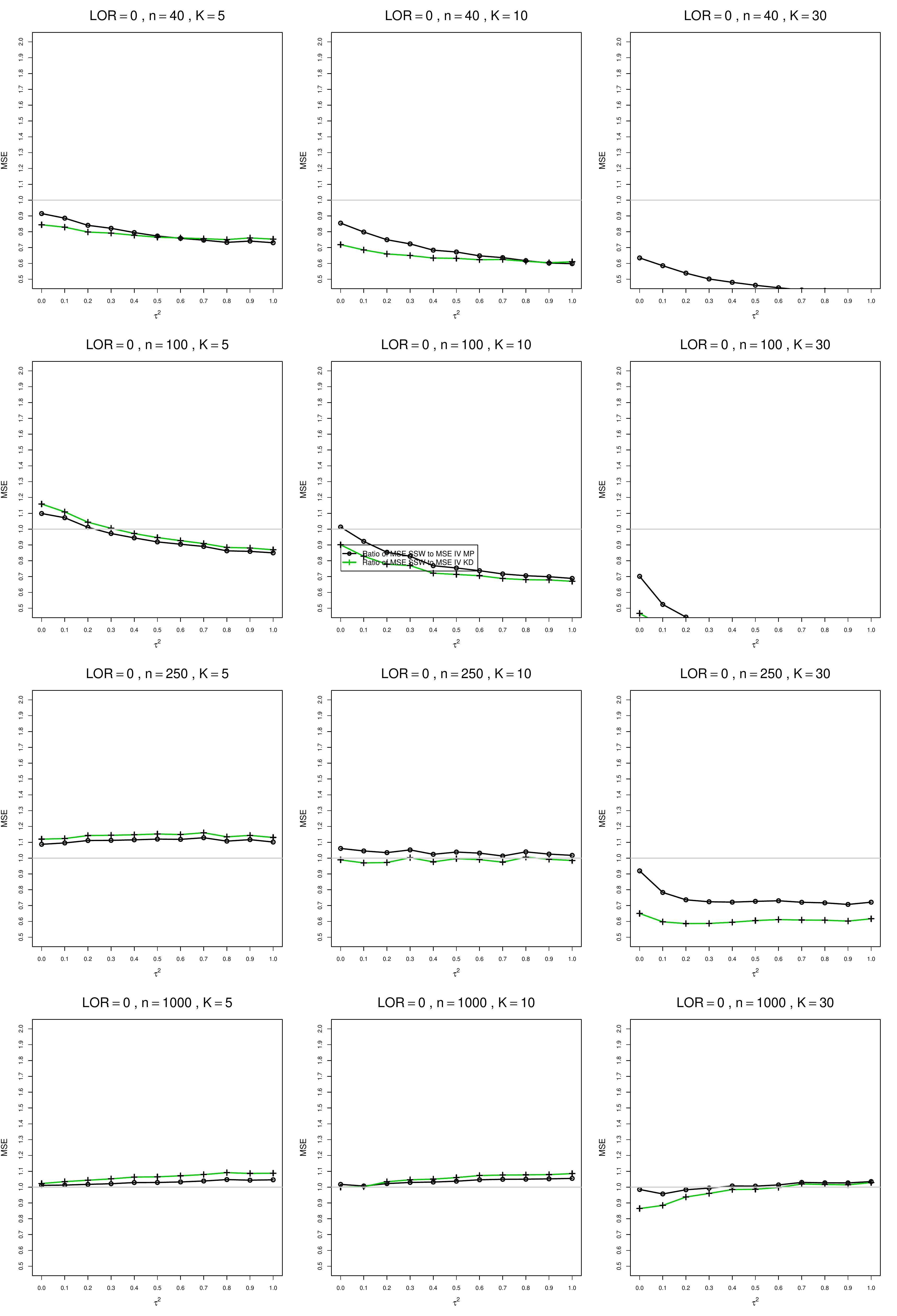}
	\caption{Ratio of mean squared errors of the fixed-weights to mean squared errors of inverse-variance estimator for $\theta=0$,$p_{iC}=0.1$, $q=0.75$, equal sample sizes $n=40,\;100,\;250,\;1000$. 
		\label{RatioOfMSEwithLOR0q075piC01fromMPandCMP}}
\end{figure}

\begin{figure}[t]\centering
	\includegraphics[scale=0.35]{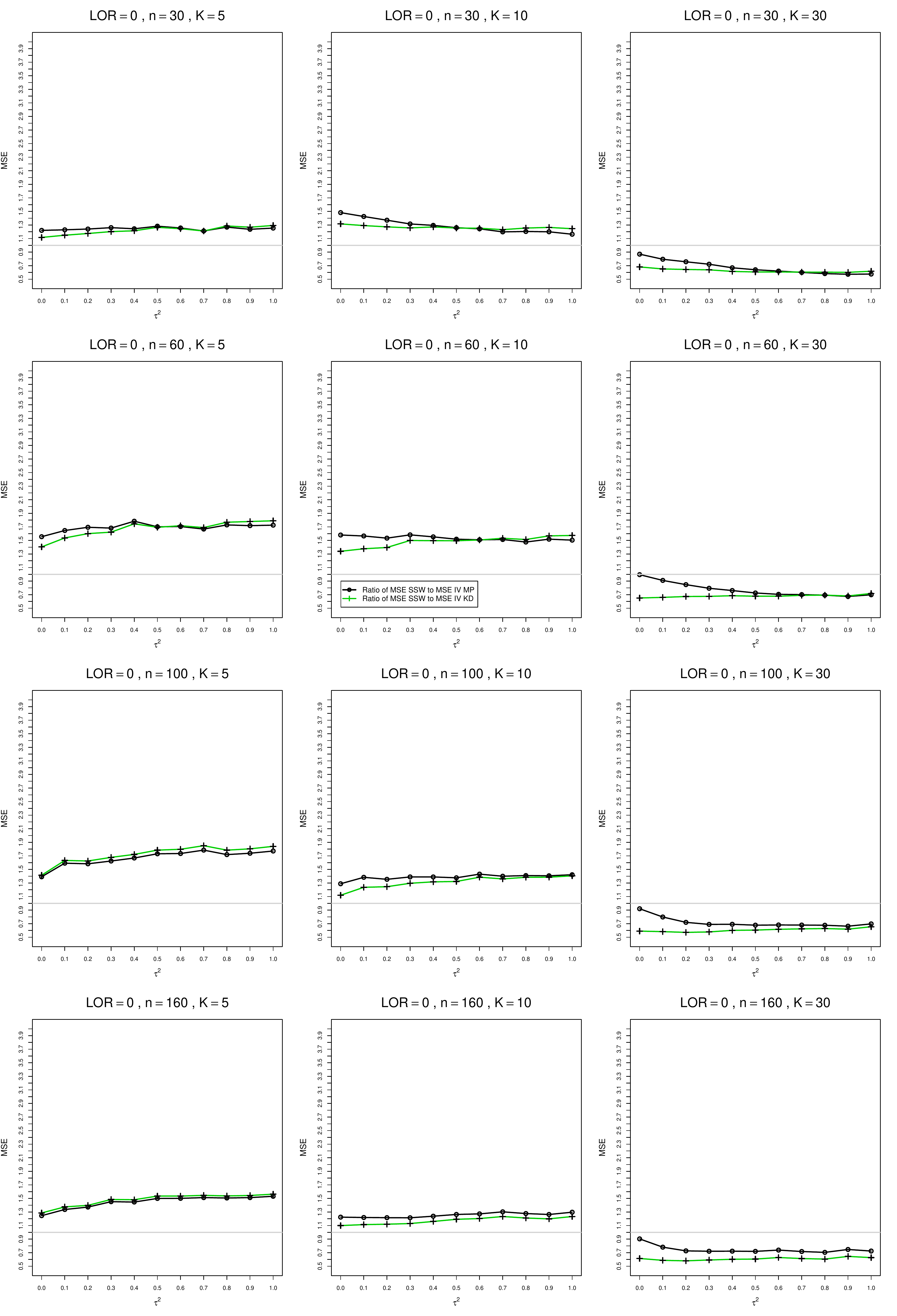}
	\caption{Ratio of mean squared errors of the fixed-weights to mean squared errors of inverse-variance estimator for $\theta=0$,$p_{iC}=0.1$, $q=0.75$, unequal sample sizes $n=30,\;60,\;100,\;160$. 
		\label{RatioOfMSEwithLOR0q075piC01fromMPandCMP_unequal_sample_sizes}}
\end{figure}


\begin{figure}[t]
	\centering
	\includegraphics[scale=0.33]{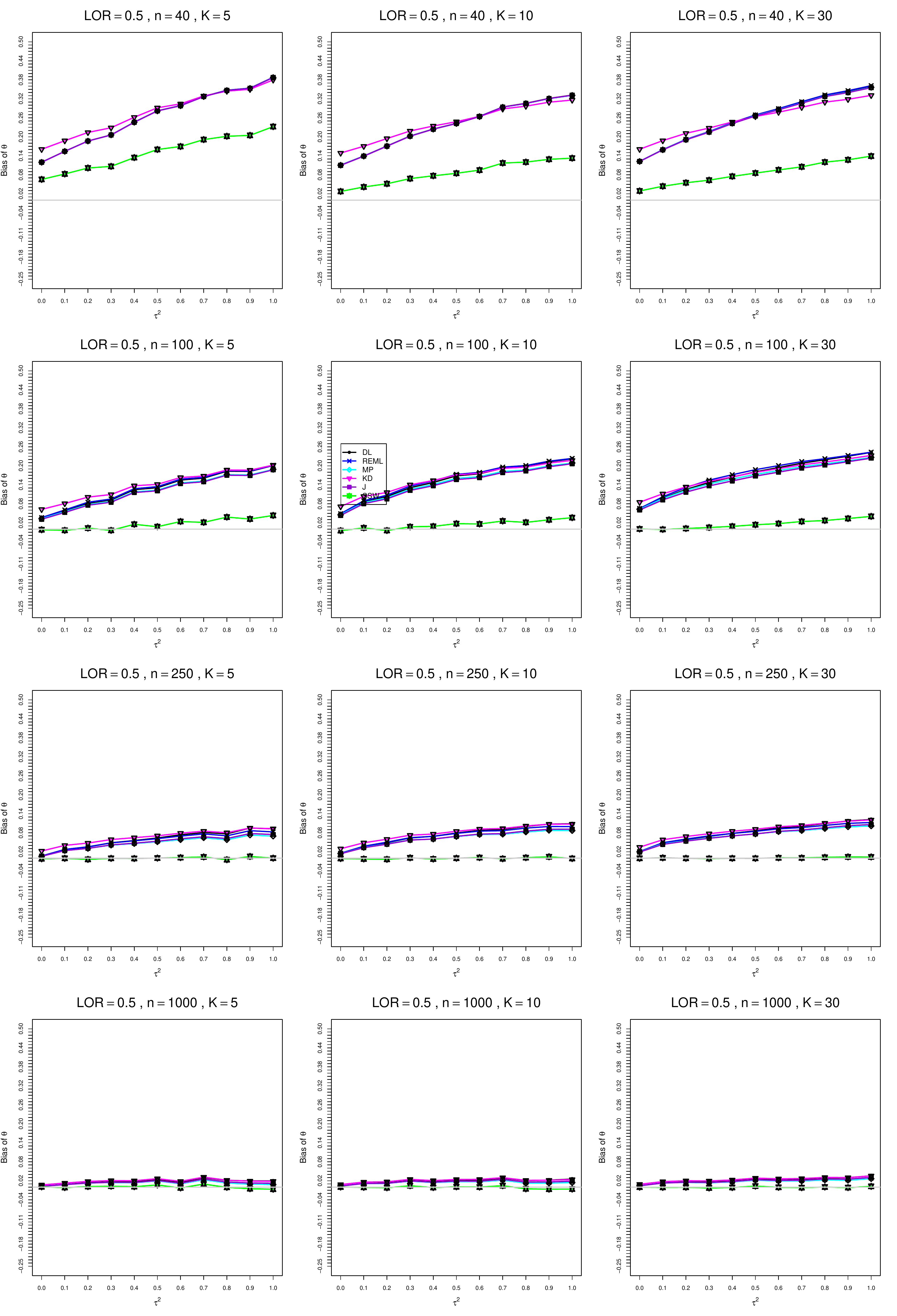}
	\caption{Bias of the estimation of  overall effect measure $\theta$ for $\theta=0.5$, $p_{iC}=0.1$, $q=0.75$, equal sample sizes $n=40,\;100,\;250,\;1000$. 
		\label{BiasThetaLOR05q075piC01}}
\end{figure}

\begin{figure}[t]
	\centering
	\includegraphics[scale=0.33]{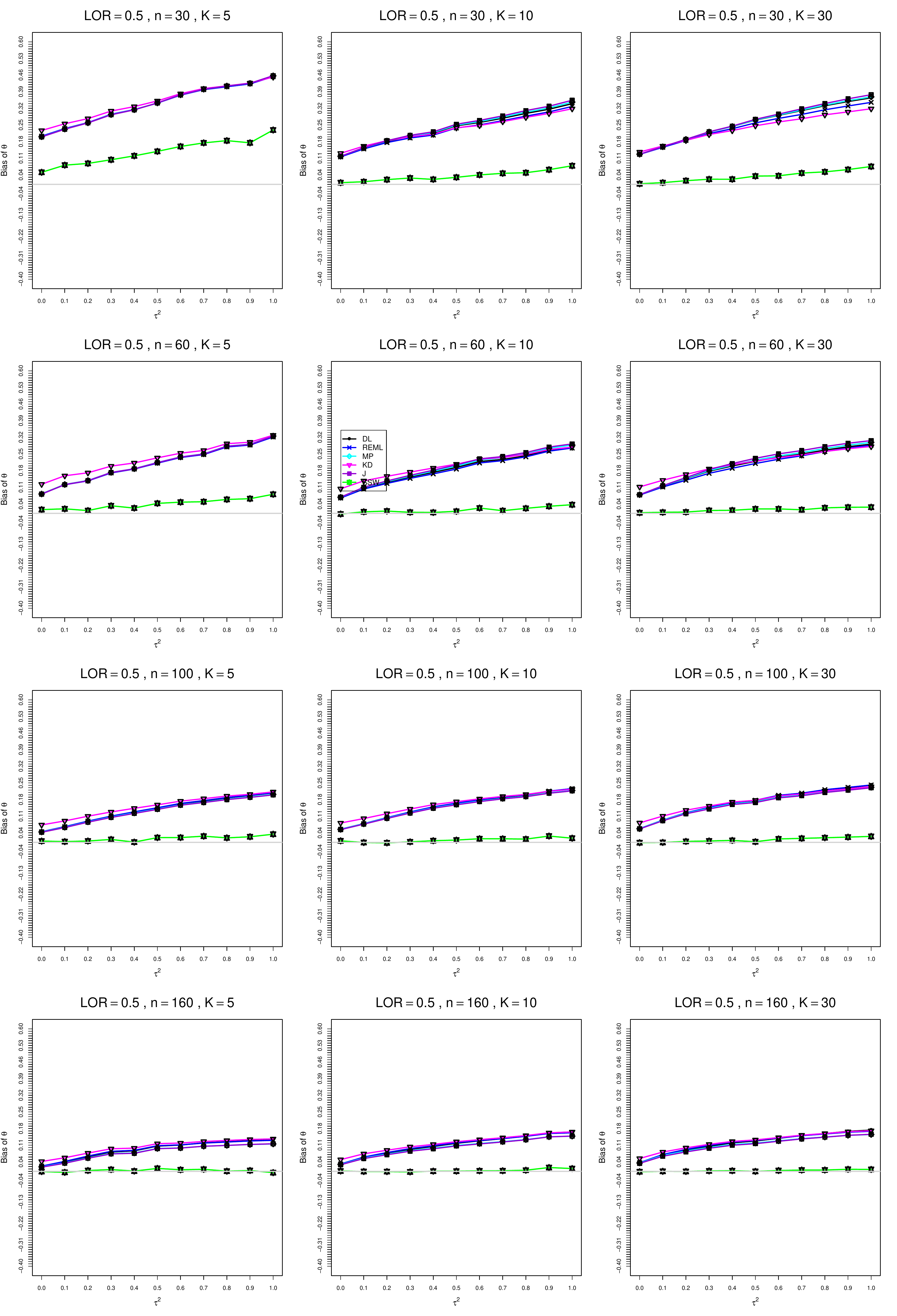}
	\caption{Bias of the estimation of  overall effect measure $\theta$ for $\theta=0.5$, $p_{iC}=0.1$, $q=0.75$, 
		unequal sample sizes $n=30,\; 60,\;100,\;160$. 
		\label{BiasThetaLOR05q075piC01_unequal_sample_sizes}}
\end{figure}

\begin{figure}[t]\centering
	\includegraphics[scale=0.35]{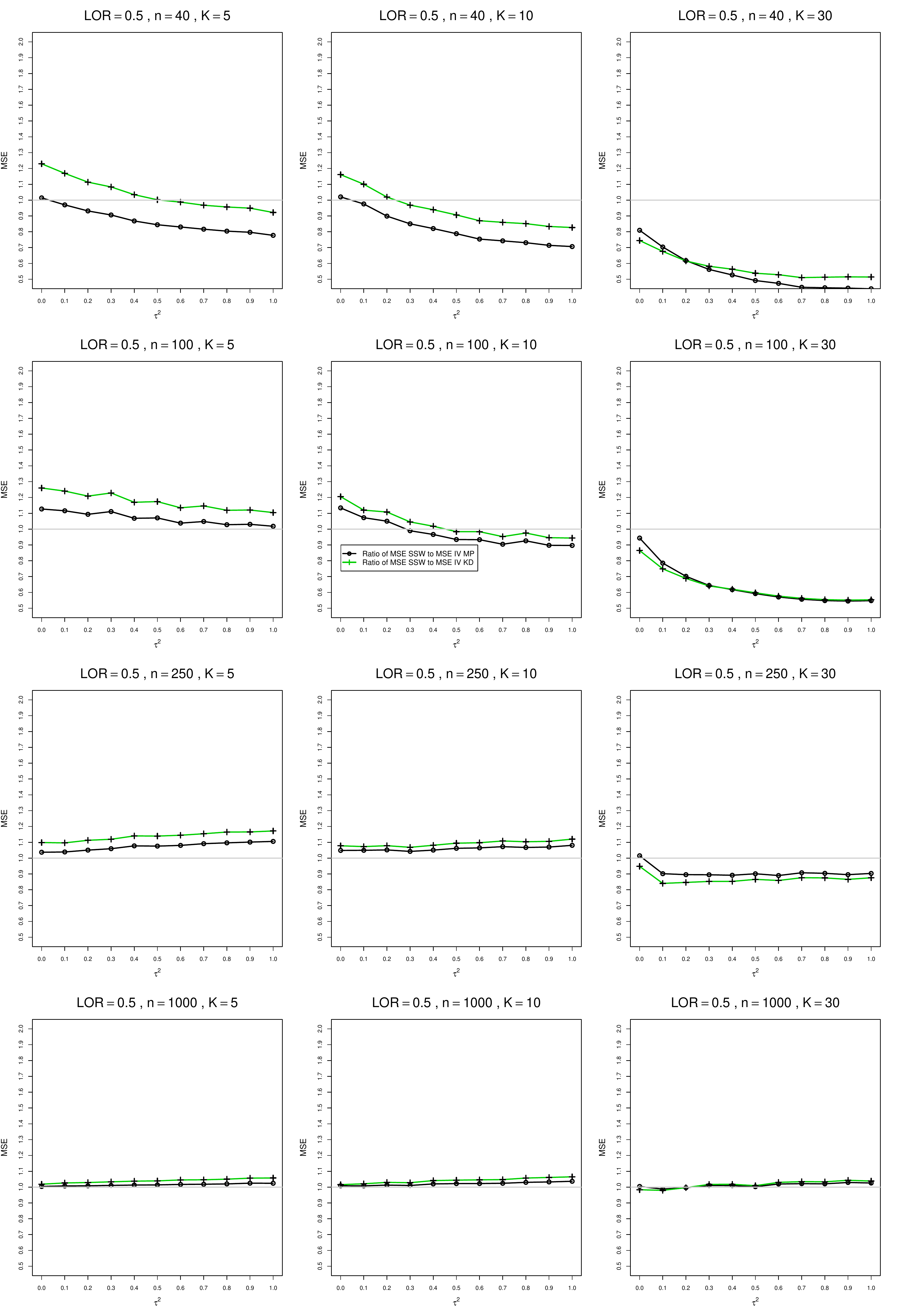}
	\caption{Ratio of mean squared errors of the fixed-weights to mean squared errors of inverse-variance estimator for $\theta=0.5$,$p_{iC}=0.1$, $q=0.75$, equal sample sizes $n=40,\;100,\;250,\;1000$. 
		\label{RatioOfMSEwithLOR05q075piC01fromMPandCMP}}
\end{figure}

\begin{figure}[t]\centering
	\includegraphics[scale=0.35]{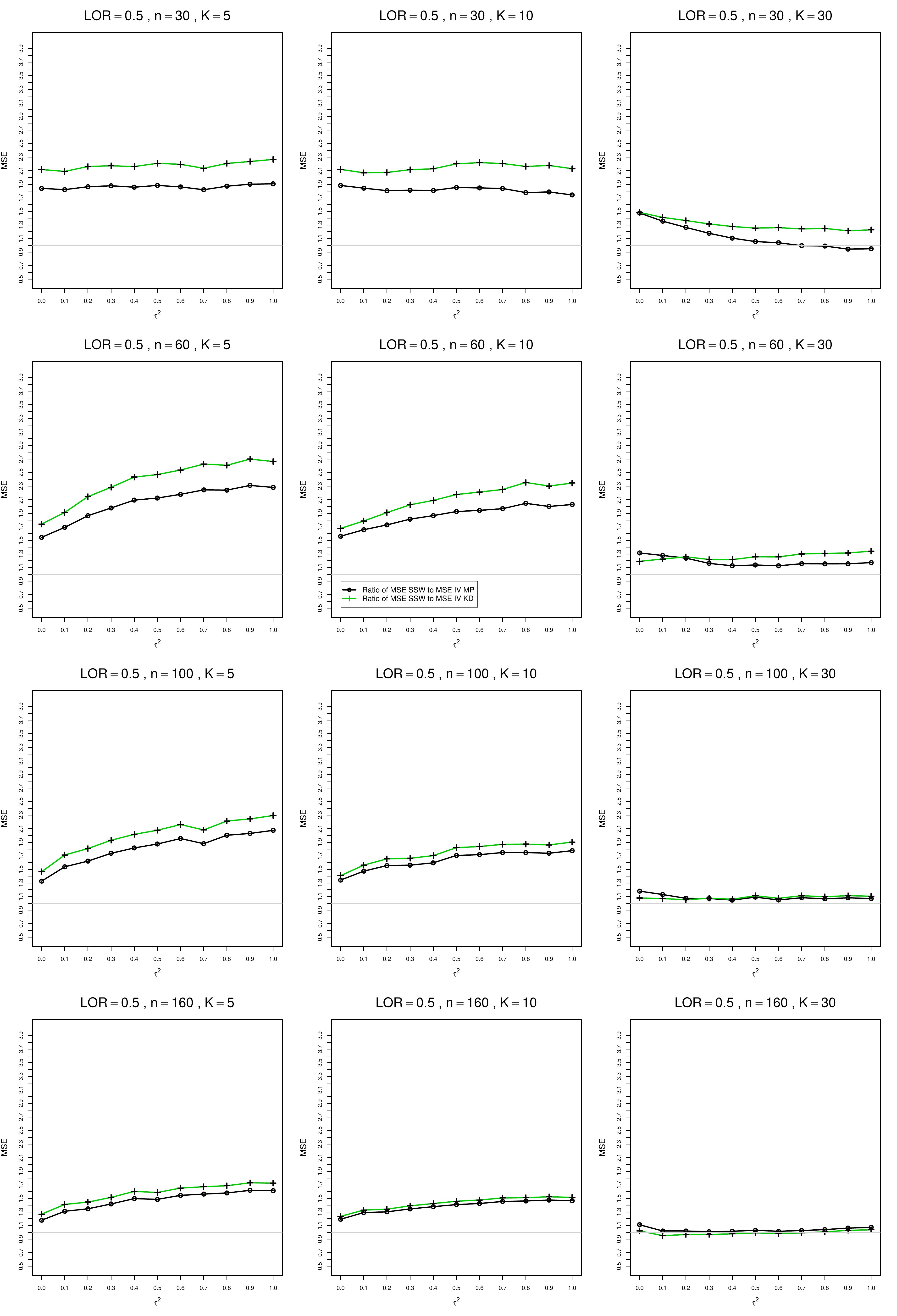}
	\caption{Ratio of mean squared errors of the fixed-weights to mean squared errors of inverse-variance estimator for $\theta=0.5$,$p_{iC}=0.1$, $q=0.75$, unequal sample sizes $n=30,\;60,\;100,\;160$. 
		\label{RatioOfMSEwithLOR05q075piC01fromMPandCMP_unequal_sample_sizes}}
\end{figure}


\begin{figure}[t]
	\centering
	\includegraphics[scale=0.33]{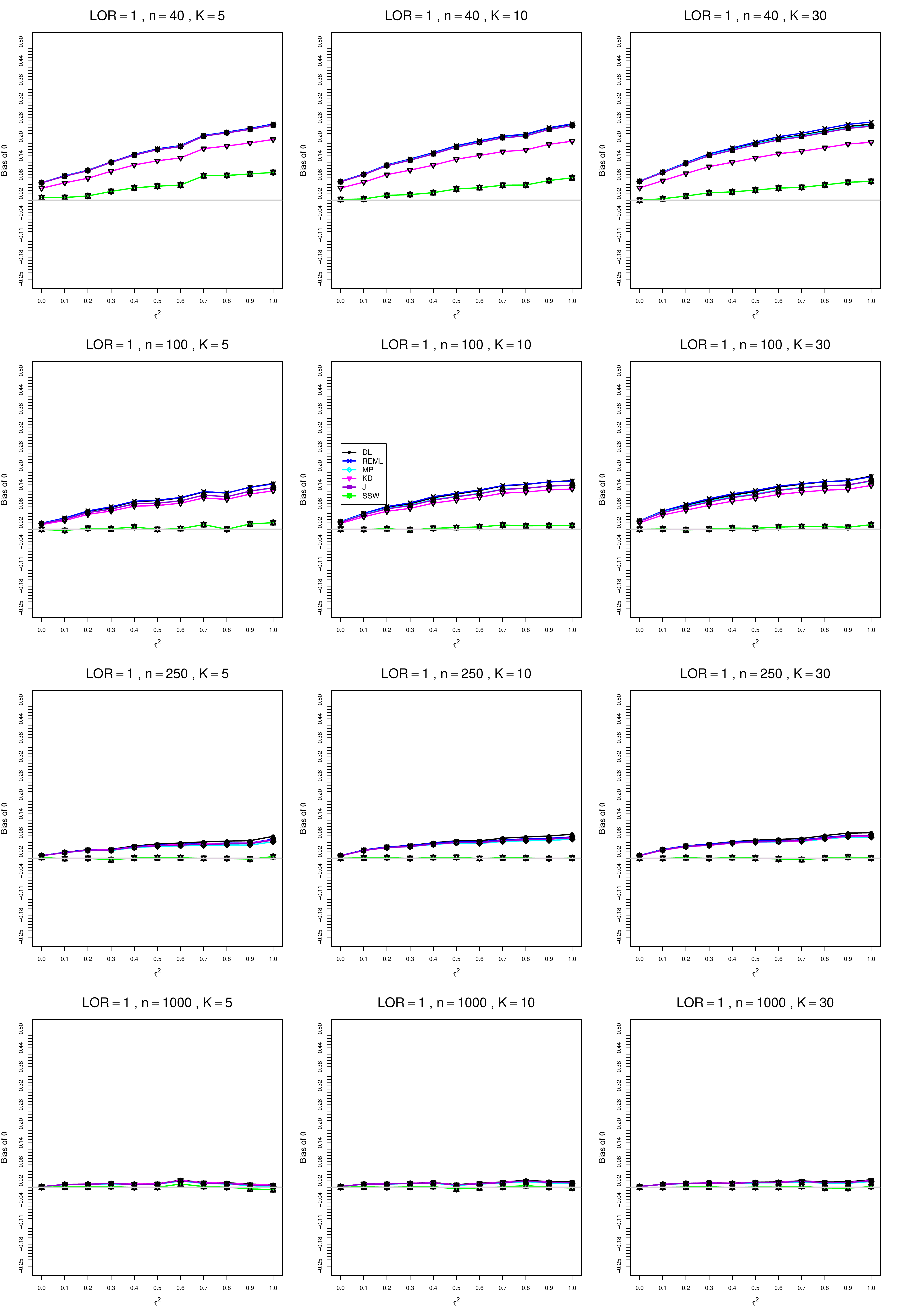}
	\caption{Bias of the estimation of  overall effect measure $\theta$ for $\theta=1$, $p_{iC}=0.1$, $q=0.75$, equal sample sizes $n=40,\;100,\;250,\;1000$. 
		\label{BiasThetaLOR1q075piC01}}
\end{figure}

\begin{figure}[t]
	\centering
	\includegraphics[scale=0.33]{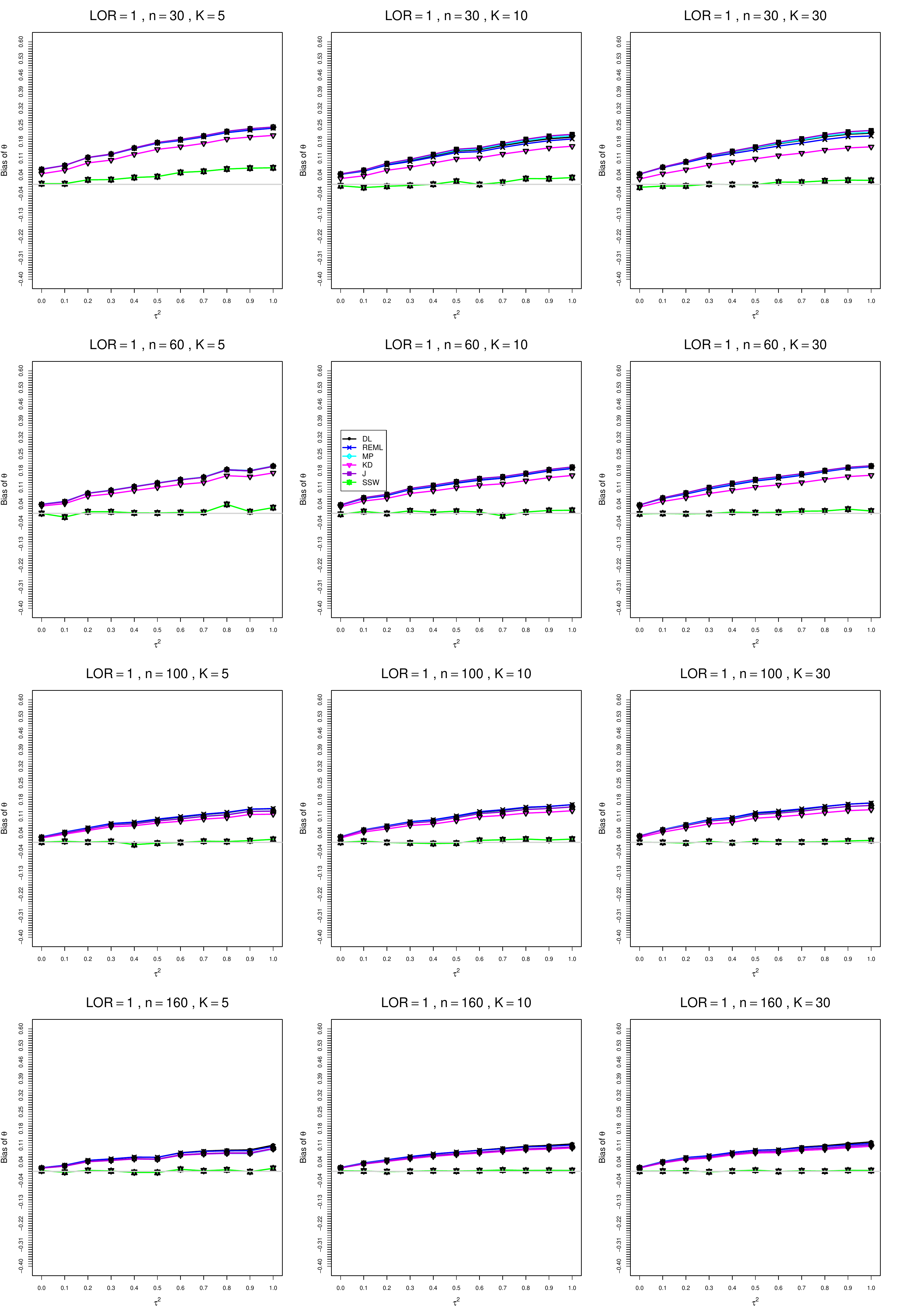}
	\caption{Bias of the estimation of  overall effect measure $\theta$ for $\theta=1$, $p_{iC}=0.1$, $q=0.75$, 
		unequal sample sizes $n=30,\; 60,\;100,\;160$. 
		\label{BiasThetaLOR1q075piC01_unequal_sample_sizes}}
\end{figure}

\begin{figure}[t]\centering
	\includegraphics[scale=0.35]{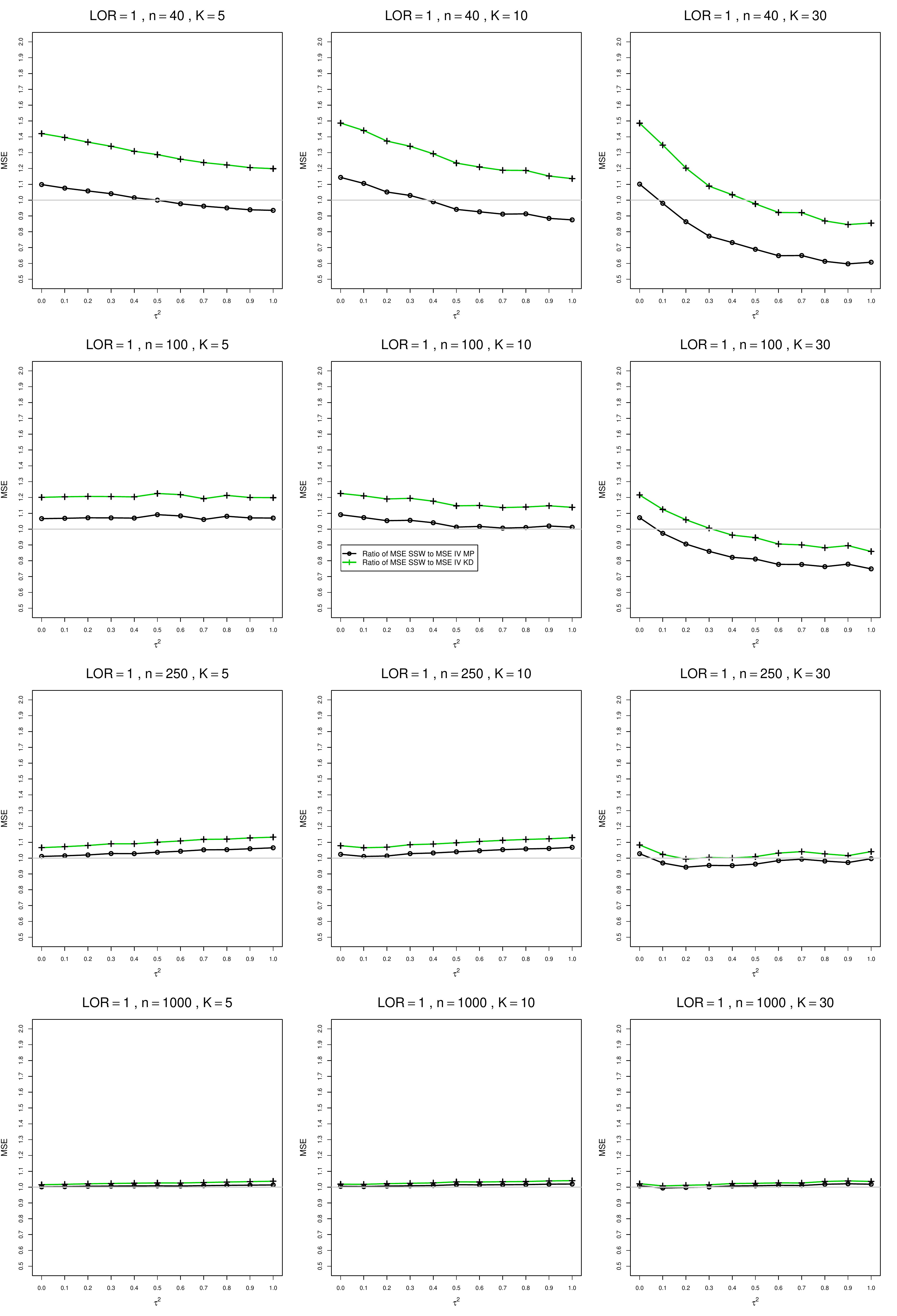}
	\caption{Ratio of mean squared errors of the fixed-weights to mean squared errors of inverse-variance estimator for $\theta=1$,$p_{iC}=0.1$, $q=0.75$, equal sample sizes $n=40,\;100,\;250,\;1000$. 
		\label{RatioOfMSEwithLOR1q075piC01fromMPandCMP}}
\end{figure}

\begin{figure}[t]\centering
	\includegraphics[scale=0.35]{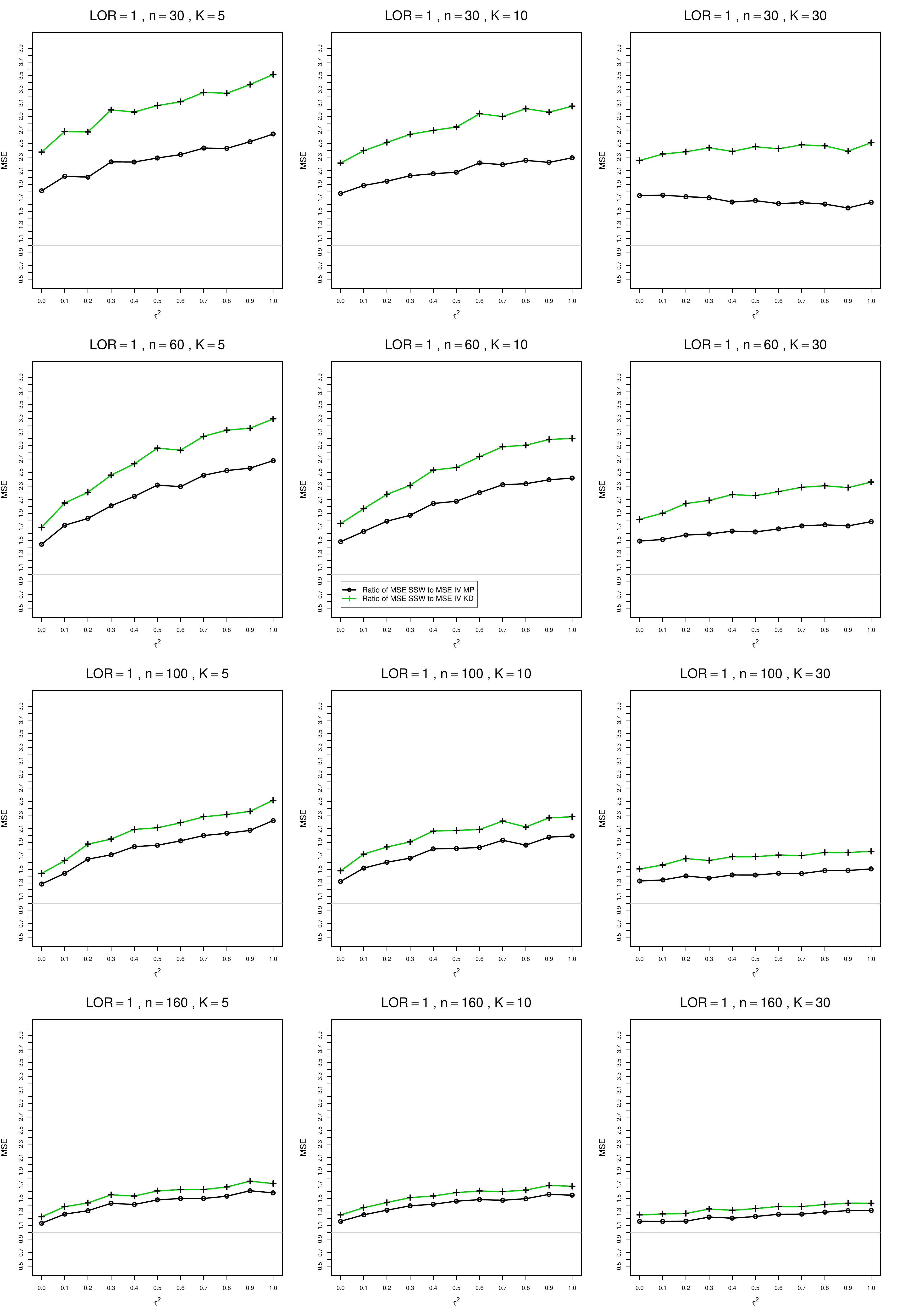}
	\caption{Ratio of mean squared errors of the fixed-weights to mean squared errors of inverse-variance estimator for $\theta=1$,$p_{iC}=0.1$, $q=0.75$, unequal sample sizes $n=30,\;60,\;100,\;160$. 
		\label{RatioOfMSEwithLOR1q075piC01fromMPandCMP_unequal_sample_sizes}}
\end{figure}


\begin{figure}[t]
	\centering
	\includegraphics[scale=0.33]{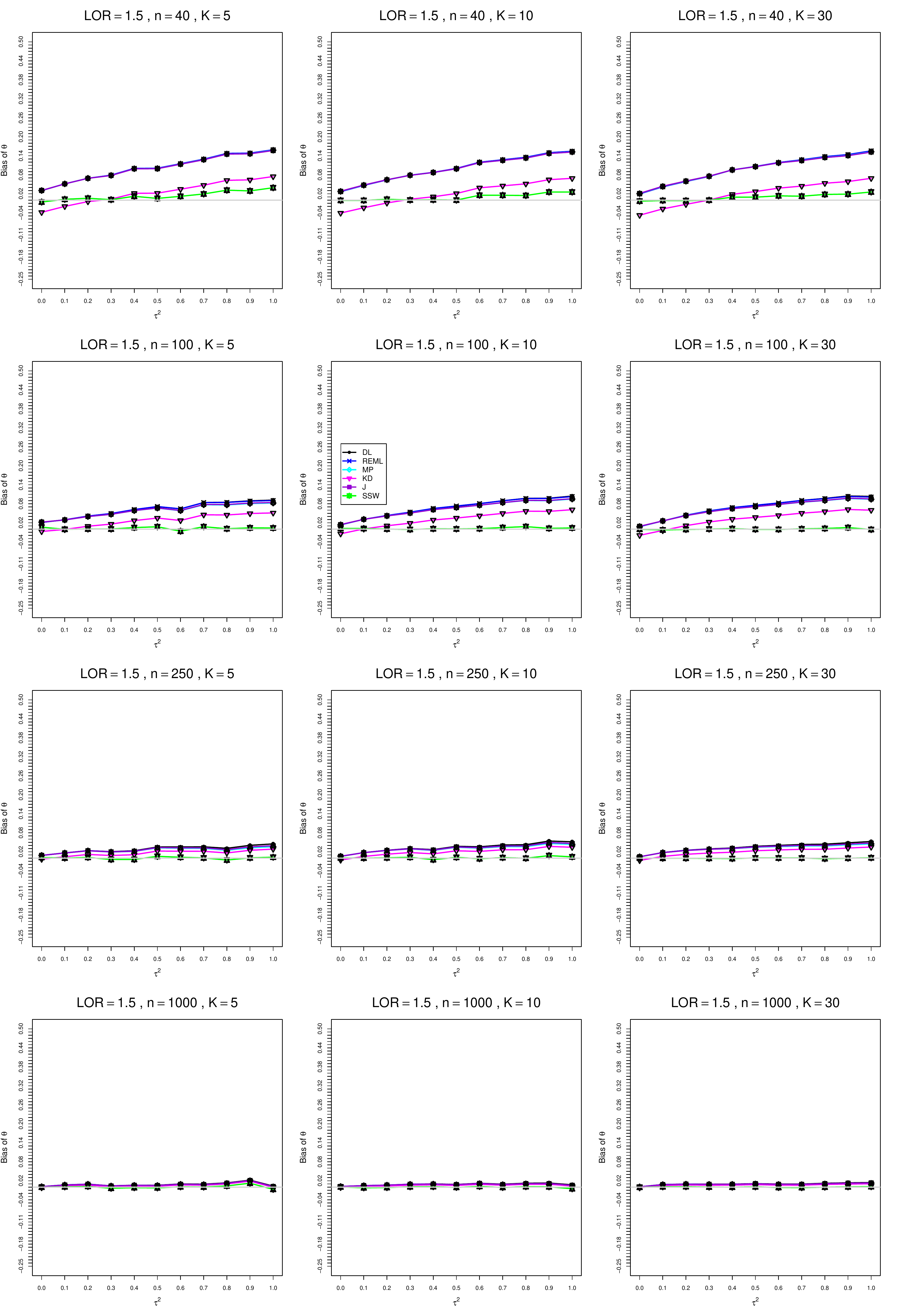}
	\caption{Bias of the estimation of  overall effect measure $\theta$ for $\theta=1.5$, $p_{iC}=0.1$, $q=0.75$, equal sample sizes $n=40,\;100,\;250,\;1000$. 
		\label{BiasThetaLOR15q075piC01}}
\end{figure}

\begin{figure}[t]
	\centering
	\includegraphics[scale=0.33]{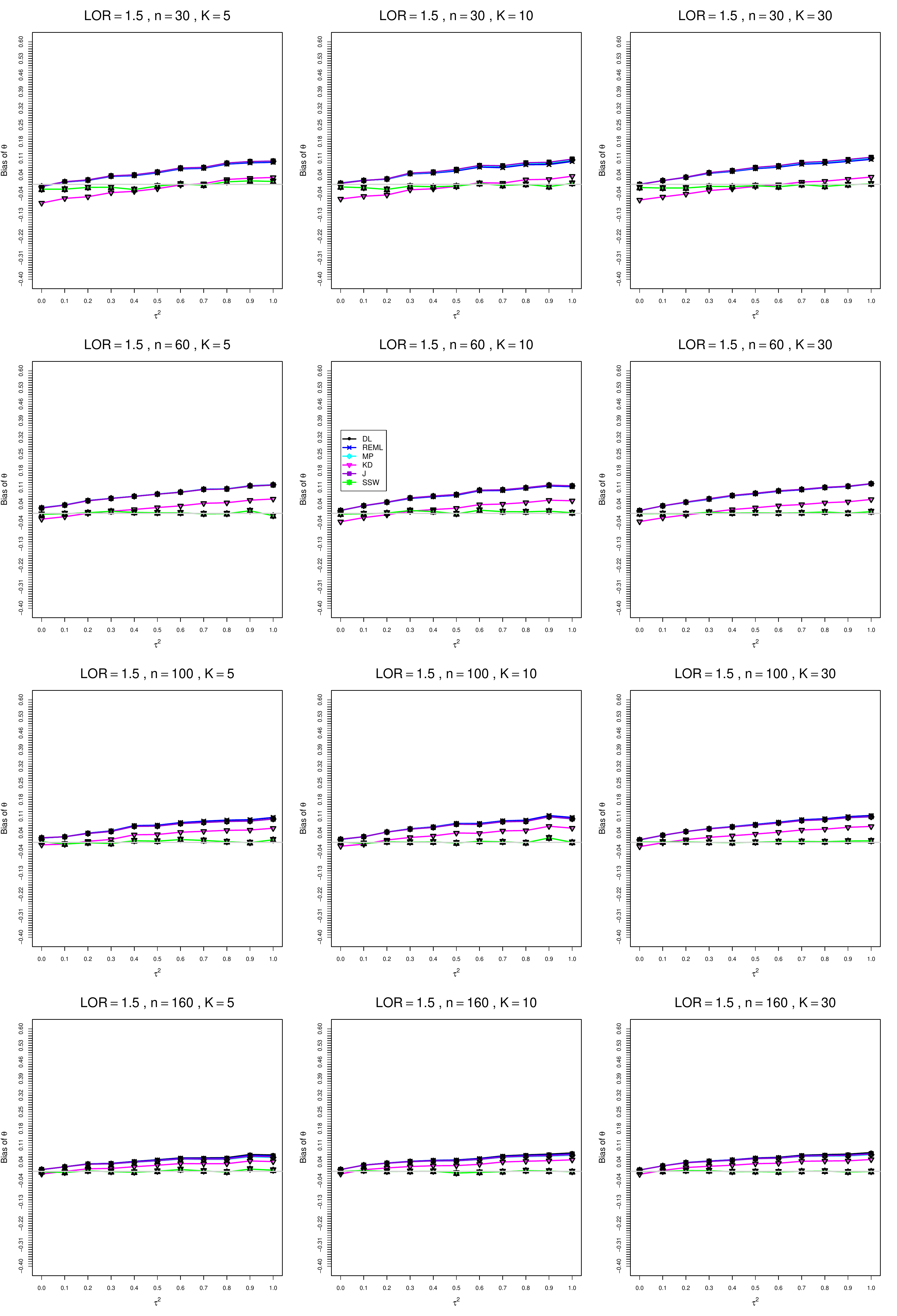}
	\caption{Bias of the estimation of  overall effect measure $\theta$ for $\theta=1.5$, $p_{iC}=0.1$, $q=0.75$, 
		unequal sample sizes $n=30,\; 60,\;100,\;160$. 
		\label{BiasThetaLOR15q075piC01_unequal_sample_sizes}}
\end{figure}

\begin{figure}[t]\centering
	\includegraphics[scale=0.35]{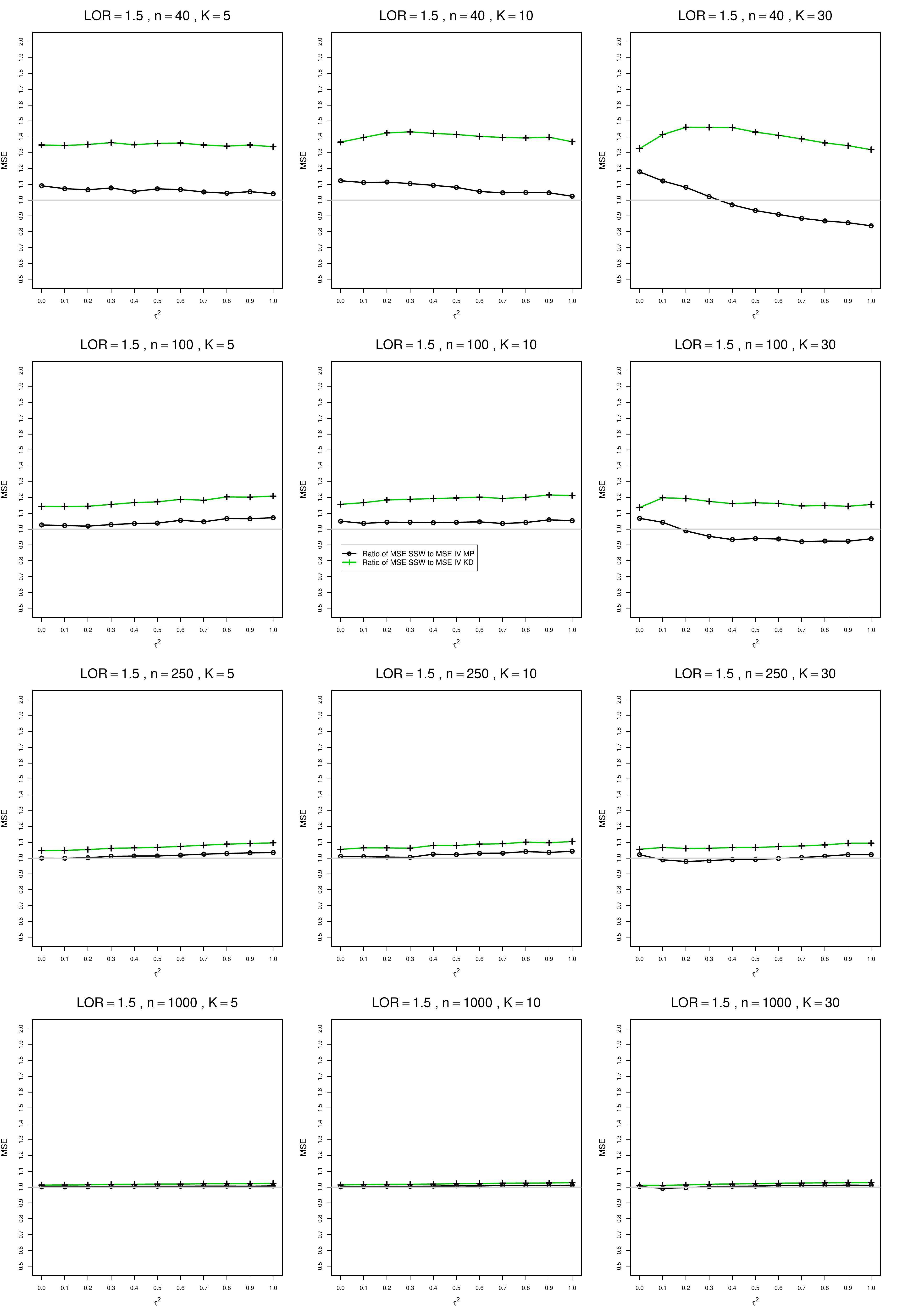}
	\caption{Ratio of mean squared errors of the fixed-weights to mean squared errors of inverse-variance estimator for $\theta=1.5$,$p_{iC}=0.1$, $q=0.75$, equal sample sizes $n=40,\;100,\;250,\;1000$. 
		\label{RatioOfMSEwithLOR15q075piC01fromMPandCMP}}
\end{figure}

\begin{figure}[t]\centering
	\includegraphics[scale=0.35]{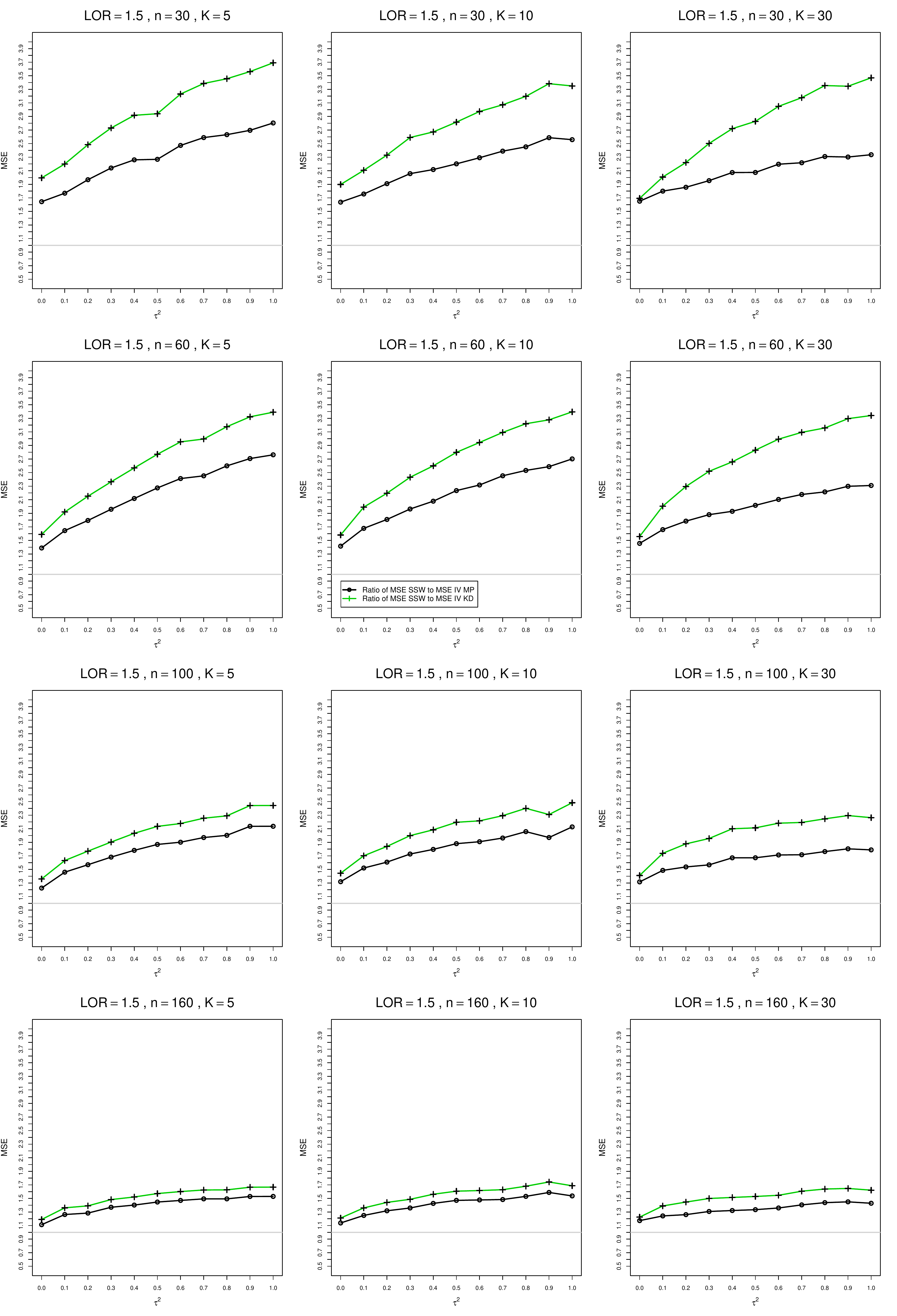}
	\caption{Ratio of mean squared errors of the fixed-weights to mean squared errors of inverse-variance estimator for $\theta=1.5$,$p_{iC}=0.1$, $q=0.75$, unequal sample sizes $n=30,\;60,\;100,\;160$. 
		\label{RatioOfMSEwithLOR15q075piC01fromMPandCMP_unequal_sample_sizes}}
\end{figure}


\clearpage
\begin{figure}[t]
	\centering
	\includegraphics[scale=0.33]{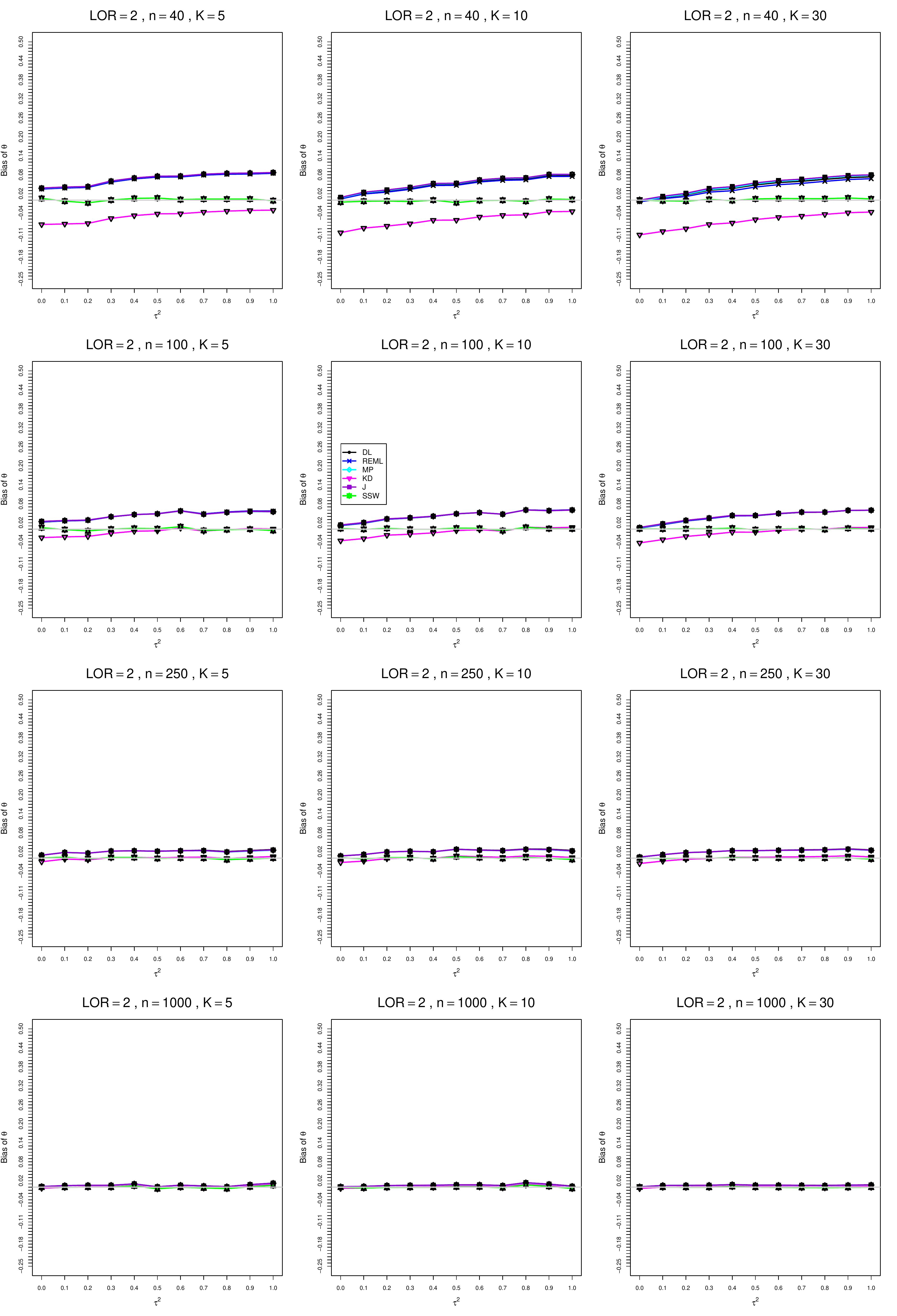}
	\caption{Bias of the estimation of  overall effect measure $\theta$ for $\theta=2$, $p_{iC}=0.1$, $q=0.75$, equal sample sizes $n=40,\;100,\;250,\;1000$. 
		\label{BiasThetaLOR2q075piC01}}
\end{figure}

\begin{figure}[t]
	\centering
	\includegraphics[scale=0.33]{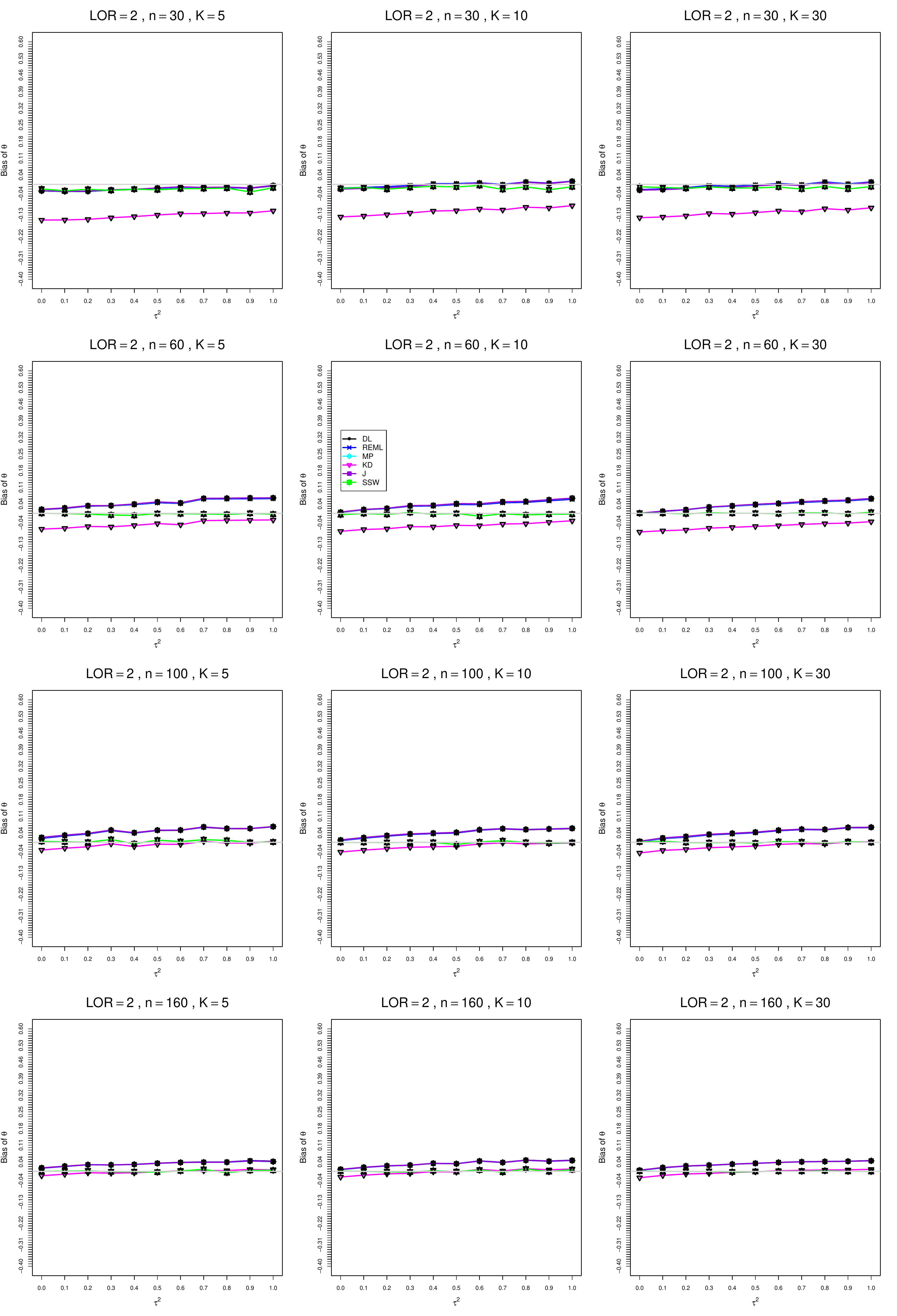}
	\caption{Bias of the estimation of  overall effect measure $\theta$ for $\theta=2$, $p_{iC}=0.1$, $q=0.75$, 
		unequal sample sizes $n=30,\; 60,\;100,\;160$. 
		\label{BiasThetaLOR2q075piC01_unequal_sample_sizes}}
\end{figure}

\begin{figure}[t]\centering
	\includegraphics[scale=0.35]{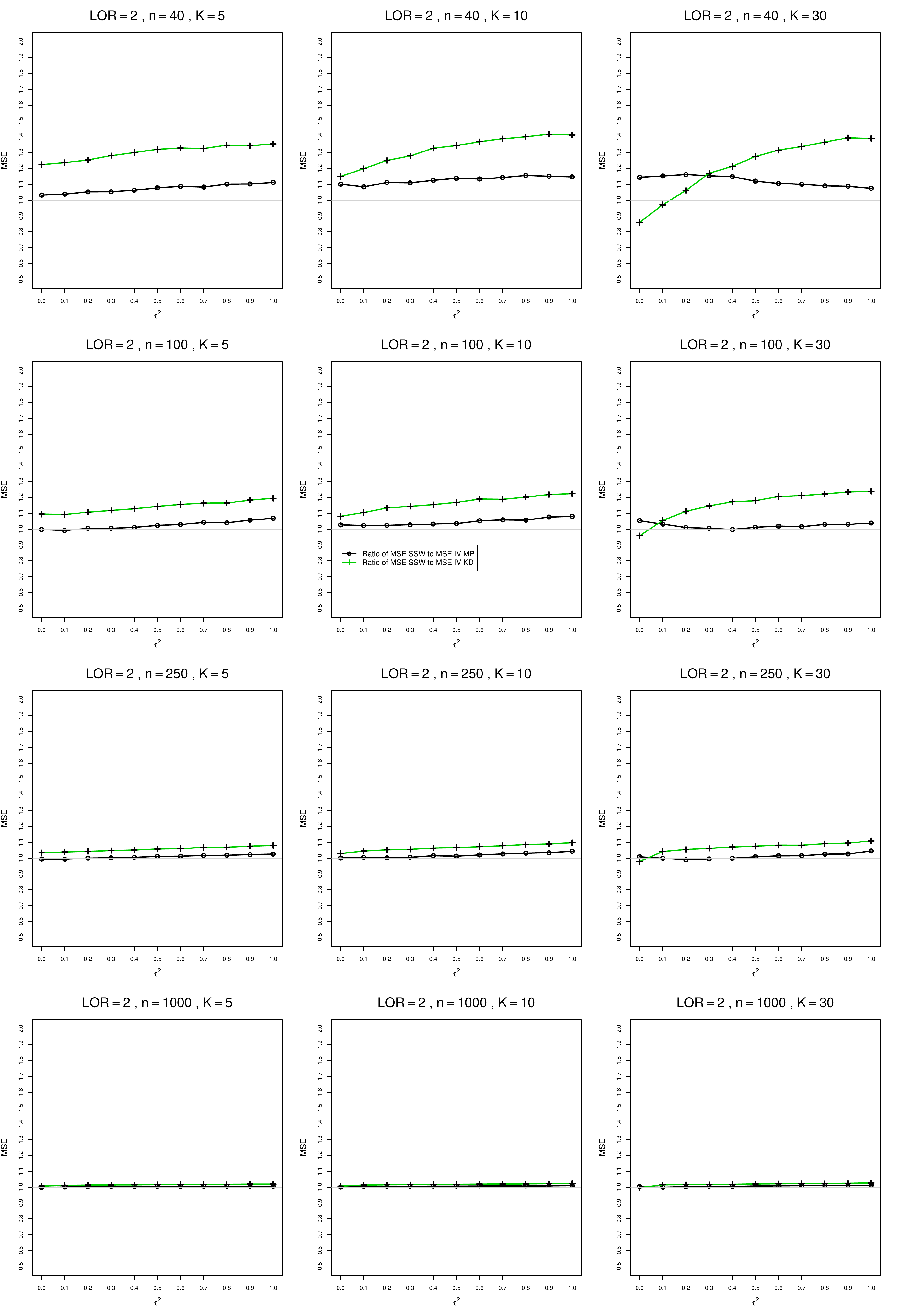}
	\caption{Ratio of mean squared errors of the fixed-weights to mean squared errors of inverse-variance estimator for $\theta=2$,$p_{iC}=0.1$, $q=0.75$, equal sample sizes $n=40,\;100,\;250,\;1000$. 
		\label{RatioOfMSEwithLOR2q075piC01fromMPandCMP}}
\end{figure}

\begin{figure}[t]\centering
	\includegraphics[scale=0.35]{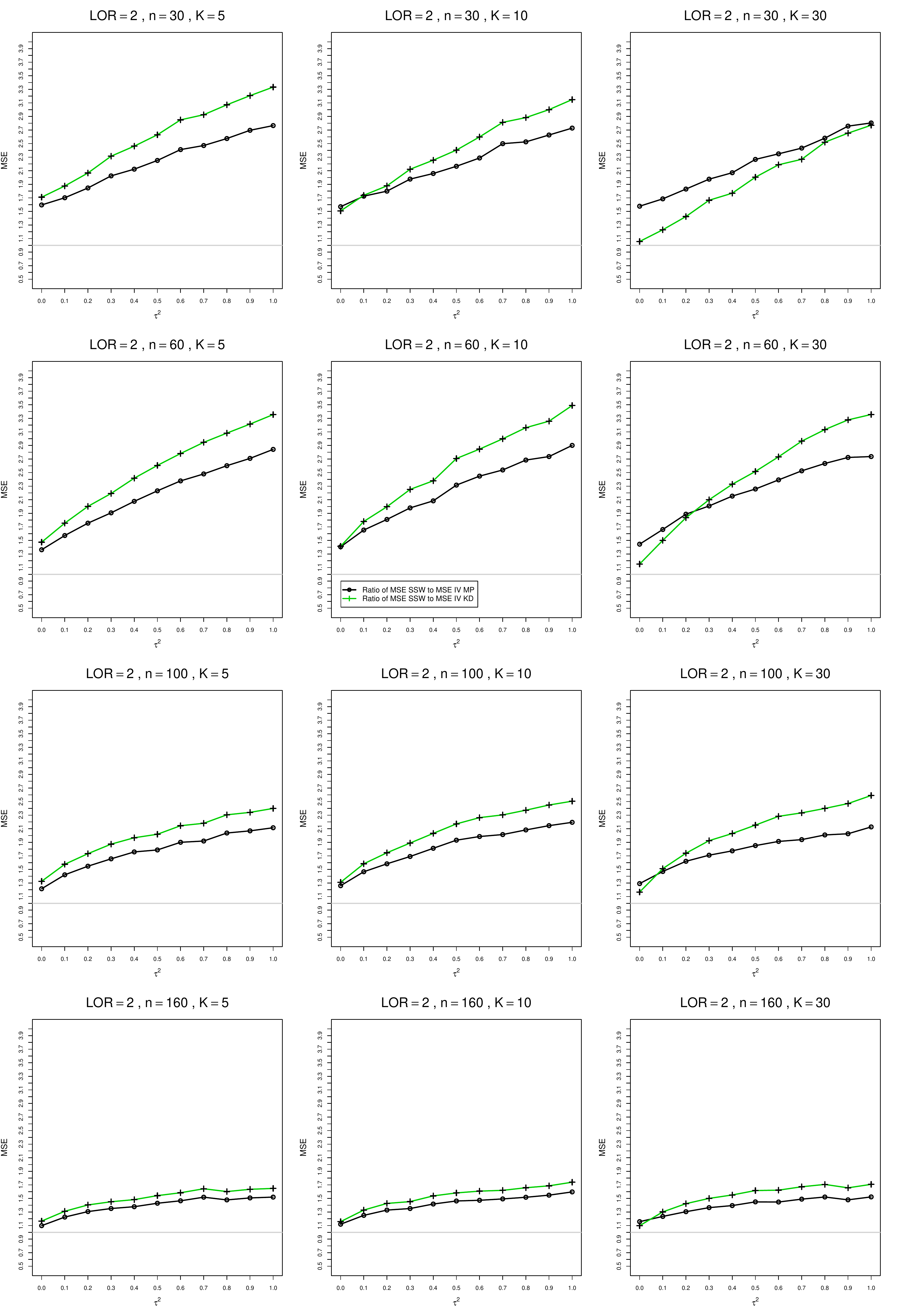}
	\caption{Ratio of mean squared errors of the fixed-weights to mean squared errors of inverse-variance estimator for $\theta=2$,$p_{iC}=0.1$, $q=0.75$, unequal sample sizes $n=30,\;60,\;100,\;160$. 
		\label{RatioOfMSEwithLOR2q075piC01fromMPandCMP_unequal_sample_sizes}}
\end{figure}


\clearpage
\renewcommand{\thefigure}{B1.2.\arabic{figure}}
\setcounter{figure}{0}
\subsection*{B1.2 Probability in the control arm $p_{C}=0.2$}

\begin{figure}[t]
	\centering
	\includegraphics[scale=0.33]{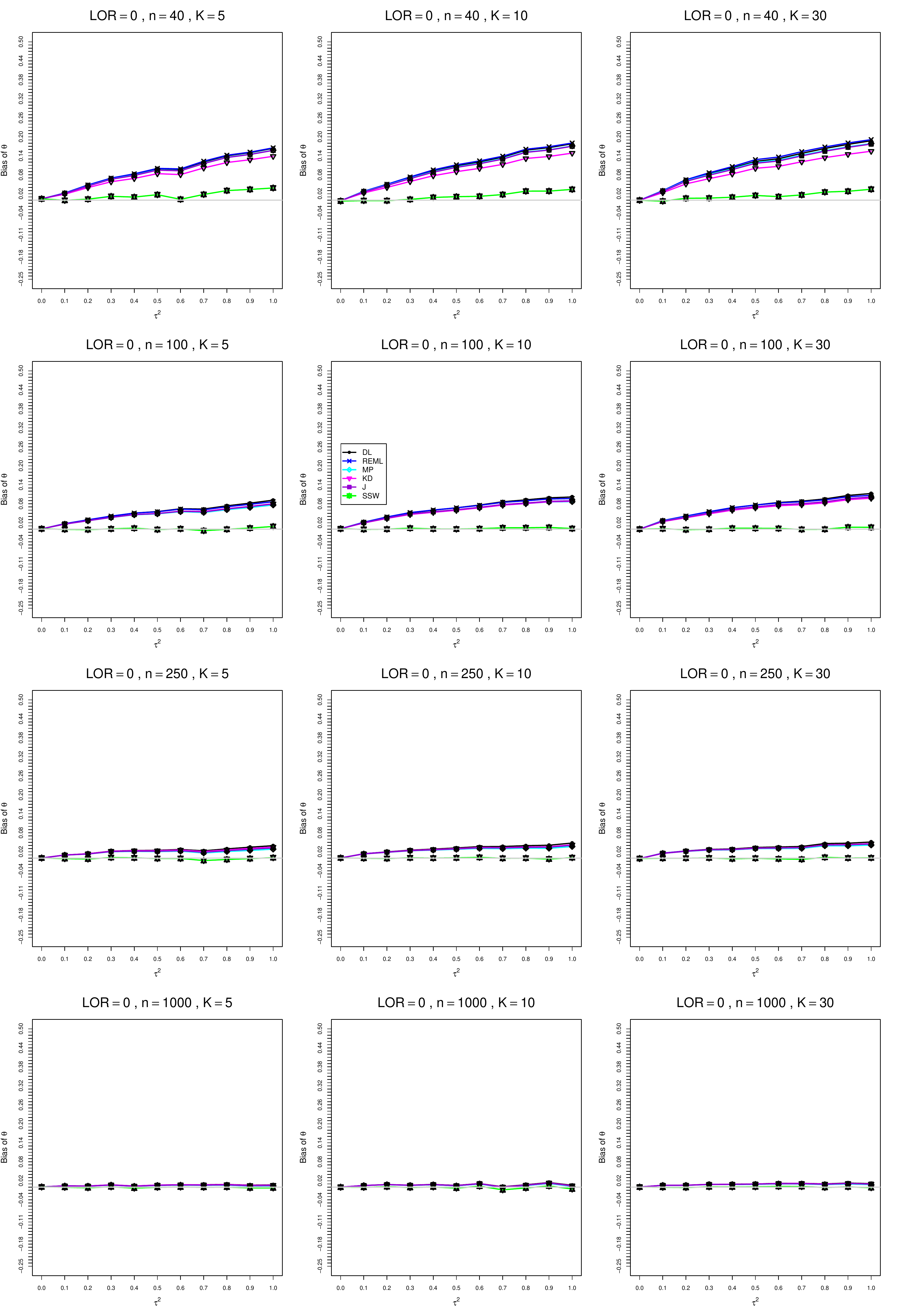}
	\caption{Bias of the estimation of  overall effect measure $\theta$ for $\theta=0$, $p_{iC}=0.2$, $q=0.5$, equal sample sizes  $n=40,\;100,\;250,\;1000$. 
		\label{BiasThetaLOR0q05piC02}}
\end{figure}

\begin{figure}[t]
	\centering
	\includegraphics[scale=0.33]{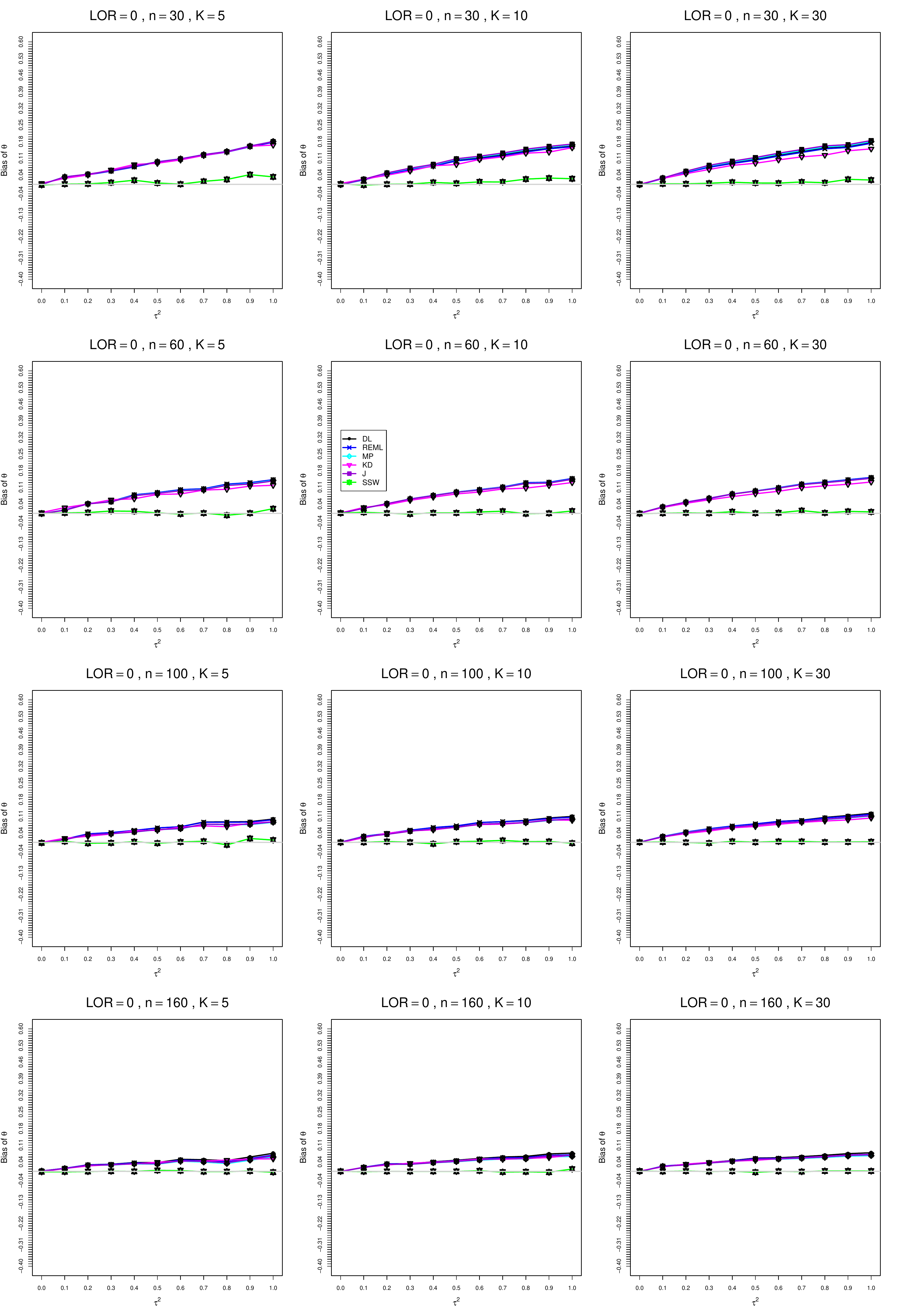}
	\caption{Bias of the estimation of  overall effect measure $\theta$ for $\theta=0$, $p_{iC}=0.2$, $q=0.5$, 
		unequal sample sizes $n=30,\; 60,\;100,\;160$. 
		\label{BiasThetaLOR0q05piC02_unequal_sample_sizes}}
\end{figure}

\begin{figure}[t]\centering
	\includegraphics[scale=0.35]{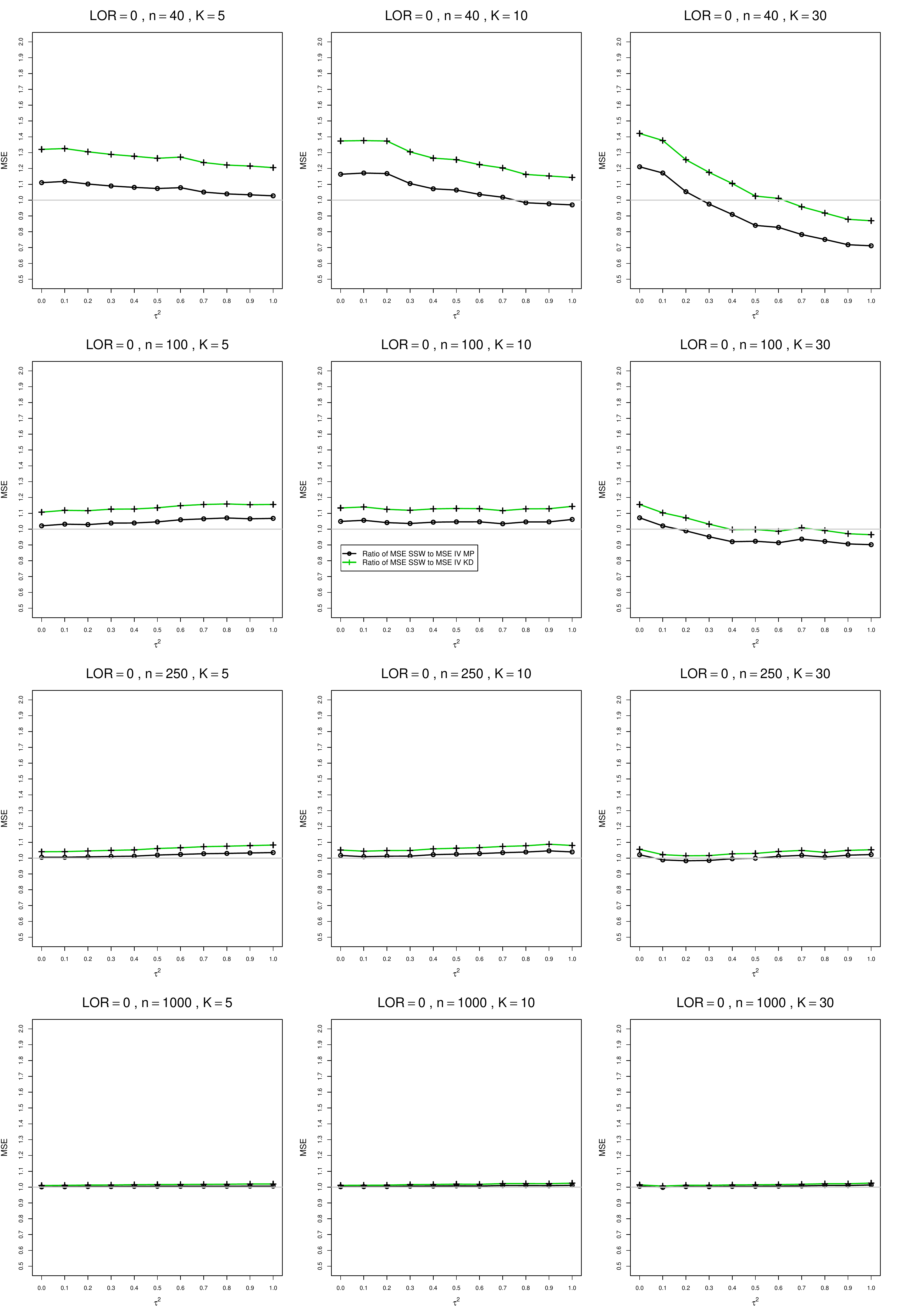}
	\caption{Ratio of mean squared errors of the fixed-weights to mean squared errors of inverse-variance estimator for $\theta=0$, $p_{iC}=0.2$, $q=0.5$, equal sample sizes $n=40,\;100,\;250,\;1000$. 
		\label{RatioOfMSEwithLOR0q05piC02fromMPandCMP}}
\end{figure}

\begin{figure}[t]\centering
	\includegraphics[scale=0.35]{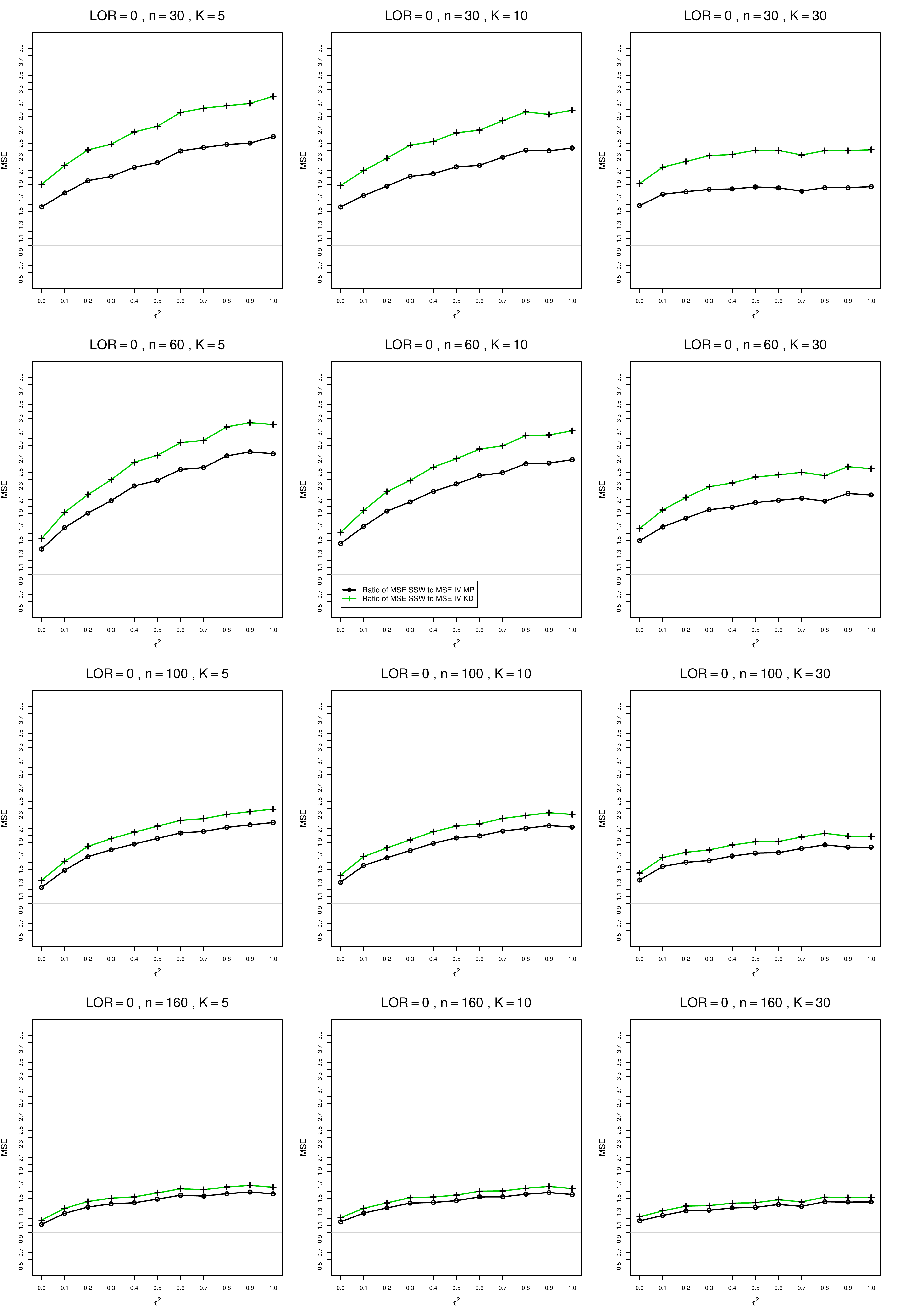}
	\caption{Ratio of mean squared errors of the fixed-weights to mean squared errors of inverse-variance estimator for $\theta=0$, $p_{iC}=0.2$, $q=0.5$, unequal sample sizes $n=30,\;60,\;100,\;160$. 
		\label{RatioOfMSEwithLOR0q05piC02fromMPandCMP_unequal_sample_sizes}}
\end{figure}


\begin{figure}[t]
	\centering
	\includegraphics[scale=0.33]{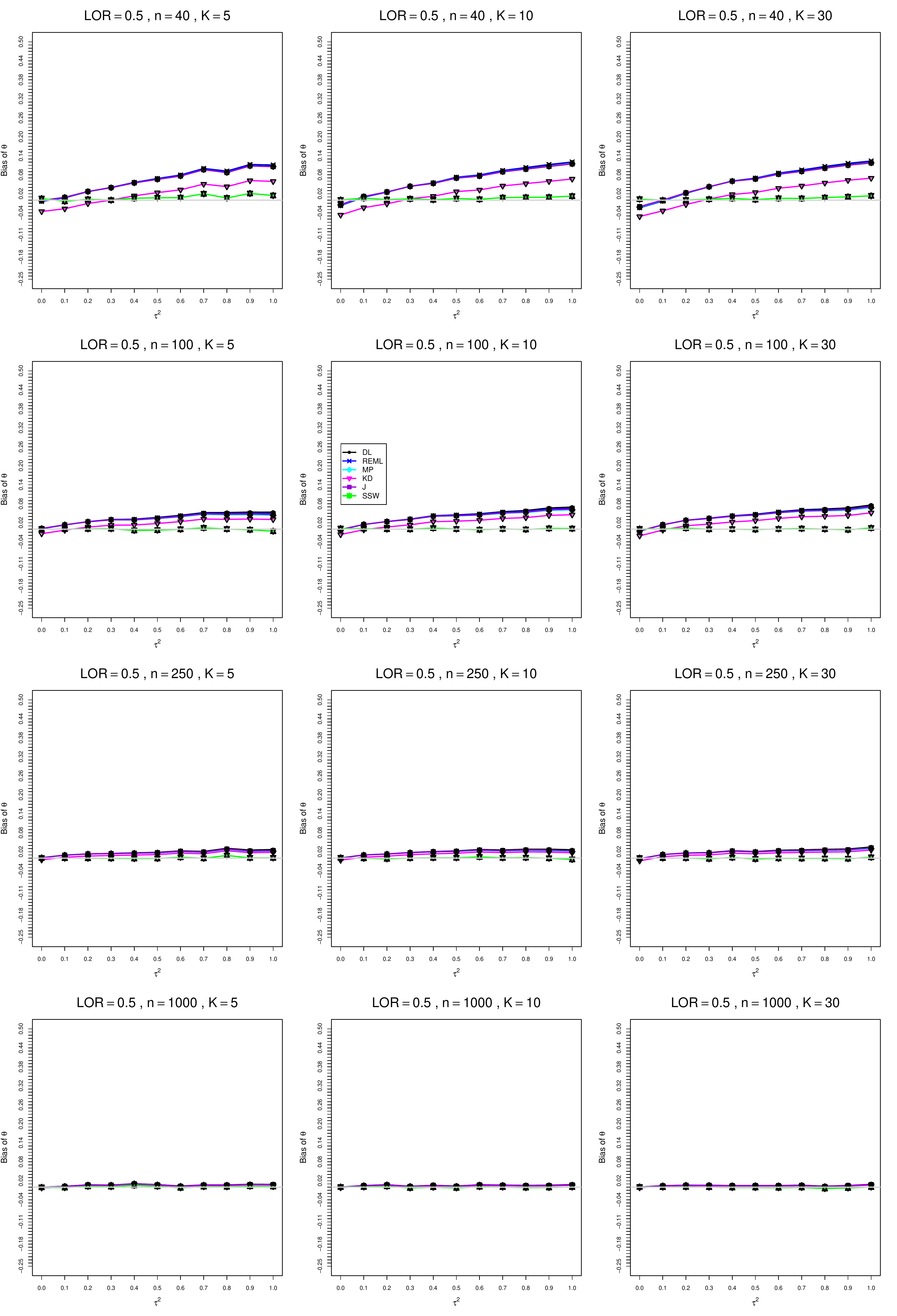}
	\caption{Bias of the estimation of  overall effect measure $\theta$ for $\theta=0.5$, $p_{iC}=0.2$, $q=0.5$, equal sample sizes $n=40,\;100,\;250,\;1000$. 
		\label{BiasThetaLOR05q05piC02}}
\end{figure}

\begin{figure}[t]
	\centering
	\includegraphics[scale=0.33]{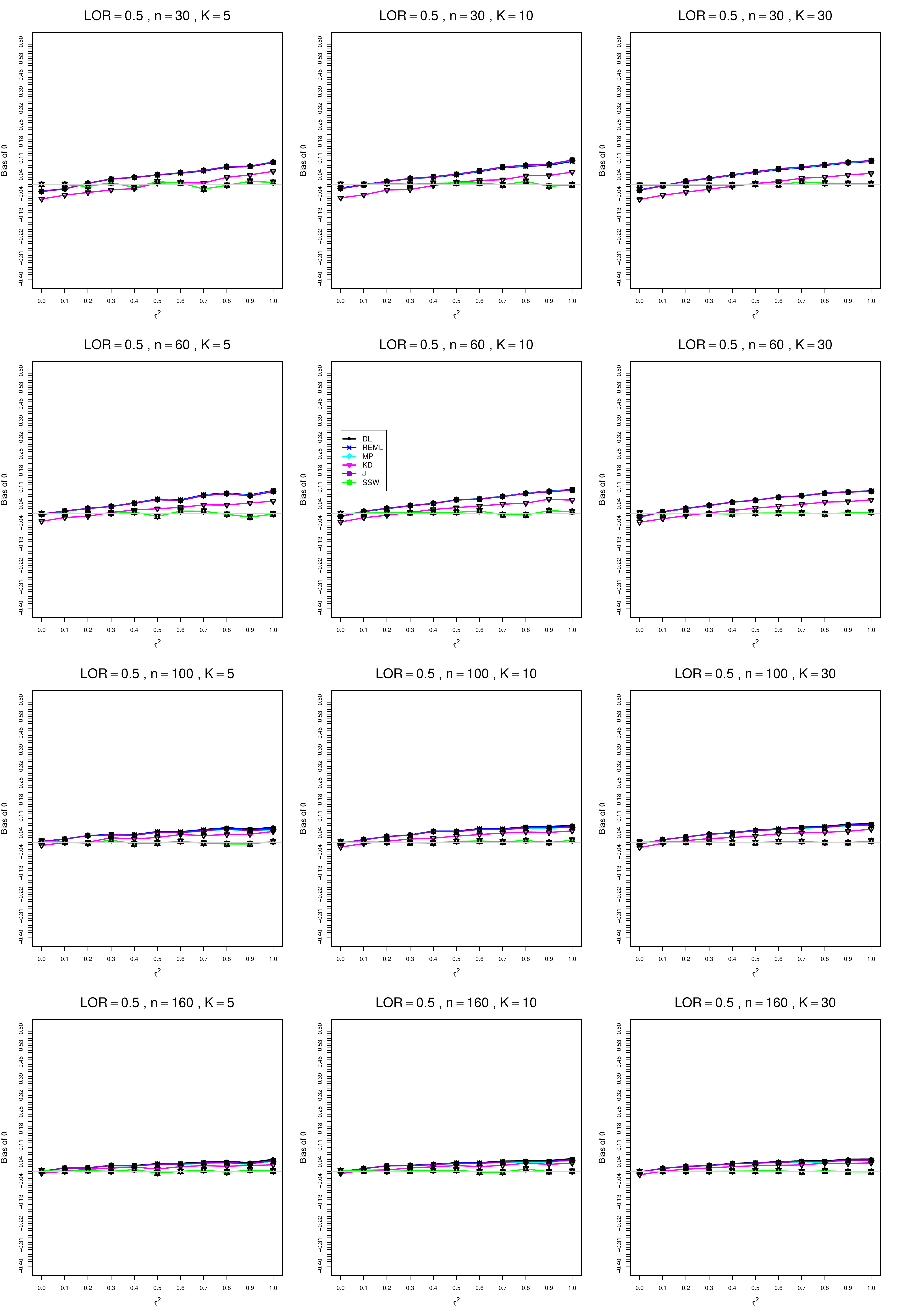}
	\caption{Bias of the estimation of  overall effect measure $\theta$ for $\theta=0.5$, $p_{iC}=0.2$, $q=0.5$, 
		unequal sample sizes $n=30,\; 60,\;100,\;160$. 
		\label{BiasThetaLOR05q05piC02_unequal_sample_sizes}}
\end{figure}

\begin{figure}[t]\centering
	\includegraphics[scale=0.35]{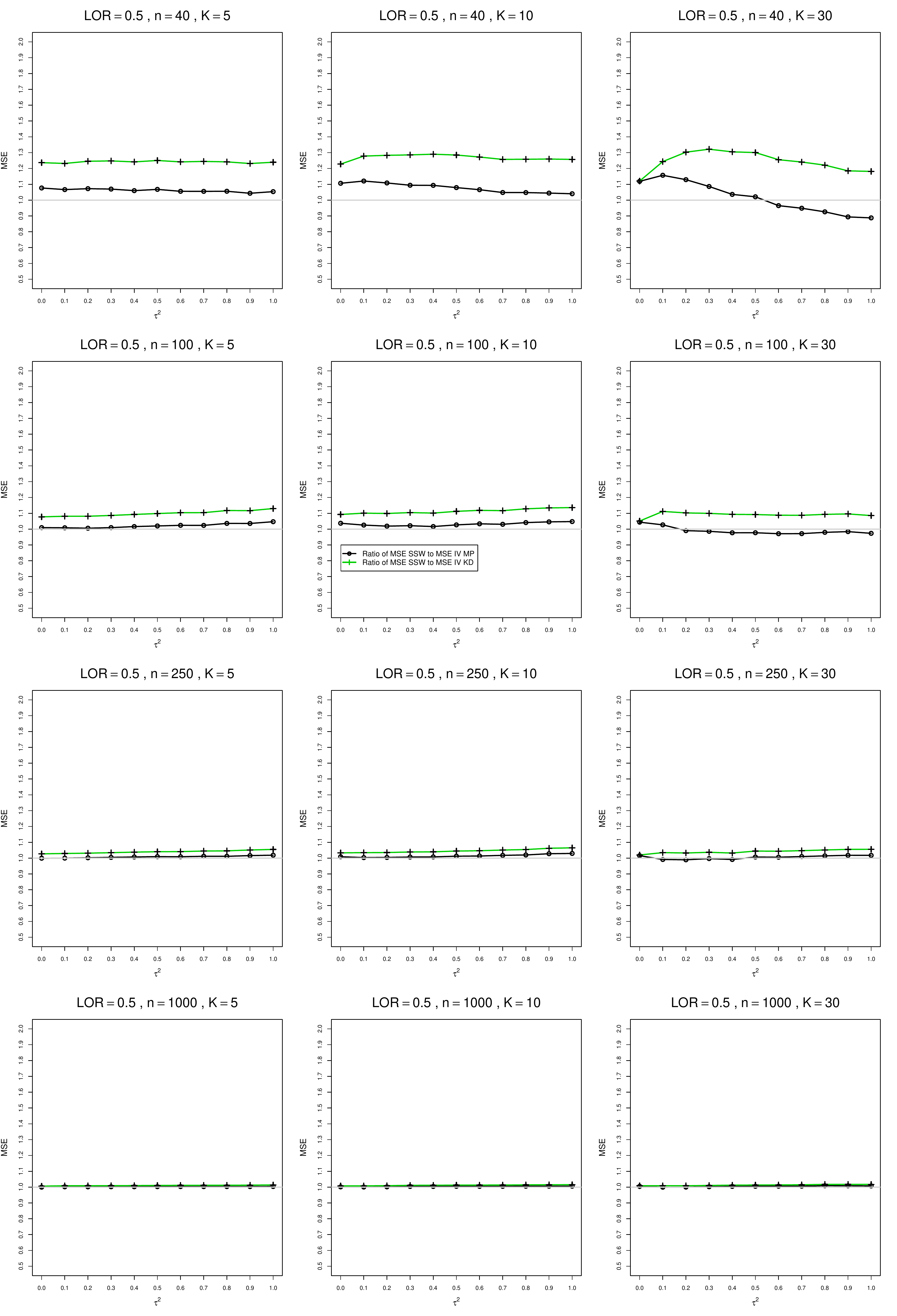}
	\caption{Ratio of mean squared errors of the fixed-weights to mean squared errors of inverse-variance estimator for $\theta=0.5$, $p_{iC}=0.2$, $q=0.5$, equal sample sizes $n=40,\;100,\;250,\;1000$. 
		\label{RatioOfMSEwithLOR05q05piC02fromMPandCMP}}
\end{figure}

\begin{figure}[t]\centering
	\includegraphics[scale=0.35]{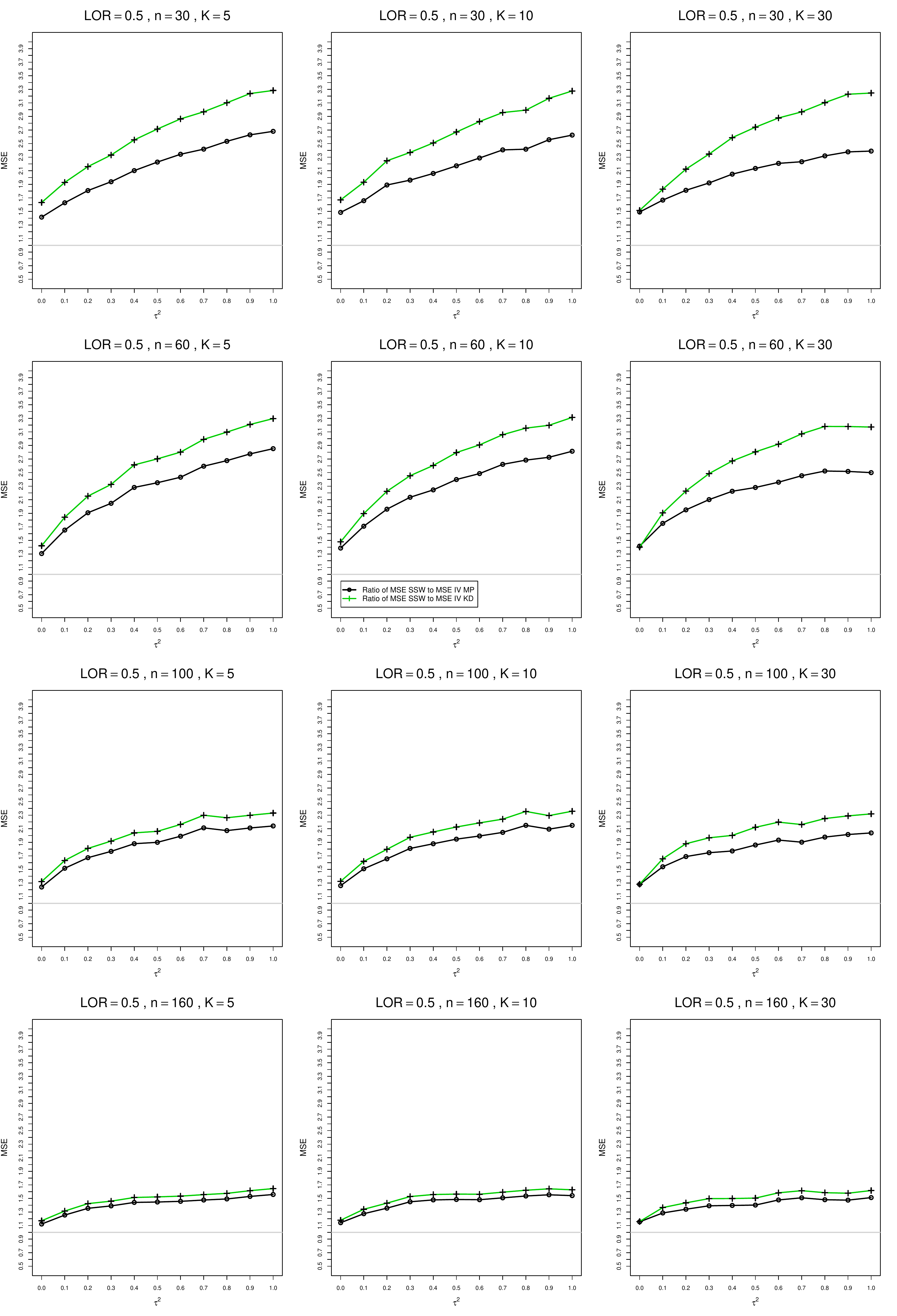}
	\caption{Ratio of mean squared errors of the fixed-weights to mean squared errors of inverse-variance estimator for $\theta=0.5$, $p_{iC}=0.2$, $q=0.5$, unequal sample sizes $n=30,\;60,\;100,\;160$. 
		\label{RatioOfMSEwithLOR05q05piC02fromMPandCMP_unequal_sample_sizes}}
\end{figure}


\begin{figure}[t]
	\centering
	\includegraphics[scale=0.33]{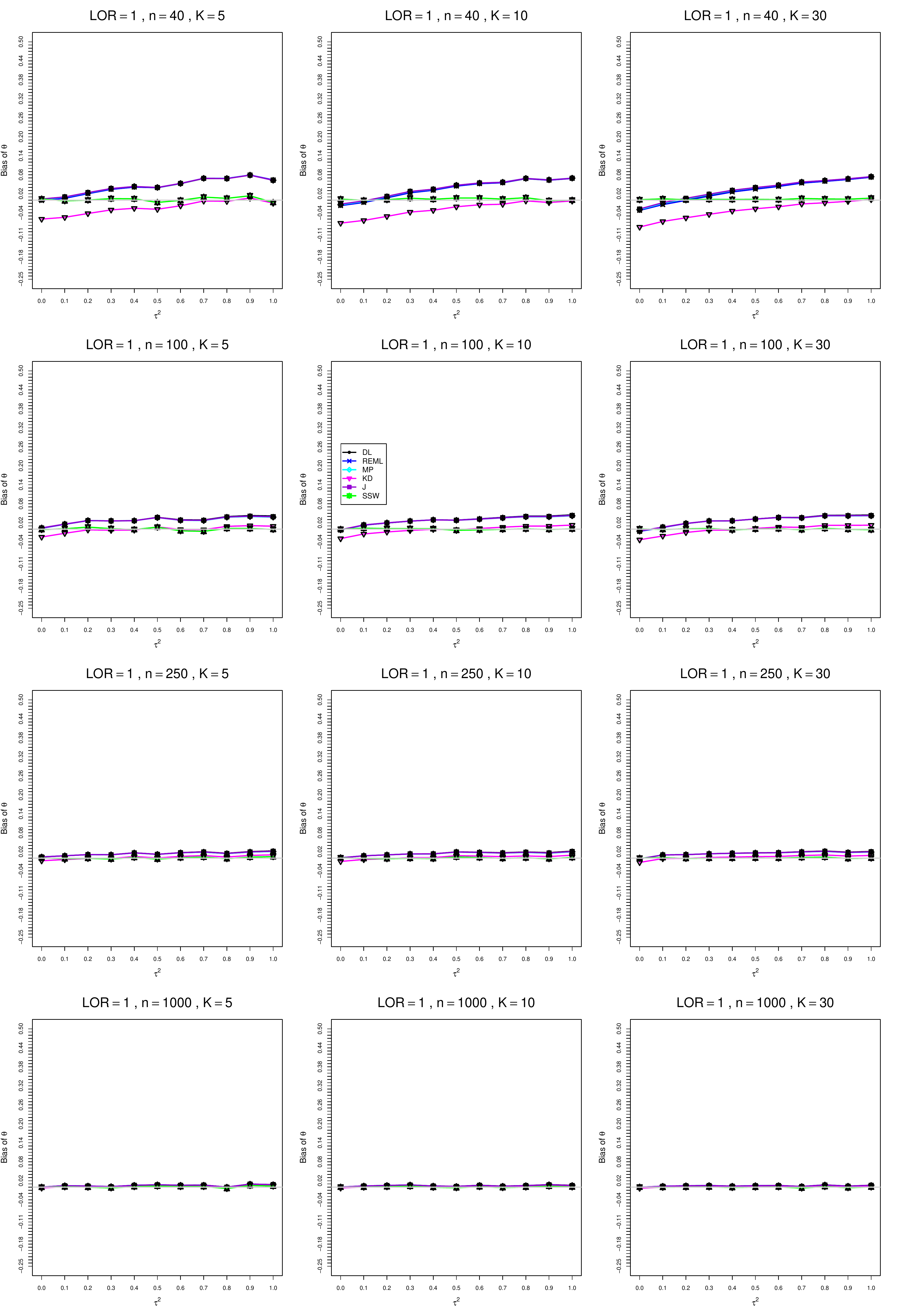}
	\caption{Bias of the estimation of  overall effect measure $\theta$ for $\theta=1$, $p_{iC}=0.2$, $q=0.5$, equal sample sizes $n=40,\;100,\;250,\;1000$. 
		\label{BiasThetaLOR1q05piC02}}
\end{figure}

\begin{figure}[t]
	\centering
	\includegraphics[scale=0.33]{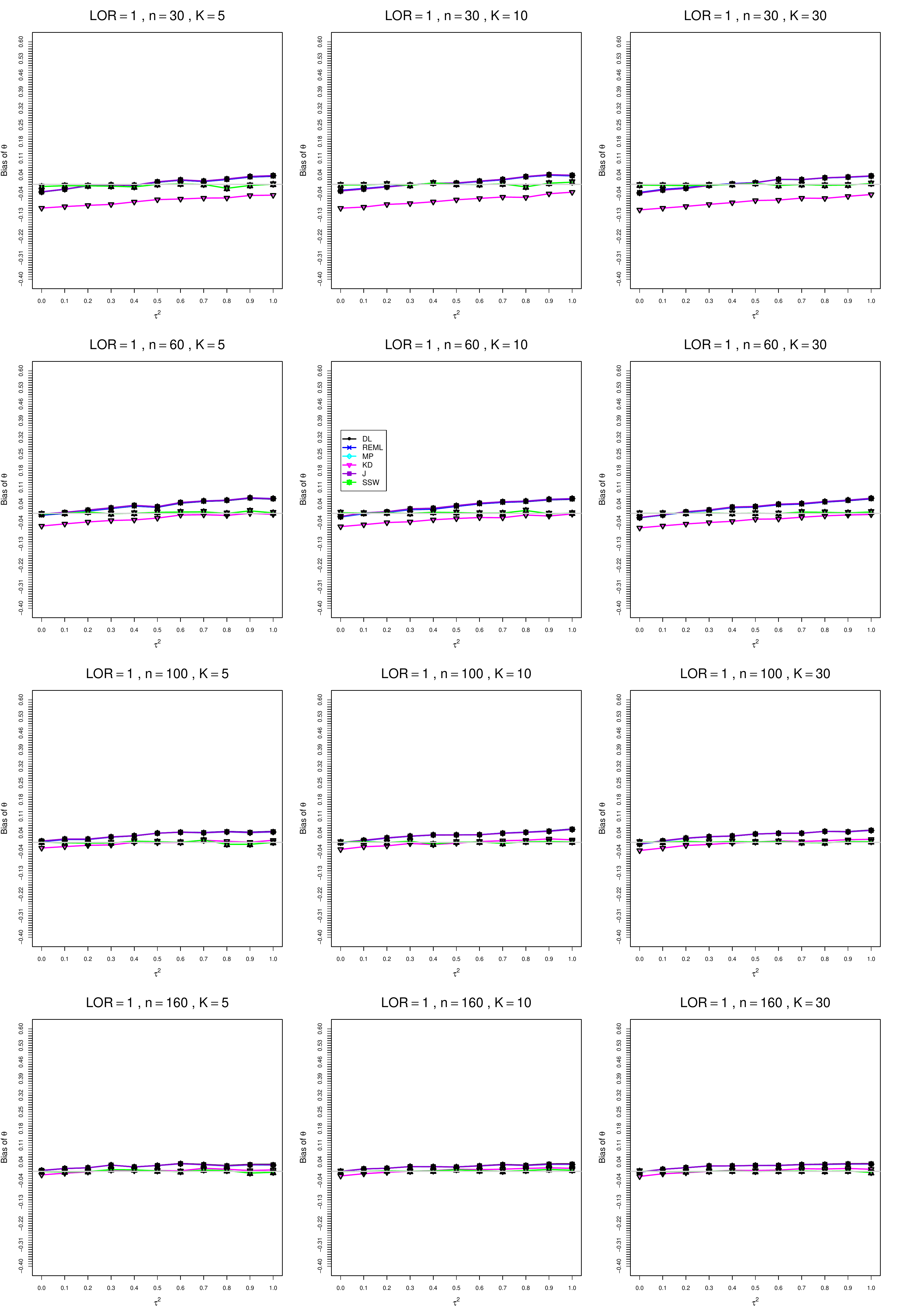}
	\caption{Bias of the estimation of  overall effect measure $\theta$ for $\theta=1$, $p_{iC}=0.2$, $q=0.5$, 
		unequal sample sizes $n=30,\; 60,\;100,\;160$. 
		\label{BiasThetaLOR1q05piC02_unequal_sample_sizes}}
\end{figure}

\begin{figure}[t]\centering
	\includegraphics[scale=0.35]{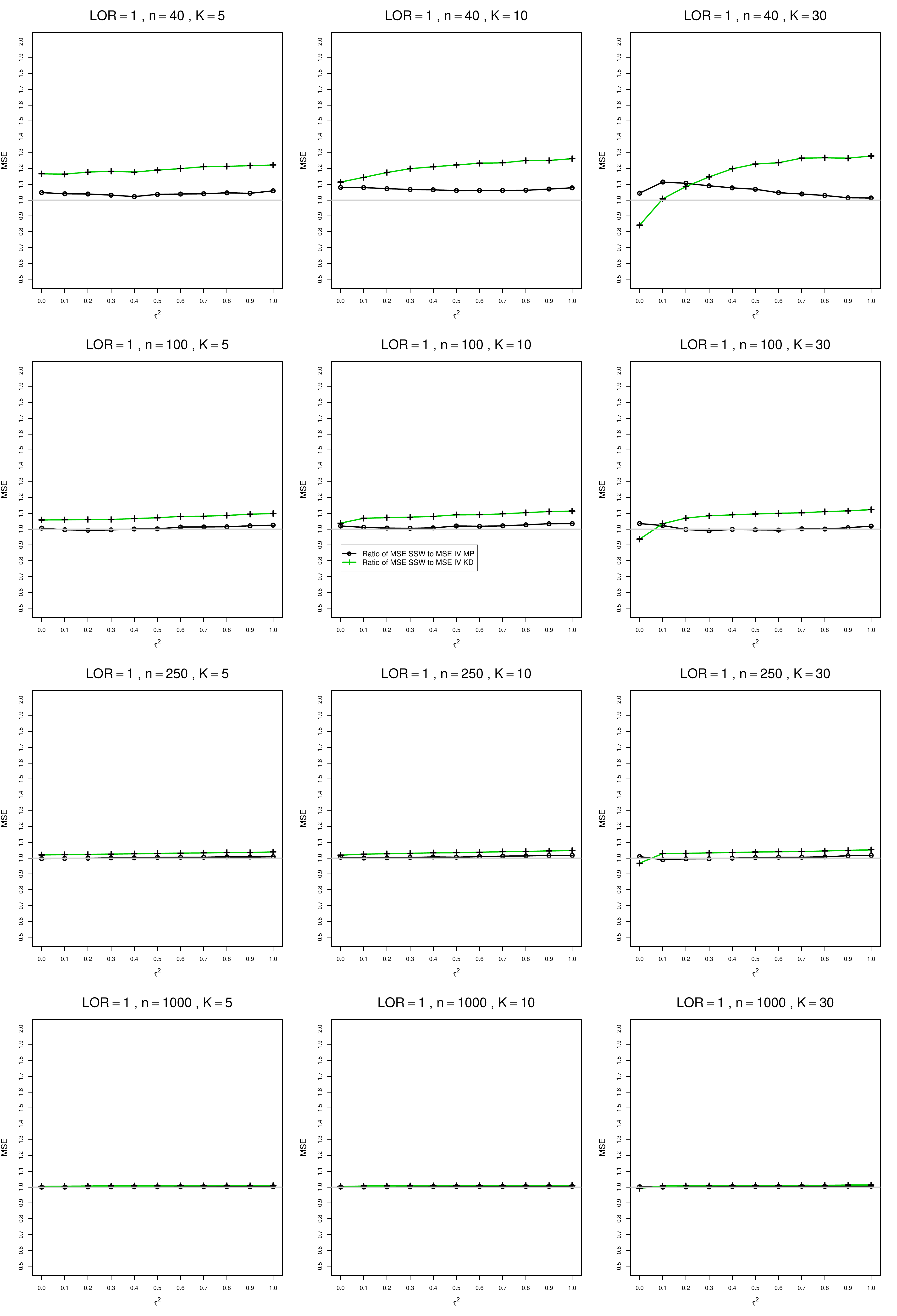}
	\caption{Ratio of mean squared errors of the fixed-weights to mean squared errors of inverse-variance estimator for $\theta=1$, $p_{iC}=0.2$, $q=0.5$, equal sample sizes $n=40,\;100,\;250,\;1000$. 
		\label{RatioOfMSEwithLOR1q05piC02fromMPandCMP}}
\end{figure}

\begin{figure}[t]\centering
	\includegraphics[scale=0.35]{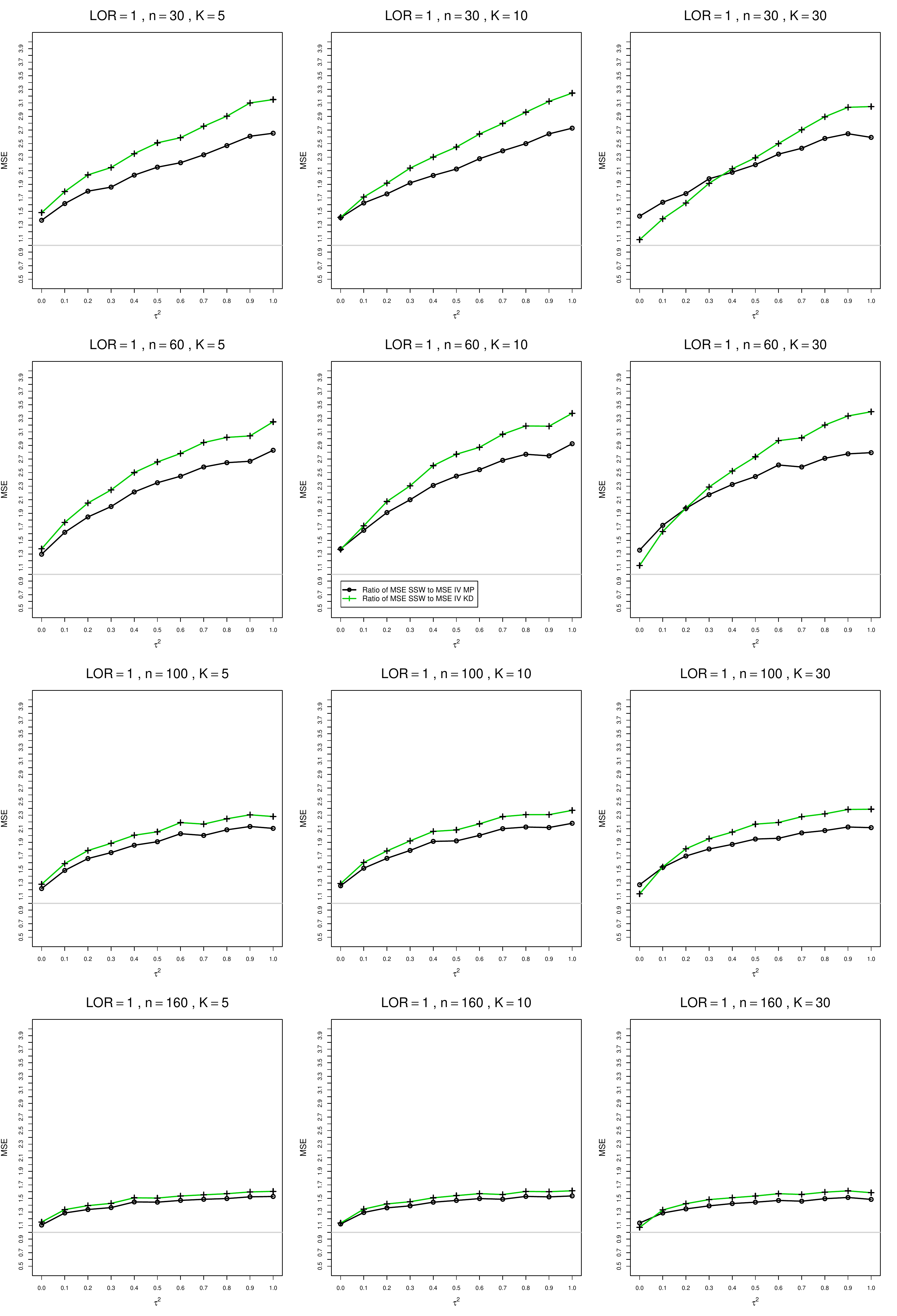}
	\caption{Ratio of mean squared errors of the fixed-weights to mean squared errors of inverse-variance estimator for $\theta=1$, $p_{iC}=0.2$, $q=0.5$, unequal sample sizes $n=30,\;60,\;100,\;160$. 
		\label{RatioOfMSEwithLOR1q05piC02fromMPandCMP_unequal_sample_sizes}}
\end{figure}

\begin{figure}[t]
	\centering
	\includegraphics[scale=0.33]{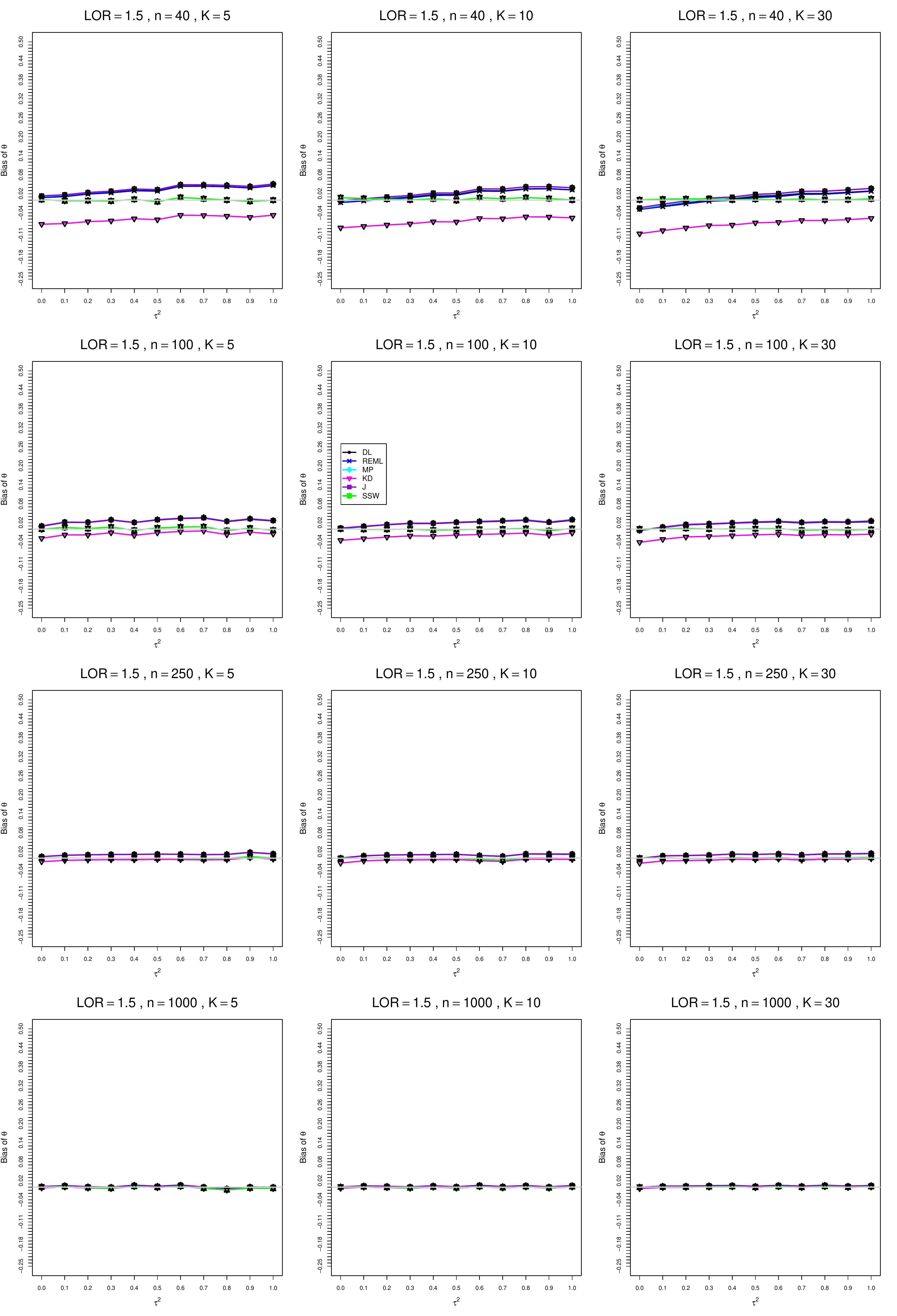}
	\caption{Bias of the estimation of  overall effect measure $\theta$ for $\theta=1.5$, $p_{iC}=0.2$, $q=0.5$, equal sample sizes $n=40,\;100,\;250,\;1000$. 
		\label{BiasThetaLOR15q05piC02}}
\end{figure}

\begin{figure}[t]
	\centering
	\includegraphics[scale=0.33]{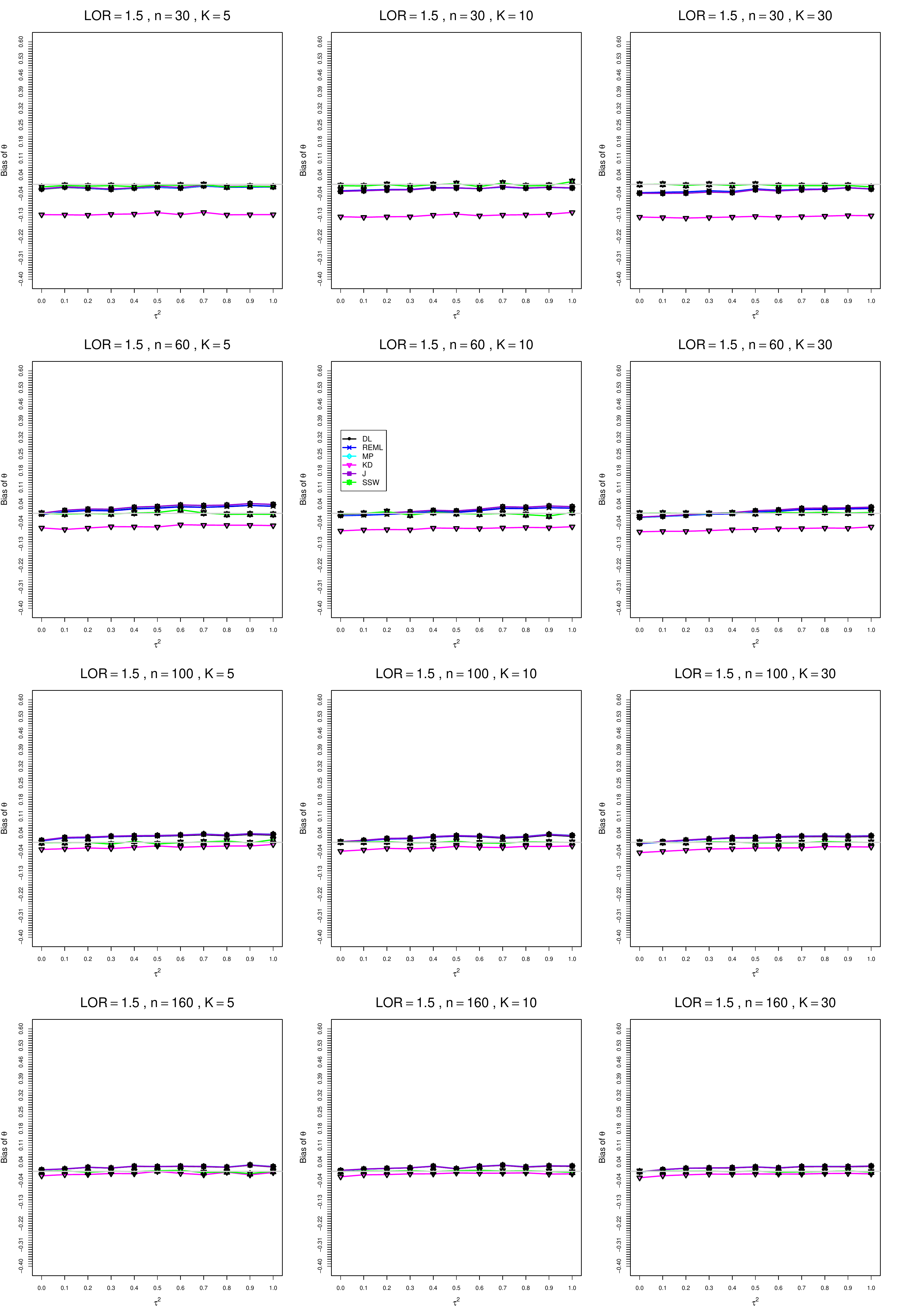}
	\caption{Bias of the estimation of  overall effect measure $\theta$ for $\theta=1.5$, $p_{iC}=0.2$, $q=0.5$, 
		unequal sample sizes $n=30,\; 60,\;100,\;160$. 
		\label{BiasThetaLOR15q05piC02_unequal_sample_sizes}}
\end{figure}

\begin{figure}[t]\centering
	\includegraphics[scale=0.35]{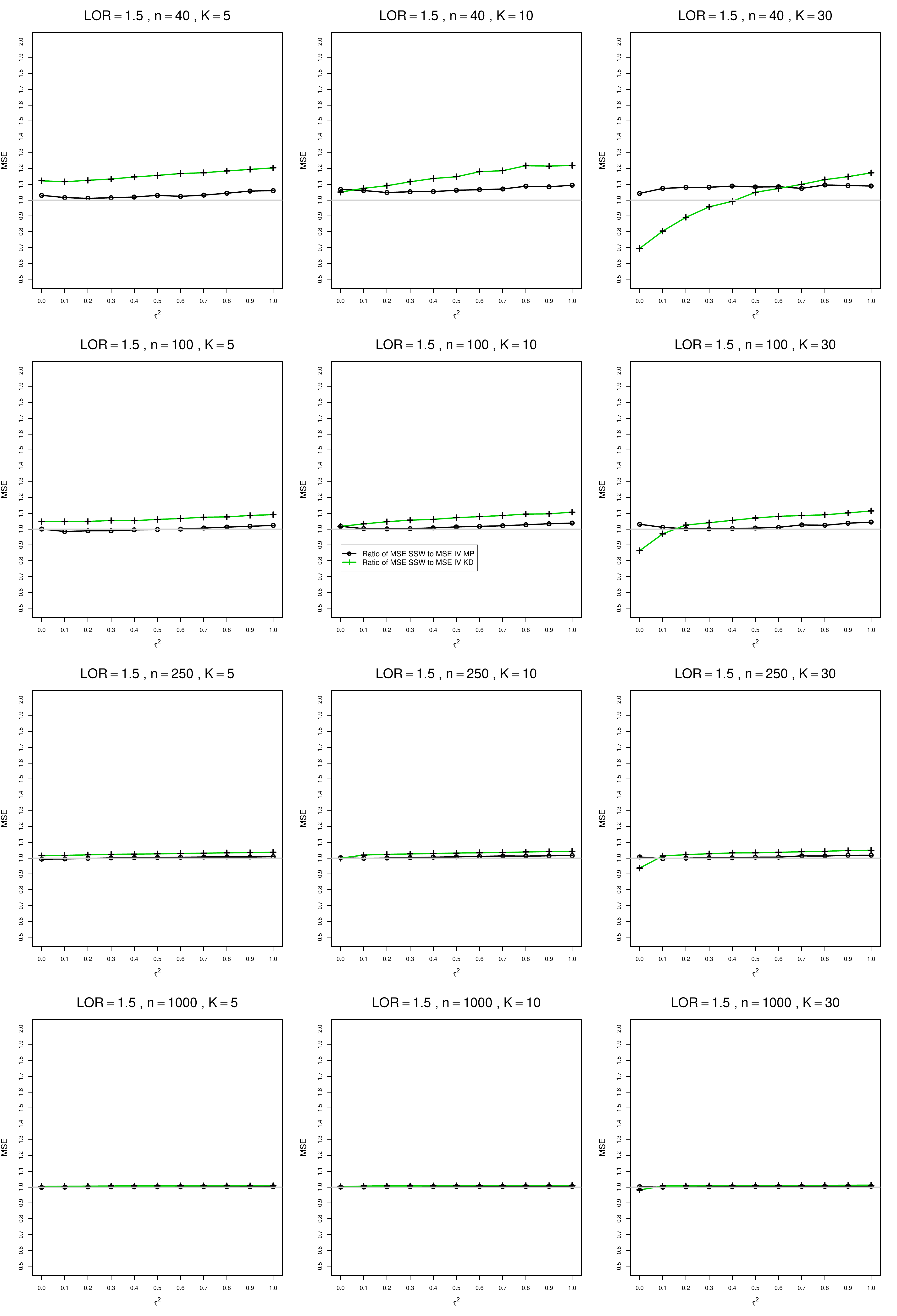}
	\caption{Ratio of mean squared errors of the fixed-weights to mean squared errors of inverse-variance estimator for $\theta=1.5$, $p_{iC}=0.2$, $q=0.5$, equal sample sizes $n=40,\;100,\;250,\;1000$. 
		\label{RatioOfMSEwithLOR15q05piC02fromMPandCMP}}
\end{figure}

\begin{figure}[t]\centering
	\includegraphics[scale=0.35]{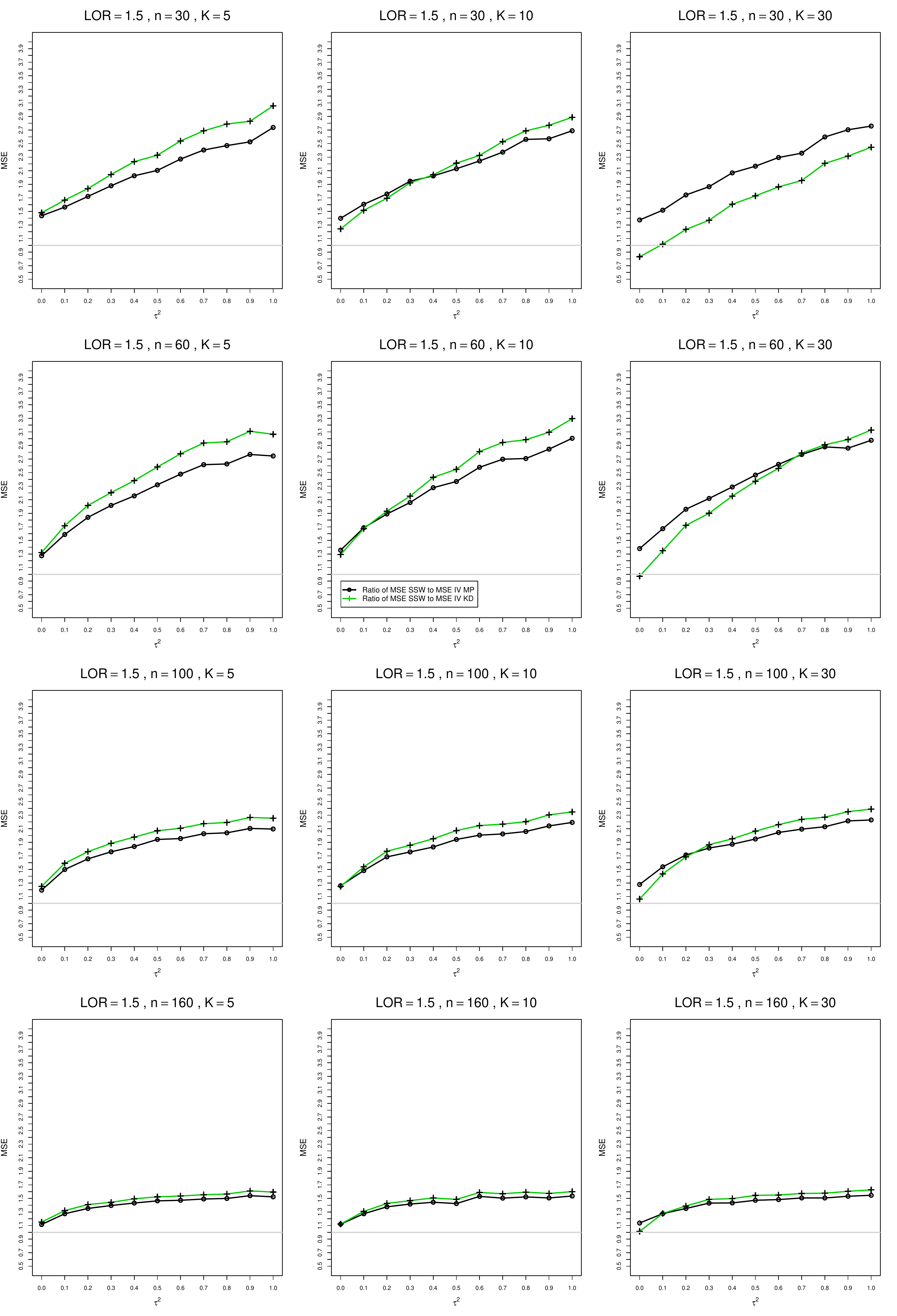}
	\caption{Ratio of mean squared errors of the fixed-weights to mean squared errors of inverse-variance estimator for $\theta=1.5$, $p_{iC}=0.2$, $q=0.5$, unequal sample sizes $n=30,\;60,\;100,\;160$. 
		\label{RatioOfMSEwithLOR15q05piC02fromMPandCMP_unequal_sample_sizes}}
\end{figure}

\begin{figure}[t]
	\centering
	\includegraphics[scale=0.33]{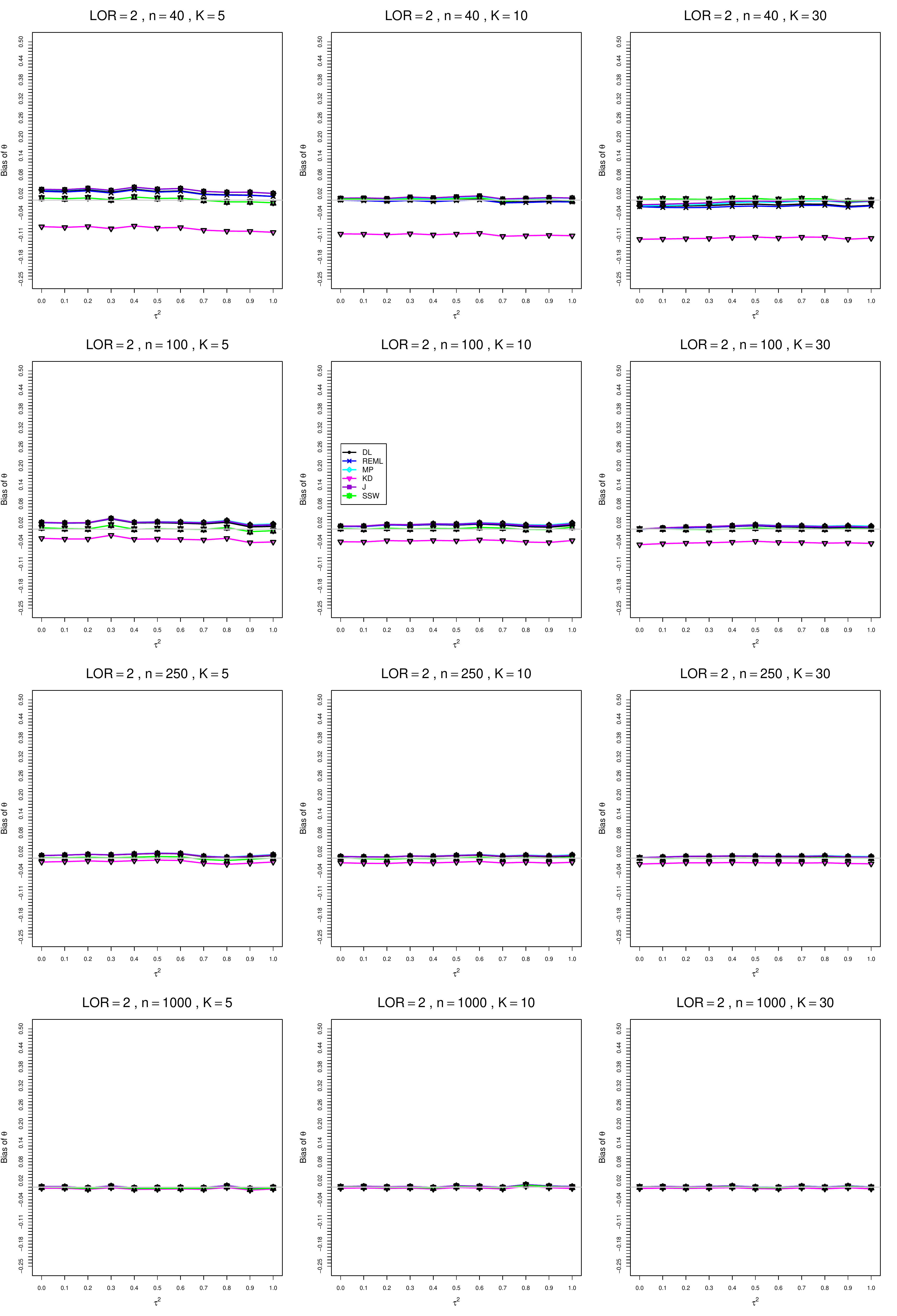}
	\caption{Bias of the estimation of  overall effect measure $\theta$ for $\theta=2$, $p_{iC}=0.2$, $q=0.5$, equal sample sizes $n=40,\;100,\;250,\;1000$. 
		\label{BiasThetaLOR2q05piC02}}
\end{figure}

\begin{figure}[t]
	\centering
	\includegraphics[scale=0.33]{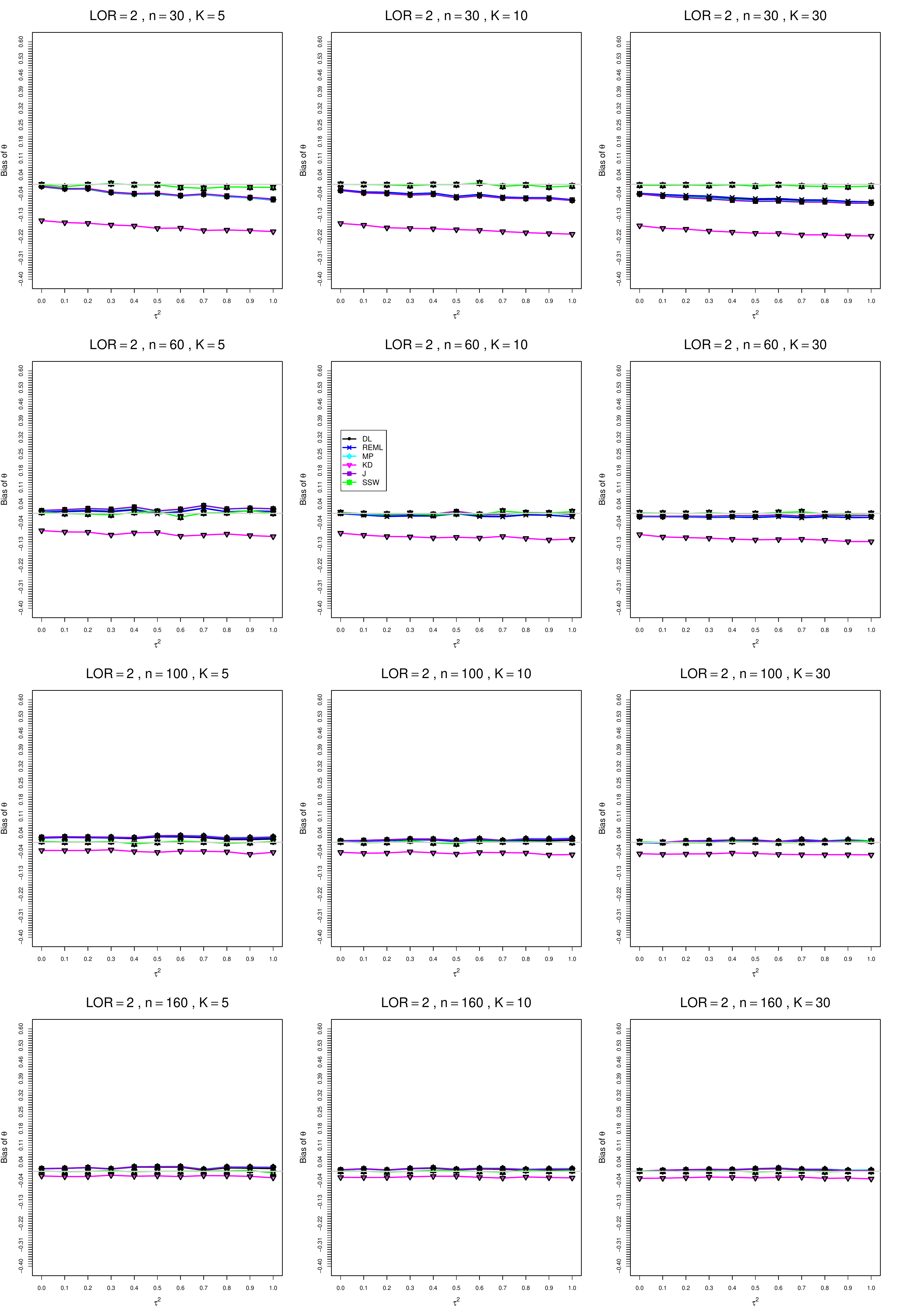}
	\caption{Bias of the estimation of  overall effect measure $\theta$ for $\theta=2$, $p_{iC}=0.2$, $q=0.5$, 
		unequal sample sizes $n=30,\; 60,\;100,\;160$. 
		\label{BiasThetaLOR2q05piC02_unequal_sample_sizes}}
\end{figure}

\begin{figure}[t]\centering
	\includegraphics[scale=0.35]{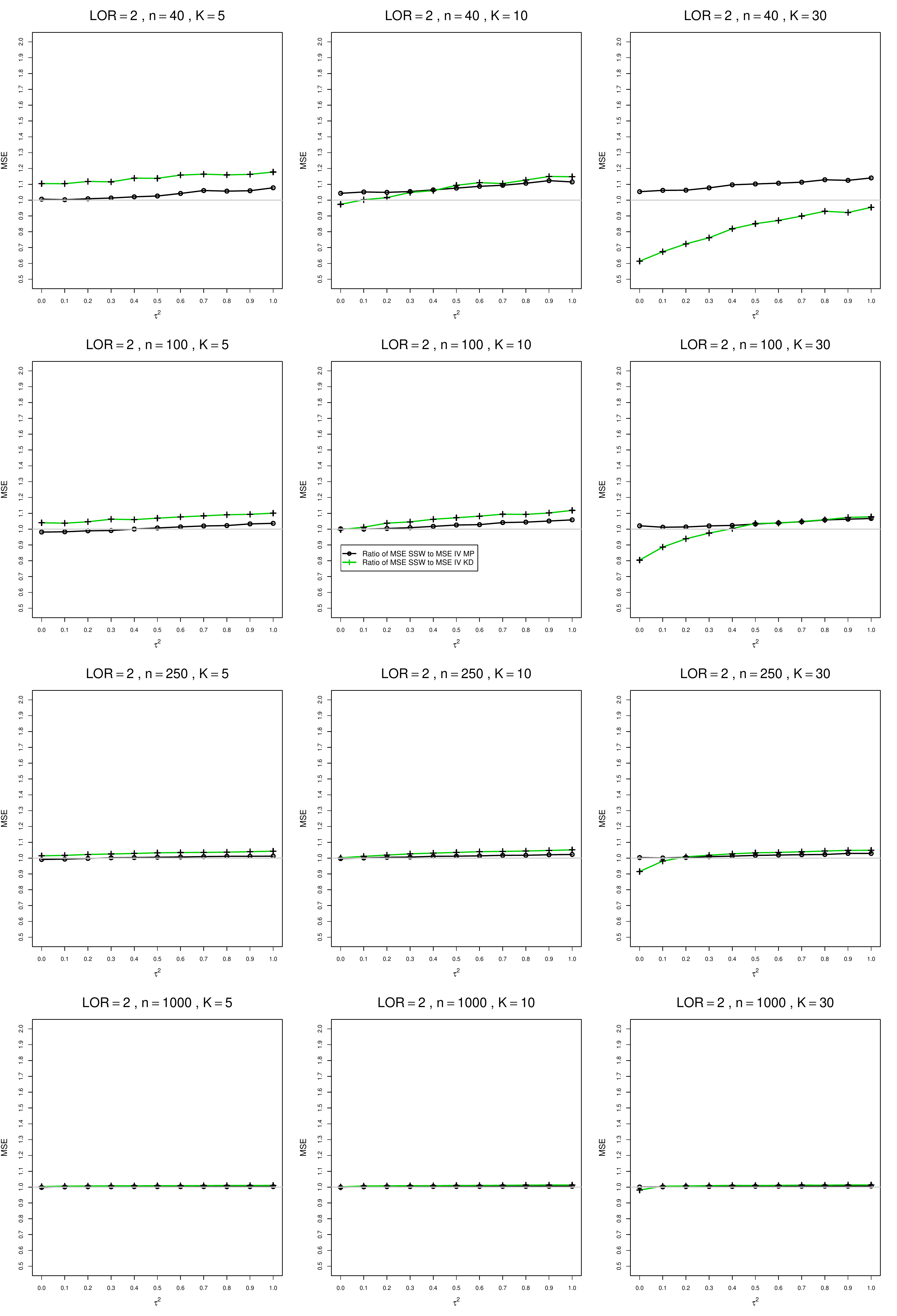}
	\caption{Ratio of mean squared errors of the fixed-weights to mean squared errors of inverse-variance estimator for $\theta=2$, $p_{iC}=0.2$, $q=0.5$, equal sample sizes $n=40,\;100,\;250,\;1000$. 
		\label{RatioOfMSEwithLOR2q05piC02fromMPandCMP}}
\end{figure}

\begin{figure}[t]\centering
	\includegraphics[scale=0.35]{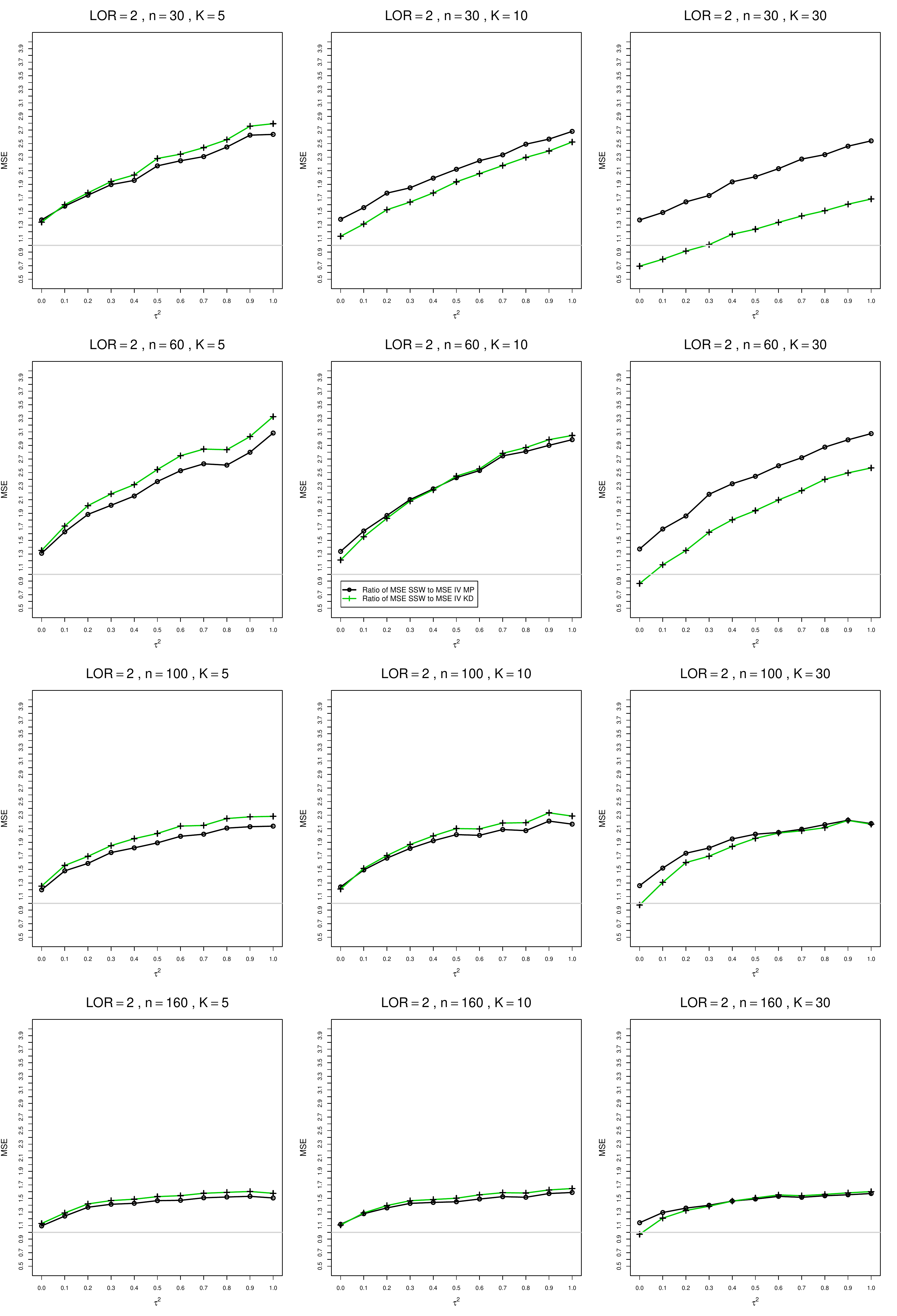}
	\caption{Ratio of mean squared errors of the fixed-weights to mean squared errors of inverse-variance estimator for $\theta=2$, $p_{iC}=0.2$, $q=0.5$, unequal sample sizes $n=30,\;60,\;100,\;160$. 
		\label{RatioOfMSEwithLOR2q05piC02fromMPandCMP_unequal_sample_sizes}}
\end{figure}


\begin{figure}[t]
	\centering
	\includegraphics[scale=0.33]{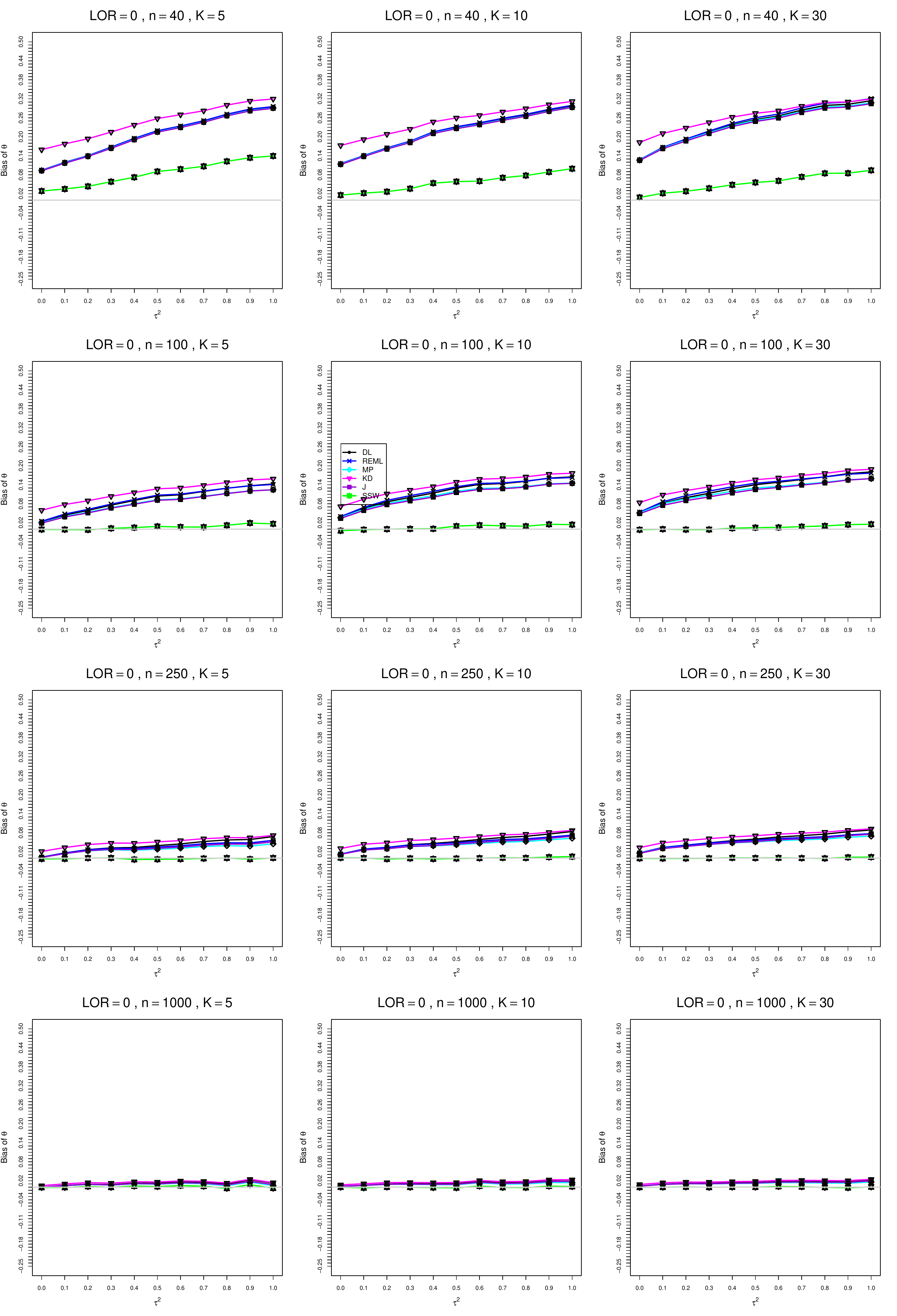}
	\caption{Bias of the estimation of  overall effect measure $\theta$ for $\theta=0$, $p_{iC}=0.2$, $q=0.75$, equal sample sizes $n=40,\;100,\;250,\;1000$. 
		\label{BiasThetaLOR0q075piC02}}
\end{figure}

\begin{figure}[t]
	\centering
	\includegraphics[scale=0.33]{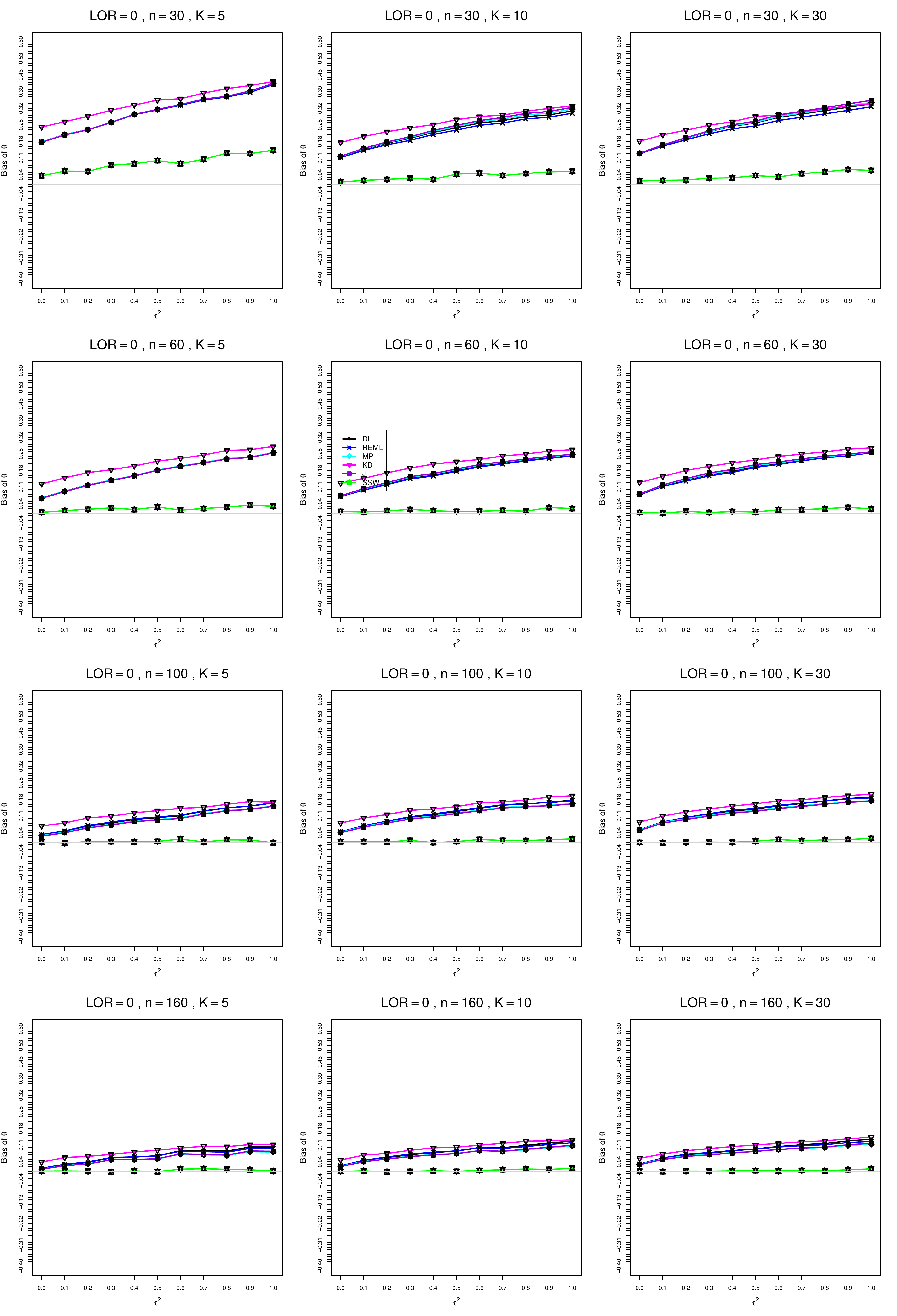}
	\caption{Bias of the estimation of  overall effect measure $\theta$ for $\theta=0$, $p_{iC}=0.2$, $q=0.75$, 
		unequal sample sizes $n=30,\; 60,\;100,\;160$. 
		\label{BiasThetaLOR0q075piC02_unequal_sample_sizes}}
\end{figure}

\begin{figure}[t]\centering
	\includegraphics[scale=0.35]{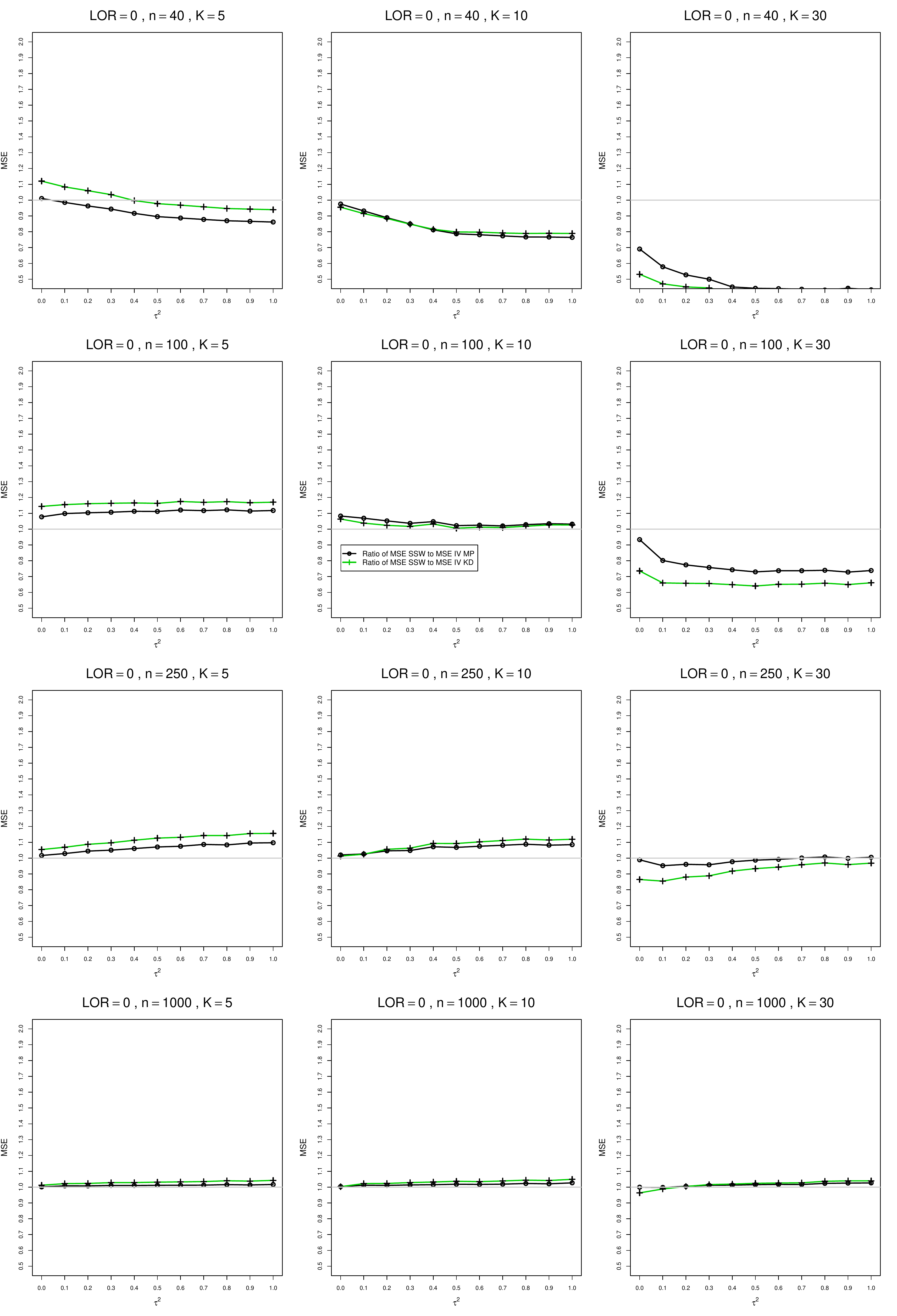}
	\caption{Ratio of mean squared errors of the fixed-weights to mean squared errors of inverse-variance estimator for $\theta=0$, $p_{iC}=0.2$, $q=0.75$, equal sample sizes $n=40,\;100,\;250,\;1000$. 
		\label{RatioOfMSEwithLOR0q075piC02fromMPandCMP}}
\end{figure}

\begin{figure}[t]\centering
	\includegraphics[scale=0.35]{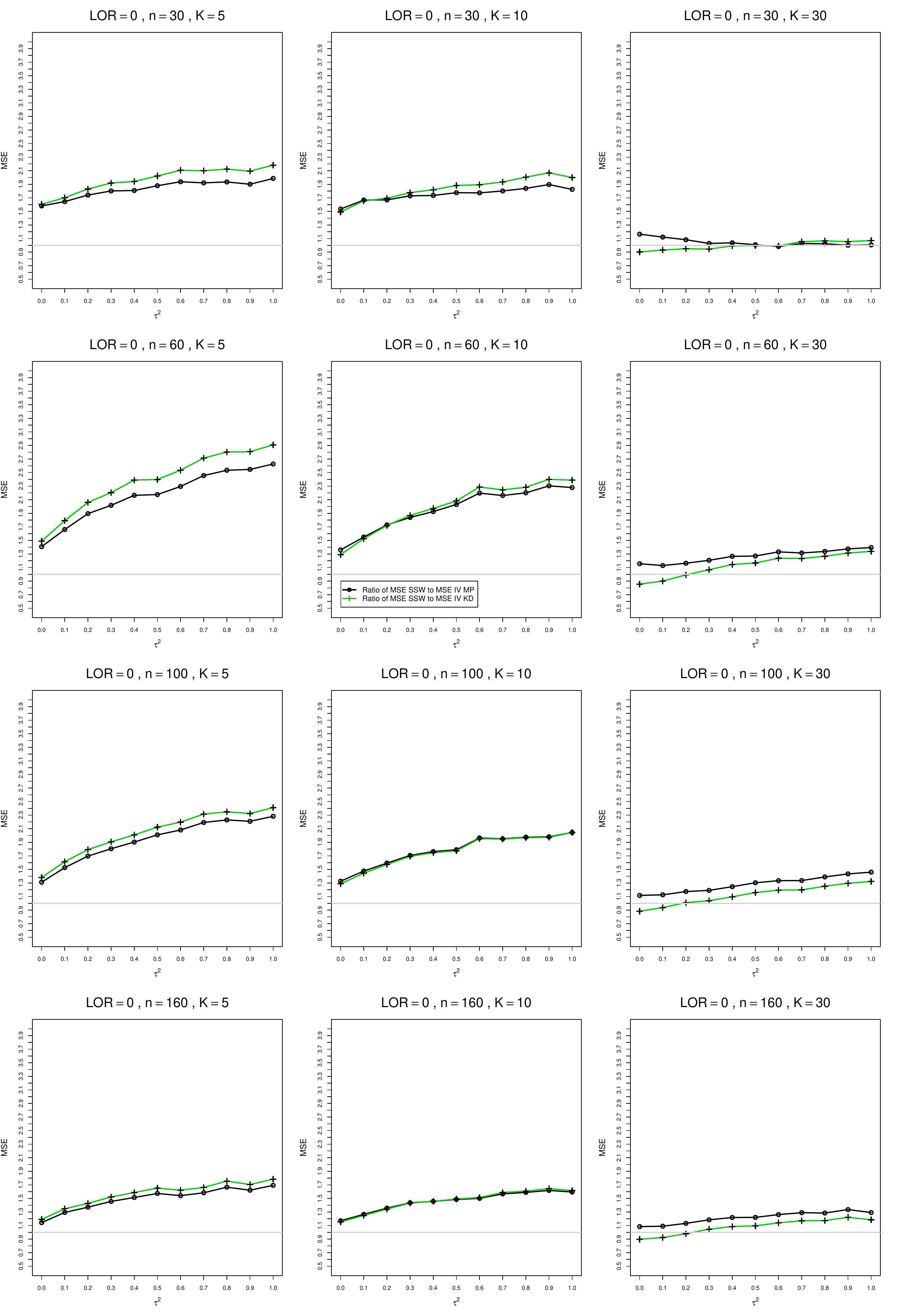}
	\caption{Ratio of mean squared errors of the fixed-weights to mean squared errors of inverse-variance estimator for $\theta=0$, $p_{iC}=0.2$, $q=0.75$, unequal sample sizes $n=30,\;60,\;100,\;160$. 
		\label{RatioOfMSEwithLOR0q075piC02fromMPandCMP_unequal_sample_sizes}}
\end{figure}

\begin{figure}[t]
	\centering
	\includegraphics[scale=0.33]{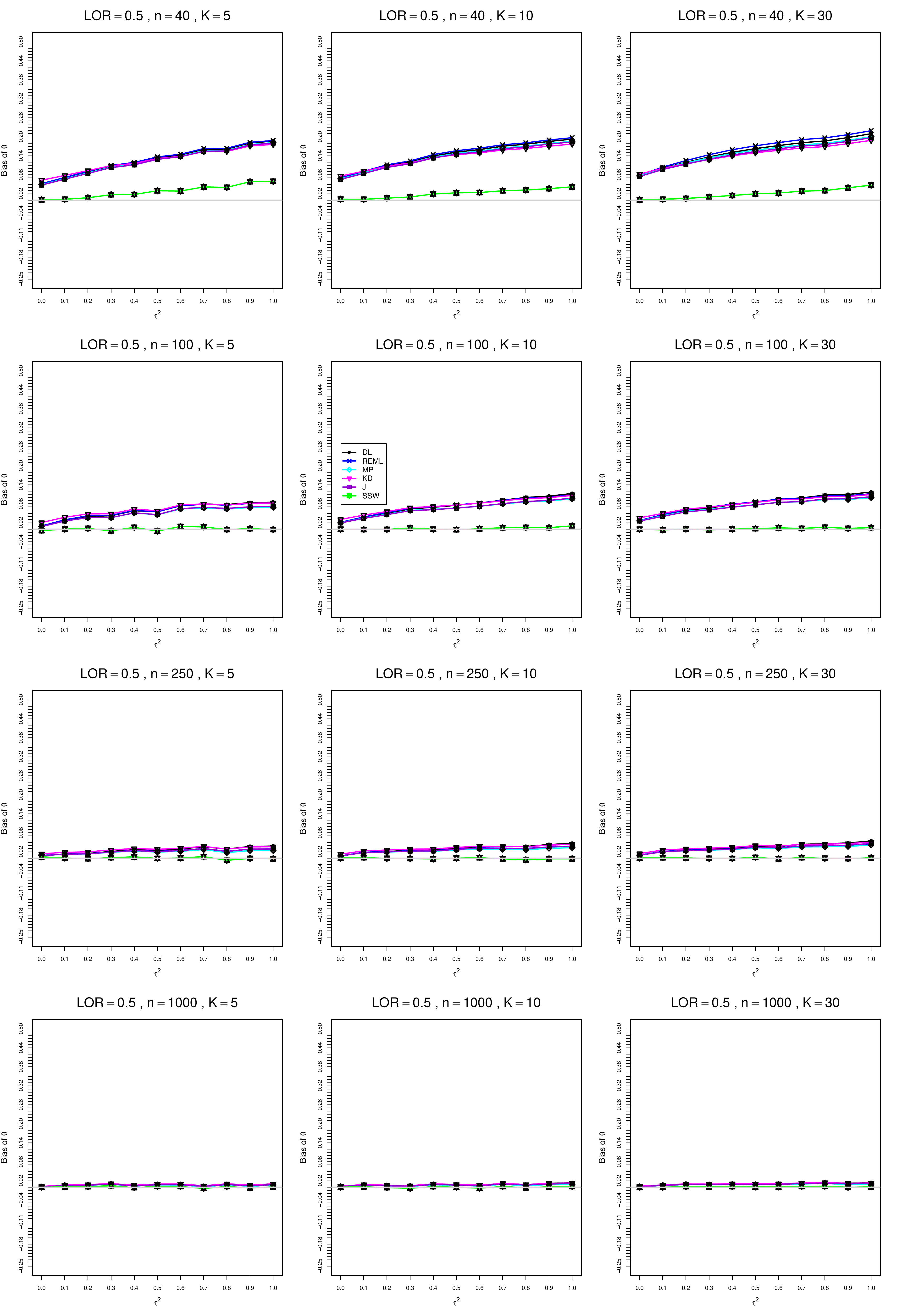}
	\caption{Bias of the estimation of  overall effect measure $\theta$ for $\theta=0.5$, $p_{iC}=0.2$, $q=0.75$, equal sample sizes $n=40,\;100,\;250,\;1000$. 
		\label{BiasThetaLOR05q075piC02}}
\end{figure}

\begin{figure}[t]
	\centering
	\includegraphics[scale=0.33]{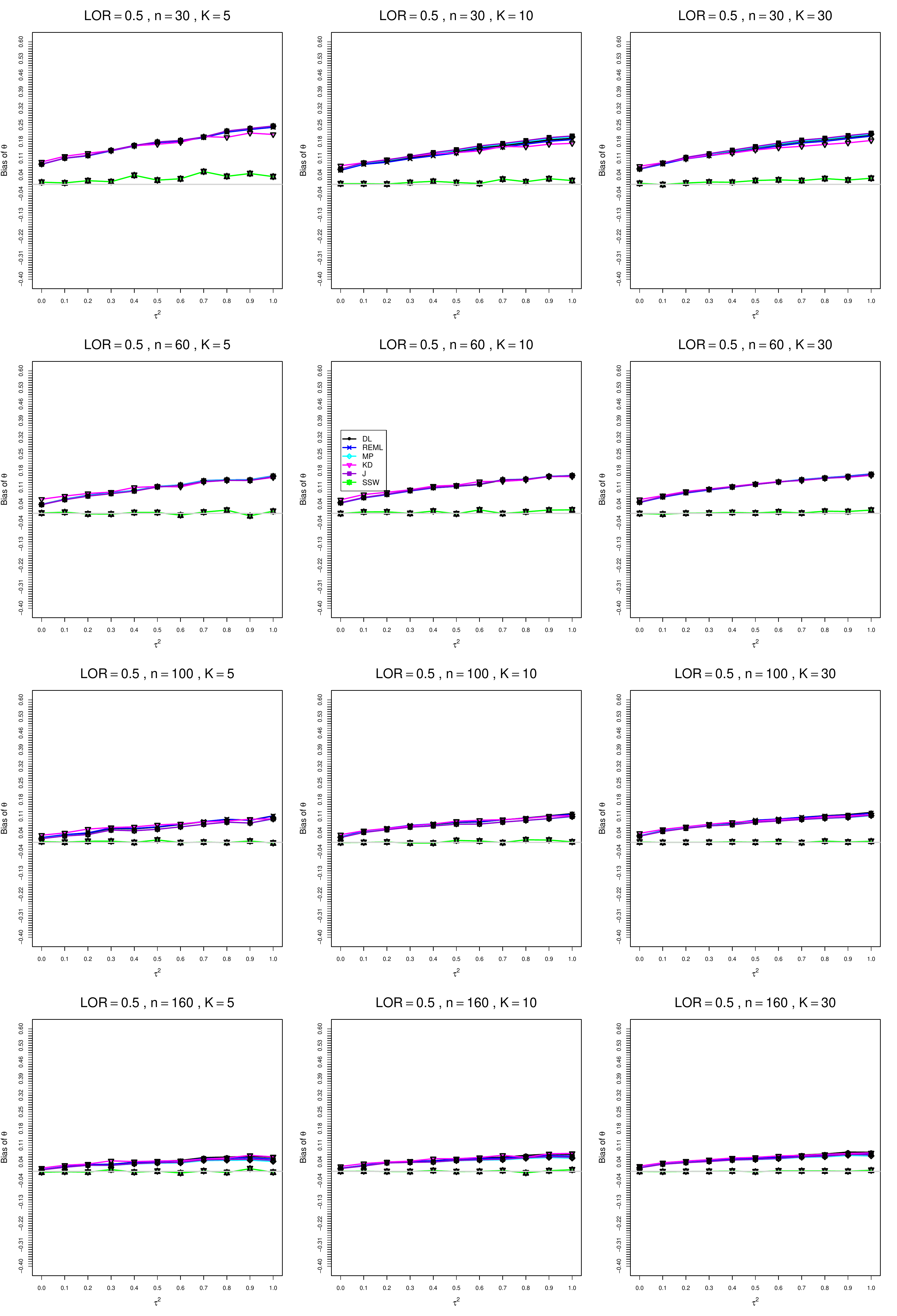}
	\caption{Bias of the estimation of  overall effect measure $\theta$ for $\theta=0.5$, $p_{iC}=0.2$, $q=0.75$, 
		unequal sample sizes $n=30,\; 60,\;100,\;160$. 
		\label{BiasThetaLOR05q075piC02_unequal_sample_sizes}}
\end{figure}
\clearpage

\begin{figure}[t]\centering
	\includegraphics[scale=0.35]{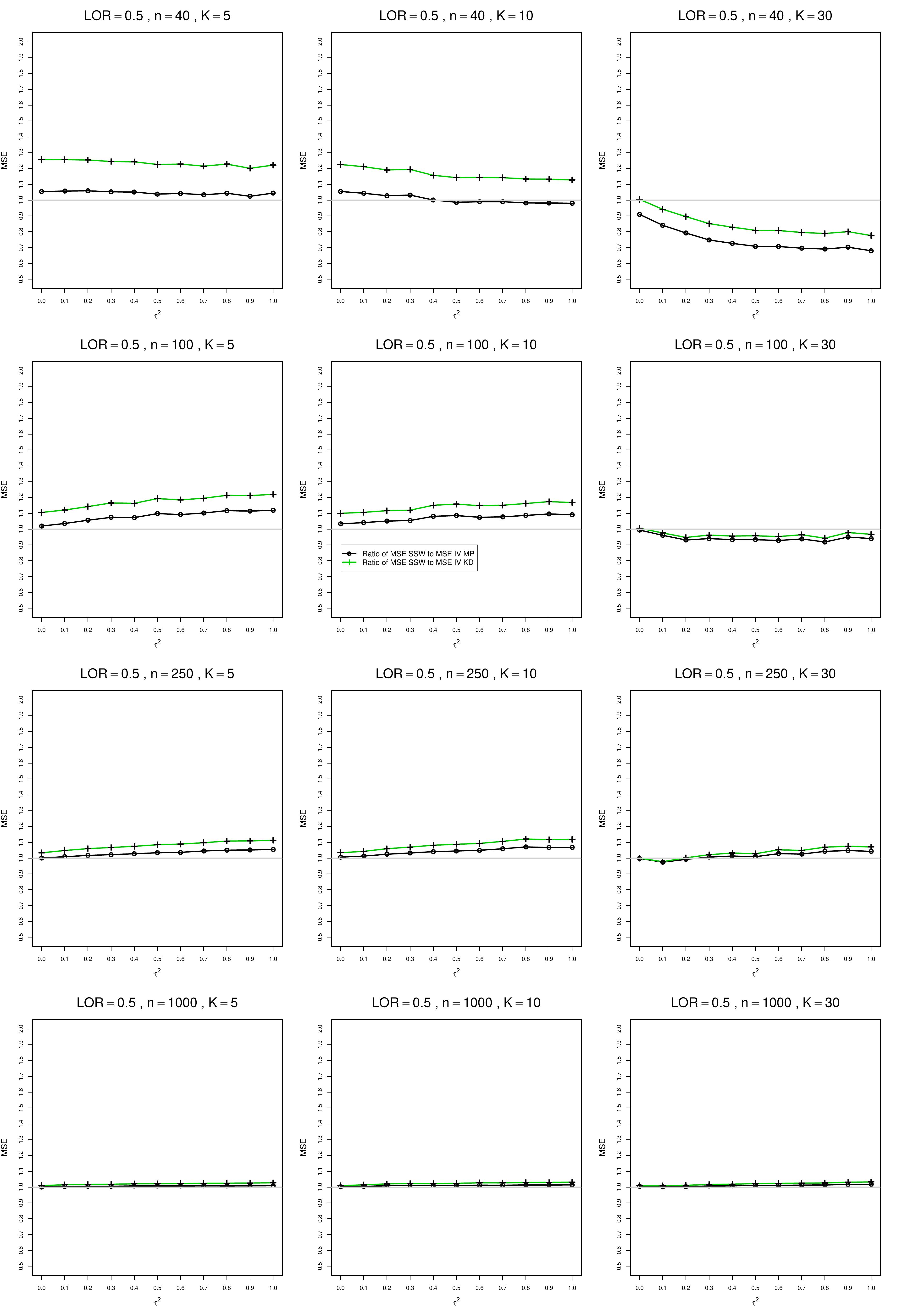}
	\caption{Ratio of mean squared errors of the fixed-weights to mean squared errors of inverse-variance estimator for $\theta=0.5$, $p_{iC}=0.2$, $q=0.75$, equal sample sizes $n=40,\;100,\;250,\;1000$. 
		\label{RatioOfMSEwithLOR05q075piC02fromMPandCMP}}
\end{figure}


\begin{figure}[t]\centering
	\includegraphics[scale=0.35]{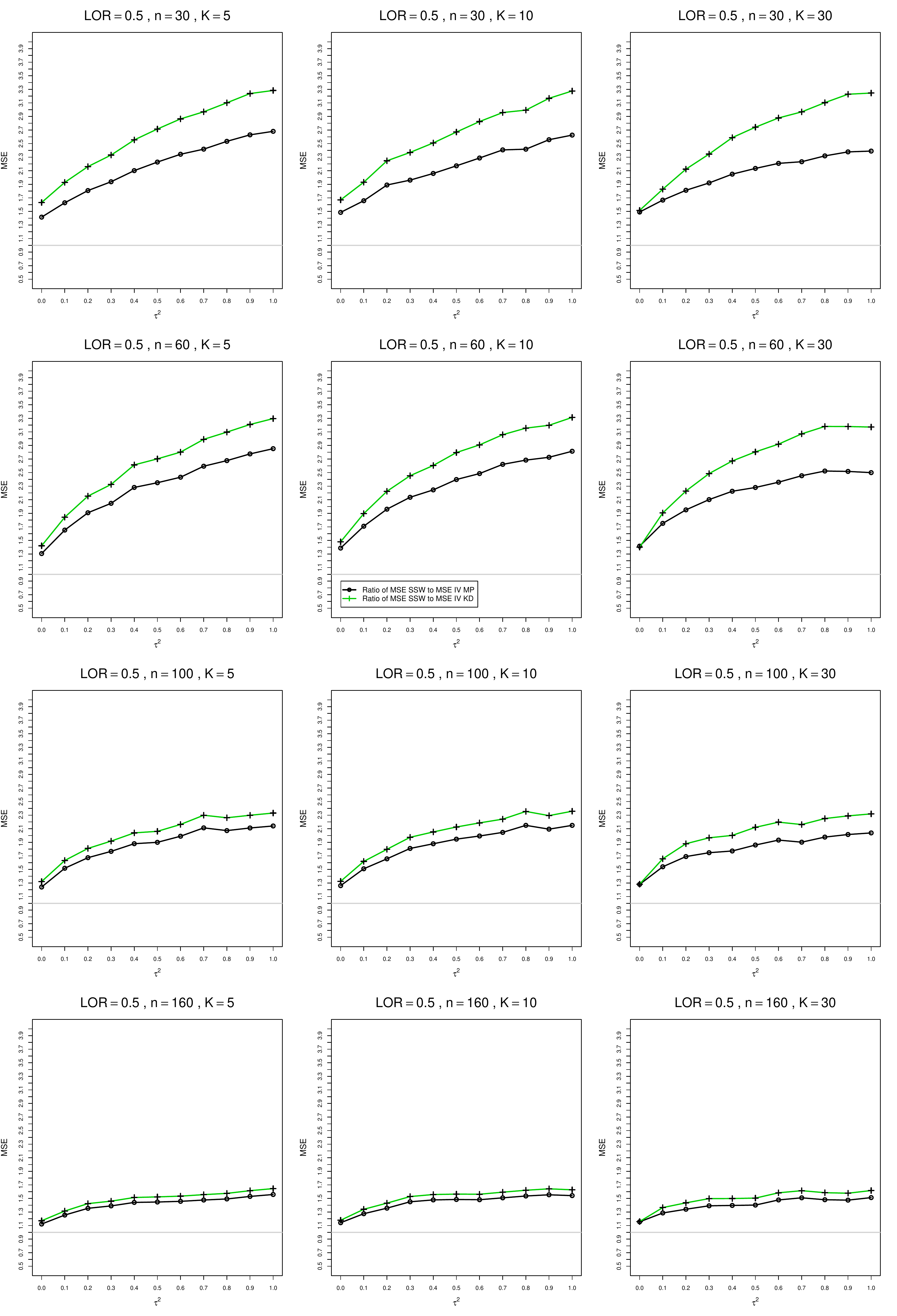}
	\caption{Ratio of mean squared errors of the fixed-weights to mean squared errors of inverse-variance estimator for $\theta=0.5$, $p_{iC}=0.2$, $q=0.75$, unequal sample sizes $n=30,\;60,\;100,\;160$. 
		\label{RatioOfMSEwithLOR05q075piC02fromMPandCMP_unequal_sample_sizes}}
\end{figure}

\begin{figure}[t]
	\centering
	\includegraphics[scale=0.33]{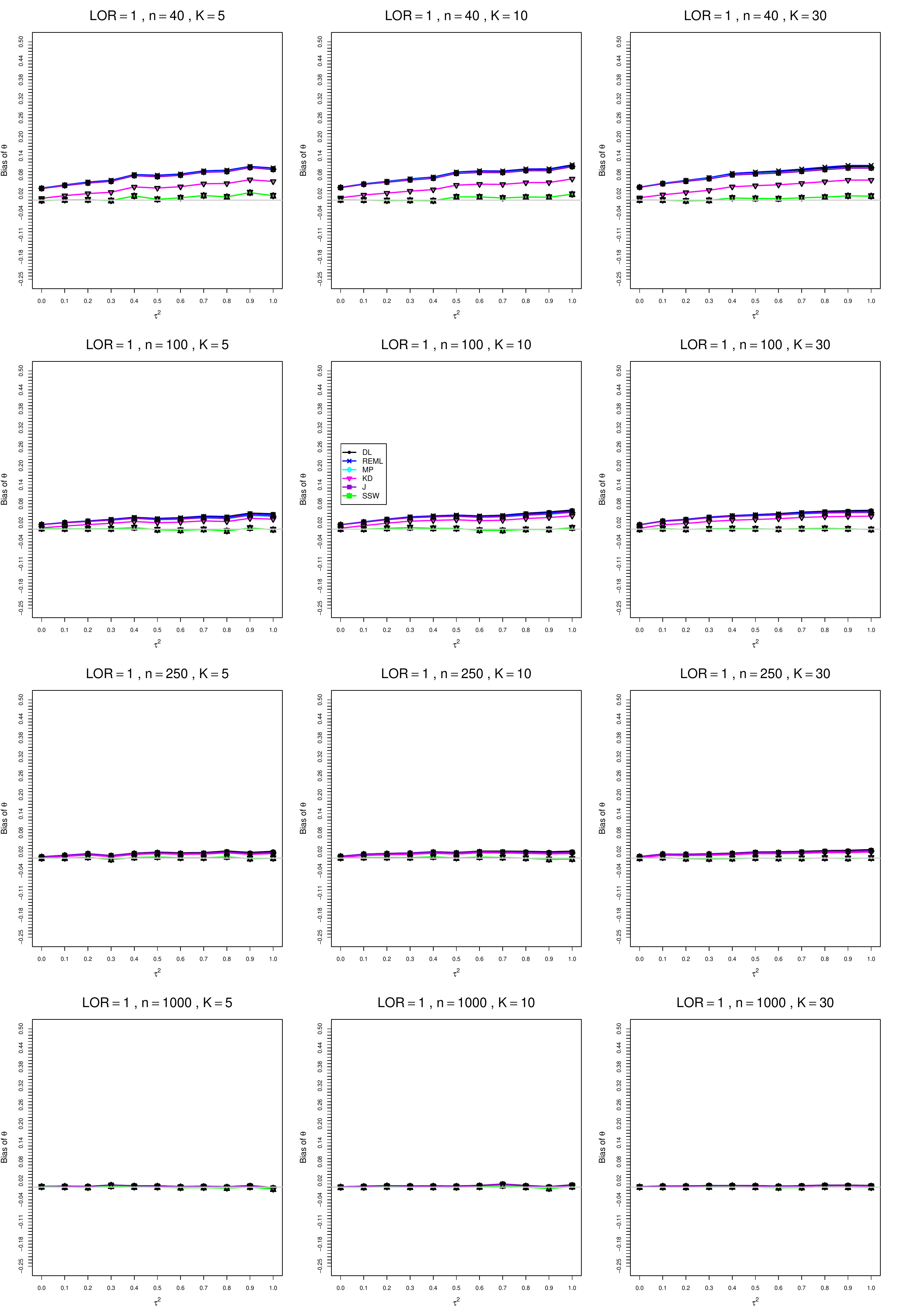}
	\caption{Bias of the estimation of  overall effect measure $\theta$ for $\theta=1$, $p_{iC}=0.2$, $q=0.75$, equal sample sizes $n=40,\;100,\;250,\;1000$. 
		\label{BiasThetaLOR1q075piC02}}
\end{figure}

\begin{figure}[t]
	\centering
	\includegraphics[scale=0.33]{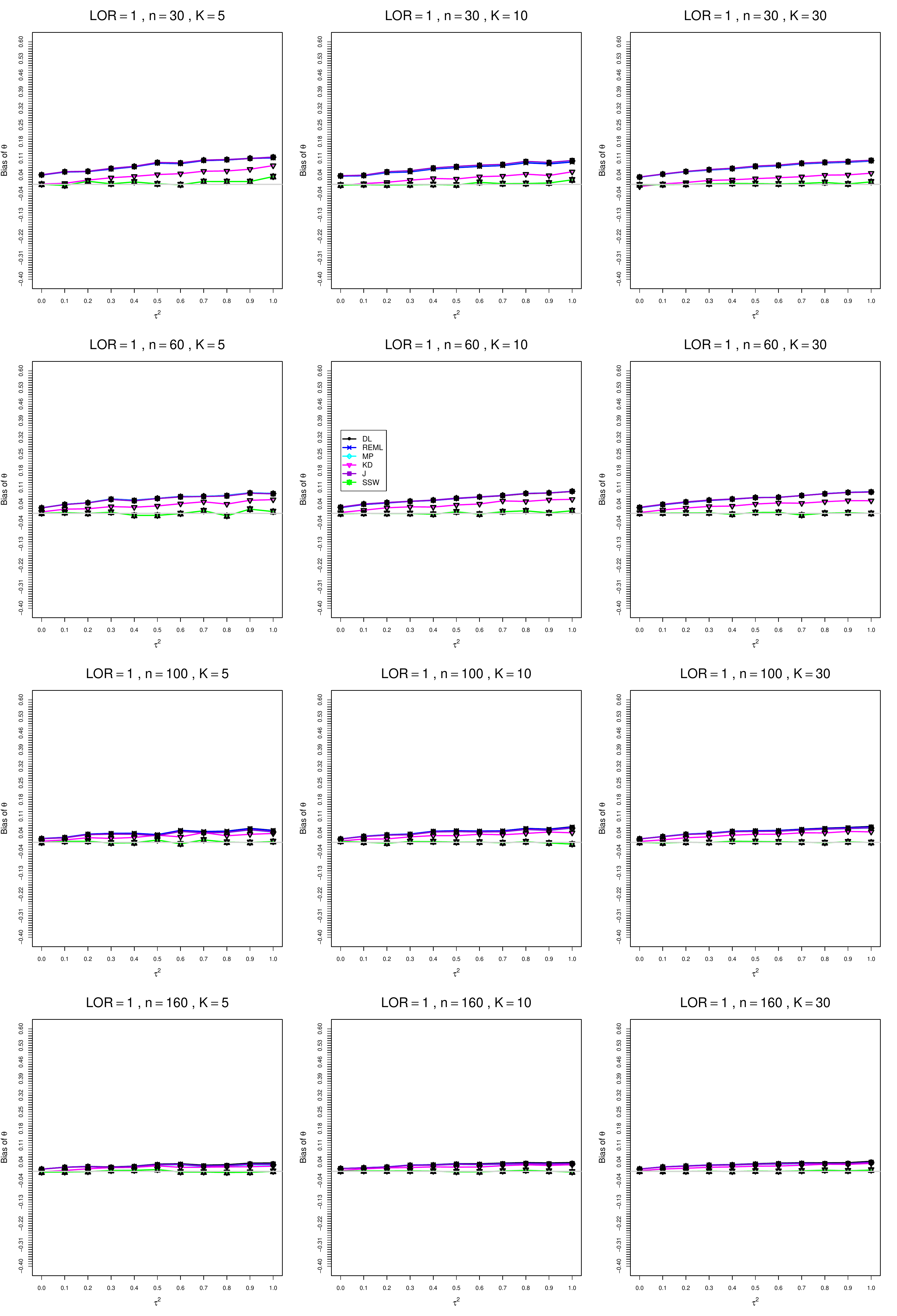}
	\caption{Bias of the estimation of  overall effect measure $\theta$ for $\theta=1$, $p_{iC}=0.2$, $q=0.75$, 
		unequal sample sizes $n=30,\; 60,\;100,\;160$. 
		\label{BiasThetaLOR1q075piC02_unequal_sample_sizes}}
\end{figure}

\begin{figure}[t]\centering
	\includegraphics[scale=0.35]{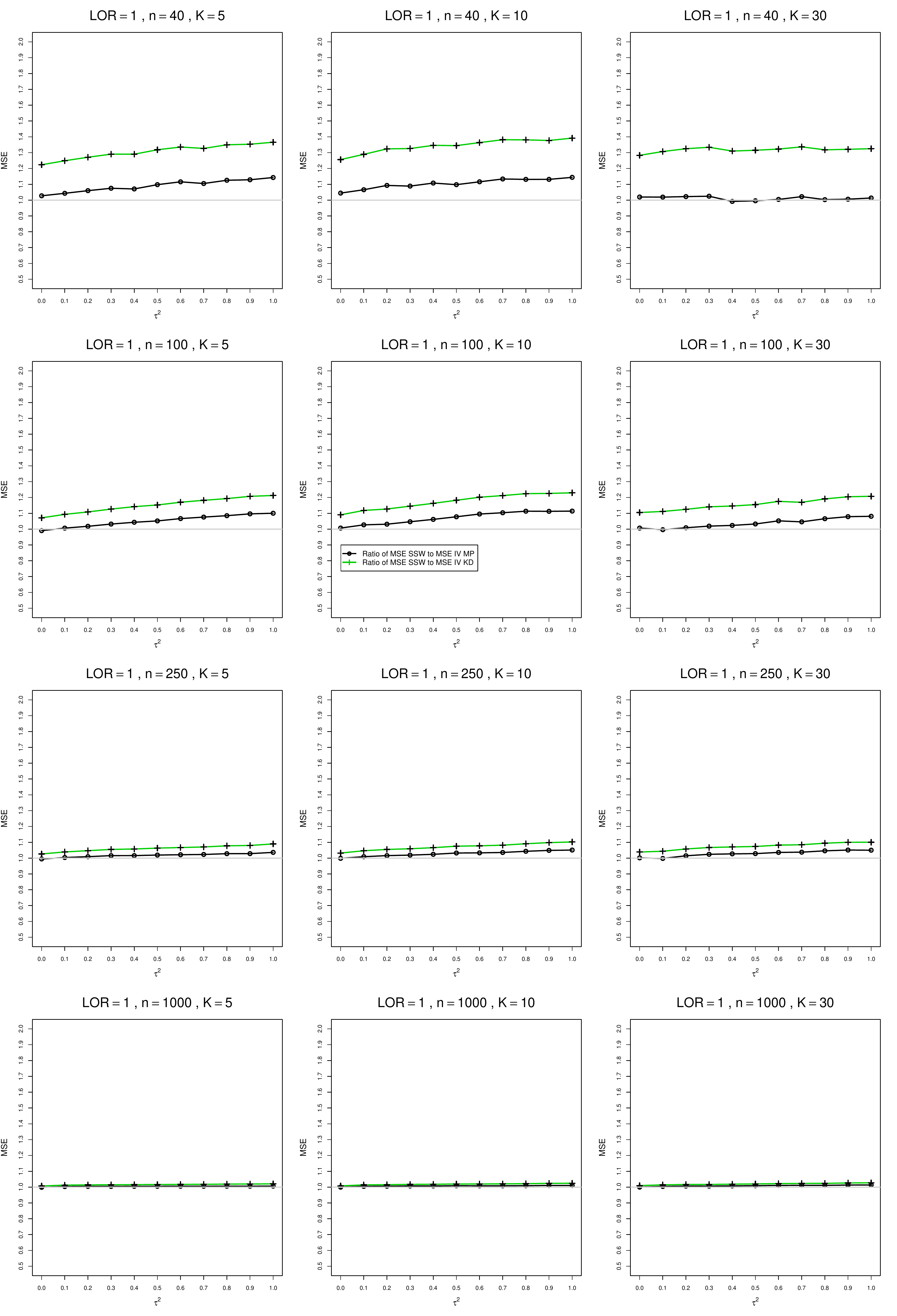}
	\caption{Ratio of mean squared errors of the fixed-weights to mean squared errors of inverse-variance estimator for $\theta=1$, $p_{iC}=0.2$, $q=0.75$, equal sample sizes $n=40,\;100,\;250,\;1000$. 
		\label{RatioOfMSEwithLOR1q075piC02fromMPandCMP}}
\end{figure}

\begin{figure}[t]\centering
	\includegraphics[scale=0.35]{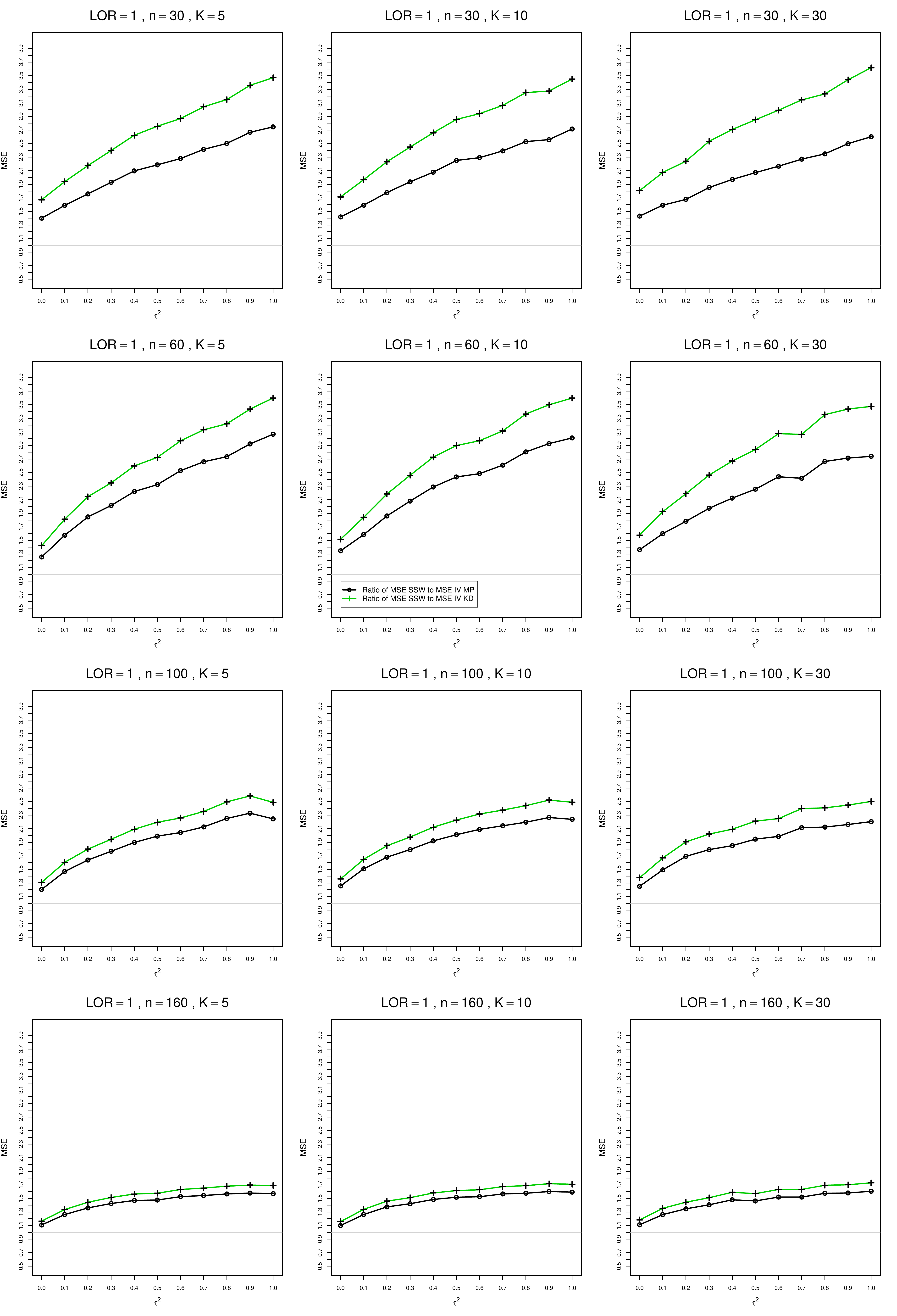}
	\caption{Ratio of mean squared errors of the fixed-weights to mean squared errors of inverse-variance estimator for $\theta=1$, $p_{iC}=0.2$, $q=0.75$, unequal sample sizes $n=30,\;60,\;100,\;160$. 
		\label{RatioOfMSEwithLOR1q075piC02fromMPandCMP_unequal_sample_sizes}}
\end{figure}

\begin{figure}[t]
	\centering
	\includegraphics[scale=0.33]{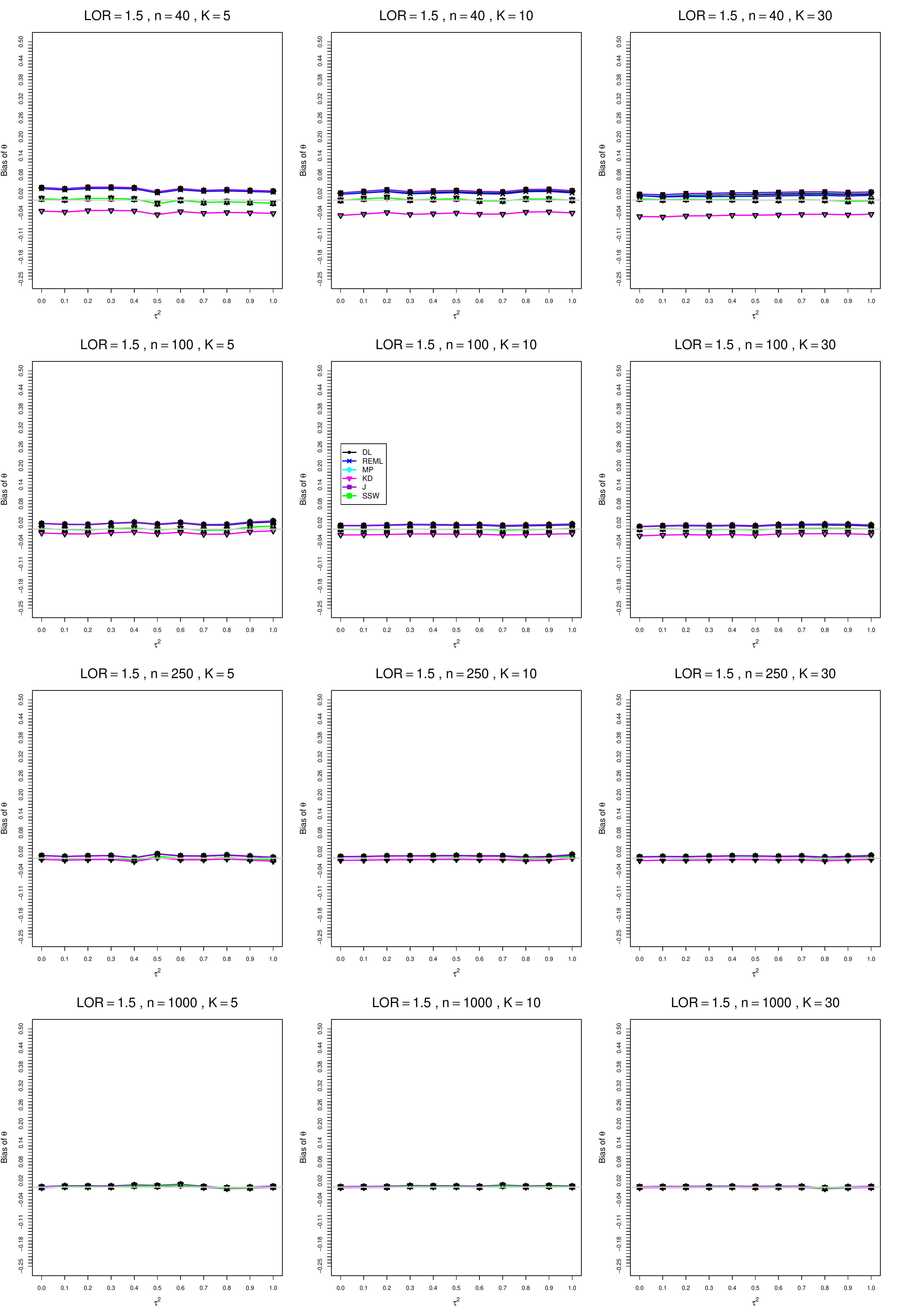}
	\caption{Bias of the estimation of  overall effect measure $\theta$ for $\theta=1.5$, $p_{iC}=0.2$, $q=0.75$, equal sample sizes $n=40,\;100,\;250,\;1000$. 
		\label{BiasThetaLOR15q075piC02}}
\end{figure}

\begin{figure}[t]
	\centering
	\includegraphics[scale=0.33]{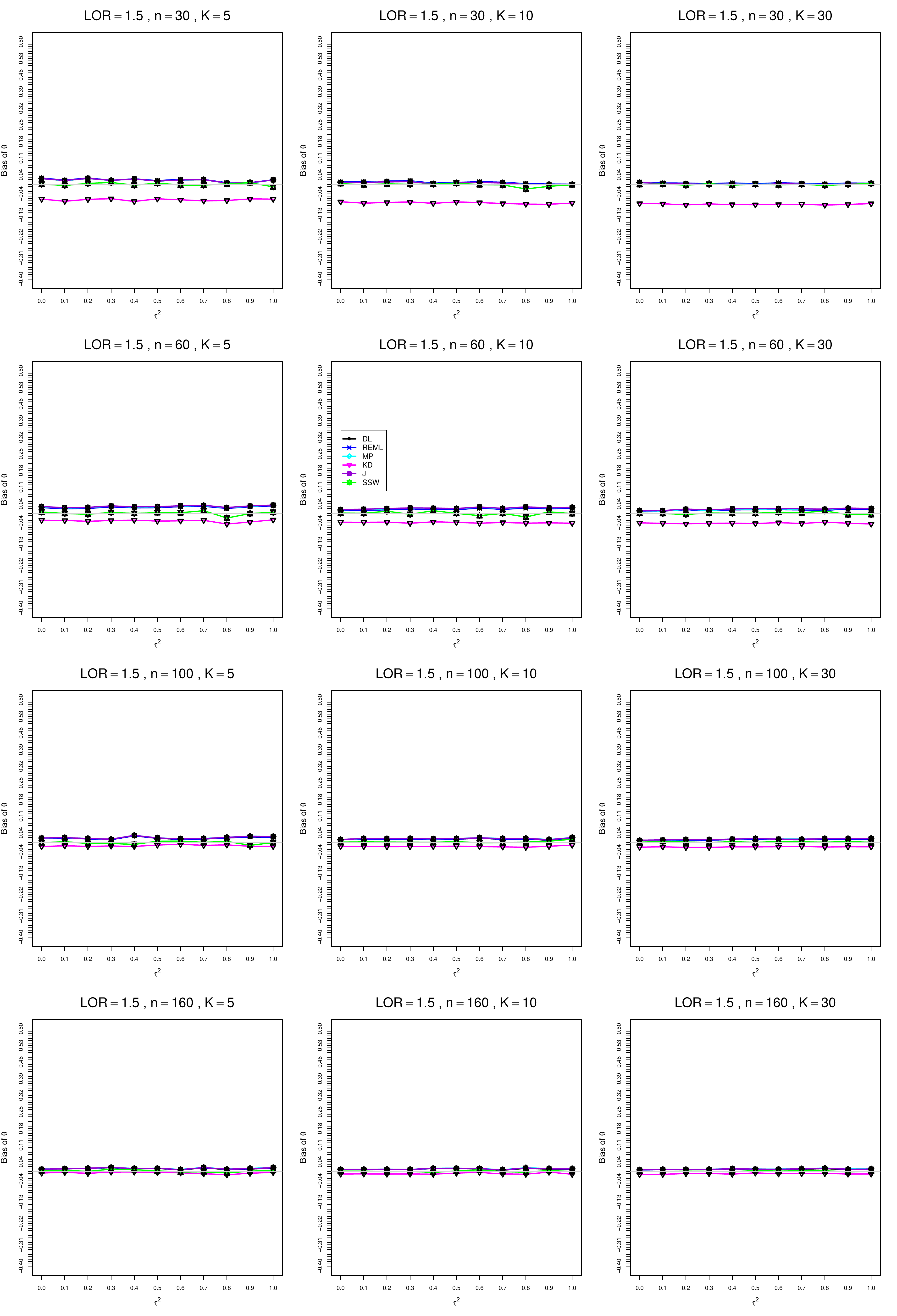}
	\caption{Bias of the estimation of  overall effect measure $\theta$ for $\theta=1.5$, $p_{iC}=0.2$, $q=0.75$, 
		unequal sample sizes $n=30,\; 60,\;100,\;160$. 
		\label{BiasThetaLOR15q075piC02_unequal_sample_sizes}}
\end{figure}

\begin{figure}[t]\centering
	\includegraphics[scale=0.35]{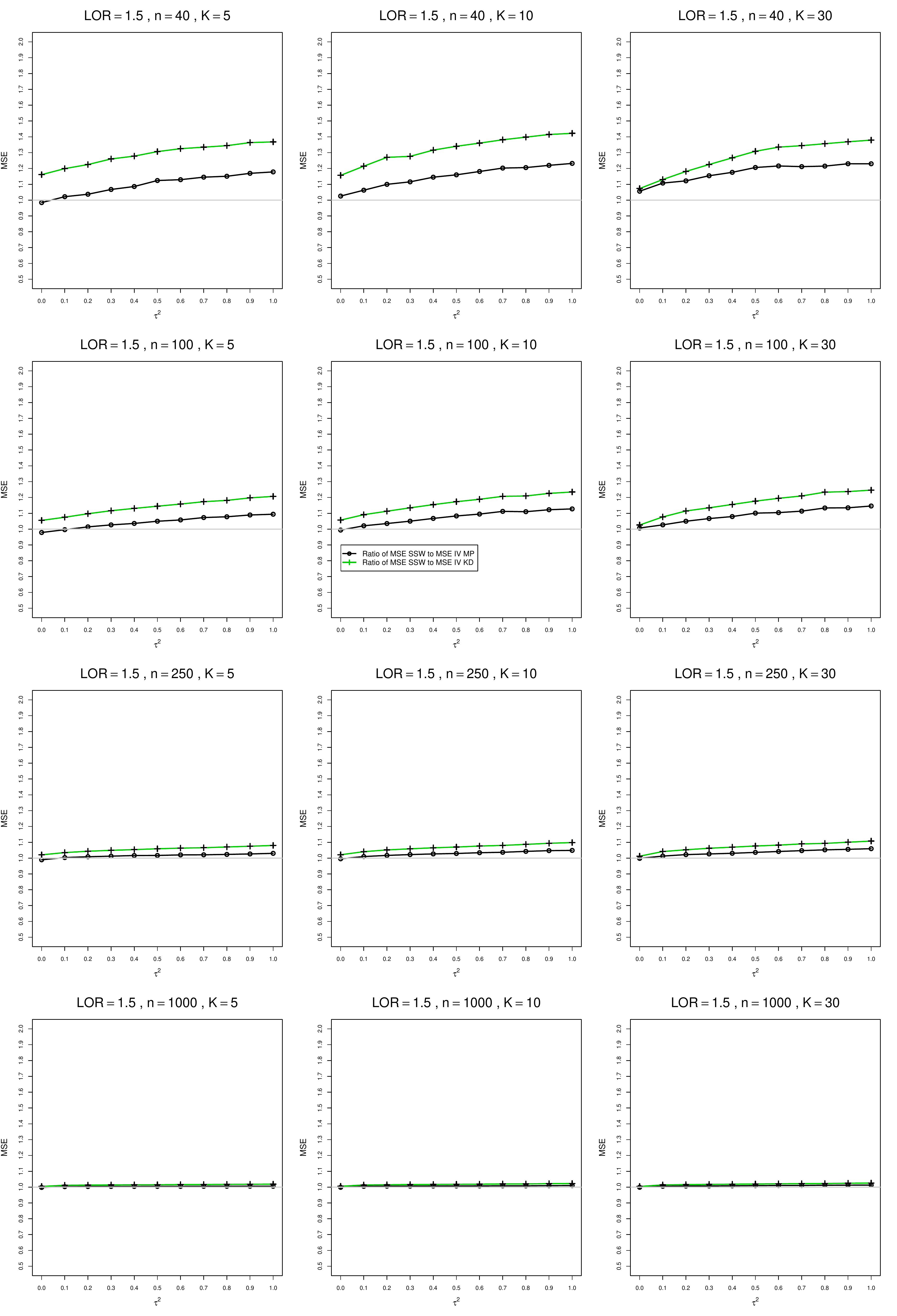}
	\caption{Ratio of mean squared errors of the fixed-weights to mean squared errors of inverse-variance estimator for $\theta=1.5$, $p_{iC}=0.2$, $q=0.75$, equal sample sizes $n=40,\;100,\;250,\;1000$. 
		\label{RatioOfMSEwithLOR15q075piC02fromMPandCMP}}
\end{figure}

\begin{figure}[t]\centering
	\includegraphics[scale=0.35]{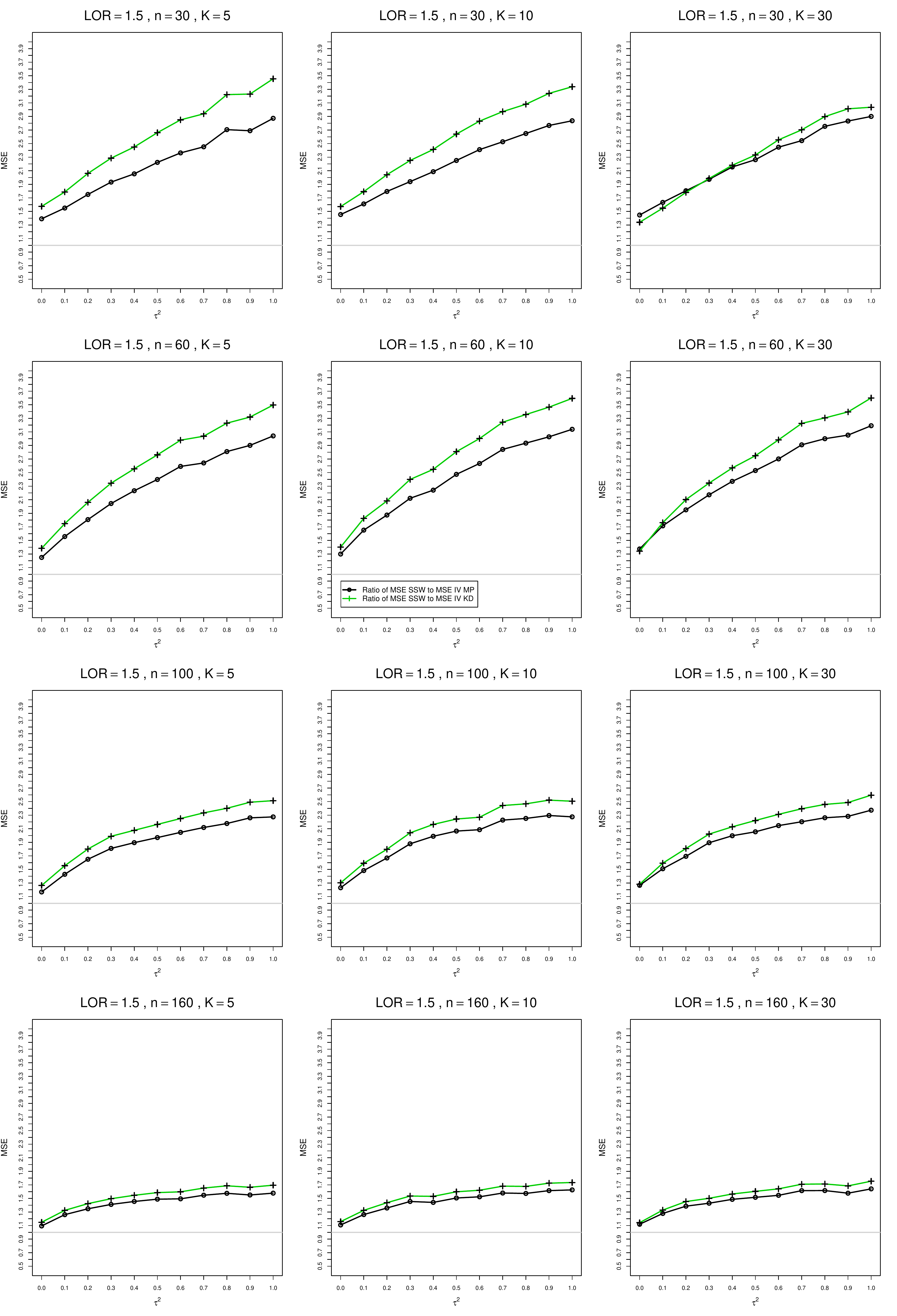}
	\caption{Ratio of mean squared errors of the fixed-weights to mean squared errors of inverse-variance estimator for $\theta=1.5$, $p_{iC}=0.2$, $q=0.75$, unequal sample sizes $n=30,\;60,\;100,\;160$. 
		\label{RatioOfMSEwithLOR15q075piC02fromMPandCMP_unequal_sample_sizes}}
\end{figure}

\begin{figure}[t]
	\centering
	\includegraphics[scale=0.33]{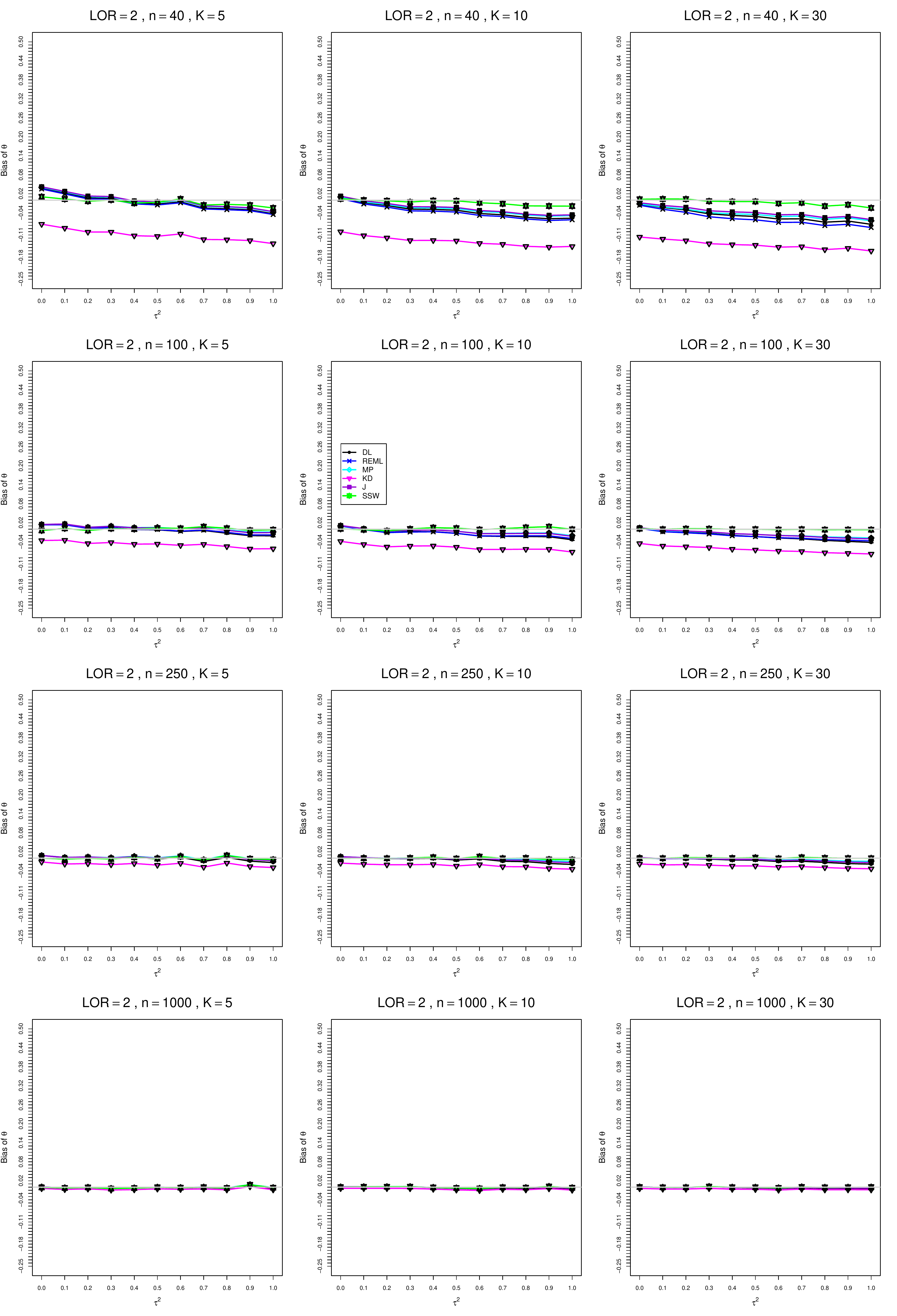}
	\caption{Bias of the estimation of  overall effect measure $\theta$ for $\theta=2$, $p_{iC}=0.2$, $q=0.75$, equal sample sizes $n=40,\;100,\;250,\;1000$. 
		\label{BiasThetaLOR2q075piC02}}
\end{figure}

\begin{figure}[t]
	\centering
	\includegraphics[scale=0.33]{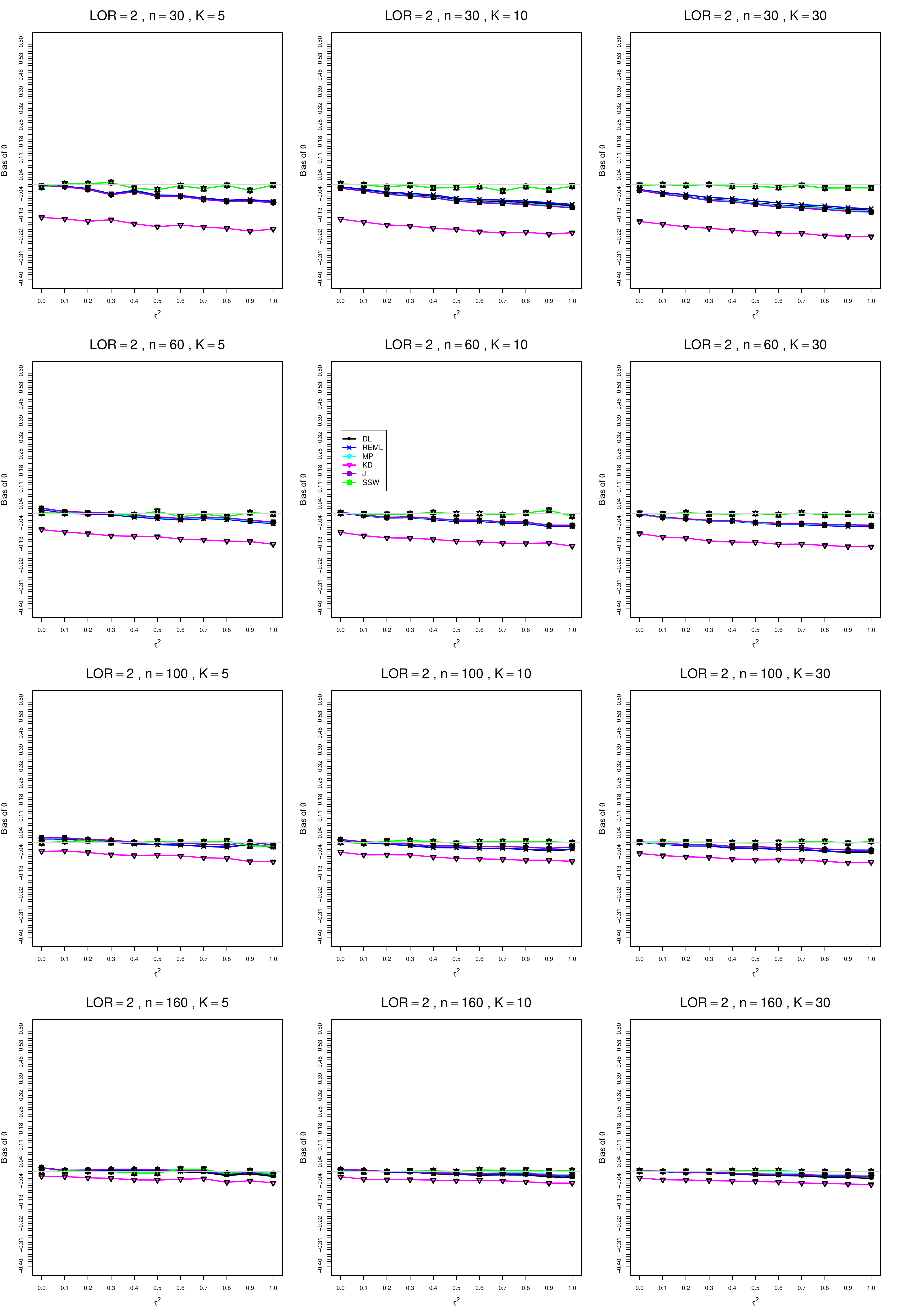}
	\caption{Bias of the estimation of  overall effect measure $\theta$ for $\theta=2$, $p_{iC}=0.2$, $q=0.75$, 
		unequal sample sizes $n=30,\; 60,\;100,\;160$. 
		\label{BiasThetaLOR2q075piC02_unequal_sample_sizes}}
\end{figure}

\begin{figure}[t]\centering
	\includegraphics[scale=0.35]{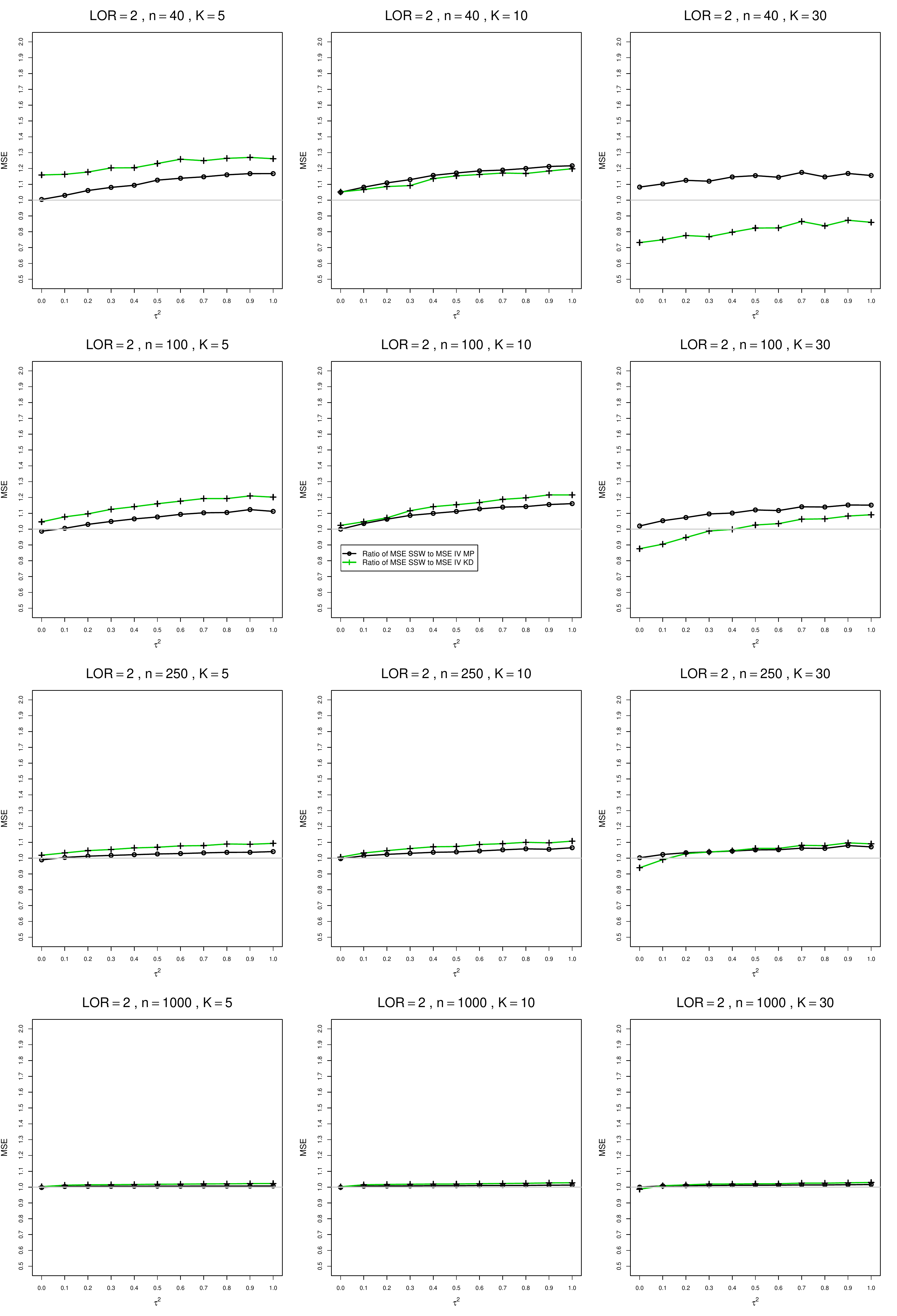}
	\caption{Ratio of mean squared errors of the fixed-weights to mean squared errors of inverse-variance estimator for $\theta=2$, $p_{iC}=0.2$, $q=0.75$, equal sample sizes $n=40,\;100,\;250,\;1000$. 
		\label{RatioOfMSEwithLOR2q075piC02fromMPandCMP}}
\end{figure}

\begin{figure}[t]\centering
	\includegraphics[scale=0.35]{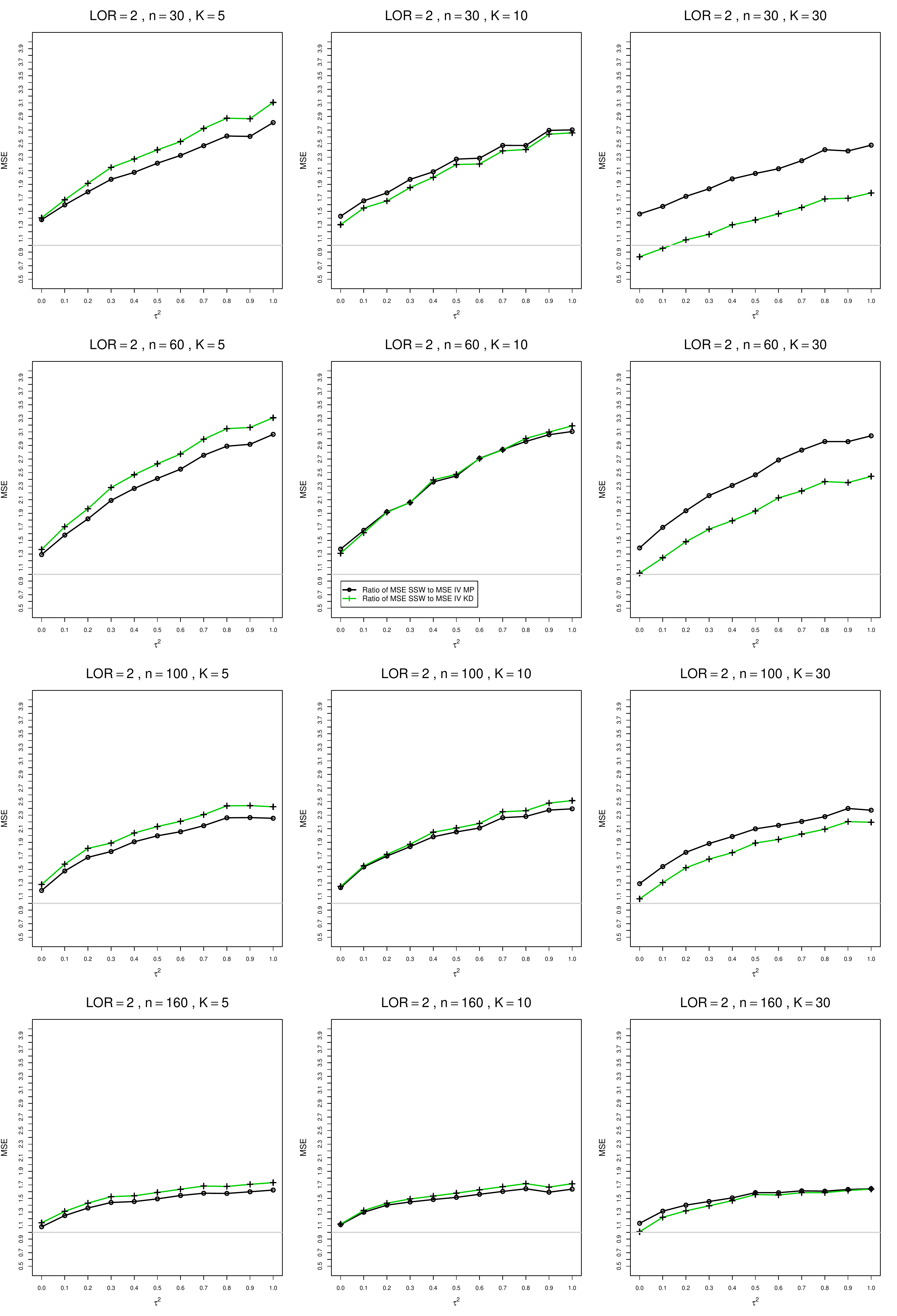}
	\caption{Ratio of mean squared errors of the fixed-weights to mean squared errors of inverse-variance estimator for $\theta=2$, $p_{iC}=0.2$, $q=0.75$, unequal sample sizes $n=30,\;60,\;100,\;160$. 
		\label{RatioOfMSEwithLOR2q075piC02fromMPandCMP_unequal_sample_sizes}}
\end{figure}

\clearpage
\renewcommand{\thefigure}{B1.3.\arabic{figure}}
\setcounter{figure}{0}
\subsection*{B1.3 Probability in the control arm $p_{C}=0.4$}
\begin{figure}[t]
	\centering
	\includegraphics[scale=0.33]{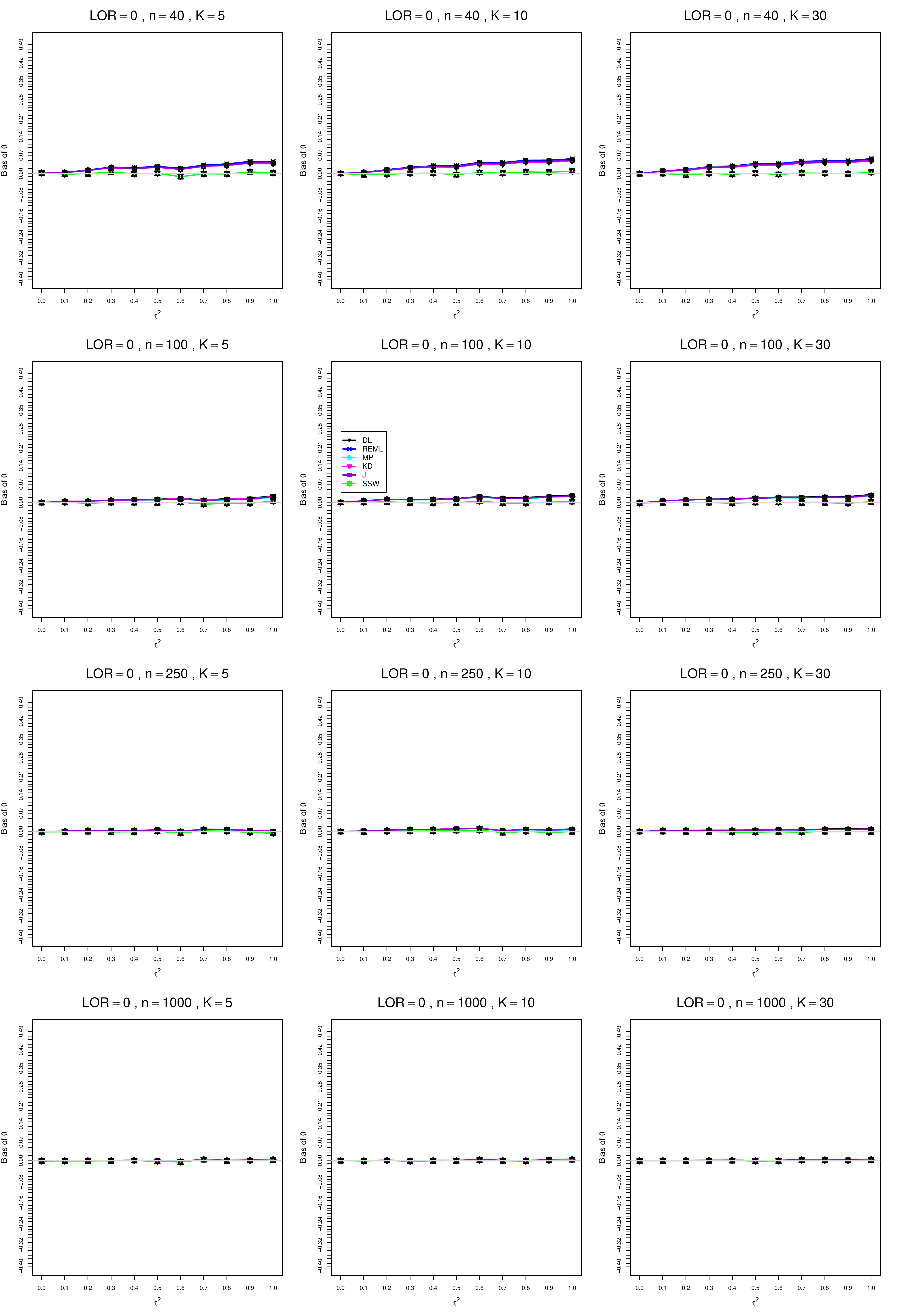}
	\caption{Bias of the estimation of  overall effect measure $\theta$ for $\theta=0$, $p_{iC}=0.4$, $q=0.5$, equal sample sizes $n=40,\;100,\;250,\;1000$. 
		\label{BiasThetaLOR0q05piC04}}
\end{figure}

\begin{figure}[t]
	\centering
	\includegraphics[scale=0.33]{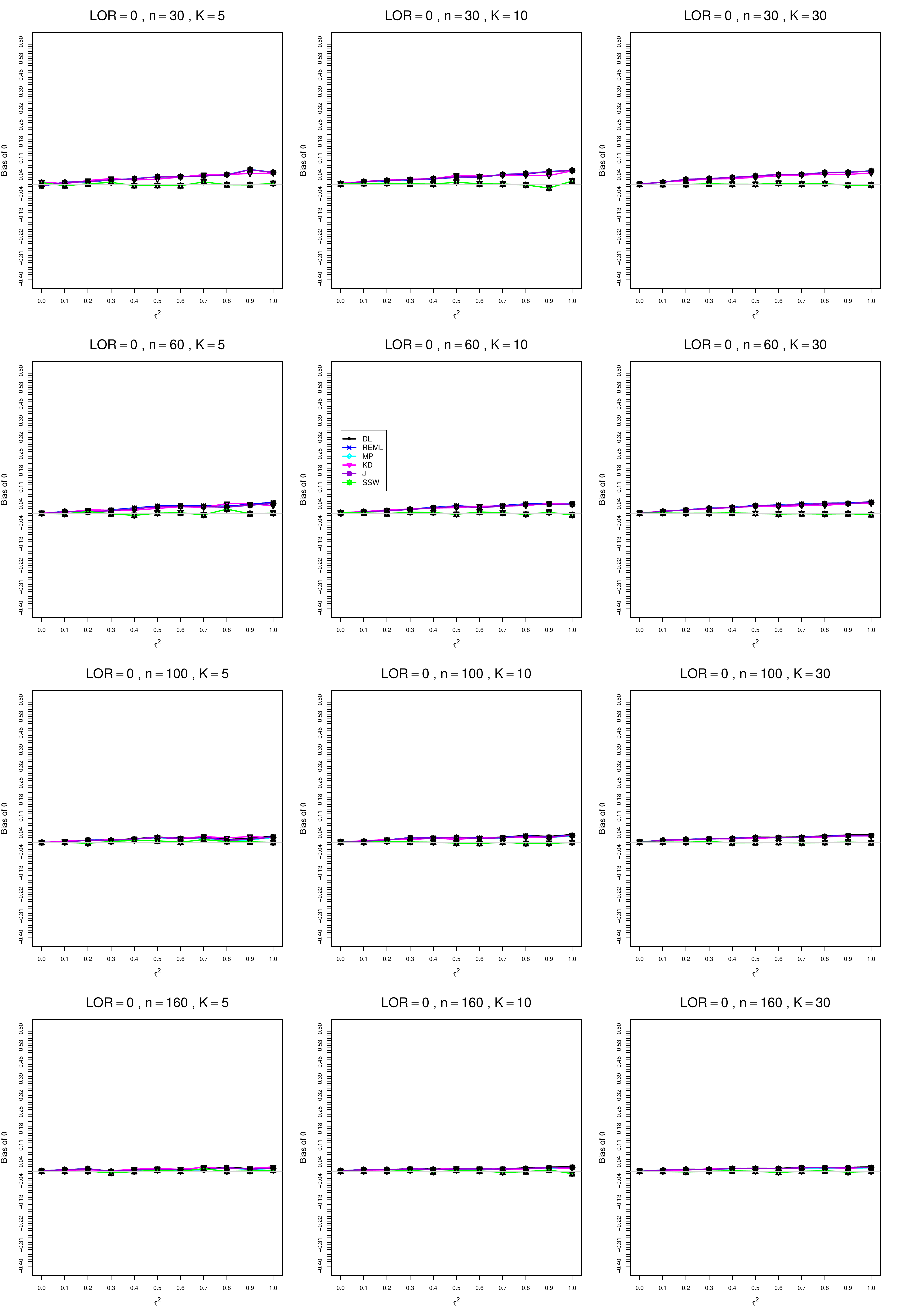}
	\caption{Bias of the estimation of  overall effect measure $\theta$ for $\theta=0$, $p_{iC}=0.4$, $q=0.5$, 
		unequal sample sizes $n=30,\; 60,\;100,\;160$. 
		\label{BiasThetaLOR0q05piC04_unequal_sample_sizes}}
\end{figure}

\begin{figure}[t]\centering
	\includegraphics[scale=0.35]{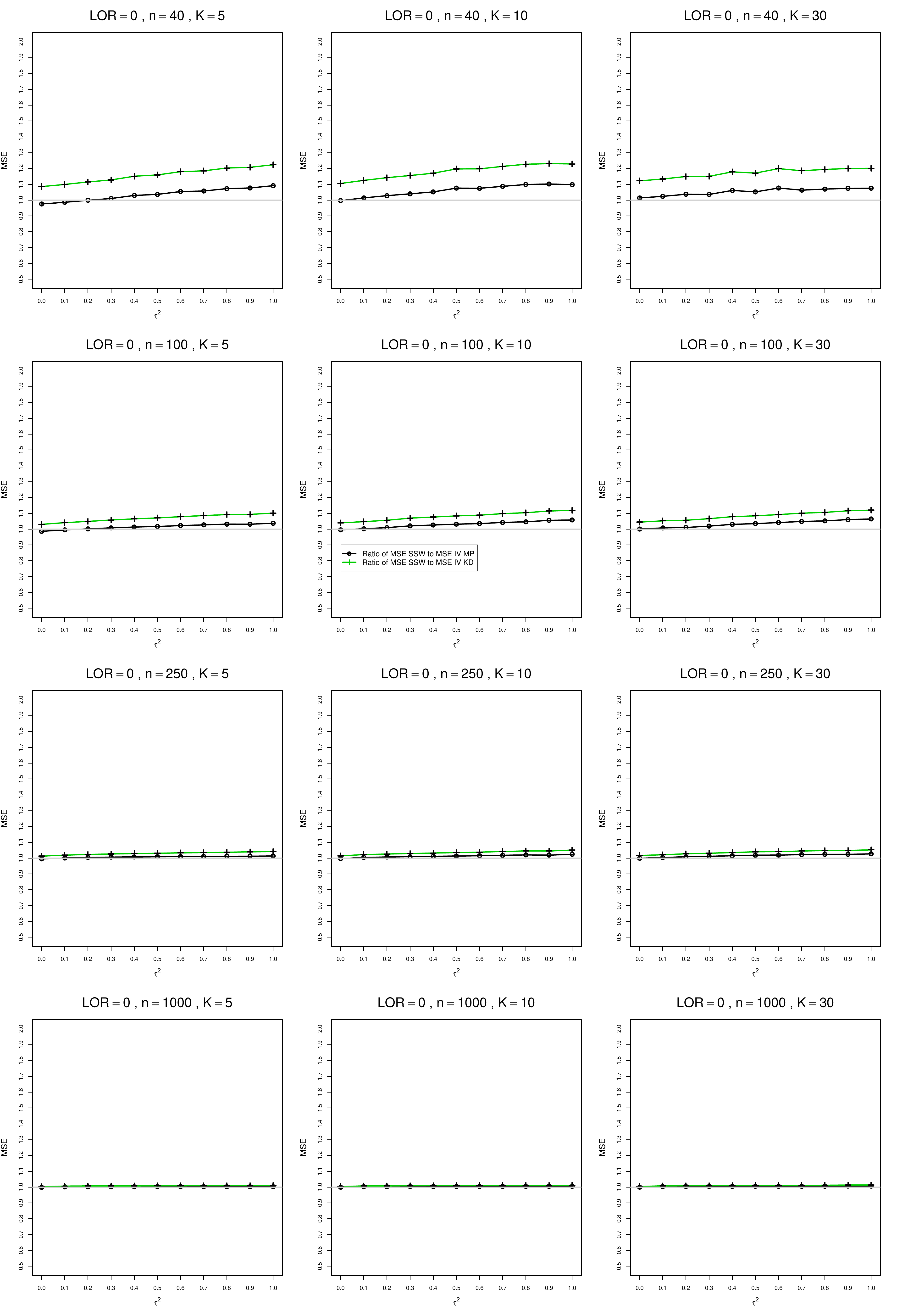}
	\caption{Ratio of mean squared errors of the fixed-weights to mean squared errors of inverse-variance estimator for $\theta=0$, $p_{iC}=0.4$, $q=0.5$, equal sample sizes $n=40,\;100,\;250,\;1000$. 
		\label{RatioOfMSEwithLOR0q05piC04fromMPandCMP}}
\end{figure}

\begin{figure}[t]\centering
	\includegraphics[scale=0.35]{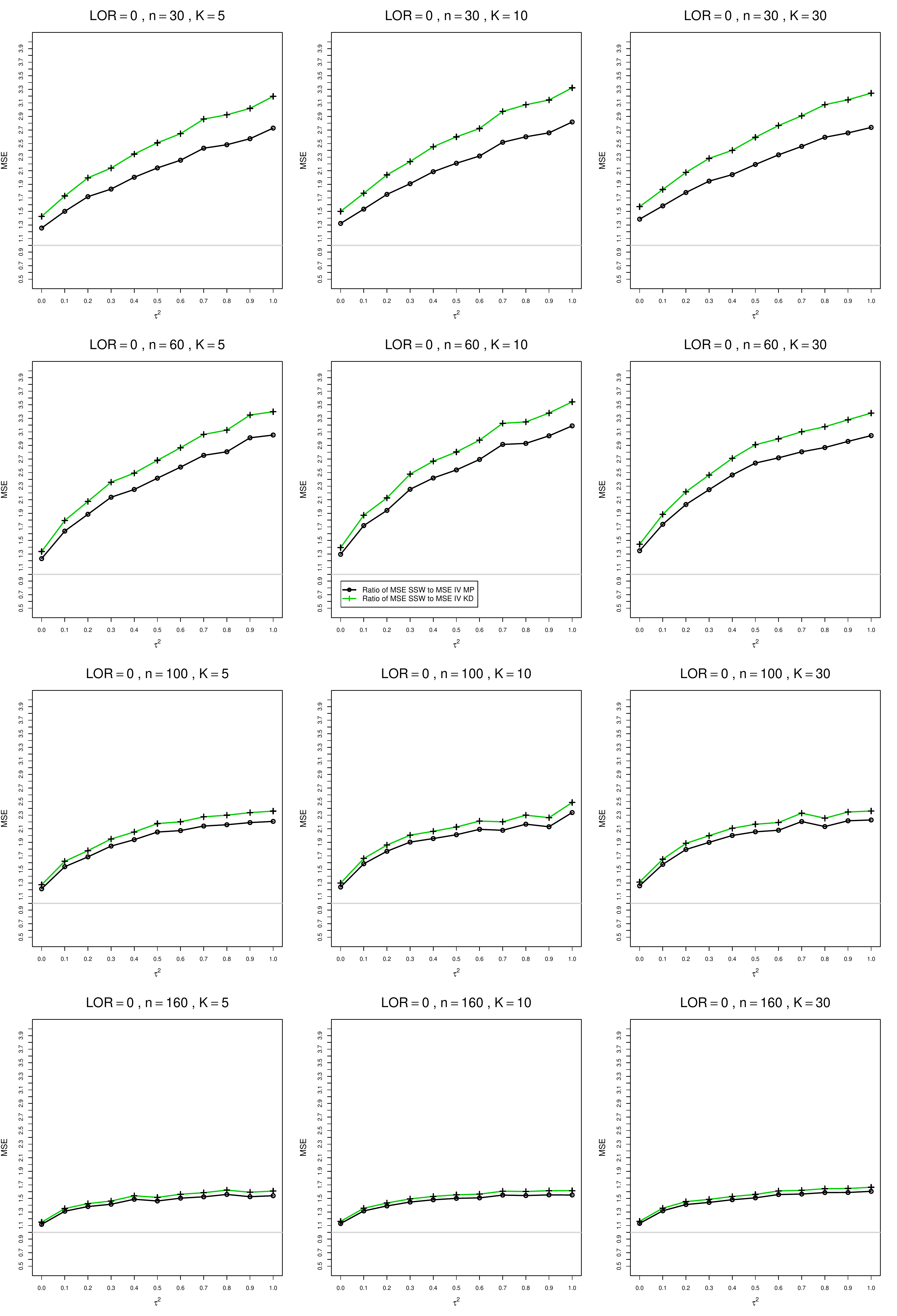}
	\caption{Ratio of mean squared errors of the fixed-weights to mean squared errors of inverse-variance estimator for $\theta=0$, $p_{iC}=0.4$, $q=0.5$, unequal sample sizes $n=30,\;60,\;100,\;160$. 
		\label{RatioOfMSEwithLOR0q05piC04fromMPandCMP_unequal_sample_sizes}}
\end{figure}


\begin{figure}[t]
	\centering
	\includegraphics[scale=0.33]{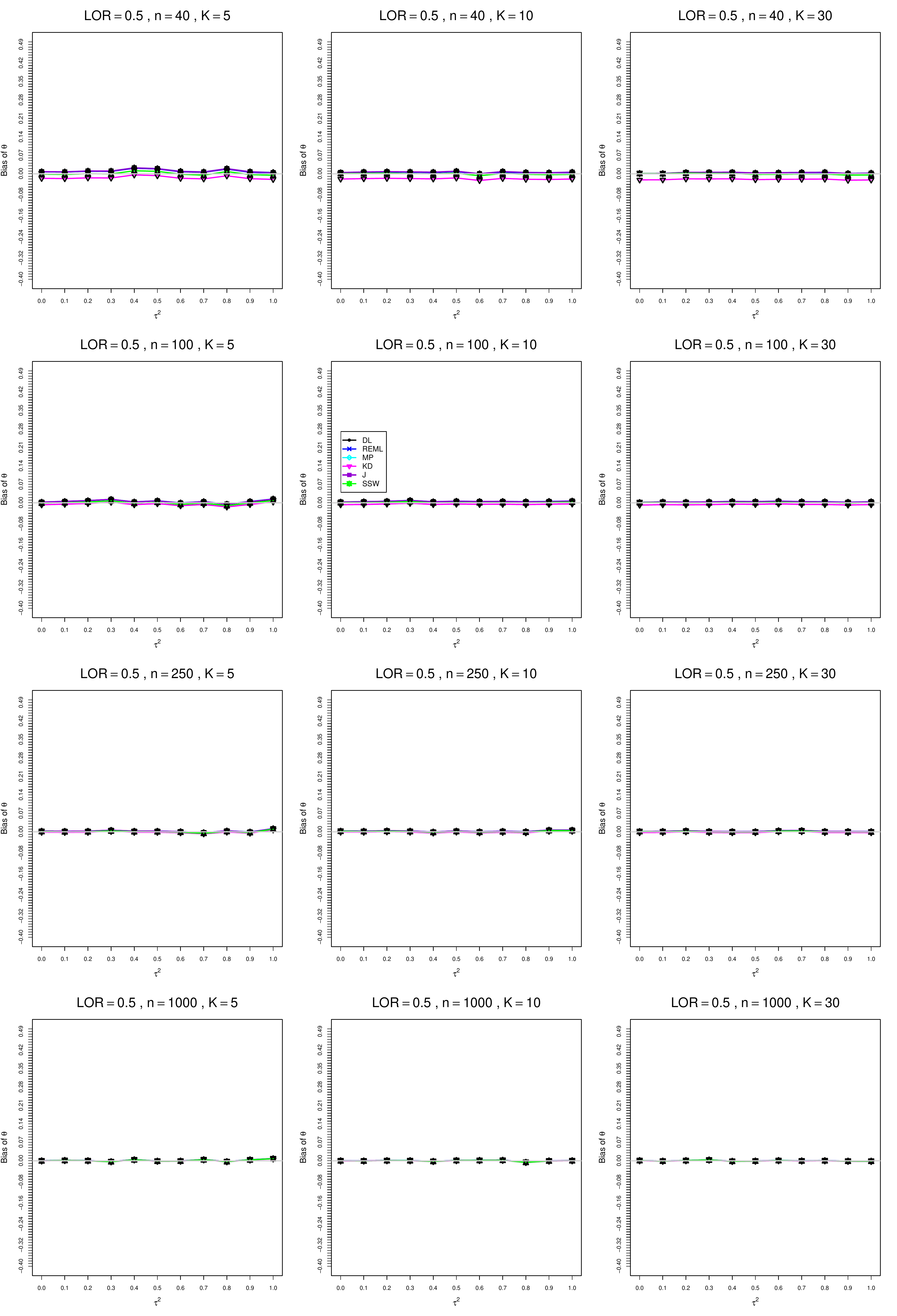}
	\caption{Bias of the estimation of  overall effect measure $\theta$ for $\theta=0.5$, $p_{iC}=0.4$, $q=0.5$, equal sample sizes $n=40,\;100,\;250,\;1000$. 
		\label{BiasThetaLOR05q05piC04}}
\end{figure}

\begin{figure}[t]
	\centering
	\includegraphics[scale=0.33]{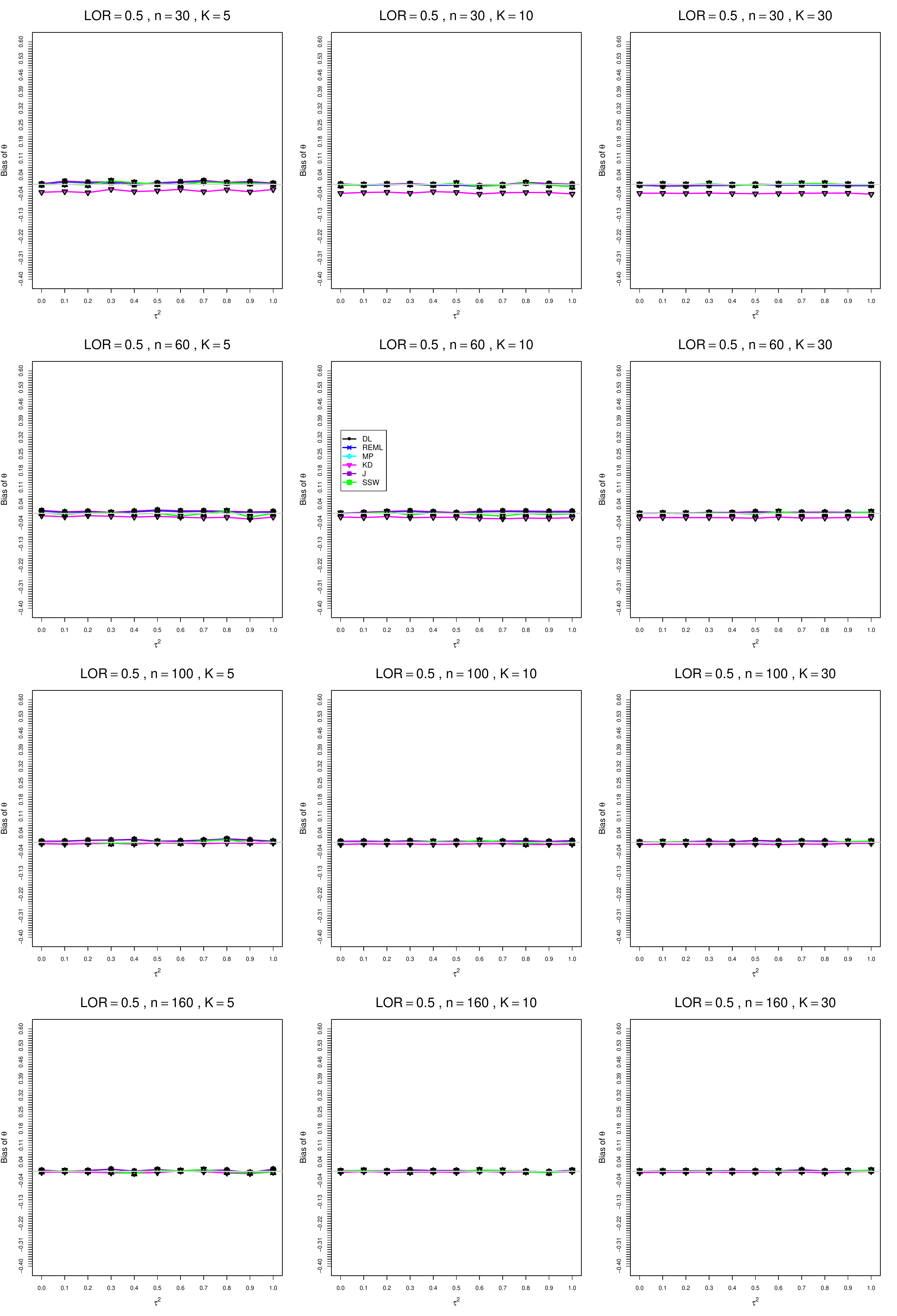}
	\caption{Bias of the estimation of  overall effect measure $\theta$ for $\theta=0.5$, $p_{iC}=0.4$, $q=0.5$, 
		unequal sample sizes $n=30,\; 60,\;100,\;160$. 
		\label{BiasThetaLOR05q05piC04_unequal_sample_sizes}}
\end{figure}

\begin{figure}[t]\centering
	\includegraphics[scale=0.35]{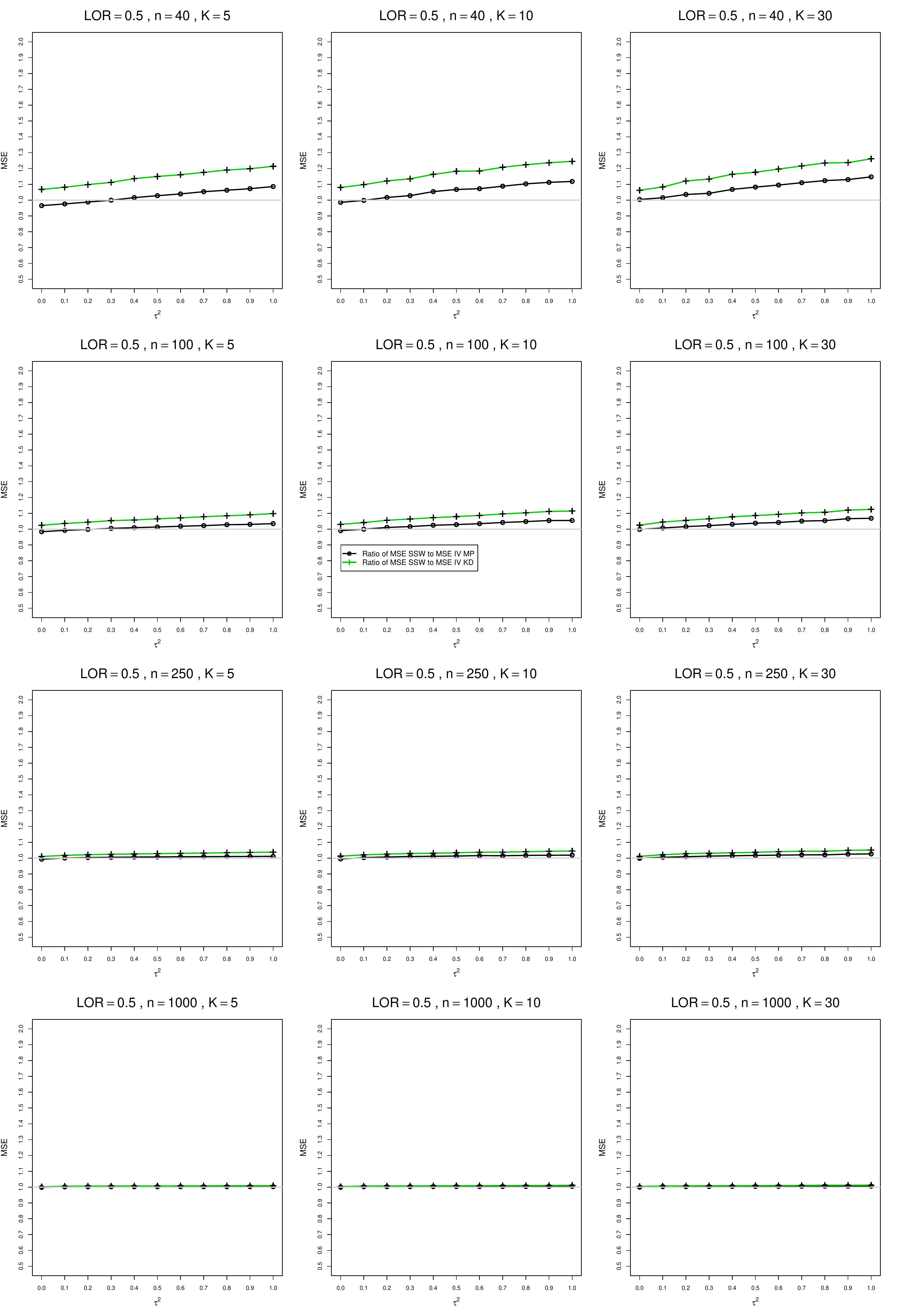}
	\caption{Ratio of mean squared errors of the fixed-weights to mean squared errors of inverse-variance estimator for $\theta=0.5$, $p_{iC}=0.4$, $q=0.5$, equal sample sizes $n=40,\;100,\;250,\;1000$. 
		\label{RatioOfMSEwithLOR05q05piC04fromMPandCMP}}
\end{figure}

\begin{figure}[t]\centering
	\includegraphics[scale=0.35]{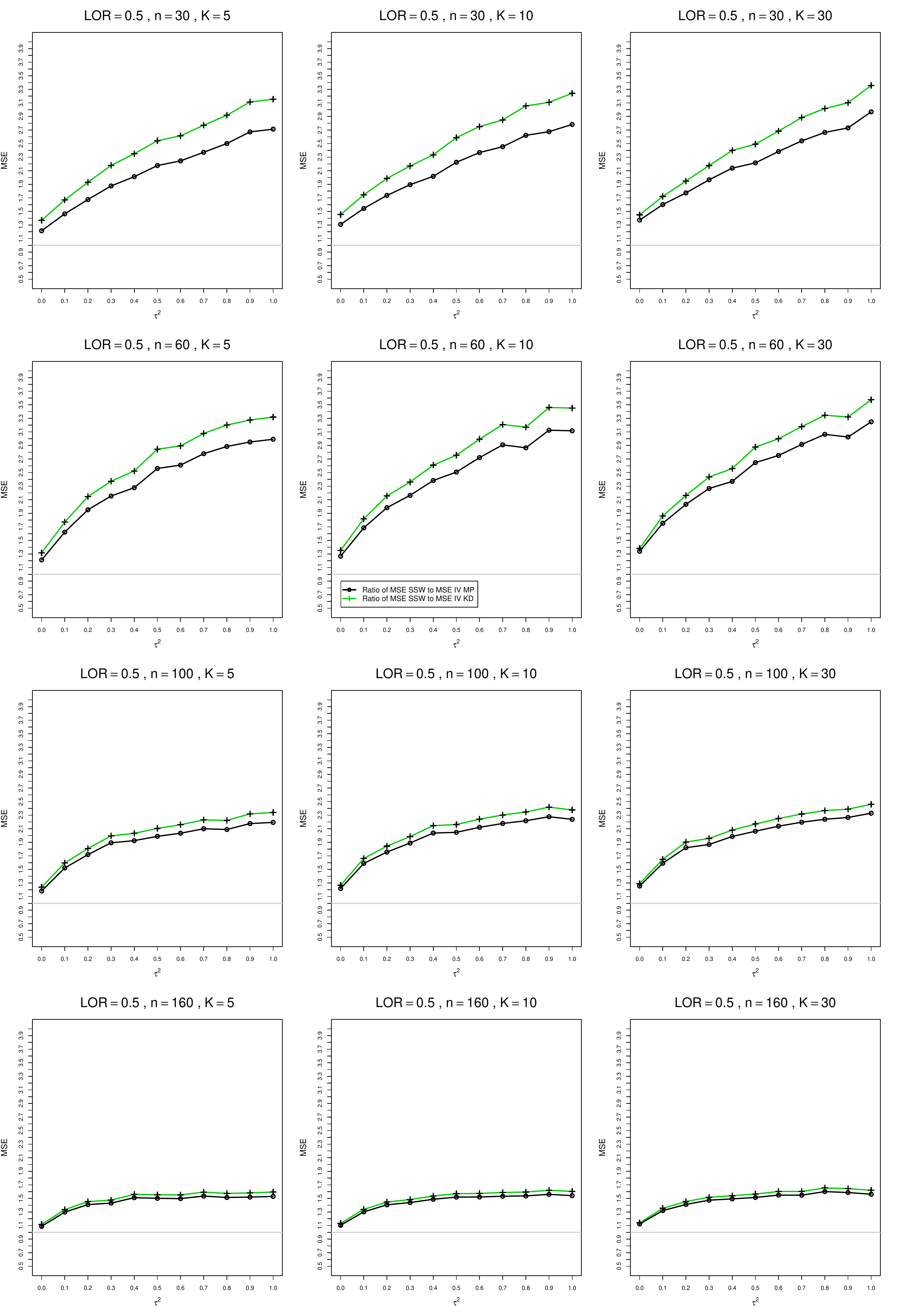}
	\caption{Ratio of mean squared errors of the fixed-weights to mean squared errors of inverse-variance estimator for $\theta=0.5$, $p_{iC}=0.4$, $q=0.5$, unequal sample sizes $n=30,\;60,\;100,\;160$. 
		\label{RatioOfMSEwithLOR05q05piC04fromMPandCMP_unequal_sample_sizes}}
\end{figure}

\begin{figure}[t]
	\centering
	\includegraphics[scale=0.33]{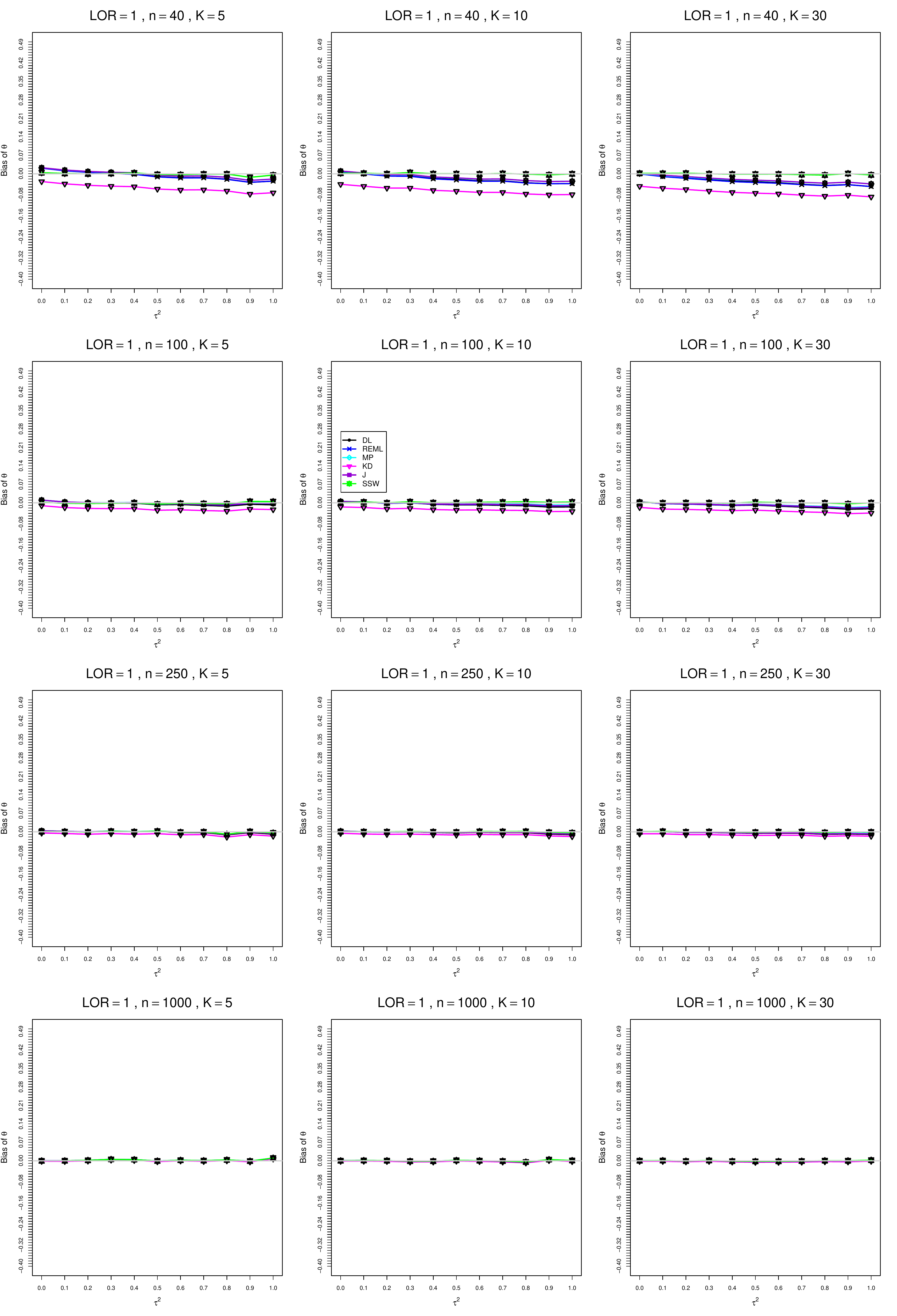}
	\caption{Bias of the estimation of  overall effect measure $\theta$ for $\theta=1$, $p_{iC}=0.4$, $q=0.5$, equal sample sizes $n=40,\;100,\;250,\;1000$. 
		\label{BiasThetaLOR1q05piC04}}
\end{figure}

\begin{figure}[t]
	\centering
	\includegraphics[scale=0.33]{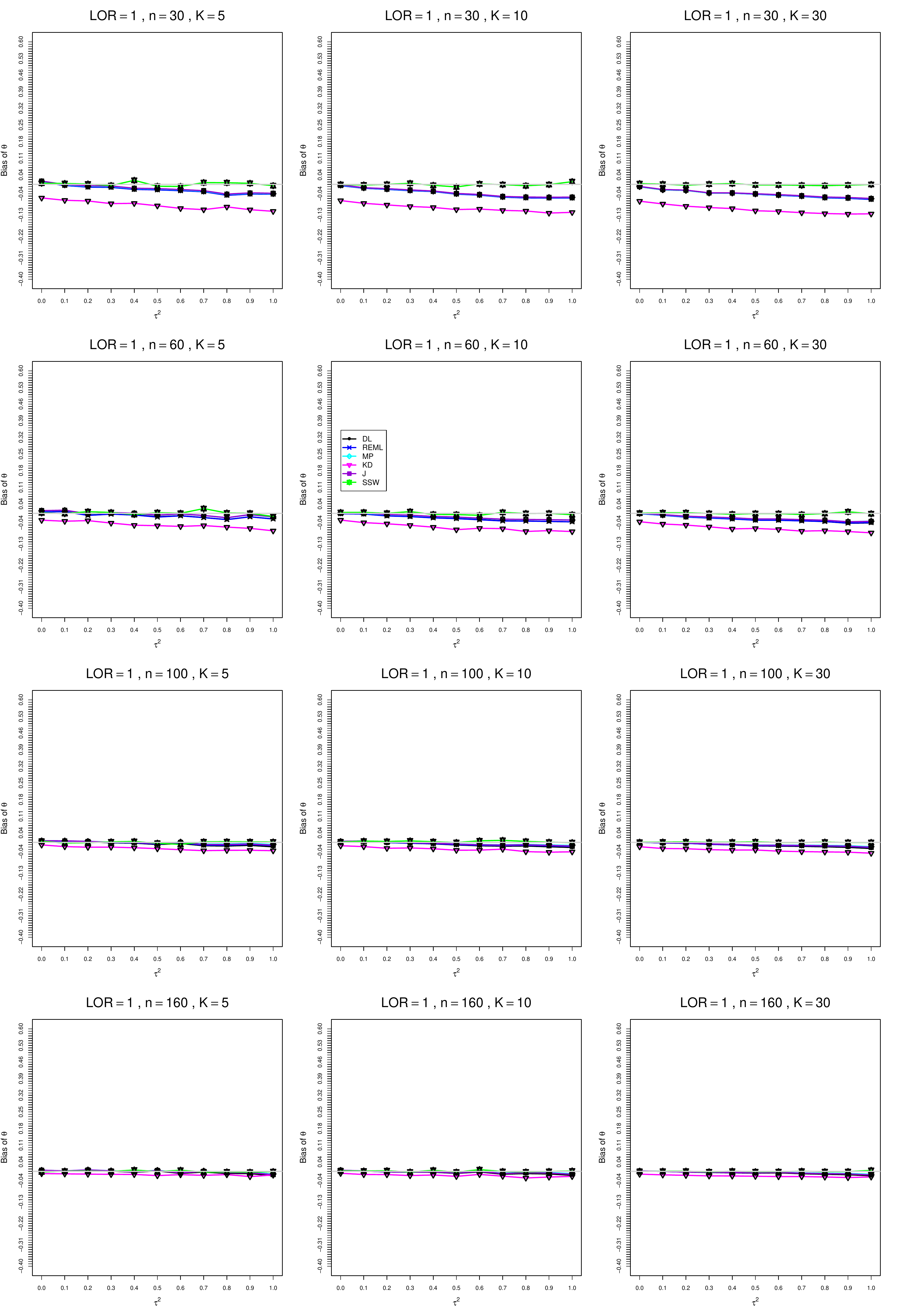}
	\caption{Bias of the estimation of  overall effect measure $\theta$ for $\theta=1$, $p_{iC}=0.4$, $q=0.5$, 
		unequal sample sizes $n=30,\; 60,\;100,\;160$. 
		\label{BiasThetaLOR1q05piC04_unequal_sample_sizes}}
\end{figure}

\begin{figure}[t]\centering
	\includegraphics[scale=0.35]{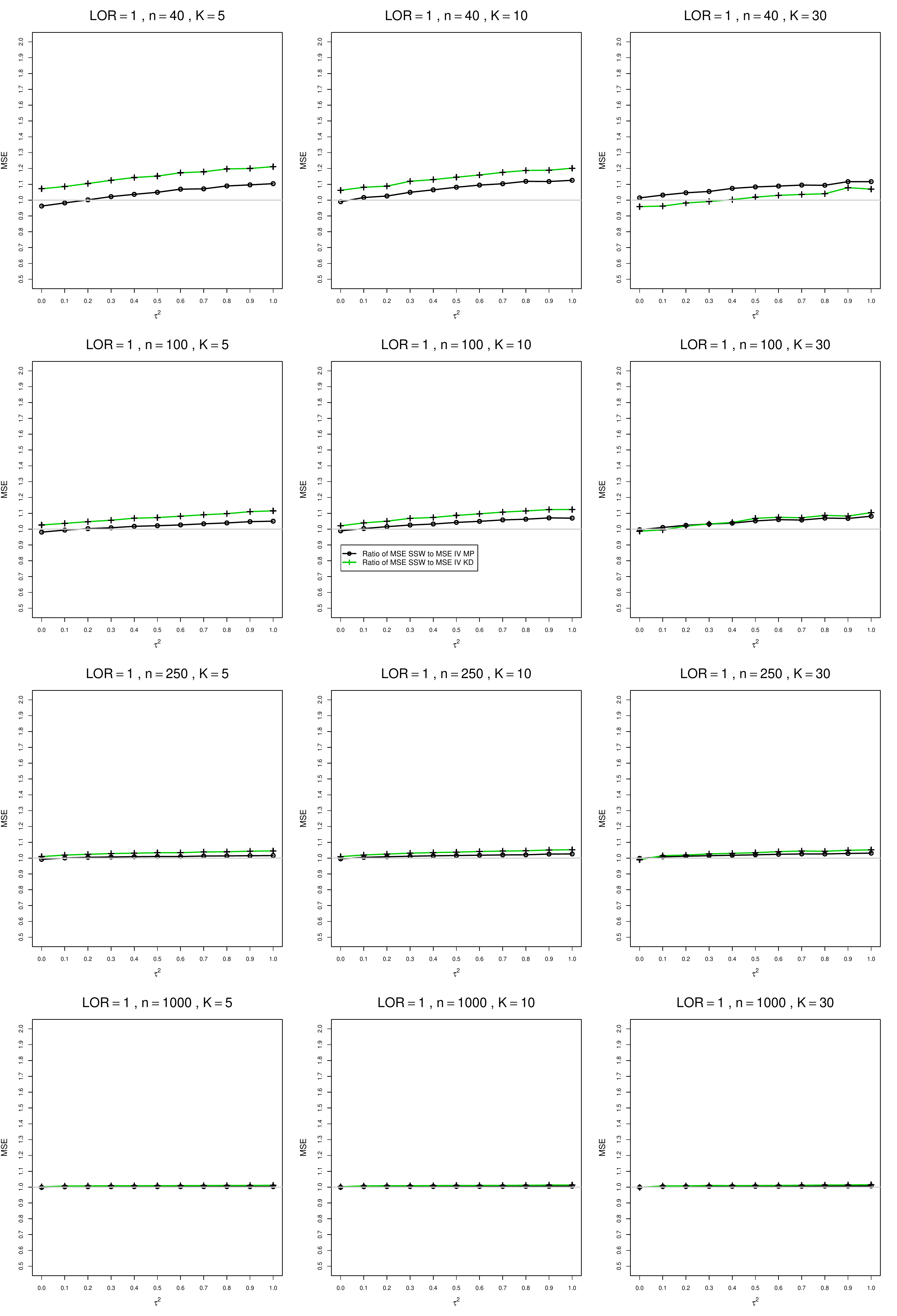}
	\caption{Ratio of mean squared errors of the fixed-weights to mean squared errors of inverse-variance estimator for $\theta=1$, $p_{iC}=0.4$, $q=0.5$, equal sample sizes $n=40,\;100,\;250,\;1000$. 
		\label{RatioOfMSEwithLOR1q05piC04fromMPandCMP}}
\end{figure}

\begin{figure}[t]\centering
	\includegraphics[scale=0.35]{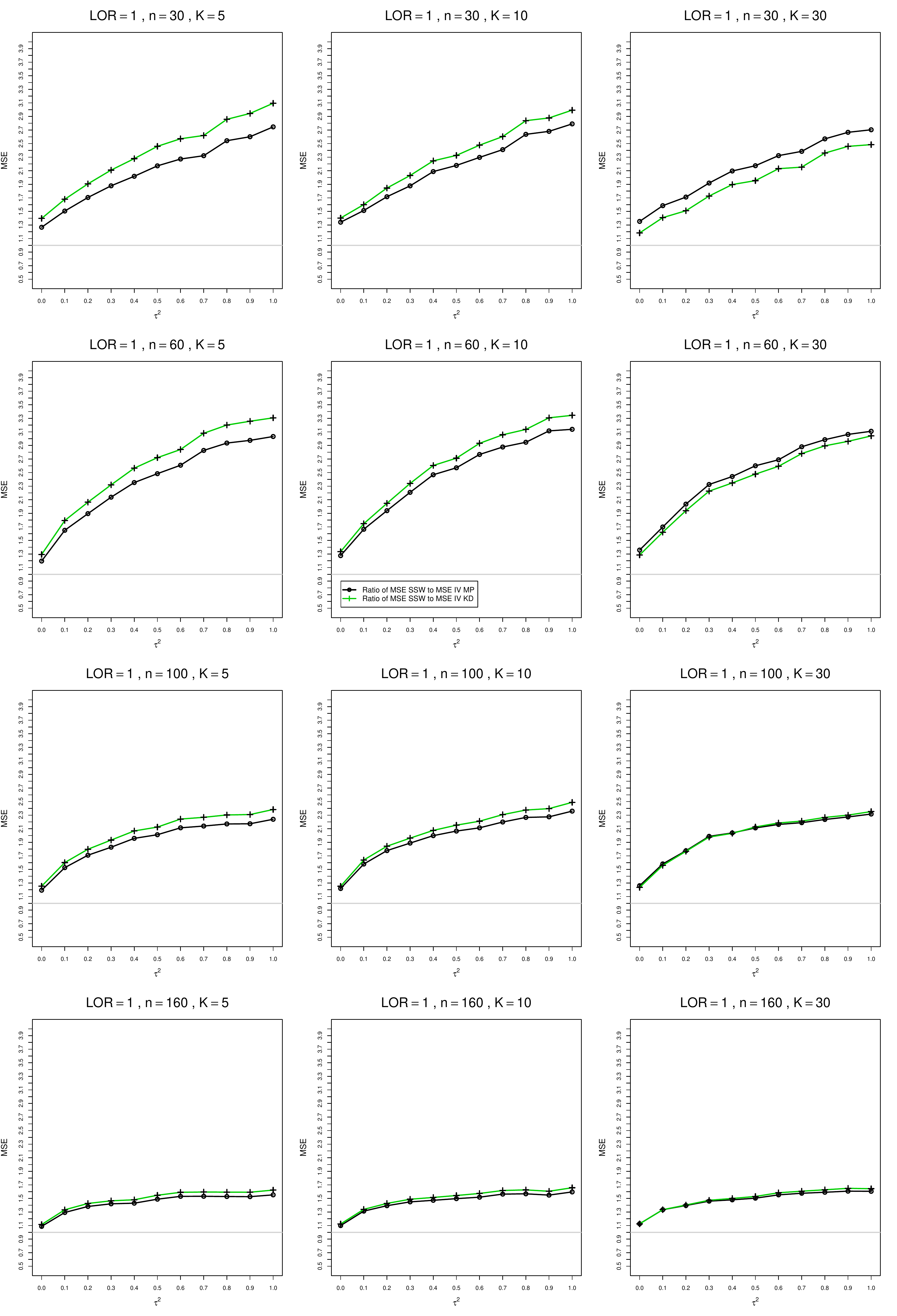}
	\caption{Ratio of mean squared errors of the fixed-weights to mean squared errors of inverse-variance estimator for $\theta=1$, $p_{iC}=0.4$, $q=0.5$, unequal sample sizes $n=30,\;60,\;100,\;160$. 
		\label{RatioOfMSEwithLOR1q05piC04fromMPandCMP_unequal_sample_sizes}}
\end{figure}


\begin{figure}[t]
	\centering
	\includegraphics[scale=0.33]{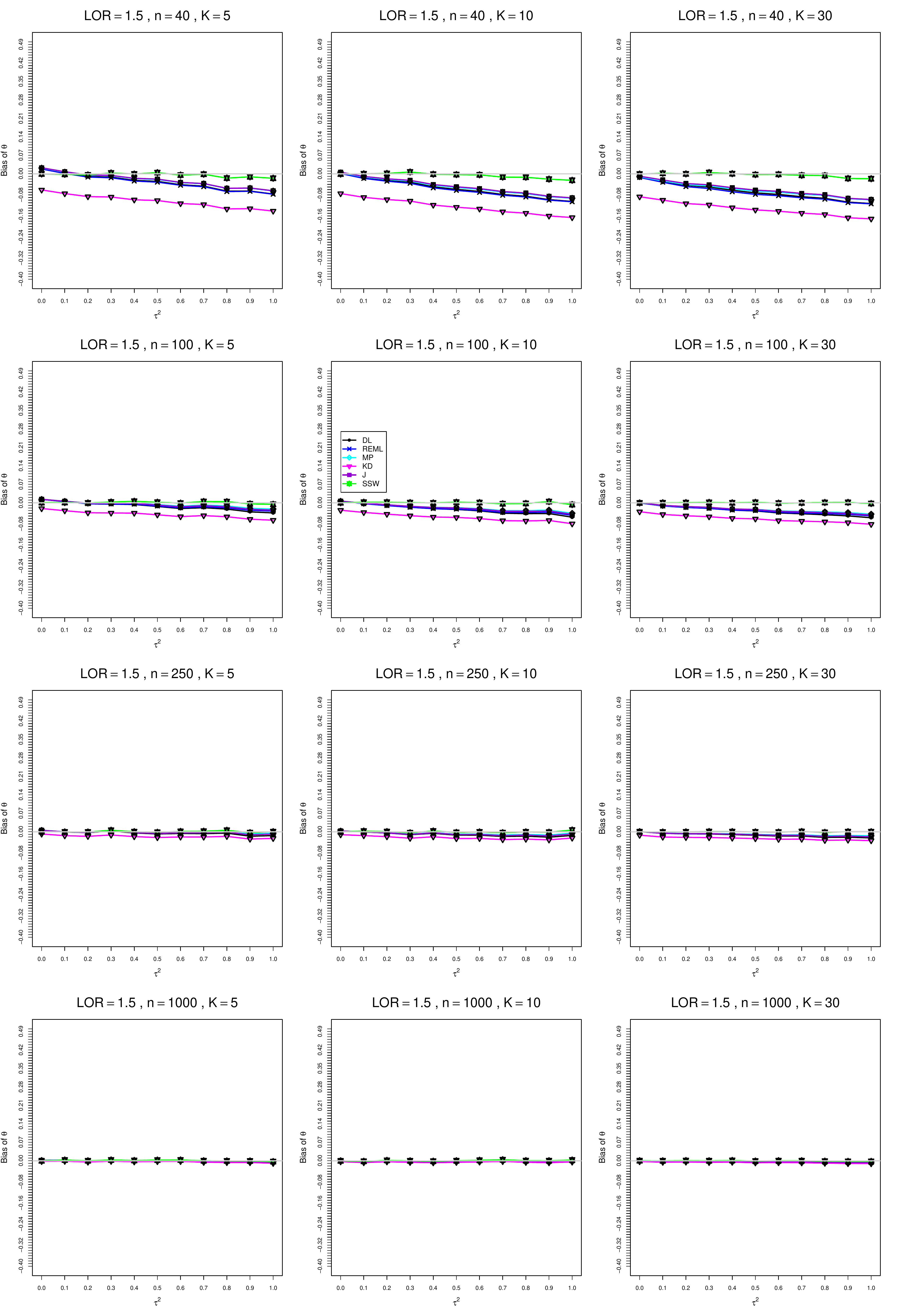}
	\caption{Bias of the estimation of  overall effect measure $\theta$ for $\theta=1.5$, $p_{iC}=0.4$, $q=0.5$, equal sample sizes $n=40,\;100,\;250,\;1000$. 
		\label{BiasThetaLOR15q05piC04}}
\end{figure}

\begin{figure}[t]
	\centering
	\includegraphics[scale=0.33]{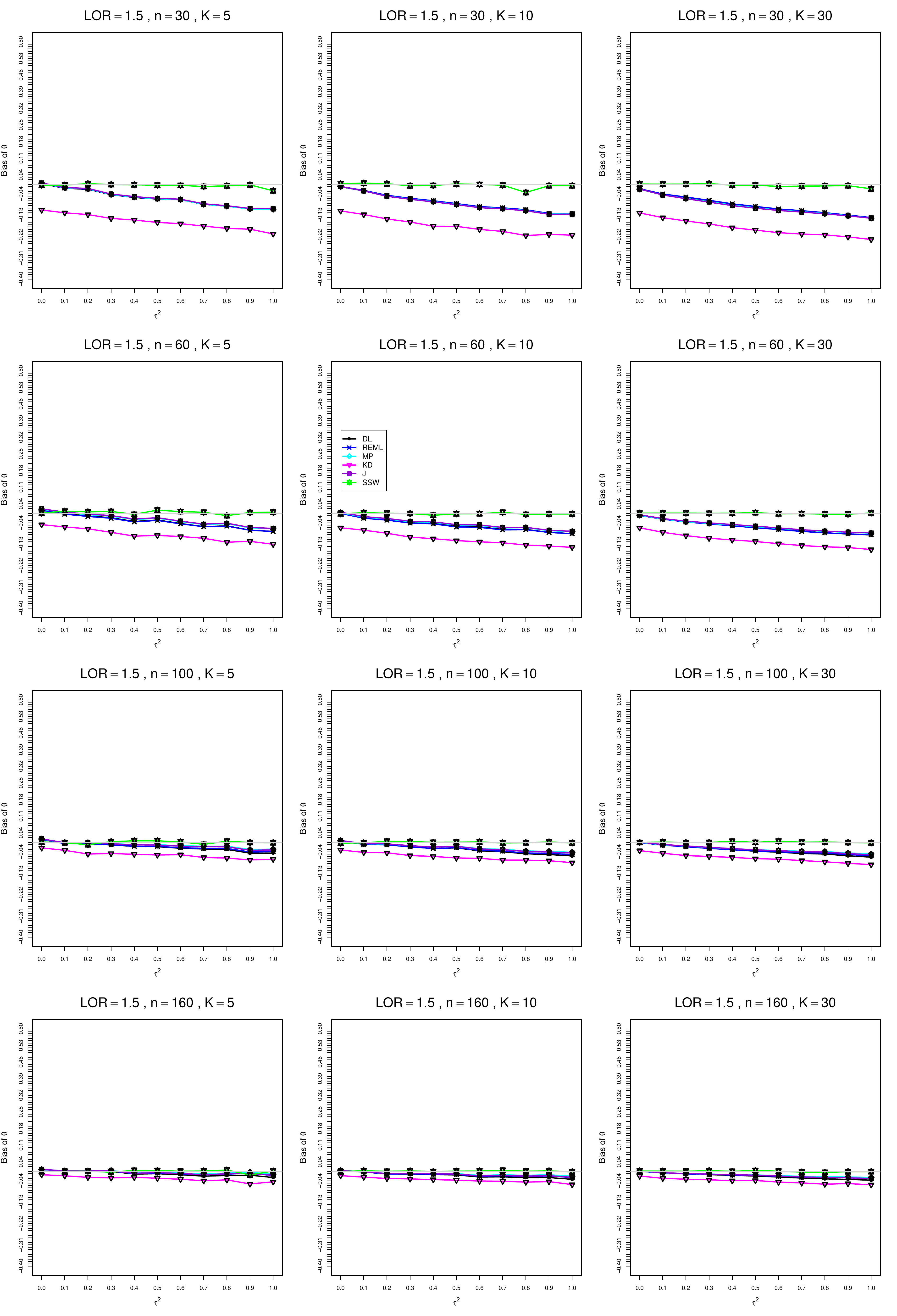}
	\caption{Bias of the estimation of  overall effect measure $\theta$ for $\theta=1.5$, $p_{iC}=0.4$, $q=0.5$, 
		unequal sample sizes $n=30,\; 60,\;100,\;160$. 
		\label{BiasThetaLOR15q05piC04_unequal_sample_sizes}}
\end{figure}

\begin{figure}[t]\centering
	\includegraphics[scale=0.35]{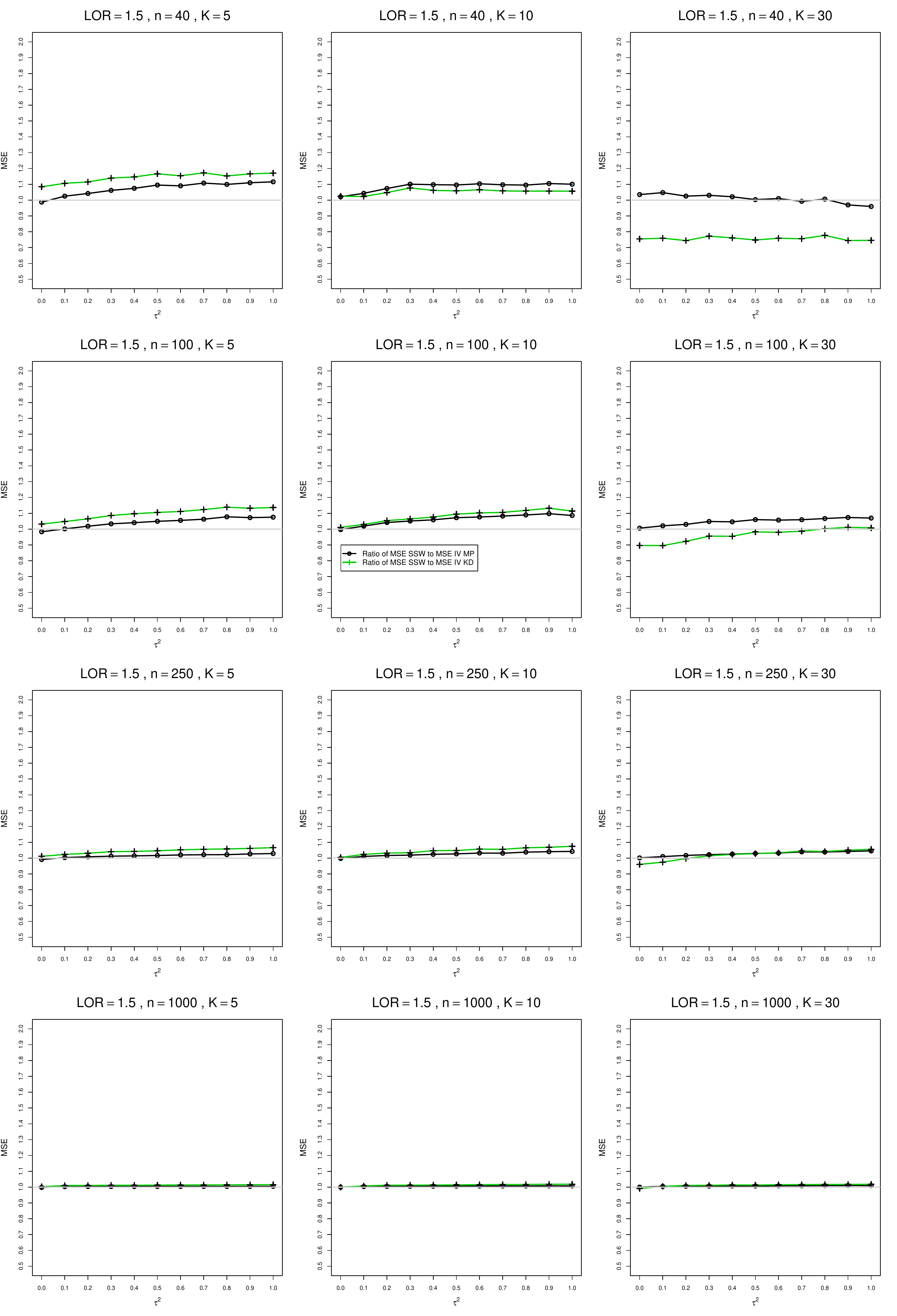}
	\caption{Ratio of mean squared errors of the fixed-weights to mean squared errors of inverse-variance estimator for $\theta=1.5$, $p_{iC}=0.4$, $q=0.5$, equal sample sizes $n=40,\;100,\;250,\;1000$. 
		\label{RatioOfMSEwithLOR15q05piC04fromMPandCMP}}
\end{figure}

\begin{figure}[t]\centering
	\includegraphics[scale=0.35]{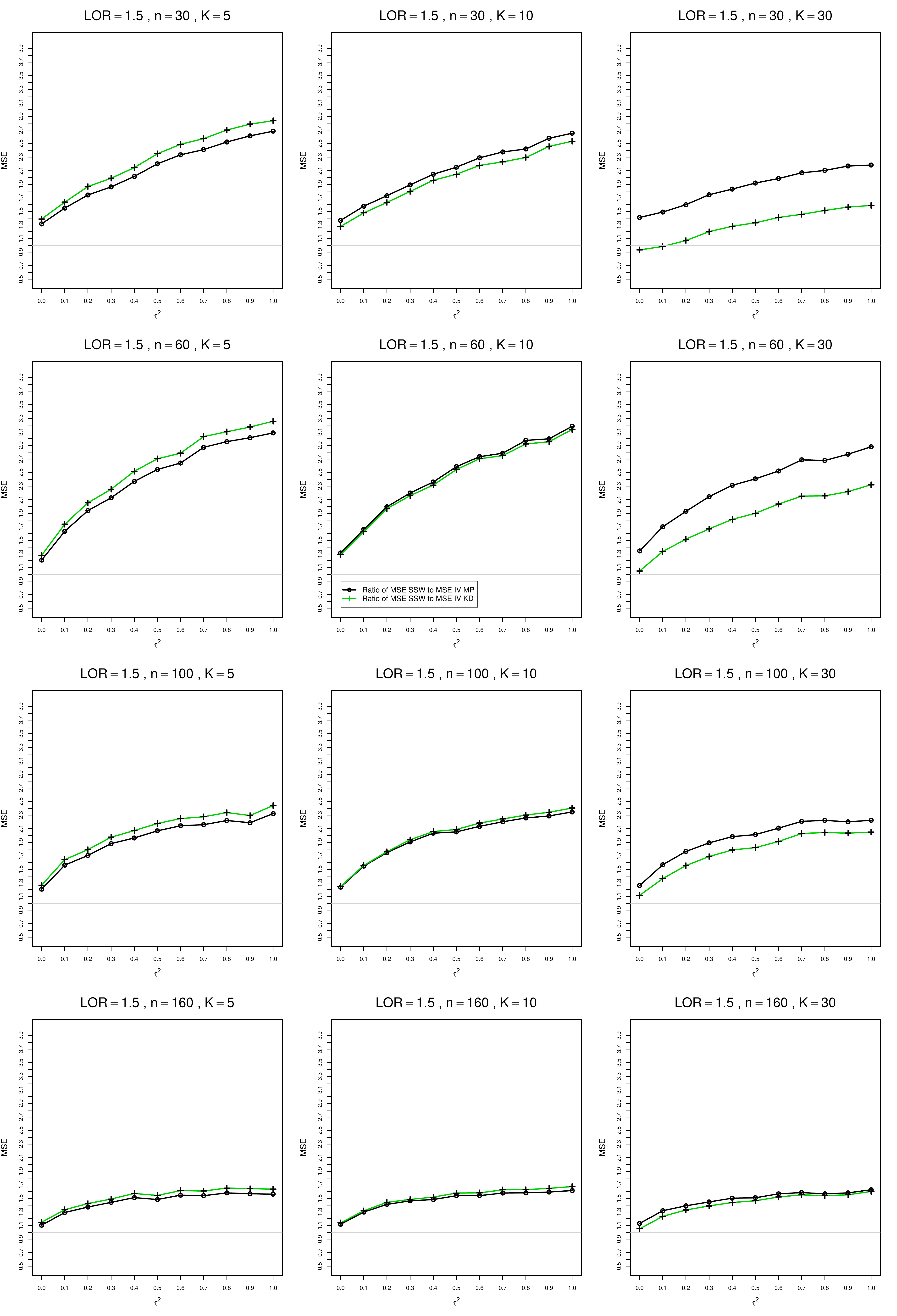}
	\caption{Ratio of mean squared errors of the fixed-weights to mean squared errors of inverse-variance estimator for $\theta=1.5$, $p_{iC}=0.4$, $q=0.5$, unequal sample sizes $n=30,\;60,\;100,\;160$. 
		\label{RatioOfMSEwithLOR15q05piC04fromMPandCMP_unequal_sample_sizes}}
\end{figure}

\begin{figure}[t]
	\centering
	\includegraphics[scale=0.33]{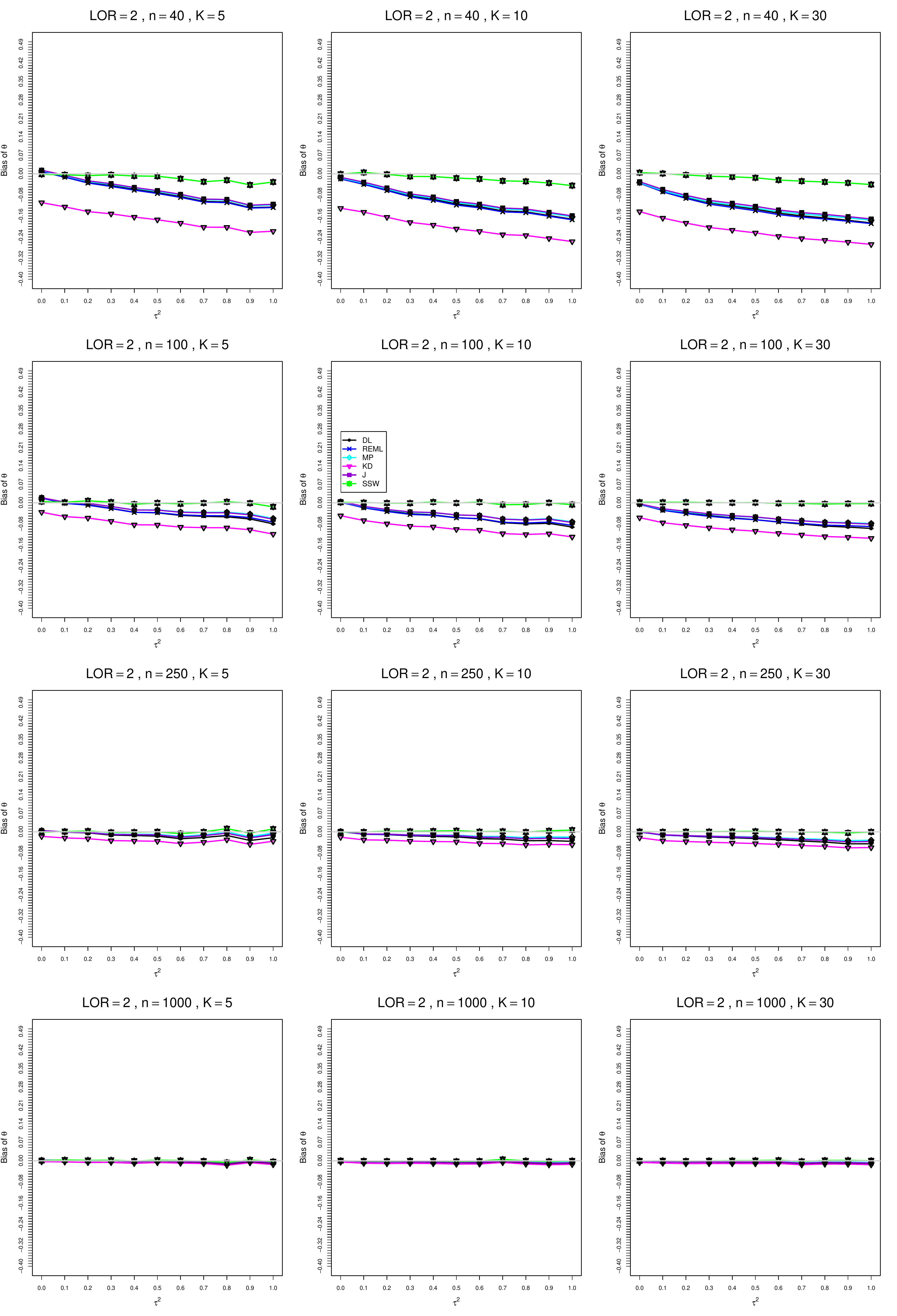}
	\caption{Bias of the estimation of  overall effect measure $\theta$ for $\theta=2$, $p_{iC}=0.4$, $q=0.5$, equal sample sizes $n=40,\;100,\;250,\;1000$. 
		\label{BiasThetaLOR2q05piC04}}
\end{figure}

\begin{figure}[t]
	\centering
	\includegraphics[scale=0.33]{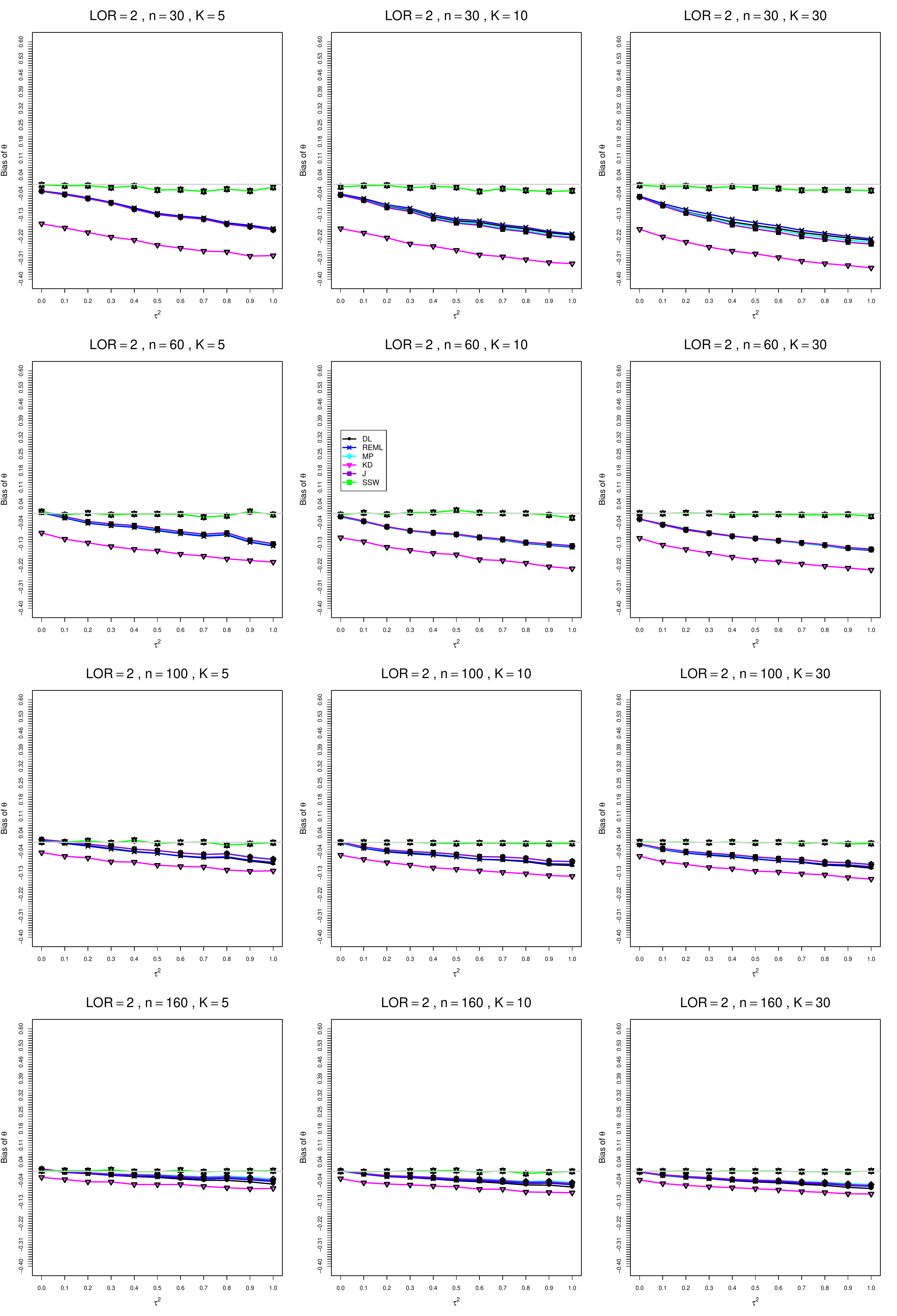}
	\caption{Bias of the estimation of  overall effect measure $\theta$ for $\theta=2$, $p_{iC}=0.4$, $q=0.5$, 
		unequal sample sizes $n=30,\; 60,\;100,\;160$. 
		\label{BiasThetaLOR2q05piC04_unequal_sample_sizes}}
\end{figure}

\begin{figure}[t]\centering
	\includegraphics[scale=0.35]{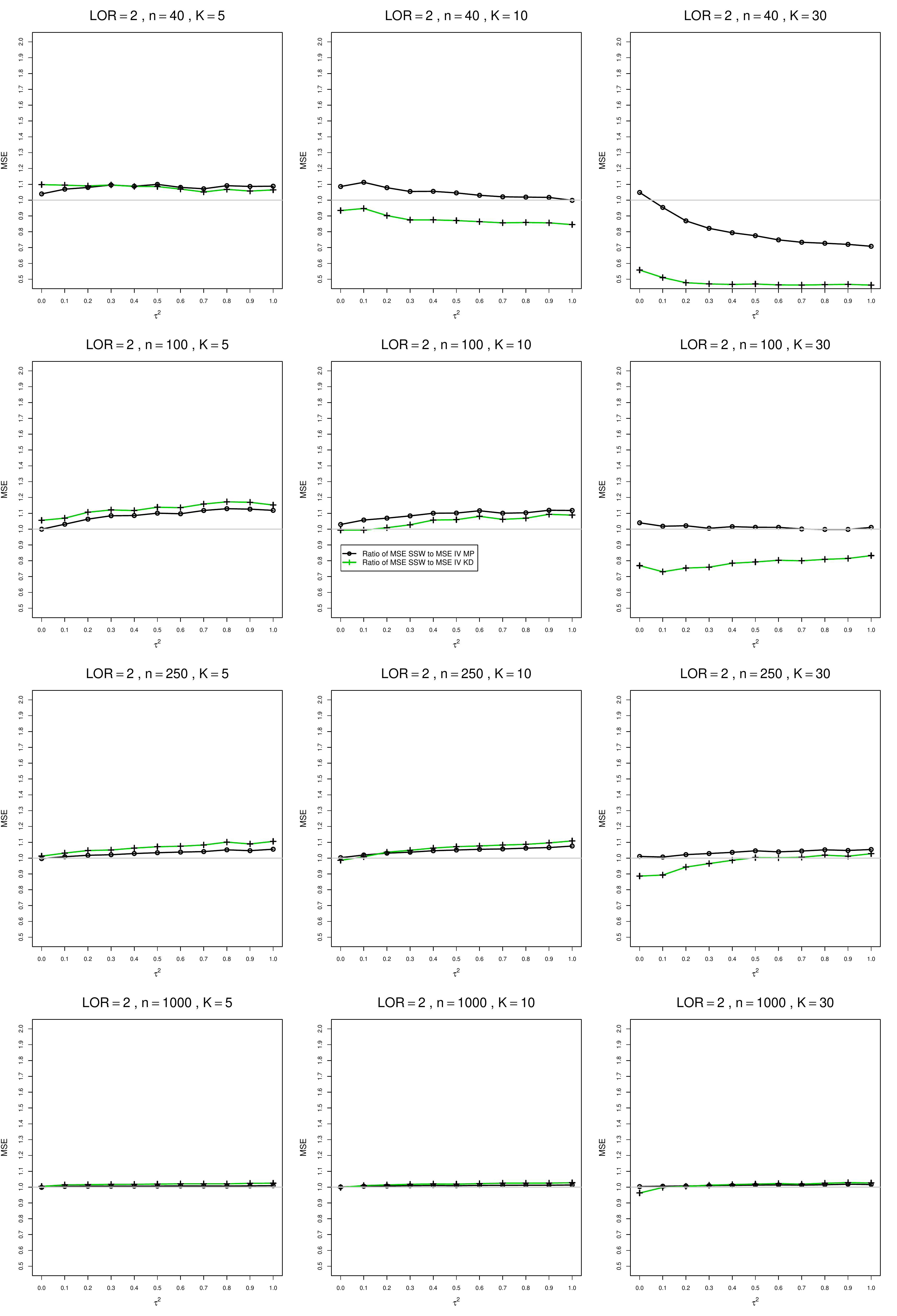}
	\caption{Ratio of mean squared errors of the fixed-weights to mean squared errors of inverse-variance estimator for $\theta=2$, $p_{iC}=0.4$, $q=0.5$, equal sample sizes $n=40,\;100,\;250,\;1000$. 
		\label{RatioOfMSEwithLOR2q05piC04fromMPandCMP}}
\end{figure}

\begin{figure}[t]\centering
	\includegraphics[scale=0.35]{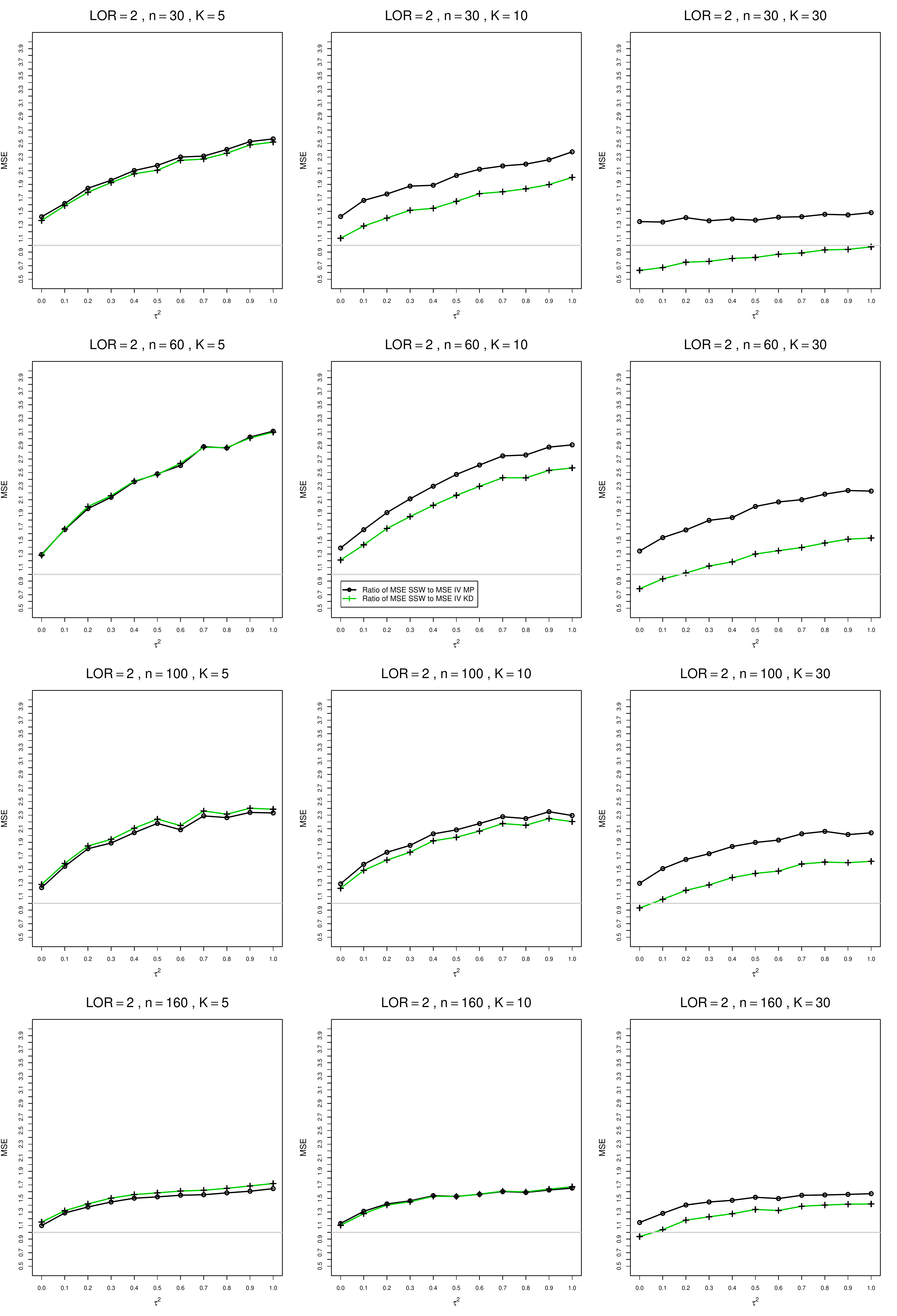}
	\caption{Ratio of mean squared errors of the fixed-weights to mean squared errors of inverse-variance estimator for $\theta=2$, $p_{iC}=0.4$, $q=0.5$, unequal sample sizes $n=30,\;60,\;100,\;160$. 
		\label{RatioOfMSEwithLOR2q05piC04fromMPandCMP_unequal_sample_sizes}}
\end{figure}


\begin{figure}[t]
	\centering
	\includegraphics[scale=0.33]{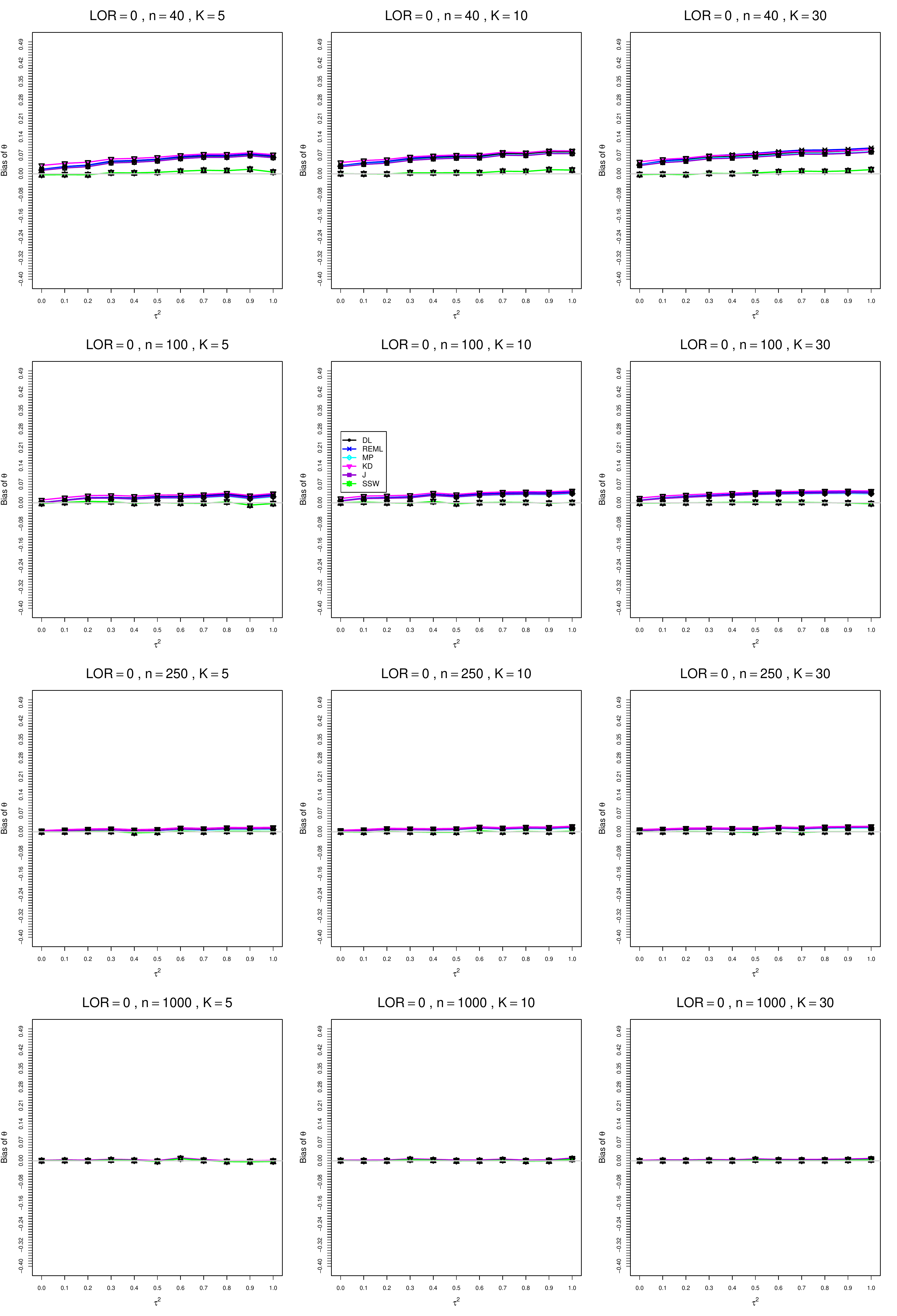}
	\caption{Bias of the estimation of  overall effect measure $\theta$ for $\theta=0$, $p_{iC}=0.4$, $q=0.75$, equal sample sizes $n=40,\;100,\;250,\;1000$. 
		\label{BiasThetaLOR0q075piC04}}
\end{figure}

\begin{figure}[t]
	\centering
	\includegraphics[scale=0.33]{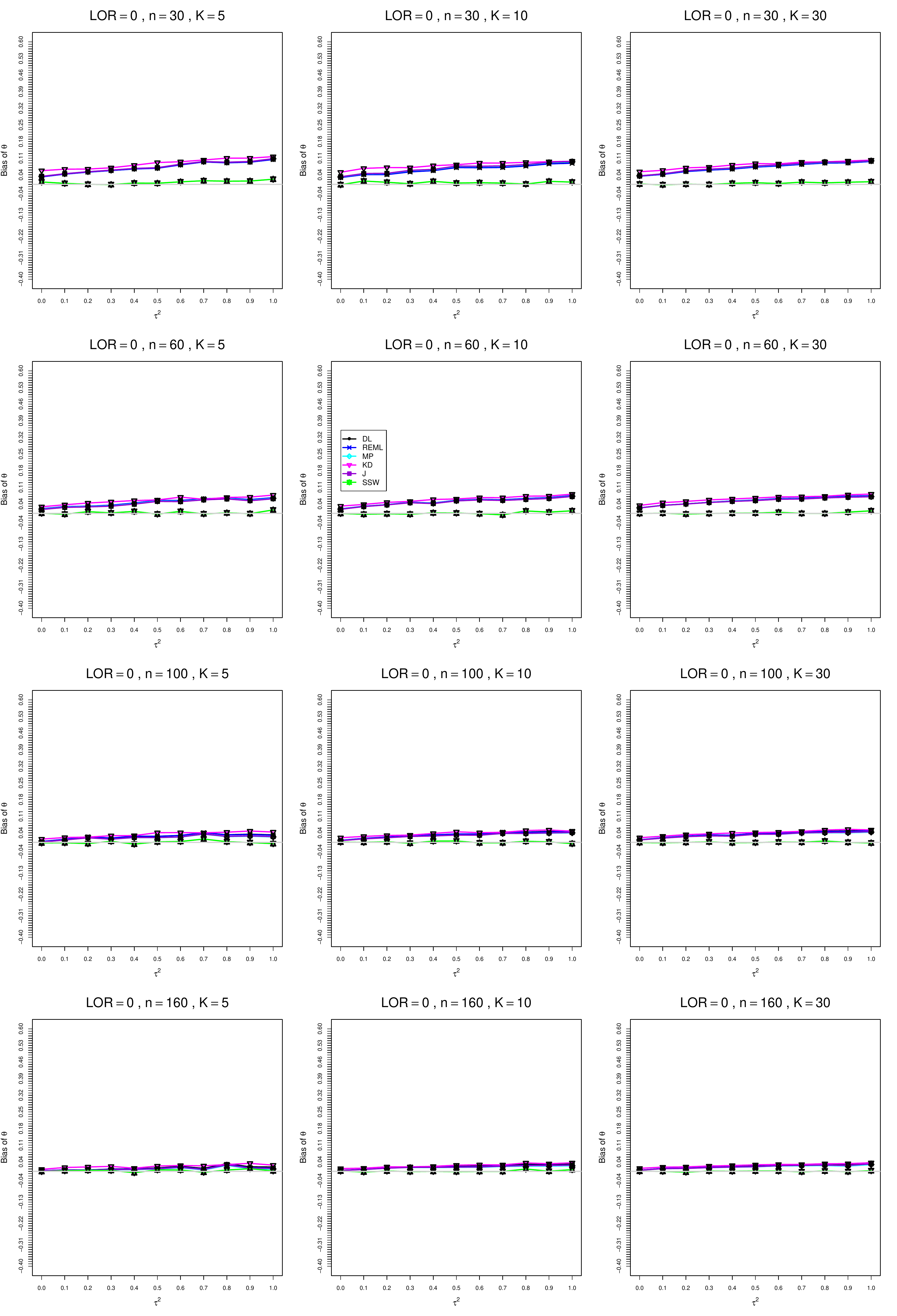}
	\caption{Bias of the estimation of  overall effect measure $\theta$ for $\theta=0$, $p_{iC}=0.4$, $q=0.75$, 
		unequal sample sizes $n=30,\; 60,\;100,\;160$. 
		\label{BiasThetaLOR0q075piC04_unequal_sample_sizes}}
\end{figure}

\begin{figure}[t]\centering
	\includegraphics[scale=0.35]{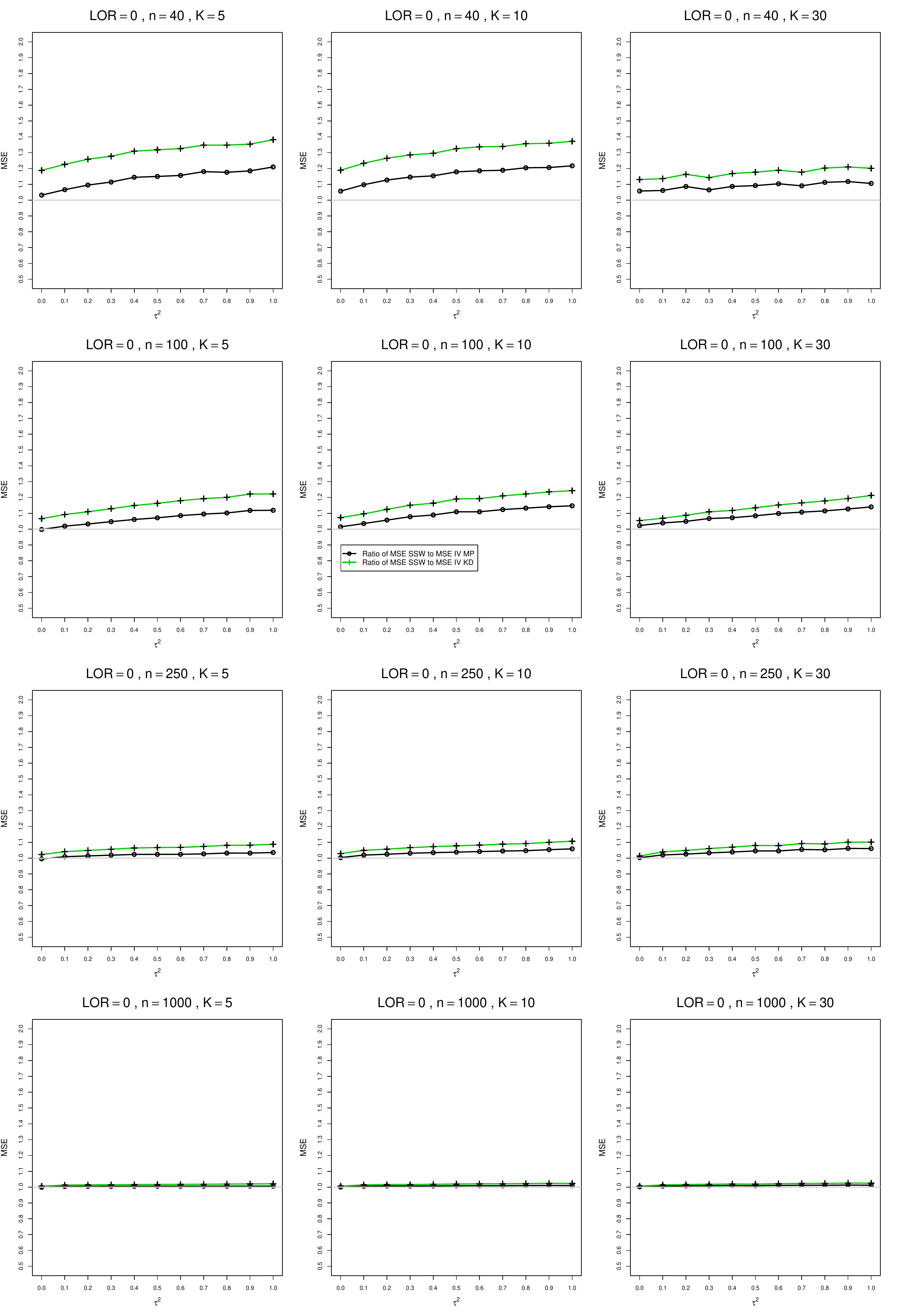}
	\caption{Ratio of mean squared errors of the fixed-weights to mean squared errors of inverse-variance estimator for $\theta=0$, $p_{iC}=0.4$, $q=0.75$, equal sample sizes $n=40,\;100,\;250,\;1000$. 
		\label{RatioOfMSEwithLOR0q075piC04fromMPandCMP}}
\end{figure}

\begin{figure}[t]\centering
	\includegraphics[scale=0.35]{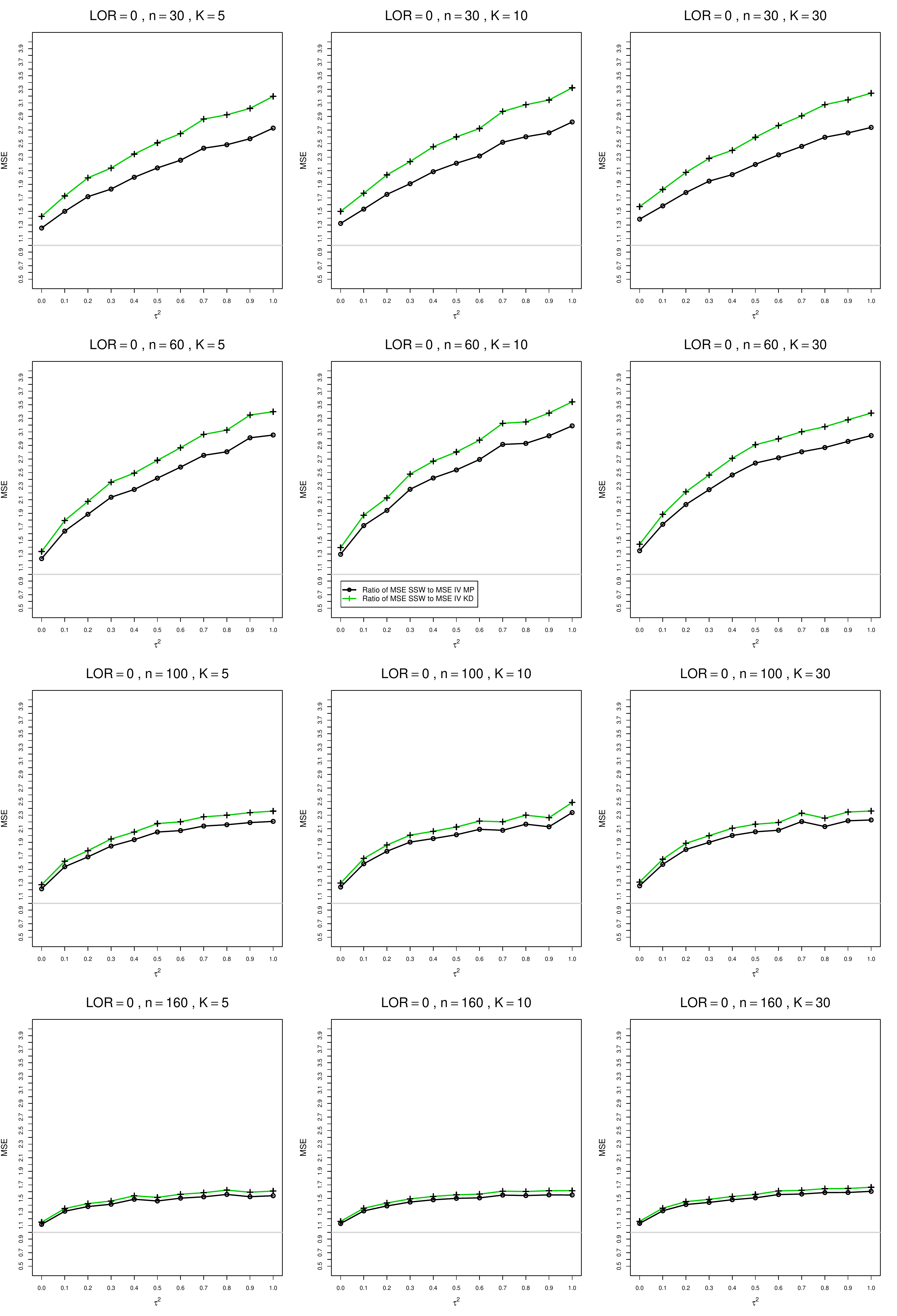}
	\caption{Ratio of mean squared errors of the fixed-weights to mean squared errors of inverse-variance estimator for $\theta=0$, $p_{iC}=0.4$, $q=0.75$, unequal sample sizes $n=30,\;60,\;100,\;160$. 
		\label{RatioOfMSEwithLOR0q075piC04fromMPandCMP_unequal_sample_sizes}}
\end{figure}

\begin{figure}[t]
	\centering
	\includegraphics[scale=0.33]{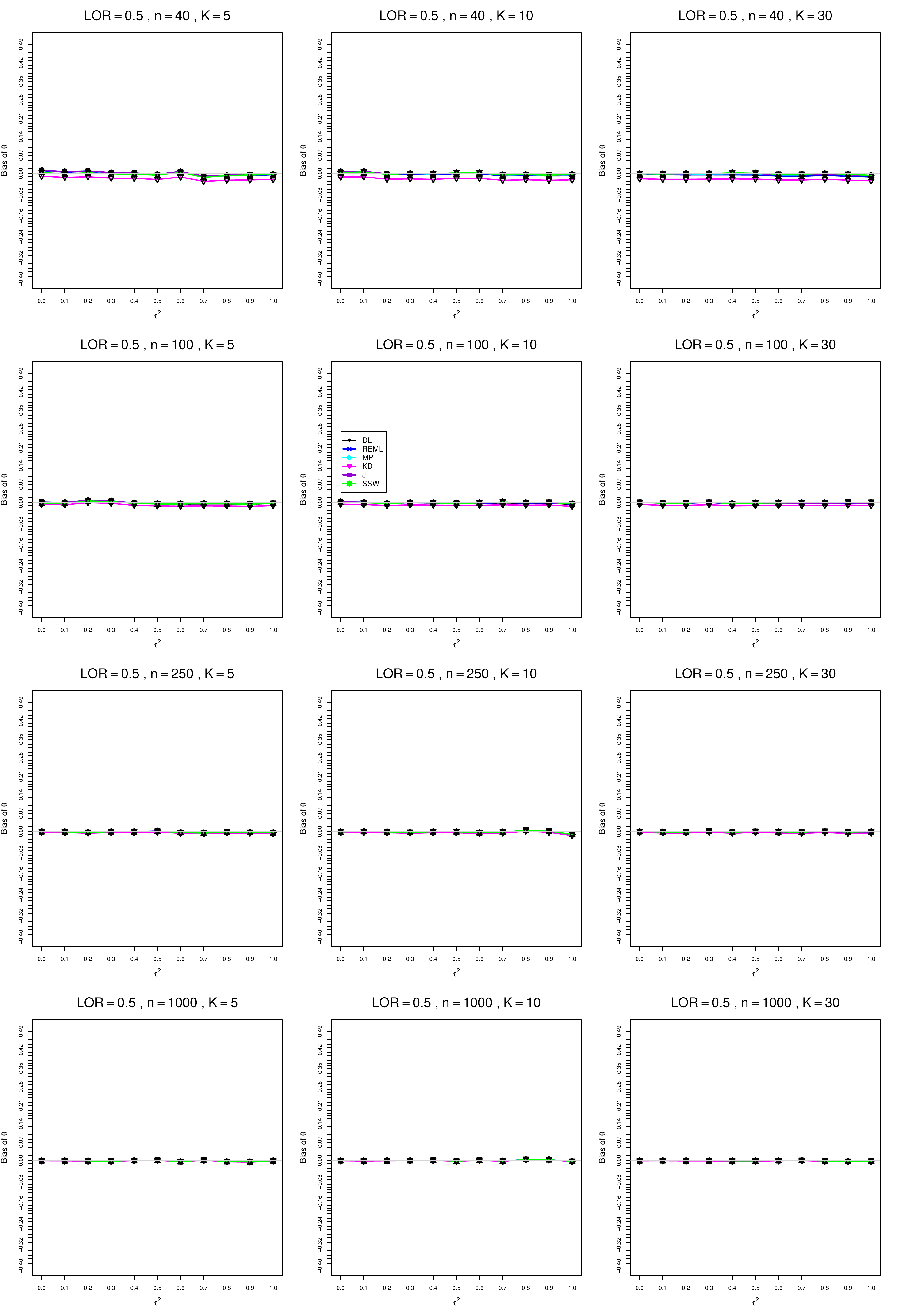}
	\caption{Bias of the estimation of  overall effect measure $\theta$ for $\theta=0.5$, $p_{iC}=0.4$, $q=0.75$, equal sample sizes $n=40,\;100,\;250,\;1000$. 
		\label{BiasThetaLOR05q075piC04}}
\end{figure}

\begin{figure}[t]
	\centering
	\includegraphics[scale=0.33]{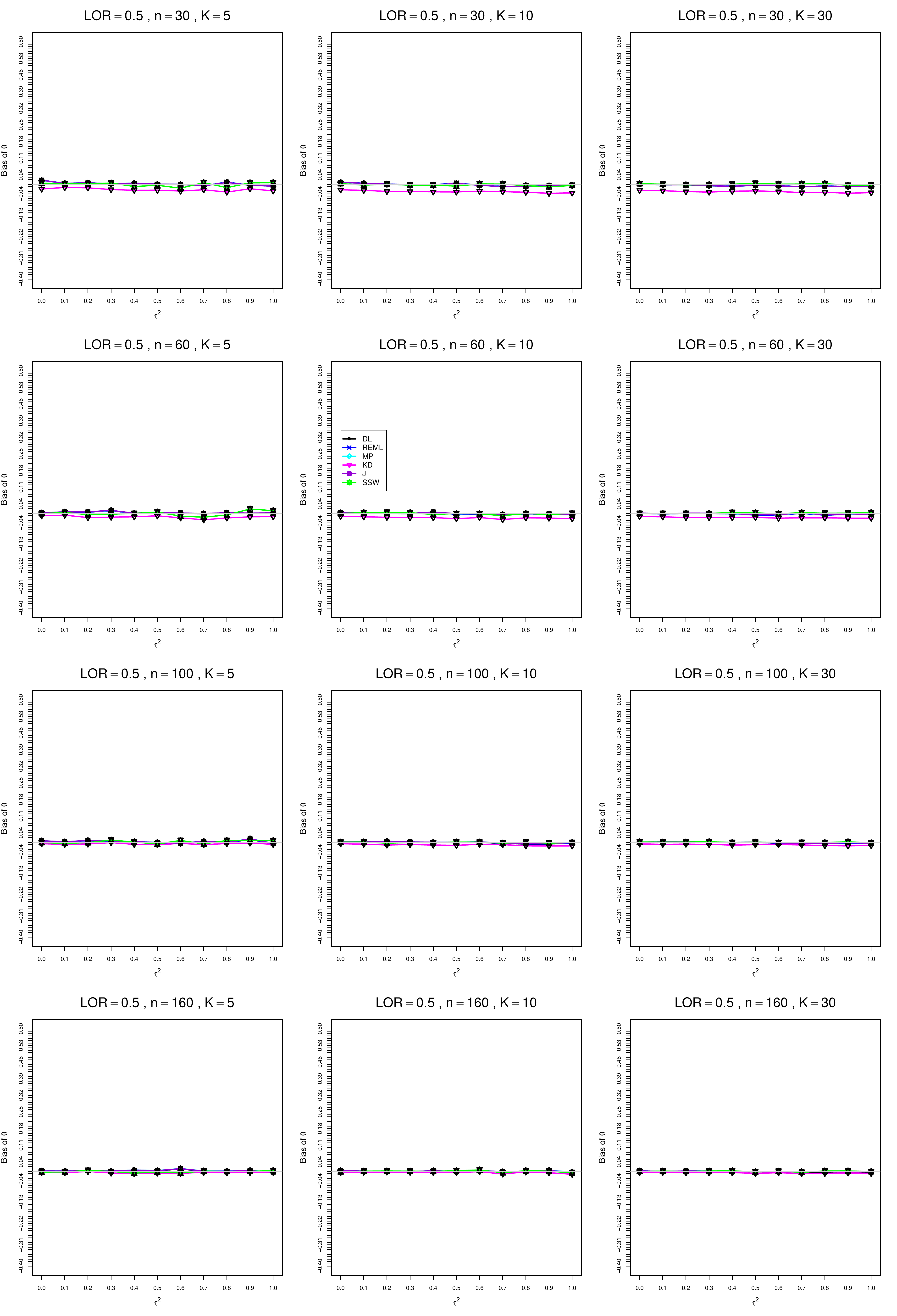}
	\caption{Bias of the estimation of  overall effect measure $\theta$ for $\theta=0.5$, $p_{iC}=0.4$, $q=0.75$, 
		unequal sample sizes $n=30,\; 60,\;100,\;160$. 
		\label{BiasThetaLOR05q075piC04_unequal_sample_sizes}}
\end{figure}

\begin{figure}[t]\centering
	\includegraphics[scale=0.35]{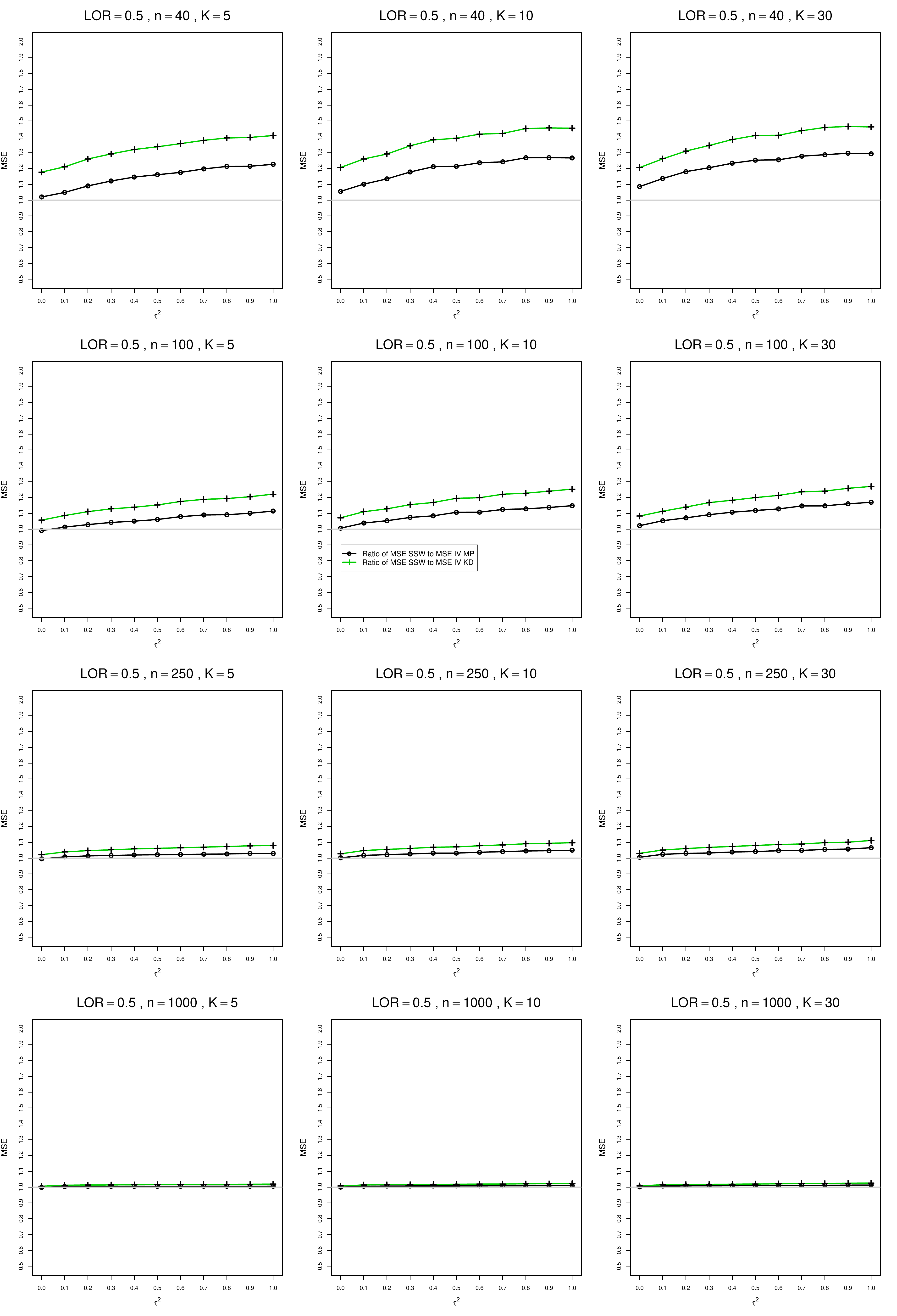}
	\caption{Ratio of mean squared errors of the fixed-weights to mean squared errors of inverse-variance estimator for $\theta=0.5$, $p_{iC}=0.4$, $q=0.75$, equal sample sizes $n=40,\;100,\;250,\;1000$. 
		\label{RatioOfMSEwithLOR05q075piC04fromMPandCMP}}
\end{figure}

\begin{figure}[t]\centering
	\includegraphics[scale=0.35]{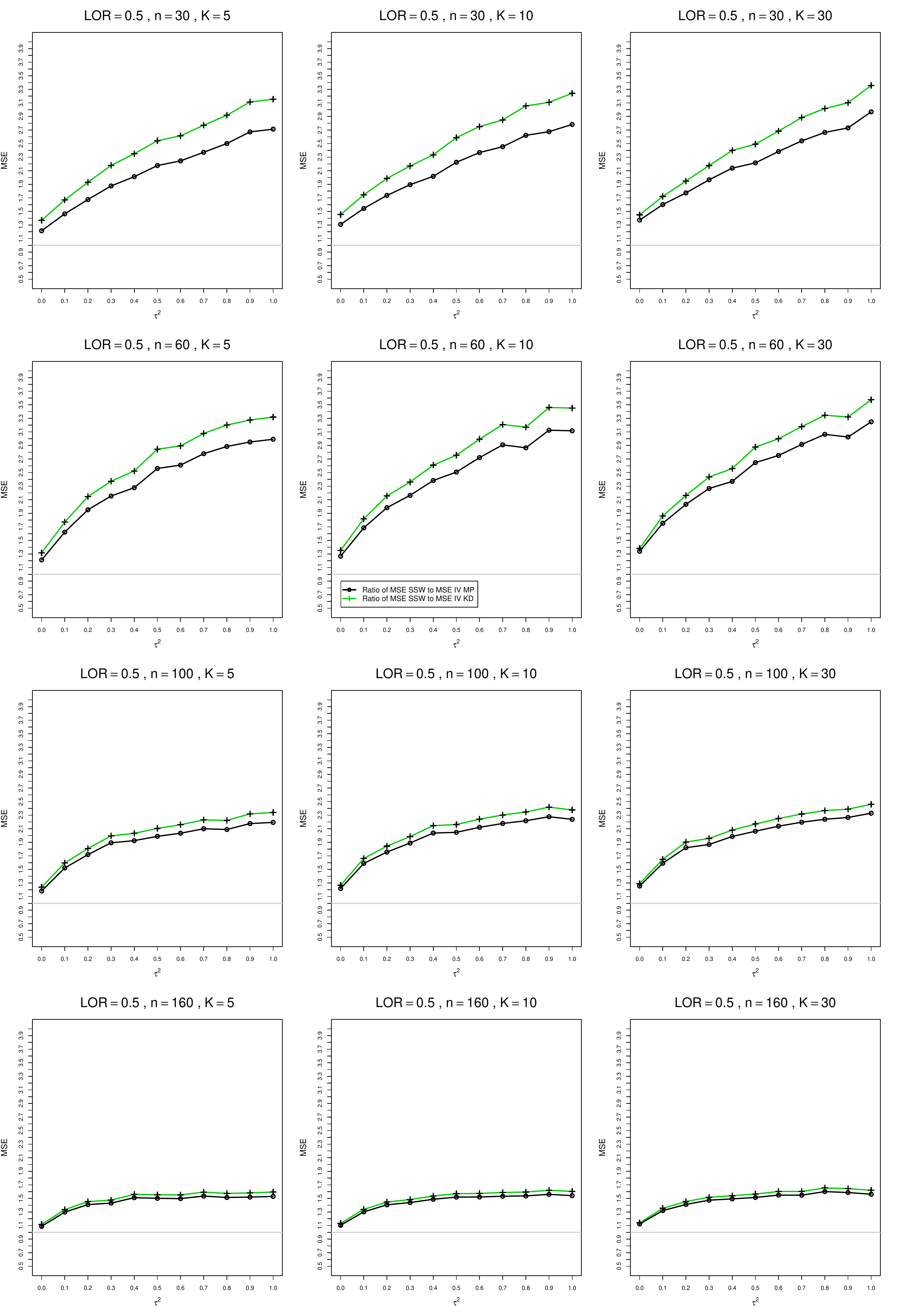}
	\caption{Ratio of mean squared errors of the fixed-weights to mean squared errors of inverse-variance estimator for $\theta=0.5$, $p_{iC}=0.4$, $q=0.75$, unequal sample sizes $n=30,\;60,\;100,\;160$. 
		\label{RatioOfMSEwithLOR05q075piC04fromMPandCMP_unequal_sample_sizes}}
\end{figure}

\clearpage

\begin{figure}[t]
	\centering
	\includegraphics[scale=0.33]{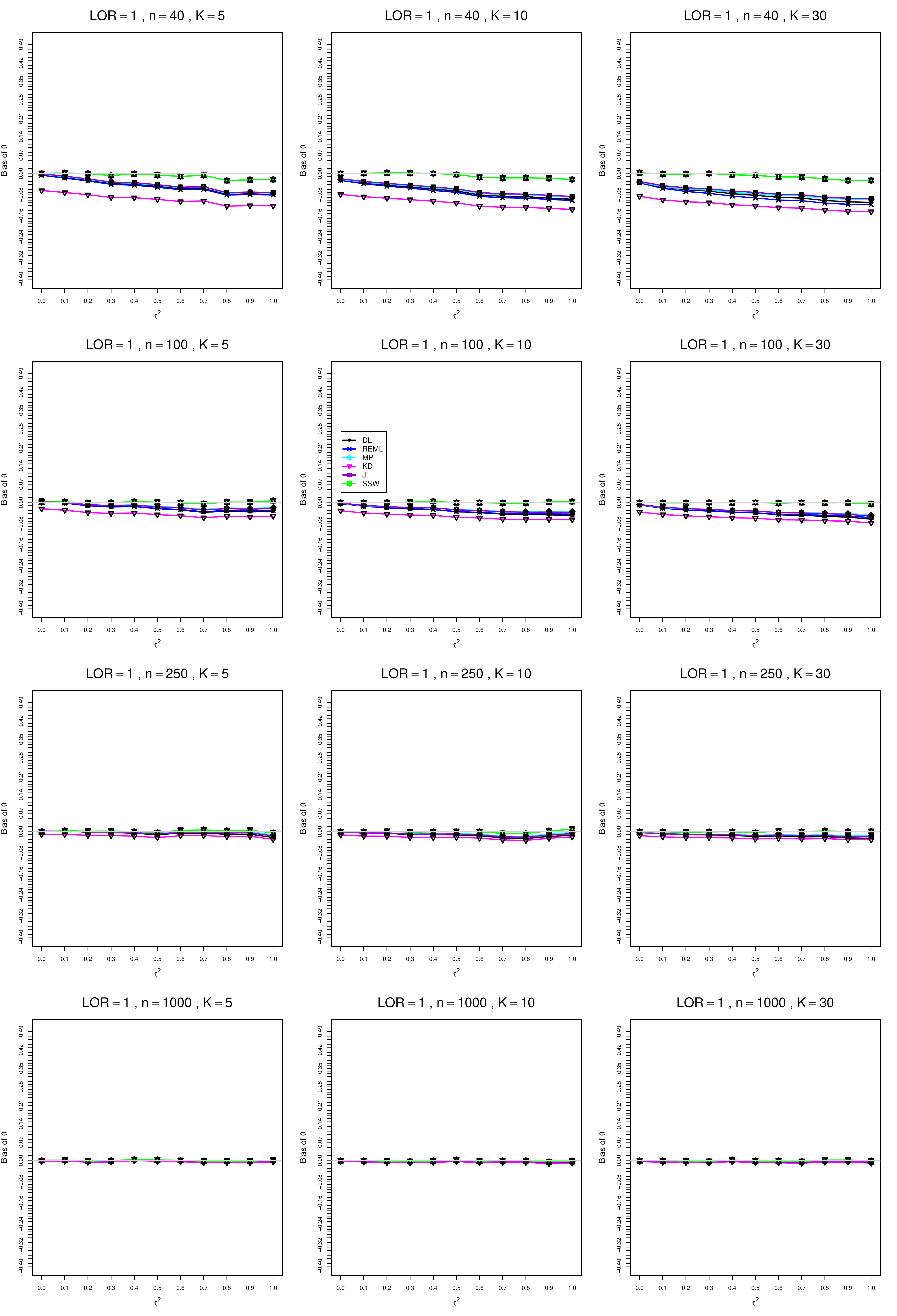}
	\caption{Bias of the estimation of  overall effect measure $\theta$ for $\theta=1$, $p_{iC}=0.4$, $q=0.75$, equal sample sizes $n=40,\;100,\;250,\;1000$. 
		\label{BiasThetaLOR1q075piC04}}
\end{figure}

\begin{figure}[t]
	\centering
	\includegraphics[scale=0.33]{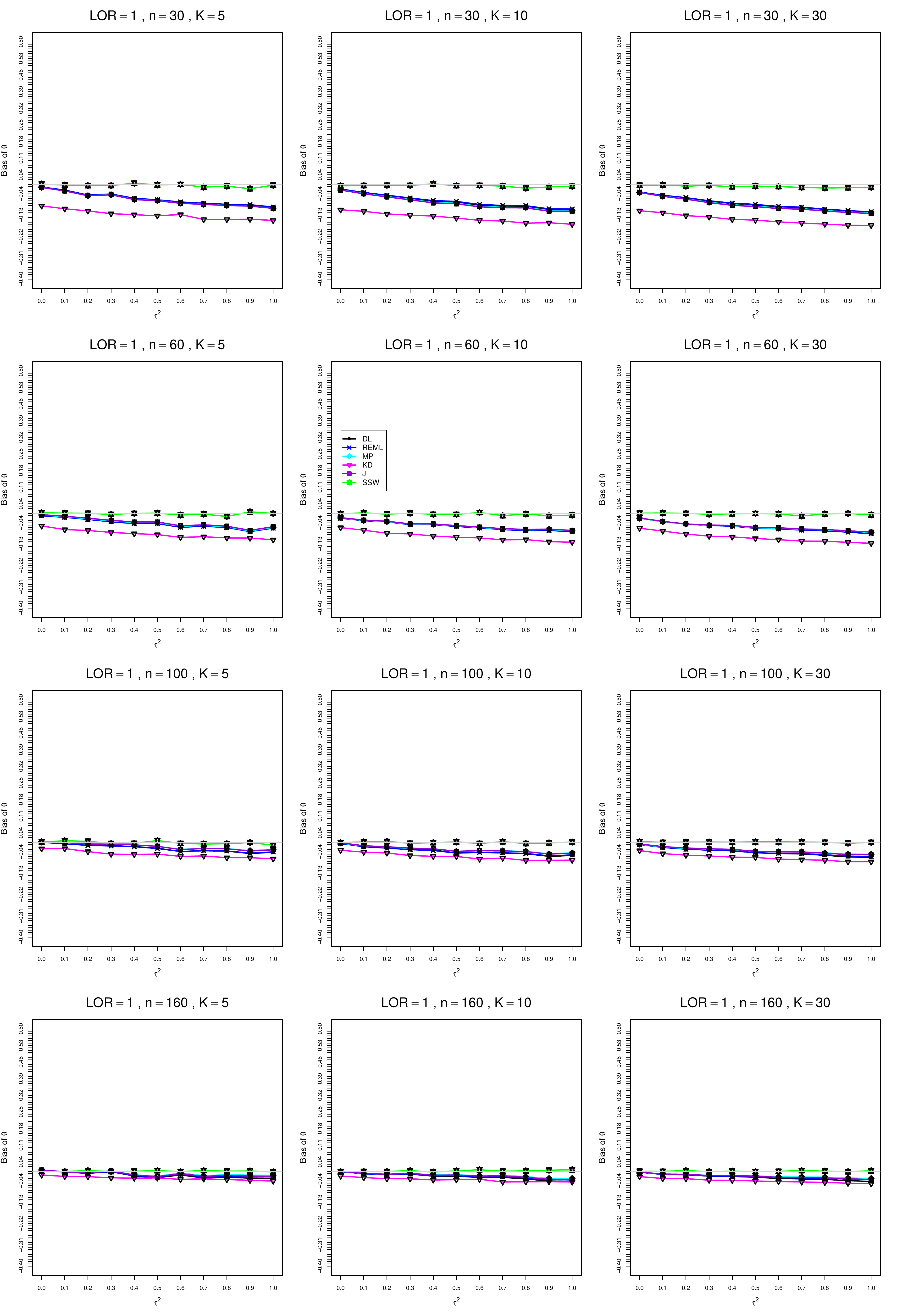}
	\caption{Bias of the estimation of  overall effect measure $\theta$ for $\theta=1$, $p_{iC}=0.4$, $q=0.75$, 
		unequal sample sizes $n=30,\; 60,\;100,\;160$. 
		\label{BiasThetaLOR1q075piC04_unequal_sample_sizes}}
\end{figure}

\begin{figure}[t]\centering
	\includegraphics[scale=0.35]{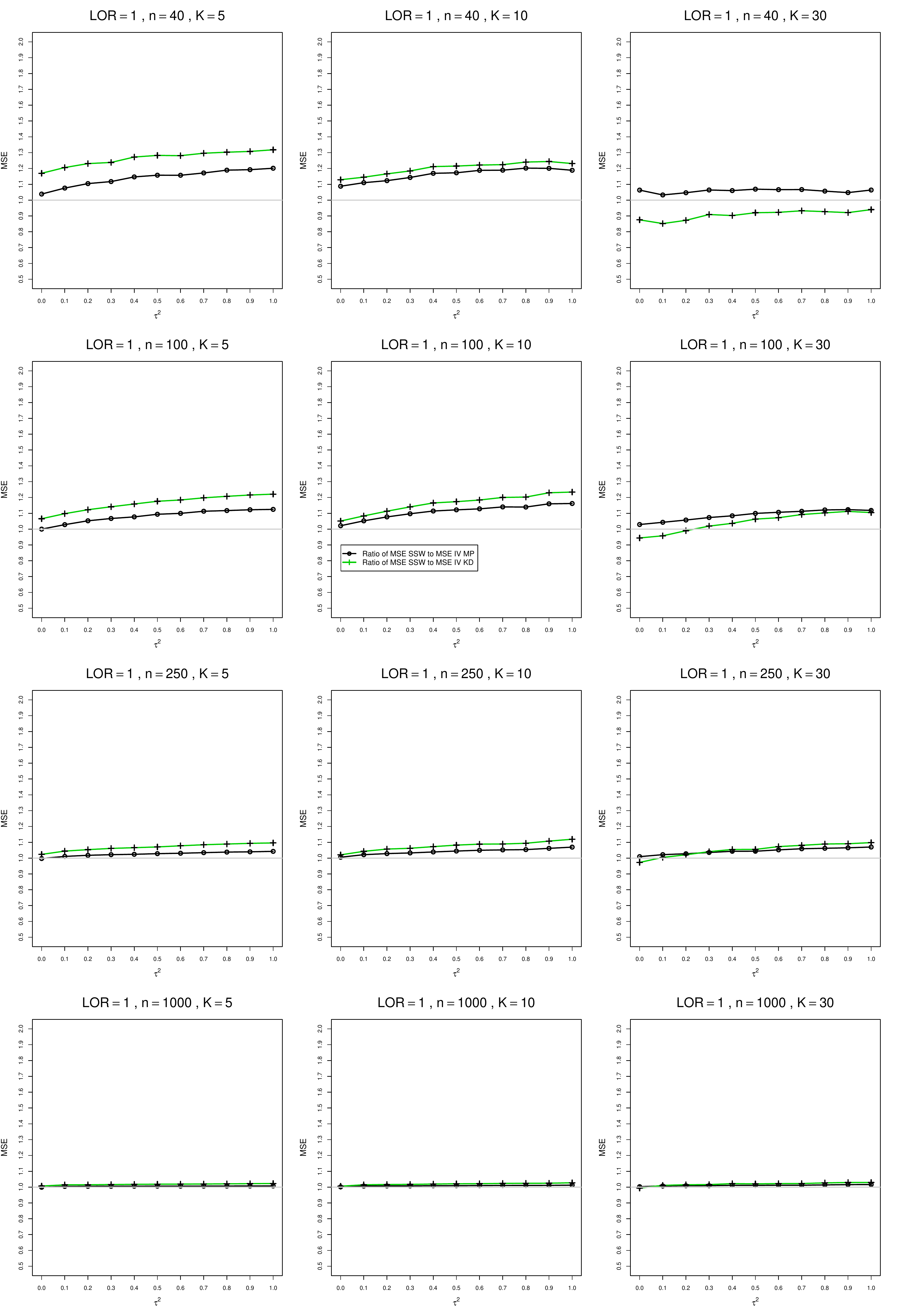}
	\caption{Ratio of mean squared errors of the fixed-weights to mean squared errors of inverse-variance estimator for $\theta=1$, $p_{iC}=0.4$, $q=0.75$, equal sample sizes $n=40,\;100,\;250,\;1000$. 
		\label{RatioOfMSEwithLOR1q075piC04fromMPandCMP}}
\end{figure}

\begin{figure}[t]\centering
	\includegraphics[scale=0.35]{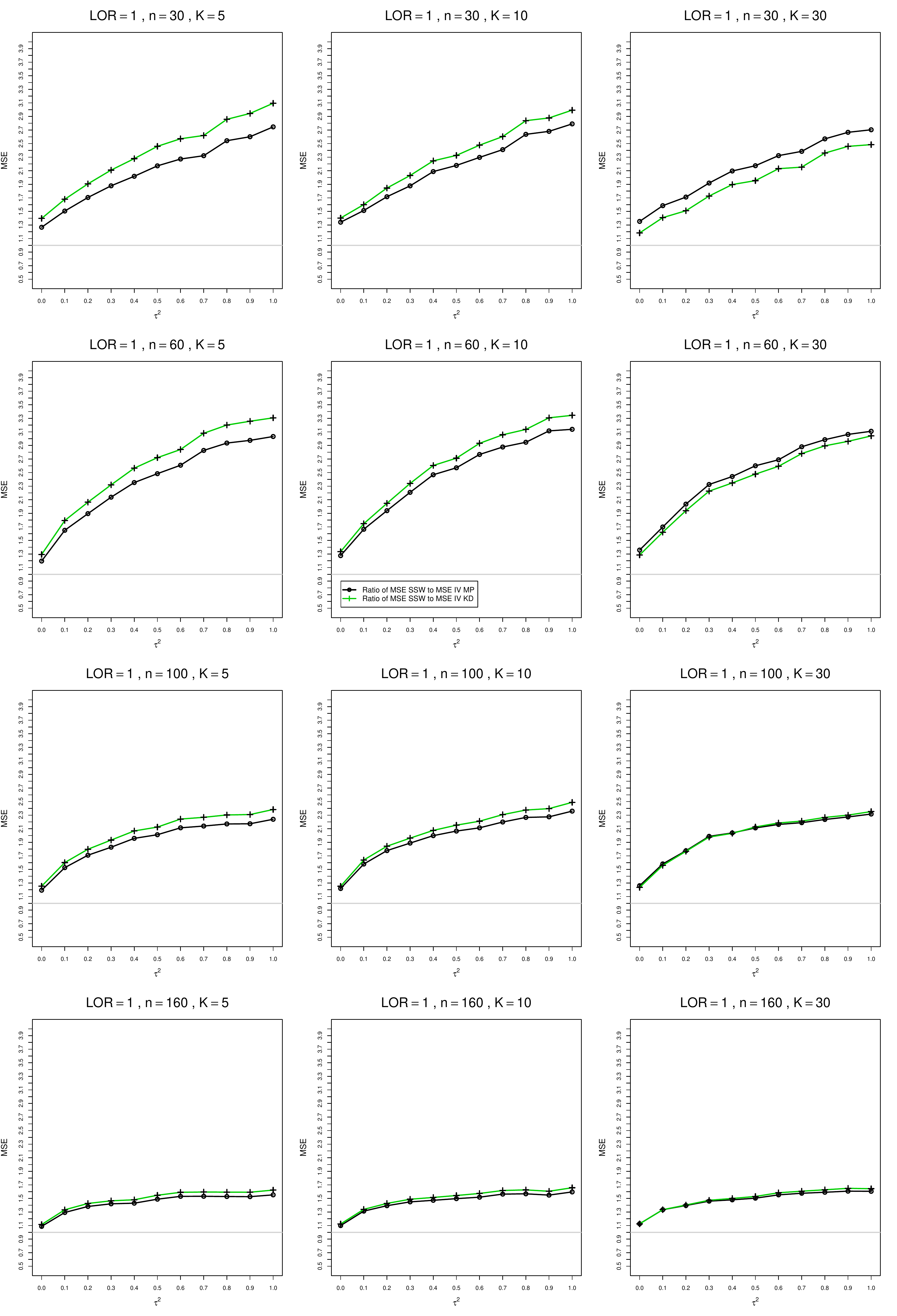}
	\caption{Ratio of mean squared errors of the fixed-weights to mean squared errors of inverse-variance estimator for $\theta=1$, $p_{iC}=0.4$, $q=0.75$, unequal sample sizes $n=30,\;60,\;100,\;160$. 
		\label{RatioOfMSEwithLOR1q075piC04fromMPandCMP_unequal_sample_sizes}}
\end{figure}


\begin{figure}[t]
	\centering
	\includegraphics[scale=0.33]{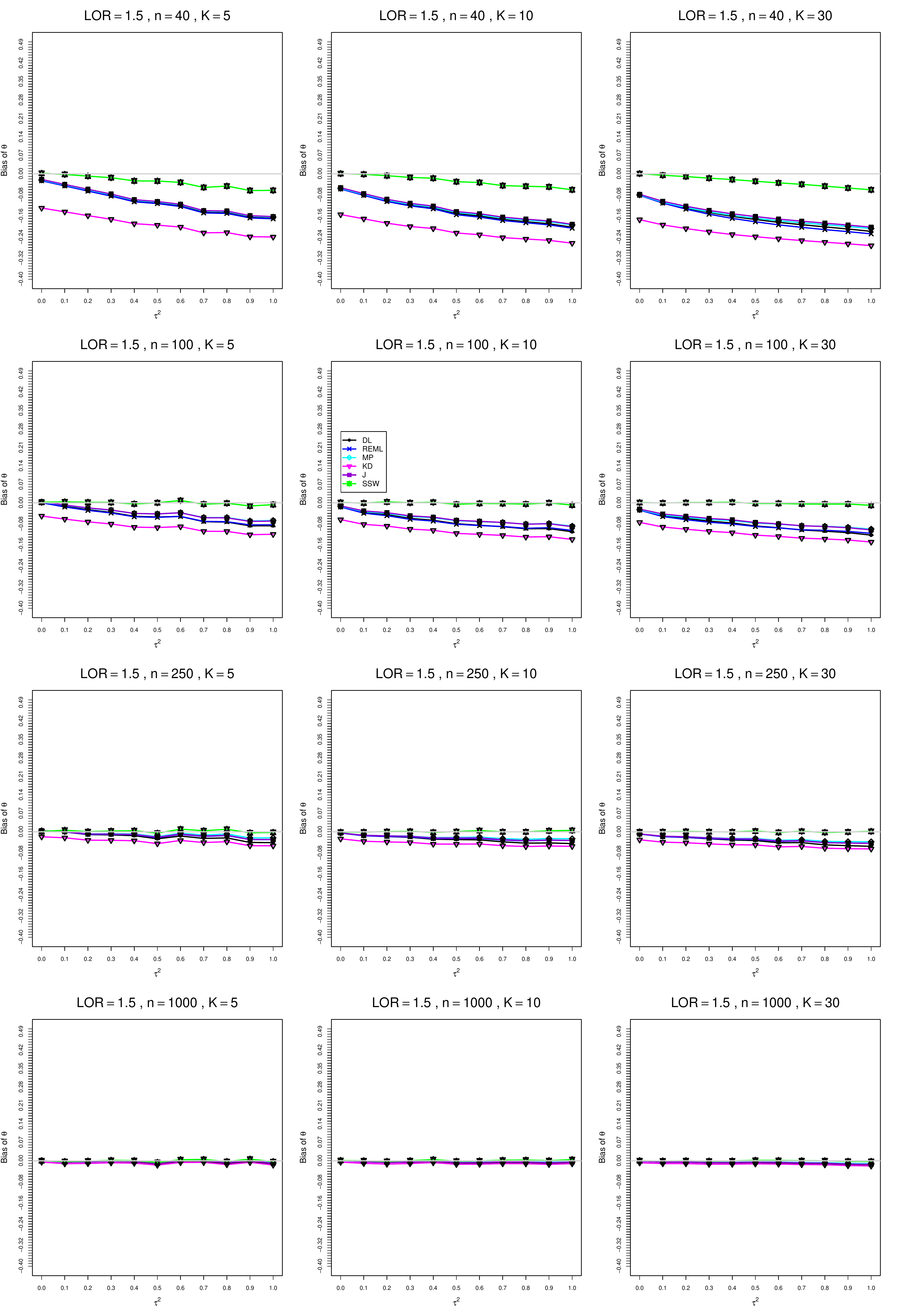}
	\caption{Bias of the estimation of  overall effect measure $\theta$ for $\theta=1.5$, $p_{iC}=0.4$, $q=0.75$, equal sample sizes $n=40,\;100,\;250,\;1000$. 
		\label{BiasThetaLOR15q075piC04}}
\end{figure}

\begin{figure}[t]
	\centering
	\includegraphics[scale=0.33]{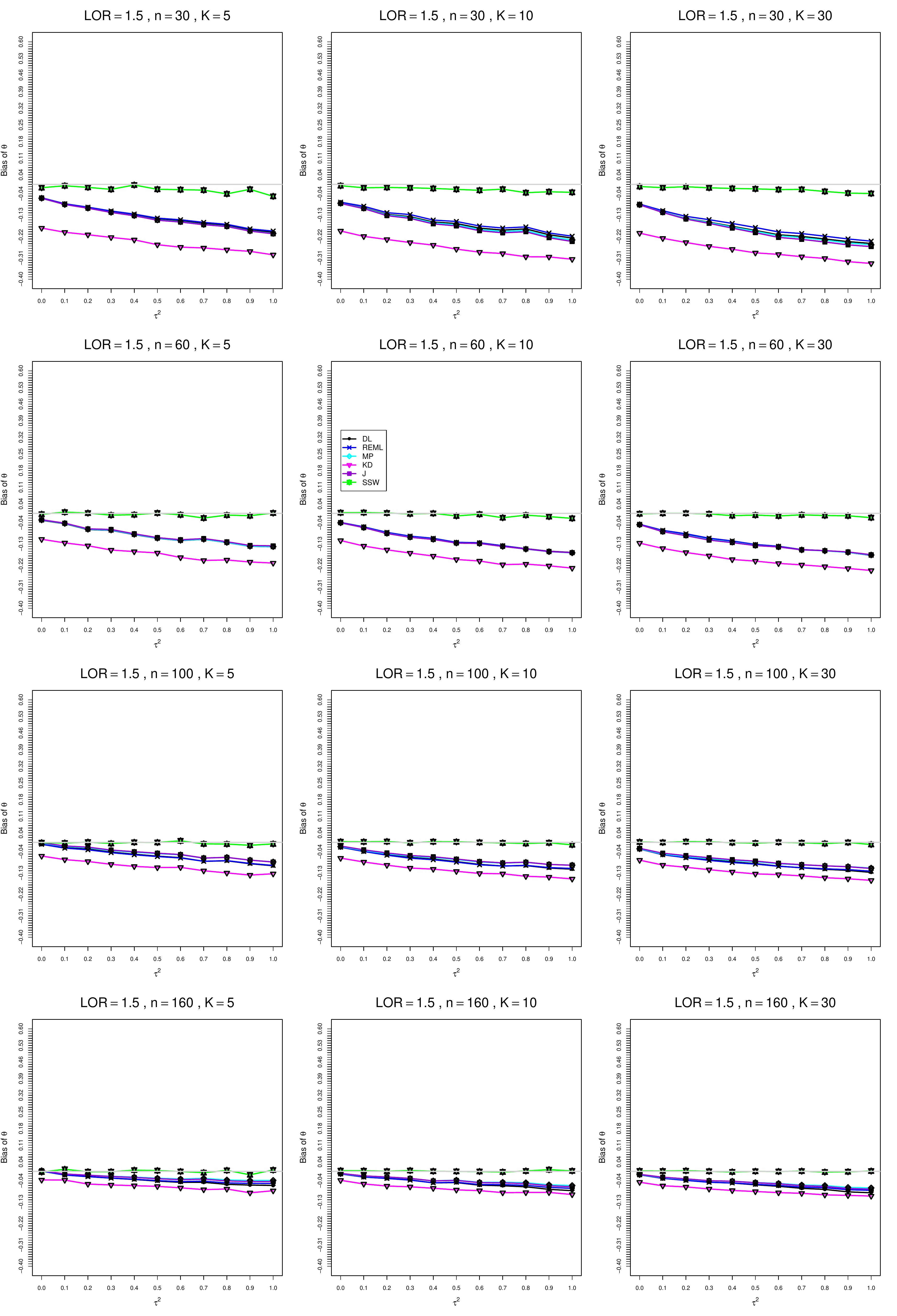}
	\caption{Bias of the estimation of  overall effect measure $\theta$ for $\theta=1.5$, $p_{iC}=0.4$, $q=0.75$, 
		unequal sample sizes $n=30,\; 60,\;100,\;160$. 
		\label{BiasThetaLOR15q075piC04_unequal_sample_sizes}}
\end{figure}

\begin{figure}[t]\centering
	\includegraphics[scale=0.35]{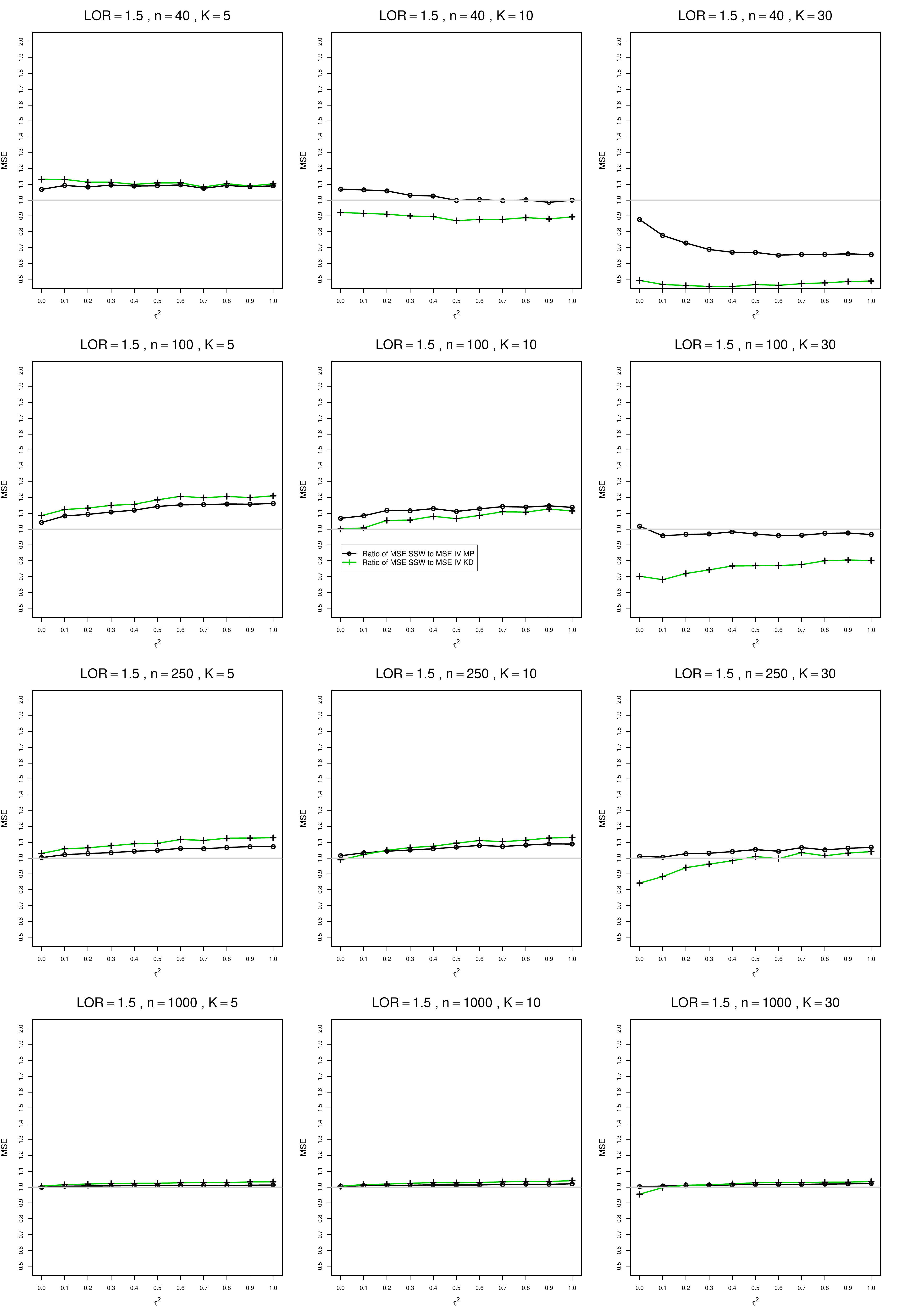}
	\caption{Ratio of mean squared errors of the fixed-weights to mean squared errors of inverse-variance estimator for $\theta=1.5$, $p_{iC}=0.4$, $q=0.75$, equal sample sizes $n=40,\;100,\;250,\;1000$. 
		\label{RatioOfMSEwithLOR15q075piC04fromMPandCMP}}
\end{figure}

\begin{figure}[t]\centering
	\includegraphics[scale=0.35]{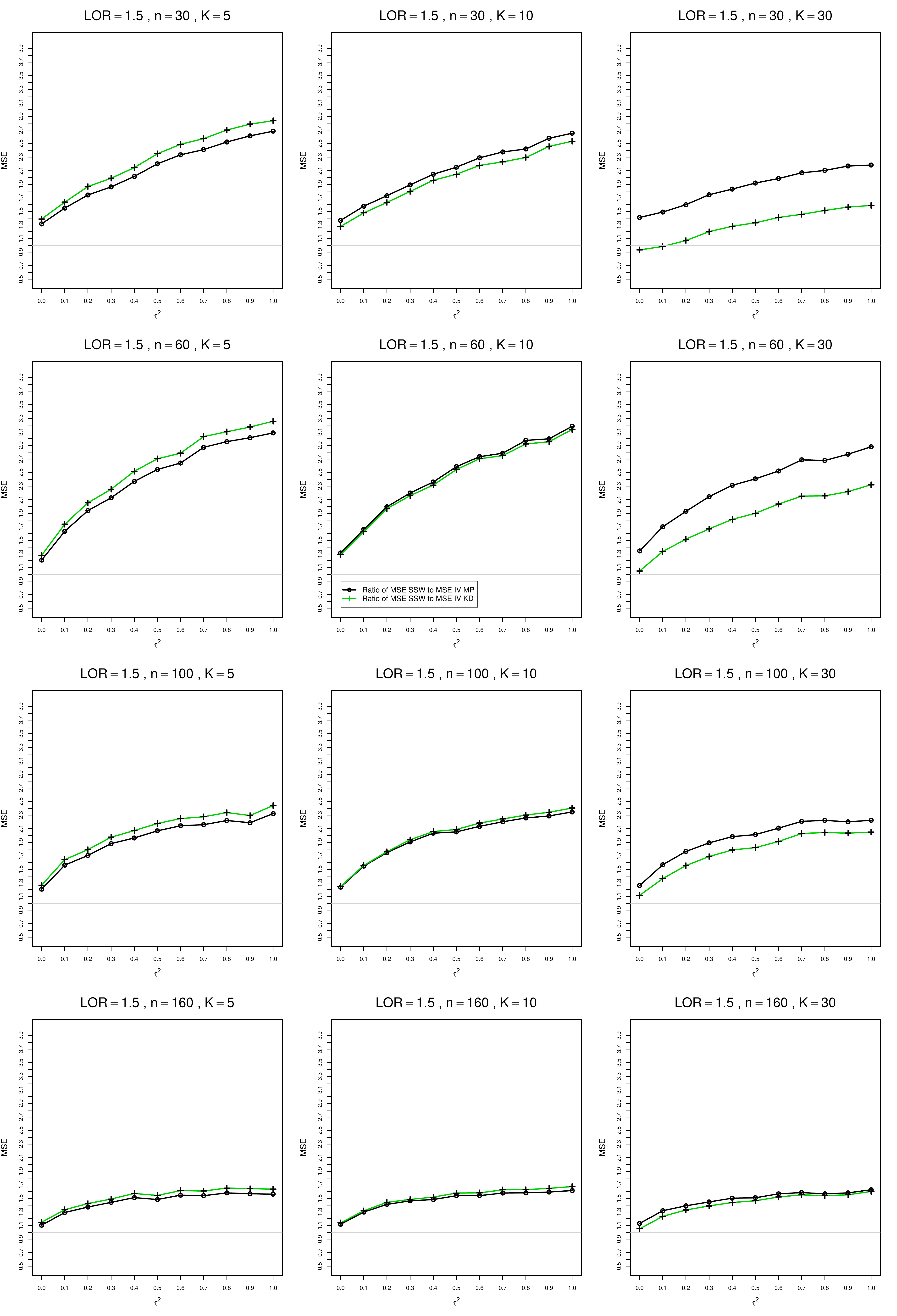}
	\caption{Ratio of mean squared errors of the fixed-weights to mean squared errors of inverse-variance estimator for $\theta=1.5$, $p_{iC}=0.4$, $q=0.75$, unequal sample sizes $n=30,\;60,\;100,\;160$. 
		\label{RatioOfMSEwithLOR15q075piC04fromMPandCMP_unequal_sample_sizes}}
\end{figure}


\begin{figure}[t]
	\centering
	\includegraphics[scale=0.33]{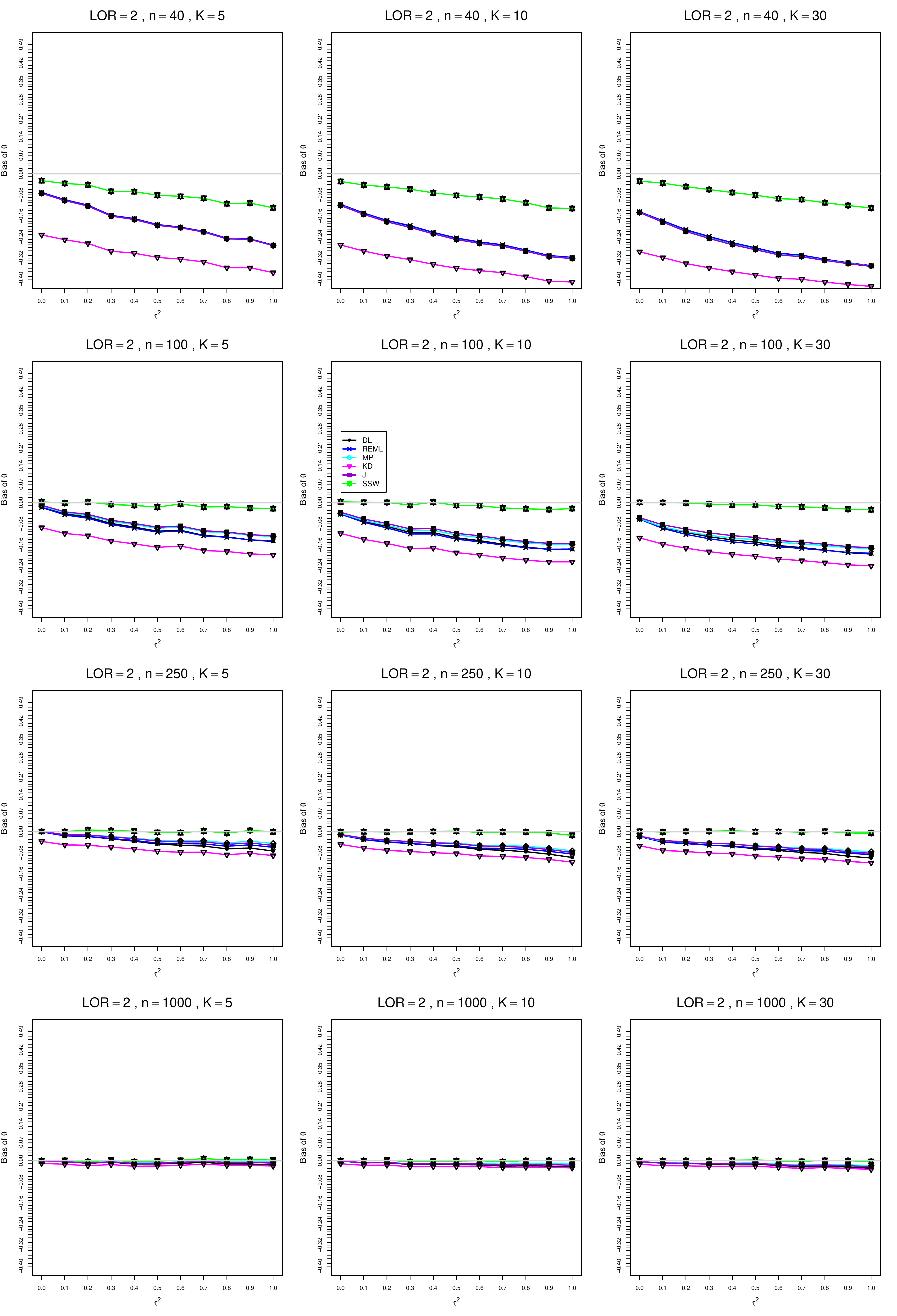}
	\caption{Bias of the estimation of  overall effect measure $\theta$ for $\theta=2$, $p_{iC}=0.4$, $q=0.75$, equal sample sizes $n=40,\;100,\;250,\;1000$. 
		\label{BiasThetaLOR2q075piC04}}
\end{figure}

\begin{figure}[t]
	\centering
	\includegraphics[scale=0.33]{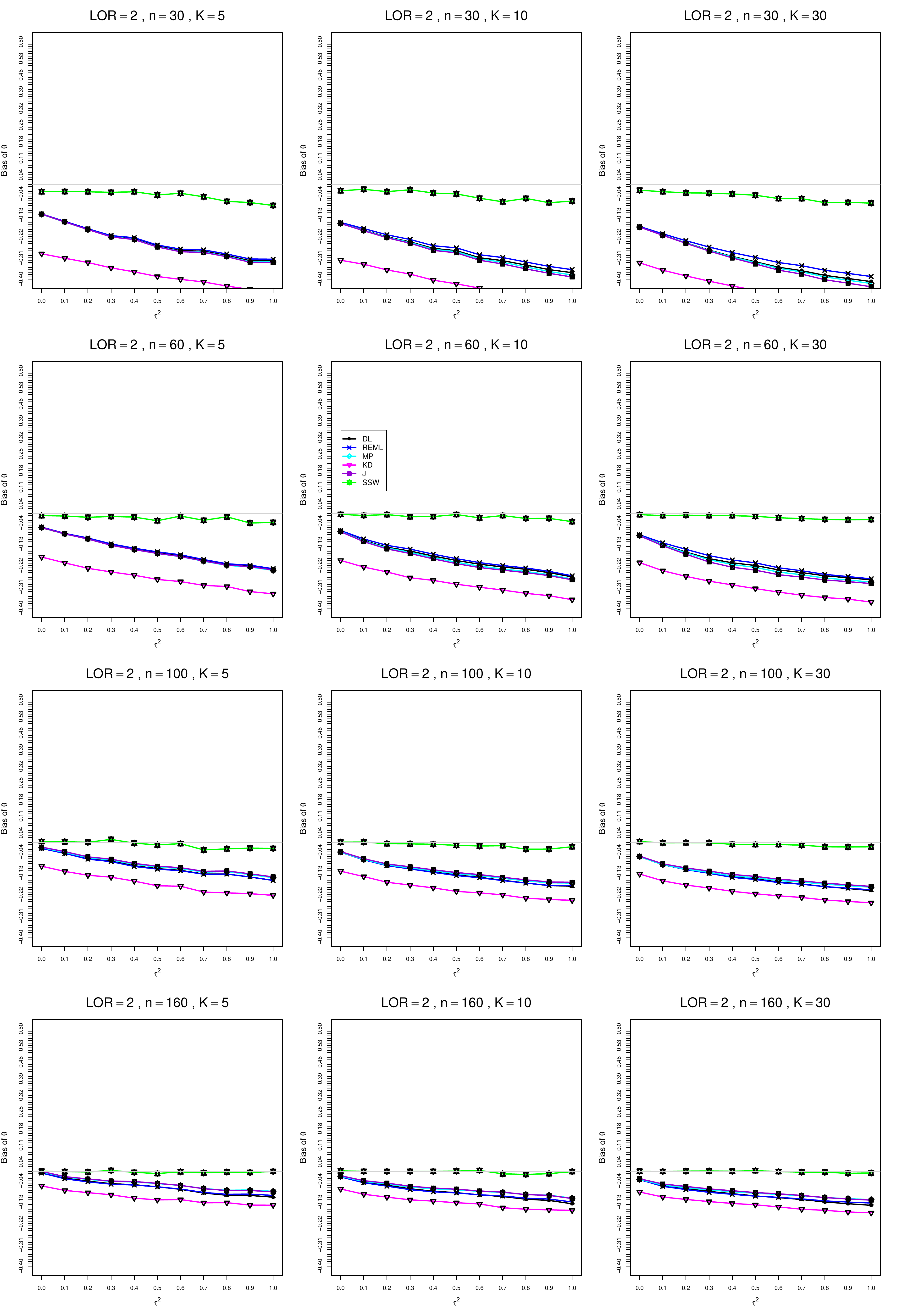}
	\caption{Bias of the estimation of  overall effect measure $\theta$ for $\theta=2$, $p_{iC}=0.4$, $q=0.75$, 
		unequal sample sizes $n=30,\; 60,\;100,\;160$. 
		\label{BiasThetaLOR2q075piC04_unequal_sample_sizes}}
\end{figure}

\begin{figure}[t]\centering
	\includegraphics[scale=0.35]{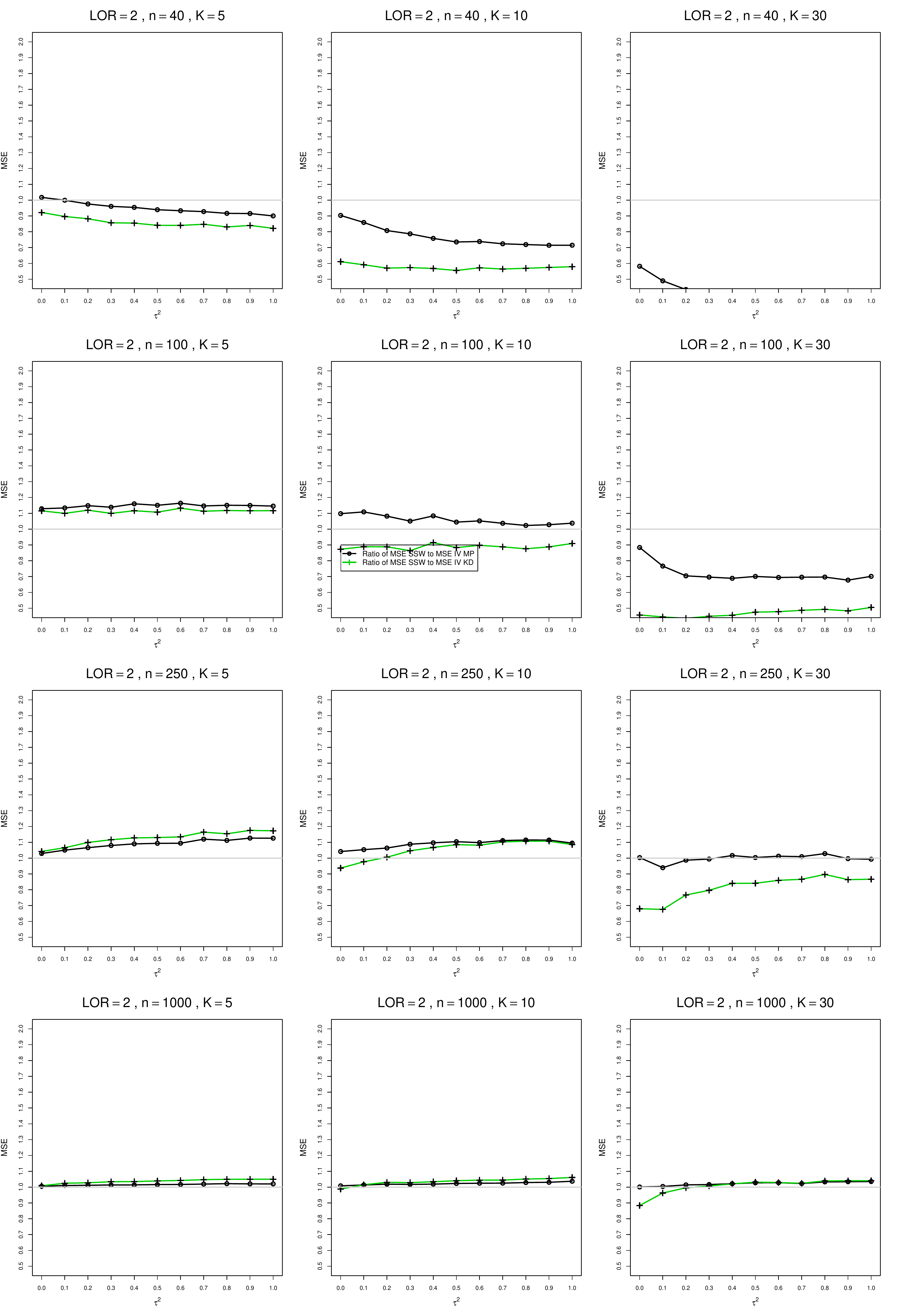}
	\caption{Ratio of mean squared errors of the fixed-weights to mean squared errors of inverse-variance estimator for $\theta=2$, $p_{iC}=0.4$, $q=0.75$, equal sample sizes $n=40,\;100,\;250,\;1000$. 
		\label{RatioOfMSEwithLOR2q075piC04fromMPandCMP}}
\end{figure}

\begin{figure}[t]\centering
	\includegraphics[scale=0.35]{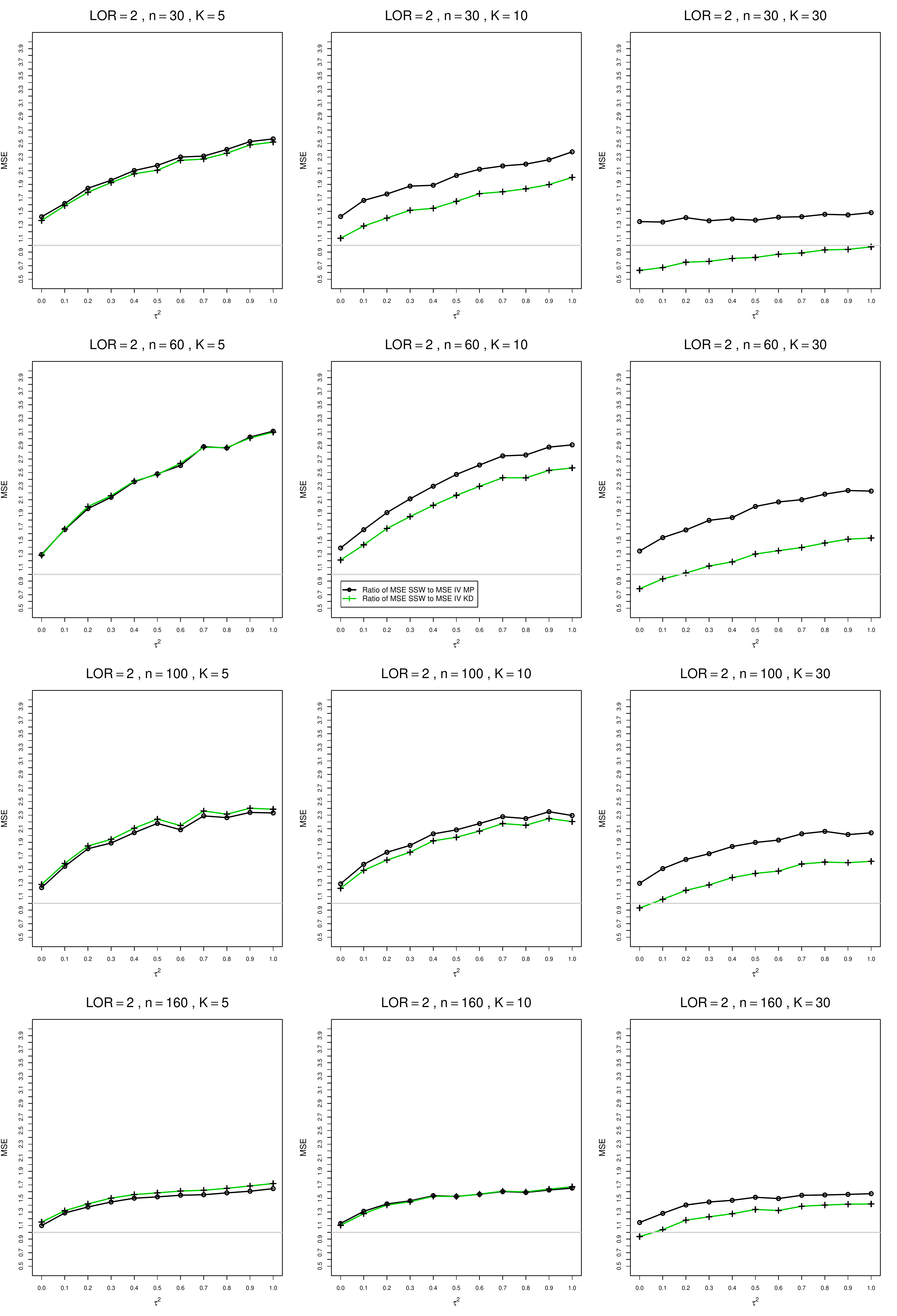}
	\caption{Ratio of mean squared errors of the fixed-weights to mean squared errors of inverse-variance estimator for $\theta=2$, $p_{iC}=0.4$, $q=0.75$, unequal sample sizes $n=30,\;60,\;100,\;160$. 
		\label{RatioOfMSEwithLOR2q075piC04fromMPandCMP_unequal_sample_sizes}}
\end{figure}


\clearpage

\renewcommand{\thefigure}{B2.1.\arabic{figure}}
\setcounter{figure}{0}
\section{Coverage of log-odds-ratio.}
Subsections B2.1, B2.2 and B2.3 correspond to $p_{C}=0.1,\; 0.2,\; 0.4$ respectively. 
For a given $p_{C}$ value, each figure corresponds to a value of $\theta (= 0, 0.5, 1, 1.5, 2)$, a value of $q (= 0.5, 0.75)$, a value of $\tau^2 = 0.0(0.1)1$, and a set of values of $n$ (= 40, 100, 250, 1000) or $\bar{n}$ (= 30, 60, 100, 160).\\
Each figure contains a panel (with $\tau^2$ on the horizontal axis) for each combination of n (or $\bar{n}$) and $K (=5, 10, 30)$.\\
The interval estimators of $\theta$ are the companions to the inverse-variance-weighted point estimators
\begin{itemize}
	\item DL (DerSimonian-Laird)
	\item REML (restricted maximum likelihood)
	\item MP (Mandel-Paule)
	\item KD (Improved moment estimator based on Kulinskaya and Dollinger (2015)) 
	\item J (Jackson)
\end{itemize}
and
\begin{itemize}
	\item HKSJ (Hartung-Knapp-Sidik-Jonkman)
	\item HKSJ KD (HKSJ with KD estimator of $\tau^2$)
	\item SSW (SSW as center and half-width equal to critical value from $t_{K-1}$
\end{itemize}
times estimated standard deviation of SSW with $\hat{\tau}^2$ = $\hat{\tau}^2_{KD}$

\clearpage

\subsection*{B2.1 Probability in the control arm $p_{C}=0.1$}
\begin{figure}[t]
	\centering
	\includegraphics[scale=0.33]{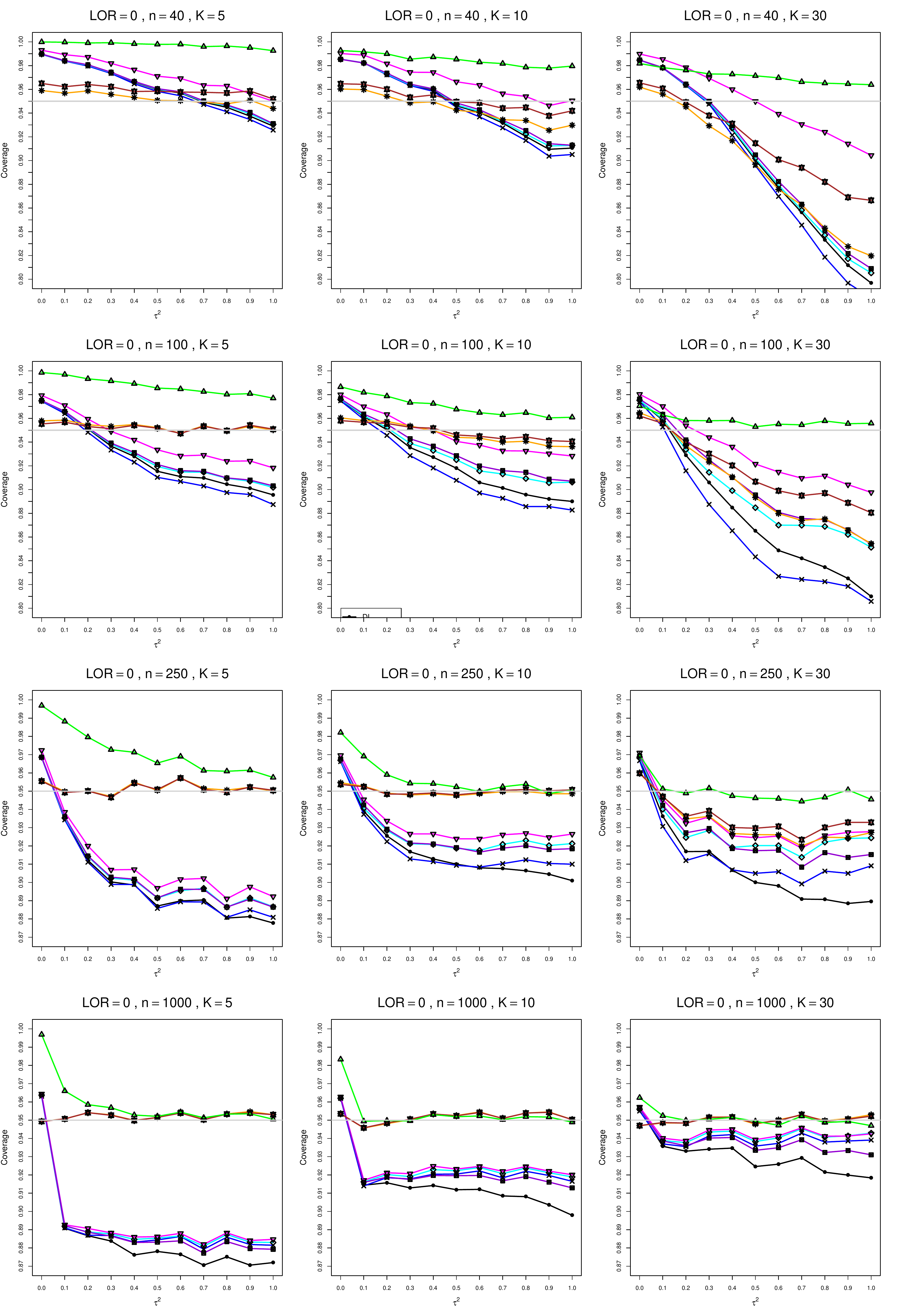}
	\caption{Coverage of  overall effect measure $\theta$ for $\theta=0$, $p_{iC}=0.1$, $q=0.5$, equal sample sizes $n=40,\;100,\;250,\;1000$. 
		\label{CovThetaLOR0q05piC01}}
\end{figure}

\begin{figure}[t]
	\centering
	\includegraphics[scale=0.33]{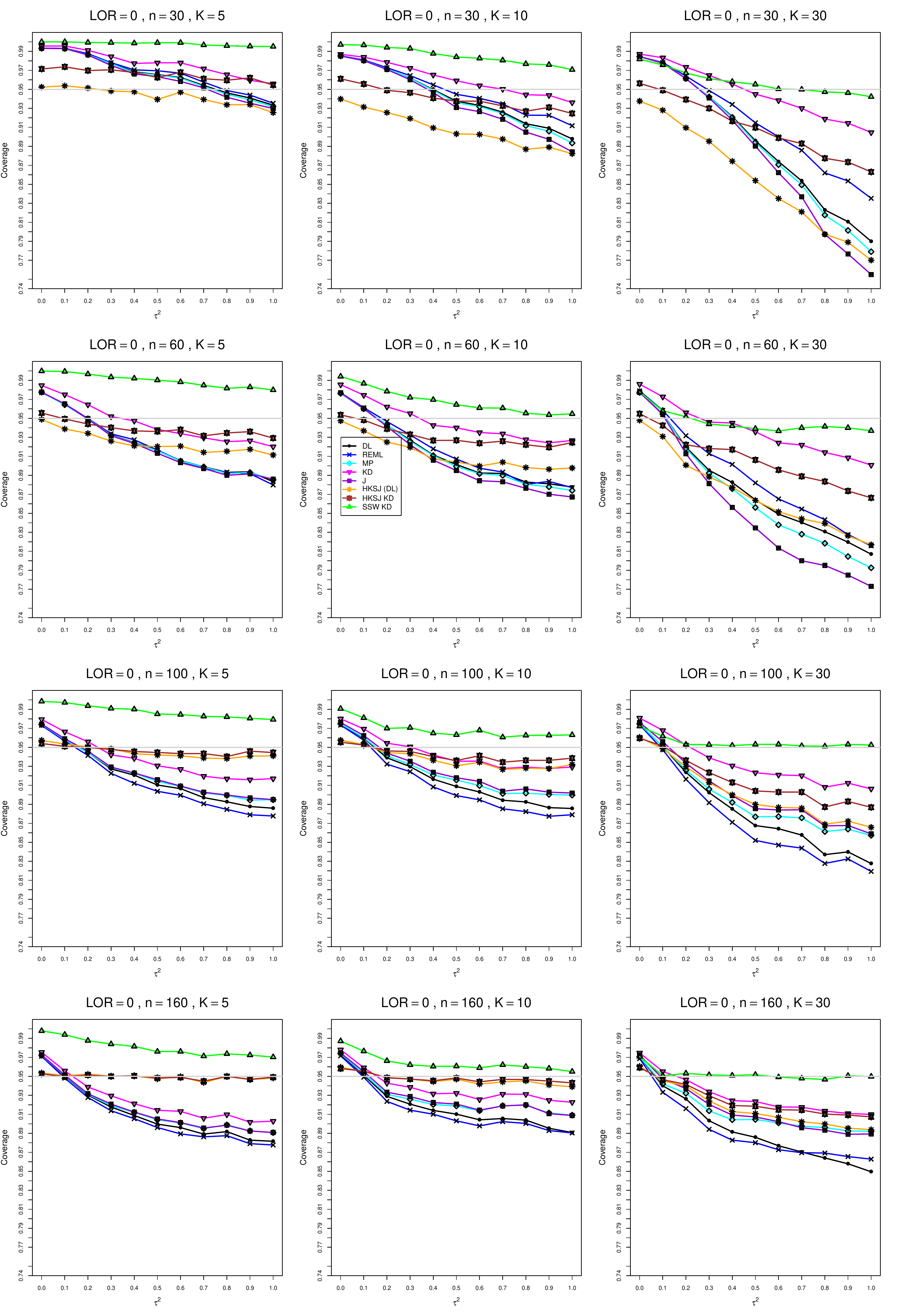}
	\caption{Coverage of  overall effect measure $\theta$ for $\theta=0$, $p_{iC}=0.1$, $q=0.5$, 
		unequal sample sizes $n=30,\; 60,\;100,\;160$. 
		\label{CovThetaLOR0q05piC01_unequal_sample_sizes}}
\end{figure}

\begin{figure}[t]
	\centering
	\includegraphics[scale=0.33]{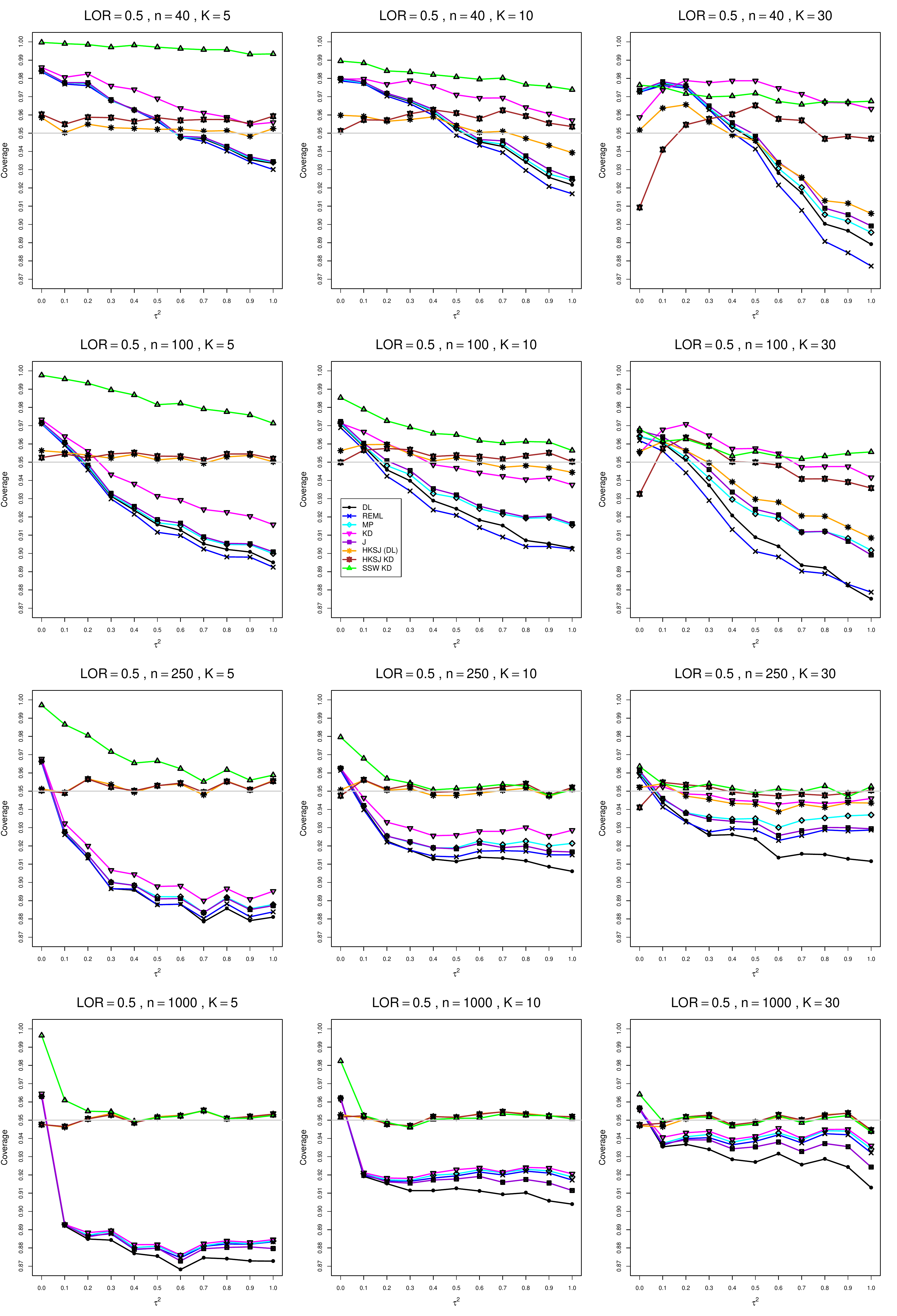}
	\caption{Coverage of  overall effect measure $\theta$ for $\theta=0.5$, $p_{iC}=0.1$, $q=0.5$, equal sample sizes $n=40,\;100,\;250,\;1000$. 
		\label{CovThetaLOR05q05piC01}}
\end{figure}

\begin{figure}[t]
	\centering
	\includegraphics[scale=0.33]{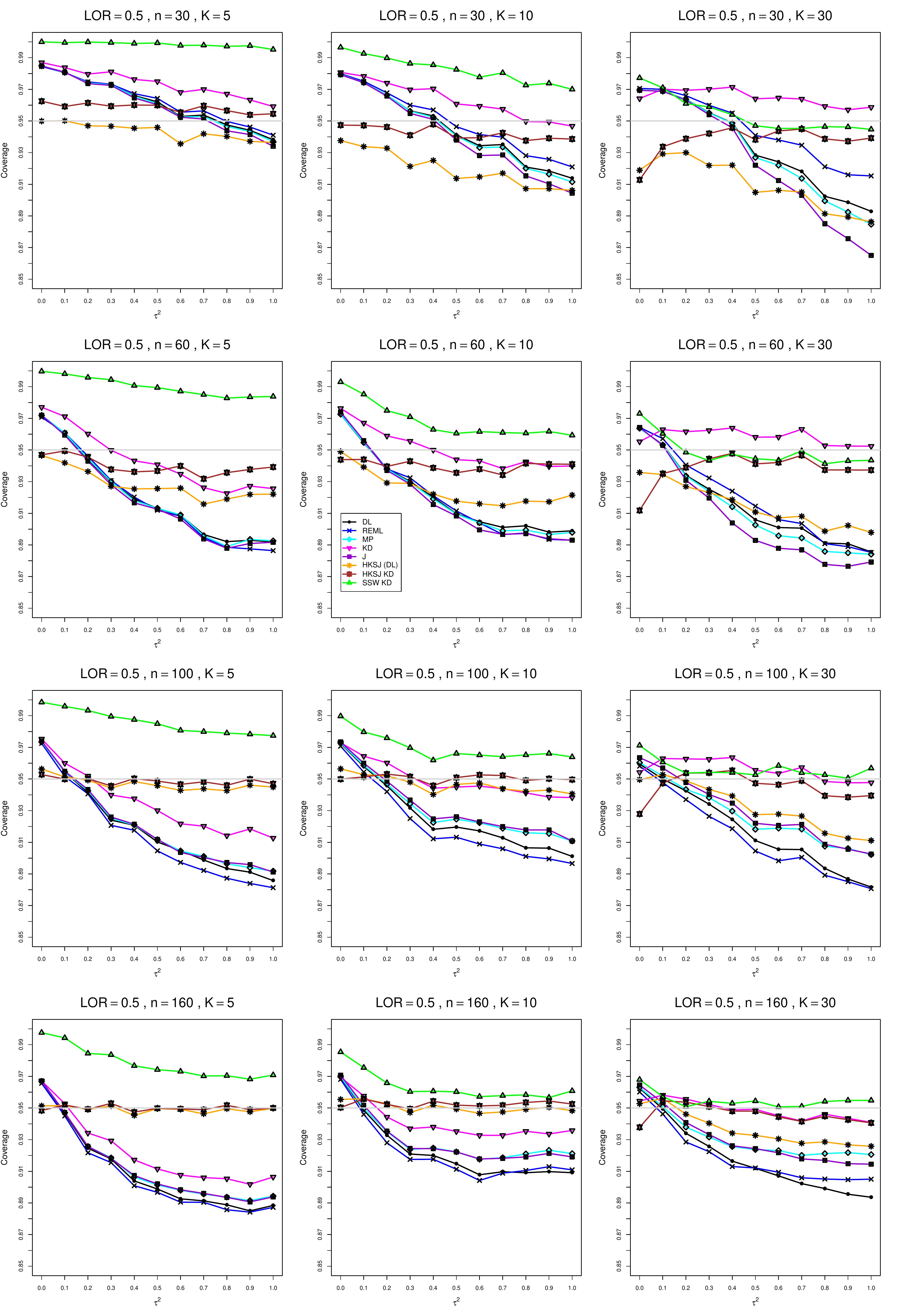}
	\caption{Coverage of  overall effect measure $\theta$ for $\theta=0.5$, $p_{iC}=0.1$, $q=0.5$, 
		unequal sample sizes $n=30,\; 60,\;100,\;160$. 
		\label{CovThetaLOR05q05piC01_unequal_sample_sizes}}
\end{figure}

\begin{figure}[t]
	\centering
	\includegraphics[scale=0.33]{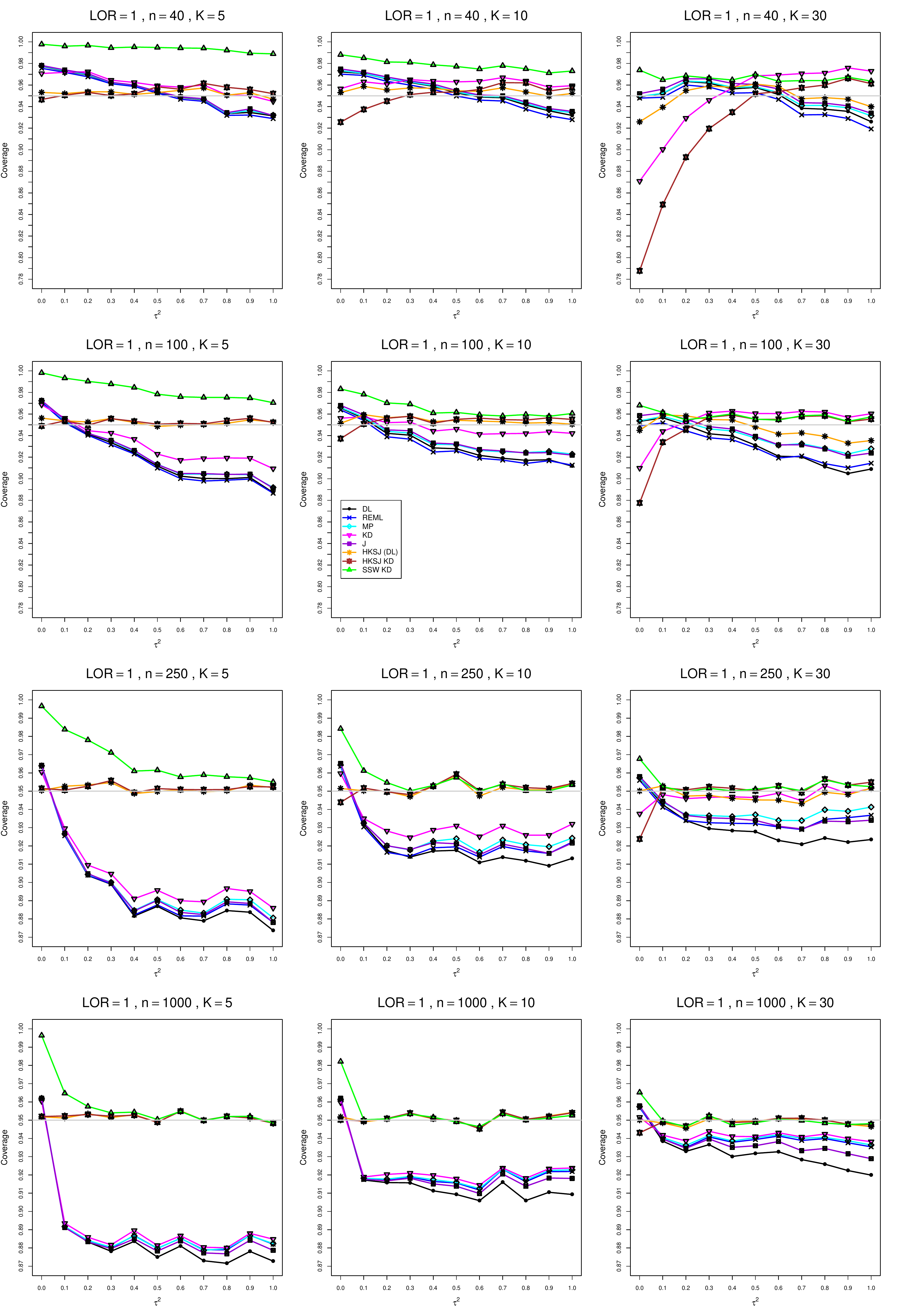}
	\caption{Coverage of  overall effect measure $\theta$ for $\theta=1$, $p_{iC}=0.1$, $q=0.5$, equal sample sizes $n=40,\;100,\;250,\;1000$. 
		\label{CovThetaLOR1q05piC01}}
\end{figure}

\begin{figure}[t]
	\centering
	\includegraphics[scale=0.33]{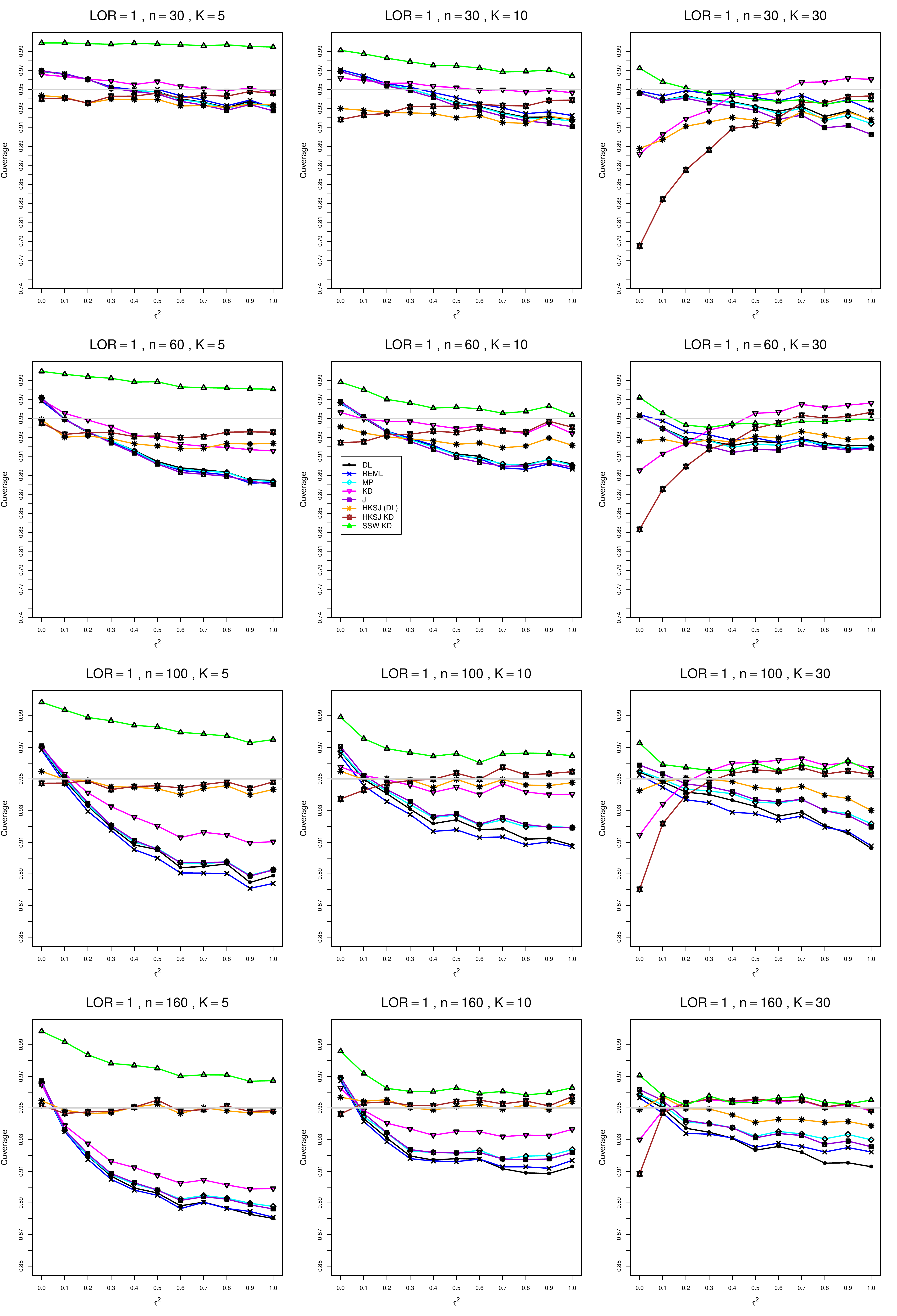}
	\caption{Coverage of  overall effect measure $\theta$ for $\theta=1$, $p_{iC}=0.1$, $q=0.5$, 
		unequal sample sizes $n=30,\; 60,\;100,\;160$. 
		\label{CovThetaLOR1q05piC01_unequal_sample_sizes}}
\end{figure}

\begin{figure}[t]
	\centering
	\includegraphics[scale=0.33]{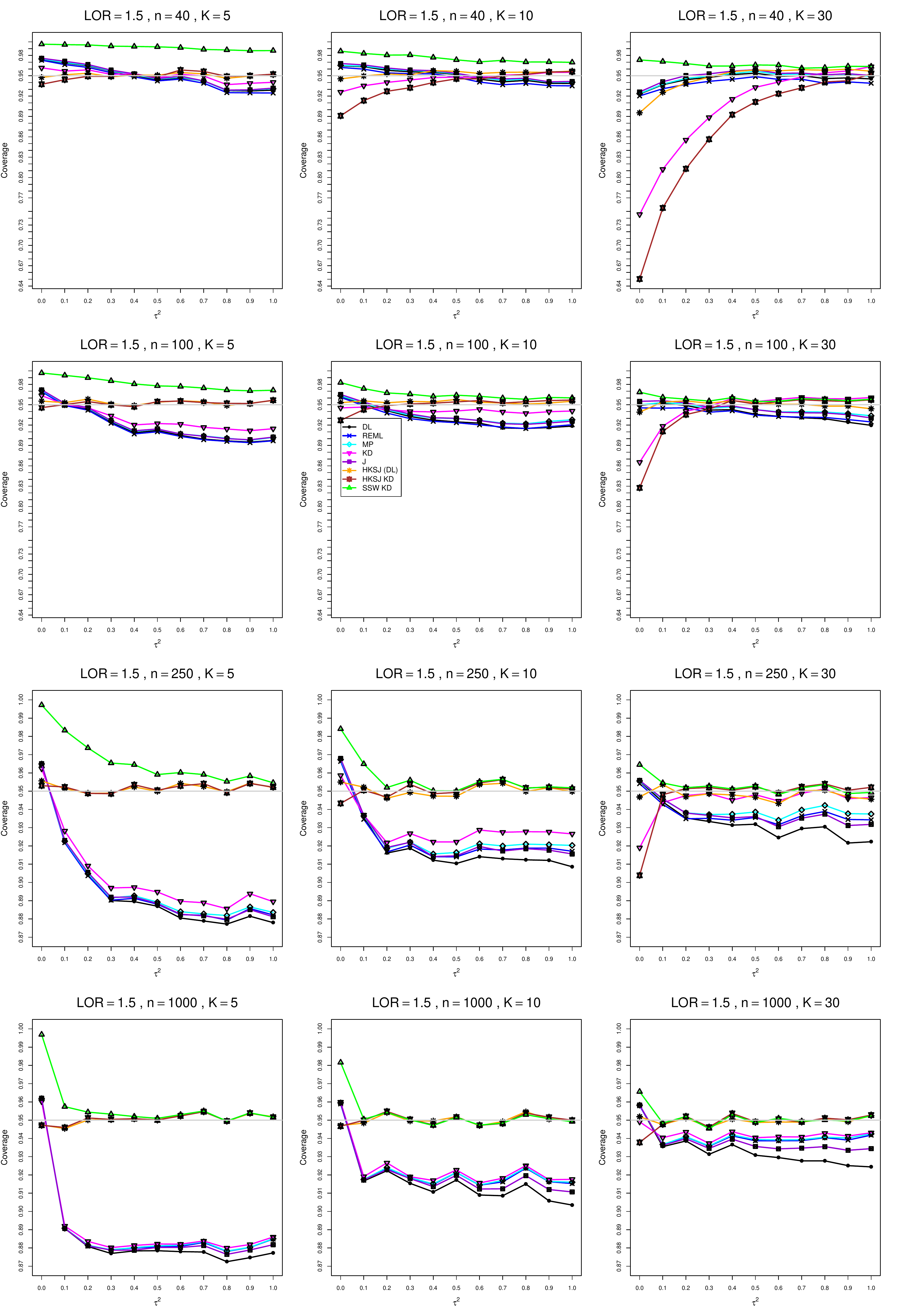}
	\caption{Coverage of  overall effect measure $\theta$ for $\theta=1.5$, $p_{iC}=0.1$, $q=0.5$, equal sample sizes $n=40,\;100,\;250,\;1000$. 
		\label{CovThetaLOR15q05piC01}}
\end{figure}

\begin{figure}[t]
	\centering
	\includegraphics[scale=0.33]{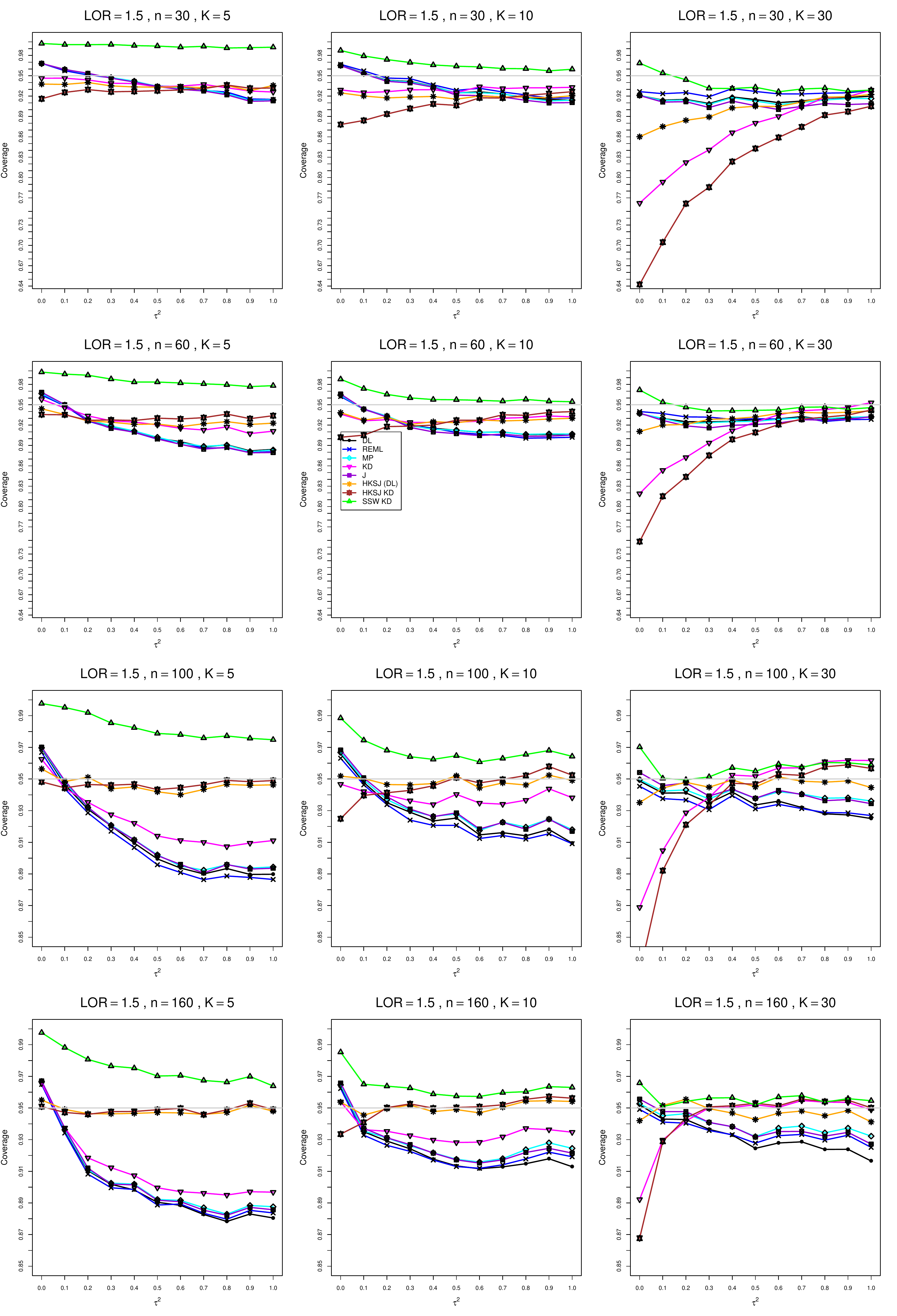}
	\caption{Coverage of  overall effect measure $\theta$ for $\theta=1.5$, $p_{iC}=0.1$, $q=0.5$, 
		unequal sample sizes $n=30,\; 60,\;100,\;160$. 
		\label{CovThetaLOR15q05piC01_unequal_sample_sizes}}
\end{figure}

\begin{figure}[t]
	\centering
	\includegraphics[scale=0.33]{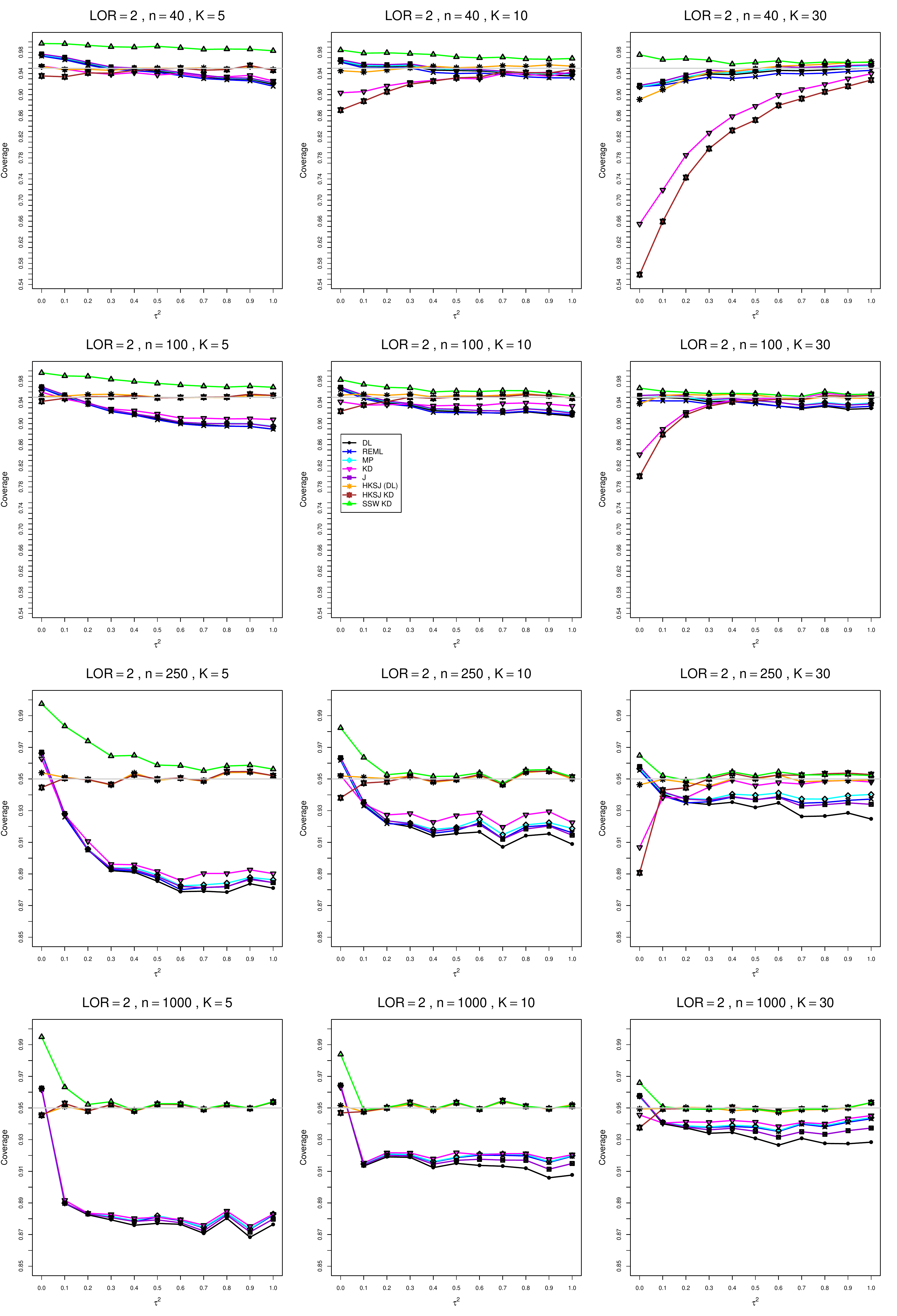}
	\caption{Coverage of  overall effect measure $\theta$ for $\theta=2$, $p_{iC}=0.1$, $q=0.5$, equal sample sizes $n=40,\;100,\;250,\;1000$. 
		\label{CovThetaLOR2q05piC01}}
\end{figure}

\begin{figure}[t]
	\centering
	\includegraphics[scale=0.33]{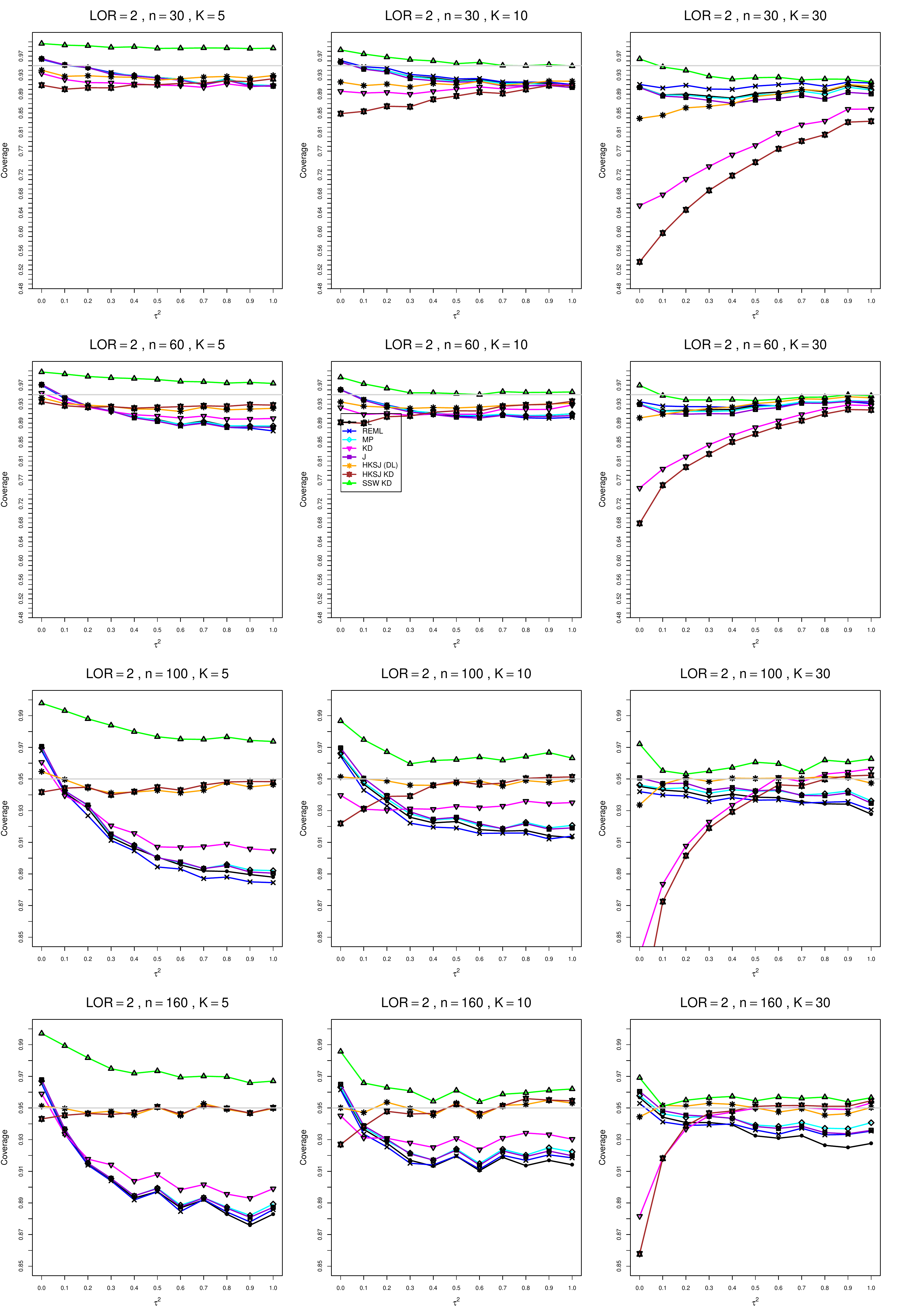}
	\caption{Coverage of  overall effect measure $\theta$ for $\theta=2$, $p_{iC}=0.1$, $q=0.5$, 
		unequal sample sizes $n=30,\; 60,\;100,\;160$. 
		\label{CovThetaLOR2q05piC01_unequal_sample_sizes}}
\end{figure}


\begin{figure}[t]
	\centering
	\includegraphics[scale=0.33]{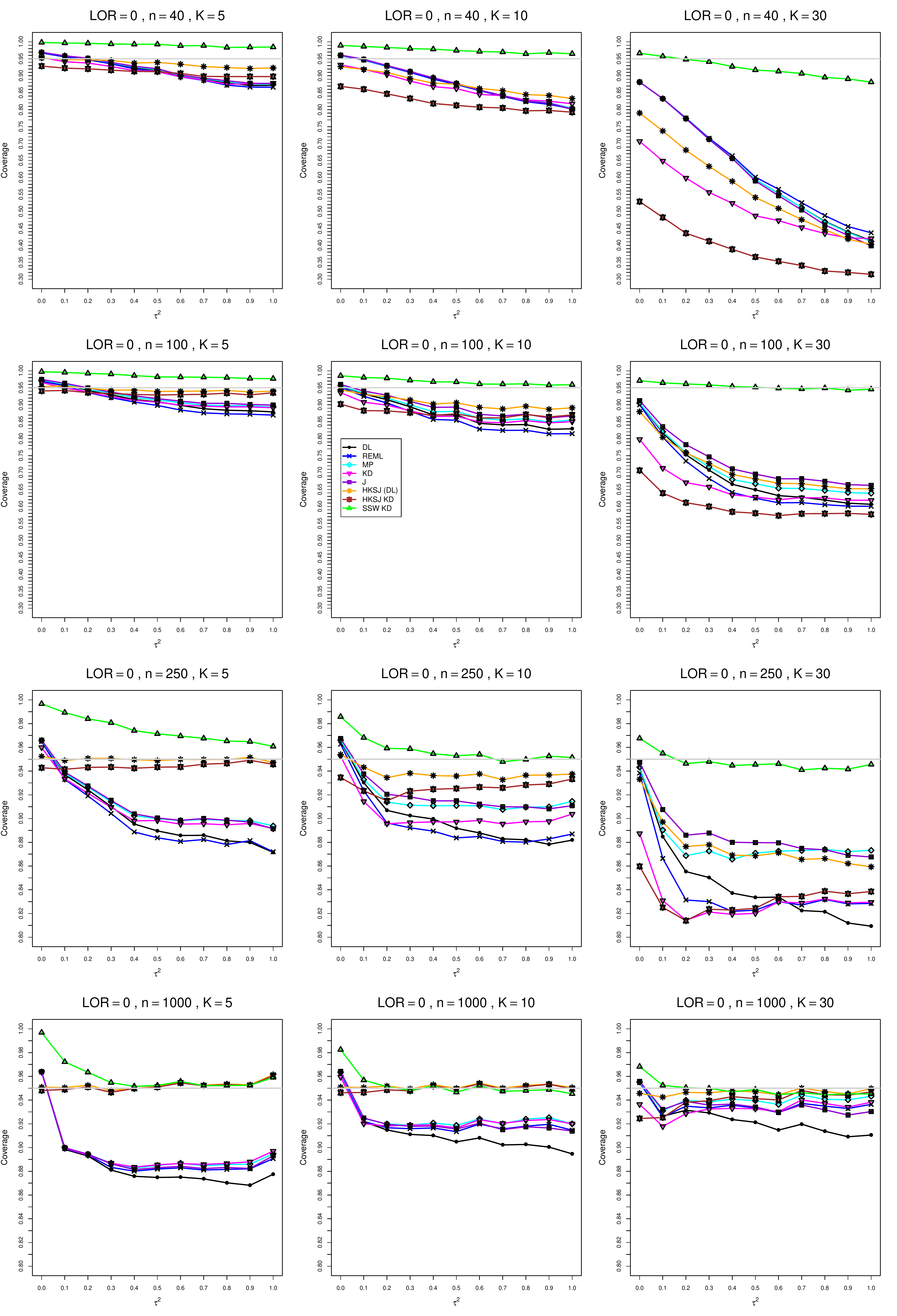}
	\caption{Coverage of  overall effect measure $\theta$ for $\theta=0$, $p_{iC}=0.1$, $q=0.75$, equal sample sizes $n=40,\;100,\;250,\;1000$. 
		\label{CovThetaLOR0q075piC01}}
\end{figure}

\begin{figure}[t]
	\centering
	\includegraphics[scale=0.33]{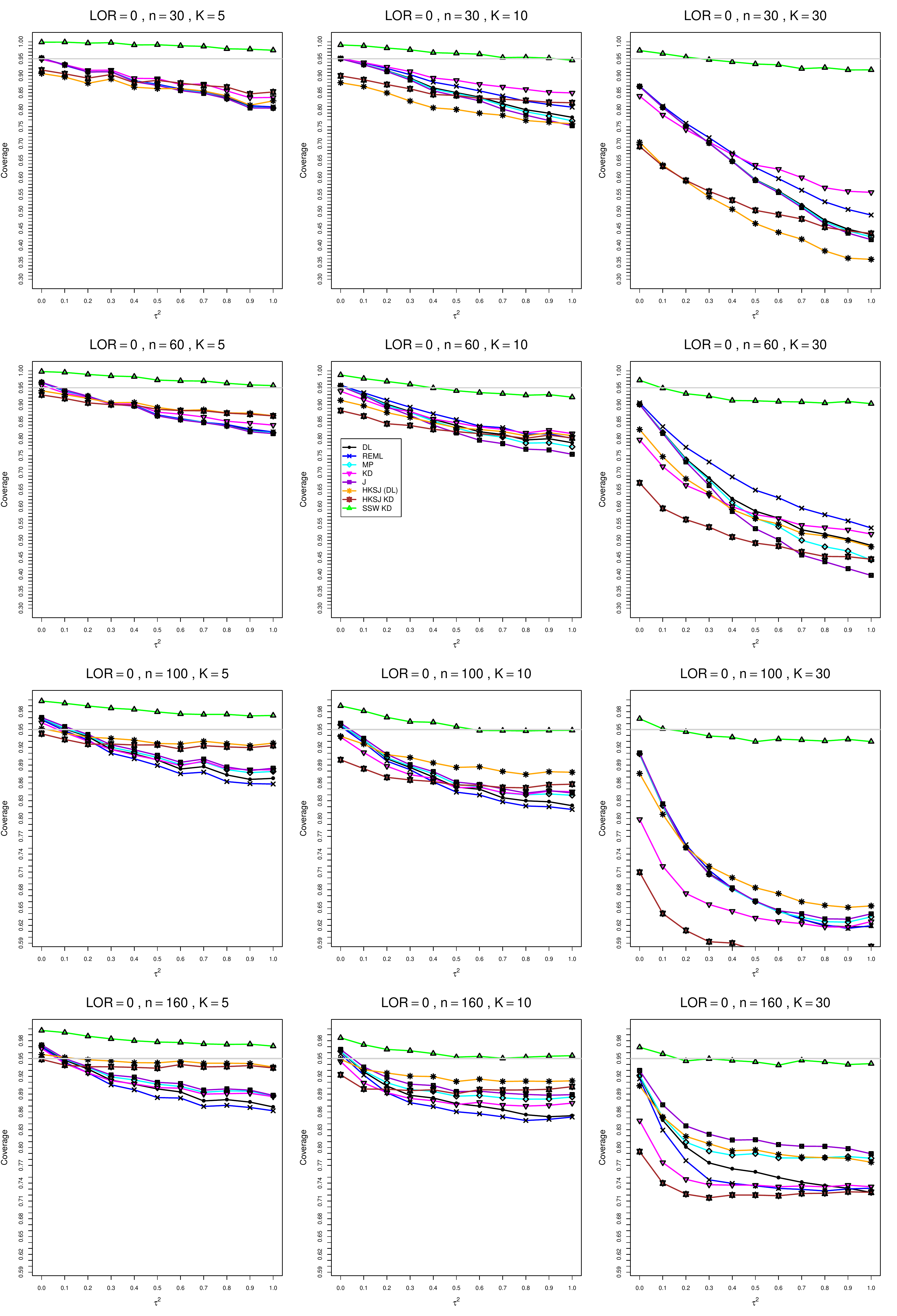}
	\caption{Coverage of  overall effect measure $\theta$ for $\theta=0$, $p_{iC}=0.1$, $q=0.75$, 
		unequal sample sizes $n=30,\; 60,\;100,\;160$. 
		\label{CovThetaLOR0q075piC01_unequal_sample_sizes}}
\end{figure}

\begin{figure}[t]
	\centering
	\includegraphics[scale=0.33]{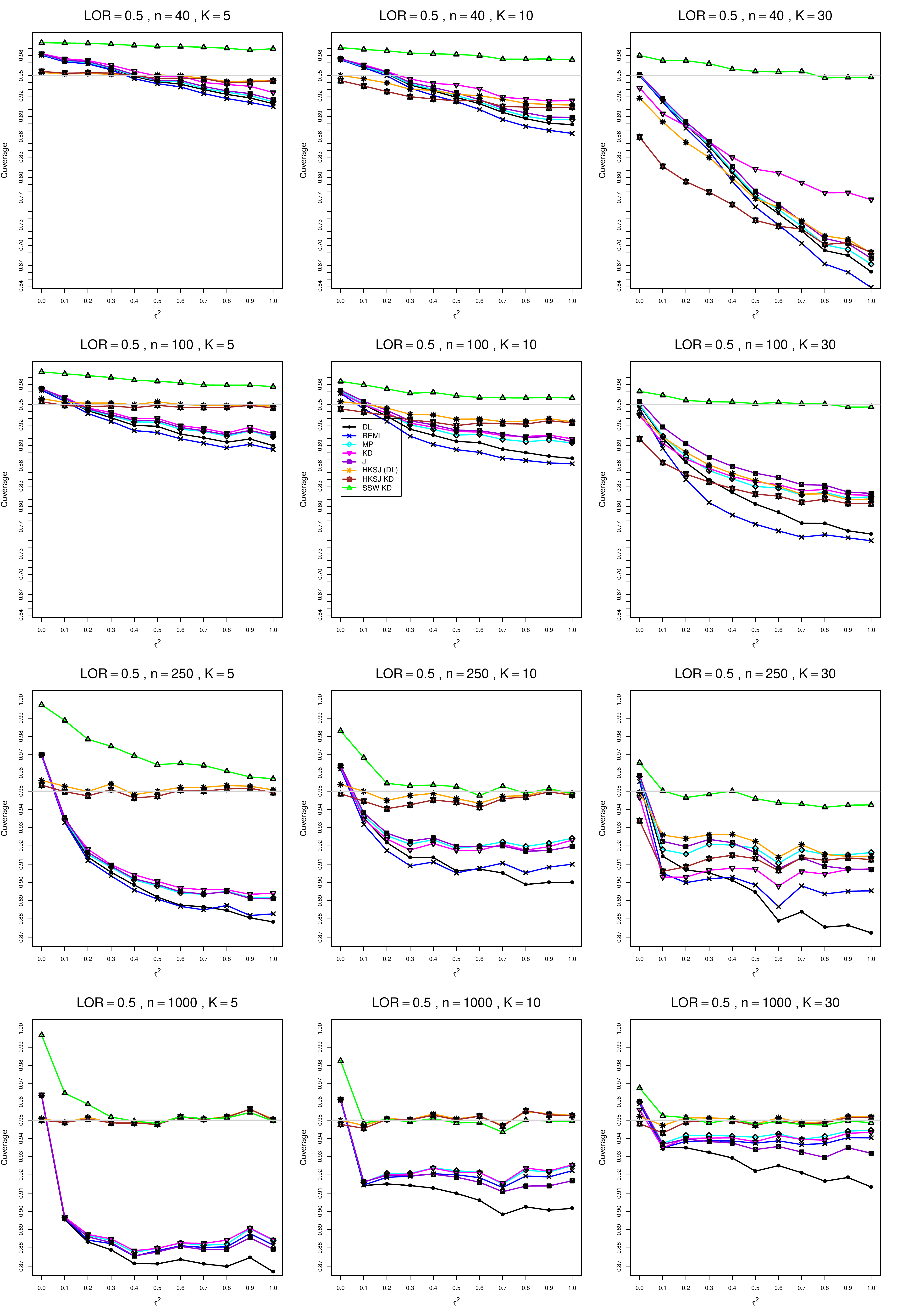}
	\caption{Coverage of  overall effect measure $\theta$ for $\theta=0.5$, $p_{iC}=0.1$, $q=0.75$, equal sample sizes $n=40,\;100,\;250,\;1000$. 
		\label{CovThetaLOR05q075piC01}}
\end{figure}

\begin{figure}[t]
	\centering
	\includegraphics[scale=0.33]{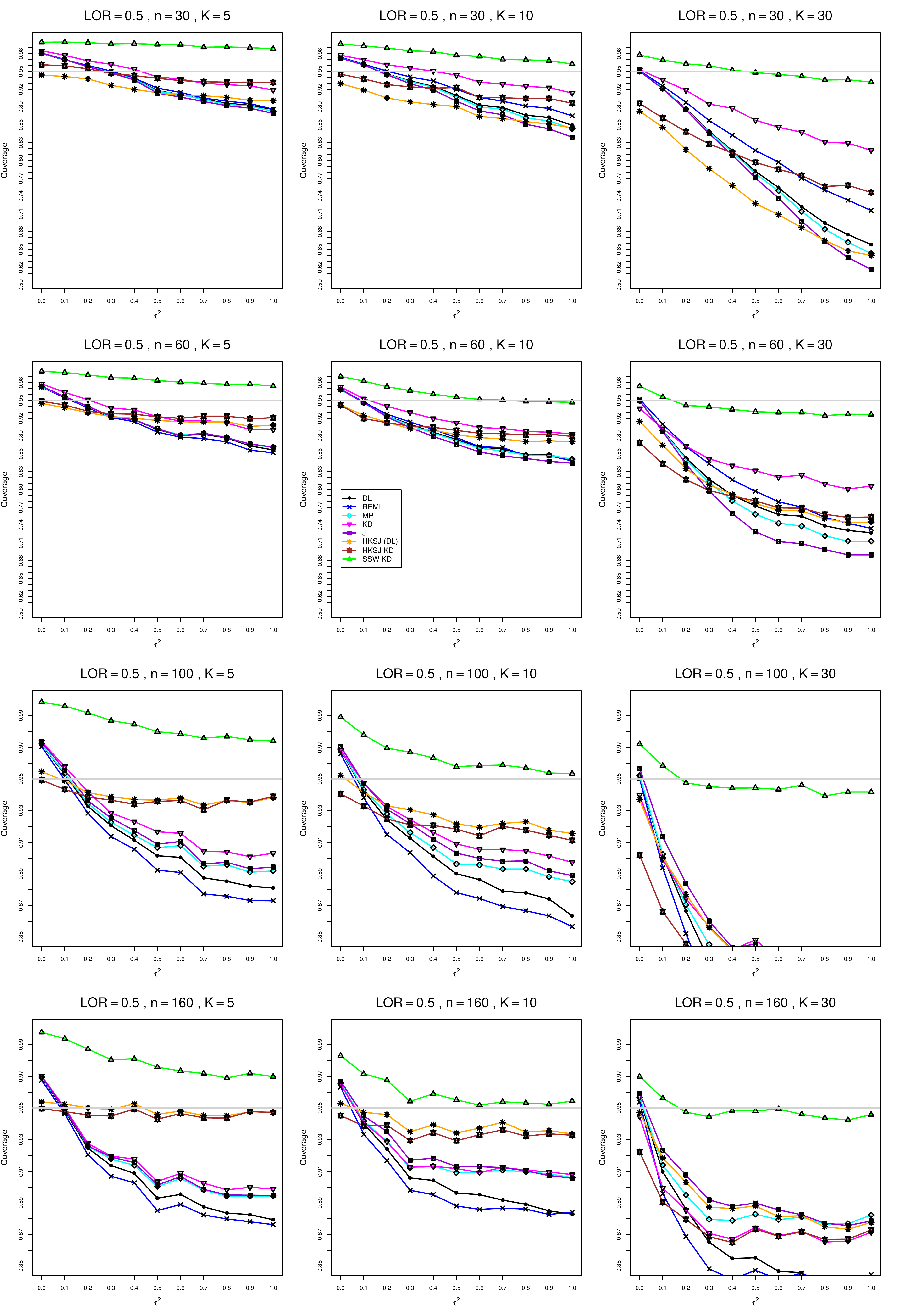}
	\caption{Coverage of  overall effect measure $\theta$ for $\theta=0.5$, $p_{iC}=0.1$, $q=0.75$, 
		unequal sample sizes $n=30,\; 60,\;100,\;160$. 
		\label{CovThetaLOR05q075piC01_unequal_sample_sizes}}
\end{figure}

\begin{figure}[t]
	\centering
	\includegraphics[scale=0.33]{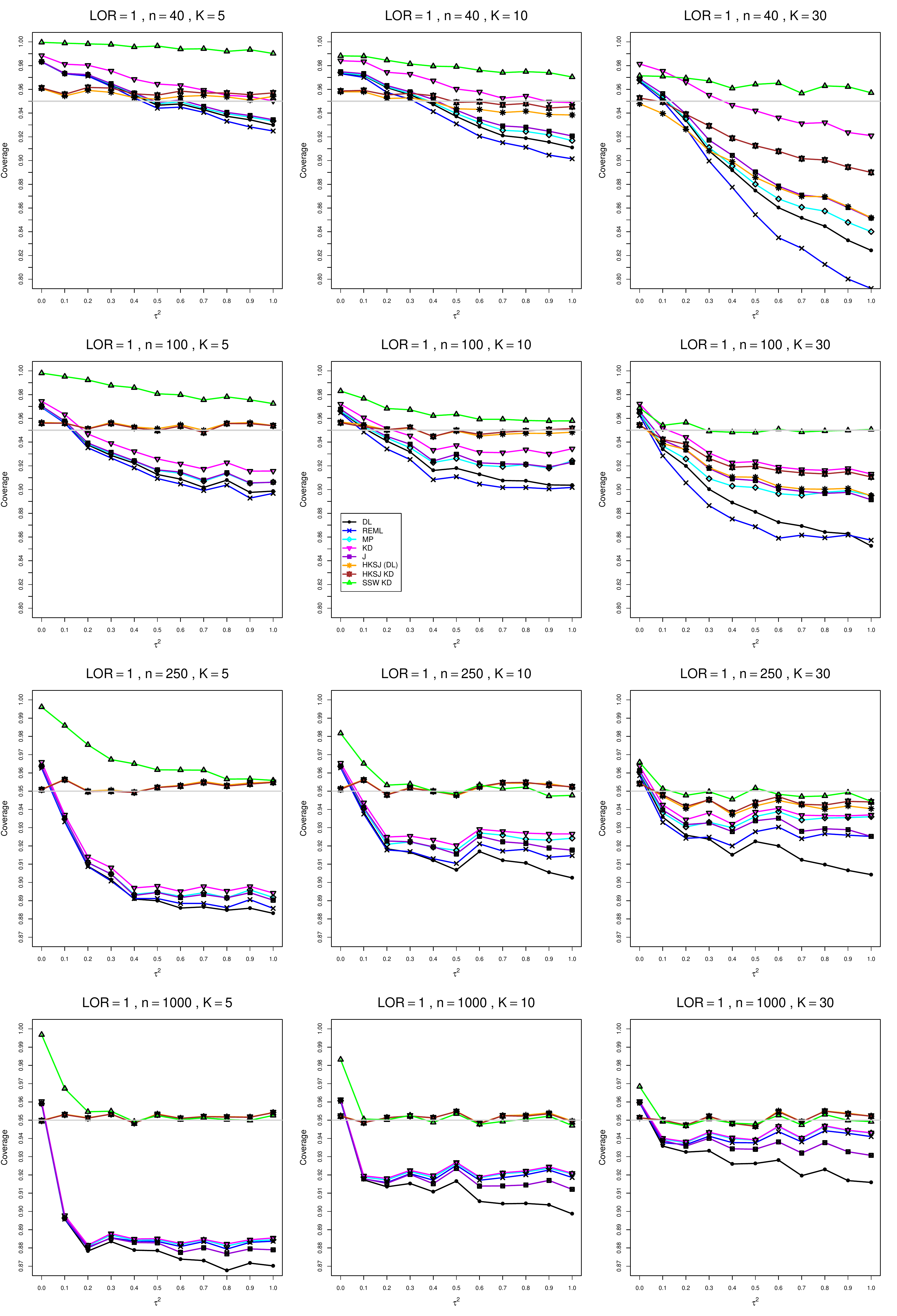}
	\caption{Coverage of  overall effect measure $\theta$ for $\theta=1$, $p_{iC}=0.1$, $q=0.75$, equal sample sizes  $n=40,\;100,\;250,\;1000$. 
		\label{CovThetaLOR1q075piC01}}
\end{figure}

\begin{figure}[t]
	\centering
	\includegraphics[scale=0.33]{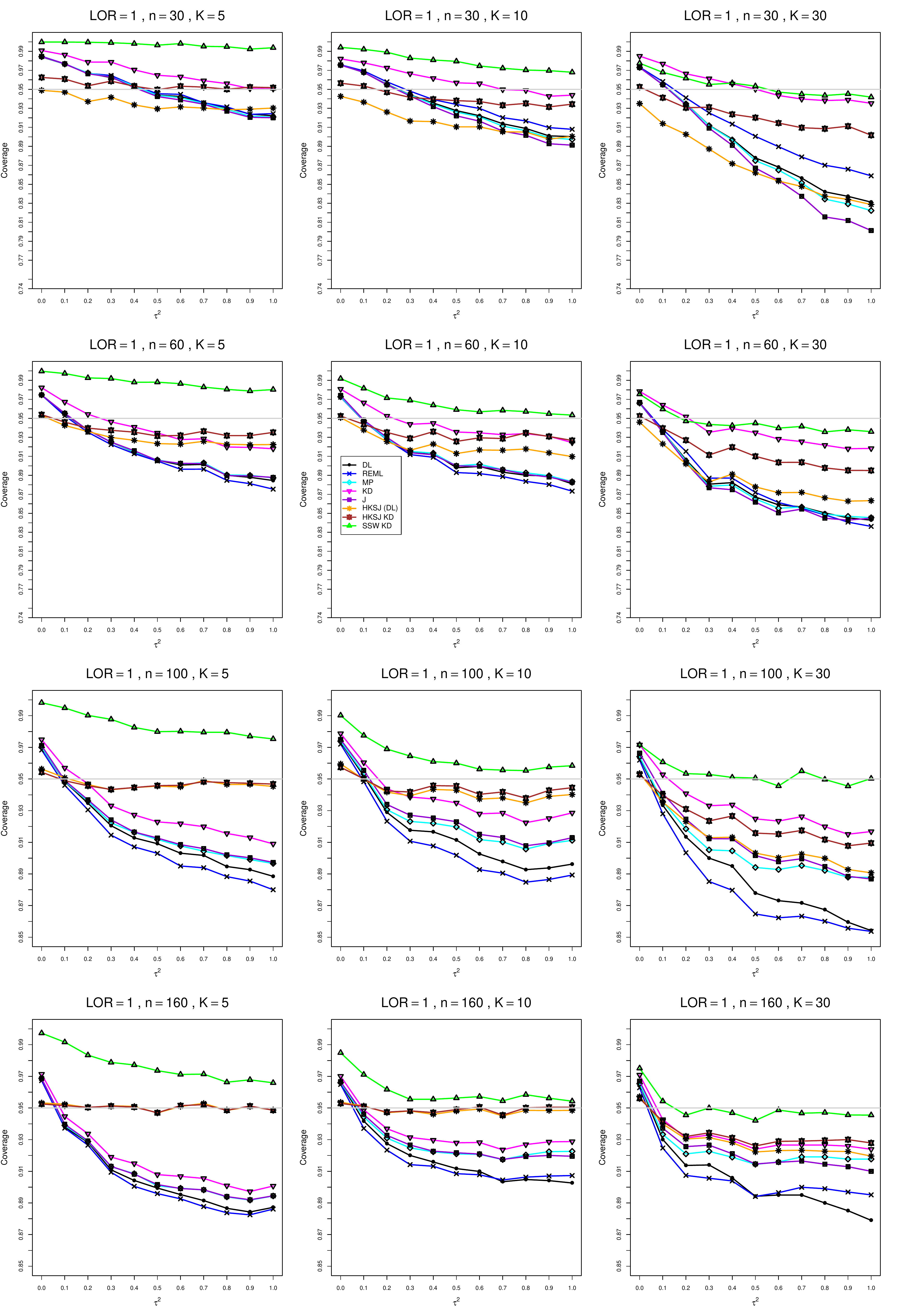}
	\caption{Coverage of  overall effect measure $\theta$ for $\theta=1$, $p_{iC}=0.1$, $q=0.75$, 
		unequal sample sizes $n=30,\; 60,\;100,\;160$. 
		\label{CovThetaLOR1q075piC01_unequal_sample_sizes}}
\end{figure}

\begin{figure}[t]
	\centering
	\includegraphics[scale=0.33]{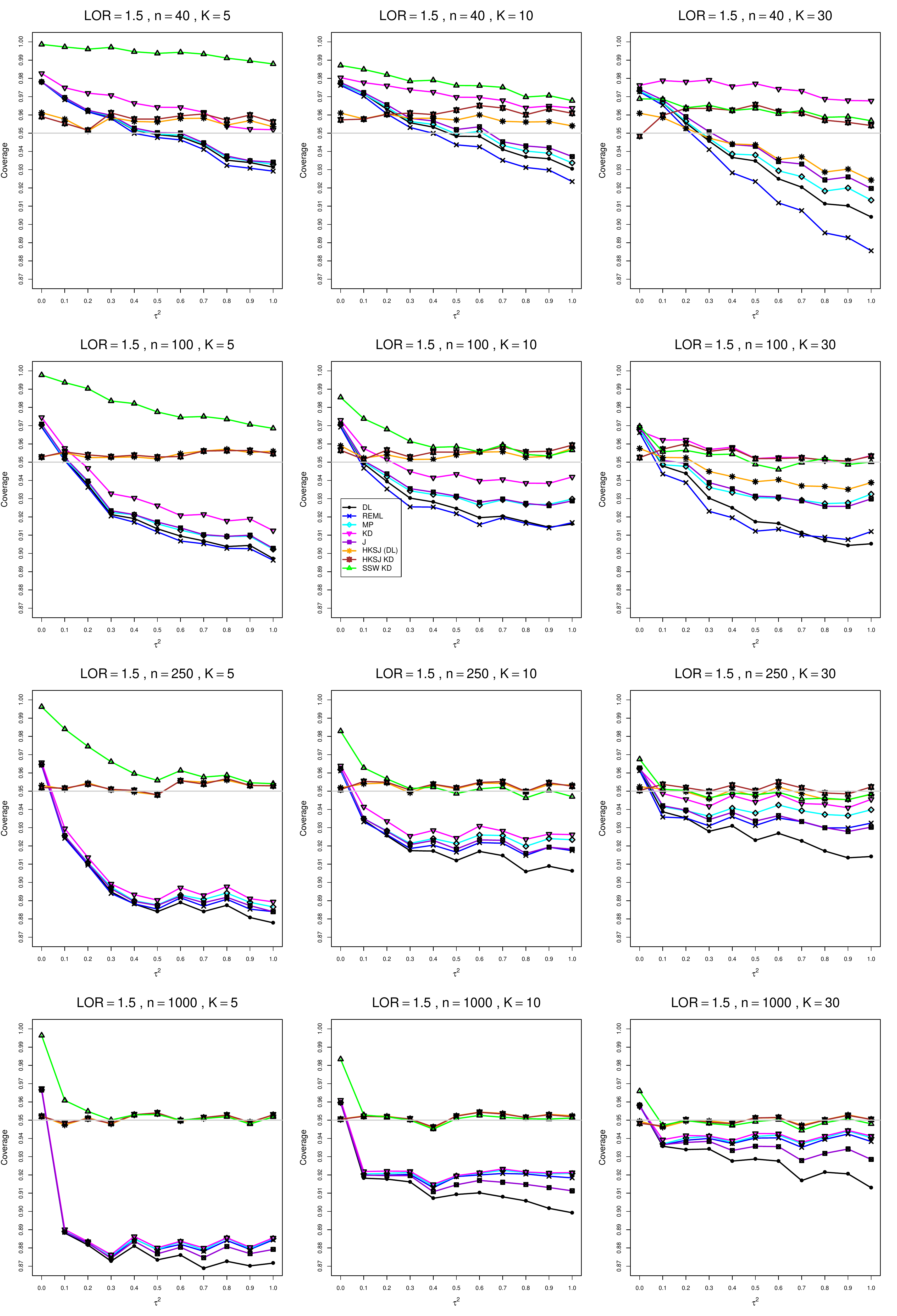}
	\caption{Coverage of  overall effect measure $\theta$ for $\theta=1.5$, $p_{iC}=0.1$, $q=0.75$, equal sample sizes $n=40,\;100,\;250,\;1000$. 
		\label{CovThetaLOR15q075piC01}}
\end{figure}

\begin{figure}[t]
	\centering
	\includegraphics[scale=0.33]{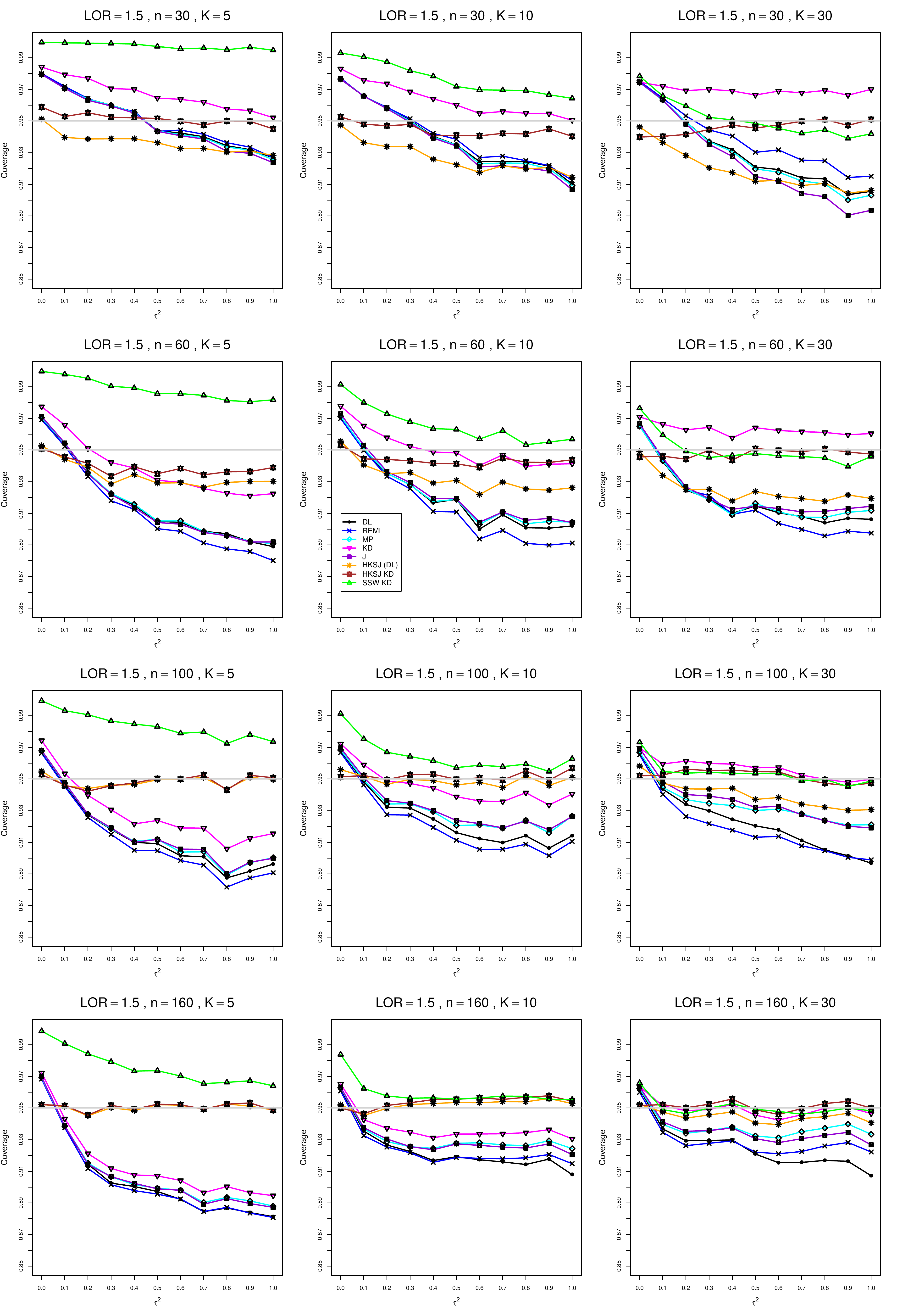}
	\caption{Coverage of  overall effect measure $\theta$ for $\theta=1.5$, $p_{iC}=0.1$, $q=0.75$, 
		unequal sample sizes $n=30,\; 60,\;100,\;160$. 
		\label{CovThetaLOR15q075piC01_unequal_sample_sizes}}
\end{figure}

\clearpage
\begin{figure}[t]
	\centering
	\includegraphics[scale=0.33]{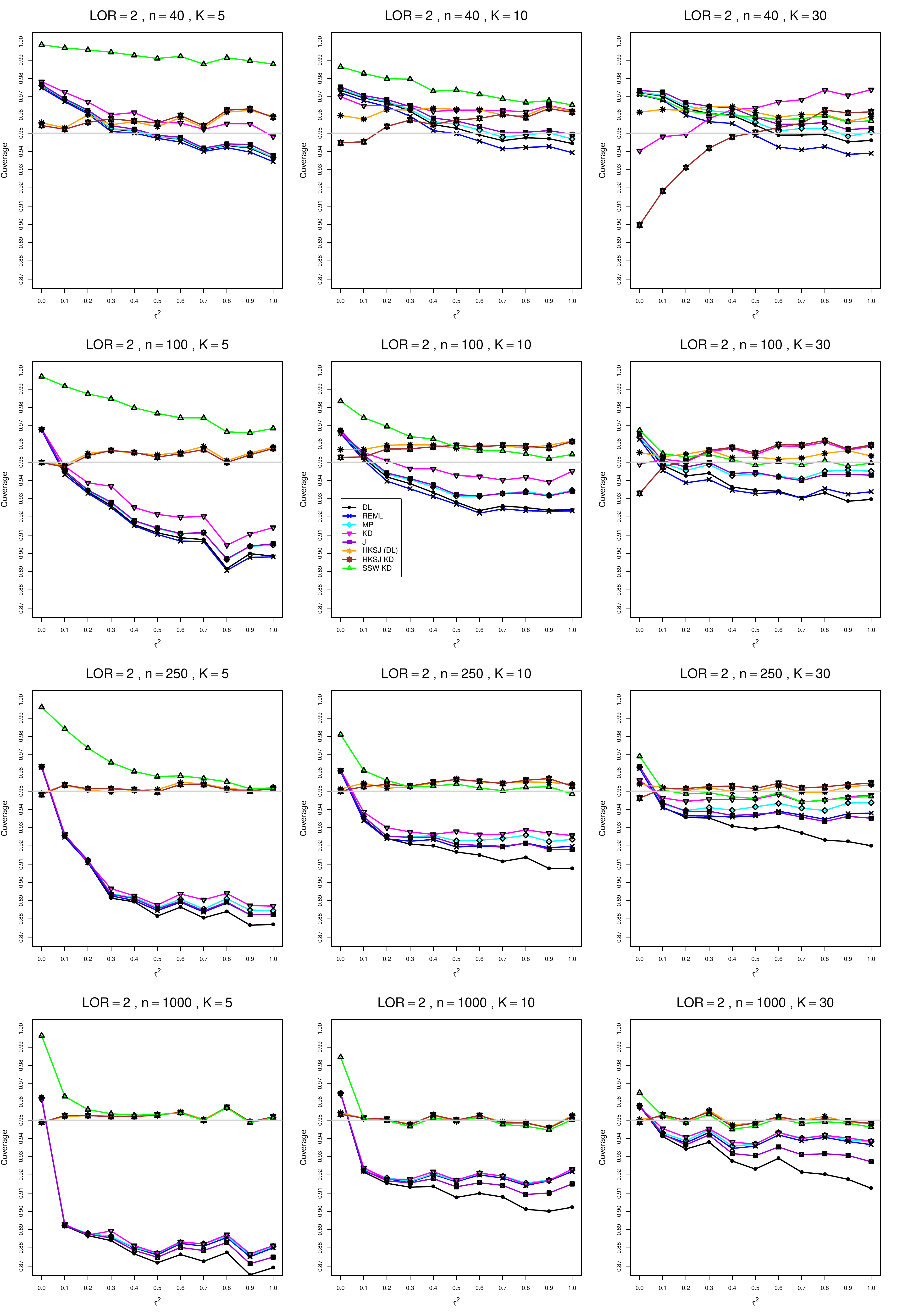}
	\caption{Coverage of  overall effect measure $\theta$ for $\theta=2$, $p_{iC}=0.1$, $q=0.75$, equal sample sizes $n=40,\;100,\;250,\;1000$. 
		\label{CovThetaLOR2q075piC01}}
\end{figure}

\begin{figure}[t]
	\centering
	\includegraphics[scale=0.33]{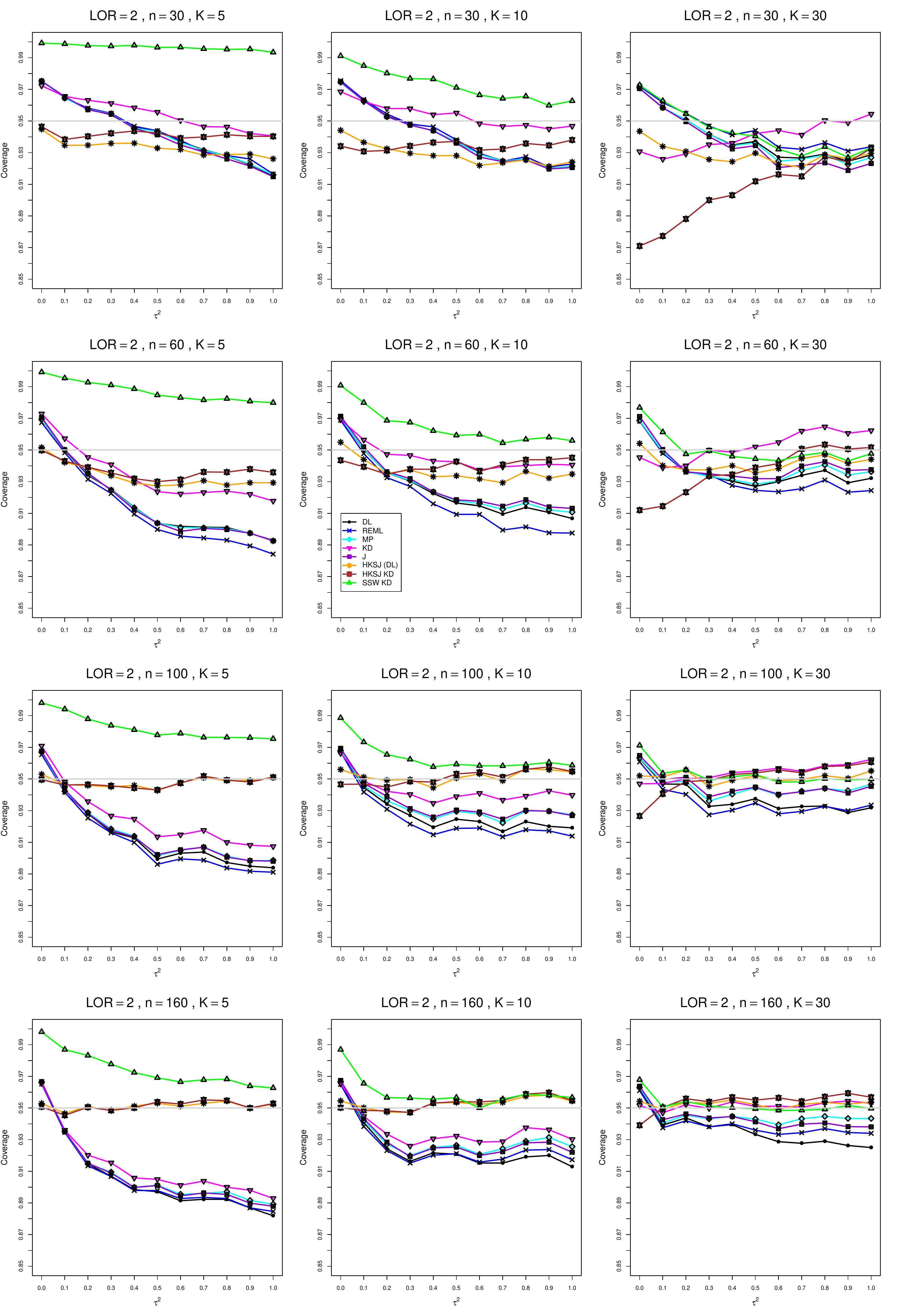}
	\caption{Coverage of  overall effect measure $\theta$ for $\theta=2$, $p_{iC}=0.1$, $q=0.75$, 
		unequal sample sizes $n=30,\; 60,\;100,\;160$. 
		\label{CovThetaLOR2q075piC01_unequal_sample_sizes}}
\end{figure}

\clearpage
\renewcommand{\thefigure}{B2.2.\arabic{figure}}
\setcounter{figure}{0}
\subsection*{B2.2 Probability in the control arm $p_{C}=0.2$}
\begin{figure}[t]
	\centering
	\includegraphics[scale=0.33]{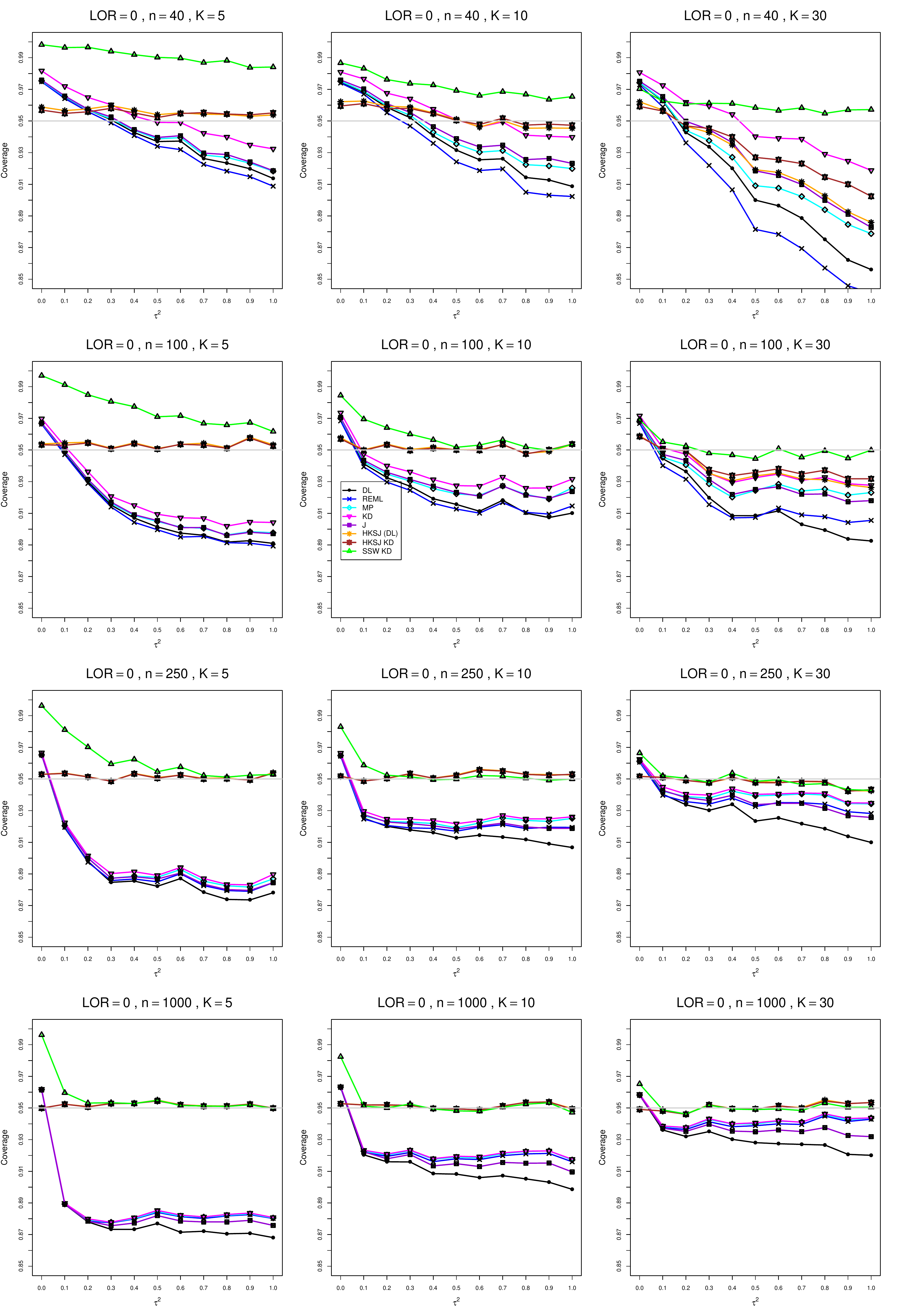}
	\caption{Coverage of  overall effect measure $\theta$ for $\theta=0$, $p_{iC}=0.2$, $q=0.5$, equal sample sizes $n=40,\;100,\;250,\;1000$. 
		\label{CovThetaLOR0q05piC02}}
\end{figure}

\begin{figure}[t]
	\centering
	\includegraphics[scale=0.33]{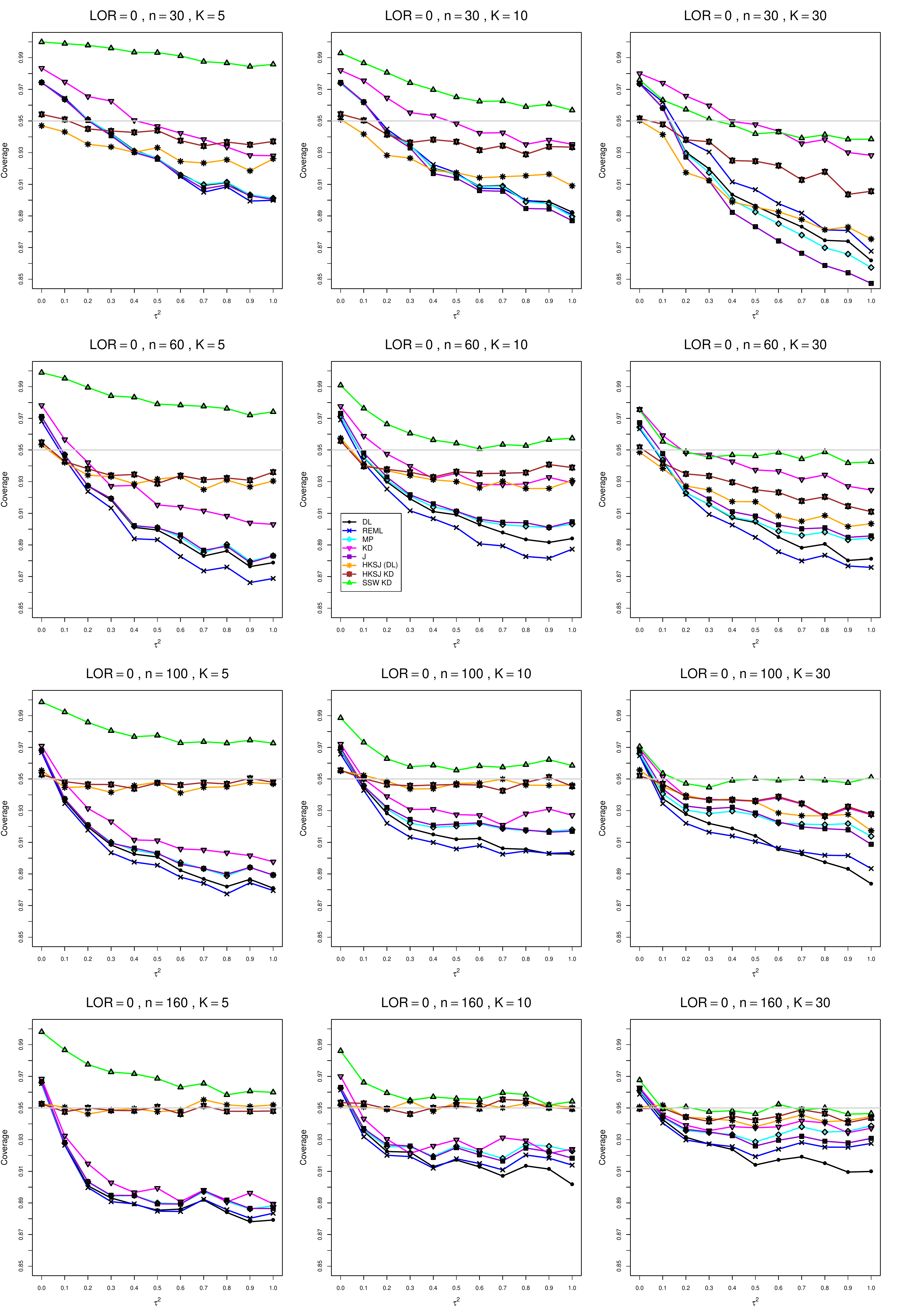}
	\caption{Coverage of  overall effect measure $\theta$ for $\theta=0$, $p_{iC}=0.2$, $q=0.5$, 
		unequal sample sizes $n=30,\; 60,\;100,\;160$. 
		\label{CovThetaLOR0q05piC02_unequal_sample_sizes}}
\end{figure}

\begin{figure}[t]
	\centering
	\includegraphics[scale=0.33]{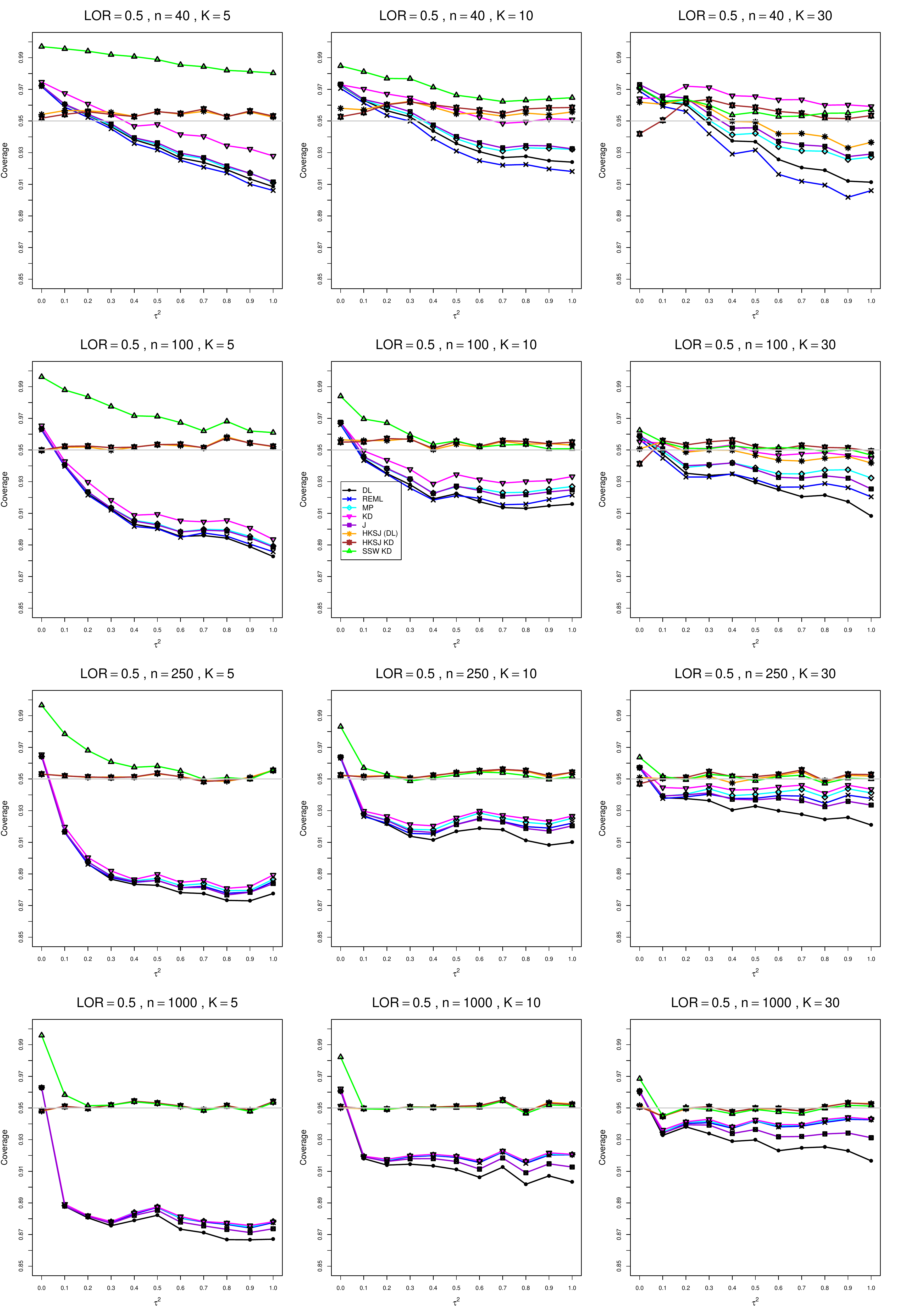}
	\caption{Coverage of  overall effect measure $\theta$ for $\theta=0.5$, $p_{iC}=0.2$, $q=0.5$, equal sample sizes $n=40,\;100,\;250,\;1000$. 
		\label{CovThetaLOR05q05piC02}}
\end{figure}

\begin{figure}[t]
	\centering
	\includegraphics[scale=0.33]{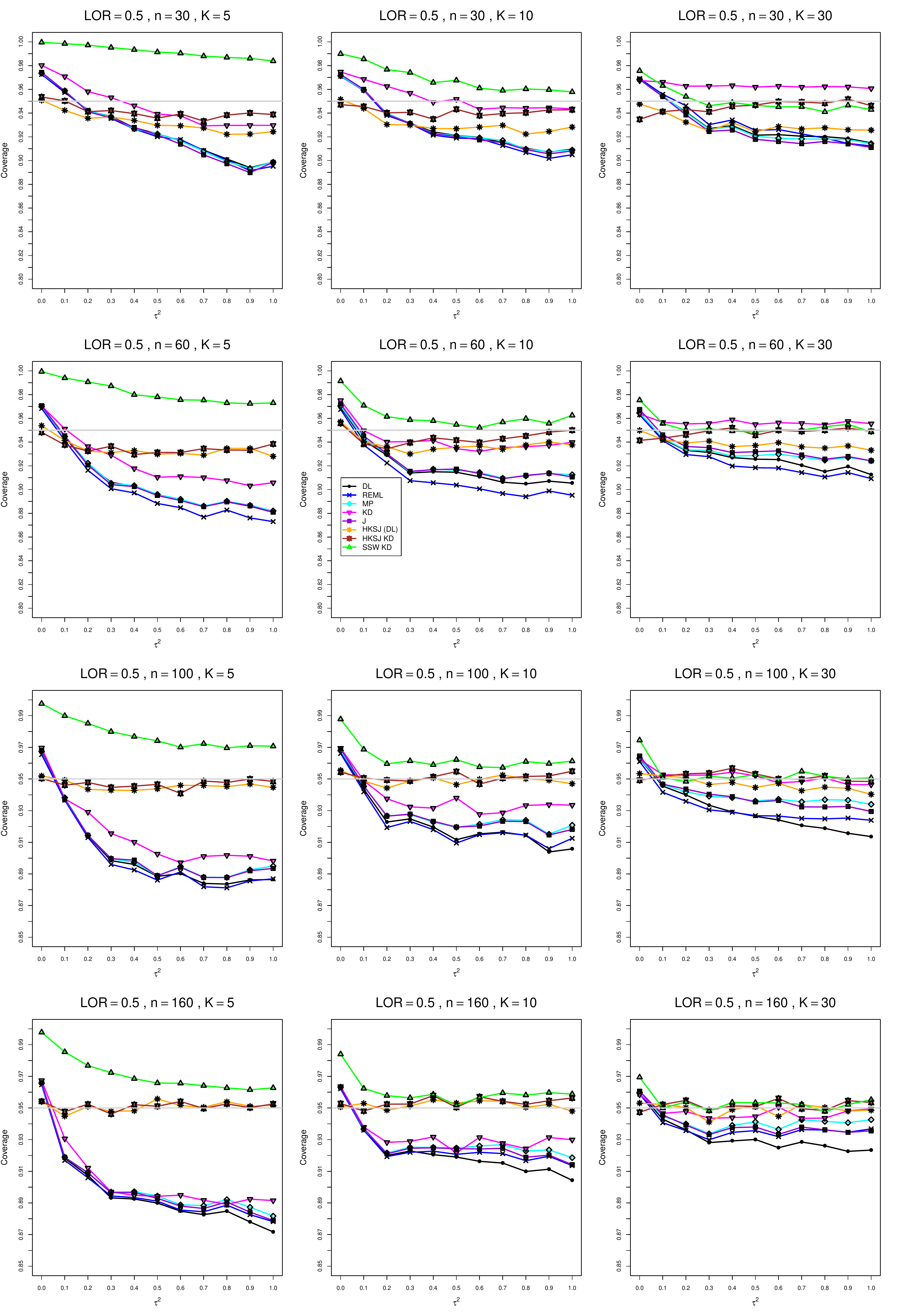}
	\caption{Coverage of  overall effect measure $\theta$ for $\theta=0.5$, $p_{iC}=0.2$, $q=0.5$, 
		unequal sample sizes $n=30,\; 60,\;100,\;160$. 
		\label{CovThetaLOR05q05piC02_unequal_sample_sizes}}
\end{figure}

\begin{figure}[t]
	\centering
	\includegraphics[scale=0.33]{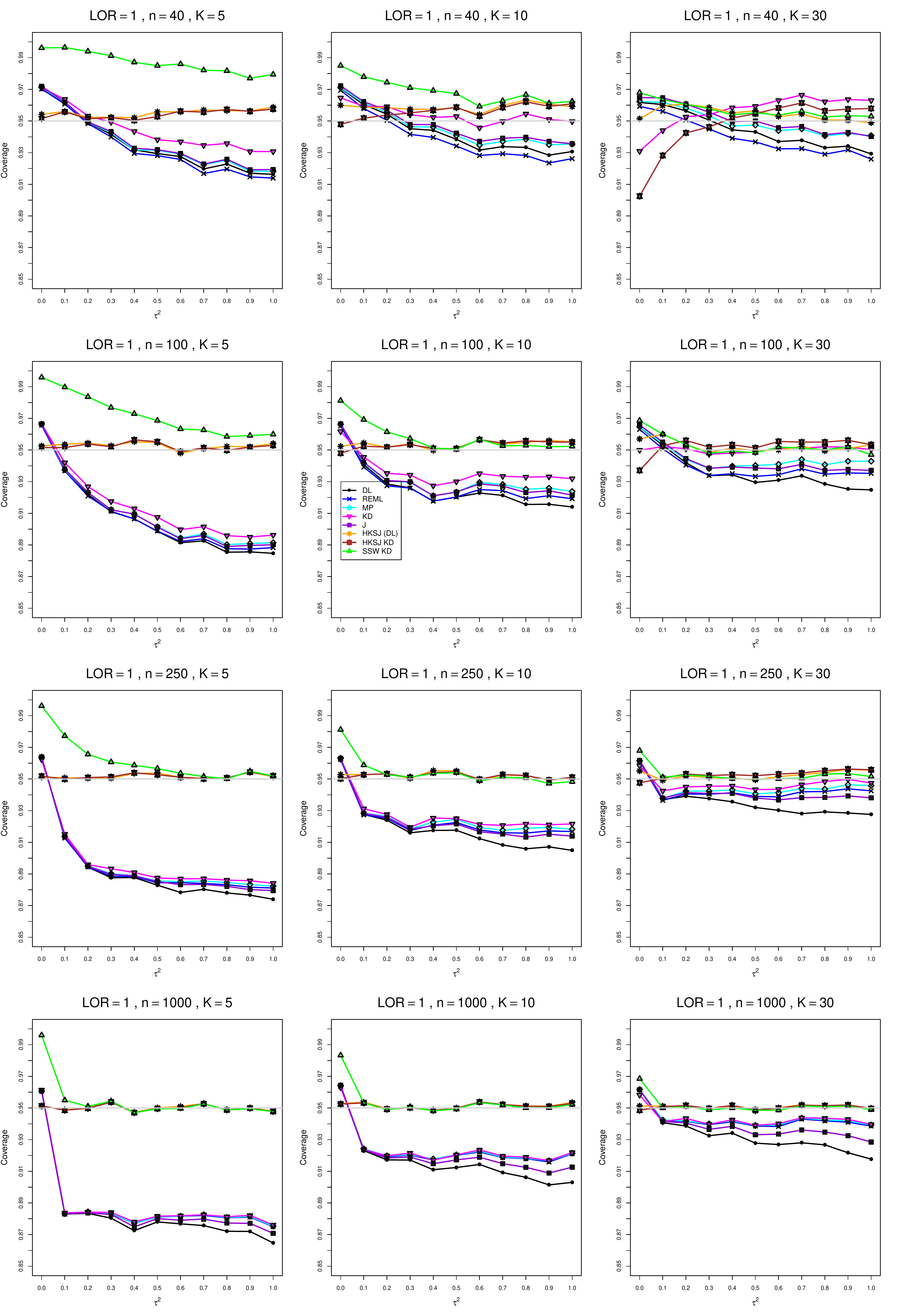}
	\caption{Coverage of  overall effect measure $\theta$ for $\theta=1$, $p_{iC}=0.2$, $q=0.5$, equal sample sizes $n=40,\;100,\;250,\;1000$. 
		\label{CovThetaLOR1q05piC02}}
\end{figure}

\begin{figure}[t]
	\centering
	\includegraphics[scale=0.33]{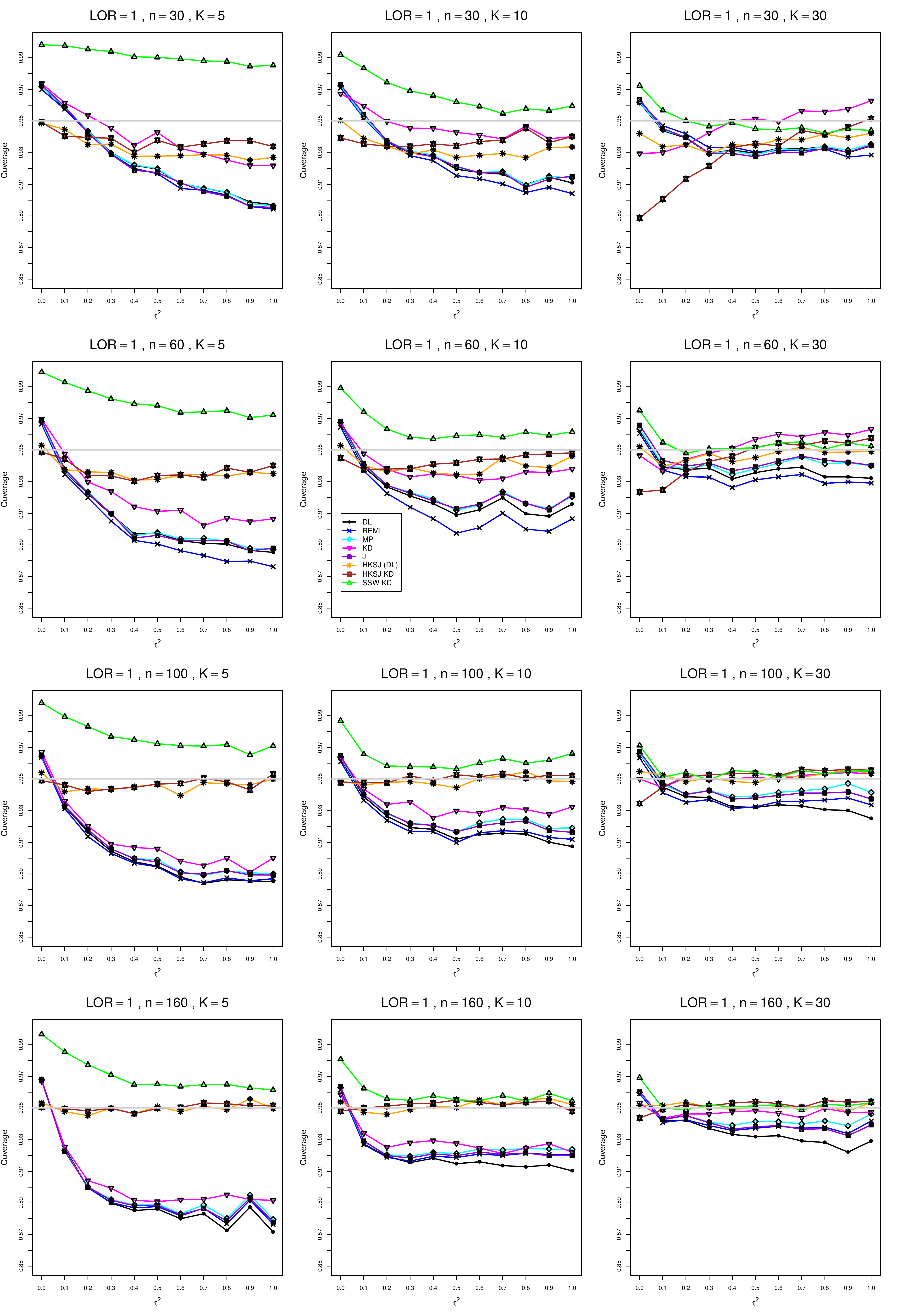}
	\caption{Coverage of  overall effect measure $\theta$ for $\theta=1$, $p_{iC}=0.2$, $q=0.5$, 
		unequal sample sizes $n=30,\; 60,\;100,\;160$. 
		\label{CovThetaLOR1q05piC02_unequal_sample_sizes}}
\end{figure}

\begin{figure}[t]
	\centering
	\includegraphics[scale=0.33]{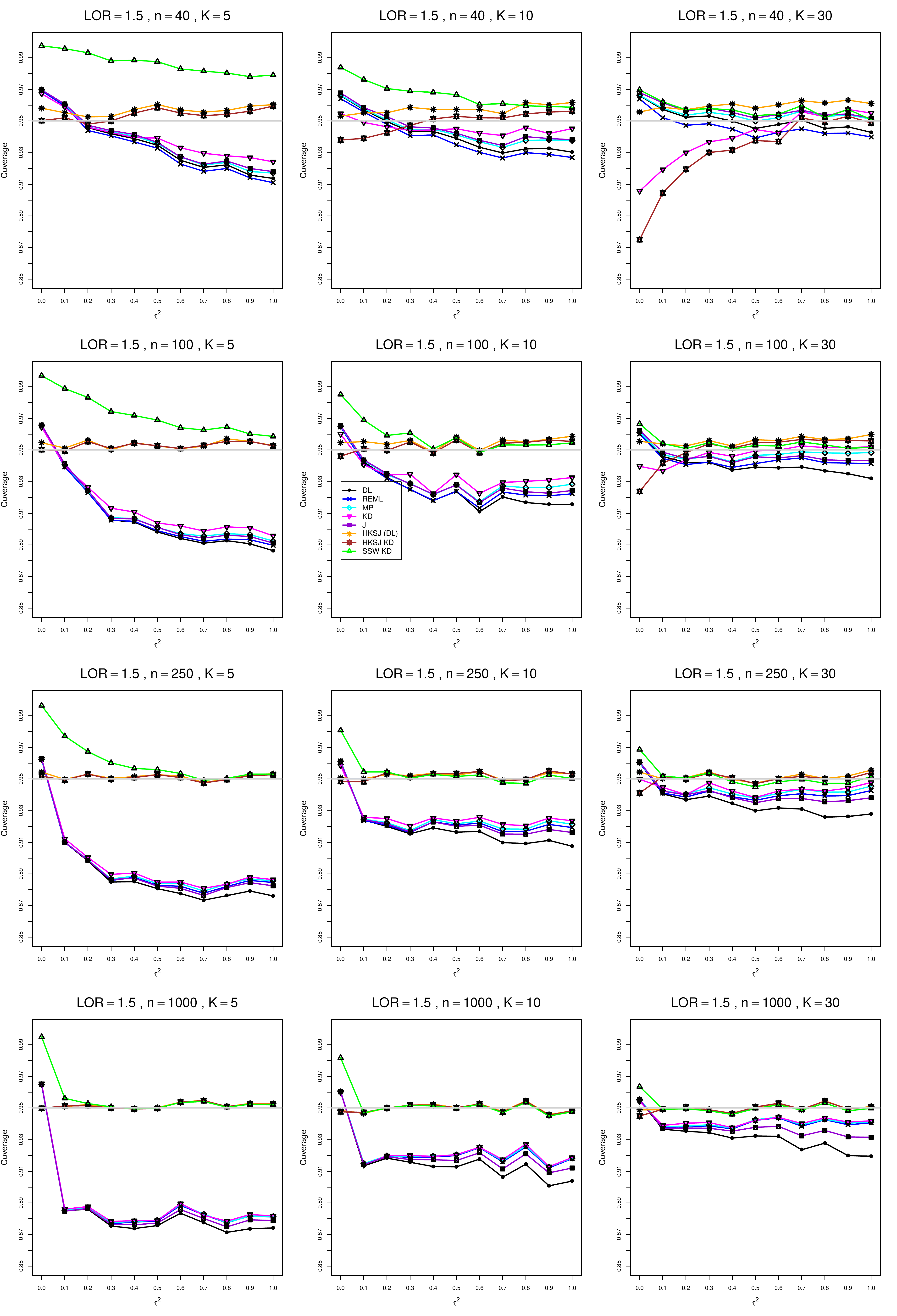}
	\caption{Coverage of  overall effect measure $\theta$ for $\theta=1.5$, $p_{iC}=0.2$, $q=0.5$, equal sample sizes  $n=40,\;100,\;250,\;1000$. 
		\label{CovThetaLOR15q05piC02}}
\end{figure}

\begin{figure}[t]
	\centering
	\includegraphics[scale=0.33]{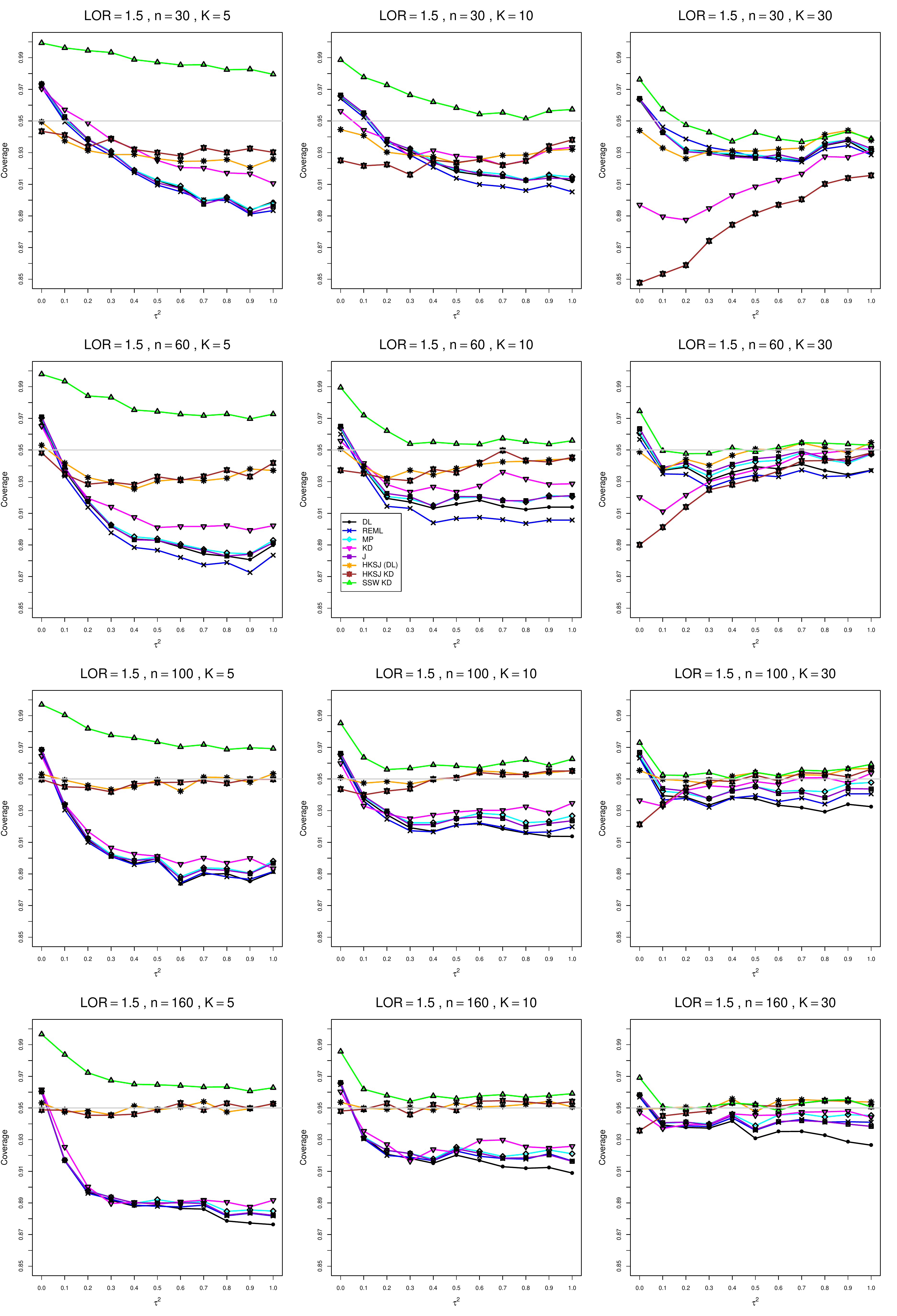}
	\caption{Coverage of  overall effect measure $\theta$ for $\theta=1.5$, $p_{iC}=0.2$, $q=0.5$, 
		unequal sample sizes $n=30,\; 60,\;100,\;160$. 
		\label{CovThetaLOR15q05piC02_unequal_sample_sizes}}
\end{figure}

\begin{figure}[t]
	\centering
	\includegraphics[scale=0.33]{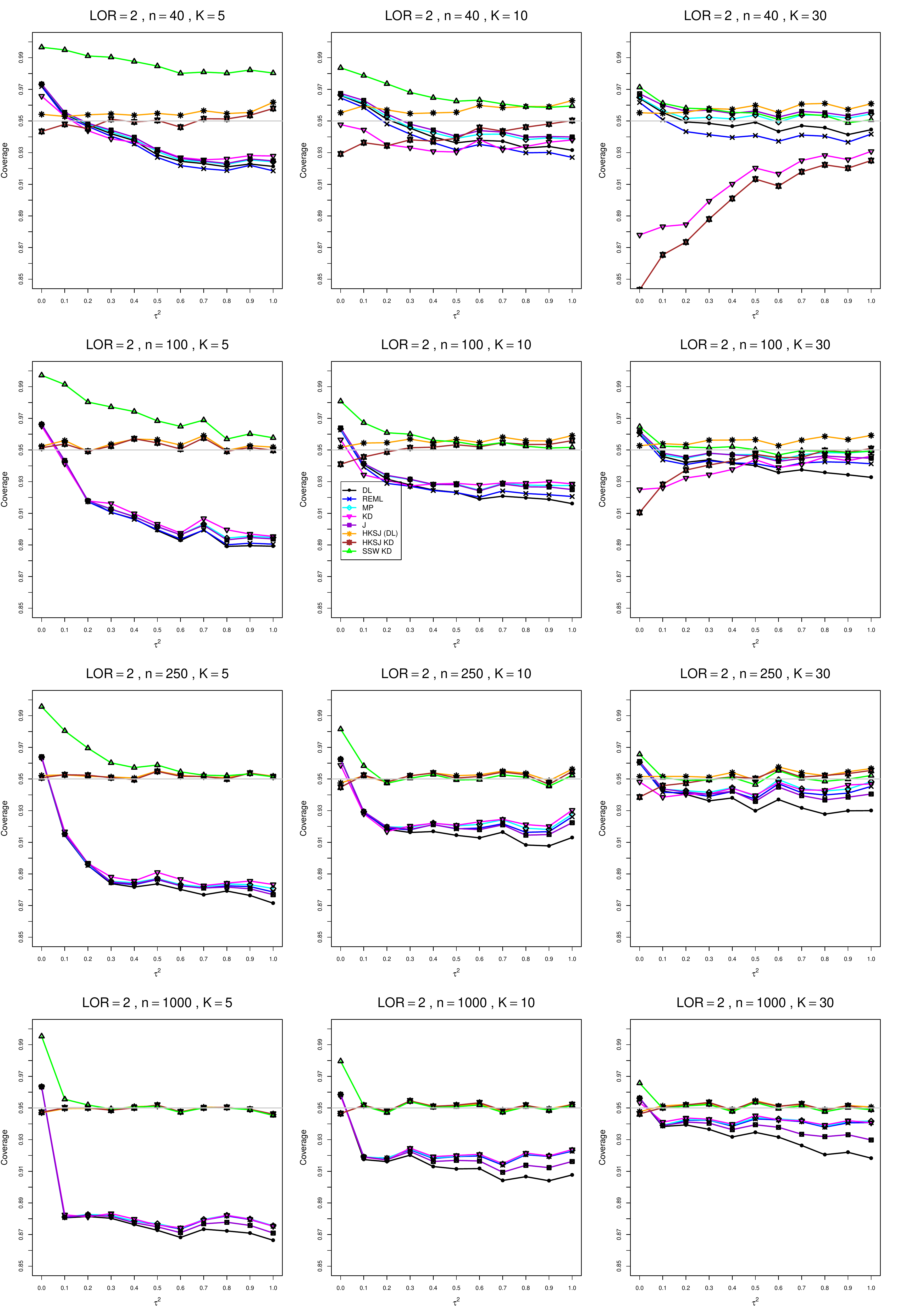}
	\caption{Coverage of  overall effect measure $\theta$ for $\theta=2$, $p_{iC}=0.2$, $q=0.5$, equal sample sizes $n=40,\;100,\;250,\;1000$. 
		\label{CovThetaLOR2q05piC02}}
\end{figure}

\begin{figure}[t]
	\centering
	\includegraphics[scale=0.33]{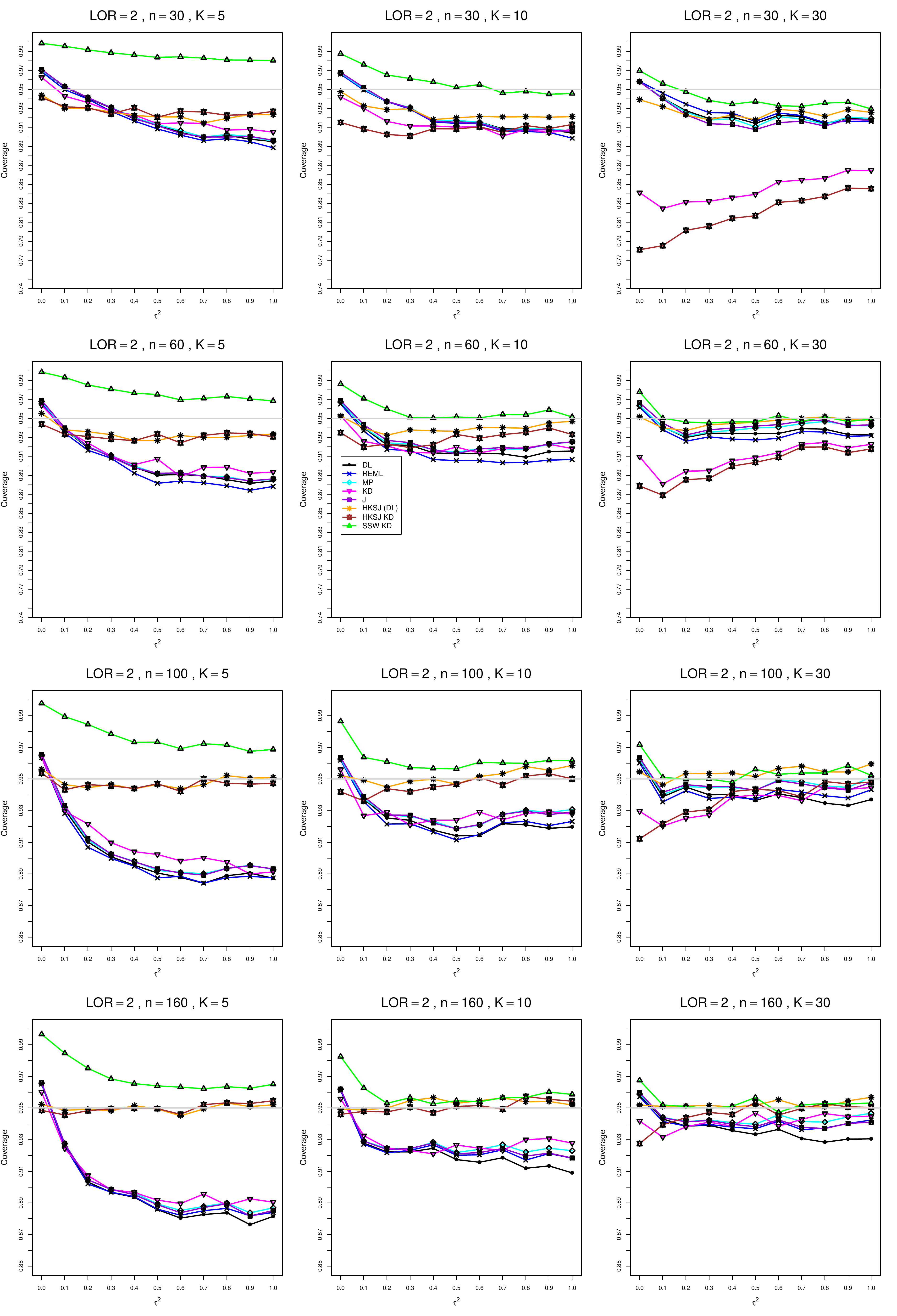}
	\caption{Coverage of  overall effect measure $\theta$ for $\theta=2$, $p_{iC}=0.2$, $q=0.5$, unequal sample sizes $n=40,\;100,\;250,\;1000$. 
		\label{CovThetaLOR2q05piC02_unequal_sample_sizes}}
\end{figure}


\begin{figure}[t]
	\centering
	\includegraphics[scale=0.33]{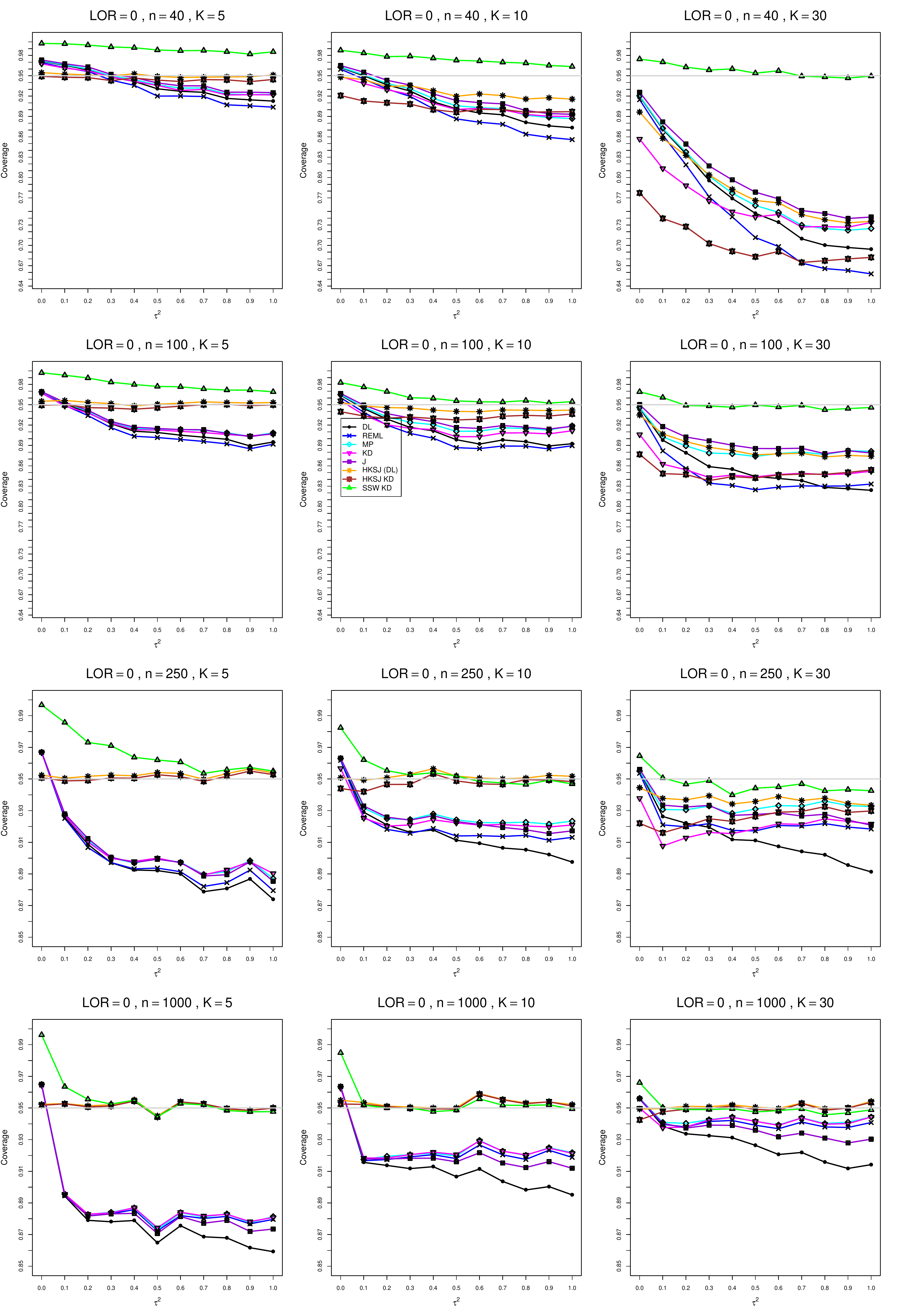}
	\caption{Coverage of  overall effect measure $\theta$ for $\theta=0$, $p_{iC}=0.2$, $q=0.75$, equal sample sizes $n=40,\;100,\;250,\;1000$. 
		\label{CovThetaLOR0q075piC02}}
\end{figure}

\begin{figure}[t]
	\centering
	\includegraphics[scale=0.33]{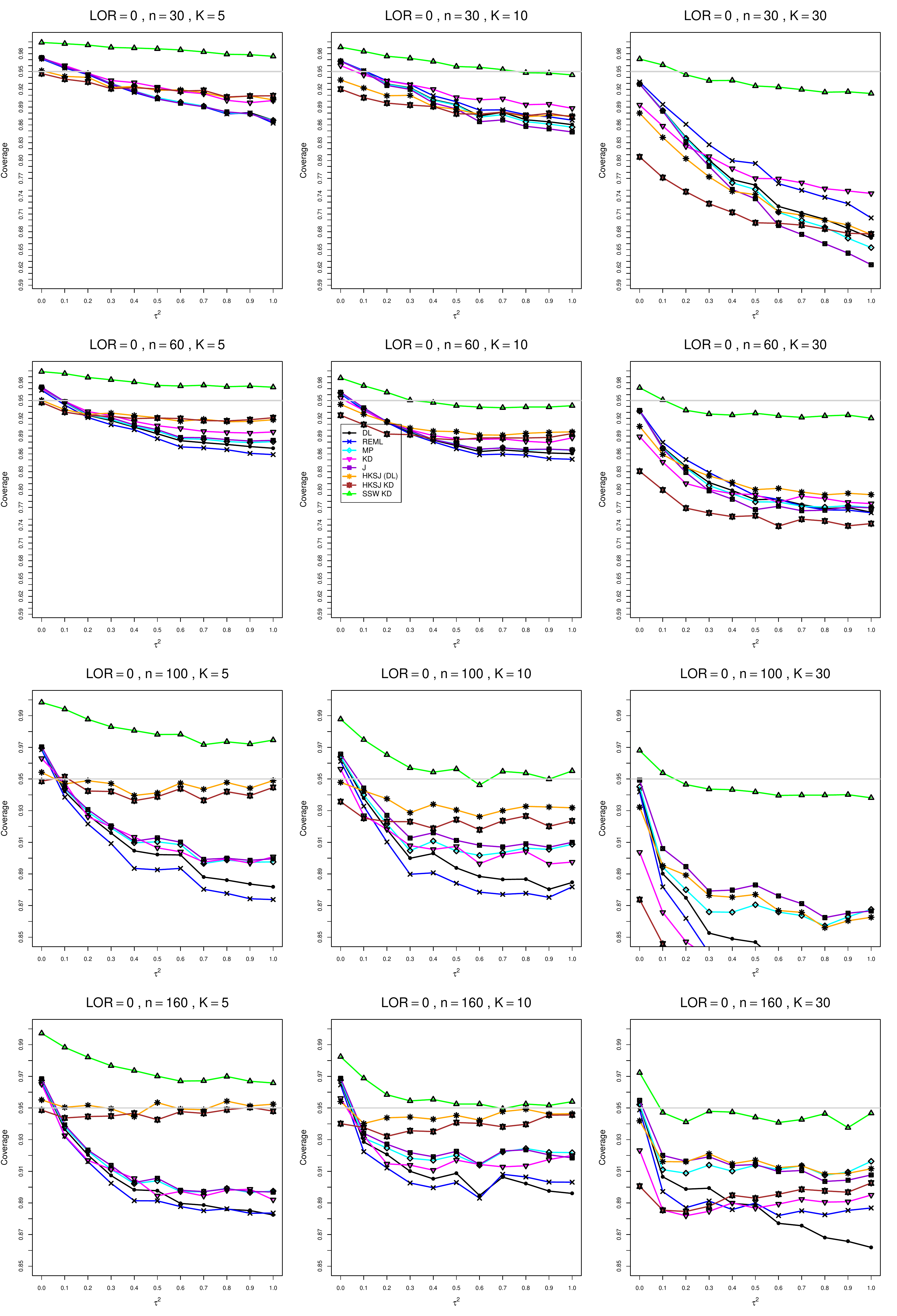}
	\caption{Coverage of  overall effect measure $\theta$ for $\theta=0$, $p_{iC}=0.2$, $q=0.75$, 
		unequal sample sizes $n=30,\; 60,\;100,\;160$. 
		\label{CovThetaLOR0q075piC02_unequal_sample_sizes}}
\end{figure}

\begin{figure}[t]
	\centering
	\includegraphics[scale=0.33]{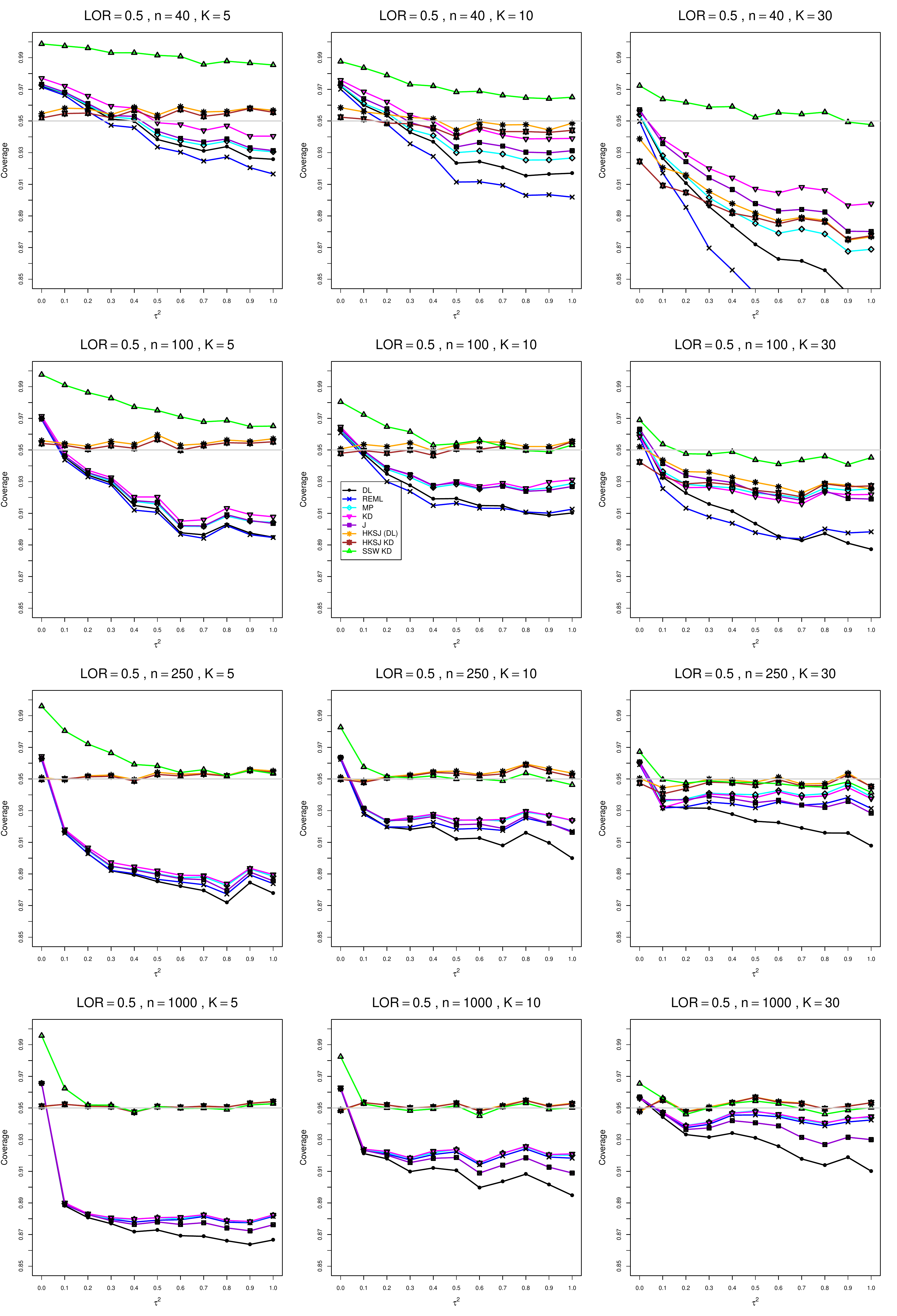}
	\caption{Coverage of  overall effect measure $\theta$ for $\theta=0.5$, $p_{iC}=0.2$, $q=0.75$, equal sample sizes $n=40,\;100,\;250,\;1000$. 
		\label{CovThetaLOR05q075piC02}}
\end{figure}

\begin{figure}[t]
	\centering
	\includegraphics[scale=0.33]{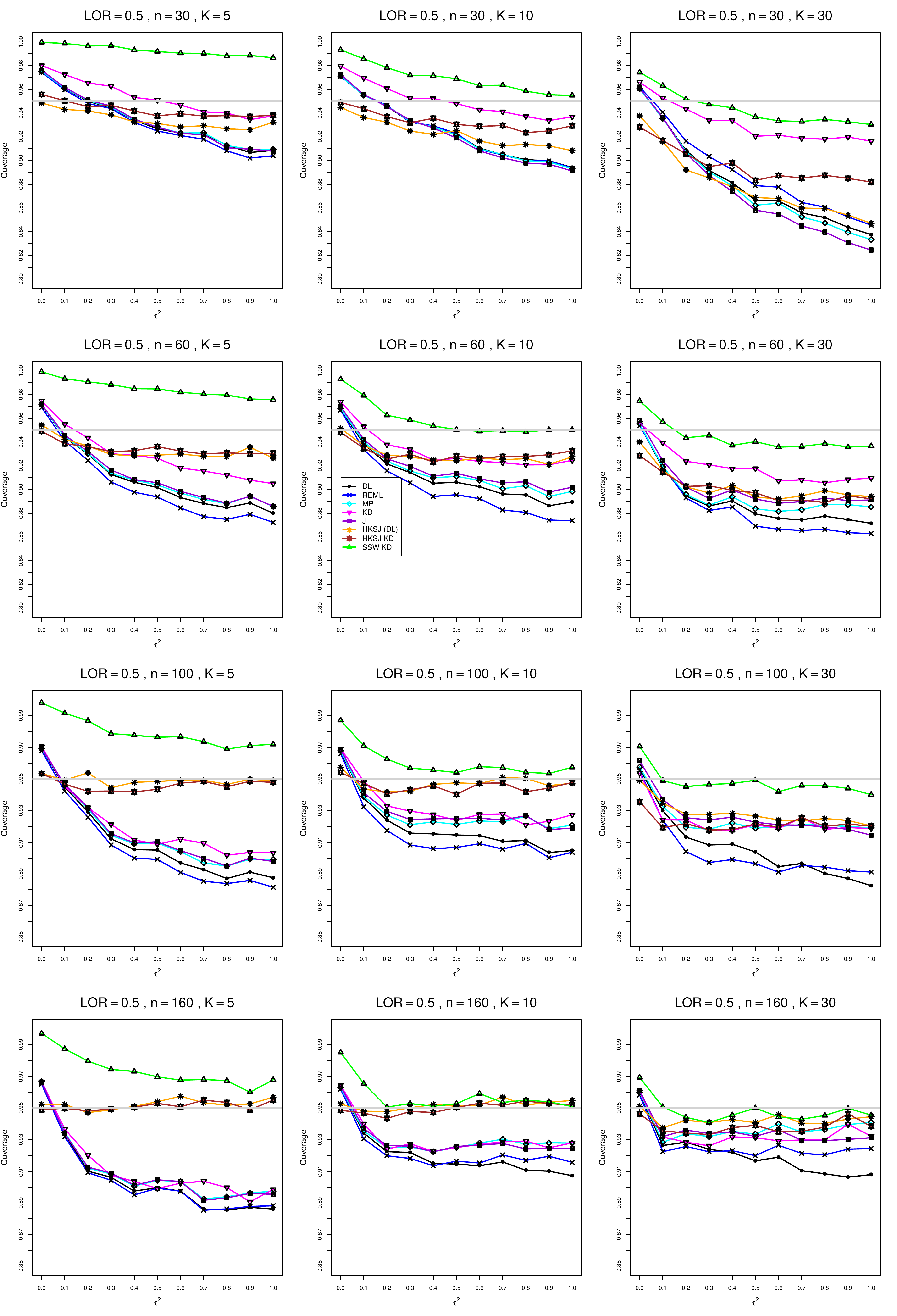}
	\caption{Coverage of  overall effect measure $\theta$ for $\theta=0.5$, $p_{iC}=0.2$, $q=0.75$, 
		unequal sample sizes $n=30,\; 60,\;100,\;160$. 
		\label{CovThetaLOR05q075piC02_unequal_sample_sizes}}
\end{figure}

\clearpage

\begin{figure}[t]
	\centering
	\includegraphics[scale=0.33]{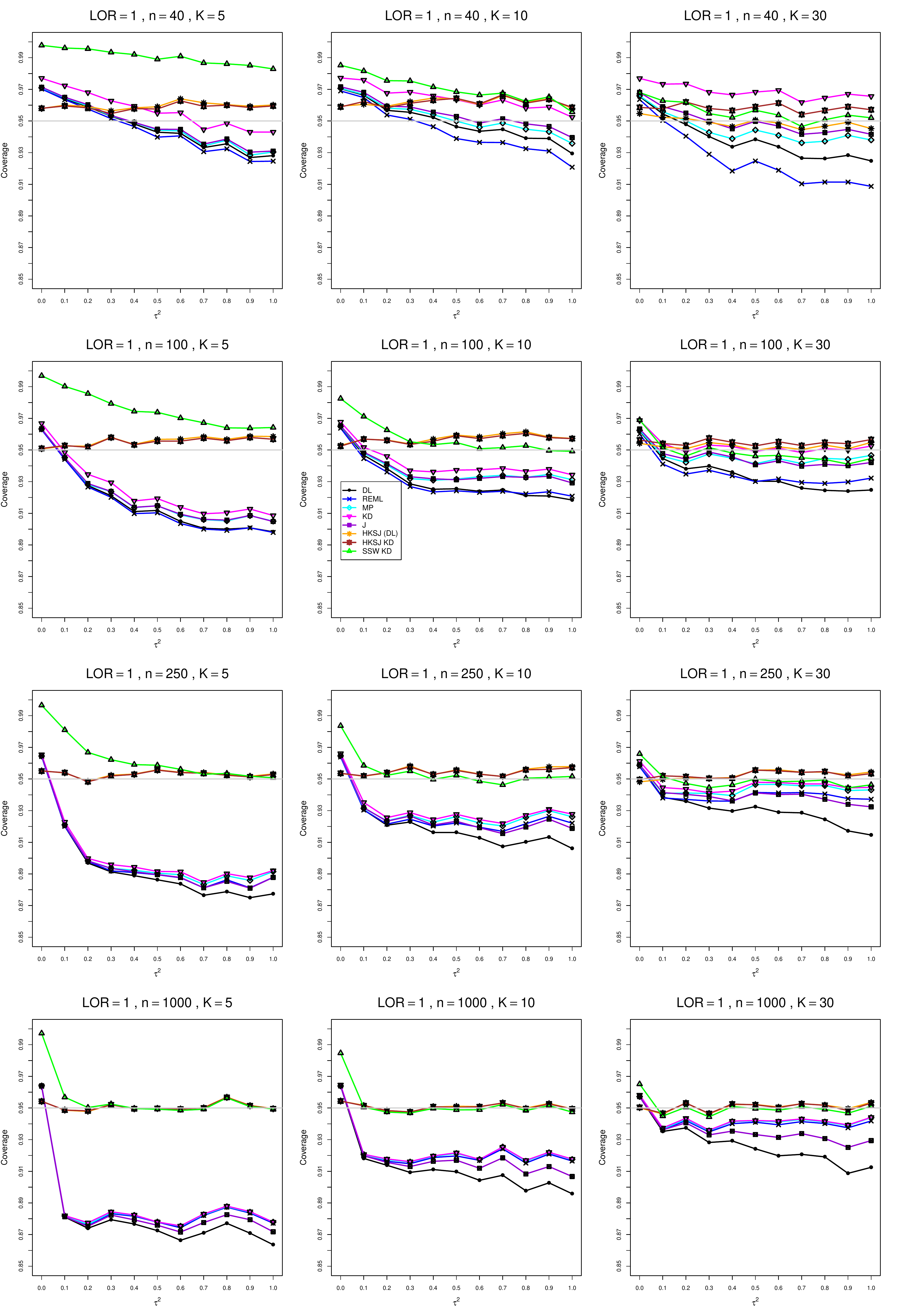}
	\caption{Coverage of  overall effect measure $\theta$ for $\theta=1$, $p_{iC}=0.2$, $q=0.75$, equal sample sizes  $n=40,\;100,\;250,\;1000$. 
		\label{CovThetaLOR1q075piC02}}
\end{figure}

\begin{figure}[t]
	\centering
	\includegraphics[scale=0.33]{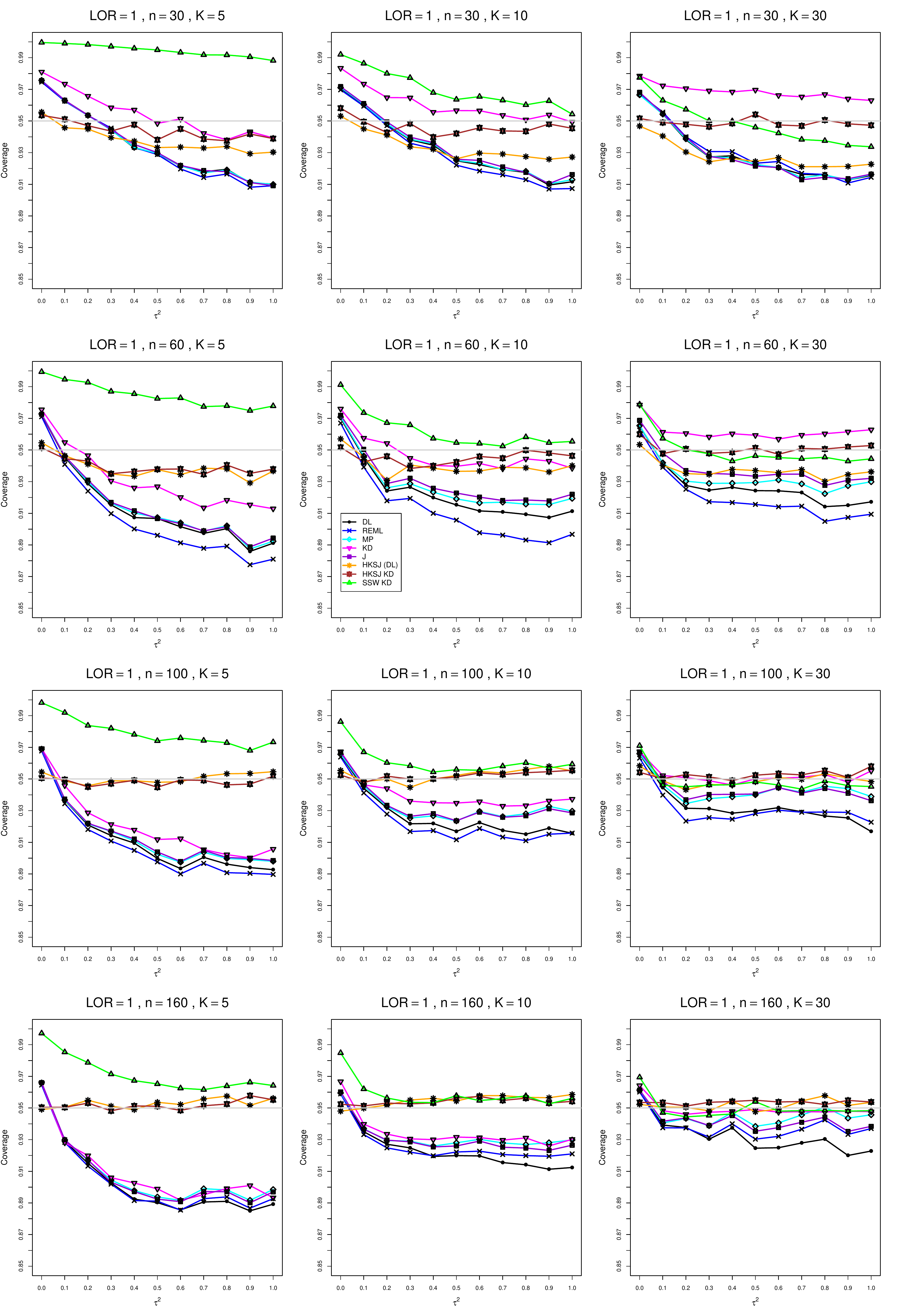}
	\caption{Coverage of  overall effect measure $\theta$ for $\theta=1$, $p_{iC}=0.2$, $q=0.75$, 
		unequal sample sizes $n=30,\; 60,\;100,\;160$. 
		\label{CovThetaLOR1q075piC02_unequal_sample_sizes}}
\end{figure}

\begin{figure}[t]
	\centering
	\includegraphics[scale=0.33]{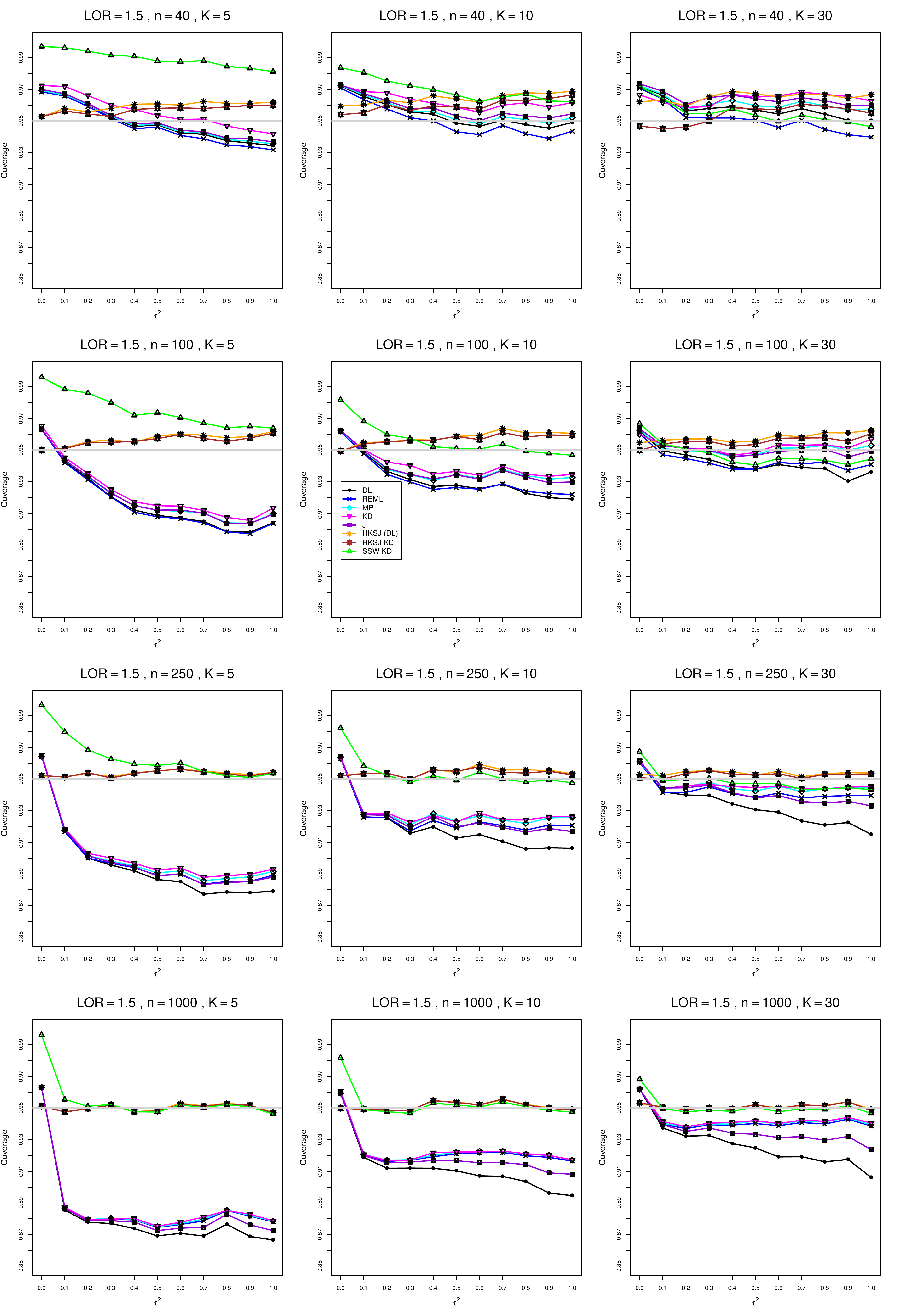}
	\caption{Coverage of  overall effect measure $\theta$ for $\theta=1.5$, $p_{iC}=0.2$, $q=0.75$, equal sample sizes $n=40,\;100,\;250,\;1000$. 
		\label{CovThetaLOR15q075piC02}}
\end{figure}

\begin{figure}[t]
	\centering
	\includegraphics[scale=0.33]{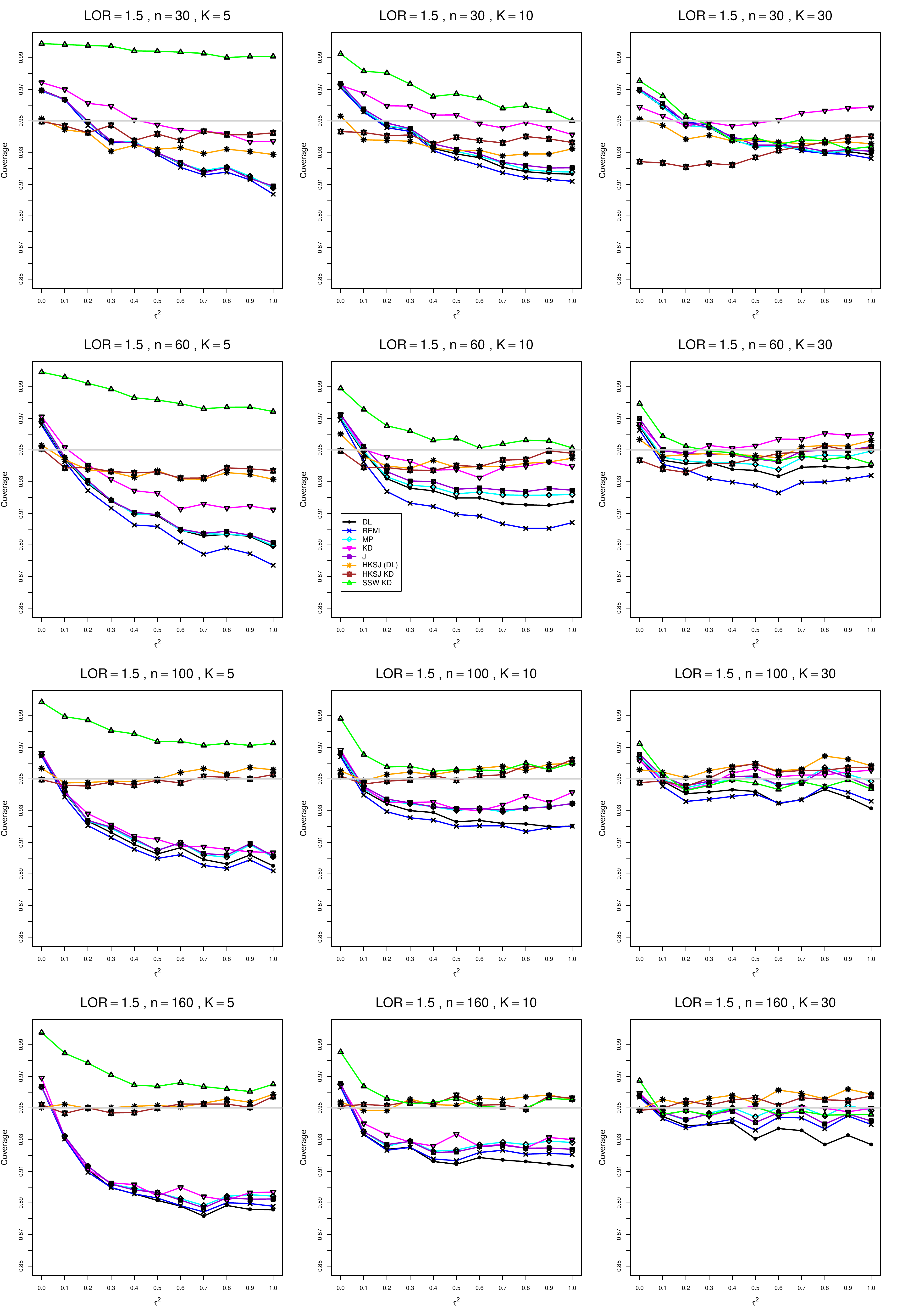}
	\caption{Coverage of  overall effect measure $\theta$ for $\theta=1.5$, $p_{iC}=0.2$, $q=0.75$, 
		unequal sample sizes $n=30,\; 60,\;100,\;160$. 
		\label{CovThetaLOR15q075piC02_unequal_sample_sizes}}
\end{figure}

\begin{figure}[t]
	\centering
	\includegraphics[scale=0.33]{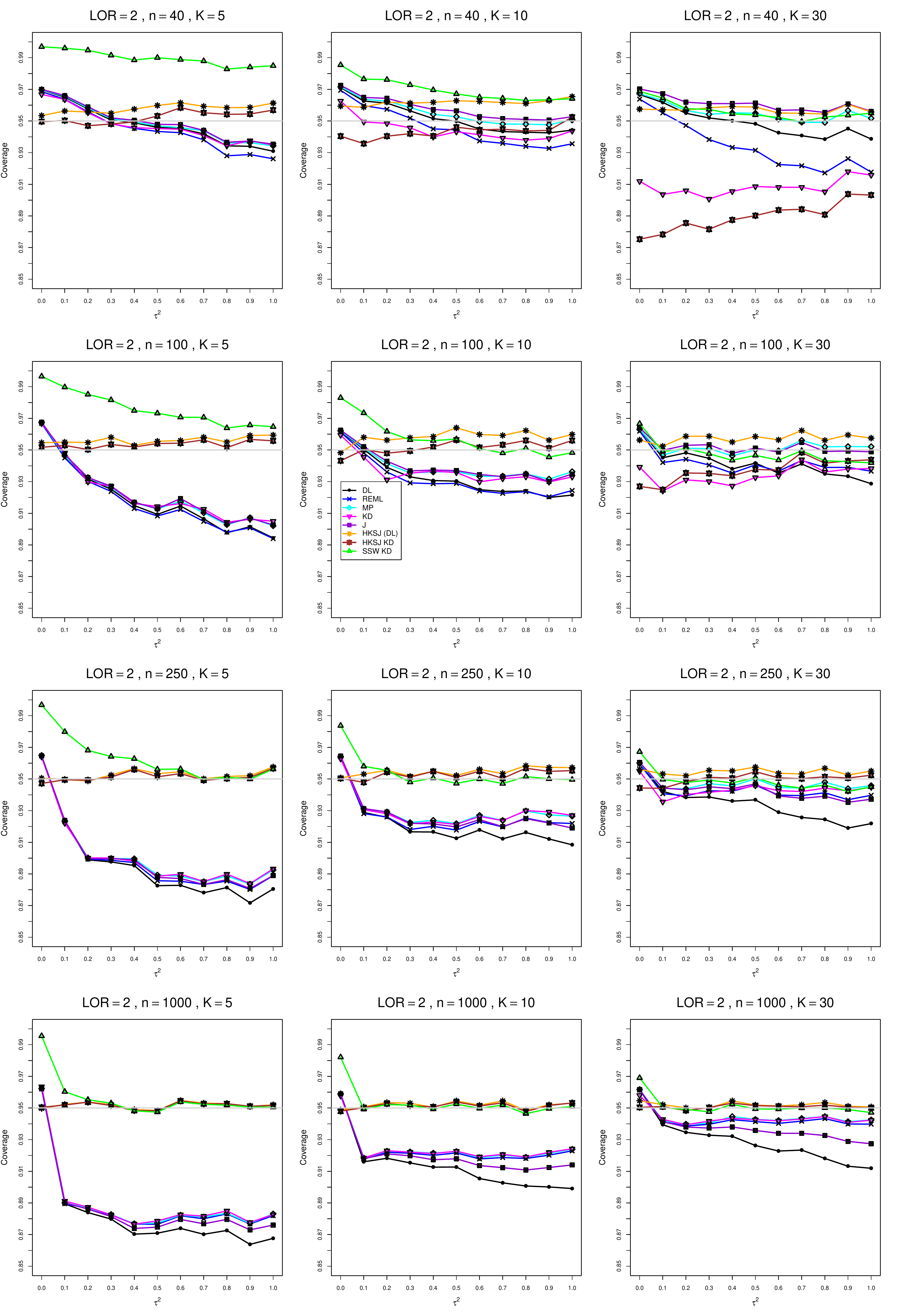}
	\caption{Coverage of  overall effect measure $\theta$ for $\theta=2$, $p_{iC}=0.2$, $q=0.75$, equal sample sizes $n=40,\;100,\;250,\;1000$. 
		\label{CovThetaLOR2q075piC02}}
\end{figure}

\begin{figure}[t]
	\centering
	\includegraphics[scale=0.33]{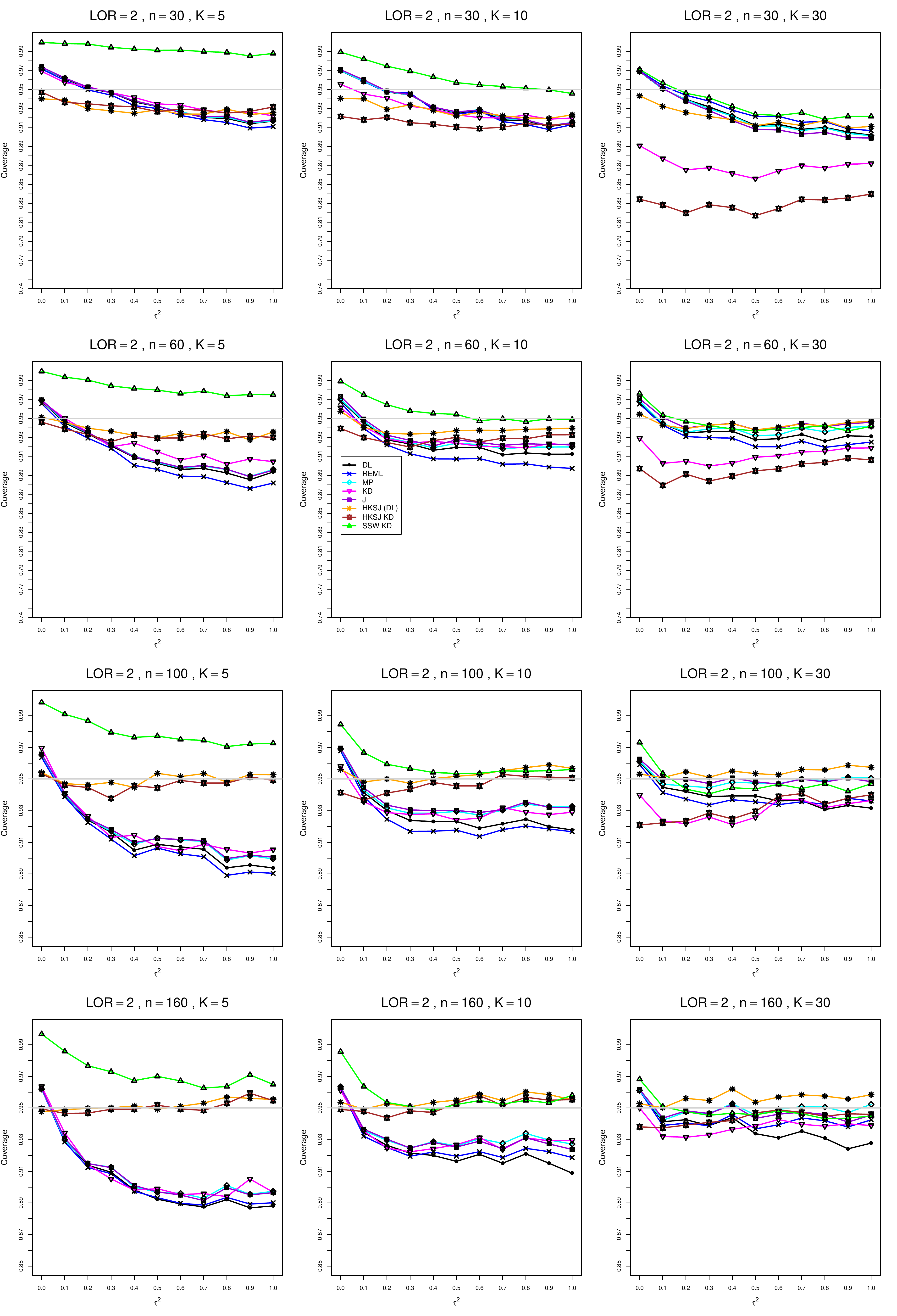}
	\caption{Coverage of  overall effect measure $\theta$ for $\theta=2$, $p_{iC}=0.2$, $q=0.75$, 
		unequal sample sizes $n=30,\; 60,\;100,\;160$. 
		\label{CovThetaLOR2q075piC02_unequal_sample_sizes}}
\end{figure}
\clearpage

\renewcommand{\thefigure}{B2.3.\arabic{figure}}
\setcounter{figure}{0}
\subsection*{B2.3 Probability in the control arm $p_{C}=0.4$}
\begin{figure}[t]
	\centering
	\includegraphics[scale=0.33]{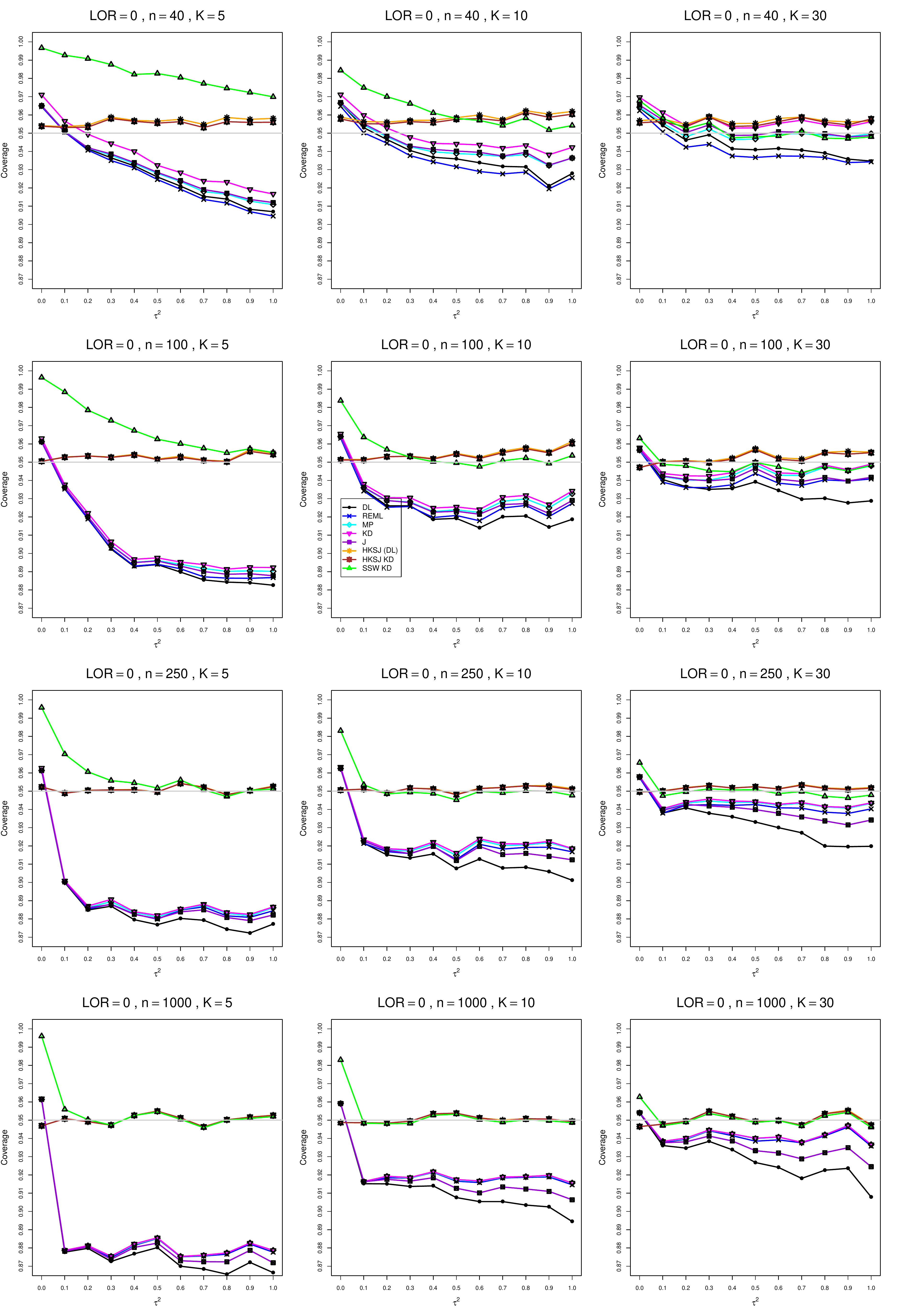}
	\caption{Coverage of  overall effect measure $\theta$ for $\theta=0$, $p_{iC}=0.4$, $q=0.5$, equal sample sizes  $n=40,\;100,\;250,\;1000$. 
		\label{CovThetaLOR0q05piC04}}
\end{figure}

\begin{figure}[t]
	\centering
	\includegraphics[scale=0.33]{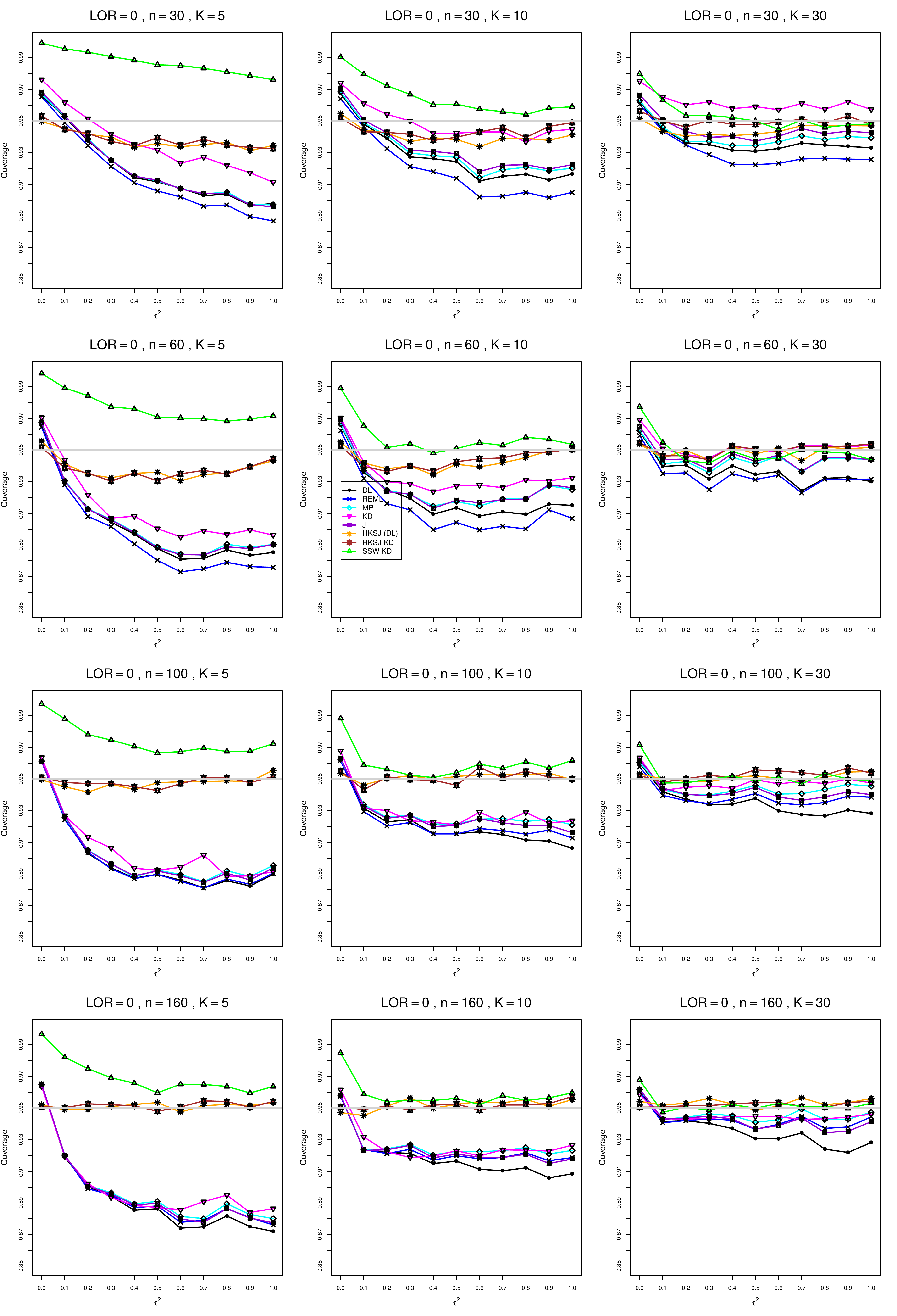}
	\caption{Coverage of  overall effect measure $\theta$ for $\theta=0$, $p_{iC}=0.4$, $q=0.5$, 
		unequal sample sizes $n=30,\; 60,\;100,\;160$. 
		\label{CovThetaLOR0q05piC04_unequal_sample_sizes}}
\end{figure}

\begin{figure}[t]
	\centering
	\includegraphics[scale=0.33]{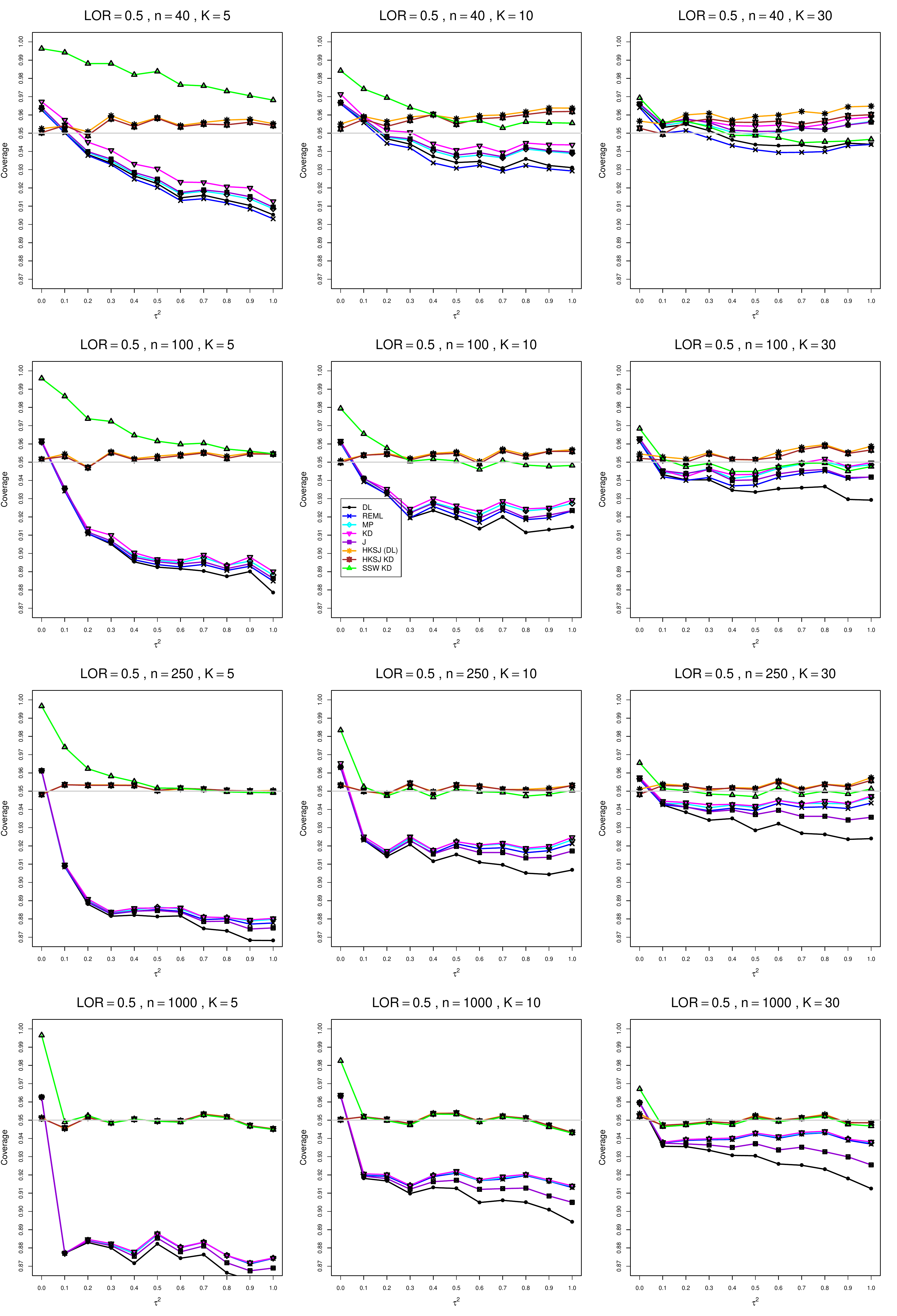}
	\caption{Coverage of  overall effect measure $\theta$ for $\theta=0.5$, $p_{iC}=0.4$, $q=0.5$, equal sample sizes $n=40,\;100,\;250,\;1000$. 
		\label{CovThetaLOR05q05piC04}}
\end{figure}

\begin{figure}[t]
	\centering
	\includegraphics[scale=0.33]{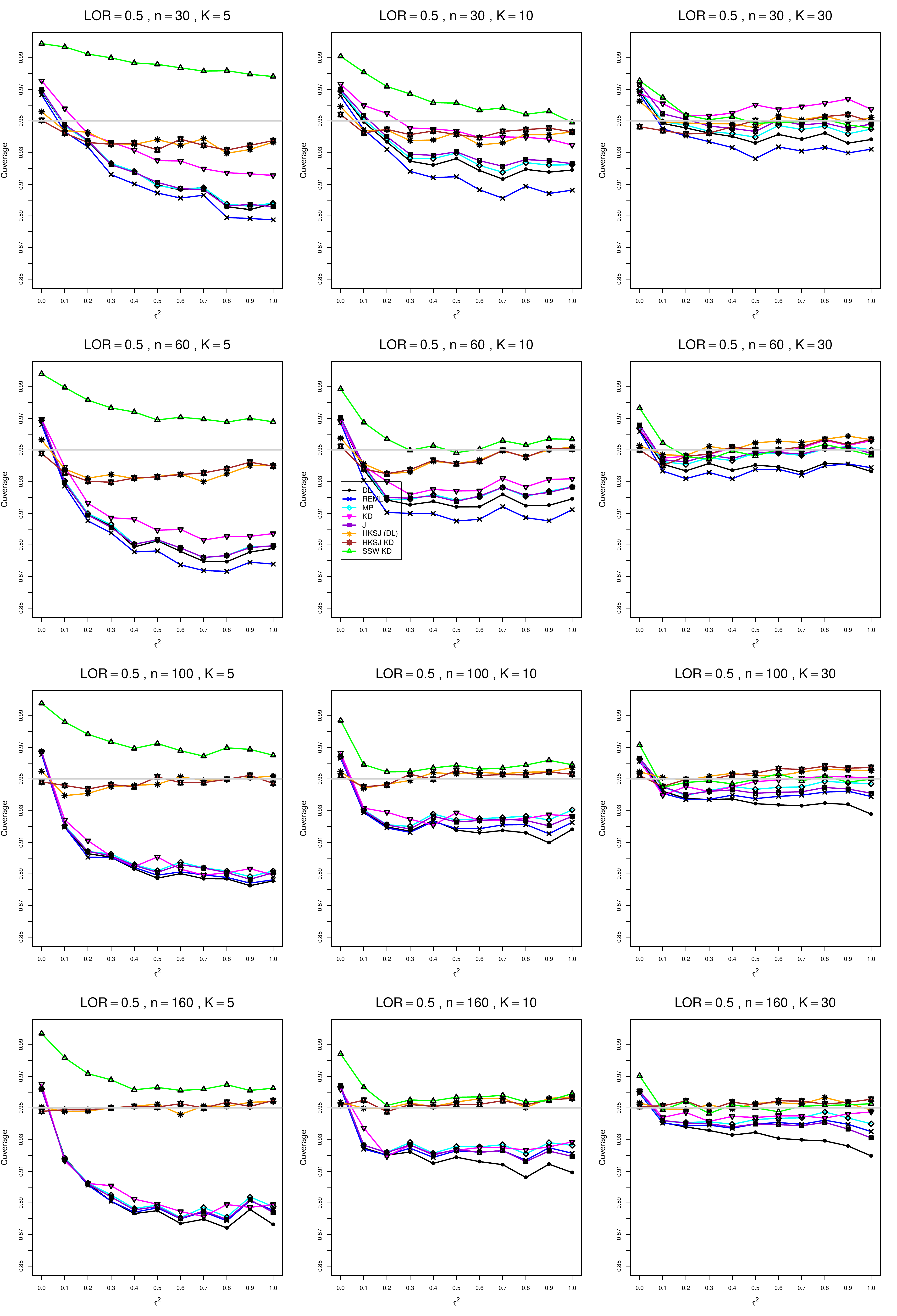}
	\caption{Coverage of  overall effect measure $\theta$ for $\theta=0.5$, $p_{iC}=0.4$, $q=0.5$, 
		unequal sample sizes $n=30,\; 60,\;100,\;160$. 
		\label{CovThetaLOR05q05piC04_unequal_sample_sizes}}
\end{figure}

\begin{figure}[t]
	\centering
	\includegraphics[scale=0.33]{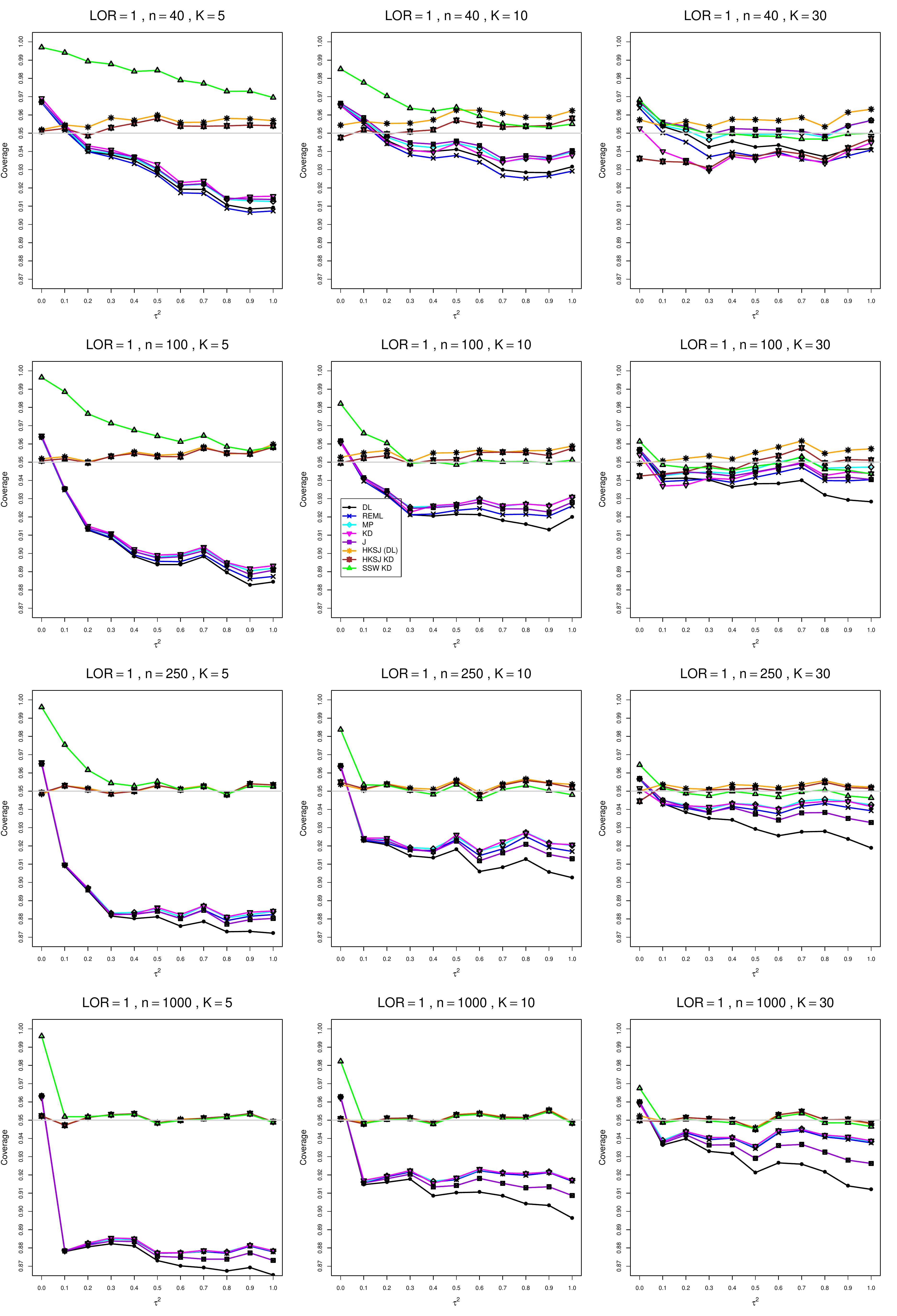}
	\caption{Coverage of  overall effect measure $\theta$ for $\theta=1$, $p_{iC}=0.4$, $q=0.5$, equal sample sizes $n=40,\;100,\;250,\;1000$. 
		\label{CovThetaLOR1q05piC04}}
\end{figure}

\begin{figure}[t]
	\centering
	\includegraphics[scale=0.33]{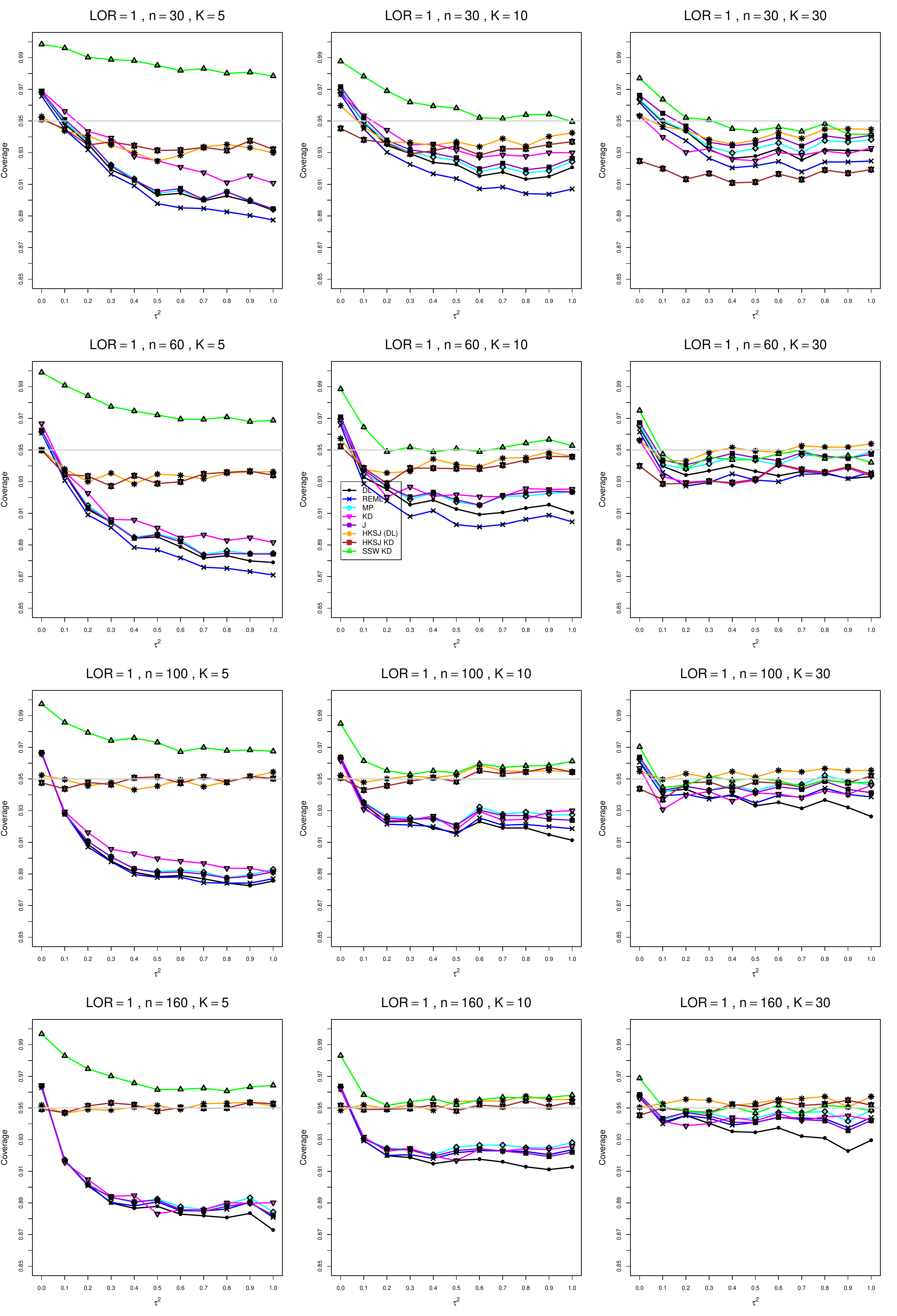}
	\caption{Coverage of  overall effect measure $\theta$ for $\theta=1$, $p_{iC}=0.4$, $q=0.5$, 
		unequal sample sizes $n=30,\; 60,\;100,\;160$. 
		\label{CovThetaLOR1q05piC04_unequal_sample_sizes}}
\end{figure}

\begin{figure}[t]
	\centering
	\includegraphics[scale=0.33]{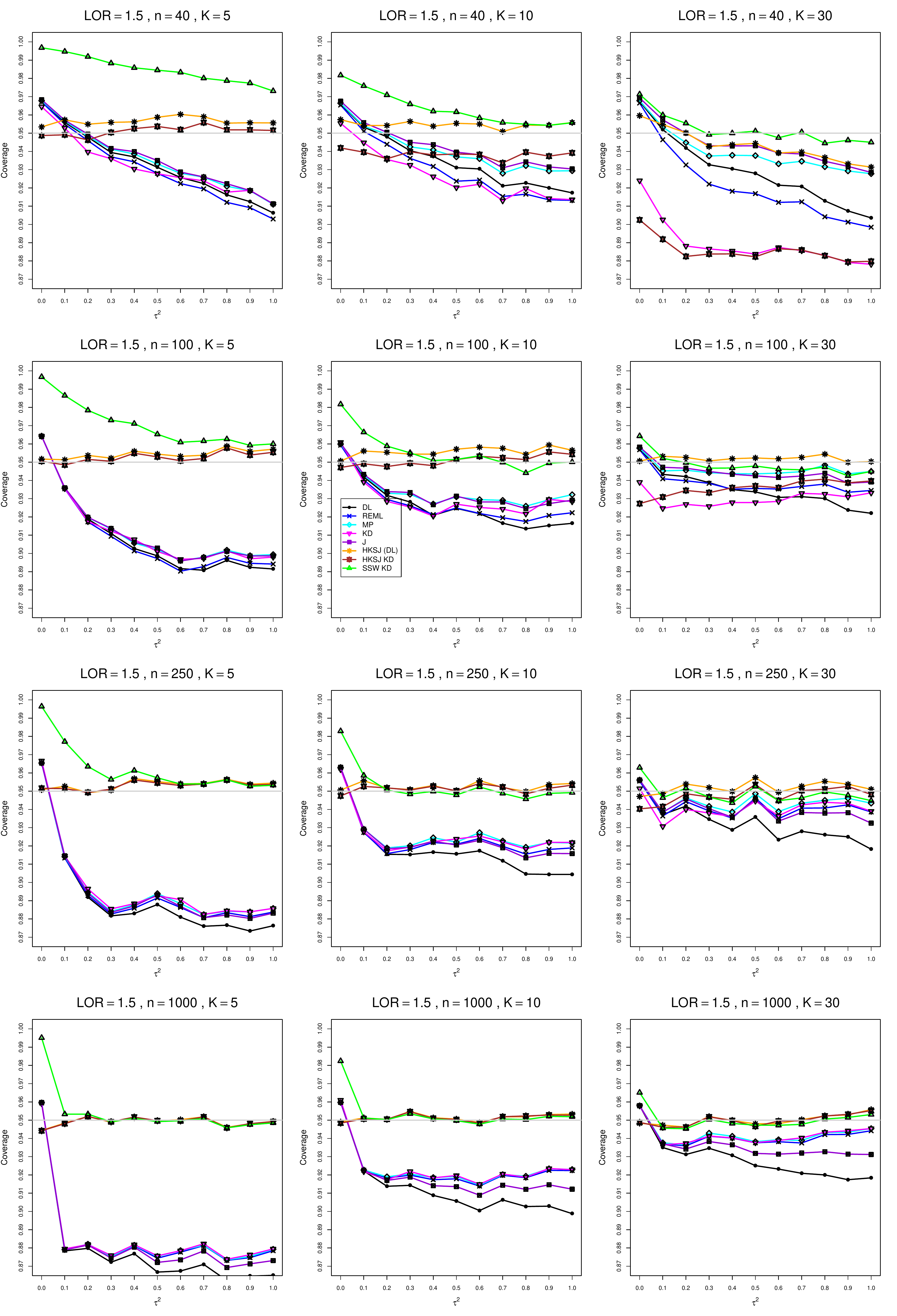}
	\caption{Coverage of  overall effect measure $\theta$ for $\theta=1.5$, $p_{iC}=0.4$, $q=0.5$, equal sample sizes $n=40,\;100,\;250,\;1000$. 
		\label{CovThetaLOR15q05piC04}}
\end{figure}

\begin{figure}[t]
	\centering
	\includegraphics[scale=0.33]{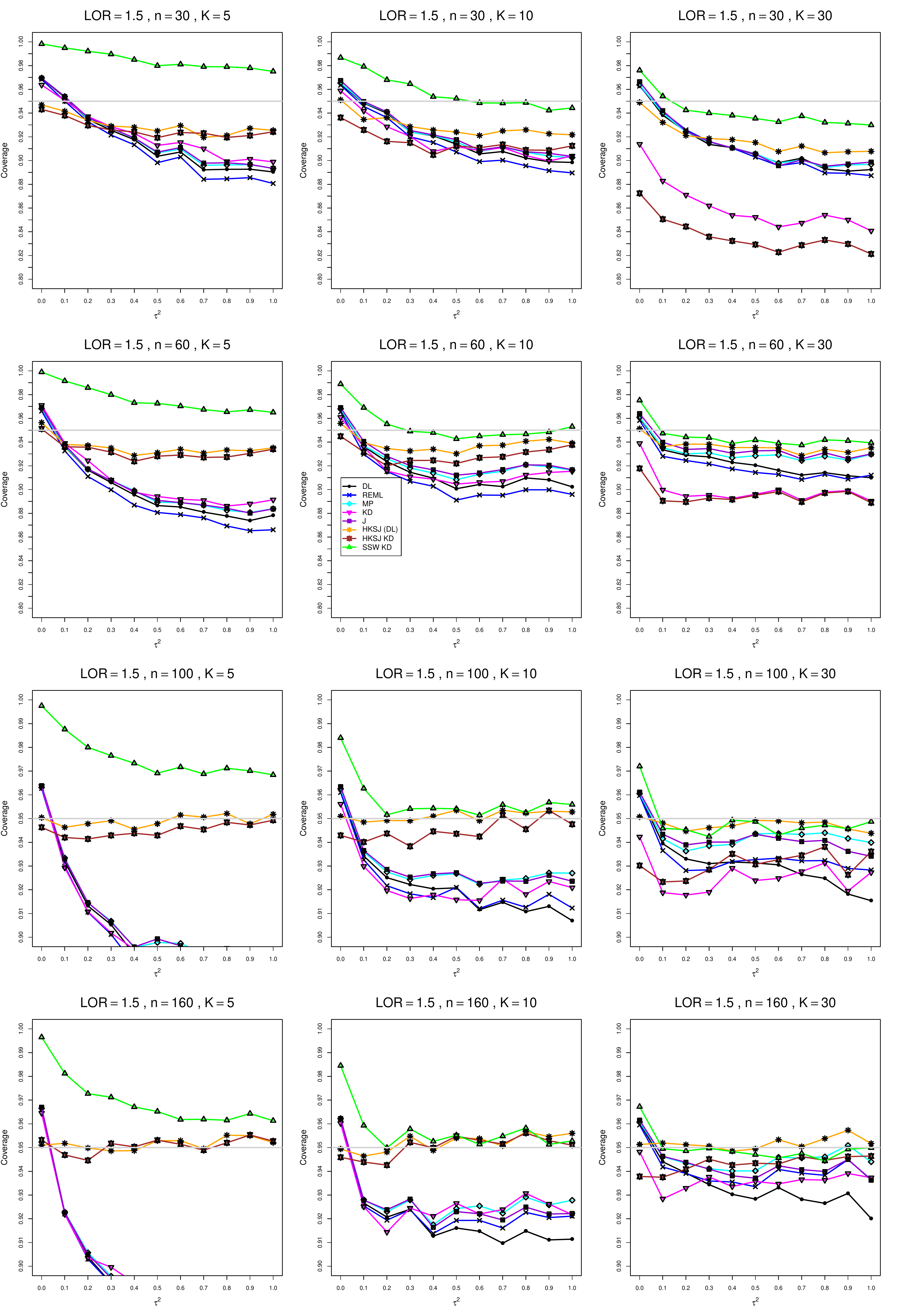}
	\caption{Coverage of  overall effect measure $\theta$ for $\theta=1.5$, $p_{iC}=0.4$, $q=0.5$, 
		unequal sample sizes $n=30,\; 60,\;100,\;160$. 
		\label{CovThetaLOR15q05piC04_unequal_sample_sizes}}
\end{figure}

\begin{figure}[t]
	\centering
	\includegraphics[scale=0.33]{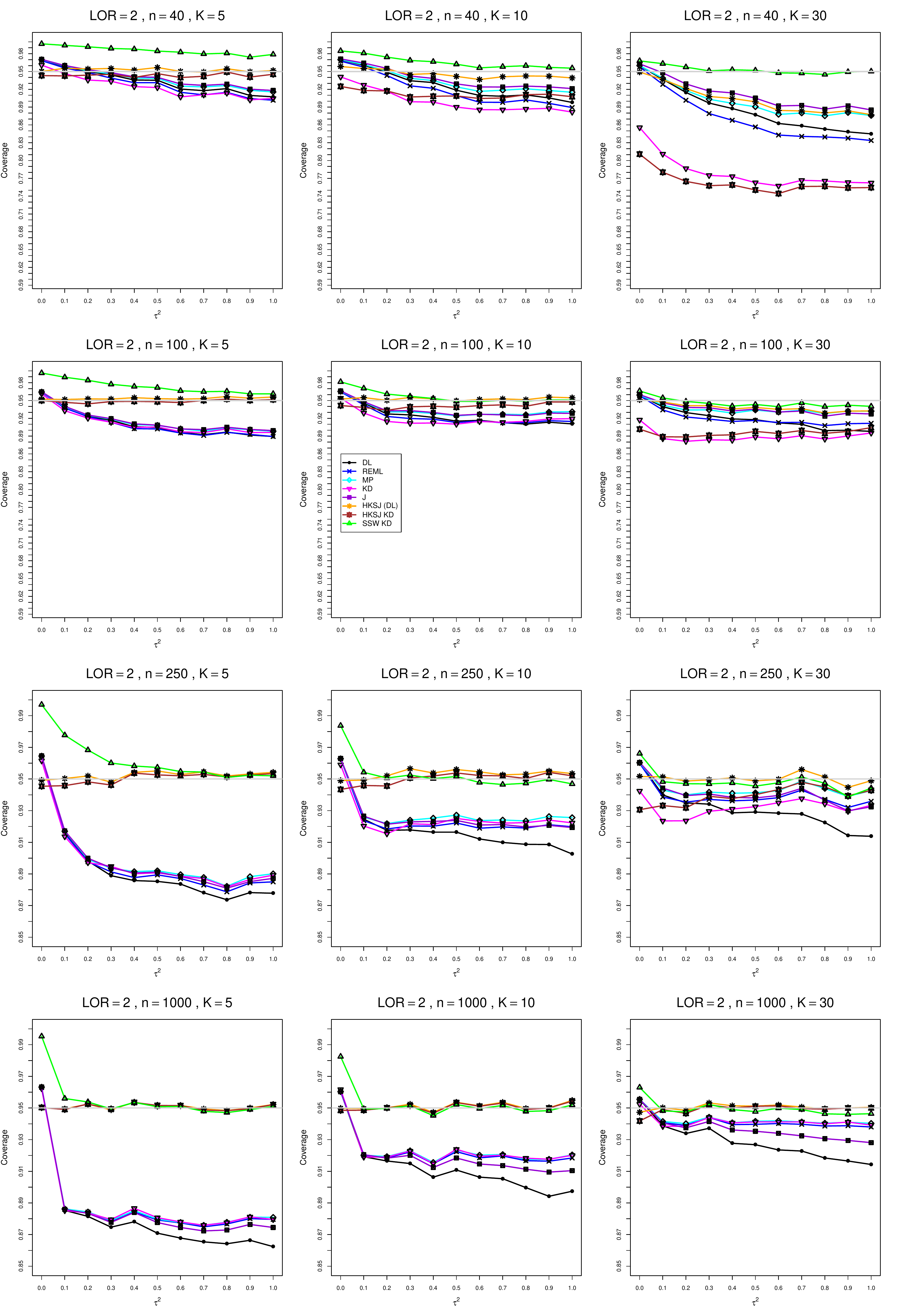}
	\caption{Coverage of  overall effect measure $\theta$ for $\theta=2$, $p_{iC}=0.4$, $q=0.5$, equal sample sizes $n=40,\;100,\;250,\;1000$. 
		\label{CovThetaLOR2q05piC04}}
\end{figure}

\begin{figure}[t]
	\centering
	\includegraphics[scale=0.33]{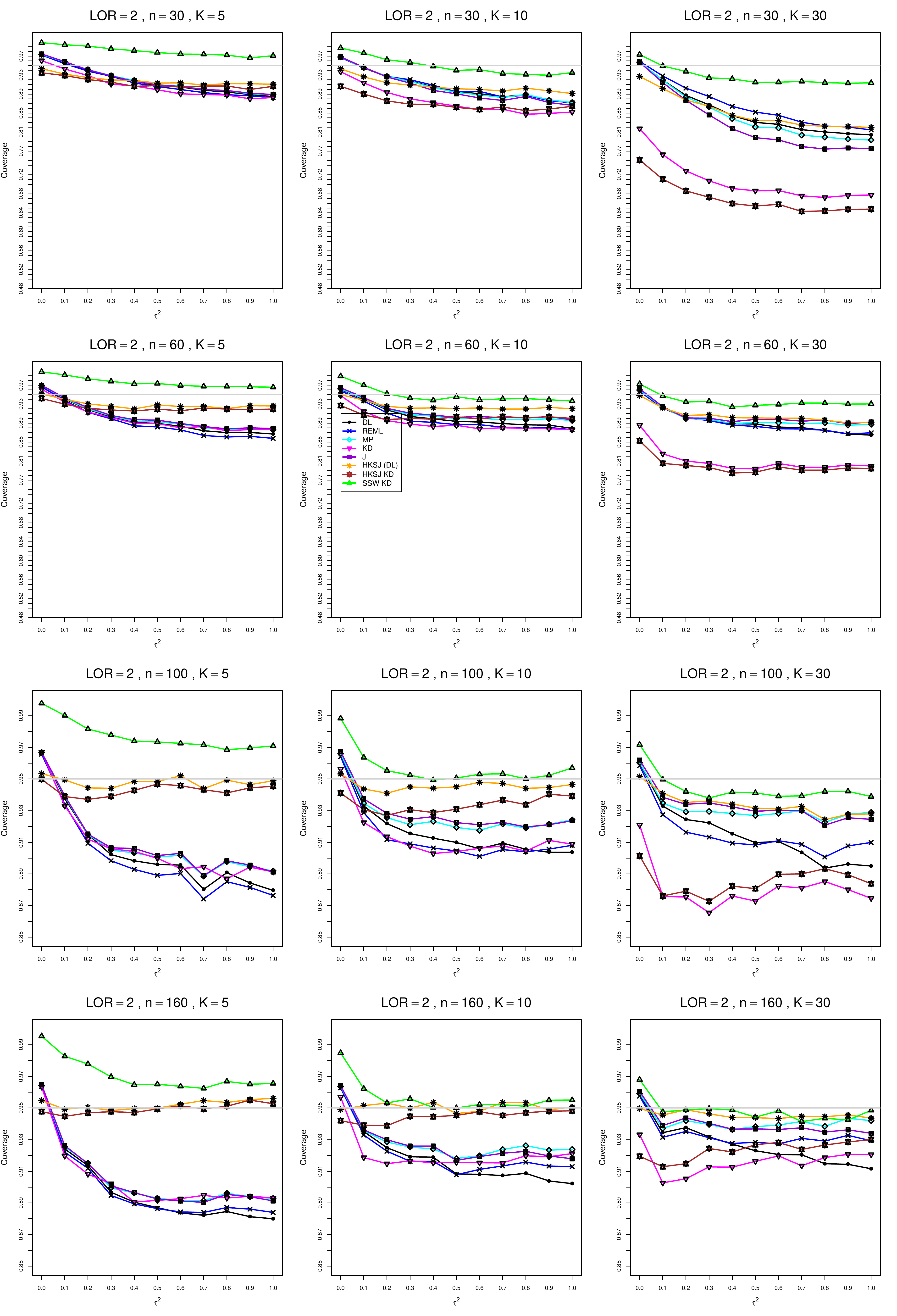}
	\caption{Coverage of  overall effect measure $\theta$ for $\theta=2$, $p_{iC}=0.4$, $q=0.5$, 
		unequal sample sizes $n=30,\; 60,\;100,\;160$. 
		\label{CovThetaLOR2q05piC04_unequal_sample_sizes}}
\end{figure}


\begin{figure}[t]
	\centering
	\includegraphics[scale=0.33]{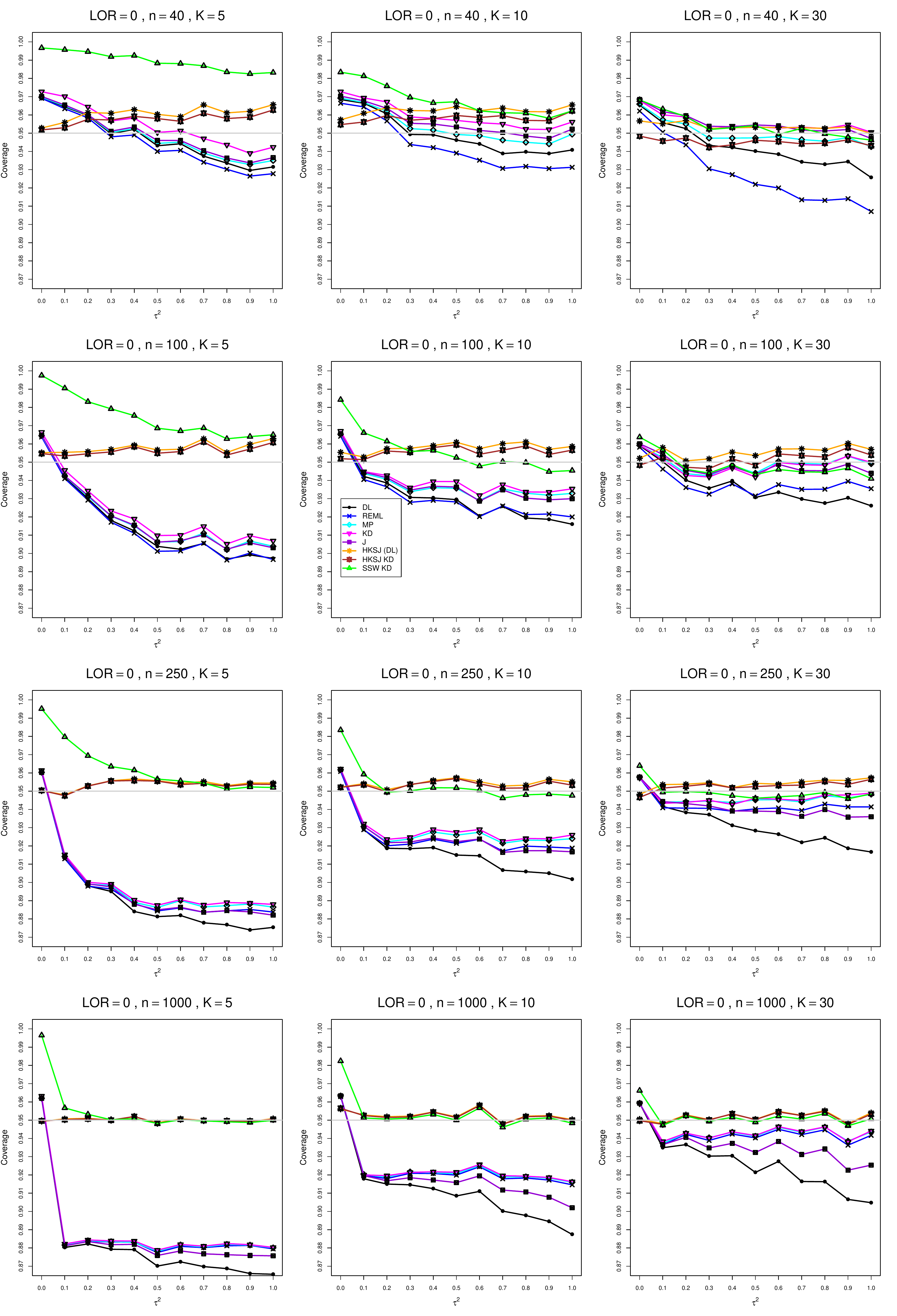}
	\caption{Coverage of  overall effect measure $\theta$ for $\theta=0$, $p_{iC}=0.4$, $q=0.75$, equal sample sizes $n=40,\;100,\;250,\;1000$. 
		\label{CovThetaLOR0q075piC04}}
\end{figure}

\begin{figure}[t]
	\centering
	\includegraphics[scale=0.33]{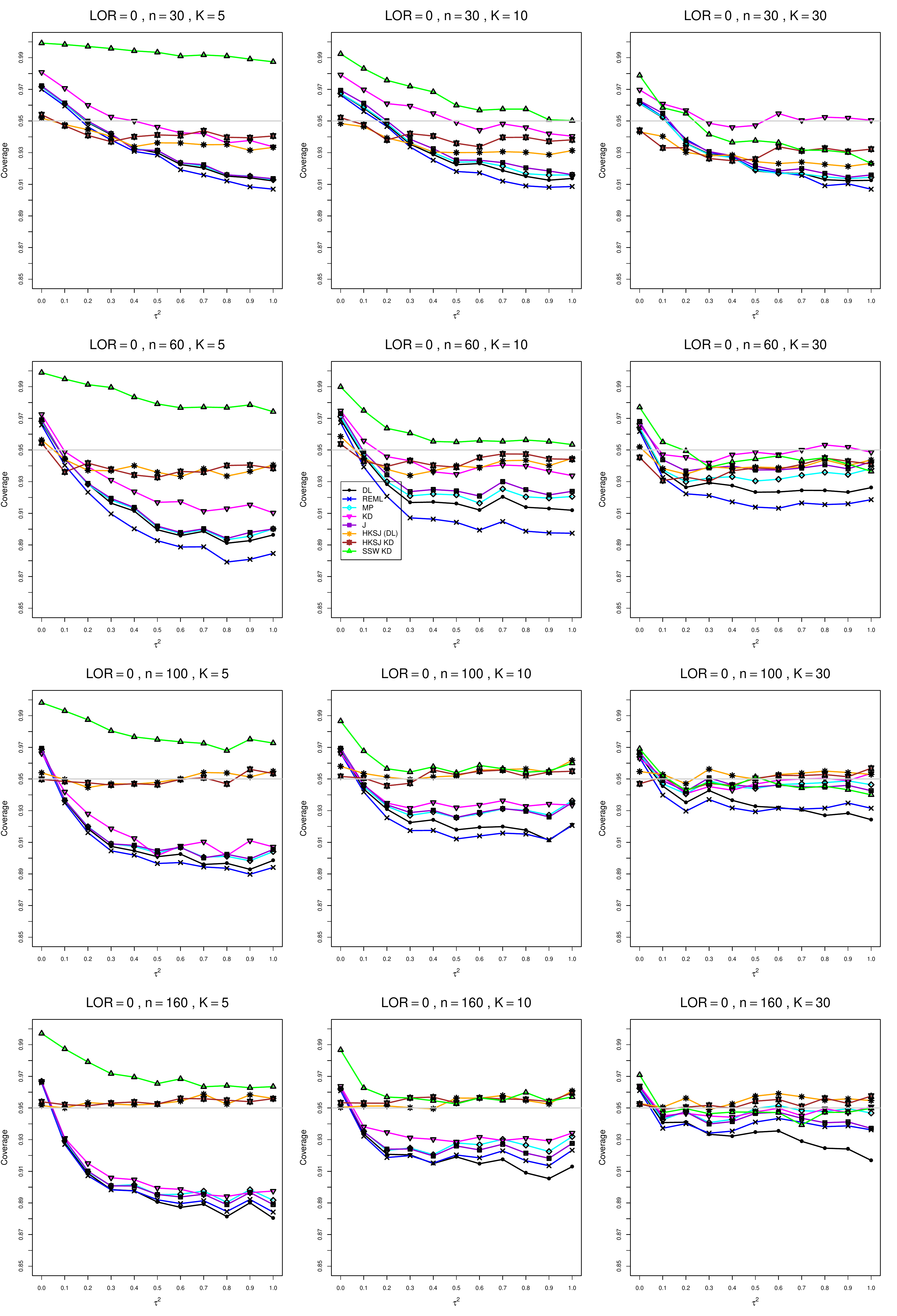}
	\caption{Coverage of  overall effect measure $\theta$ for $\theta=0$, $p_{iC}=0.4$, $q=0.75$, 
		unequal sample sizes $n=30,\; 60,\;100,\;160$. 
		\label{CovThetaLOR0q075piC04_unequal_sample_sizes}}
\end{figure}

\begin{figure}[t]
	\centering
	\includegraphics[scale=0.33]{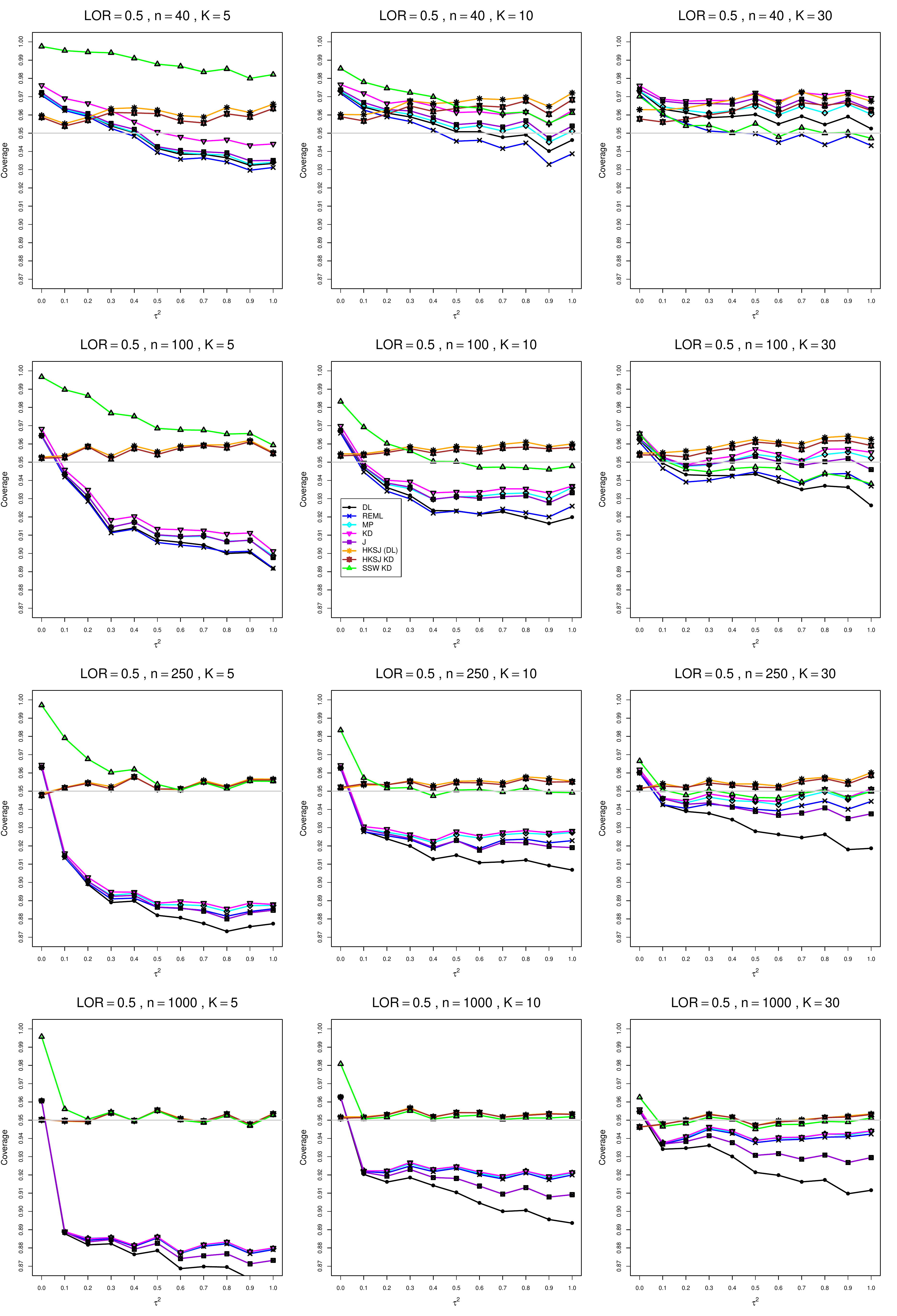}
	\caption{Coverage of  overall effect measure $\theta$ for $\theta=0.5$, $p_{iC}=0.4$, $q=0.75$, equal sample sizes $n=40,\;100,\;250,\;1000$. 
		\label{CovThetaLOR05q075piC04}}
\end{figure}

\begin{figure}[t]
	\centering
	\includegraphics[scale=0.33]{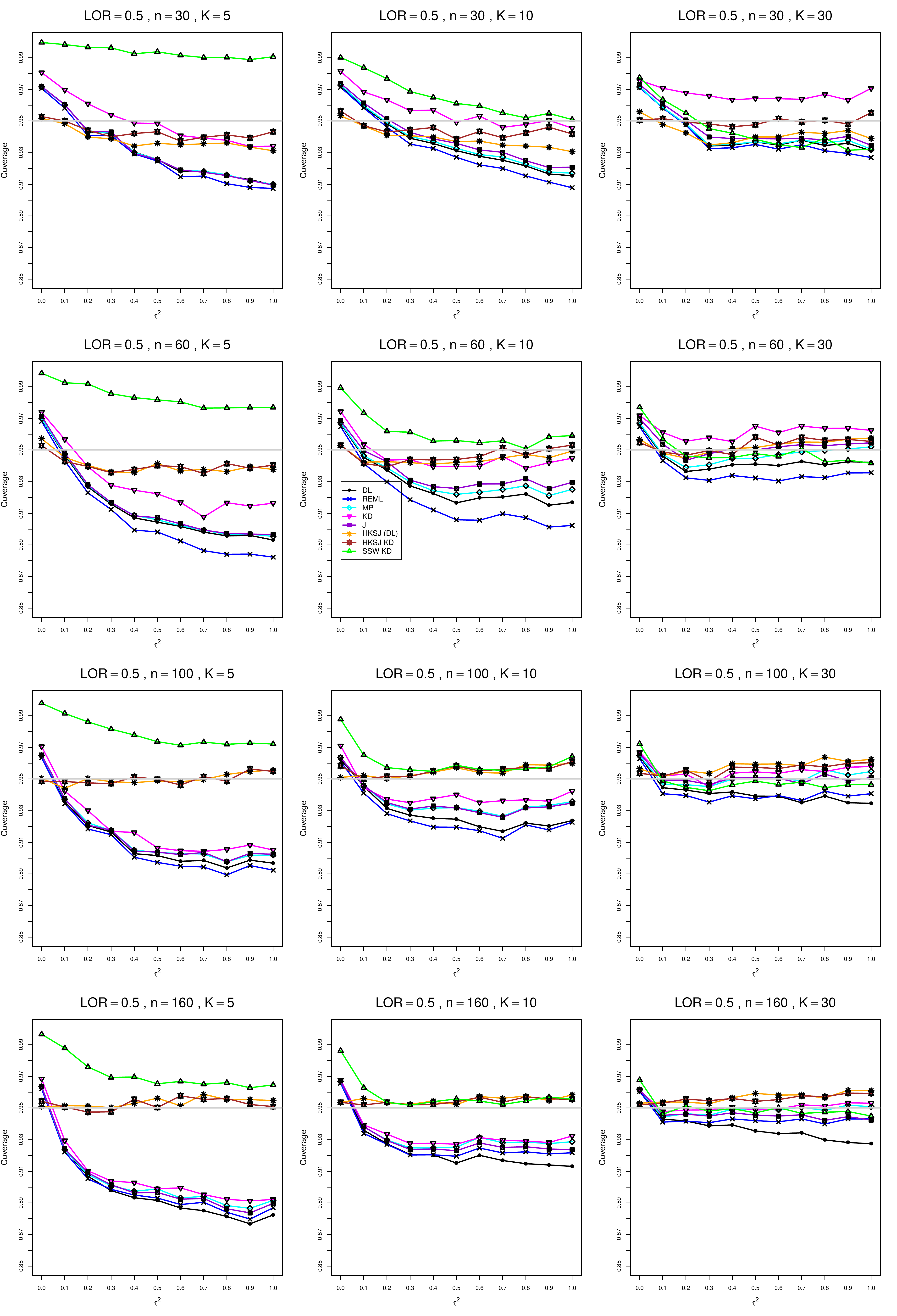}
	\caption{Coverage of  overall effect measure $\theta$ for $\theta=0.5$, $p_{iC}=0.4$, $q=0.75$, 
		unequal sample sizes $n=30,\; 60,\;100,\;160$. 
		\label{CovThetaLOR05q075piC04_unequal_sample_sizes}}
\end{figure}
\clearpage

\begin{figure}[t]
	\centering
	\includegraphics[scale=0.33]{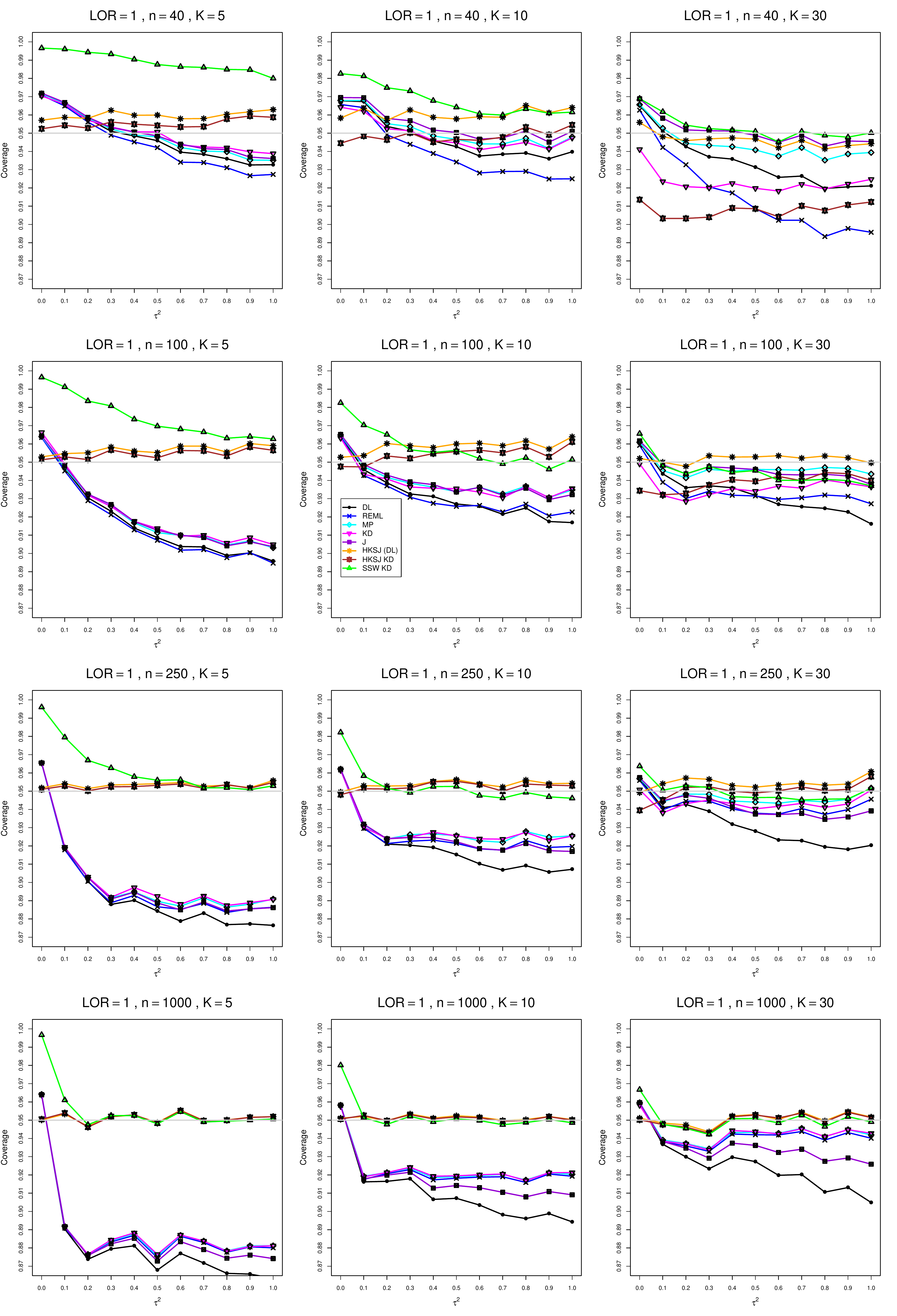}
	\caption{Coverage of  overall effect measure $\theta$ for $\theta=1$, $p_{iC}=0.4$, $q=0.75$, equal sample sizes  $n=40,\;100,\;250,\;1000$. 
		\label{CovThetaLOR1q075piC04}}
\end{figure}

\begin{figure}[t]
	\centering
	\includegraphics[scale=0.33]{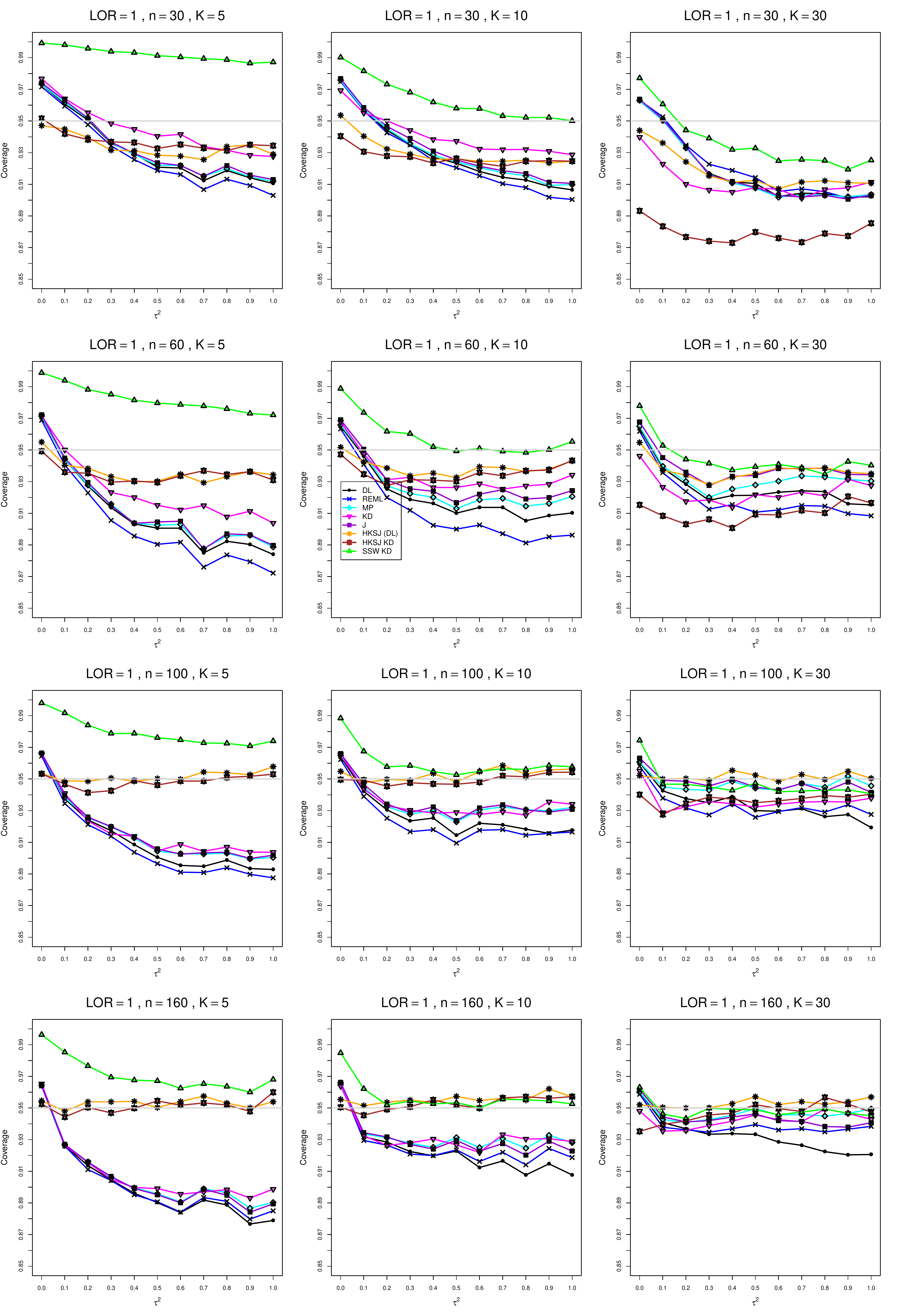}
	\caption{Coverage of  overall effect measure $\theta$ for $\theta=1$, $p_{iC}=0.4$, $q=0.75$, 
		unequal sample sizes $n=30,\; 60,\;100,\;160$. 
		\label{CovThetaLOR1q075piC04_unequal_sample_sizes}}
\end{figure}

\begin{figure}[t]
	\centering
	\includegraphics[scale=0.33]{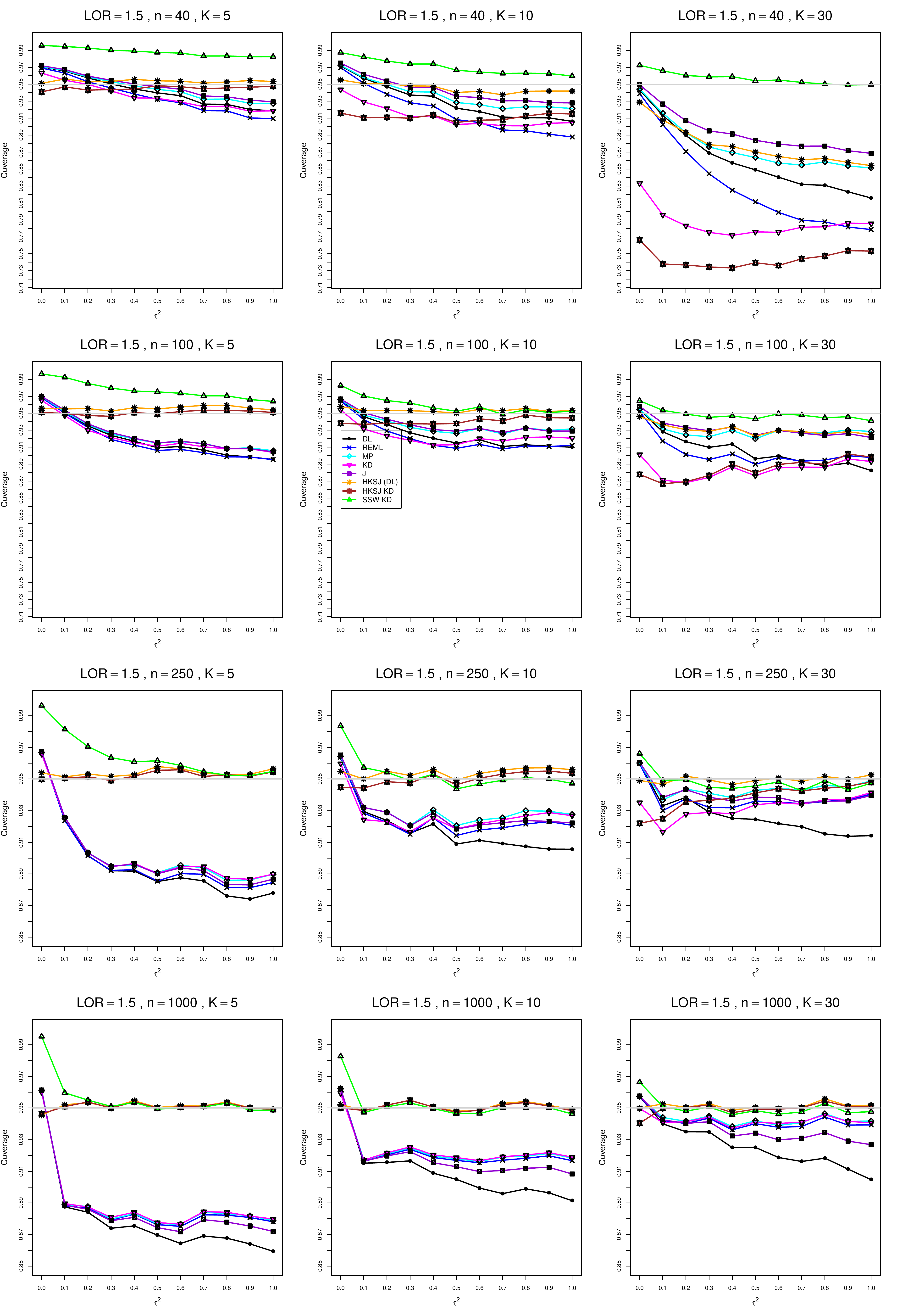}
	\caption{Coverage of  overall effect measure $\theta$ for $\theta=1.5$, $p_{iC}=0.4$, $q=0.75$, equal sample sizes $n=40,\;100,\;250,\;1000$. 
		\label{CovThetaLOR15q075piC04}}
\end{figure}

\begin{figure}[t]
	\centering
	\includegraphics[scale=0.33]{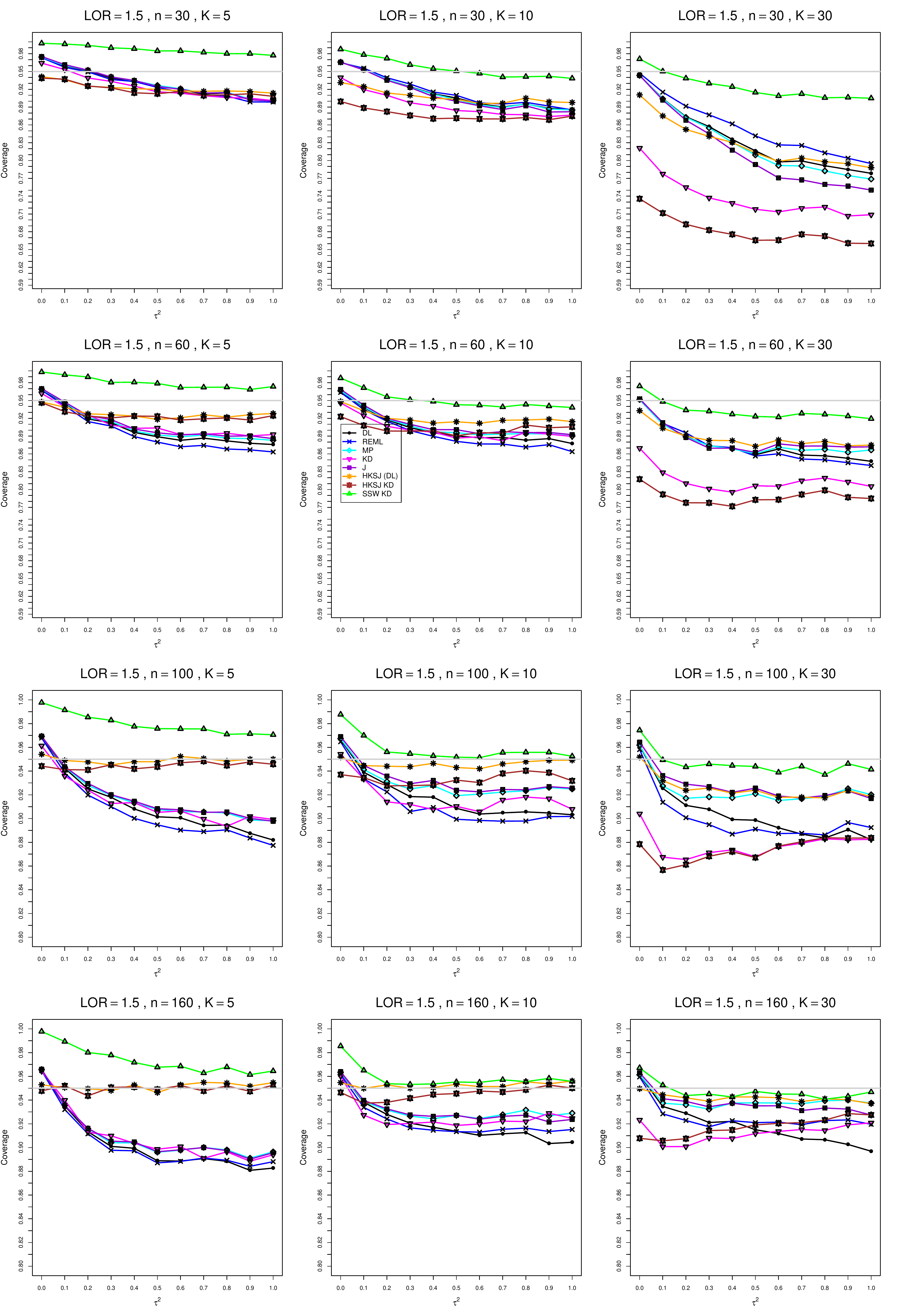}
	\caption{Coverage of  overall effect measure $\theta$ for $\theta=1.5$, $p_{iC}=0.4$, $q=0.75$, 
		unequal sample sizes $n=30,\; 60,\;100,\;160$. 
		\label{CovThetaLOR15q075piC04_unequal_sample_sizes}}
\end{figure}

\begin{figure}[t]
	\centering
	\includegraphics[scale=0.33]{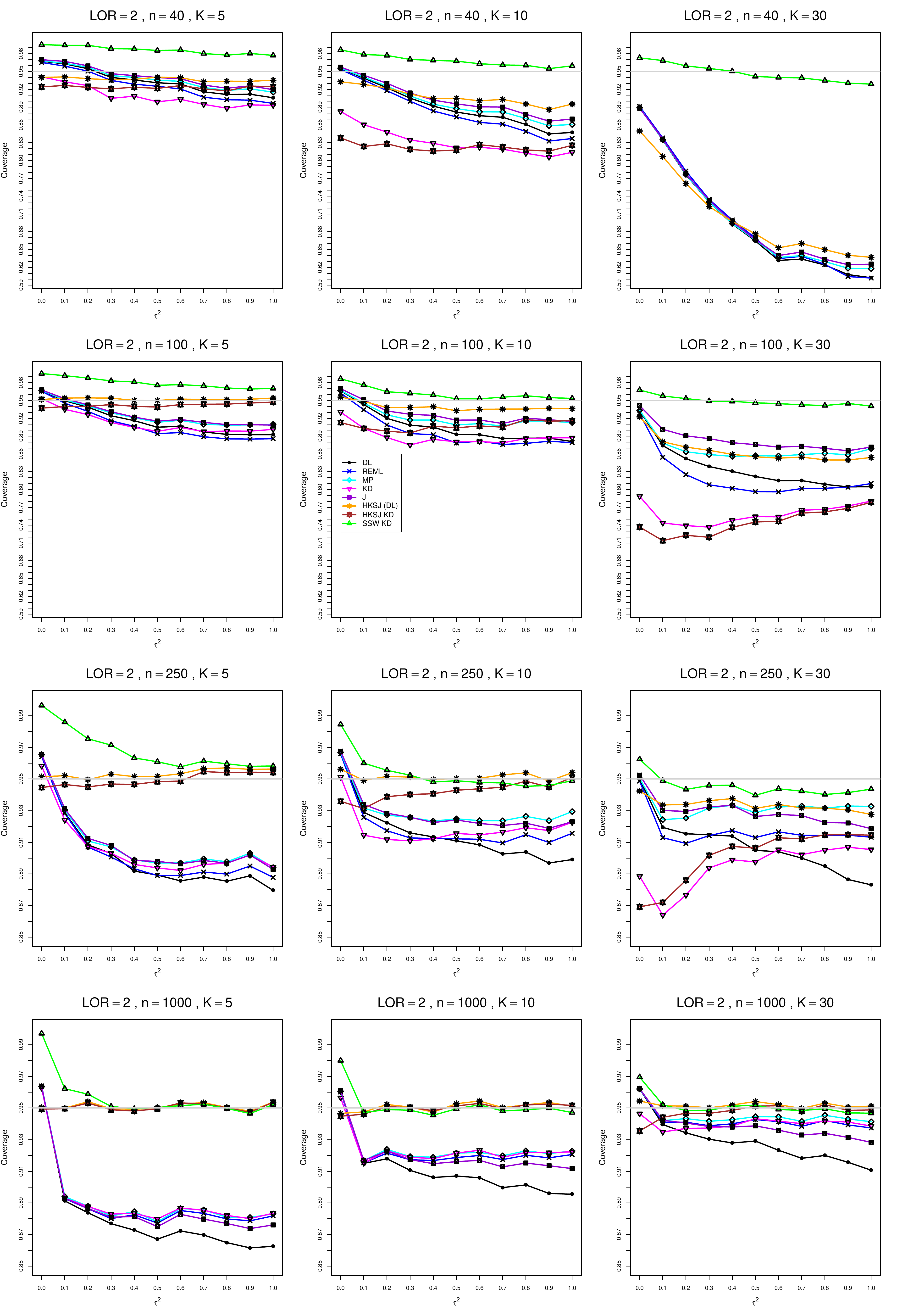}
	\caption{Coverage of  overall effect measure $\theta$ for $\theta=2$, $p_{iC}=0.4$, $q=0.75$, equal sample size $n=40,\;100,\;250,\;1000$. 
		\label{CovThetaLOR2q075piC04}}
\end{figure}

\begin{figure}[t]
	\centering
	\includegraphics[scale=0.33]{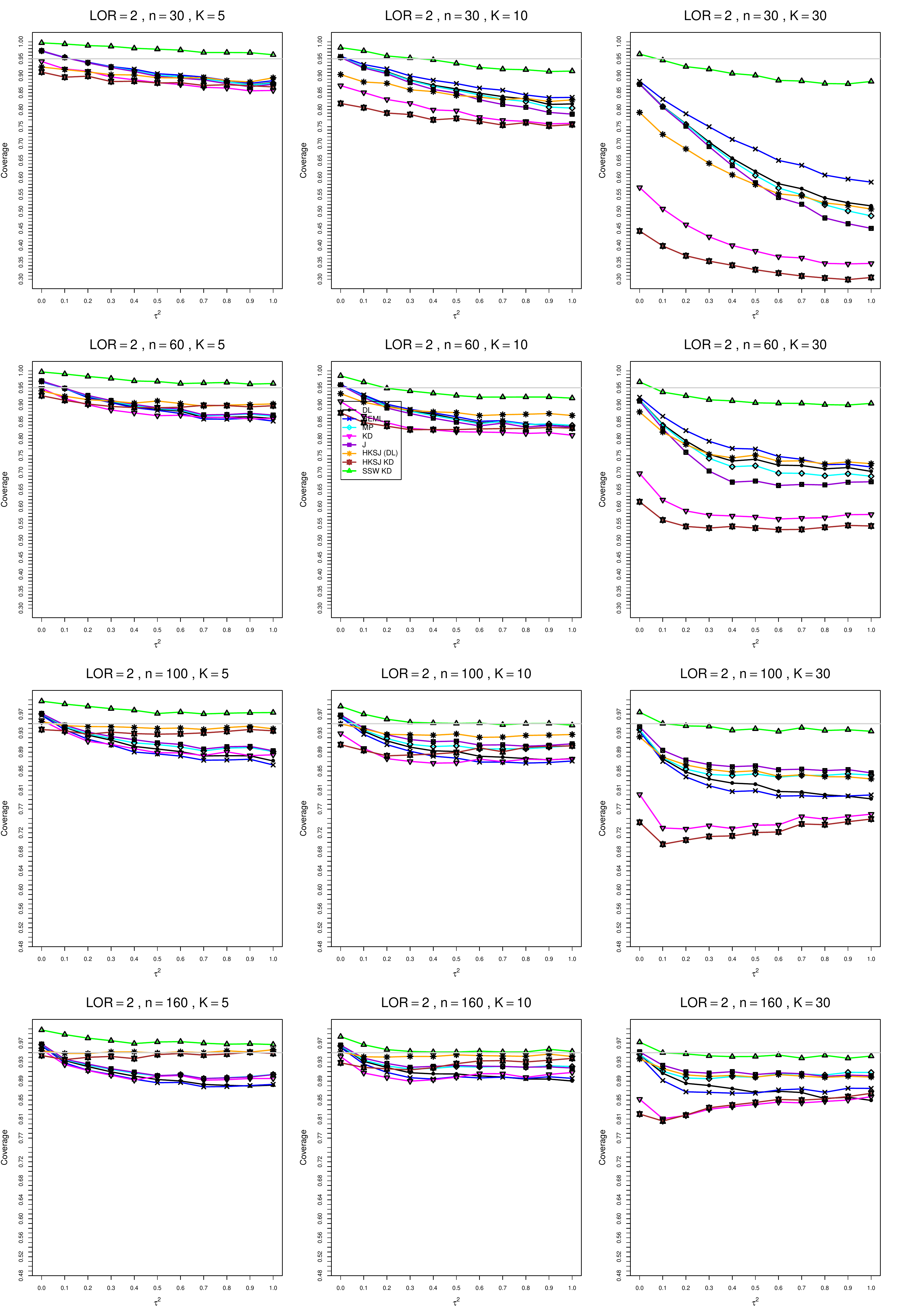}
	\caption{Coverage of  overall effect measure $\theta$ for $\theta=2$, $p_{iC}=0.4$, $q=0.75$, 
		unequal sample sizes $n=30,\; 60,\;100,\;160$. 
		\label{CovThetaLOR2q075piC04_unequal_sample_sizes}}
\end{figure}
\end{document}